\documentclass[review]{elsarticle}

\usepackage{hyperref}
\hypersetup{colorlinks,citecolor=blue,linktoc=all,linkcolor=cyan}

\usepackage[T1]{fontenc}
\usepackage{dsfont}               
\usepackage{mathrsfs}             
\usepackage{slashed}              
\usepackage{amsmath}
\usepackage{amssymb}
\usepackage{amsbsy}
\usepackage{amsfonts}

\usepackage{pdfpages} 
\usepackage{graphicx}
\usepackage{mathtools,slashed}
\usepackage[scale=2]{ccicons}
\usepackage{epstopdf}

\usepackage{bm,array,caption}

\usepackage{float,subfig,physics,siunitx,combelow}
\usepackage{placeins}

\numberwithin{equation}{section}
\numberwithin{table}{section}
\numberwithin{figure}{section}

\journal{Progress in Particle and Nuclear Physics}

\usepackage{titlesec}
\usepackage{sectsty}
\titleformat{\section}{\normalfont\Large\bfseries}{\thesection}{1em}{}
\titleformat{\subsection}{\normalfont\large\bfseries}{\thesubsection}{1em}{}
\titleformat{\subsubsection}{\normalfont\normalsize\bfseries}{\thesubsubsection}{1em}{}

\DeclareMathOperator*{\sumint}{%
\mathchoice%
  {\ooalign{$\displaystyle\sum$\cr\hidewidth$\displaystyle\int$\hidewidth\cr}}
  {\ooalign{\raisebox{.14\height}{\scalebox{.7}{$\textstyle\sum$}}\cr\hidewidth$\textstyle\int$\hidewidth\cr}}
  {\ooalign{\raisebox{.2\height}{\scalebox{.6}{$\scriptstyle\sum$}}\cr$\scriptstyle\int$\cr}}
  {\ooalign{\raisebox{.2\height}{\scalebox{.6}{$\scriptstyle\sum$}}\cr$\scriptstyle\int$\cr}}
}

\newcommand{\bs}{\boldsymbol}

\newcommand{\snn}[1]{$\sqrt{s_{_{NN}}}$ = #1\,GeV}
\newcommand{\nn}{\nonumber}
\newcommand{\p}{\parallel}

\newcommand{\eB}{\left|eq_f B\right|}
\newcommand{\dpi}{(2\pi)}

\newcommand{\qp}{q_\parallel}
\newcommand{\qt}{q_\perp}

\def\bra{\langle}
\def\ket{\rangle}
\newcommand{\bea}{\begin{eqnarray}}
\newcommand{\eea}{\end{eqnarray}}
\newcommand{\be}{\begin{equation}}
\newcommand{\ee}{\end{equation}}

\newcommand{\sgn}{{\rm sign}}

\renewcommand\Im{\text{Im}}
\newcommand{\tb}[1]{t_B({#1})}
\newcommand{\qB}{eB}

\newcommand{\A}{\mathcal{A}}
\renewcommand\vec{\bm}

\renewcommand\Im{\text{Im}}
\newcommand{\ii}{\mathrm{i}}
\newcommand{\qfB}{|q_f B|}

\newcommand{\pt}{\mathbf{p}_\perp}

\newcommand{\gt}{g_\perp}
\newcommand{\gp}{g_\parallel}
\newcommand{\gn}{\gamma^\nu}
\newcommand{\gm}{\gamma^\mu}

\newcommand{\ppar}{p_\parallel}

\renewcommand{\Tr}[1]{\text{Tr}\left\{{#1}\right\}}

\usepackage{bbm}

\DeclareMathSymbol{\mhyphen}{\mathord}{AMSa}{"39}

\def\gcoups{g_S}
\def\gcoupr{g_{V}}
\def\gcoupw{g_{V_0}}
\def\gcoupa{g_{A_0}}

\def\cc{\alpha}
\def\gmatrix{{\cal G}}
\def\jmatrix{{\cal J}}

\def\trmin{{\rm tr}}
\def\mf{{\mbox{\tiny MF}}}

\def\xi{v}

\def\taua{{  \tau_b}}
\def\sigmaa{{  \sigma_b}}
\def\aamu{{  a_b^\mu}}

\def\sigmac{{  \sigma_0}}
\def\pia{{  \pi_b}}
\def\pic{{  \pi_0}}
\def\rhocmu{{  \rho_0^\mu}}
\def\rhocnu{{  \rho_0^\nu}}
\def\rhocmud{{  \rho_{0\mu}}}
\def\rhoc{{  \rho_{0}}}
\def\acmud{{  a_{0\mu}}}
\def\acmu{{  a_0^\mu}}
\def\acnu{{  a_0^\nu}}
\def\ac{{  a_{0}}}
\def\sigmat{{  \sigma_3}}
\def\pit{{  \pi_3}}
\def\rhotmu{{  \rho_3^\mu}}
\def\rhotnu{{  \rho_3^\nu}}
\def\rhot{{  \rho_3}}
\def\atmu{{  a_3^\mu}}
\def\atnu{{  a_3^\nu}}
\def\at{{  a_3}}

\def\rhoamu{{  \rho_b^\mu}}

\begin{document}
	
	\begin{frontmatter}

		\title{Strongly interacting matter in extreme magnetic fields}
		

        \author[b]{Prabal Adhikari}
        \author[jena]{Martin Ammon}
        \author[ssa]{Sidney S. Avancini}
        \author[a]{Alejandro Ayala\corref{mycorrespondingauthor}}
		\cortext[mycorrespondingauthor]{Corresponding author}
		\ead{ayala@nucleares.unam.mx}
        \author[heidelberg]{Aritra Bandyopadhyay}
        \author[blaschke1,blaschke2,blaschke3]{\\ David Blaschke}
        \author[c]{Fabio L. Braghin}
        \author[mymainaddresspb]{Pavel Buividovich}
        \author[ufsc]{Rafael P. Cardoso}
        \author[utrecht]{Casey Cartwright}
        \author[d]{\\ Jorge David Casta\~no-Yepes}
        \author[uniTours,uniUVT]{Maxim Chernodub}
		\author[e]{M. Coppola}
        \author[zuraiq1]{Mayusree Das}
        \author[lourenco_add1,dutra2]{Mariana Dutra}
        \author[uniELTE,uniBielefeld]{\\ Gergely Endr\H{o}di}
        \author[pais2]{Jianjun Fang}
        \author[ufsm]{Ricardo L. S. Farias}
		\author[f]{Eduardo S. Fraga}
        \author[krein]{Arthur Frazon}
        \author[tokyo]{Kenji Fukushima}
        \author[Cinvestav]{\\ Juan D. Garc\'ia-Mu\~noz}
        \author[uniBielefeld]{Eduardo Garnacho-Velasco}
		\author[g]{D. Gomez Dumm}
        \author[stonybrook]{Sebastian Grieninger}
        \author[pais3]{\\ Francesca Gulminelli}

        \author[hernandez]{Juan Hernandez}
        \author[cai_add1]{Chowdhury Aminul Islam}
        \author[tuscaloosa]{Matthias Kaminski}
        \author[juelich]{Andrey Kotov}
        \author[krein]{Gast\~ao Krein}
        \author[pais2]{Jing Li}
        \author[wroclaw]{Pok Man Lo}
		\author[h,j]{Marcelo Loewe}
        \author[lourenco_add1,dutra2]{Odilon Louren\c{c}o}
        \author[uniBielefeld]{Gergely Mark\'o}
        \author[marquez]{Kau D. Marquez}
        \author[krein,saopaulo2,chile]{\\ Ana Mizher}
        \author[zuraiq1]{Banibrata Mukhopadhyay}
		\author[d,i]{Enrique Mu\~noz}
		\author[l]{S. Noguera}
        \author[ufsm]{Rodrigo M. Nunes}
        \author[pais1]{Helena Pais}
		\author[m]{Let\'\i cia F. Palhares}
        \author[pais1]{Constança Providência}
        \author[ifm,chile]{Alfredo Raya}
		\author[f]{Tulio Restrepo}
		\author[n]{Juan Crist\'obal Rojas}
		\author[e,Scoccola2]{N.N. Scoccola}
        \author[pais1]{Luigi Scurto}
        \author[blaschke1,sedrakian2]{Armen Sedrakian}
        \author[gsi]{Dominik Smith} 
        \author[m]{William Rafael Tavares}
		\author[o]{Maria E. Tejeda-Yeomans}
        \author[vst]{Varese S. Tim\'oteo}
        \author[tolos1,tolos2,sedrakian2]{Laura Tolos}
        \author[chile]{Cristian Villavicencio}
        \author[zuraiq2,zuraiq3]{Fridolin Weber}
        \author[yasui1,yasui2,yasui3]{\\Shigehiro Yasui}
		\author[p,q]{Renato Zamora}
        \author[zuraiq1]{Zenia Zuraiq}

		\address[b]{Physics Department,
        Faculty of Natural Sciences and Mathematics,
        St.~Olaf College,
        Northfield, MN 55057, USA.}
        \address[jena]{Theoretisch-Physikalisches Institut, Friedrich-Schiller-Universität Jena, Max-Wien-Platz 1, D-07743 Jena, Germany.}
        \address[ssa]{Departamento de F\'{\i}sica, Universidade Federal de Santa Catarina, 88040-900 Florian\'{o}polis, Santa Catarina, Brazil.}
        \address[a]{Instituto de Ciencias Nucleares, Universidad Nacional Aut\'onoma de M\'exico,\\ Circuito Ext. S.N., Ciudad Universitaria, CdMx 04510, Mexico.}
        \address[heidelberg]{Institut fur Theoretische Physik, Universitat Heidelberg, Philosophenweg 16, 69120 Heidelberg, Germany.}
          \address[blaschke1]{Institute of Theoretical Physics, University of Wroc\l{}aw, pl. M. Borna 9, 50-204 Wroc\l{}aw, Poland.}
        \address[blaschke2]{Helmholtz-Zentrum Dresden-Rossendorf (HZDR), Bautzener Landstrasse 400, 01328 Dresden, Germany.}
        \address[blaschke3]{Center for Advanced Systems Understanding (CASUS), Untermarkt 20, 02826 G\"orlitz, Germany.}
        \address[c]{Instituto de Fisica - Federal University of Goias, Av. Esperança, s/n, 74690-900, Goiania, GO, Brazil.}
        \address[mymainaddresspb]{Department of Mathematical Sciences, University of Liverpool, UK.}
        \address[ufsc]{Departamento de F\'{\i}sica, Universidade Federal de Santa
        Catarina, 88040-900 Florian\'{o}polis, SC, Brazil.}
        \address[utrecht]{Institute for Theoretical Physics, Utrecht University, Princetonplein 5, 3584 CC Utrecht, The Netherlands.}
        \address[d]{Facultad de F\'isica, Pontificia Universidad Cat\'olica de Chile, Vicu\~{n}a Mackenna 4860, Santiago, Chile.}
        \address[uniTours]{Institut Denis Poisson, Universit\'e de Tours, Tours 37200, France.}
		\address[uniUVT]{Department of Physics, West University of Timi\cb{s}oara, Bd.~Vasile P\^arvan 4, Timi\cb{s}oara 300223, Romania.}
        \address[e]{Departamento de F\'isica Te\'orica, Comisi\'{o}n Nacional de Energ\'{\i}a At\'{o}mica, Av. Libertador 8250, 1429 Buenos Aires, Argentina.}
        \address[zuraiq1]{Department of Physics, Indian Institute of Science, Bengaluru 560012, India.}

        \address[lourenco_add1]{Departamento de F\'isica e Laborat\'orio de Computa\c c\~ao Cient\'ifica Avan\c cada e Modelamento (Lab-CCAM), Instituto Tecnol\'ogico de Aeron\'autica, DCTA, 12228-900, S\~ao Jos\'e dos Campos, SP, Brazil.}
        \address[dutra2]{Institut de Physique des 2 infinis de Lyon, CNRS/IN2P3, Universit\'e de Lyon, Universit\'e Claude Bernard Lyon 1,\\ F-69622 Villeurbanne Cedex, France.}
        \address[uniELTE]{Institute of Physics and Astronomy, ELTE E\"otv\"os Lor\'and University, P\'azm\'any P.\ s\'et\'any 1/A, H-1117 Budapest, Hungary.}
        \address[uniBielefeld]{Faculty of Physics, Universit\"{a}t Bielefeld, Bielefeld, Germany.}
        \address[pais2]{School of Physics and Physical Engineering, Qufu Normal University, 273165 Qufu, China.}
        \address[ufsm]{Departamento de F\'{\i}sica, Universidade Federal de Santa Maria, zero5-900, Santa Maria, RS, Brazil.}
        \address[f]{Instituto de F\'isica, Universidade Federal do Rio de Janeiro, Caixa Postal 68528, 21941-972, Rio de Janeiro, RJ, Brazil.}
		\address[krein]{Instituto de F\'{\i}sica Te\'orica, Universidade Estadual Paulista,  Rua Dr. Bento Teobaldo Ferraz, 271 - Bloco II\\ 01140-070 S\~ao Paulo, SP, Brazil.}
        \address[tokyo]{Department of Physics, The University of Tokyo, 7-3-1 Hongo,
        Bunkyo-ku, Tokyo, 113-0033, Japan.}
        \address[Cinvestav]{Physics Department, CINVESTAV, P.O.B. 14-740, 0700, Mexico City, Mexico.}
        \address[g]{IFLP, CONICET $-$ Departamento de F\'{\i}sica, Facultad de Ciencias Exactas, Universidad Nacional de La Plata, 1900 La Plata, Argentina.}
        \address[stonybrook]{Center for Nuclear Theory, Department of Physics and Astronomy, Stony Brook University, Stony Brook, New York 11794–3800, USA.}
        \address[pais3]{Normandie Univ., ENSICAEN, UNICAEN, CNRS/IN2P3, LPC Caen, F-14000 Caen, France.}
        \address[hernandez]{Theoretische Natuurkunde, Vrije Universiteit Brussel (VUB) and The International Solvay Institutes,\\ Pleinlaan 2, B-1050 Brussels, Belgium.}
        \address[cai_add1]{Institut f\"{u}r Theoretische Physik, Johann Wolfgang Goethe–Universit\"{a}t, Max-von-Laue-Str. 1, D–60438 Frankfurt am Main, Germany.}
        \address[tuscaloosa]{Department of Physics and Astronomy, University of Alabama, 514 University Boulevard, Tuscaloosa, AL 35487, USA.}
        \address[juelich]{J\"ulich Supercomputing Centre, Forschungszentrum J\"ulich, D-52428 J\"ulich, Germany.}
        \address[wroclaw]{Institute of Theoretical Physics, University of Wroclaw,
        plac Maksa Borna 9, PL-50204 Wroclaw, Poland.}
        \address[h]{Centre for Theoretical and Mathematical Physics, and Department of Physics, University of Cape Town, Rondebosch 7700, South Africa.}
		\address[j]{Facultad de Ingeniería, Arquitectura y Diseño, Universidad San Sebastián, Santiago, Chile.}
		\address[k]{Center for Nanotechnology and Advanced Materials CIEN-UC, Avenida Vicuña Mackenna 4860, Santiago, Chile.}
        
        \address[marquez]{Departamento de F\'isica e Laborat\'orio de Computação Científica Avançada e Modelamento (Lab-CCAM),
        Instituto Tecnológico de Aeronáutica, DCTA, 12228-900 São José dos Campos/SP, Brazil.}
		\address[saopaulo2]{Laborat\'orio de F\'isica Te\'orica e Computacional, Universidade Cidade de S\~ao Paulo, R. Galv\~ao Bueno,\\ 868, Liberdade, 01506-000, S\~ao Paulo, Brazil.}
		\address[chile]{Centro de Ciencias Exactas, Universidad del B\'io-B\'io, Casilla 447, Chill\'an, Chile.}
        \address[l]{Departamento de F\'isica Te\'orica and IFIC, Centro Mixto Universidad de Valencia-CSIC, E-46100 Burjassot (Valencia), Spain.}
        \address[pais1]{CFisUC, Department of Physics, University of Coimbra, 3004-516 Coimbra, Portugal.}
        \address[m]{Universidade do Estado do Rio de Janeiro, Instituto de F\'isica, Departamento de F\'isica Te\'orica, Rua S\~ao Francisco Xavier 524, 20550-013 Maracan\~a, Rio de Janeiro, Brasil.}
        \address[ifm] {Instituto de F\'{\i}sica y Matem\'aticas, Universidad Michoacana de San Nicol\'as de Hidalgo, Edificio C-3, Ciudad Universitaria. Francisco J. M\'ujica s/n. Col. Fel\'{\i}citas del R\'{\i}o. 58040 Morelia, Michoac\'an, M\'exico.}
        \address[n]{Departamento de Física, Universidad Cat\'olica del Norte, Angamos 610, Antofagasta,Chile.}
        \address[Scoccola2]{CONICET, Godoy Cruz 2290 (C1425FQB) ,  Buenos Aires, Argentina}
        \address[sedrakian2]{Frankfurt Institute for Advanced Studies, Ruth-Moufang-Str. 1, 60438 Frankfurt am Main, Germany.}
        \address[gsi]{Facility for Antiproton and Ion Research in Europe GmbH (FAIR GmbH), 64291 Darmstadt, Germany.}
        \address[o]{Facultad de Ciencias-CUICBAS, Universidad de Colima, Bernal Díaz del Castillo Número 340,\\ Colonia Villas San Sebasti\'an, Colima, 28045, Mexico.}
		\address[vst]{Grupo de \'Optica e Modelagem Num\'erica - GOMNI, Faculdade de Tecnologia - FT,\\ Universidade Estadual de Campinas - UNICAMP, 13484-332 Limeira, SP , Brazil.}
        \address[tolos1]{Institute of Space Sciences (ICE, CSIC), Campus UAB, Carrer de Can Magrans, 08193 Barcelona, Spain.} 
        \address[tolos2]{Institut d'Estudis Espacials de Catalunya (IEEC), 08860 Castelldefels (Barcelona), Spain.}
        \address[zuraiq2]{Department of Physics, San Diego State University, 5500 Campanile Drive, San Diego, California 92182, USA.}
        \address[zuraiq3]{Department of Physics, University of California at San Diego, La Jolla, California 92093, USA.}
\address[yasui1]{Nishogakusha University, 6-16, Sanbancho, Chiyoda, Tokyo 102-8336, Japan.}
        \address[yasui2]{International Institute for Sustainability with Knotted Chiral Meta Matter (WPI-SKCM$^{2}$), Hiroshima University,\\ Higashi-Hiroshima, Hiroshima 739-8526, Japan.}
        \address[yasui3]{Research and Education Center for Natural Sciences, Keio University, Hiyoshi 4-1-1, Yokohama, Kanagawa 223-8521, Japan.}
        \address[p]{Centro de Investigaci\'on y Desarrollo en Ciencias Aeroespaciales (CIDCA), Fuerza A\'erea de Chile, Casilla 8020744, Santiago, Chile.}
		\address[q]{Instituto de Ciencias B\'asicas, Universidad Diego Portales, Casilla 298-V, Santiago, Chile.}



        




\begin{abstract}
Magnetic fields are ubiquitous across different physical systems of current interest; from the early Universe, compact astrophysical objects and heavy-ion collisions to condensed matter systems. A proper treatment of the effects produced by magnetic fields during the dynamical evolution of these systems, can help to understand observables that otherwise show a puzzling behavior. Furthermore, when these fields are comparable to or stronger than $\Lambda_{\mbox{\tiny{QCD}}}$, they serve as excellent probes to help elucidate the physics of strongly interacting matter under extreme conditions of temperature and density. In this work we provide a comprehensive review of recent developments on the description of QED and QCD systems where magnetic field driven effects are important. These include the modification of meson static properties such as masses and form factors, the chiral magnetic effect, the description of anomalous transport coefficients, superconductivity in extreme magnetic fields, the properties of neutron stars, the evolution of heavy-ion collisions, as well as effects on the QCD phase diagram. We describe recent theory and phenomenological developments using effective models as well as LQCD methods. The work represents a state-of-the-art review of the field, motivated by presentations and discussions during the \lq\lq Workshop on Strongly Interacting Matter in Strong Electromagnetic Fields\rq\rq\ that took place in the European Centre for Theoretical Studies in Nuclear Physics and Related Areas (ECT*) in the city of Trento, Italy, September 25-29, 2023.\\
\end{abstract}
		
\end{frontmatter}
	
	\newpage
	
	\thispagestyle{empty}
	\setcounter{tocdepth}{2}
	\tableofcontents
	

    \newpage
    \part{Introduction}

The properties of physical systems influenced by the presence of strong magnetic fields have become a subject of intense research over the recent past. It has been known already for a long time, that the magnetic field leads to dimensional reduction in the description of particle dynamics, which is at the core of a myriad of important effects involving the physical properties of systems subject to the presence of magnetic fields. Among these, an important effect is magnetic catalysis, namely, the dynamical generation of mass and the enhancement of the chiral condensate in the presence of a magnetic field. Nevertheless, it is by now known that in the simultaneous presence of a heat bath and a magnetic field, strongly interacting matter exhibits an interesting phenomenon, whereby the chiral condensate melts for temperatures close to the chiral transition, and the transition temperature itself also decreases. This effect is dubbed inverse magnetic catalysis and was discovered about a decade ago by means of lattice QCD simulations. It is fair to say that this unexpected phenomenon sparked a great deal of activity, mainly among practitioners performing research on the properties of strongly interacting matter under extreme conditions, since it was realized that a magnetic field of a strength similar or larger than $\Lambda_{\mbox{\tiny{QCD}}}$ offered an excellent probe to explore QCD matter in this regime. The exploration includes studying the pattern of modifications of mass and other static properties, such as form factors, in the meson and baryon sectors. On the other hand, the intense magnetic fields present in semi-central heavy-ion collisions at large energies, opened up the possibility to produce and detect chirally imbalanced matter in this kind of systems, a phenomenon dubbed the Chiral Magnetic Effect (CME). This has further motivated exploring the effects of these fields not only during the hot and dense phases of the reaction but also at the beginning, when matter is not yet equilibrated but the field is the strongest. Although the CME has proven difficult to be experimentally confirmed, an analogous effect has been successfully detected in condensed matter systems. This finding has also motivated another intense line of research aiming to explore two-dimensional systems of electromagnetically interacting particles, subject to the presence of magnetic fields. In another front of research, the advent of gravitational wave detection has strengthened the era of precision cosmology and astrophysics, motivating more detailed studies to elucidate the role played by the intense magnetic fields, present in compact astrophysical objects, on the QCD equation of state. The goal in this context is to find possible explanations for the anomalous mass-radius relations found for some of these objects, or the role that these fields may play as catalyzers of a rocket effect to help understand the neutron star kick velocities.

In this work we present a comprehensive review of the state of the art of research aimed to understand the properties of a range of physical systems subject to the influence of strong magnetic fields. The work was  motivated by presentations and discussions  during the ``Workshop on Strongly Interacting Matter in Strong
Electromagnetic Fields'' that took place in the European Centre for Theoretical Studies in Nuclear Physics and
Related Areas (ECT*) in the city of Trento, Italy, September 25-29, 2023. The work is organized as follows: In Part~\ref{II} we discuss recent developments in the theory front, covering a wide range of topics, from the description of mixing of charged states and the proper treatment  of Schwinger phases to account for the magnetic field driven mass and form factor modifications for mesons and baryons, photon production during pre-equilibrium in heavy-ion collisions, the thermomagnetic properties of QCD, including finite volume effects, and finally the description of magnetic field effects in QED matter beyond the constant field approximation, considering the field as a background of white noise around an average value. Part~\ref{III} is devoted to describing recent developments in the description of non-perturbative phenomena in  QCD and the electroweak theory with magnetic fields. In particular, we study anomalous transport phenomena in QCD as well as in low-dimensional QED, a subject that is relevant for certain condensed matter systems. Given the non-perturbative nature of these phenomena, they require either first-principles lattice field theory simulations or effective approaches, for instance hydrodynamics, which are also discussed in this part. In Part~\ref{Eff_Mod} we present several effective models that have been developed for the description of few-nucleon systems, nuclear matter, nuclei and QCD matter, extended to describe strongly interacting magnetized matter. Although these models were conceived to address specific features, they can still be used to gain insight into the properties of strong interactions
in the presence of magnetic fields or nonzero densities, which otherwise cannot be obtained from first-principles calculations. In Part~\ref{densemagnetizedmatter}  we present recent theoretical advancements in the exploration of magnetic field effects on the QCD phase diagram and dense magnetized matter. We discuss a new mechanism for accelerating proto-neutron stars involving the chiral separation effect. We examine the impact of incorporating ring diagrams into the four-quark interaction on the quark gap equation within a chiral model. Moreover, we use finite energy sum rules in dense nuclear matter and in the presence of a constant and uniform external magnetic field to obtain information on different hadronic and QCD parameters. We also study the momentum diffusion coefficients of heavy quarks in a hot, magnetized medium, examining a wide spectrum of external magnetic field strengths. In addition, we address the effects of considering a quark anomalous magnetic moment to probe new phenomena in the context of the magnetized QCD phase diagram. 
Part~\ref{VI} continues the discussion with neutron stars, which are among the densest objects in astrophysics, formed from massive stars after a supernova explosion. They possess a large mass in a relatively small volume as well as strong magnetic fields in their interior. These objects provide laboratories for nuclear physics under extreme conditions and therefore in this Part we discuss the their physics and the way magnetic fields can influence their properties. Finally in Part~\ref{concl} we summarize and conclude.

	\newpage

\graphicspath{{./Figures_Theory_Dev/} }

	\part{Theory Developments}\label{II}

	\section{Introduction}\label{intro}

Electromagnetic fields are a unique tool to explore the properties of the QCD vacuum, thermodynamics and phase structure. When the energy associated to the field strength is larger than $\Lambda_{\mbox{\small{QCD}}}$, the field is able to probe length scales smaller than the hadron size, potentially revealing aspects of the dynamics associated to confinement and chiral symmetry breaking. For example, it is by now well known that at zero temperature magnetic fields catalyze the breaking of chiral symmetry, producing a stronger light quark-antiquark condensate~\cite{Semenoff:1999xv,Leung:1996poh,Lee:1997zj,Ayala:2006sv, Rojas:2008sg}. However, contrary to the first predictions from chiral models \cite{Fraga:2008qn,Mizher:2010zb,Gatto:2010pt} and preliminary results from LQCD \cite{DElia:2010abb}, at finite temperature magnetic fields inhibit the condensate formation, reducing the critical temperature for chiral symmetry restoration and thus giving rise to inverse magnetic catalysis (IMC), as found in LQCD analysis and described using effective models with magnetic field-dependent couplings~\cite{Bali:2011qj,Bali:2012zg,Bali:2014kia,Bruckmann:2013oba,Farias:2014eca,Ferreira:2014kpa,Ayala:2014gwa,Ayala:2015lta,Ayala:2014iba,Farias:2016gmy,Ayala:2016bbi,Ferrer:2014qka,Ayala:2015bgv,Ayala:2014uua,Ayala:2018wux,Mueller:2014tea,Mueller:2015fka,Bandyopadhyay:2020zte,Pagura:2016pwr}. This behavior has motivated an intense activity aimed to search for the influence of magnetic fields on hadron dynamics~\cite{Bali:2018sey,Simonov:2015xta,Aguirre:2018fbo,Yoshida:2016xgm,Dudal:2014jfa,Dudal:2018rki,Marasinghe:2011bt,Gubler:2015qok,Machado:2013yaa,Cho:2014exa,Cho:2014loa,Ghosh:2016evc,Bandyopadhyay:2016cpf,Ayala:2014mla,Avancini:2015ady,Wang:2017vtn,Braghin2018c,Coppola:2018vkw,Braghin2020a,Li:2020hlp,Liu:2018zag,GomezDumm:2017jij,GomezDumm:2020bxj,Andreichikov:2016ayj,Avancini:2018svs,Bandyopadhyay:2019pml,Mao:2018dqe,Mukherjee:2017dls,Ghosh:2017rjo,Zhang:2016qrl,Arias:2016zqu,Liu:2014uwa,Andreichikov:2018wrc,Colucci:2013zoa,Aguirre:2017dht,Taya:2014nha,Kawaguchi:2015gpt,Andersen:2012dz,Hattori:2015aki,Loewe:2022aaw,Loewe:2021ekj,Ayala:2020muk,Andreichikov:2013zba,Ghosh:2019fet,Chaudhuri:2019lbw,Ghosh:2020qvg,Chaudhuri:2020lga,Avila:2019pua,Avila:2020ved,Callebaut:2011ab,Andreichikov:2013pga,He:2015zca,Mukherjee:2018ebw,Endrodi:2019whh,Dominguez:2020sdf,Villavicencio:2022gbr,Das:2022mic,Ayala:2023llp,Parui:2023lyo,Coppola:2018ygv,Coppola:2019idh,Coppola:2019wvh,Coppola:2020mon}. The possibility to generate large intensity, albeit short-lived, magnetic fields in peripheral heavy-ion collisions at high energies~\cite{Skokov:2009qp,Voronyuk:2011jd,McLerran:2013hla,Bzdak:2011yy,Zhang:2023ppo,Sun:2023rhh} has further stimulated the experimental search for signals associated to the presence of these fields in that environment. In fact, experimentally, a peak intensity of order $B\sim 10^{19}$ G has been inferred at RHIC energies~\cite{STAR:2019wlg,Brandenburg:2021lnj}. The strength of these fields is usually taken as uniform, although it can in principle be also treated as white noise around a peak strength~\cite{Castano-Yepes:2023brq,Castano-Yepes:2022luw,Castano-Yepes:2024ctr}.

In this Part, we describe recent theoretical developments that contribute to the ongoing exploration of the properties of the QCD vacuum using magnetic fields as a probe. We approach the study both from the point of view of hadron, as well as from fundamental QCD degrees of freedom. Hadron properties are addressed from the perspective of effective models, in particular the Nambu--Jona-Lasinio (NJL) model and the constituent quark model (CQM). In Sec.~\ref{Scoccola}, the mass spectrum of light mesons in the presence of a constant and homogeneous magnetic field is described. The importance of including the mixing between the pseudoscalar, vector and axial-vector channels is emphasized. The description of this mixing is made using a two-flavor NJL-like model which includes isoscalar and
isovector couplings together with a flavor mixing 't Hooft-like term. Attention is also payed to the proper account of the effects of the Schwinger phase charged particles. In Sec.~\ref{Braghin}, the magnetic field modification of constituent quark-meson form factors and of the one-meson exchange amplitude in the weak field limit, are explored. In Sec.~\ref{Tejeda}, photon production from gluon fusion and splitting induced by magnetic fields in relativistic heavy-ion collisions at pre-equilibrium is studied as a possible venue to solve the photon puzzle. Next, in Sec.~\ref{Fraga}, thermo-magnetic properties of QCD are explored, first from a perturbative computation of the pressure, the chiral condensate and the strange quark number susceptibility up to two-loops for very large values of the field strength and for physical quark masses. The convergence of the perturbative series for the pressure for different choices of the renormalization scale in the running coupling is also studied and results are compared with recent LQCD data, away from the chiral transition. Effects of a finite baryon density and of the renormalization scale in the running coupling and strange quark mass are also studied and the results are used to build a simple analytical model for the equation of state of pure quark magnetars. In Sec.~\ref{Adhikari}, the impact of a uniform magnetic field on the lowest two QCD topological cumulants is explored using chiral perturbation theory. The nature of the corrections is discussed and low energy theorems that connect the topological susceptibility with the chiral condensate and the fourth cumulant with the chiral condensate and susceptibilities, are deduced. The nature of finite volume thermodynamics with a particular focus on the chiral condensate, renormalized magnetization and the anisotropic pressure, is investigated. It is found that the local chiral condensate is doubly periodic on the plane transverse to the magnetic field. Both the renormalized magnetization and (spatially averaged) chiral condensate exhibit significant corrections, with the former being particularly sensitive to finite volume corrections due to its small size. 

The effects of magnetic fields on QCD dynamics are usually explored assuming that the field intensity is constant and uniform. Nevertheless this assumption is an idealization of physical systems where the magnetic fields may at least fluctuate in intensity and even in direction in an extended space-time region. In order to include space and time fluctuations into a field theoretical framework, Secs.~\ref{Munoz} ,~\ref{Loewe} and~\ref{Castano} are devoted to the description of the properties of fermions and gauge fields immersed in a noisy magnetic field background. In Sec.~\ref{Munoz}, the effective quasi-particle fermion propagator, dressed with magnetic noise effects, is obtained as a solution of the corresponding Dyson equation. The perturbative results at first-order in the magnetic noise indicate a renormalization of charge as well as the emergence of an effective index of refraction. The picture that emerges corresponds to quasiparticles propagating in a dispersive medium induced by the fluctuations of the background field. In Sec.~\ref{Loewe}, a vector order parameter,  representing an ensemble average over magnetic noise of the fermion currents, is introduced together with the replica formalism to study the modification of the fermion propagator induced by white noise at mean field level. The region where this order parameter becomes finite, thus breaking the $U(1)$-symmetry due to the magnetic noise, is identified. In Sec.~\ref{Castano}, the one-loop polarization tensor for photons and gluons in a magnetized medium, where the magnetic field experiences fluctuations, is investigated. These fluctuations are described using the replica method. The resulting polarization tensor is not transverse and this property is shown to be connected with the dynamical generation of a magnetic mass for otherwise massless gauge bosons (photons or gluons). 

\section{Magnetized mesons in the Nambu--Jona-Lasinio model}\label{Scoccola}

As already mentioned, the effects caused by magnetic fields, $|eB|$ larger than $\sim\Lambda_{QCD}^2$, on the properties of strong-interacting matter
have attracted a lot of attention along the past decades. In this
context, the understanding of the way in which the properties of
light hadrons are modified by the presence of an intense magnetic
field becomes a very relevant task. Clearly, this is a nontrivial
problem, since first-principle theoretical calculations require to
deal in general with QCD in a low energy non-perturbative regime.
As a consequence, the corresponding theoretical analyses have been
carried out using a variety of approaches. In particular, in the framework of Nambu-Jona-Lasinio (NJL)-like
models the effect of intense external magnetic fields on $\pi$ meson properties has been
extensively studied~\cite{Fayazbakhsh:2012vr,Fayazbakhsh:2013cha,Liu:2014uwa,Avancini:2015ady,
Zhang:2016qrl,Avancini:2016fgq,Mao:2017wmq,GomezDumm:2017jij,Wang:2017vtn,Liu:2018zag,
Coppola:2018vkw,Mao:2018dqe,Avancini:2018svs,Coppola:2019uyr,Cao:2019res,Cao:2021rwx,Sheng:2021evj,
Avancini:2021pmi,Xu:2020yag,Lin:2022ied}. In addition, several results
for the $\pi$ meson spectrum in the presence of background
magnetic fields have been obtained from lattice QCD (LQCD)
calculations~\cite{Bali:2011qj,Luschevskaya:2015bea,Luschevskaya:2014lga,Bali:2017ian,Ding:2020hxw,Ding:2022tqn}.
Regarding the $\rho$ meson sector, studies of magnetized $\rho$
meson masses in the framework of NJL-like models and LQCD can be
found in
Refs.~\cite{Zhang:2016qrl,Liu:2018zag,
Cao:2019res,Xu:2020yag,Chernodub:2011mc,Ghosh:2020qvg,Avancini:2022qcp}
and
Refs.~\cite{Luschevskaya:2015bea,Luschevskaya:2014lga,
Bali:2017ian,Luschevskaya:2012xd,Hidaka:2012mz,Luschevskaya:2016epp}, respectively. In
most of the existing calculations the mixing
between states of different spin/isospin has been neglected.
Although such mixing contributions are usually forbidden by
isospin and/or angular momentum conservation, they can be nonzero
in the presence of the external
magnetic field. In the framework of
NJL-like models effects of this kind have been studied recently in Refs.~\cite{Carlomagno:2022inu,
Carlomagno:2022arc,Coppola:2023mmq}. The aim of the present
contribution is to review the main results of those works.

    \subsection{Effective Lagrangian and mean field quantities}\label{second}

    We consider the Lagrangian for an extended
SU(2) NJL model in the presence of an electromagnetic field.
We have
\begin{eqnarray}
{\cal L}\!\! & \! = \! & \! \! \bar{\psi}(x)\left(i\,\rlap/\!D-m_{c}\right)\psi(x)
+\gcoups\sum_{a=0}^{3}\Big[\left(\bar{\psi}(x)\taua\psi(x)\right)^{2}+\left(\bar{\psi}(x)\,i\gamma_{5}\taua\psi(x)\right)^{2}\Big]
 -\gcoupr\left[\left(\bar{\psi}(x)\,\gamma_{\mu}\boldsymbol{\tau}\,\psi(x)\right)^{2}+\left(\bar{\psi}(x)\,\gamma_{\mu}\,\gamma_{5}\,\boldsymbol{\tau}\,\psi(x)\right)^{2}\right]\nonumber \\
 &  & -\,\gcoupw\left(\bar{\psi}(x)\,\gamma_{\mu}\,\psi(x)\right)^{2}-\,\gcoupa\left(\bar{\psi}(x)\,\gamma_{\mu}\,\gamma_{5}\,\psi(x)\right)^{2}
 +\,2g_{D}\,\left({\rm det}[\bar{\psi}(x)(1+\gamma_{5})\psi(x)]+{\rm det}[\bar{\psi}(x)(1-\gamma_{5})\psi(x)]\right)\ ,\label{lagrangian}
\end{eqnarray}
where $\psi=(u\ d)^{T}$, $\taua=(\mathbbm{1},\boldsymbol{\tau})$, $\boldsymbol{\tau}$
being the usual Pauli-matrix vector, and $m_{c}$ is the current quark
mass, which is assumed to be equal for $u$ and $d$ quarks. The interaction
between the fermions and the electromagnetic field ${\cal A}_{\mu}$
is driven by the covariant derivative
$
D_{\mu}\ =\ \partial_{\mu}+i\,\hat{Q}\mathcal{A}_{\mu}\ ,\label{covdev}
$
where $\hat{Q}=\mbox{diag}(Q_{u},Q_{d})$, with $Q_{u}=2e/3$ and
$Q_{d}=-e/3$, $e$ being the proton electric charge.
We consider the case of a homogeneous
stationary magnetic field $\boldsymbol{B}$ orientated along the axis 3,
or $z$. To write down the explicit form of $\mathcal{A}^{\mu}$
one has to choose a specific gauge. Some commonly used (\lq\lq standard'') gauges are
the symmetric gauge, in which ${\cal A}^{\mu}(x)=(0,-B\,x^{2}/2,B\,x^{1}/2,0)$,
the Landau gauge 1, in which ${\cal A}^{\mu}(x)=\left(0,-B\,x^{2},0,0\right)$
and the Landau gauge 2, in which ${\cal A}^{\mu}(x)=\left(0,0,B\,x^{1},0\right)$. To test
the gauge independence of our results, all these gauges will be considered
in our analysis.

Since we are interested in studying meson properties, it is convenient
to bosonize the fermionic theory, introducing scalar, pseudoscalar,
vector and axial vector fields $\sigmaa(x)$, $\pia(x)$, $\rhoamu(x)$,
$\aamu(x)$, with $b=0,1,2,3$, and integrating out the fermion fields.
The bosonized action can be written as
\begin{eqnarray}
S_{\mathrm{bos}} & = & -i\ln\det\!\big(i\mathcal{D}\big)-\frac{1}{4g}\int d^{4}x\ \Big[\sigmac(x)\,\sigmac(x)+\boldsymbol{\pi}(x)\cdot\boldsymbol{\pi}(x)\Big]
 -\,\frac{1}{4g(1-2\cc)}\int d^{4}x\ \Big[\boldsymbol{\sigma}(x)\cdot\boldsymbol{\sigma}(x)+\pic(x)\,\pic(x)\Big]\nonumber \\
 &  & +\,\frac{1}{4\gcoupr}\int d^{4}x\ \left[\boldsymbol{\rho}_{\mu}(x)\cdot\boldsymbol{\rho}^{\,\mu}(x)+\boldsymbol{a}_{\mu}(x)\cdot\boldsymbol{a}^{\,\mu}(x)\right]
 +\frac{1}{4\gcoupw}\int d^{4}x\ \rhocmud(x)\,\rhocmu(x)+\frac{1}{4\gcoupa}\int d^{4}x\ \acmud(x)\,\acmu(x)\ ,\qquad \label{sbos}
\end{eqnarray}
with
\begin{equation}
i\mathcal{D}_{x,x'}\ =\ \delta^{(4)}(x-x')\,\Big\{i\,\rlap/\!D-m_{0}-\taua\left[\sigmaa(x)+i\,\gamma_{5}\,\pia(x)+\gamma_{\mu}\,\rhoamu(x)+\gamma_{\mu}\gamma_{5}\,\aamu(x)\right]\Big\}\ .\label{dxx}
\end{equation}
In Eq.~(\ref{sbos}) we have introduced $g \equiv \gcoups + g_{D}$ and $\cc \equiv g_{D}/(\gcoups+g_{D})$,
in such a way that the flavor mixing turns out to be regulated
by the parameter $\cc$.

We proceed by expanding $S_{\mathrm{bos}}$ in powers of the fluctuations
of the boson fields around the corresponding mean field (MF) values.
We assume that the fields $\sigma_b(x)$ have nontrivial translational
invariant MF values given by $\tau_b\,\bar{\sigma}_b=\mbox{diag}(\bar{\sigma}_{u},\bar{\sigma}_{d})$,
while vacuum expectation values of other boson fields are zero;
thus, we write
\begin{equation}
\mathcal{D}_{x,x'}\ =\ \ {\rm diag}\big(\mathcal{D}_{x,x'}^{\mf,\,u}\,,\,\mathcal{D}_{x,x'}^{\mf,\,d}\big)+\delta\mathcal{D}_{x,x'}\ ,
\end{equation}
where $\mathcal{D}_{x,x'}^{\mf,\,f}\ =\ -i\,\delta^{(4)}(x-x')\left(i\rlap/\partial+Q_{f} \ \rlap/\!\!{\cal A}(x)   -M_{f}\right)\,$, with $f=u,d$. Here
$M_{f}=m_{c}+\bar{\sigma}_{f}$ is the quark effective
mass for each flavor $f$.

The MF action per unit volume is given by
\begin{equation}
\frac{S_{\mathrm{bos}}^{\mbox{\tiny MF}}}{V^{(4)}}\ =-\ \frac{(1-\cc)(\bar{\sigma}_{u}^{2}+
\bar{\sigma}_{d}^{2})-2\,\cc\,\bar{\sigma}_{u}\bar{\sigma}_{d}}{8g(1-2\,\cc)}-\frac{iN_{c}}{V^{(4)}}\sum_{f=u,d}\int d^{4}x\,d^{4}x'
\ \trmin_{D}\,\ln\left(\mathcal{S}_{x,x'}^{\mf,\,f}\right)^{-1}\ ,
\label{seff}
\end{equation}
where $\trmin_{D}$ stands for the trace over Dirac space and $\mathcal{S}_{x,x'}^{\mf,\,f}=\big(i\mathcal{D}_{x,x'}^{\mf,\,f}\big)^{-1}$
is the MF quark propagator in the presence of the magnetic field.
Its explicit expression can be written as
\begin{equation}
\mathcal{S}_{x,y}^{\mf,\,f}=e^{i\Phi_{Q_{f}}(x,y)}\int\frac{d^{4}p}{(2\pi)^{4}}\ e^{-i\,p(x-y)}
\,\bar{S}^{f}(p_{\parallel},p_{\perp})\ ,
\label{uno}
\end{equation}
where the function $\Phi_{Q}(x,y)$ is the
so-called Schwinger phase, which is shown to be a gauge dependent quantity. In addition, introducing $B_f = |B Q_f|$ , $s_f={\rm sign}(B Q_f)$ and $t_f = \tan (\sigma B_f)$,
one has
\begin{eqnarray}
\bar{S}^{f}(p_{\parallel},p_{\perp}) = -\,i\int_{0}^{\infty}d\sigma\
\exp\!\bigg[\!-i\sigma\Big(M_{f}^{2}-p_{\parallel}^{2}+ \,\dfrac{t_f \, \boldsymbol{p}_{\perp}^{\ 2}}{\sigma B_{f}}-i\epsilon\Big)\bigg]
\left[\left(p_{\parallel}\cdot\gamma_{\parallel}+M_{f}\right)(1-s_f\, t_f\,\gamma^{1}\gamma^{2} )- (1-t_f^2)\, \boldsymbol{p}_{\perp}\cdot\boldsymbol{\gamma}_{\perp}\right]\ .
\label{sfp_schw}
\end{eqnarray}
Here, we have defined $p_{\parallel}^{\mu}=(p^{0},0,0,p^{3})$, $p_{\perp}^{\mu}=(0,p^{1},p^{2},0)$,
and equivalent definitions have been used for $\gamma_\parallel$,
$\gamma_\perp$.

The gap equations needed to determine the effective masses $M_f$
are obtained from $\partial S_{\mathrm{bos}}^{\mbox{\tiny
MF}}/\partial\bar{\sigma}_{f}=0$. It is seen that they can be
written in terms of the quark condensates
$\phi_{f}\equiv\langle\bar{\psi}_{f}\psi_{f}\rangle$. As usual in
NJL-like models, the expressions for the condensates involve
divergent loop integrals that have to be properly regularized. We
use here the magnetic field independent regularization (MFIR)
scheme~\cite{Menezes:2008qt,Avancini:2019wed}: for a given
unregularized quantity, the corresponding (divergent) $B\to0$
limit is subtracted and then it is added in a regularized form.
Thus, the quantities can be separated into a (finite) \lq\lq $B=0$''
part and a \lq\lq magnetic'' piece. To regularize the \lq\lq $B=0$'' terms
(which still keep an implicit dependence on the magnetic field,
e.g.~through the masses $M_f$) we use here a proper time (PT)
scheme. This requires the introduction of a dimensionful parameter
$\Lambda$, which plays the role of an ultraviolet cutoff.

    \subsection{Meson masses in the NJL model at finite magnetic field}\label{third}

The quadratic piece of the bosonized action corresponding to the neutral mesons can be
written as
\begin{equation}
S_{\mathrm{bos}}^{{\rm quad,\,neutral}}\ =-\ \frac{1}{2}\int d^{4}x\;d^{4}x'\sum_{M,M'}\delta M(x)^{\dagger}
\ {\gmatrix}_{MM'}(x,x')\ \delta M'(x')\ .
\end{equation}
Here $M.M'=\sigmac,\pic,\rhocmu,\acmu,\sigmat,\pit,\rhotmu,\atmu$. Note that $\sigmac$,
$\pic$, $\rhoc$ and $\ac$ correspond to the isoscalar states $\sigma$,
$\eta$, $\omega$ and $f_{1}$, while $\sigmat$, $\pit$, $\rhot$ and $\at$
stand for the neutral components of the isovector triplets $\boldsymbol{
a}_{0}$, $\boldsymbol{\pi}$, $\boldsymbol{\rho}$ and $\boldsymbol{a}_{1}$, respectively.
The meson indices $M,M'$, as well as the functions
${\gmatrix}_{MM'}$, include Lorentz indices in the case of vector mesons.
This also holds for the functions $\delta_{MM'}$, ${\cal J}_{MM'}$,
$c_{MM'}^{ff'}$, etc., introduced below. In the corresponding
expressions, a contraction of Lorentz indices is understood when
appropriate. The functions ${\gmatrix}_{MM'}(x,x')$ can be
separated in two terms, namely
\begin{equation}
{\gmatrix}_{MM'}(x,x')\ =\ \frac{1}{2g_{M}}\ \delta_{MM'}\ \delta^{(4)}(x-x')-\,{\jmatrix}_{MM'}(x,x')\ ,
\label{gmat}
\end{equation}
where
\begin{equation}
\frac{1}{g_{M}}\ \delta_{MM'}\ =\ \left\{ \begin{array}{cl}
1/g & \ \ \ \ \mbox{for}\ \ M=M'=\sigmac,\pit\\
1/[g(1-2\cc)] & \ \ \ \ \mbox{for}\ \ M=M'=\sigmat,\pic\\
-\eta^{\mu\nu}/\gcoupr & \ \ \ \ \mbox{for}\ \ MM'=\rhotmu\rhotnu,\atmu\atnu\\
-\eta^{\mu\nu}/\gcoupw & \ \ \ \ \mbox{for}\ \ MM'=\rhocmu\rhocnu\\
-\eta^{\mu\nu}/\gcoupa & \ \ \ \ \mbox{for}\ \ MM'=\acmu\acnu
\end{array}\right.
\label{valgm}
\end{equation}
and $\delta_{MM'}=0$ otherwise. Here $\eta^{\mu\nu}$ is the Minkowski metric
tensor, $\eta^{\mu\nu} = \mbox{diag}(1,-1,-1,-1)$.
In turn, the polarization functions ${\jmatrix}_{MM'}(x,x')$ can be
separated into $u$ and $d$ quark pieces,
\begin{equation}
{\jmatrix}_{MM'}(x,x')\ =\ c_{MM'}^{uu}(x,x')\,+\,\varepsilon_{M}\,\varepsilon_{M'}\,c_{MM'}^{dd}(x,x')\ .
\label{jotas}
\end{equation}
Here, $\varepsilon_{M}=1$ for $M=\sigmac,\pic,\rhocmu,\acmu$
and $\varepsilon_{M}=-1$ for $M=\sigmat,\pit,\rhotmu,\atmu$, while
the functions $c_{MM'}^{ff'}(x,x')$ are given by
\begin{eqnarray}
c_{MM'}^{ff'}(x,x')\ =-i\,N_{c}\ \trmin_{D}\bigg[i\,\mathcal{S}_{x,x'}^{\mf,\,f}\,\Gamma^{M'}i\,\mathcal{S}_{x',x}^{\mf,\,f'}\,\Gamma^{M}\,\bigg]\ ,
\label{polfun}
\end{eqnarray}
with $\Gamma^{\sigma}=\mathbbm{1}$, $\Gamma^{\pi}=i\gamma^{5}$,
$\Gamma^{\rho^{\mu}}=\gamma^{\mu}$ and
$\Gamma^{a^{\mu}}=\gamma^{\mu}\gamma^{5}$.

We now address the charged mesons case, i.e.\ the states
$s^{\pm}=(s_{1}\mp is_{2})/\sqrt{2}$ and $v^{\pm\mu}=(v_{1}^{\mu}\mp iv_{2}^{\mu})/\sqrt{2}$,
with $s=\sigma,\pi$ and $v=\rho,a$. We concentrate on the positive
charge sector, noticing that the analysis of negatively charged mesons
is completely equivalent. The corresponding quadratic piece of the
bosonized action can be written as
\begin{equation}
S_{\mathrm{bos}}^{{\rm quad,+}}\ =\ -\frac{1}{2}\int d^{4}x\;d^{4}x'\sum_{M,M'}\,\delta M(x)^{\dagger}\ {\cal G}_{MM'}(x,x')\ \delta M'(x')\ ,\label{quadbosch}
\end{equation}
where, for notational convenience, we simply denote the positively
charged states by $M,M'=\sigma,\pi,\rho^{\mu},a^{\mu}$ (a proper
contraction of Lorentz indices of vector mesons is understood).
Once again, the inverse propagator ${\cal G}_{MM'}(x,x')$ can be separated in a form as that given by Eq.~(\ref{gmat}). In this case one has
\begin{eqnarray}
{\cal J}_{MM'}(x,x')= 2\ c_{MM'}^{ud}(x,x')
\label{jcharged}
\end{eqnarray}
 and
\begin{equation}
\frac{1}{g_{M}}\ \delta_{MM'}\ =\ \left\{ \begin{array}{cl}
1/g & \ \ \ \ \mbox{for}\ \ M=M'=\pi\\
1/[g(1-2\cc)] & \ \ \ \ \mbox{for}\ \ M=M'=\sigma\\
-\eta^{\mu\nu}/\gcoupr & \ \ \ \ \mbox{for}\ \ MM'=\rho^{\mu}\rho^{\nu},a^{\mu}a^{\nu}
\end{array}\right.\qquad,
\end{equation}
with $\delta_{MM'}=0$ otherwise.

To proceed we have to evaluate the functions
$c_{MM'}^{ff'}(x,x')$ in Eqs.~(\ref{jotas}) and (\ref{jcharged}).
From Eqs.~(\ref{polfun}) and (\ref{uno}) we have
\begin{equation}
c_{MM'}^{ff'}(x,x')\ =\
e^{i\Phi_{M}(x,x')}\int\frac{d^{4}t}{(2\pi)^{4}}\;e^{-it(x-x')}\,
\bar c_{MM'}^{ff'}(t_\parallel,t_\perp)\ ,
\label{nueve}
\end{equation}
where
\begin{equation}
\bar c_{MM'}^{ff'}(t_\parallel,t_\perp)\ =\ -iN_{c}\int\frac{d^{4}p}{(2\pi)^{4}}\;
\mbox{tr}_{D}\left[i\bar{S}^{f}({p_{\parallel}^{+}},{p_{\perp}^{+}})\,
\Gamma^{M'}\,i\bar{S}^{f'}({p_{\parallel}^{-}},{p_{\perp}^{-}})\,\Gamma^{M}\right]\ .
\label{jmmtt}
\end{equation}
Here we have defined $p_{a}^{\pm}=p_{a}\pm t_{a}/2$, where $a=\parallel,\perp$.
In addition, we have used $\Phi_{M}(x,x')=\Phi_{Q_{f}}(x,x')+\Phi_{Q_{f'}}(x',x)$.
It can be seen that $\Phi_{M}$ vanishes in the case of the neutral mesons, while it does not for
charged mesons.

In what follows we consider the particular case $M=M'=\pi$. The procedure is similar for the all the other cases, although, in general, the Lorentz structure is more complicated. Details can be found in Ref.~\cite{Coppola:2023mmq}. We introduce
\begin{eqnarray}
c_{\pi\pi}^{ff'}(\bar{q},\bar{q}') & = & \int d^{4}x\,d^{4}x'\ {\cal F}_{Q}(x,\bar{q})^{\ast}\,
c_{\pi\pi}^{ff'}(x,x')\,{\cal F}_{Q}(x',\bar{q}')\ ,
\label{diez}
\end{eqnarray}
where the functions ${\cal F}_{Q}(x,\bar{q})$ are solutions of the
meson field equations in the presence of an external constant and
homogenous magnetic field. For the neutral pion  $\bar q$ stands
for the usual four-momentum, $\bar q = q =(q^0,\vec q\,)$, while
${\cal F}_{Q}(x,\bar{q})=\exp[ -i q\cdot x]$. On the other hand,
for charged pions ${\cal F}_{Q}(x,\bar{q})$ is a gauge dependent
Ritus-like function; in this case one has
$\bar{q}=(q^{0},\ell,\chi,q^{3})$, where $\ell$ is an integer
related with the so-called Landau level, and the fourth quantum
number $\chi$ can be chosen according to the gauge fixing.
Details on this issue can be found in
Ref.~\cite{GomezDumm:2023owj}, where the explicit form of ${\cal
F}_{Q}(x,\bar{q})$ for the standard gauges is given. Replacing
Eq.~(\ref{nueve}) into Eq.~(\ref{diez}) we find
\begin{eqnarray}
c_{\pi\pi}^{ff'}(\bar{q},\bar{q}') & = & \int\frac{d^{4}t}{(2\pi)^{4}}\ \bar c_{\pi\pi}^{ff'}(t_\parallel,t_\perp)\ h_Q(\bar{q},\bar{q}', t)\ ,
\end{eqnarray}
where $h_{Q}(\bar{q},\bar{q}',t)$ is a gauge invariant quantity given by
\begin{equation}
h_{Q}(\bar{q},\bar{q}',t)\ =
\ \int d^{4}x\,d^{4}x'\,{\cal F}_{Q}(x,\bar{q})^{\,\ast}\,{\cal F}_{Q}(x',\bar{q}')\,e^{i\Phi_{Q}(x,x')}\,e^{-it(x-x')}\ .
\label{hPiInt}
\end{equation}

For neutral pions it is easy to show that $h_{Q=0}(q,q',t) =
(2\pi)^4\, \delta^{(4)}(q-q')\,  (2\pi)^4 \, \delta^{(4)} (q-t)$.
Thus, one has $c_{\pi\pi}^{ff,0}(q,q') = (2\pi)^4 \delta^{(4)}
(q-q') \ C^{ff,0}_{\pi\pi}(q_\parallel,q_\perp)\,$, where $
C^{ff,0}_{\pi\pi}(q_\parallel,q_\perp) =  \bar
c^{ff}_{\pi\pi}(q_\parallel,q_\perp)$. Therefore, from
Eqs.~(\ref{gmat}), (\ref{jotas}) and (\ref{diez}) we see that for
neutral mesons ${\cal G}_{MM'}(x,x')$ becomes diagonal when
transformed into ``Fourier space''. In turn, for charged pions
it is possible to show~\cite{GomezDumm:2023owj} that for all the
standard gauges one has
\begin{equation}
h_{Q=e}(\bar{q},\bar{q}',t)\ =\
\delta_{\chi\chi'}\left(2\pi\right)^{4}\,\delta^{\left(2\right)}(q_{\parallel}-q_{\parallel}^{\prime})
\,\left(2\pi\right)^{2}\,\delta^{\left(2\right)}(q_{\parallel}-t_{\parallel})\,f_{\ell\ell\,'}(t_{\perp})\ ,
\label{hPiGauge}
\end{equation}
where
\begin{equation}
f_{\ell\ell\,'}(t_{\perp}) \ = \
\frac{4\pi(-i)^{\ell+\ell'}}{B_{e}}\,\sqrt{\frac{\ell!}{\ell^{\prime}!}}\,
\left(\frac{2\,t_{\perp}^{\;2}}{B_{e}}\right)^{\frac{\ell'-\ell}{2}}
L_{\ell}^{\ell^{\prime}-\ell}\Big(\frac{2\,t_{\perp}^{\;2}}{B_{e}}\Big)\;
e^{-t_{\perp}^{\;2}/B_{e}}\;e^{is(\ell-\ell')\varphi_{\perp}}\ .
\label{fkkp}
\end{equation}
Here $B_{e}=|eB|$, $s=\mbox{sign}(eB)$, $L_k^m(x)$ are generalized
Laguerre polynomials, and $t_\perp^\mu = |t_\perp| (0,\cos
\varphi_{\perp}, \sin \varphi_{\perp},0)$. We can take these
expressions and integrate over $\varphi_{\perp}$, using the fact
that, due to invariance under rotations around the $3$-axis and
boosts in the 0-3 plane, $c_{\pi\pi}^{ff'}(t_\parallel,t_\perp)$
can only depend on $t_\parallel^2$ and $t_\perp^2$. One gets in
this way  $c_{\pi\pi}^{ud,+}(\bar q,\bar q') = (2\pi)^4
\delta_{\chi\chi'} \delta_{\ell\ell'} \delta^{(2)}
(q_\parallel-q_\parallel') \
C^{ud,+}_{\pi\pi}(\ell,q_\parallel)\,$, where
\begin{equation}
 C^{ud,+}_{\pi\pi}(\ell,q_\parallel) =  \int_0^\infty\, dt_\perp^2 \, \bar c^{ud}_{\pi\pi}(q_\parallel,t_\perp) \ \rho_\ell(t_\perp^2)\ ,
\end{equation}
with $\rho_\ell(t_\perp^2) = (-1)^\ell\, e^{-t_\perp^2/B_e} \,
L_\ell(2 t_\perp^2/B_e)/B_e\,$. Thus, from Eqs.~(\ref{gmat}),
(\ref{jcharged}) and (\ref{diez}) we see that for charged mesons
${\cal G}_{MM'}(x,x')$ becomes diagonal when transformed into
``Ritus space''. In this case, even for $\ell =0$, we cannot set
$t_\perp=0$, i.e. we cannot take a charged meson to be rest.

As already mentioned, when vector or axial vector mesons are
involved, the Lorentz structure of the polarization function is
more complicated. For example, in the case of the pion-axial
vector mixing polarization functions one gets
\begin{equation}
C_{\pi a^{\mu}}^{ff,0}(q_\parallel,q_\perp) \ =
d_{\pi a,1}^{ff,0}(q_\parallel,q_\perp)\ q_{\parallel}^{\mu}+d_{\pi a,2}^{ff,0}(q_\parallel,q_\perp) \ q_{\perp}^{\mu}
\end{equation}
for the neutral $\pi^0-{\rm a}_1^0$ meson mixing, and
\begin{equation}
C_{\pi a^{\mu}}^{ud,+}(\ell,q_{\parallel}) = \, d_{\pi a,1}^{ud,+}(\ell,q_{\parallel})\, \Pi_{\parallel}^{\mu\,\ast}+d_{\pi a,2}^{ud,+}(\ell,q_{\parallel})\, \Pi_{\perp}^{\mu\,\ast}-
d_{\pi a,3}^{ud,+}(\ell,q_{\parallel})\, is\, {{F}}^{\mu\nu}\,\Pi_{\perp,\nu}^{\ast}/B
\end{equation}
for the $\pi^+-{\rm a}_1^+$ mixing, where
$\Pi_{\parallel}^{\mu} = q_\parallel^\mu$ and $\Pi_{\perp}^{\mu} =
\left(0,
i\sqrt{\frac{B_{M}}{2}}\left(\sqrt{\ell+1}-\sqrt{\ell}\right),
-s\sqrt{\frac{B_{M}}{2}}\left(\sqrt{\ell+1}+\sqrt{\ell}\right),0\right)$.
The tensor structure of all possible polarization functions, as
well as the explicit form of the corresponding coefficients, can
be found in Ref.~\cite{Coppola:2023mmq}. It should be noted that
some of these coefficients involve divergent integrals. To
regularize them we use the MFIR method discussed in the previous
section.

To proceed we have to contract the tensors ${\cal G}_{MM'}$ that
involve vector or axial vector mesons with the appropriate
polarization vectors. In the case of neutral mesons there are
4 independent polarization vectors: one of them, that we
indicate by $L$, is timelike, while the other three are spacelike
and will be labelled with $c=1,2,3$. For charged mesons the
number of independent polarization vectors depends on the Landau
level $\ell$. For $\ell=-1$ there is no timelike vector,
whereas only one spacelike vector exists (
$c=1$); for $\ell=0$ we have one timelike vector and two spacelike
vectors ($c=1,2$), while for $\ell > 0$ all four polarization
vectors exist. The explicit expressions of these vectors can be
found in Ref.~\cite{Coppola:2023mmq}.  After performing the
contractions, one is faced with a neutral meson inverse propagator
given by a $20 \times 20$ matrix $G_{N,N'}(q_\parallel,q_\perp)$,
where $N,N'=\sigma_b, \pi_b, \rho_{b,L}, \rho_{b,c},a_{b,L},
a_{b,c}$, $b=0,3$. For the positively charged meson sector,  the size of
the corresponding matrix depends on $\ell$, since, as stated, not
all polarizations are available for $\ell=-1,0$. In any case, one
finds at most a $10 \times 10$ matrix with elements
$G_{N,N'}(\ell,q_\parallel)$, where $N,N'=\sigma,\pi, \rho_{L},
\rho_{c},a_{L}, a_{c}$ with $c=1,2,3$. Once the above described
matrices are obtained, the meson masses can be determined by
solving the equations
\begin{eqnarray}
&&\det G(q_\parallel,q_\perp) \Big|_{q^0=m, \vec q =0} = 0          \qquad \mbox{for neutral mesons\ ,}
\label{eqsuno}\\
&&\det G(\ell,q_\perp) \Big|_{q^0=E, q^3=0} = 0        \qquad \mbox{for charged mesons\ ,}
\label{eqsdos}
\end{eqnarray}
where $E=\sqrt{m^2 + (2 \ell + 1) B_e}$.

At this point we note that the matrices become simplified due
to an extra symmetry of the system, which corresponds to a parity
operation followed by a $\pi$-rotation around the direction of the
$\vec B$~\cite{Coppola:2023mmq}. For charged mesons with
$\ell=-1$, this symmetry implies that $\rho_1$ and $a_1$ (the only
states in this sector) do not mix, while for $\ell=0$ it is seen
that the lowest energy charged pion state only mixes with $\rho_2$, $a_L$ and $a_1$. In the case of neutral mesons, the $20\times 20$
matrix is given by a direct sum of four submatrices. One of them
corresponds to the states $\pi_b, \rho_{b,2},a_{b,L}$, one to
$\sigma_b, \rho_{b,L},a_{b,2}$, one to the (degenerated) $S_z=\pm
1$ $\rho$ meson states and, finally, one to the (degenerated)
$S_z=\pm 1$ $a$ meson states.

\hfill

To obtain numerical results it is necessary to fix
the model parameters.  As stated, in
our framework divergent quantities are regularized using the MFIR scheme,
with a proper time cutoff $\Lambda$. Within this scenario, we take
$m_c = 7.01$~MeV, $\Lambda = 842$~MeV, $g=5.94/\Lambda^2$, $\gcoupr =
3.947/\Lambda^2$ and
$\alpha=0.114$. At $B=0$, this leads to
 $M_{u,d} = 400$~MeV,
$\phi_{u,d} = (-227$~MeV$)^3$, $m_\pi = 140$~MeV, $f_\pi=92.2$~MeV, $m_\eta=548$~MeV, $m_\rho=775$~MeV
 and to a phenomenologically acceptable value of about
1020~MeV for the a$_1$ mass (in fact, as usual in this type of model,
the a$_1$ mass is found to lie above the $q \bar q$ production
threshold and can be determined only after some extrapolation). For simplicity, the remaining coupling constants are taken to be $\gcoupw=\gcoupa=\gcoupr$, which leads to
$m_\omega=m_\rho$ and $m_{f_1}=m_{{\rm a}_1}$.

Now, while most NJL-like models are able to reproduce the
effect of magnetic catalysis at vanishing temperature, they fail
to describe the inverse magnetic catalysis effect observed in
lattice QCD at finite temperature. One of the simplest approaches
to partially cure this behavior consists of allowing the model
couplings to depend on the magnetic
field~\cite{Ayala:2014iba,Farias:2014eca,Ferreira:2014kpa}. Thus,
we consider here both the situation in which the couplings are
constant and the one in which they vary with $B$. For
definiteness, we adopt for $g(B)$ the form proposed in
Ref.~\cite{Avancini:2016fgq}, namely
\begin{equation}
g(B) \ =\ g\,\mathcal{F}(B) \ , \qquad \qquad
\mathcal{F}(B) \ =\ \kappa_1 + (1-\kappa_1)\,e^{-\kappa_2(eB)^2} \ ,
\end{equation}
with $\kappa_1= 0.321$ and $\kappa_2= 1.31$~GeV$^{-2}$. Concerning the
vector couplings, given the common gluonic origin of $g$ and $\gcoupr$, we
assume that they get affected in a similar way by the magnetic field; hence,
we take $\gcoupr(B)=\gcoupr \mathcal{F}(B)$.

\begin{figure}[h]
\centering{}\includegraphics[width=0.45\textwidth]{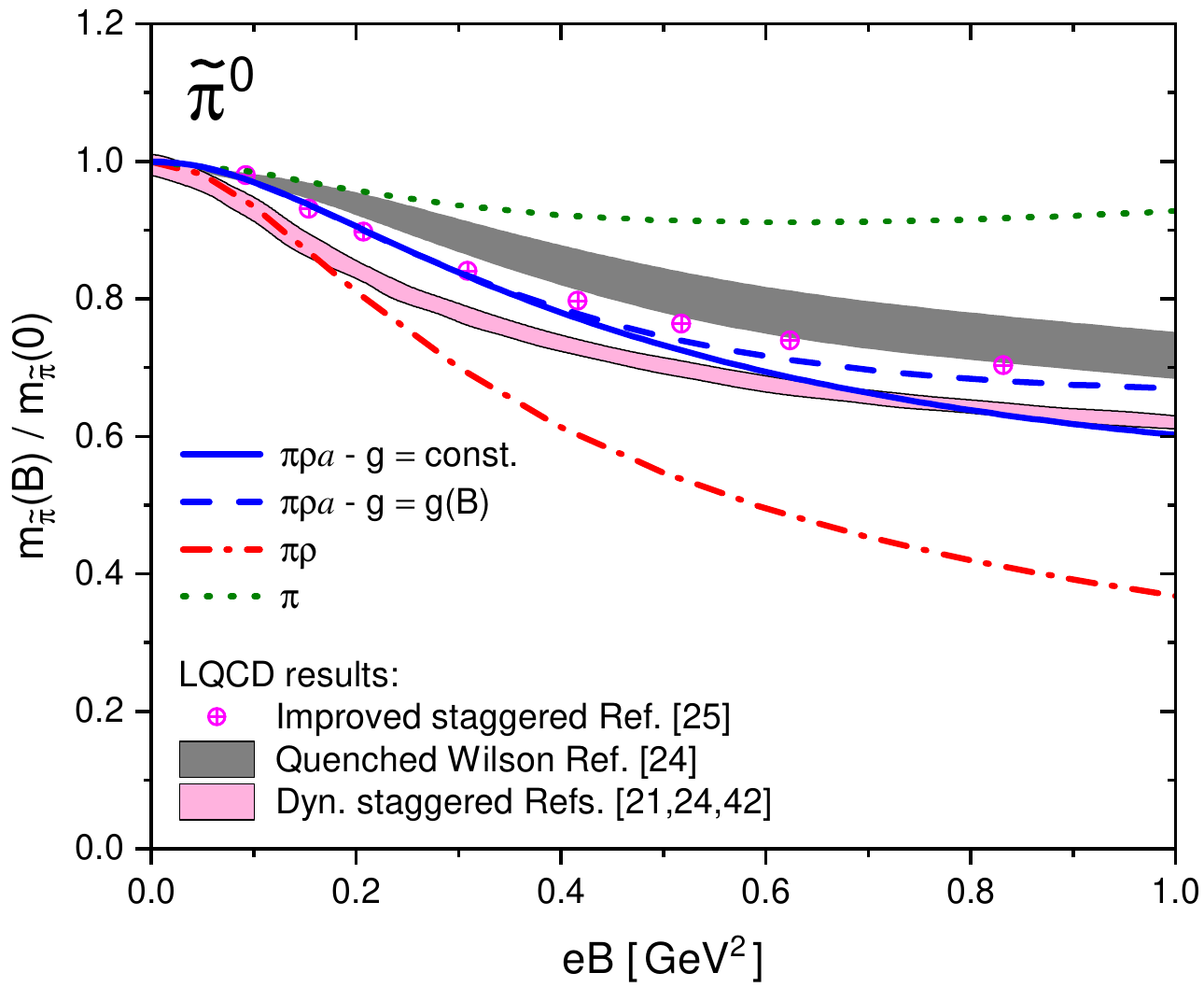}
\vspace*{-0.5cm} \caption{Normalized mass of the $\tilde{\pi}^0$
meson as a function of $eB$.} \label{fig:neutralpion}
\end{figure}

In what follows we present and analyze some selected results for
the effect of the magnetic field on meson masses. Let us start
with the neutral sector. As well known, for $B=0$ pseudoscalar
mesons mix with ``longitudinal'' axial vector mesons. For nonzero
$B$ the mixing also involves neutral vector mesons with spin
projection $S_z=0$ (corresponding to the polarization state
$c=2$). The four lowest mass states of this sector are to be
identified with the physical states $\tilde\pi^0$, $\tilde\eta$,
$\tilde\rho^{\,0}$ and $\tilde\omega$, where the particle names
are chosen according to the spin-isospin composition of the states
in the limit of vanishing external field. In addition, the neutral
sector includes states with spin projections $S_z = \pm 1$, i.e.,
spin parallel to the direction of the magnetic field. Here we
concentrate only on the mass of the lowest state, $\tilde\pi^0$.
Results for other states can be found in
Ref.~\cite{Coppola:2023mmq}. In Fig.~\ref{fig:neutralpion} we
compare our predictions with LQCD results that have been
obtained using different methods and various values of the $B=0$
pion mass. For comparison, we include the results obtained for
NJL-like models in which different meson sectors have been taken
into account. In the case where one considers just the
pseudoscalar sector, the result is quite different depending on
whether $g$ is kept constant or not. In the first case, the
behavior of $m_{\tilde\pi^0}$ with the magnetic field is found to
be non-monotonic, deviating just slightly from its value at $B=0$
(green dotted line). In contrast, for a $B$-dependent $g$ one
can get a monotonic decrease, in good agreement with LQCD
results~\cite{Avancini:2016fgq}. When the mixing with the vector
sector is considered, the results for both constant and
$B$-dependent couplings are similar to each other (red dash-dotted
line) and monotonically decreasing, lying however quite below LQCD
predictions~\cite{Carlomagno:2022inu}. Finally, if the mixing with ${\rm a}_1$ mesons is
also included, we obtain, for both constant  (blue solid line) and
$B$-dependent couplings (blue dotted line), a monotonic decrease
which is in good qualitative agreement with LQCD calculations. One
may infer that the incorporation of axial vector mesons, being the
chiral partners of vector mesons, leads to cancellations that help
to alleviate the magnitude of the neutral pion mass suppression.
Their inclusion into the full picture leads to relatively more
robust results and improves the agreement with LQCD
predictions.

\begin{figure}[h]
\centering{}\includegraphics[width=0.85\textwidth]{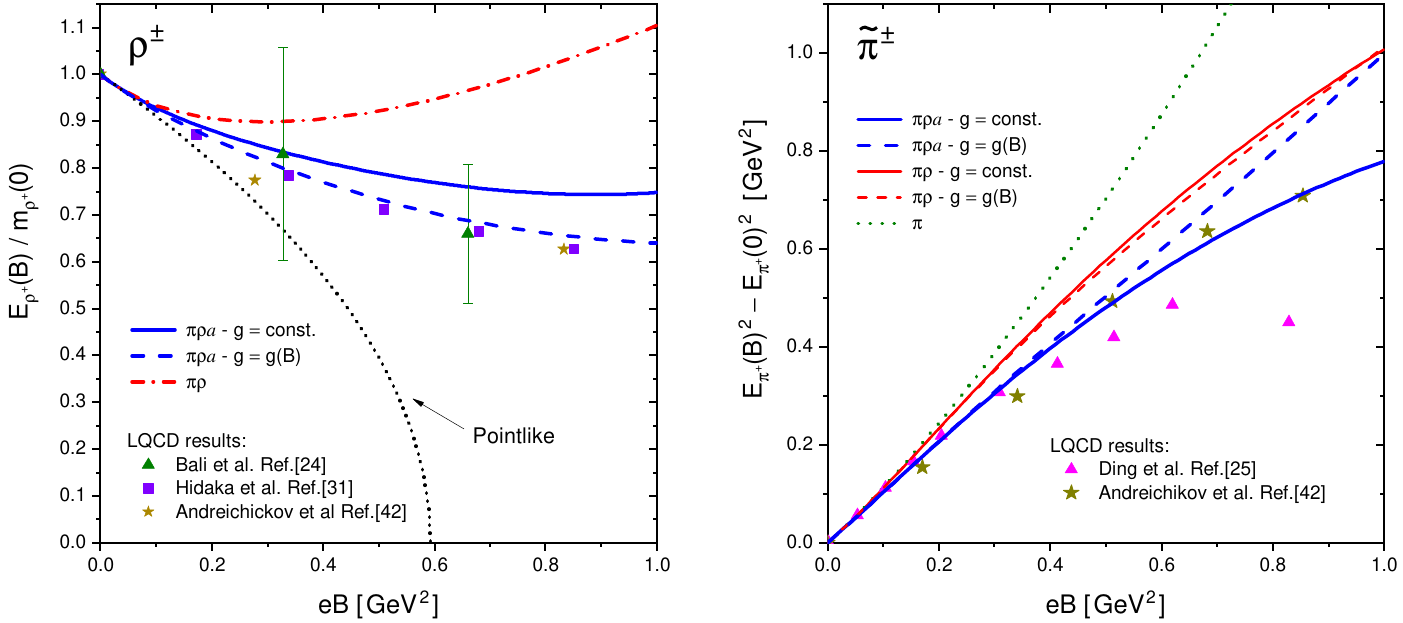}
 \caption{Left panel:  Normalized energy of the
lowest $\rho^+$ meson state as a function of $eB$. Right panel:
Difference $E_{\pi^+}(B)^2-E_{\pi^+}(0)^2$ for the lowest $\pi^+$
meson satate as a function of $eB$.} \label{fig:chargedrho}
\end{figure}

We turn now to the charged meson sector. To study the lowest lying
states one has to consider the Landau modes $\ell=-1$ and $\ell =
0$. For $\ell = -1$, the lowest mass state is the one that we have
denoted as $\rho_1$ and its energy is given by $E_{\rho^+} =
\sqrt{m_{\rho^+}^2-B_e}$, where $m_{\rho^+}$ is obtained from
Eq.~(\ref{eqsdos}). The left panel of Fig.~\ref{fig:chargedrho}
displays
 our numerical results for $E_{\rho^+}$
as a function of $eB$, normalized by $m_{\rho^+}|_{B=0}$.
%
It is seen that the results for constant $g$ (blue solid line)
and for $g(B)$ (blue dashed line) are quite similar. In turn,
they differ considerably from those obtained in a similar
model~\cite{Carlomagno:2022arc} that does not take into account
the presence of axial vector mesons (red dash-dotted line).
In fact, here the differences between models that include or not
axial vector mesons do not arise from direct mixing effects but
from the fact that axial vector states mix with pions already for
$B=0$; this leads to some change in the model parameters so as to
get consistency with the phenomenological inputs. In any case, it
is found that ---as in the case of neutral mesons--- the results
from the full model (blue solid and dashed lines) appear to be
rather robust: they show a similar behavior either for constant or
$B$-dependent couplings, and this behavior is shown to be in good
agreement with LQCD calculations of
Refs.~\cite{Bali:2017ian,Hidaka:2012mz,Andreichikov:2016ayj}. We
point out that our results, as those from LQCD, are not consistent
with $\rho^+$ condensation for the considered range of values of
$eB$.  The curve corresponding to a pointlike $\rho^+$ meson is
shown for comparison. It is worth mentioning that our results are
qualitatively different from those obtained in other
works~\cite{Liu:2014uwa,Cao:2019res}, which do find $\rho^+$ meson
condensation for $eB \sim 0.2$ to $0.6$~GeV$^2$ in the framework
of two-flavor NJL-like models. Contrary to what is done here, in
those works Schwinger phases are neglected and it is assumed that
charged $\pi$ and $\rho$ mesons lie in zero three-momentum states.

In the case of the mode $\ell=0$,  the
lowest mass state $\pi^+$ is given in general by a mixing between the states
that we have denoted as $\pi$, $\rho_2$, $a_L$ and $a_1$. Its energy is given by $E_{\pi^+} =
\sqrt{m_{\pi^+}^2+B_e}$ where $m_{\pi^+}$ is obtained
from the lowest $\ell=0$ solution of Eq.~(\ref{eqsdos}). Our numerical results
are shown in the right panel of Fig.~\ref{fig:chargedrho} where we plot the values of the difference
$E_{\pi^+}(B)^2-E_{\pi^+}(0)^2$ as a function of $eB$. Blue
solid and dashed lines correspond to the cases of constant and $B$-dependent
couplings, respectively. We also include for comparison the results obtained
from similar NJL-like models that just include the pseudoscalar meson sector
(green dotted line), or just include the mixing between the pseudoscalar and
vector meson sectors (red full and dashed line), neglecting the effect of the
presence of axial vector mesons. It can be seen that the inclusion of the
axial vector meson sector leads to an improvement of the agreement with LQCD
data.
 We
remark, however, that our numerical results indicate a monotonic enhancement of the
charged pion energy with the magnetic field, in contrast with the nonmonotic
behavior found by recent LQCD simulations, see Ref.~\cite{Ding:2020hxw}. It would be interesting to get more insight on
this open issue from other effective models and further LQCD calculations.

    \subsection{Summary}\label{sec:sum}

We have considered the effect of a constant magnetic field $\vec B$ on light meson masses, taking into account the various possible mixing effects. This has been done in the framework of a two-flavor
NJL-like model that includes isoscalar and isovector couplings in the
scalar-pseudoscalar and vector-axial vector sector, as well as a flavor
mixing term in the scalar-pseudoscalar sector.
The divergences associated with the non-renormalizability of the
model have been regularized using the \lq\lq magnetic field independent
regularization" method. We have also explored the
possibility of using $B$-dependent couplings to
account for the effect of the magnetic field on sea quarks.
As is well known, for $B=0$ pseudoscalar mesons mix with
\lq\lq longitudinal'' axial vector mesons. The presence of an external
uniform $B$ breaks isospin and full rotational symmetry, allowing for a more complex meson
mixing pattern. The latter is constrained by the
remaining unbroken symmetries, in such a way that the mass matrices can be separated into
several \lq\lq boxes''.

In the case of neutral mesons, Schwinger phases cancel and the polarization
functions become diagonal in the usual momentum basis.
We have found that when both vector and axial-vector interactions are
taken into account, the mass
of the lightest state in this sector, $\tilde\pi^0$, displays a monotonic
decreasing behavior with $B$ in the studied range $eB<1$~GeV$^2$, which is
in good qualitative agreement with LQCD calculations for both $g={\rm
constant}$ and $g(B)$. This represents an improvement
over the results that take into account just the mixing with the
vector meson sector, or no mixing at all.

In the case of charged mesons, the polarization functions are
diagonalized by expanding meson fields in appropriate Ritus-like bases, so
as to account for the effect of non-vanishing Schwinger phases. Once again,
the symmetries of the system constrain the mixing matrices, which
also depend on the value of the meson Landau level $\ell$. For $\ell=-1$ one
has only one vector and one axial vector polarization states that do not mix with each other. Thus, in this case the effect of
the inclusion of axial vector mesons on the $\rho^+$ mass comes solely from
the model parametrization, which is affected by the presence of $\pi$-a$_1$
mixing at $B=0$. Our results show that when the axial-vector sector is
included, the energy $E_{\rho^+}$ of this state
undergoes a considerable reduction, leading to a decreasing behavior that is in qualitative agreement with LQCD predictions, independently of the
assumption of a $B$-dependent coupling constant. We notice that, in accordance with LQCD calculations,
$E_{\rho^+}$ does not vanish for any considered value of $B$.
For $\ell=0$,
the pion mixing subspace is given by $\pi^+\mhyphen\rho^+\mhyphen {\rm
a}_1^+$ for only certain polarizations states. Here, the lowest mass state
can be identified with the $\pi^+$. We find that, even though
vector mixing already induces a softening in the enhancement of the $E_{\pi^+}$ with $B$,
the inclusion of the axial vector meson sector reinforces
this softening, leading to an improved agreement with LQCD predictions.

\section{
Magnetic field effects in form factors and in the meson exchange potential }\label{Braghin}

Hadron interactions have been shown to present different types of modifications in background  magnetic fields, mostly due to the non-trivial hadron structure with quarks and gluons~\cite{Braghin2018c,Braghin2020a,Ferrer2014,Ferrer2015,Braghin2018a,Villavicencio2023,WorkshopTrento,FPV2024,Dominguez2023}. Experimental measurements and theoretical descriptions of form factors provide a powerful tool to understand hadron structure and interactions as described for example in Ref.~\cite{DrechselWalcher}.
Currently, there are several quite refined calculations for  different types of form factors, mostly for the nucleon, based in {\it ab initio} or continuous in-vacuum methods.
However, it may be useful, and conceptually enriching, to decompose such type of calculations in a CQM framework, such that
the role of each degree of freedom may be tested or further understood~\cite{Lavelle,Plessas}. Besides its historical importance, the CQM  is usually very good to describe global hadron properties, such as masses, overall interactions and decay constants. In the CQM, in different versions, hadrons are built out from colorless states of valence quarks. A dynamical derivation and investigation of constituent quark and quark-antiquark meson form factors in constant magnetic fields was initiated in the last years, starting with a one-gluon exchange 
quark-antiquark interaction as tbe leading QCD quark effective action~\cite{Braghin2018c,Braghin2020a,Braghin2018a}. By applying the background field method and the auxiliary field method, it is possible to describe the formation of quark-antiquark meson multiplets and the chiral condensate,
and the development of constituent quark currents that emerge as a dressing of quark currents by a gluon cloud due to components of a non-perturbative gluon propagator.
The resulting one-loop form factors, that may be considered for constant or running quark masses, have a structure akin to the Schwinger-Dyson equations (SDE) in the rainbow ladder approach~\cite{Tandy1997,Binosi2015}. This becomes more appealing when embedding hadrons in an external magnetic field, in which case, the description of hadron structure and dynamics become more intricate. Expectations for the strength of magnetic fields in  relativistic heavy ion collisions 
and in magnetars
may be of the order of  $eB \sim 10^{12}-10^{13}$ T
that may be weak with respect to quark or hadron masses, $eB \sim 0.1 m_\pi^2 - 0.1 M^2$, where $m_\pi$ and $M \sim 0.35$ GeV are the pion and a constituent quark masses  ~\cite{WorkshopTrento,magnetarsB,rhicB1,PhysRevX.14.011028}. The weak magnetic field offers the possibility to allow (semi)analytical calculations since, in this limit,  quark and meson  propagators usually have well established
analytical expressions. In the CQM, mesons couple to baryons via the baryon constituent quark that in turn couples to the magnetic field by means of insertions. In a slightly larger distance scale, the knowledge of hadron interactions and, consequently nuclear interactions, has been the subject of a wide and extensive investigation, under the assumptions of  meson exchange dominated processes from which nucleon and nuclear potentials can be parameterized~\cite{Yukawa1935,Machleidt}.
In meson exchange potentials, the meson propagator also couples to the magnetic field,
which can easily lead to non-trivial contributions for nucleon and nuclear interactions. In this section, some results obtained in Refs.~\cite{Braghin2020a,Braghin2024} are reviewed and  highlighted: the leading {\it weak} magnetic field corrections for 
the one-loop axial and pseudoscalar  pion-constituent form factors are discussed for space-like momenta. With the magnetic field contribution for the pseudoscalar form factor and  the corresponding leading contribution for a (pseudoscalar) meson field propagator,
the Fourier transform of the meson exchange process is performed for 
off-shell pions and on-shell constituent quarks.

\subsection{Meson and constituent quark form factors
in constant magnetic fields}
\label{sub:formfactors}

The one-gluon exchange quark-antiquark interaction
is a leading term of the QCD quark effective action. It is convenient to describe the structure of an effective gluon propagator in terms of longitudinal and transverse components:
$\tilde{R}^{\mu\nu}_{ab}(k) \equiv \delta_{ab} \left[
 \left( g^{\mu\nu} - \frac{k^\mu k^\nu}{k^2}
\right)   R_T  (k)
+ \frac{k^\mu k^\nu}{k^2} R_L (k) \right]$.
The one loop background field method (BFM) can be applied
such that the quantum quark currents can be arranged into multiplets with the same quantum numbers as the quark-antiquark mesons. The auxiliary field method (AFM) makes it possible to introduce  multiplets of quark-antiquark local meson fields, although non-local meson fields can also be defined. By integrating out the quantum  quark field, a quark determinant is obtained with background quark currents and local meson fields arranged in flavor multiplets. Given that the dressed gluon propagator and corresponding running quark-gluon coupling constant
has enough strength, Dynamical Chiral Symmetry Breaking (DChSB) takes place with the formation of a scalar quark-antiquark chiral condensate in the ground state. This leads to a large correction of the Lagrangian quark mass leading to an effective quark mass that can be considered at the  basis of the CQM.

A large quark mass expansion  leads to one-loop equations for meson-constituent quark current couplings that are well established in phenomenological investigations. In the presence of the magnetic field, leading effects in the quark propagator can be computed. Hereby,
effects in the gluon propagator and in the quark-gluon running coupling constant will not be discussed. Magnetic field effects in the propagator require numerical calculations, which make it more difficult to be analyzed. However,  by resorting to the weak magnetic field limit, weak with respect to the constituent 
quark mass, $M \sim 0.30-40$ GeV, analytical or semi-analytical results can be obtained.
\begin{figure}[t!]
\centering
\includegraphics[width=150mm]{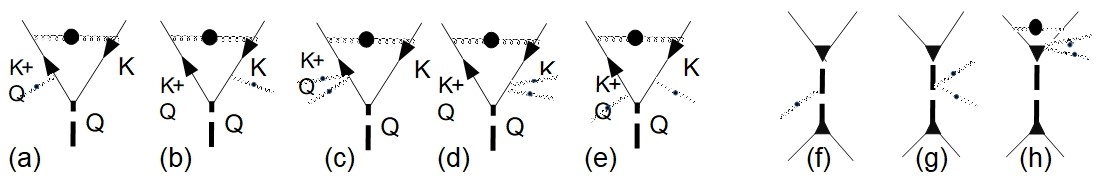}
 \caption{ 
\small
 (a)  - (e): One loop meson-constituent quark
 form factor with leading magnetic field insertions in the quark propagator.
\\
 (f) -  (h): Meson exchange process, with
the leading one or two weak magnetic field insertions in the meson propagator and the leading insertion in the pseudoscalar form factor (h). Dashed lines stand for meson, solid (wiggly with a big dot) lines for quarks (gluons) and small wavy lines, with a small dot, for magnetic field insertions.
}
\label{fig:FFMesB}
\end{figure}

The computation of weak magnetic field effects in the one-loop quark determinant, including in the resulting form factors, can be done by considering at least two methods: The first one, more convenient for weak constant background magnetic fields, is the expansion based in the Fock-Schwinger gauge~\cite{Schwinger1951}. A more commonly used approach is to consider the weak magnetic field limit of the quark (or meson) propagator, usually calculated in the proper time Schwinger representation~\cite{Chyi2000}. These contributions can be translated into the corresponding diagrams, with the 
leading ones containing one or two magnetic field insertions, as shown in
diagrams (a)-(e) of Fig.~\eqref{fig:FFMesB}.
Notice however, that the meson coupling to quark in these diagrams happens through the flavor-Dirac current with the same quantum numbers as the meson. Accordingly, the wiggly line with a big dot is not a full (dressed) gluon propagator but instead only a component of it that can be written in terms of $R_L(k)$
and $R_T(k)$. Each magnetic field insertion
is of order $(eB/M^2)$, with $M$ being the constituent quark mass.

The dynamically generated terms for the leading magnetic field corrections for the dimensionless axial and pseudoscalar form factors, with  canonical normalization of the isotriplet pion local field $P_i$, can be written in momentum space as
\begin{eqnarray} \label{Lcoupling}
{\cal L}_{\pi}^B =
 c_{i}   \; 
\left( \frac{e B_0}{M^2}
\right)^2 \left[ F^{B}_{ps}  (Q , K) 
+ F^{B,a}_{ps} (Q , K) \right]
P_i (Q)  (j_{ps}^i (K,Q) )^\dagger
+ \; 
d  \;
\epsilon^{0 \rho \mu 3} 
\left( \frac{e B_0}{M^2} \right)
 \frac{F^{B}_{A}  (Q , K) }{F}
\partial_\rho P_i (Q) \;  (j^{A,\mu}_i  (K,Q) )^\dagger,
\end{eqnarray}
where
$F^{B}_{ps}  (Q , K)$  ($F^{B,a}_{ps} (Q , K)$) is the isotropic (anisotropic)  contribution for the pseudoscalar form factor and
$F^{B}_{A}  (Q , K)$ the one for the axial form factor. $Q$ and $K$ are the pion and incoming constituent quark momenta, respectively.
$F=92$ MeV is the pion normalization constant,  namely, the pion decay constant, and 
the constituent quark currents are
$j_{ps}^i = \bar{\psi} i \gamma_5 \lambda_i \psi$ and  $j_{A,\mu}^i = \bar{\psi} \gamma_\mu \gamma_5 
\lambda_i \psi$
with $i = 1,2$ for charged and $i=3$ neutral pions. The following  numerical coefficients were defined: $c_1= c_2 = - 4/9$ and $c_3=5/9$, 
respectively, for the charged and neutral pion pseudoscalar couplings, and unique $d = 1/3$ for the axial coupling. $\epsilon^{0 \rho \mu 3}$ stands for some of the components of the Levi-Civita tensor that picks up the transverse pion and axial current components.
The leading contributions for the pseudoscalar  form factor is that of one magnetic field insertion for each of the internal quark lines, which is of order $(eB)^2$, in diagram (e). 
The two magnetic field insertions in a single quark propagator, diagrams (c) and (d), 
and one single magnetic field insertion in only one quark propagator, diagrams (a) and (b),
are either vanishing or sub-leading and will not be taken into account. The axial form factor
goes with $(eB)$ corresponding to diagrams (a) and (b), leading to a linear behavior of the axial coupling with magnetic fields~\cite{Braghin2020a} that is seemingly present in the pion-nucleon axial coupling~\cite{Villavicencio2023}. However, it turns out that diagram (a) provides a vanishing contribution and diagram (b) is responsible for the above result.

\begin{figure}[t!]
\begin{center}
\includegraphics[width=100mm]{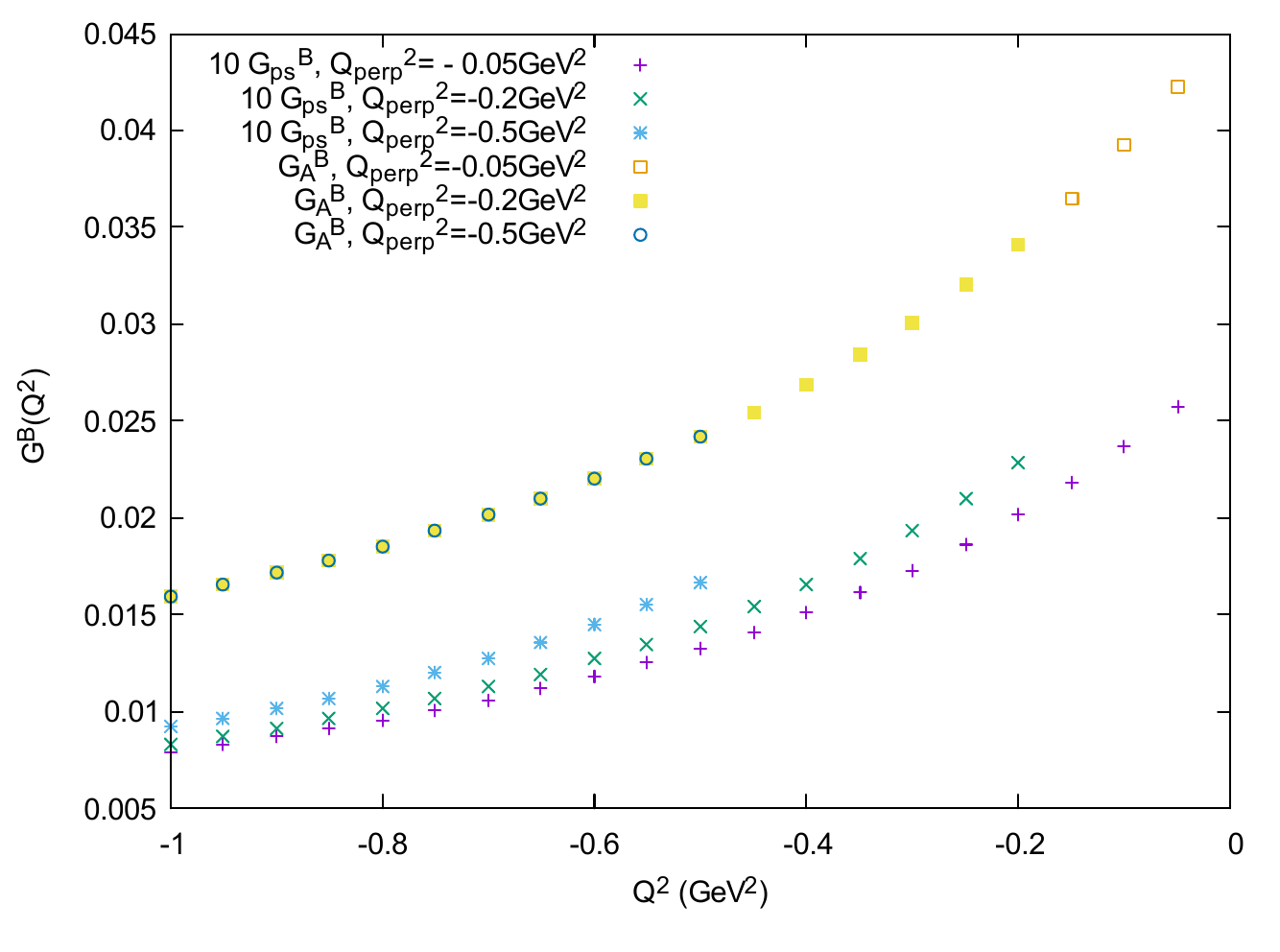}
\end{center}
 \caption{ 
\small
Ratio of leading magnetic field corrections to the pseudoscalar and axial form factors
$ 10 \times \beta^2 F_{ps}^B(Q^2)/F_{ps}(Q^2)$  and
$\beta F_{A}^B(Q^2)/F_{A}(Q^2)$  in the pion space-like region with $K^2 = M_q^2$.
}
\label{fig:FF}
\end{figure}

The  leading magnetic field corrections to the corresponding form factors can be written as
\begin{eqnarray}  
\label{FF-B}
\left[ F_{ps}^{B}   (Q , K) \; ; \;  F_{ps}^{B,a}   (Q , K) \right]
&=&   
i  C_{PS}^B    {M^*}^4
  \; 
 \int \frac{ d^4 k}{(2\pi)^4}   
\frac{ \left[
 -  k \cdot (k +Q) + M^2 
 \right) ; \left( -  k_\perp \cdot ( k_\perp + Q_\perp) \right]
}{[ k^2 - M^2
 ]^2 [  (k+Q)^2 - M^2]^2} 
 \frac{R_L(-k-K)}{K_g}
\nonumber
\\
\label{FF-ani}
F_{A}^{B}  (Q , K)
&=& i C_{PS}^B   F   {M}^3 \frac{g_{\mu\nu}}{10}
\int  \frac{ d^4 k}{(2\pi)^4}   
 \frac{  1
}{
[ k^2 - M^2 ] [  (k+Q)^2 - M^2 ]^2
}  \frac{ \bar{R}^{\mu\nu}(-k-K)}{K_g},
\end{eqnarray}
where for the pseudoscalar coupling
$R (-k)= 3 R_T (-k) + R_L (-k)$, and for the axial coupling $\bar{R}^{\mu\nu} (-k)  
=  g^{\mu\nu} (R_T (-k) +R_L (-k) ) + 
2 \frac{k^{\mu} k^{\nu}}{k^2} (R_T (-k)  - R_L )$, where $R(-k)$ is the effective gluon propagator (EGP), with a normalization $K_g$ discussed below, and the constant 
$C_{PS}^B  = 8 N_c \alpha K_g$ was defined.
Hereby, an effective confining gluon propagator inspired in Ref.~\cite{cornwall} is considered. This leads to Dynamical Chiral Symmetry Breaking, given by
$R_L(k)  = R(k)  = 
\frac{ K_g}{ (k^2 - M_G^2 + i \epsilon)^2 }$,
where $K_g, M_G$ are a dimensionful normalization constant, that incorporates the running quark gluon coupling constant, and an effective (constant) mass, respectively.
The normalization of the effective gluon propagator was fixed by the pseudoscalar pion coupling constant in the vacuum, such that it reproduces the phenomenological value $G_{ps} \simeq 13$~\cite{Schindler2007} for a reasonably large gluon effective mass $M_G=0.5$ GeV~\cite{Braghin2024}. With this EGP, it is possible to carry out overall analytical calculations, yielding form factors free of UV divergences, in spite of the need of a renormalization, that settles the scale of $K_g$. Besides that, it provides numerical results for mesons-constituent quark form factors that are similar to results obtained using other effective gluon propagator 
extracted from SDE. These diagrams are in the class of triangle diagrams analyzed in~\cite{Ayala2015a} for which the Schwinger phase does not contribute. In Fig.~\eqref{fig:FF} the ratios of the magnetic field correction to the pseudoscalar and axial form factors, with respect to their values in the vacuum, calculated by the same method~\cite{Braghin2019}, are presented in the low momentum space-like region for different values of $Q_{perp}^2$ that manifests in the anisotropic part of $G_{ps}^B$ where in the figure caption $\beta = (eB)/M^2$. The magnetic field correction $G_{ps}^B(Q^2) \equiv 
\beta^2 F_{ps}^B(Q^2)/ F_{ps}(Q^2)$ (multiplied by a factor 10) is of order $(eB)^2$  and smaller than  $ G_A^B(Q^2)  \equiv \beta F_A^B(Q^2)/F_A(Q^2)$. The anisotropy  of $G_{ps}^B$ for different values of $Q_{perp}^2$
was not exploited before, and it is small. 

\subsection{One-pion exchange potential}

The leading magnetic field effects will be considered for the Feynman diagrams of Figs.~\eqref{fig:FFMesB}  (f), (g) and (h). The magnetic field contributions to the pion-constituent quark (nucleon) coupling is accounted for by means of its pseudoscalar form factor, shown above, with the pion propagator obtained in Ref.~\cite{Ayala2005a}.
In the weak field limit, the leading terms of the meson exchange amplitude in momentum space can be written as
\begin{eqnarray}  \label{VYB}
\tilde{V}(Q, Q_z) 
&\simeq&
\left[g_{ps}   + \beta^2 F_{ps}^B   (Q, Q_\perp)    \right]
\left[ D_0  (Q)  + \beta^2 D_1^{B} (Q, Q_\perp)  \right]
\left[ g_{ps}   + \beta^2 F_{ps}^B  (Q, Q_\perp)     \right]\nonumber\\
&=&
V_0 (Q^2) + V_{\pi}^B (Q,Q_z) + V_{FF}^B (Q^2,Q_z) +...
\end{eqnarray}
where it has been assumed $K^2=M^2$, $D_0(Q^2)$ is the pion propagator in vacuum
and $D_1^B(Q,Q_\perp)$ the  leading magnetic field correction
that contains an anisotropy expressed as a dependence on $Q_\perp$. The first term, $V_0(Q^2)$, yields the (isotropic) Yukawa potential, 
and the two types of corrections considered here are due to the pion propagator
($V_{\pi}^B (Q,,Q_z)$) and to the form factor ($V_{FF}^B (Q,Q_z)$), respectively, 
which contain anisotropic terms expressed as a dependence on $Q_z$. Due to the linearization above, these two terms add to each other. The Fourier
transform of the four-point Feynman diagrams amplitude,
with off-shell meson propagation, $Q^2= -\boldsymbol{Q}^2$,
can be performed semi-analytically and is given by
\begin{eqnarray}
V(R,R_z) &=&
\int \frac{ d^3 k}{(2\pi)^3} 
e^{- i \boldsymbol{Q} \cdot  \boldsymbol{R}}  \;
 \left[
V_0 (Q^2) + V_{\pi}^B (Q,Q_z) + V_{FF}^B (Q,Q_z) + ...
\right], 
\end{eqnarray}
where, due to the Fourier transform,
the anisotropic can be written
in terms of the distance between two sources and
the longitudinal distance  between sources along 
the magnetic field direction $R_z$.
\begin{figure}[t!]
\begin{center}
\includegraphics[width=100mm]{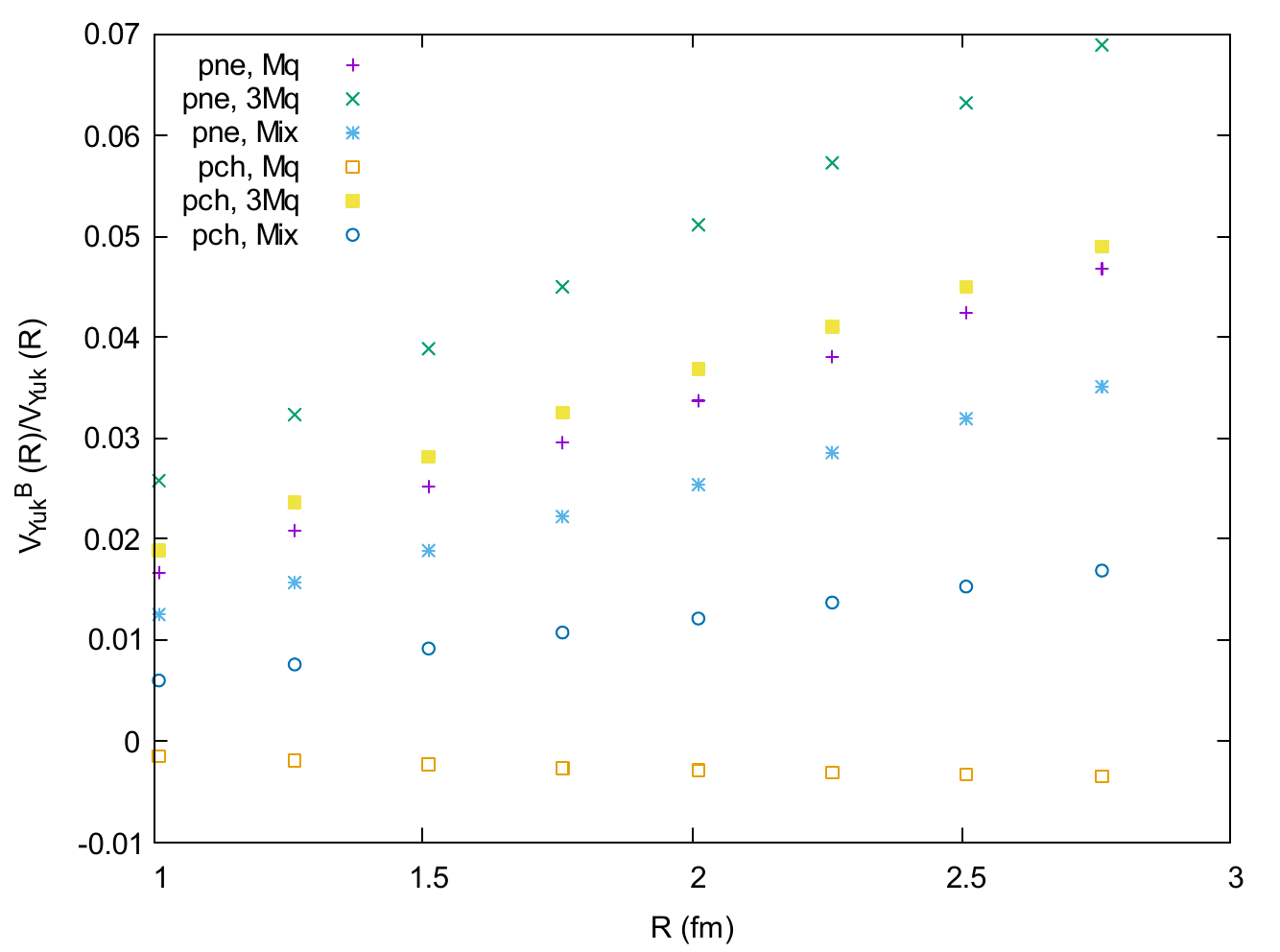}
\end{center}
\caption{\small
Ratio of magnetic field corrections to the Yukawa potential as a function of distance $R$, for the neutral and charged pions with corresponding magnetic 
field corrections to their masses and three different constituent quark masses as described in the 
text.
(eB) = 0.1 $M^2 \simeq$  0.01 GeV$^2$ 
and
$R_z=1$ fm.
}
\label{fig:YUK}
\end{figure}

The (semi)analytical result of this Fourier transformation was exhibited in~\cite{Braghin2024} and it will not be written here. There are mostly terms with two residual integrations in two Feynman
parameters to be performed numerically, although some anisotropic terms need a numerical integration in the pion transversal momentum $Q_{\perp}$. Overall, the anisotropies in all contributions are very small.
In Fig.~\eqref{fig:YUK} the resulting ratio of the  magnetic field corrections to the Yukawa potential with respect to the Yukawa potential
is presented for $R_z=1$ fm for the anisotropic terms of the potential. The anisotropy noted   for the pseudoscalar form factor in Fig.~\eqref{fig:FF} is washed out in the calculation of the potential such that, by 
fixing $R_z\neq 1$ fm, the difference with respect to the results with another value
is hardly noted in this figure. It can be seen that results for neutral (pne) and charged  (pch) pions are non-degenerate, and there are  two different reasons for that: First, the magnetic field contributions for the  neutral (charged) pions  in the range analyzed $(eB) \sim 0.01$ GeV$^2$, are of the order of $-20$ MeV ($+20$ MeV)~\cite{LQCDBfield1,LQCDBfield2}.
Second, Eq.~\eqref{Lcoupling} showed that the neutral and charged pion form factors have different coefficients $c_i$. Three different cases for the constituent quark masses were considered. First, a typical constituent  quark mass with a correction due to the magnetic field $M = (M_q + 0.02) GeV = (0.35 + 0.02)$ GeV~\cite{NJLB3}. Second, a larger mass, as if the constituent quark carried all the nucleon mass, of order $M \sim 3 M_q$. The overall contributions are amplified for large quark masses due to the structure of the magnetic field contributions to the form factor.
Finally, a mixed case in which the normalization of the gluon propagator
was considered to reproduce $G_{ps} = 13$ with $M=M_q$, as described above. However, in  the magnetic field correction to the Yukawa potential, the quark mass was considered to be three times larger so that the role of the form factors (the part that depends on the 
constituent quark mass) was reduced, in agreement with expectations that an infinite quark mass leads to point-like interaction.
The leading magnetic field contributions for the case of neutral pion exchange
leads to an increase of the strength of the potential with distance 
and consequently a larger range of the  interaction. For the case of the charged pion, results are more ambiguous since they strongly depend on the (considered) value of the constituent quark mass. This is mostly due to a possible cancellation of different contributions from different effects, since the charged pion form factor receives a contribution with opposite sign from 
the one of the neutral pion.

\subsection{Discussion}

The limit of constant magnetic fields,  weak with respect to a hadronic mass scale such 
as the pion mass ($m_\pi$), or the constituent quark mass ($M$), was shown to 
make possible to develop analytical or semi-analytical calculations for intricate hadron processes from the point of view of the fundamental  constituents, quarks and gluons. The strength of such magnetic fields is nearly in the range 
expected to appear in relativistic heavy-ion collisions and magnetars. For that, a leading term of the QCD (quark) effective action, that is the quark-antiquark interaction mediated by one (non-perturbative) gluon exchange,
is shown to produce, dynamically, meson-constituent quark 
interactions  described by one loop form factors. The form factor was calculated and used in the one-loop level, 
being akin to the rainbow ladder approximation of the SDE.
The pion pseudoscalar and axial form factors receive corrections with (leading) strengths
that are of order $(eB/M^2)^2$ and $(eB/M^2)$, respectively.
Whereas the pseudoscalar form factor presents some sizeable anisotropy, manifested in the dependence on the perpendicular component of the pion momentum, 
$Q_{\perp}$, the leading anisotropy of the axial form factor is mostly present in the 
shape of the interaction. Each of these pion couplings occur in color-singlet quark currents that, nevertheless, were shown to be extracted from a color-quark current interaction.
The pseudoscalar form factor was employed in the calculation of magnetic field corrections to the one-pion exchange potential
which requires an off-shell pion.
The limit of weak magnetic field allowed to linearize the potential in two different contributions: one from the magnetic field dependent meson propagator and one from  the
form factor, since both are of order of magnitude $(eB/M^2)^2$.
If a deuteron is formed in a region with such weak magnetic field, when leaving this 
region it would go into an excited state or even disintegrate. It should be possible to detect both processes by means of either the excitation emission spectrum or the outgoing neutron and proton.
The CQM offers a framework for the prediction of observable  hadron dynamics that must be tested and whose precise role must be understood in the more complete picture of baryon
interactions. The use of the meson form factor in the meson exchange process also, hopefully, will make possible to probe nucleon and meson quark and gluon structures as participants of the nucleon and nuclear interactions. A more complete analysis of all the ingredients of these calculations is to be done
with more complete account of the dependence on the magnetic field, for example for 
gluon propagator and running coupling constant. It is therefore possible to use magnetic fields (even relatively weak ones) as probes of the quark-gluon structure of hadrons
and of the nucleon (and consequently nuclear) interaction.

\section{
Magnetic fields in heavy-ion collisions: a catalyst of photo-production via gluon fusion}\label{Tejeda}

\subsection{Introduction}\label{sec:intro}

Most of the photons produced in heavy-ion collisions are \textit{decay} photons, a by-product of the fireball expansion, cooling and eventual hadronization. The rest of the photons produced in heavy-ion collisions are the so called \textit{direct} photons: \textit{prompt} photons, just as the ones produced in proton-proton collisions, so we expect they scale as the number of binary collisions and \textit{non-prompt} photons which can come from pre-equilibrium, jet-medium interactions or they can be purely thermal from the QGP phase or the hadron gas phase. We have learned from recent experiments that direct photons produced in heavy-ion collisions, have a large elliptic flow coefficient $v_2$, in some cases as large as that of hadrons~\cite{PHENIX:2011oxq,ALICE:2018dti,PHENIX:2015igl}.

The evidence from data could be interpreted as direct photons being produced in the later stages of the collision, since it is then when flow is already leaving an imprint in hadrons. But, looking closely at the photon yields, we can see that there is a large thermal component that dominates over the
prompt one, for low values of $p_T$, the transverse momentum. Now the $p_T$ dependence of $v_2$ seems to confirm that most of the direct photons are really \textit{early} photons, since $v_2$ for large $p_T$ is consistent with zero~\cite{David:2019wpt}. These observations initiated a discussion in the community a few years ago, that was named as the \textit{direct photon puzzle}. 

On top of this puzzle, the excess of low $p_T$ photons with respect to
known sources, as reported by the PHENIX and the STAR Collaborations, has eluded explanations and models that work well for other electromagnetic probes. 
In summary, these recent data analysis show that the yield of these low $p_T$ photons scales with a given power of the number of binary collisions,
both in large and small colliding systems, which suggests that
the source of these photons is similar for different colliding
species and beam energies. The data also shows tension between the yields measured by PHENIX and STAR, so further measurements of photon-production in the low energy regime will complement these and will push the discussion towards elucidating the source of these soft photons or making the effect disappear altogether~\cite{PHENIX:2014nkk,ALICE:2015xmh,STAR:2016use,PHENIX:2018che,PHENIX:2022qfp}.  

\subsection{New photo-production channel available in pre-equilibrium stage}

\begin{figure}
	\centering\includegraphics[scale=0.3]{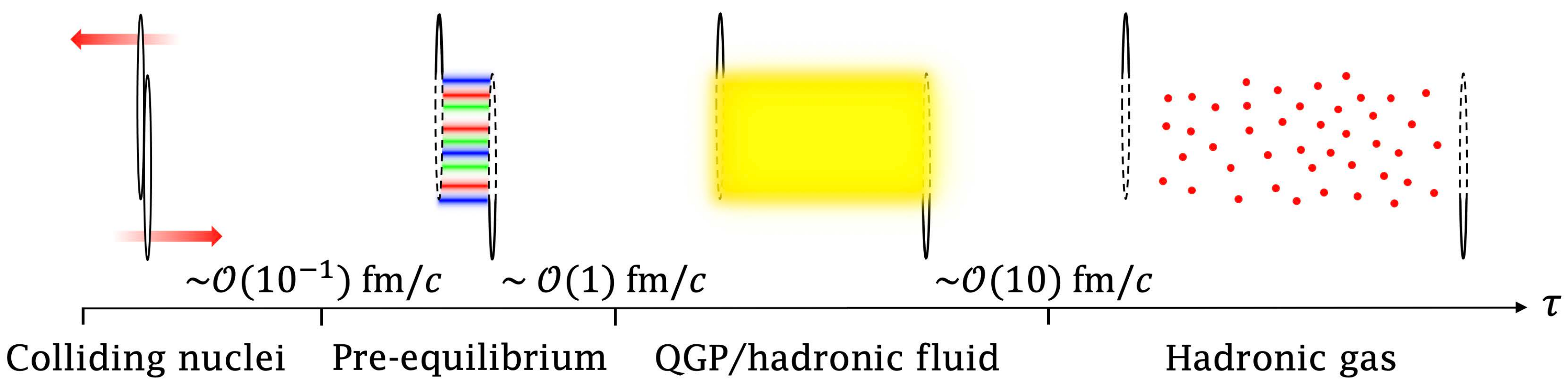}
	\caption{Distinct epochs in a heavy-ion collision and its evolution~\cite{Monnai:2022hfs}.} \label{f1}
\end{figure}

\begin{figure}
	\centering\includegraphics[scale=1.2]{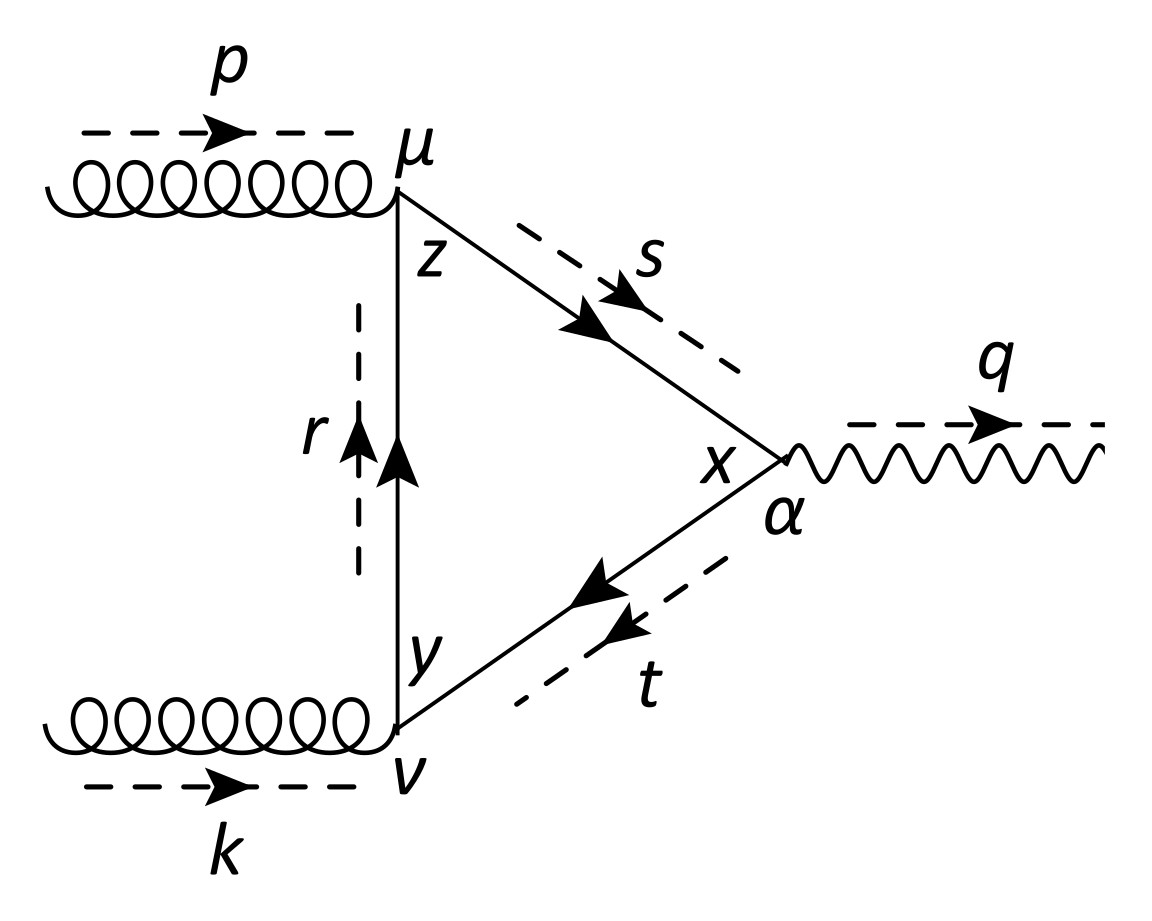}
	\caption{New photo-production channel available in pre-equilibrium stage: gluon fusion into a photon, mediated by a quark loop in the presence of an external magnetic field $B$.} \label{f2}
\end{figure}

\noindent Figure~\ref{f1} shows a diagram of the system created in heavy-ion collisions and its evolution in time. In the pre-equilibrium stage, there are large number of gluons (glasma) available, unlike quarks which are suppressed due to the parametric proportionality in $\alpha_s$. 

In Refs.~\cite{Ayala:2017vex,Ayala:2019jey,Ayala:2022zhu}, we have reported how fusion and splitting processes involving quarks and gluons at this stage, can give rise to an excess (over conventional sources) of photons. The posited source of this excess of photons is a new production channel via gluon fusion which is possible by the presence of an intense external magnetic field $B$. Figure~\ref{f2} shows the Feynman diagram for this process where all quark propagators include the effect of the external magnetic field.

The magnetic field acts as a catalyst of the newly open channel that would otherwise not be available, since in the $B \to 0$ limit we recover Furry’s theorem and QED and QCD are back to having charge conjugation symmetry. Altogether the idea we put forward is to focus on processes that are catalysed by these external magnetic fields in a heavy-ion collision epoch where there is a large abundance of soft pre-equilibrium gluons and test weather this allows an enhancement of the photon yield and a large $v_2$.

An immediate consequence of the presence of an external magnetic field in this pre-equilibrium stage is the breaking of Lorentz invariance, so calculation of observables in space-time have to be done with tensorial structures that are now differentiated: parallel and perpendicular with respect the external magnetic field.

Now, there is an initial limitation to this approach: the presence of an intense external magnetic field means that $qB$ is the dominant scale, in particular it is larger than the loop momenta, so the analysis of results should be constrained to the low photon momenta regime.\\

\subsubsection{Estimation and simulation of magnetic fields created in heavy-ion collisions}\label{sec:mag}

\begin{figure}
	\centering\includegraphics[scale=0.4]{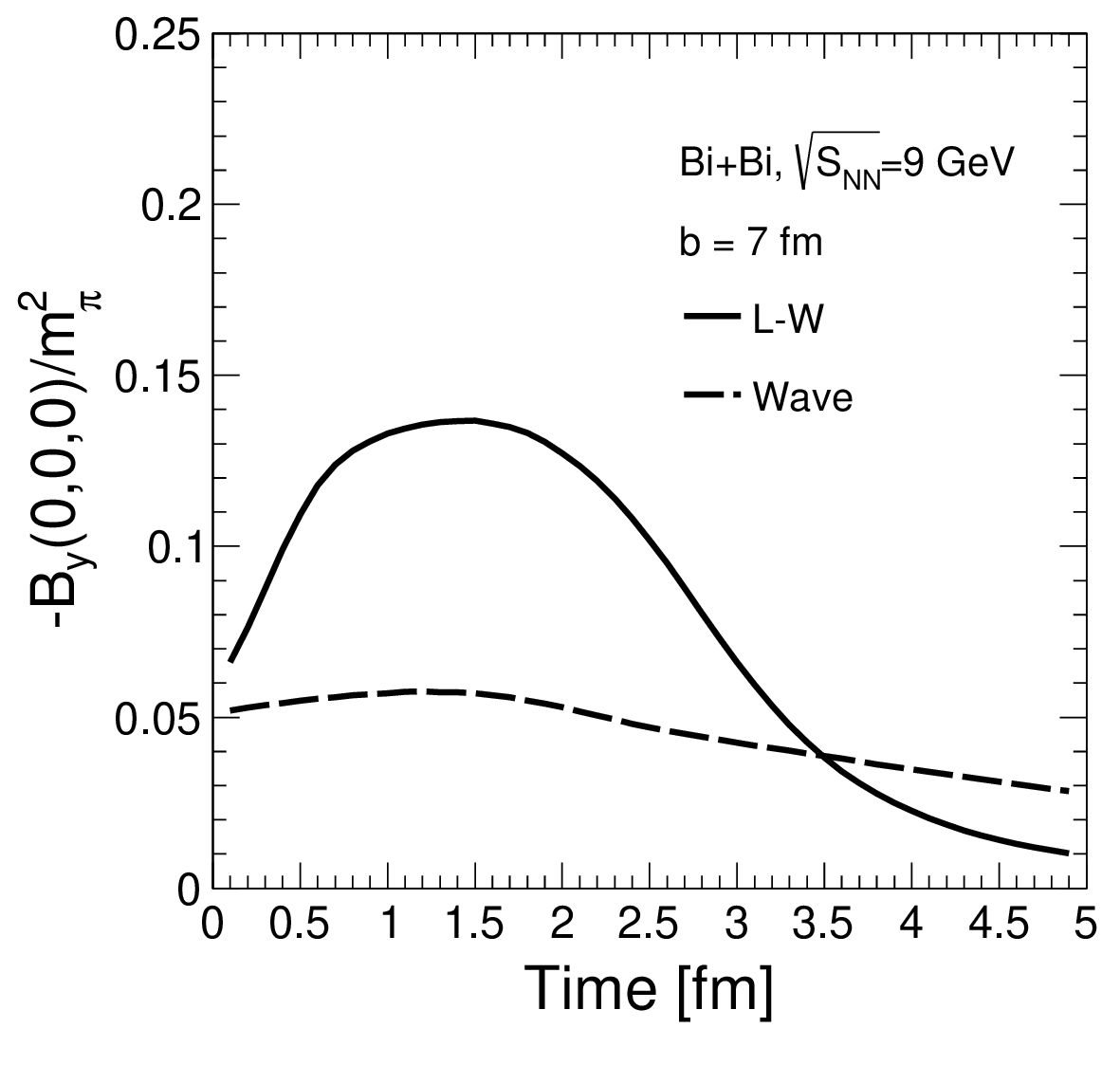}\includegraphics[scale=0.4]{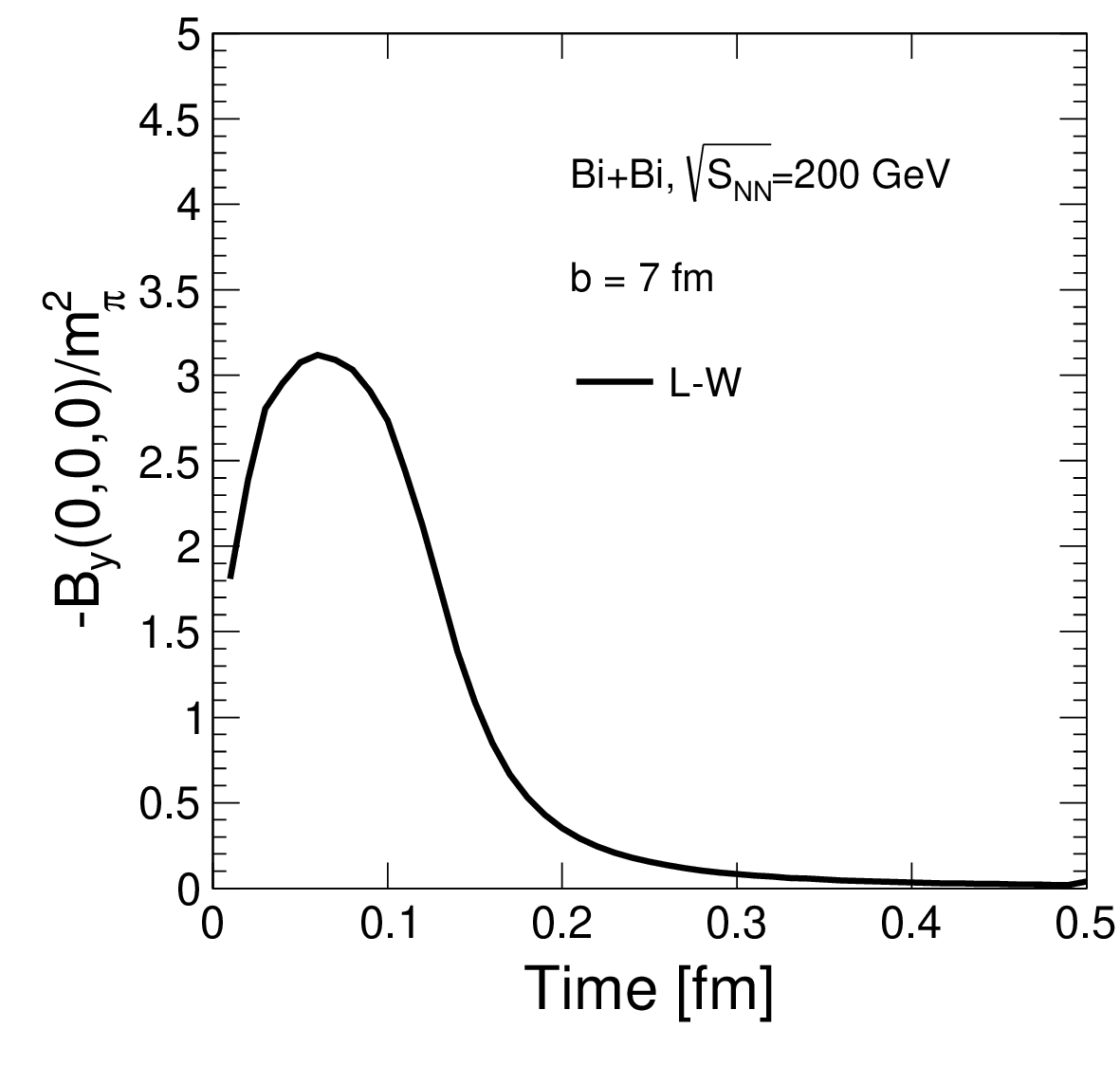}
	\caption{Evolution in time of the average dominant component of the magnetic field pulse, that would be created in heavy-ion collisions~\cite{DGNT}.}\label{f3}
\end{figure}	

\begin{figure}
	\centering\includegraphics[scale=0.3]{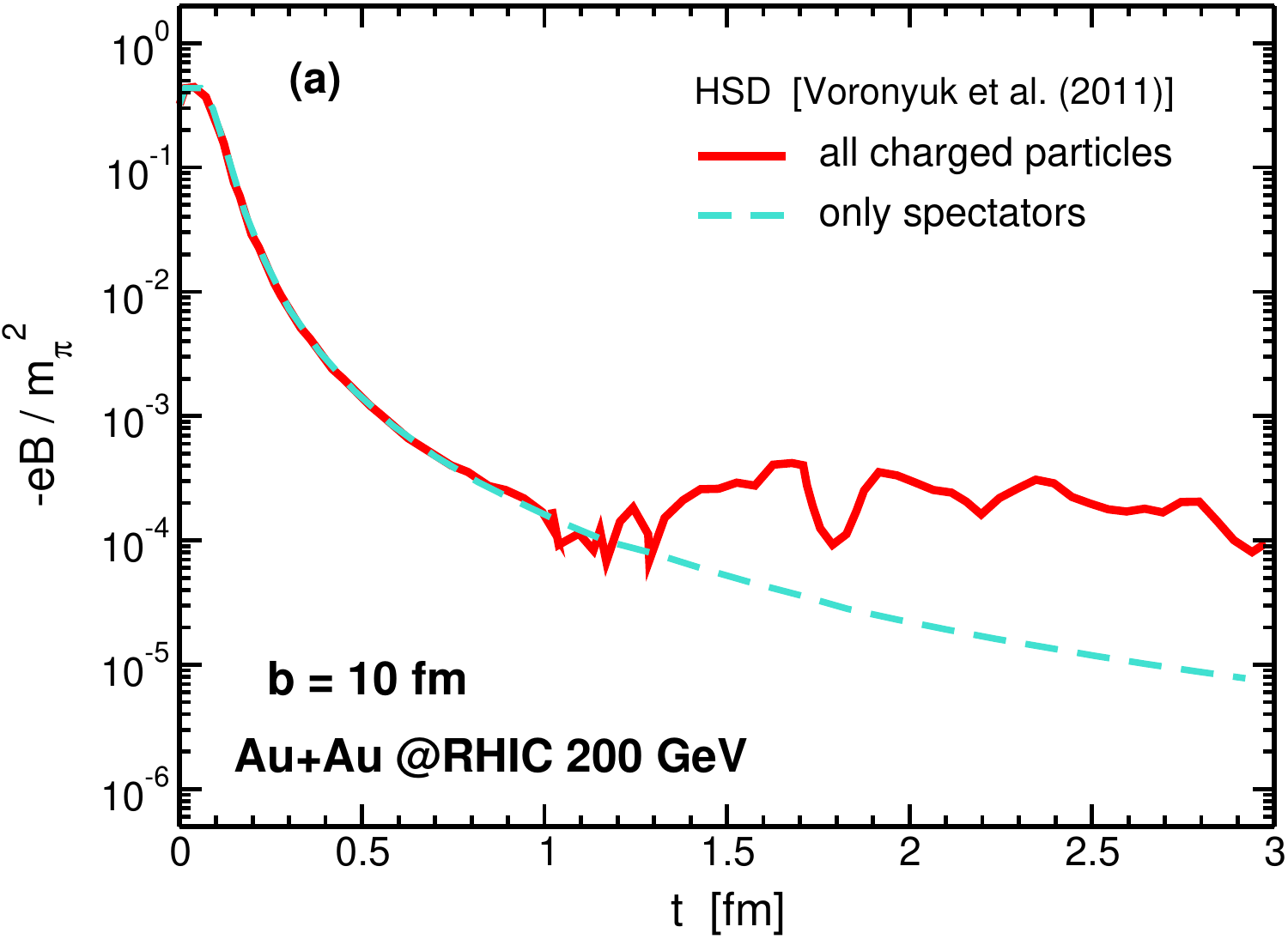}\includegraphics[scale=0.3]{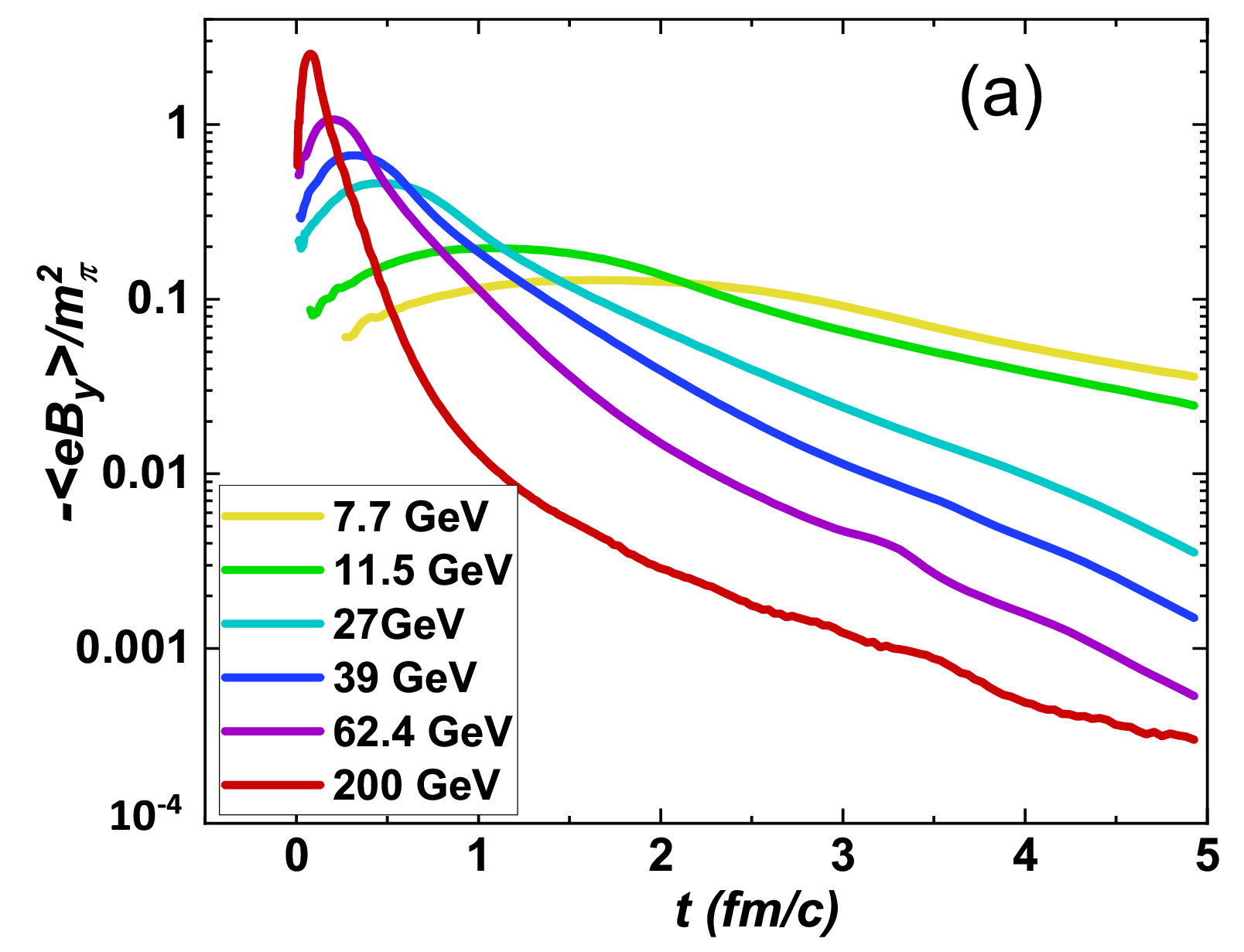}
	\caption{Evolution in time of the average dominant component of the magnetic field pulse, that would be created under diverse heavy-ion collision conditions (colliding energy, impact parameter or centrality), as reported in Refs.~\cite{Oliva:2020mfr,Siddique:2021smf}.}\label{f4}
\end{figure}

The classical approach to estimate the magnetic field $\mathbf{B}(\mathbf{r}_i,t)$ produced by non-accelerated charges moving along the beam direction in a heavy-ion collision event, can be done using the Liénard-Wiechert potential as ~\cite{Landau:1975pou}	
			\begin{equation}
			e\boldsymbol{\mathbf{B}}(\mathbf{r}_i,t) = \alpha_{em} \sum_{j}  
			\frac{(1-v_j^{2})~\boldsymbol{\mathbf{v}_j}\times \boldsymbol{\mathbf{R}_j}}{ R_j^{3}\left[1-\frac{(\boldsymbol{\mathbf{v}_i}\times\boldsymbol{\mathbf{R}_j})^{2}}{R_j^{2}}\right]^{3/2}},\nonumber
			\end{equation}
			where $\boldsymbol{\mathbf{R}}_j = \boldsymbol{\mathbf{r}} - \boldsymbol{\mathbf{r}}_j(t)$, $\boldsymbol{\mathbf{x}}_j(t)$ is the position of the $j$-th charge moving with velocity $\boldsymbol{\mathbf{v}}_j$ and the sum runs over charged particles in each event. 
   Figure~\ref{f3} shows the magnetic field pulse using the Liénard-Wiechert approach for Bi+Bi collisions at both \snn{9} and \snn{200} with an impact parameter $b=7$ fm. We can appreciate how the intensity of the pulse is of the order of $3m_\pi^2$ at RHIC energies and about $m_\pi^2/10$ for smaller collision energies. This is contrast with the duration of the pulse which is approximately $3$ fm in the low energy collision and approximately $1/10$ fm for the RHIC energy collision. Figure~\ref{f4} shows similar results reported in some of the recent reviews devoted to the topic of understanding the emergence and effects on observables of electromagnetic fields produced in heavy-ion collisions~\cite{Oliva:2020mfr,Siddique:2021smf,Giacalone:2021bzr}.

Clearly the interplay of the height and width of the magnetic field is key to have in mind when implementing phenomenological test to look for such effects. Using simulations, we have been able to include the magnetic field effects into relevant observables for photo-production through gluon fusion so that effects of participant or spectator nucleons from the centrality of the collision, can also be taken into account~\cite{Ayala:2019jey,Ayala:2017vex,ayala2022anisotropic}.   

\subsubsection{Magnetic-field induced gluon-fusion to produce photons in the early stages of the collision}
	
The lowest order gluon fusion process comes from an amplitude made out of a quark triangle diagram with two gluons and one photon attached each at one of the vertices of the triangle, as shown in Figure~\ref{f2}. The over occupied gluon state can be modeled as a quasi-particles system and perturbative methods are applicable~\cite{McLerran:2014hza}. In this context, the coordinate representation of the quark Schwinger propagator is
\begin{figure}
	\centering\includegraphics[scale=0.4]{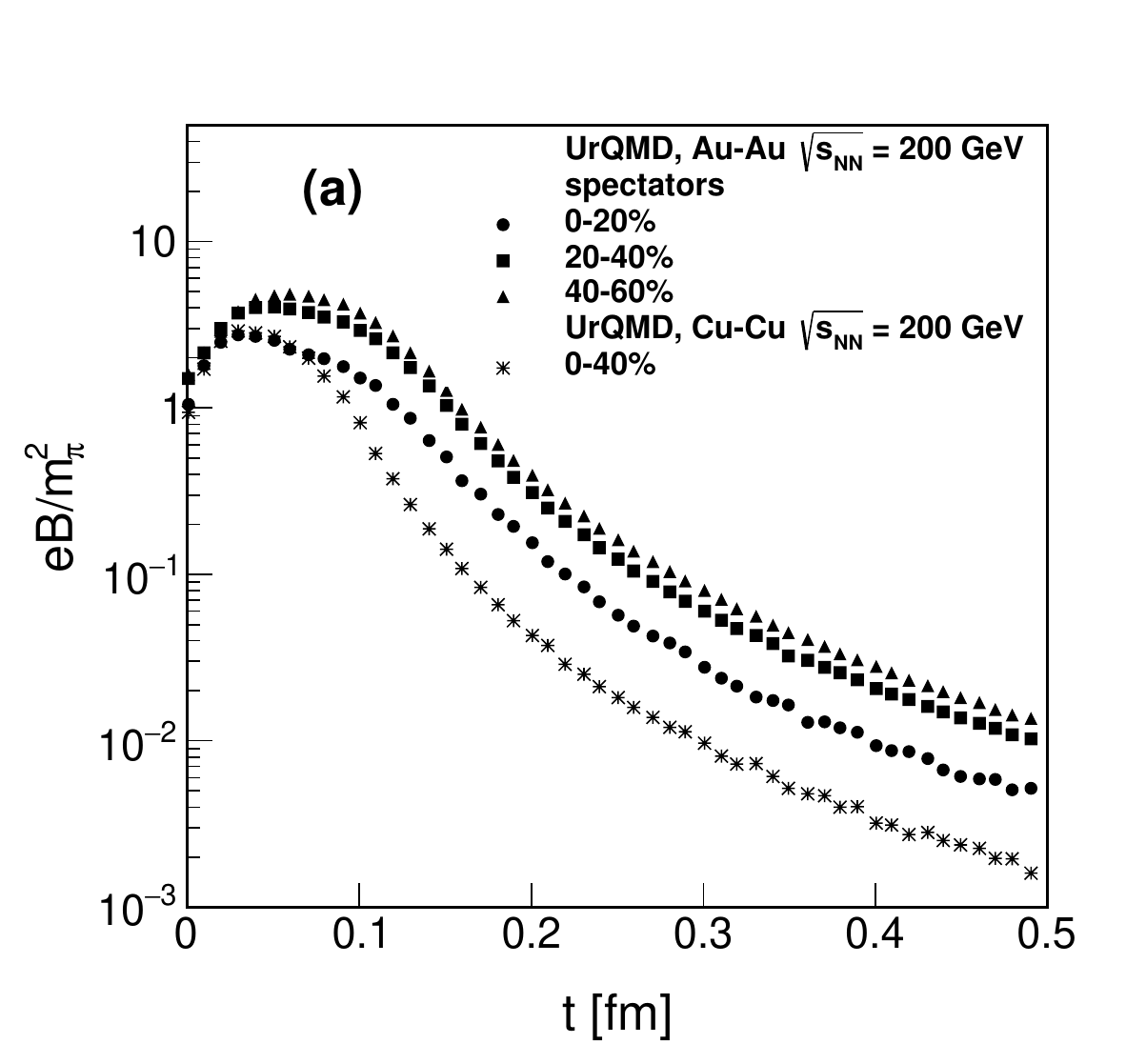}\includegraphics[scale=0.4]{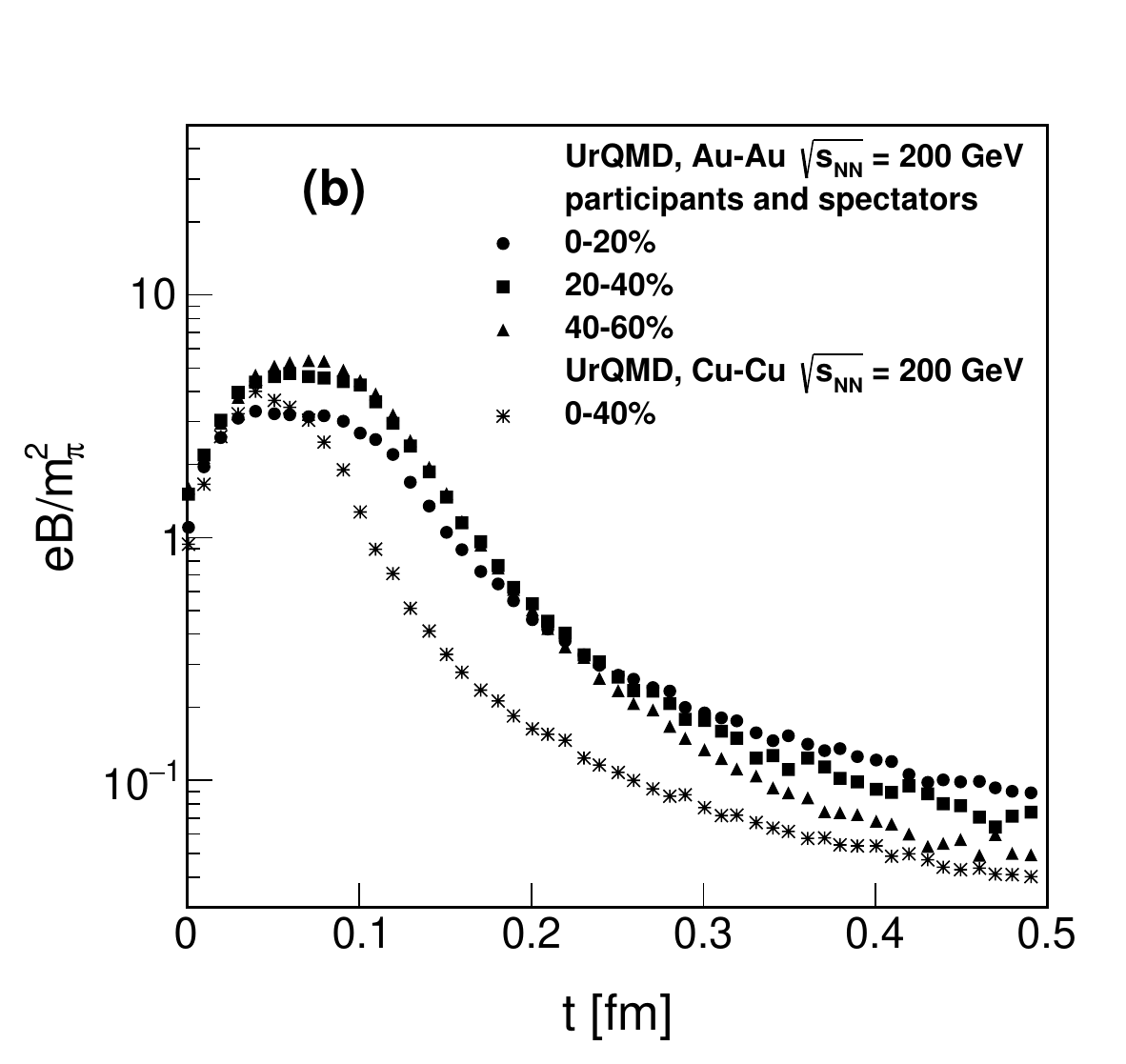}
	\caption{Mean magnetic field strength produced by (a) spectators and (b) spectators and participants at the middle of the interaction region as a function of time for three centrality classes 0-20\%, 20-40\% and 40-60\% in
		Au+Au collisions and one centrality class 0-40\% in
		Cu+Cu collisions at \snn{200}, as reported in Refs.~\cite{Ayala:2017vex,Ayala:2019jey}.}\label{f5}
\end{figure} 
\bea
S(x,x')=\Phi(x,x')\int\frac{d^4p}{\dpi^4}e^{-p\cdot(x-x')}S(p),
\label{propagator}
\eea
where $S(p)$ is the Fourier transform to momentum space of the translational invariant part of the propagator, given by
		\bea	iS(p)&=&\int_{0}^{\infty}\frac{d\tau}{\cos(\eB\tau)}e^{i\tau\left[p_{\parallel}^2-p_{\perp}^2\frac{\tan\left(\eB\tau\right)}{\eB\tau}-m_f^2+i\epsilon\right]} \\
		&\times&\Bigg{\{}\left[\cos(\eB\tau)+\gamma_1\gamma_2\sin(\eB\tau)\right](m_f+\slashed{p}_{\parallel}) - \frac{\slashed{p}_{\perp}}{\cos(\eB\tau)}\Bigg{\}} \nn ,
		\eea
		and $\Phi(x,x')$ is the phase factor with exclusive coordinate dependence
		\bea
		\Phi(x,x')=\exp\left\{ieq_f\int_{x'}^x d\xi^\mu\left[A_\mu+\frac{1}{2}F_{\mu\nu}(\xi-x')^\nu\right]\right\},
		\label{phasefactor}
		\eea
		with $eq_f$ the quark charge in units of $e$, the absolute value of the electron charge. We assume a constant magnetic field that points in the $z$-direction obtained from a vector potential $A^\mu$ in the symmetric gauge $A^\mu=\frac{B}{2}(0,-y,x,0)$. Note that as we announced previously, Lorentz invariance breaks and so from a four-momentum $p^\mu$, we define $p_{\perp}^\mu=(0,p_1,p_2,0)$, $p_{\parallel}=(p_0,0,0,p_3)$, $p_{\perp}^2=p_1^2+p_2^2$, $p_{\parallel}^2=p_0^2-p_3^2$, and therefore $p^2=p_{\parallel}^2-p_{\perp}^2$. Then we can write the expression for the amplitude for the gluon fusion channel $\widetilde{{\mathcal{M}}}_{gg\rightarrow\gamma}$ as
		\bea
		\widetilde{{\mathcal{M}}}_{gg\rightarrow\gamma}&=&-\int \!d^4xd^4yd^4z\int\!\frac{d^4r}{(2\pi)^4}
		\frac{d^4s}{(2\pi)^4}\frac{d^4t}{(2\pi)^4}
		e^{-it\cdot (y-x)}e^{-is\cdot (x-z)}e^{-ir\cdot (z-y)}e^{-ip\cdot z}e^{-ik\cdot y}e^{iq\cdot x}\nonumber\\
		&\times&
		\Big\{
		{\mbox{Tr}}\left[ ieq_f\gamma_\alpha iS_{ac}(s) ig\gamma_\mu t^c iS_{cd}(r) ig\gamma_\nu t^d iS_{da}(t) \right]
		+
		{\mbox{Tr}}\left[ ieq_f\gamma_\alpha iS_{ad}(t) ig\gamma_\nu t^d iS_{dc}(r) ig\gamma_\mu t^c iS_{ca}(s) \right]
		\Big\}
		\nonumber\\
		&\times&\Phi(x,y)\Phi(y,z)\Phi(z,x)\epsilon^\mu(\lambda_p)\epsilon^\nu(\lambda_k)\epsilon^{\alpha}(\lambda_q),
		\label{amplitude}
		\eea
		where $p$ and $k$ are the gluon and $q$ the photon four-mo\-men\-ta, $t^c, t^d$ are Gell-Mann matrices, and the polarization vectors for the gluons and the photon are $\epsilon^\mu(\lambda_p)$, $\epsilon^\nu(\lambda_k)$, $\epsilon^\alpha(\lambda_q)$, respectively. The Lorentz indices $\mu, \nu,\alpha$ and the space-time coordinates $x,y,z$ associated to each vertex, are also depicted in Fig.~\ref{f2}. The product of phase factors is
		$\Phi(x,y)\Phi(y,z)\Phi(z,x)=\exp\left\{i\frac{\eB}{2}\epsilon_{mj}(z-x)_m(x-y)_j\right\}$,
		where the indices $m,j=1,2$ and $\epsilon_{mj}$ is the Levi-Civita tensor. 
  
  If we pursue the idea that at the earliest times, when thermalization has not been achieved, the magnetic field is the dominant internal energy scale in the process. This means that in the absence of thermal effects, and with $|eq_fB|>>m_f^2$, we can work using the quark propagators in the lowest Landau level. Notice that the first non-vanishing term is where one of the quarks in the loop is in the first excited Landau level~\cite{Ayala:2017vex,Ayala:2019jey,ayala2022anisotropic}. \begin{figure}
    \begin{center}
	\includegraphics[scale=0.9]{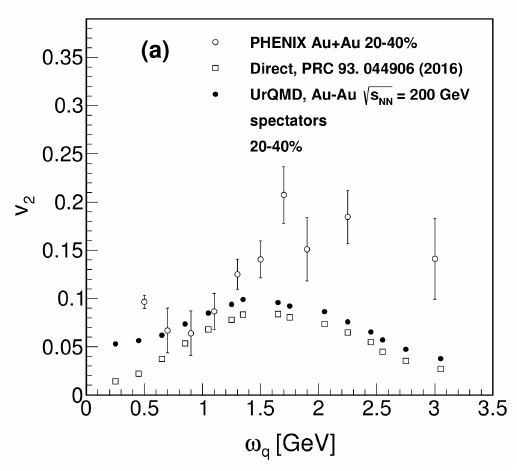}\includegraphics[scale=0.9]{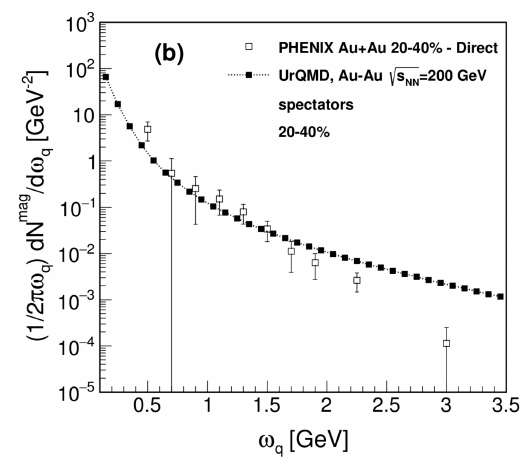}
    \end{center}
	\caption{Second harmonic coefficient and invariant photon spectra for Au+Au collisions at $\sqrt{s_{NN}}=200$ GeV with 20-40\% centrality events compared with PHENIX data, as reported in Ref.~\cite{Ayala:2019jey}.} \label{f6}
\end{figure}
Once we have the matrix elements for both gluon fusion and splitting we can build the invariant photon momentum distribution as follows
	\bea
	\omega_q\frac{dN^{\mbox{\tiny{mag}}}}{d^3q}&=&\frac{{\mathcal{V}}\Delta \tau}{2(2\pi)^3}
	\int\frac{d^3p}{\dpi^32\omega_p}\int\frac{d^3k}{\dpi^32\omega_k}\dpi^4\Bigg{\{}\delta^{(4)}\left(q-k-p\right)n(\omega_p)n(\omega_k){\overline{\sum_{\mbox{\small{c,p,f}}}}}|{\mathcal{M}}_
	{gg\rightarrow\gamma}|^2 \nonumber \\
	&& +\delta^{(4)}\left(q+k-p\right)n(\omega_p)\left[1+n(\omega_k)\right]{\overline{\sum_{\mbox{\small{c,p,f}}}}}|{\mathcal{M}}_
	{g\rightarrow g\gamma}|^2\Bigg{\}} 
	\label{invdist}
	\eea
where $\mathcal{V}\Delta\tau$ is the space-time volume where the reaction takes place, spatial volume of the nuclear overlap region ${\mathcal{V}}(t)$ at time $t$ and the time interval $\Delta\tau$ is where the magnetic field is taken as having a constant intensity $B(t)$ as shown in Fig.~\ref{f5}. As we commented back in Sec.~\ref{sec:mag}, the overall lifetime $\Delta{\mathcal{T}}$ can be estimated calculating the duration of the magnetic pulse using simulations. Once we calculate the invariant momentum distribution we can extract the second harmonic coefficient $v_2$ taking into account the contribution from direct photons reported in the literature~\cite{Paquet:2015lta} as a weighted average 
\bea
v_2(\omega_q)&=&\frac{
	\sum_{i=1}^m \left[
	\frac{dN}{d\omega_q}
	\right]_i 
	[v_2^{\text{mag}}(\omega_q)]_i
	+
	\frac{dN^{\mbox{\tiny{direct}}}}{d\omega_q}(\omega_q)\
	v_2^{\mbox{\tiny{direct}}}(\omega_q)} 
{\sum_{i=1}^m \left[
	\frac{dN}{d\omega_q}
	\right]_i 
	+ 
	\frac{dN^{\mbox{\tiny{direct}}}}{d\omega_q}(\omega_q)}
\eea

We can finally compare with data and analyze the effectiveness of our model in explaining both the excess of low $p_T$ photon yield and large $v_2$. Figure~\ref{f6} shows the second harmonic coefficient and invariant photon spectra for Au+Au collisions at \snn{200} with 20-40\% centrality events compared with PHENIX data, as reported in Ref.~\cite{Ayala:2019jey}. Our work indicates that the magnetic field contribution improves the agreement with experimental data for the low part of the spectrum helping to describe the rise of $v_2$ as the photon energy decreases, but requires improvements as a systematic effect for the behavior of flow at larger $p_T$. At this point, the limitation is the kinematic hierarchy imposed by us, when we claim that the magnetic field drives the dynamics and dominates the loop momenta in gluon fusion. So we embarked on an approach where we can improve our calculation and extend its range of applicability.

\subsubsection{Recent improvements}
	
\noindent In order to go beyond the access only to low $p_T$ photons and to push back on implementing the hierarchy between the magnetic field and the loop momenta as $2|eq_fB|\gg t_\p^2,\ s_\p^2,\ r_\p^2$, we pursue a calculation where now we do a fully strong $\boldsymbol{B}$-field approximation, as reported in Ref.~\cite{ayala2022anisotropic}. The physical picture that emerges is the following: under a strong magnetic field two of the vector particles occupy parallel polarization states~\cite{Hattori:2017xoo,Fukushima:2011nu}. So, when the vertex involves a third vector particle, invariance under charge conjugation and conservation of angular momentum require that its polarization is transverse. In a practical sense, using the Schwinger propagators this time, one of the quarks that make up the loop needs to be placed not in the lowest Landau level but instead in the first Landau level. There is plenty of work in the literature where the tensor and analytic structure of magnetized vertices is well understood, so that properties and identities need to be satisfied under controlled conditions of the external strong magnetic field~\cite{Hattori:2022hyo,Hattori:2017xoo,Shovkovy:2022bnd,Wang:2022jxx}. In the approximation where the magnetic field is the largest of the kinematical energy (squared) variables, the tensor structure can be expressed as $\Gamma^{\mu\nu\alpha}_{ab}=\delta_{ab}\Gamma^{\mu\nu\alpha}$, where 
\bea
\Gamma^{\mu\nu\alpha}&\equiv&\sum_{n=1}^3 \Gamma_n(\omega_q,\omega_k,q^2)\Gamma^{\mu\nu\alpha}_n\nn\\
&=&\Gamma_1(\omega_q,\omega_k,q^2)\frac{\epsilon_{ij} q_\perp^i g_\perp^{j\mu}}{\sqrt{q_\perp^2}}\left(g_\parallel^{\nu\alpha}-\frac{q_\parallel^\nu q_\parallel^\alpha}{q_\parallel^2}\right)
+\Gamma_2(\omega_q,\omega_k,q^2)\frac{\epsilon_{ij} q_\perp^i g_\perp^{j\nu}}{\sqrt{q_\perp^2}}\left(g_\parallel^{\mu\alpha}-\frac{q_\parallel^\mu q_\parallel^\alpha}{q_\parallel^2}\right)\nn\\
&&+\Gamma_3(\omega_q,\omega_k,q^2)\frac{\epsilon_{ij} q_\perp^i g_\perp^{j\alpha}}{\sqrt{q_\perp^2}}\left(g_\parallel^{\mu\nu}-\frac{q_\parallel^\mu q_\parallel^\nu}{q_\parallel^2}\right)
\label{vertex}
\eea
where $\Gamma_n(\omega_q,\omega_k,q^2)$, $n=1,2,3$, are scalar coefficients and $\epsilon_{ij}$ is the Levy-Civita symbol in the transverse components: $\epsilon_{12}=-\epsilon_{21}=1$. The contributing Feynman diagrams, obtained when placing two fermion propagators in the LLL and the other in the 1LL, after integration of the configuration space variables, the vertex is given by
\begin{equation}
\Gamma^{\mu\nu\alpha}_{ab}=-\delta^{(4)}(q-k-p)\text{Tr}\left[t_at_b\right]\frac{8\pi^4 q_f g^2}{\eB}\qp^2e^{f\left(p_{\perp}, k_{\perp}\right)}\sum_{i=1}^3D_{i}^{\mu\nu\alpha}.
\end{equation}
We calculate $\Gamma_n(\omega_q,\omega_k,q^2)$ for $n=1,2,3$ by projecting onto the basis
\bea
\left\{v_\perp^\mu\Pi_\parallel^{\nu\alpha},v_\perp^\nu\Pi_\parallel^{\mu\alpha}, v_\perp^\alpha\Pi_\parallel^{\mu\nu}\right\}.
\label{basis}
\eea
Using these results, we can go back to the vertex and collect the contributions that correspond to each coefficient $\Gamma_n(\omega_q,\omega_k,q^2)$. The projection with $v_\perp^\mu \Pi_\parallel^{\nu\alpha}$ contributes to 
\bea
\Gamma_n&\equiv&\frac{8\pi^4 q_f g^2}{\eB}e^{f\left(p_{\perp}, k_{\perp}\right)}\left|\qt\right|\widetilde{\Gamma}_n(\omega_p,\omega_q,\theta).
\label{eq:Gamma1}
\eea
As we reported in Ref.~\cite{ayala2022anisotropic}, we perform the calculation of $\sum|\widetilde{\Gamma}_n|^2$ and with a thorough numerical analysis we find that it is larger for a photon propagation within the reaction plane ($\theta = \pi/2$) and that for $\theta=\pi/2$ it is also larger for smaller field intensities. We also note that $|\qt|e^{f\left(p_{\perp}, k_{\perp}\right)}$ also depends on $\theta$ and peaks at small angles for large gluon energies, but here the square amplitude is highly suppressed close to the reaction plane. So for small gluon energies and/or a large field strength, the pre-factor is dominated by emission angles close to the reaction plane and vanishes for $\theta=0$, which prevents photons from being emitted along the direction of the magnetic field. Since at pre-equilibrium the largest gluon abundance happens for small energies, a positive $v_2$ coefficient may be expected. The final steps to verify this result are work in progress.

\subsection{Concluding remarks}

Here we have reviewed the photon puzzle in the context of a new approach to understanding photon production at the earliest stages of a heavy-ion collision, where intense magnetic fields could be present. We summarize the results in the literature where we have computed the contribution to the photon yield and $v_2$ from gluon fusion by a magnetic field during the early stages of a relativistic heavy-ion collision, where there is a large gluon occupation number below the saturation scale. We note an agreement of the calculation with data in the lowest part of the spectrum: observed experimental fall between 0.5 and 1 GeV. Above 1 GeV the calculation overshoots the data spectrum and $v_2$ peaks for energy values $\sim\sqrt{eB}$. We further analyze how we have improved this approach by constructing a one-loop two-gluon and one-photon vertex where we fully include the effects of the magnetic field in the tensor structure: we construct the vertex as $\Pi_\parallel^{\mu\nu} \times \tilde{\epsilon}_\perp^{\alpha\beta}$ which is required to have components in the transverse plane with respect to the $\boldsymbol{B}$-field and we place two of the loop quarks in the lowest Landau level and the other one in the first Landau level. We are in the process of obtaining and updated result for both the photon yield and the second harmonic coefficient, to further test the range of our model in describing pre-equilibrum photon production in heavy-ion collisions.

\section{Hot and dense perturbative QCD in a very strong magnetic background}\label{Fraga}

\subsection{Introduction}

One of the systematic approaches to describe QCD matter at high temperatures and densities and extremely large magnetic fields is perturbative QCD (pQCD) in a non-perturbative magnetic background.
Thermodynamic quantities of magnetic QCD like the pressure, chiral condensate and strange quark number susceptibility can be computed from first principles within perturbative QCD at finite temperature and very high magnetic fields. This indeed has been accomplished up to two-loop order and for physical quark masses~\cite{Blaizot:2012sd,Fraga:2023cef,Fraga:2023lzn}, including a full implementation of the effects of the renormalization scale in the running coupling and running strange quark mass. Within this framework, the one-loop QCD correction to the photon-quark-antiquark vertex has also been computed, yielding a significantly altered contribution to the quark anomalous magnetic moment \cite{Fraga:2024klm}. 

In this section, we summarize the technical approach and discuss the main findings for observables. In particular, these perturbative magnetic QCD results for the chiral condensate and strange quark number susceptibility can be directly compared to recent lattice QCD data away from the chiral transition. Furthermore,  the calculation of the pressure at finite baryon density is discussed, as well as the simplifications that come about in the chiral limit. In fact, the effectively negligible contribution of the exchange diagram allows for building a simple analytic model for the equation of state for pure quark magnetars. The latter, however valid only for very large baryon densities, provides constraints on the behavior of the maximum mass and radius directly from in-medium perturbative QCD. 

\subsection{Thermal pQCD in a strong magnetic background}

To compute the thermal pressure and chiral condensate in the lowest Landau level approximation up to two loops in perturbative QCD, we assume that the system is embedded in a uniform, {\it very} large magnetic field ${\bf B}=B \hat{\bf z}$, where the field strength $B$ is much larger than the temperature and all masses.

The one-loop (free) contribution to the pressure from the quark sector is given by the following renormalized expression, where the pure vacuum term has been subtracted \cite{Fraga:2012rr,Kharzeev:2013jha,Andersen:2014xxa,Miransky:2015ava}:

%
\begin{align}
\begin{split}
 \frac{P_{\rm free}^q}{N_c}=&\sum_f\frac{(q_f B)^2}{2\pi^2}\left[\zeta^\prime(-1,x_f)-\zeta^\prime(-1,0)+\frac{1}{2}\left(x_f-x_f^2\right)\ln x_f+\frac{x_f^2}{2}\right]\\
 &+T\sum_{n,f}\frac{q_f B}{\pi}\left(1-\delta_{n,0}/2\right)\int \frac{dp_z}{2\pi}\bigg \{\ln\left(1+e^{-\beta\left[E(n,p_z)-\mu_f\right]}\right)+\ln\left(1+e^{-\beta\left[E(n,p_z)+\mu_f\right]}\right)\bigg \} \, ,
 \end{split}\label{P0}
 \end{align}
%
where $E^2(n,p_z)=p_z^2+m_f^2+2q_f B n$, $x_f\equiv m_f^2/2q_f B$, $T=1/\beta$ is the temperature, $\mu$ is the quark chemical potential, $N_c$ is the number of colors, $f$ labels quark flavors, $q_f$ is the quark electric charge, and $n=0, 1, 2, \cdots$ stands for the Landau levels. In this expression, Matsubara sums have already been performed in the medium contribution. One should notice that there is an inherent arbitrariness in the renormalization procedure (see Refs. \cite{Fraga:2008qn,Mizher:2010zb,Fraga:2012fs,Endrodi:2013cs,Haber:2014ula,Avancini:2020xqe,Tavares:2021fik,Farias:2021fci} for further discussion). Taking the limit of very high magnetic fields ($m_s \ll T \ll \sqrt{eB}$), one ends up with the lowest Landau level (LLL) expression
%
\begin{align}
\begin{split}
 \frac{P_{\rm free}^{\rm LLL}}{N_c}=&
 -\sum_f\frac{(q_fB)^2}{2\pi^2}\left[x_f\ln\sqrt{x_f}\right]+T\sum_{f}\frac{q_fB}{2\pi}\int \frac{dp_z}{2\pi}\bigg \{\ln\left(1+e^{-\beta\left[E(0,p_z)-\mu_f\right]}\right)+\ln\left(1+e^{-\beta\left[E(0,p_z)+\mu_f\right]}\right)\bigg \} \, .
 \end{split} \label{Pfree}
\end{align}
%
Since gluons do not carry electric charge, their one-loop contribution has the usual Stefan-Boltzmann form \cite{Kapusta:2006pm}
\begin{equation}
P_{\rm free}^G=2(N_c^2-1)\frac{\pi^2 T^4}{90} \,.
\end{equation}

Strong interactions may then be introduced perturbatively, with the first corrections appearing for the pressure at two loops. The purely gluonic contribution at this order is given by the well-known formula \cite{Kapusta:2006pm}:
\begin{equation}
P_{\rm 2}^G=-N_c (N_c^2-1)\frac{g^2 T^4}{144} \,.
\end{equation}
On the other hand, the exchange diagram for the pressure in the quark sector was first computed in Ref. \cite{Blaizot:2012sd} in the lowest Landau level approximation, displaying a dimensionally reduced structure for the matter part. For numerical purposes, however, it is convenient to recast the result found in that reference in a different fashion, where one first evaluates the momentum integrals and then carries out the Matsubara sums numerically at $\mu =0$, as detailed in Ref. \cite{Fraga:2023cef}. The final expression for the exchange pressure has the form
\begin{align}
\begin{split}
\frac{P_{\rm exch}^{\rm LLL}}{N_c}=&\frac{1}{2}g^2 \left(\frac{N_c^2-1}{2N_c}\right)T^2 \sum_f m_f^2\left(\frac{q_fB}{2\pi}\right)\sum_{\ell,n_2}\int \frac{dm_k}{2\pi}m_k
 e^{-\frac{m_k^2}{2 q_f B}}\frac{\mathcal{E}_\ell-\mathcal{E}_{n_2}}{\mathcal{E}_\ell \mathcal{E}_{n_1} \mathcal{E}_{n_2} \left|\mathcal{E}_\ell-\mathcal{E}_{n_2}\right| \left(\left|\mathcal{E}_\ell-\mathcal{E}_{n_2}\right|+\mathcal{E}_{n_1}\right)} \, ,
\end{split}\label{Pexch}
\end{align}
where 
$\mathcal{E}_\ell=\sqrt{\omega_\ell^2+m_k^2}$, $\mathcal{E}_{n_1}=\sqrt{(\omega_{n_2}+\omega_\ell)^2+m_f^2}$, and $\mathcal{E}_{n_2}=\sqrt{\omega_{n_2}^2+m_f^2}$, with the  Matsubara frequencies given by $\omega_\ell = 2\pi \ell T$ and $\omega_{n_2}=(2n_2+1)\pi T$.
This expression has the advantage of being numerically simple. Its downside, however, is that it only holds for $\mu=0$ and can not be used for cold and dense QCD, as will be discussed in the sequel.

Besides the pressure, one may also extract other observables from these calculations.
The chiral condensate represents a very relevant physical quantity in the investigation of the phase diagram for strong interactions, since it can be considered a pseudo order parameter for the chiral transition in this case. Of course, a perturbative analysis is reliable only for very large temperatures and even larger magnetic fields, so that it can not bring information on the region near the phase transition or crossover. Nevertheless, since there are lattice results for high  temperatures and magnetic fields, the comparison of these two first-principle calculations in this regime is certainly relevant.
The condensate is obtained from the pressure as a derivative with respect to the quark mass. So, the $f$-flavor condensate is given by
\begin{align}
\begin{split}
 \left\langle\bar\psi_f\psi_f\right\rangle=-\frac{\partial P_f}{\partial m_f}
 =-\frac{\partial P_{\rm free}^{\rm LLL}}{\partial m_f}-\frac{\partial P_{\rm exch}^{\rm LLL}}{\partial m_f} \, .
 \end{split} 
\end{align}
On the lattice, one computes the $f$-flavor renormalized condensate
\begin{align}
  \Sigma_f^{r}(B,T)=\frac{m_f}{m_{\pi}^2 f_{\pi}^2}\left[\left\langle\bar\psi_f\psi_f\right\rangle_{B,T}-\left\langle\bar\psi_f\psi_f\right\rangle_{0,0}\right] \, ,
\end{align}
which eliminates additive and multiplicative divergences. Here, $m_\pi=135$ MeV, $f_\pi=86$ MeV, and $m_f=5$ MeV for the light quarks. Within perturbative QCD in a nonperturbative magnetic background, however, one can not simply take the zero-field limit and obtain the vacuum condensate, since very large fields are assumed from the outset \cite{Blaizot:2012sd}.

A different observable that can also be computed and directly compared to available lattice data is the strange quark number susceptibility
\begin{equation}
\chi^s=\frac{1}{T^2}\frac{\partial^2 P} {\partial\mu_s^2} \,.
\label{chi-s}
\end{equation}
Given the presence of a derivative with respect to the chemical potential, pure vacuum terms are excluded. This presents an advantage when comparing lattice results to pQCD, even if the temperature range in the simulations is still far from optimal for this purpose \cite{Endrodi:2015oba,DElia:2021yvk}.

The pressure and chiral condensate to this order depend also on the renormalization scale  $\bar{\Lambda}$ via the scale dependence of the strong coupling $\alpha_s(\bar{\Lambda})$ and the strange quark mass $m_s(\bar{\Lambda})$. Computing the pressure and chiral condensate consistently up to order $\alpha_s$, 
the running coupling to be adopted reads \cite{Fraga:2004gz}
     \begin{equation}
     \alpha_{s}(\bar{\Lambda})=\frac{4\pi}{\beta_{0}L}\left(
     1-\frac{2\beta_{1}}{\beta^{2}_{0}}\frac{\ln{L}}{L}\right) \,,
     \label{eq:alphas}
     \end{equation}
where $\beta_{0}=11-2N_{f}/3$, $\beta_{1}=51-19N_{f}/3$, $L=2\ln\left(\bar{\Lambda}/\Lambda_{\rm \overline{MS}}\right)$ and $N_f$ is the number of quark flavors. Since $\alpha_{s}$ depends on $N_{f}$, fixing the mass of the quark at some energy scale also depends on the number of flavors.
For the strange quark mass, one has
	\begin{eqnarray}
	\begin{aligned}
	m_{s}(\bar{\Lambda})=\hat{m}_{s}\left(\frac{\alpha_{s}}{\pi}\right)^{4/9}
	\left[1+0.895062\left(\frac{\alpha_{s}}{\pi}\right)  
 \right] \;,
	\end{aligned}
	\label{eq:smass}
	\end{eqnarray}
with $\hat{m}_{s}$ being the renormalization group invariant strange quark mass, i.e. $\bar{\Lambda}$ independent. Since Eq. (\ref{eq:alphas}) for $\alpha_{s}$ implies that different values of $N_{f}$ yield different values of $\Lambda_{\overline{\rm MS}}$, by choosing  $\alpha_{s}(\bar{\Lambda}=1.5~{\rm GeV},~N_{f}=3)=0.336^{+0.012}_{-0.008}$ \cite{Bazavov:2014soa}, one obtains $\Lambda^{2+1}_{\overline{\rm MS}}=343^{+18}_{-12}~$MeV. Fixing the strange quark mass at $m_{s}(2~{\rm GeV}, N_{f}=3)=92.4(1.5)~$MeV \cite{Chakraborty:2014aca} gives $\hat{m}^{2+1}_{s}~{\approx}~248.7~$MeV when using $\alpha^{2+1}_{s}$ in Eq. (\ref{eq:smass}).

As usual, there is arbitrariness in the way one should connect the renormalization scale $\bar{\Lambda}$ to a physical mass scale of the system under consideration \cite{Kapusta:2006pm}. In thermal QCD where, besides quark masses, the only scale is given by the temperature, and $T\gg m_f$, the usual choice is the Matsubara frequency $2\pi T$ with a band around it, i.e. $\pi T < \bar{\Lambda} < 4\pi T$. In the present case, where the magnetic field also provides a relevant mass scale given by $\sqrt{eB}$, the choice becomes more ambiguous. Therefore, one can find a few different assumptions for the form of the running coupling in this case, ranging from different choices of the renormalization scale to entirely modified running coupling expressions \cite{Ayala:2018wux,Bandyopadhyay:2017cle,Karmakar:2019tdp}. The case with the standard pQCD running and $\bar{\Lambda}=\sqrt{(2\pi T)^2+eB}$ seems to be the most physical, but results for a few relevant choices will also be presented below.

Fig. \ref{fig:P_tot_band} displays the full pressure as a function of the magnetic field for two different values of the temperature, and as a function of the temperature for two different values of the magnetic field. The bands, as usual, correspond to increasing/decreasing the central renormalization running scale by a factor of $2$. Notice that the green band in the last panel is divergent for small temperatures. Deviations between the computations within different choices of the renormalization scale are large in the behavior of $\alpha_s$, $m_s$ and the exchange term \cite{Fraga:2023cef}. Here they seem to be small because the exchange correction becomes negligible for very high magnetic fields, except in the case of the nonstandard $\alpha_s(|eB|)$ that increases with $B$ \cite{Ayala:2018wux}.


\begin{figure*}[!ht]
 \centering
 \includegraphics[width=0.45\textwidth]{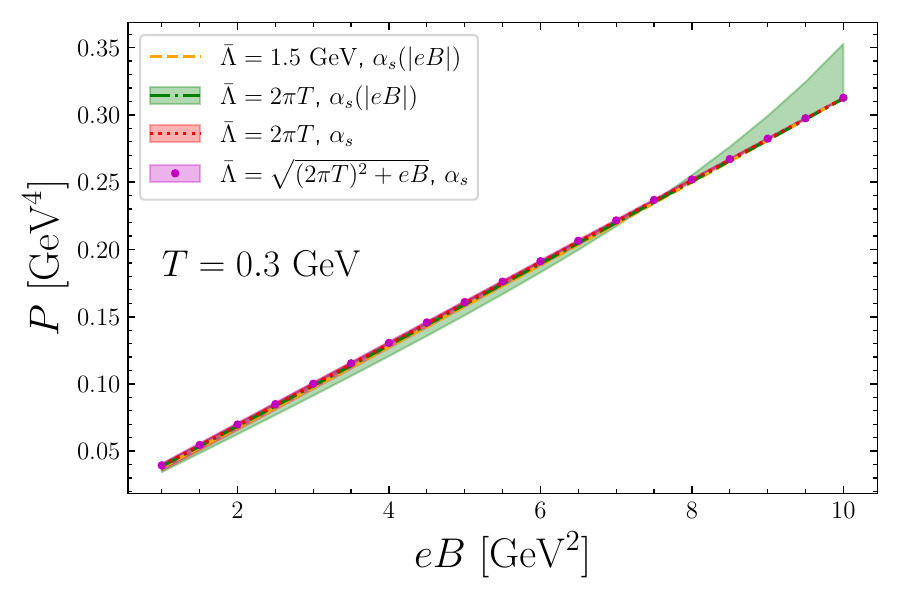} 
  \centering
 \includegraphics[width=0.45\textwidth]{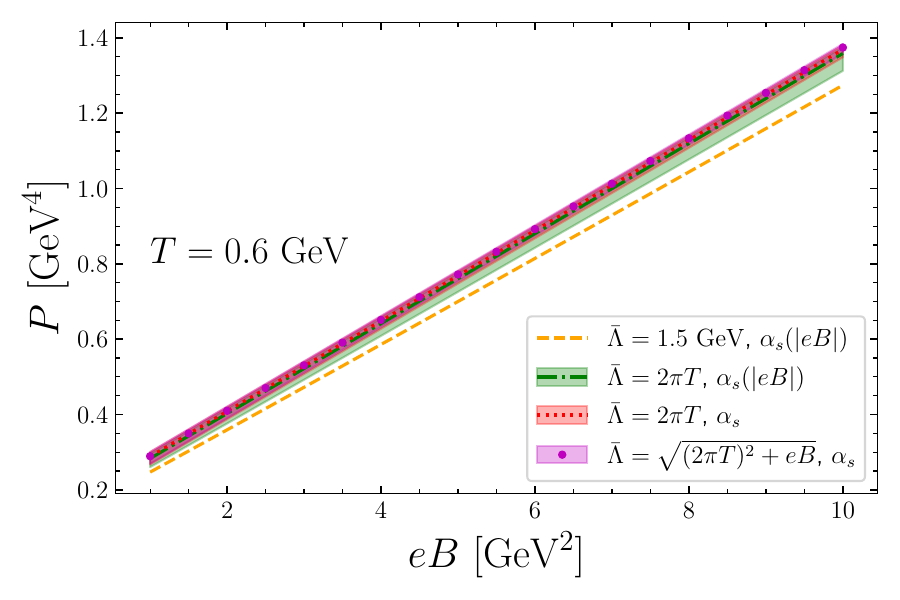}
  \centering
 \includegraphics[width=0.45\textwidth]{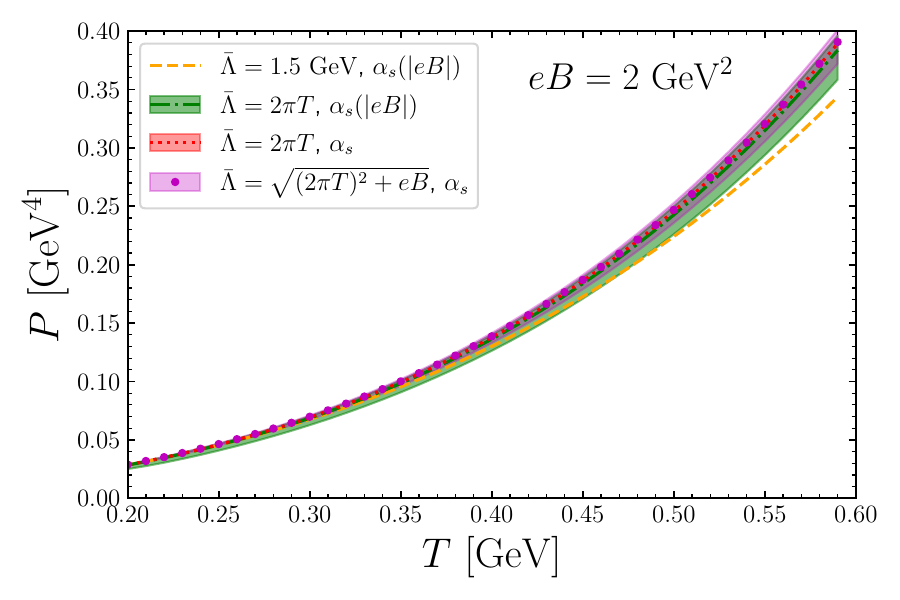}
 \centering
 \includegraphics[width=0.45\textwidth]{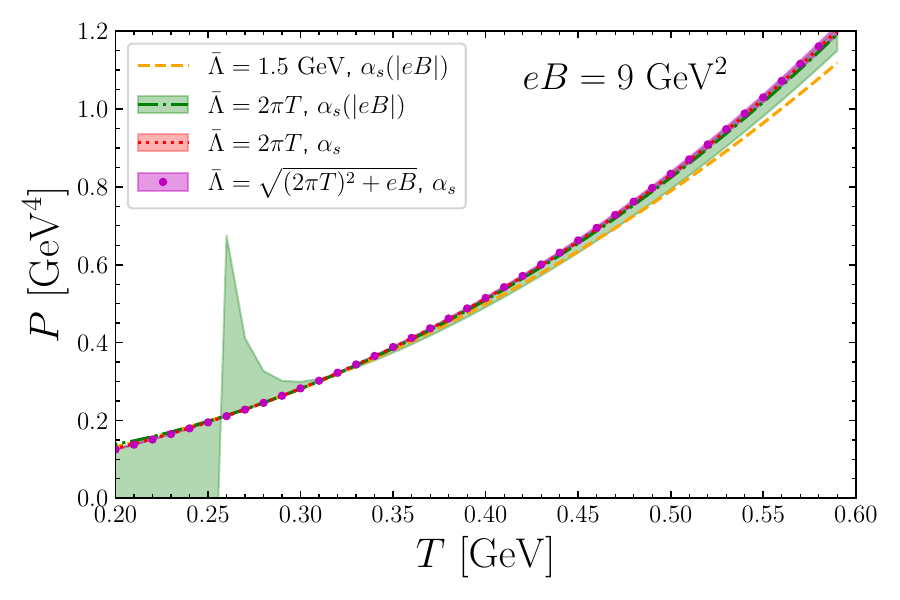} 
\caption{Full pressure as function of the magnetic filed at $T=0.3$ GeV (top-left) and $T=0.6$ GeV (top-right), and as function of temperature at $eB=2$ $\rm{GeV}^2$ (bottom-left) and $eB=9$ $\rm{GeV}^2$ (bottom-right). The bands correspond to changes in the central scale by a factor of two. The running $\alpha_s(|eB|)$ is not the standard pQCD one and has been proposed in Ref. \cite{Ayala:2018wux}. Figure from Ref. \cite{Fraga:2023cef}.}
\label{fig:P_tot_band}
\end{figure*}

\begin{figure}[!ht]
 \centering
 \includegraphics[width=0.45\textwidth]{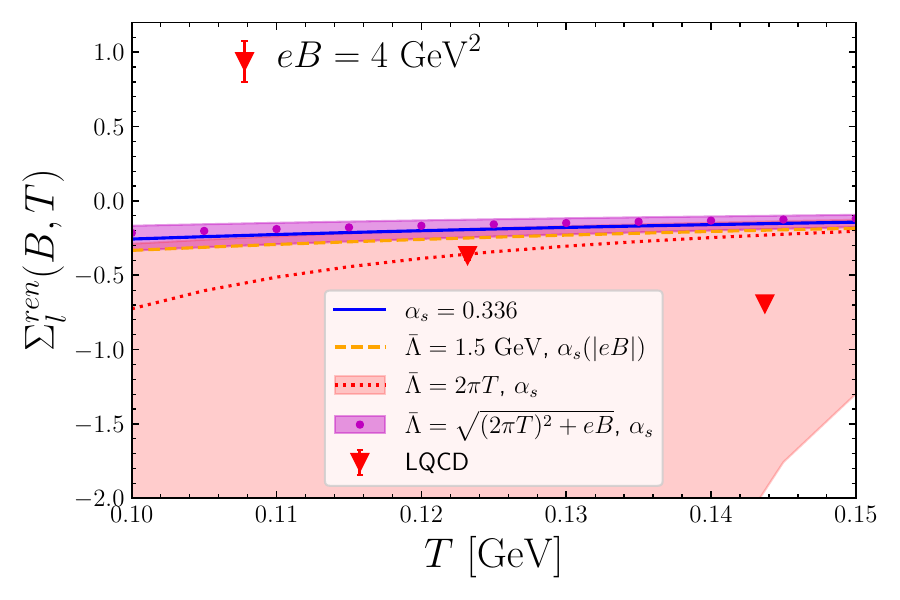} 
  \centering
 \includegraphics[width=0.45\textwidth]{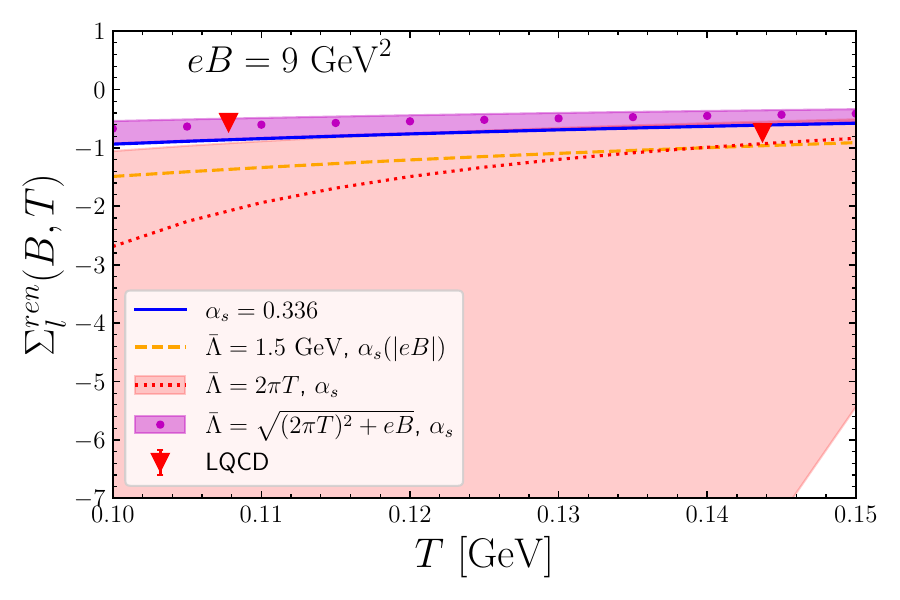}

\caption{Renormalized light quark chiral condensate as a function of the temperature for $eB=4$ $\rm{GeV}^2$ (left) and $eB=9$ $\rm{GeV}^2$ (right). Figure from Ref. \cite{Fraga:2023cef}.}
\label{Cond_Sig_B4_9}
\end{figure}

\begin{figure}[!ht]
 \centering
 \includegraphics[width=0.45\textwidth]{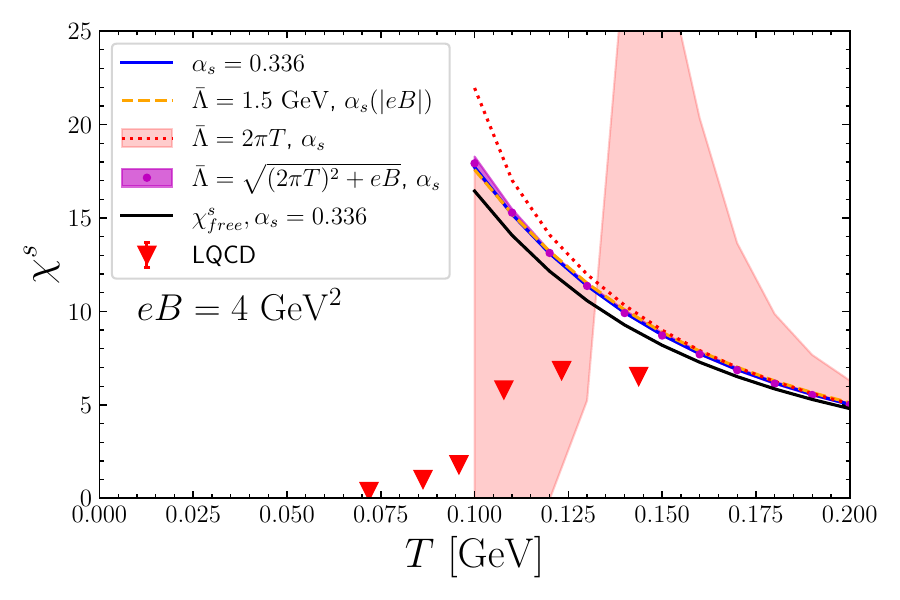} 
  \centering
 \includegraphics[width=0.45\textwidth]{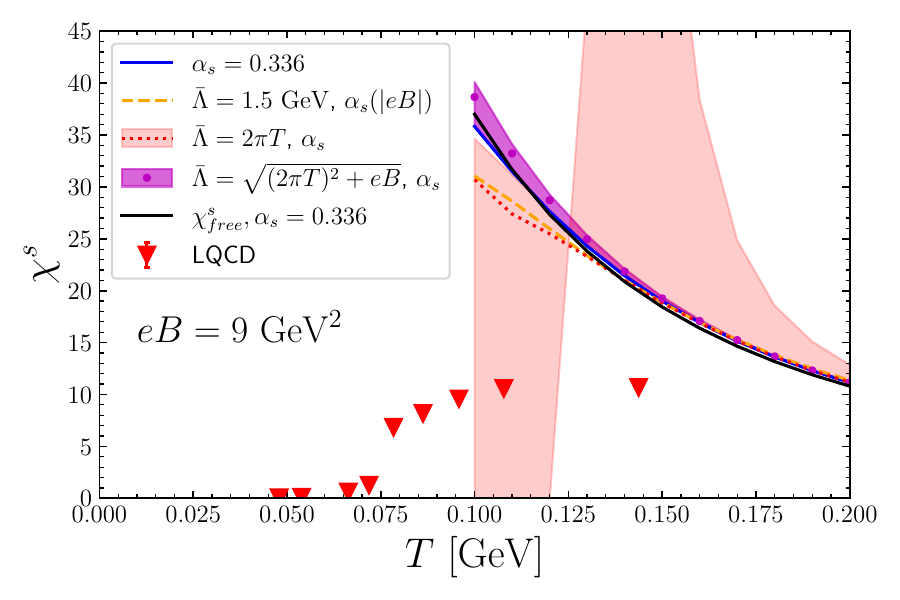}

\caption{Strange quark number susceptibility as a function of the temperature for $eB=4$ $\rm{GeV}^2$ (left) and $eB=9$ $\rm{GeV}^2$ (right). Figure from Ref. \cite{Fraga:2023cef}.}
\label{Suscep_B4_9}
\end{figure}

In Fig. \ref{Cond_Sig_B4_9} the renormalized light quark chiral condensate is shown as a function of the temperature for $eB=4$ $\rm{GeV}^2$ and $eB=9$ $\rm{GeV}^2$ computed using pQCD \cite{Fraga:2023cef}. We also show points obtained via lattice simulations for comparison \cite{DElia:2021yvk}. In Fig. \ref{Suscep_B4_9} the same is shown for the strange quark number susceptibility.
Even though the temperature range for lattice results is still well below the ideal for a fair comparison to pQCD, one can see that perturbative results are in the right ballpark for the upper end of temperatures. In order to verify whether the perturbative calculations capture the qualitative trend at high temperatures, Lattice results for higher temperatures, and even higher magnetic fields, are necessary. Since the strange quark number susceptibility represents a better observable for our comparison, one can see from the figures that the results display a more promising trend for temperatures above the ones currently simulated. One should recall that this framework is valid only if the hierarchy of scales $m_s \ll T \ll \sqrt{eB}$ is satisfied. 

\subsection{Cold and dense pQCD in a strong magnetic background}

One can follow an analogous procedure to study the complementary sector of the phase diagram, namely the one of vanishing temperature and high baryon density \cite{Fraga:2023lzn}. Cold and dense magnetic QCD, however, has a major difference with respect to its counterpart at finite temperature and zero net baryon density: the Sign Problem \cite{Aarts:2015tyj}. Monte Carlo simulations in the parameter regime that is relevant for compact star physics -- corresponding to large values of the baryon chemical potential -- are hindered. Besides magnetic pQCD, one can approach the equation of state in the limit of a large number of colors $N_c$ \cite{Fraga:2012ev}, via chiral perturbation theory \cite{Colucci:2013zoa}, using holographic models \cite{Preis:2010cq,Preis:2011sp} and, of course, within effective models. For a detailed discussion and list of references, see Refs. \cite{Fraga:2012rr,Kharzeev:2013jha,Andersen:2014xxa,Miransky:2015ava}. Effects from color superconductivity \cite{Alford:2007xm} in the presence of a strong magnetic field \cite{Ferrer:2005vd,Ferrer:2006vw,Ferrer:2007iw,Fukushima:2007fc,Noronha:2007wg}, which might be relevant to transport phenomena in the core of neutron stars and magnetars, are not discussed here.

The equation of state for a system composed by up, down and strange quarks at zero temperature and nonzero baryon chemical potential, i.e. cold and dense quark matter, was first obtained within perturbative QCD more than four decades ago by Freedman and McLerran \cite{Freedman:1976ub,Freedman:1977gz}, and also by Baluni \cite{Baluni:1977ms} and Toimela \cite{Toimela:1984xy}. Since then, it has been systematically improved \cite{Blaizot:2000fc,Fraga:2001id,Fraga:2004gz,Kurkela:2009gj,Fraga:2013qra,Kurkela:2014vha,Fraga:2015xha,Ghisoiu:2016swa,Annala:2017llu,Gorda:2018gpy,Annala:2019puf,Gorda:2021kme}. One relevant feature of perturbative QCD for cold and dense quark matter is that it seems to be much better behaved, compared to its thermal analog, as far as the convergence of the series is concerned \cite{Braaten:2002wi}. Nevertheless, if one considers the exchange (two-loop) contribution to the pressure, including the renormalization group running of $\alpha_s$ and $m_s$, it brings corrections $\sim 30\%$ for $\mu_q \sim 600$ MeV \cite{Fraga:2004gz}. Therefore, the enormous reduction in the size of the exchange term observed in thermal {\it magnetic} QCD \cite{Blaizot:2012sd,Fraga:2023cef} has remarkable effects on the perturbative series. In fact, for cold magnetic QCD this feature is even more pronounced, making the exchange diagram effectively negligible \cite{Fraga:2023lzn}. Therefore, one can build a simple, and yet nontrivial, (analytic) description for the high-density sector of the equation of state, provided that the magnetic background is strong enough to justify the lowest-Landau level description. 

In the case with zero temperature, there is only the contribution coming from the quark sector. The free part contribution is given by the following renormalized expression (subtracting the pure vacuum term) in the lowest Landau level (LLL) \cite{Palhares:2012fv,Fraga:2012rr,Fraga:2012fs,Blaizot:2012sd,Kharzeev:2013jha,Andersen:2014xxa,Miransky:2015ava,Fraga:2023cef}:
%
%
\begin{align}
 \frac{P_{\rm free}^{\rm LLL}}{N_c}
=& -\sum_f\frac{(q_fB)^2}{2\pi^2}\left[x_f\ln\sqrt{x_f}\right]+\sum_{f}\frac{(q_fB)}{4\pi^2}\left[ \mu_f P_F - m_f^2 \log\left( \frac{\mu_f+P_F}{m_f} \right) \right] 
 \, ,
  \label{Pfree_T0}
\end{align}
%
where $P_F=\sqrt{\mu_f^2-m_f^2}$ is the Fermi momentum. The exchange contribution reads \cite{Fraga:2023lzn}
\begin{align}
\begin{split}
 \frac{P_{\rm exch}^{\rm LLL}}{N_c}=&-\frac{1}{2}g^2 \left(\frac{N_c^2-1}{2N_c}\right) m_f^2\left(\frac{q_fB}{2\pi}\right)\int \frac{dm_k}{2\pi}m_ke^{-\frac{m_k^2}{2 q_f B}}\int \frac{dp_zdq_zdk_z}{(2\pi)^3}(2\pi)\delta(k_z-p_z+q_z)\\
 &\times\frac{1}{\omega E_p E_q}\Bigg\{\frac{\omega}{E_-^2-\omega^2}\Theta(\mu_f-E_{\bf{p}})\Theta(\mu_f-E_{\bf{q}}) -\left[\frac{2\left(E_{\bf{q}}+\omega\right)}{\left(E_--\omega\right)\left(E_++\omega\right)}\right]\Theta(\mu_f-E_{\bf{p}})-\frac{1}{E_{+}+\omega}\Bigg\} \, .
\end{split}
\label{P_exch_T_finite}
\end{align}
%
where
\begin{align}
\begin{split}
     E_\pm\equiv& E_{\bf{p}}\pm E_{\bf{q}}=\sqrt{{\bf{p}}^2+m_f^2}\pm\sqrt{{\bf{q}}^2+m_f^2},
\end{split}
\end{align}
and $\omega=\sqrt{k_z^2+m_k^2}$, where $m_k^2$ accounts for the gluon momentum components transverse to the magnetic field \cite{Blaizot:2012sd,Palhares:2012fv}.

As before, there is arbitrariness in the way one should connect the renormalization scale $\bar{\Lambda}$ to a physical mass scale of the system under consideration \cite{Kapusta:2006pm}. In cold and dense QCD, besides quark masses, the only available scale is given by the quark chemical potential $\mu_f\gg m_f$. The usual choice, then, is $2\mu_f$ with a band around it, i.e. $\mu_f < \bar{\Lambda} < 4\mu_f$. Following the discussion in Refs. \cite{Fraga:2023cef,Fraga:2023lzn}, the most physical case is given by $\bar{\Lambda}=\sqrt{(2\mu_f)^2+eB}$.

\begin{figure*}[!ht]
 \centering
 \includegraphics[width=0.45\textwidth]{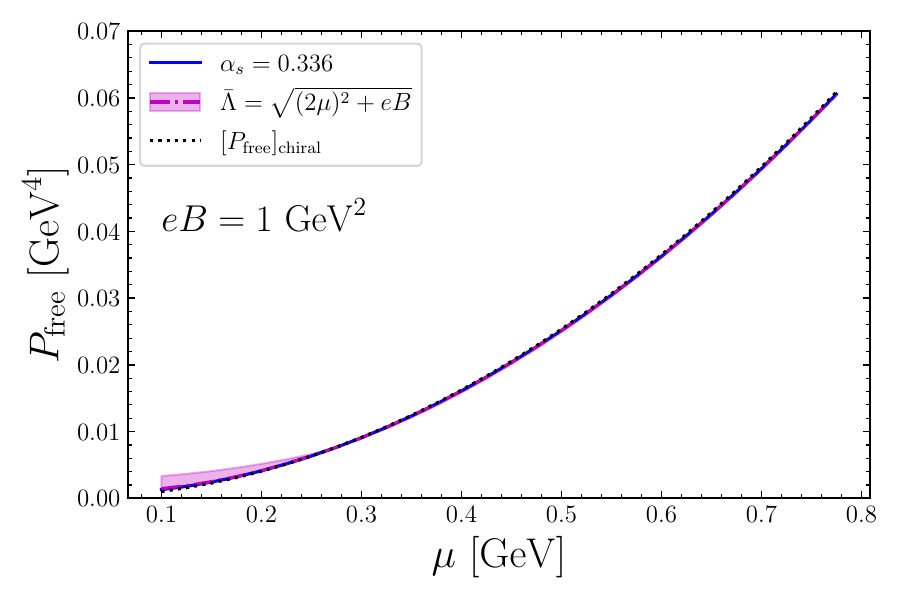} 
 \centering
 \includegraphics[width=0.45\textwidth]{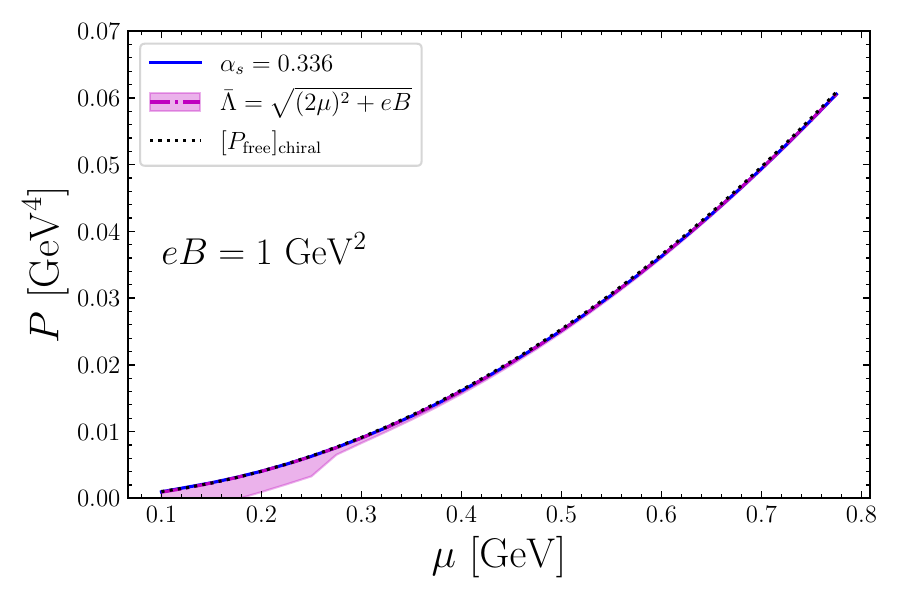} 
\caption{$P_{\rm free}$ (left) and total pressure $P$ (right) as functions of the chemical potential for $eB=1$ $\rm{GeV}^2$.  The bands correspond to changes in the central scale by a factor of $2$. For comparison, we also display the pressure in the chiral limit. Figure from Ref. \cite{Fraga:2023lzn}.}
\label{fig:Ptot_B1}
\end{figure*}

Figure \ref{fig:Ptot_B1} exhibits the free and full pressures, including their renormalization group running bands, for $eB=1$ GeV$^2$. For comparison, the pressure in the chiral limit is also displayed. One can see that, for large magnetic fields, all cases are essentially indistinguishable unless one goes to values of the chemical potential below $\mu_s=300~$ MeV. This fact suggests that one could build a pQCD-based simple analytic model for the equation of state of cold and dense quark matter under very large magnetic fields. 

\begin{figure*}[!t]
 \centering
 \includegraphics[width=0.45\textwidth]{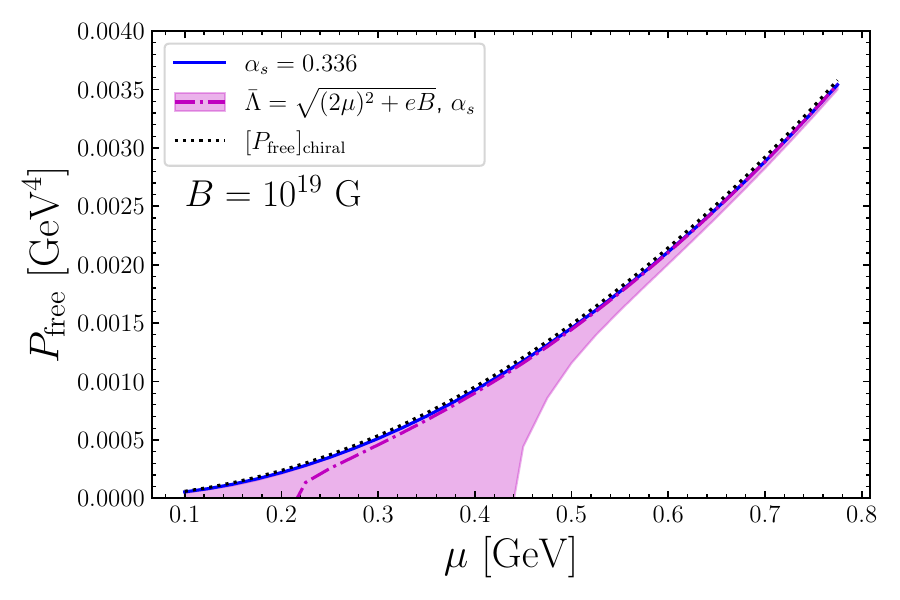} 
 \centering
 \includegraphics[width=0.45\textwidth]{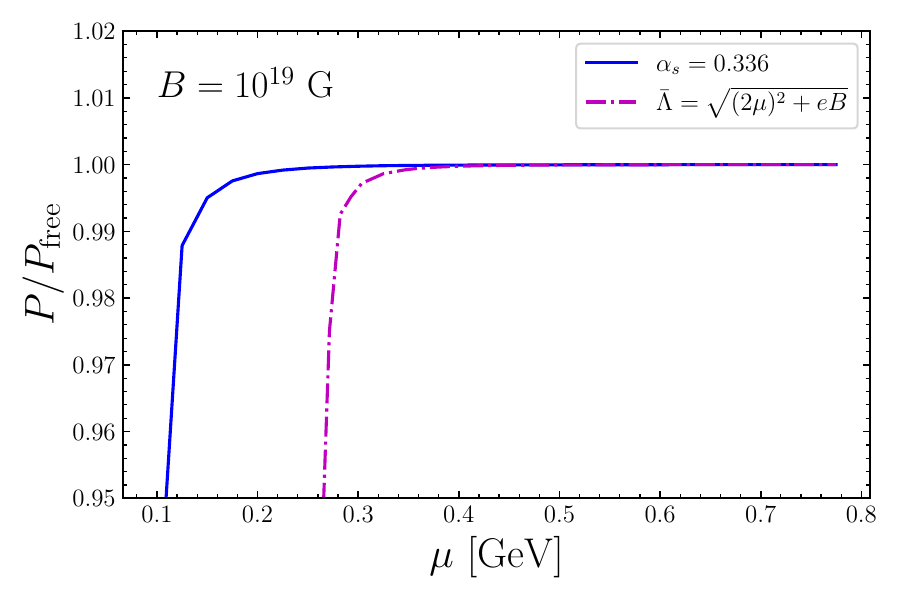} 
\caption{$P_{\rm free}$ (left) and $P/P_{\rm free}$ (right) as functions of the quark chemical potential for $B=10^{19}$ Gauss ($eB=0.059~$GeV$^2$). The bands correspond to changes in the central scale by a factor of $2$. For comparison, we also display the pressure in the chiral limit. Figure from Ref. \cite{Fraga:2023lzn}.}
\label{fig:Ptot_B00195}
\end{figure*}

As discussed previously, the contribution of the two-loop exchange diagram is negligible when compared to the one-loop free term in the case of very high magnetic fields. For magnetic fields expected to be found in the core of magnetars and higher, one can show that the (two-loop) exchange contribution remains small, but renormalization group running effects are significant. This point is illustrated in Figure \ref{fig:Ptot_B00195}, that shows the free pressure $P_{\rm free}$ and how it compares to the total pressure $P$ as functions of the chemical potential for $B=10^{19}$ Gauss. One should notice the thicker running bands that appear for such values of magnetic field, indicating the increasing error in the pQCD prediction at low densities. 

Hence, as discussed in Ref. \cite{Fraga:2023lzn}, one can build a simple analytic pQCD-based model for the equation of state by using
\begin{equation}
 \frac{P_{\rm eff}}{N_c}=
 -\sum_f\frac{(q_fB)^2}{2\pi^2}\left[x_f\ln\sqrt{x_f}\right]+\sum_{f}\frac{(q_fB)}{4\pi^2}\left[ \mu_f P_F - m_f^2 \log\left( \frac{\mu_f+P_F}{m_f} \right) \right] \,,
  \label{eos-eff}
\end{equation}
including the renormalization group running. This simple expression represents an excellent approximation to the two-loop pQCD result at very large magnetic fields and intermediate to high densities. Once again, one will notice the larger bands for the pressure as pQCD is applied for smaller values of the magnetic field, essentially leaving the strict region of validity for the full calculation. Nevertheless, these bands provide a measure of the uncertainty in the perturbative calculation, a useful feature that is usually lacking in model calculations. In fact, one can use the perturbative band at a given chemical potential to restrict possible equations of state, in the same fashion as performed in Ref. \cite{Kurkela:2014vha} for neutron star matter\footnote{Even though the exchange contribution is small for $B=10^{19}$ Gauss, such field is not high enough to justify the LLL approximation. Thus, the equation of state above should not be naively applied to magnetars. Reliable results in the current description are only obtained for higher values of the magnetic field.}.

\begin{figure*}[!ht]
 \centering
 \includegraphics[width=0.45\textwidth]{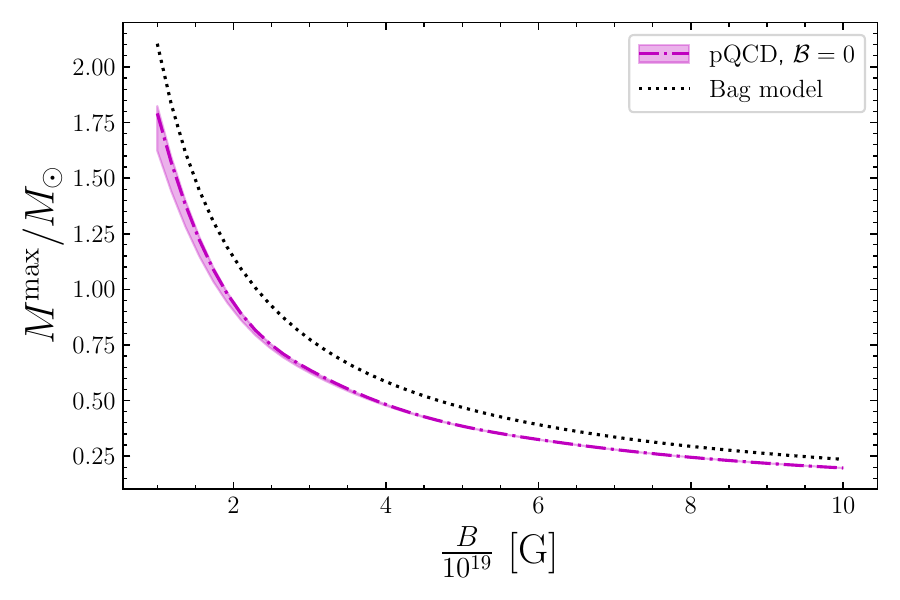} 
 \centering
 \includegraphics[width=0.45\textwidth]{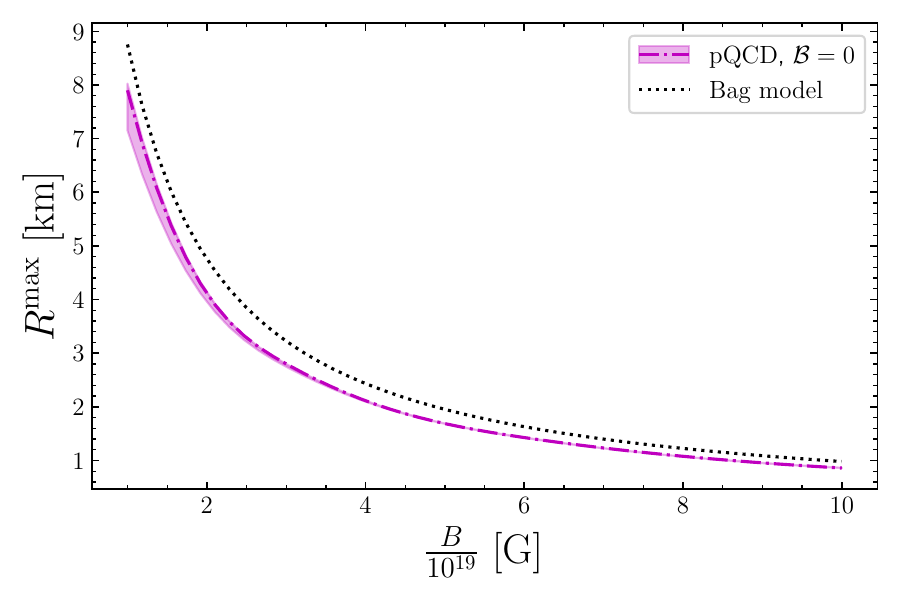} 
\caption{Maximum mass (left) and its respective radius (right) as functions of the magnetic field $B$ obtained from perturbative QCD. The different curves correspond to changes in the central scale by a factor of $2$. We also show results from the bag model for comparison. Figure from Ref. \cite{Fraga:2023lzn}.}
\label{fig:Mvsr_B0195}
\end{figure*}

The utility of this formulation can be illustrated by the computation of the maximum mass and total radius of pure quark magnetars (strange magnetars) as functions of the magnetic field $B$. These results are shown in Figure \ref{fig:Mvsr_B0195}. The different curves correspond to changes in the central scale by a factor of $2$. The figure also shows results from the magnetic bag model for comparison. These results place constraints on the
behavior of the maximum mass and associated radius from perturbative QCD coming down from very high values of the magnetic field. Any given model description should, ideally, 
approach these constraints for high enough values of $B$.

It is clear that the equation of state extracted from perturbative QCD can not provide a good description of the low- and intermediate-density sectors all by itself. However, this equation of state is obtained from first principles and, thus, should guide the description of the high-density regime. For instance, model calculations of the equation of state for quark matter in the presence of strong magnetic fields should approach this behavior for high $\mu_B$ and very high $B$. A more realistic description of the equation of state in magnetar matter, even for very large magnetic fields, can only be achieved by matching a low-density equation of state onto the equation of state from perturbative QCD.

Even though the window of applicability of hot and dense pQCD in a very strong magnetic field is still narrow, this framework yields results  obtained from a clean first-principle calculation, one that can be systematically improved and used to provide solid bounds for other approaches. Furthermore, medium loop corrections seem to become essentially negligible as compared to the free term for very high magnetic fields and physical choices of the renormalization running scale. This fact can, in principle, improve the convergence of the theory in this limit and suggest simple effective models.

\section{Strongly Interacting Matter in Magnetic Fields}\label{Adhikari}

In this section, we review recent developments characterizing the impact of a uniform magnetic field, assumed to point in the longitudinal direction, on the lowest two topological cumulants in a model independent approach using chiral perturbation theory. We identify low energy theorems that connect the topological susceptibility to the chiral condensate and the fourth cumulant to chiral condensates and susceptibilities. We also investigate the nature of finite volume thermodynamics with a particular focus on the chiral condensate, renormalized magnetization and the pressure, which becomes anisotropic. The local chiral condensate is doubly periodic on the plane transverse to the magnetic field, while the renormalized magnetization and (spatially averaged) chiral condensate exhibit significant finite volume corrections. The former is particularly sensitive to finite volume effects due to its small size.

\subsection{Chiral perturbation theory}\label{sub:chiral-perturbation-theory}
	
	Due to its versatility, chiral perturbation theory allows for model-independent investigations of a wide range of physical phenomena. In the presence of magnetic fields, these include the study of pion polarizabilities~\cite{bijnens1988two,donoghue1988reaction,holstein1990comments}, magnetic catalysis~\cite{Agasian:1999sx,shushpanov1997quark,Cohen:2007bt,Werbos:2007ym,Andersen:2012zc,andersen2012thermal,Adhikari:2021bou,tiburzi2008hadrons}, vacuum magnetization~\cite{cohen2009magnetization,kabat2002qcd}, pion superconductivity~\cite{Son:2000xc,Son:2000by,Splittorff:2000mm}, magnetic vortices~\cite{Adhikari:2015wva,Adhikari:2018fwm,adhikari2022phonon} and chiral soliton lattices~\cite{Brauner:2016pko,Brauner:2021sci,gronli2022competition,Gronli:2022cri,Evans:2022hwr}. Since the primary objective of this section is to illustrate theory developments and the focus of this section is on recent developments of the impact of magnetic fields on topological cumulants and finite volume thermodynamics in a magnetic background, we refer readers interested in a thorough review of chiral perturbation theory to seminal papers~\cite{Gasser:1983yg,Gasser:1984gg} and review articles~\cite{scherer2011primer,lin2015lattice}. 
	
Chiral perturbation theory is an effective theory of pseudo-Goldstone modes associated with chiral symmetry breaking by the QCD vacuum. The resulting Goldstone manifold characterizes field fluctuations of these modes, which in two-flavor QCD are pions. Observables and condensates constructed from chiral perturbation theory are model-independent and valid in the low energy regime of QCD characterized by $p\ll \Lambda_{\chi}$. The mass dimension one quantity, $p$, in the context of a magnetic background is either the pion mass, or the renormalization invariant quantity, $\sqrt{eB}$. Throughout the discussion, we assume $e>0$ and use $B$ to denote the external magnetic field. $\Lambda_{\chi}\equiv 4\pi F_{\pi}$ is a characteristic hadronic scale that is a function of the pion decay constant, $F_{\pi}$, and emerges in one-loop calculations. In the context of finite volume thermodynamics~\cite{gasser1988spontaneously}, a further scale, the box size, $L$, is required. Since modern lattice calculations are conducted in the $p$-regime, for which box sizes fit a large number of pions, $m_{\pi}L\gg 1$, standard power counting rules of chiral perturbation theory apply.
	 
\subsection{Topological cumulants in a magnetic field}\label{sub:topological-cumulants}
	
The QCD vacuum is characterized by $CP$-even topological cumulants constructed out of the gluon field tensor, $G_{\mu\nu}^{a}$ and its dual, $\widetilde{G}_{\mu\nu}^{a}=\tfrac{1}{2}\epsilon_{\mu\nu\alpha\beta}G^{\alpha\beta}$ . The lowest topological cumulant, the topological susceptibility, is a time-ordered correlation function of $\widetilde{G}G$
	\begin{align}
	\chi_{t}\,(0)&=-i\int d^{4}x\left\langle \mathcal{T}\frac{g^{2}\widetilde{G}G(x)}{32\pi^{2}}\frac{g^{2}\widetilde{G}G(0)}{32\pi^{2}}\right\rangle
	\end{align}
that is a measure of the susceptibility of the QCD-vacuum to second order changes in $\theta$ near the QCD ground state. Assuming a spacetime invariant angle $\theta$, the topological susceptibility can be expressed as a partial derivative, $\chi_{t}=\left.\frac{\partial^{2}\mathcal{F}}{\partial\theta^{2}}\right|_{\theta=0}$. The fourth cumulant, on the other hand, is a four-point correlation function of $\widetilde{G}G$ and a measure of fourth order changes to the free energy as a function of the vacuum angle. 
\begin{figure*}[t!]
\centering
\includegraphics[width=0.48\textwidth]{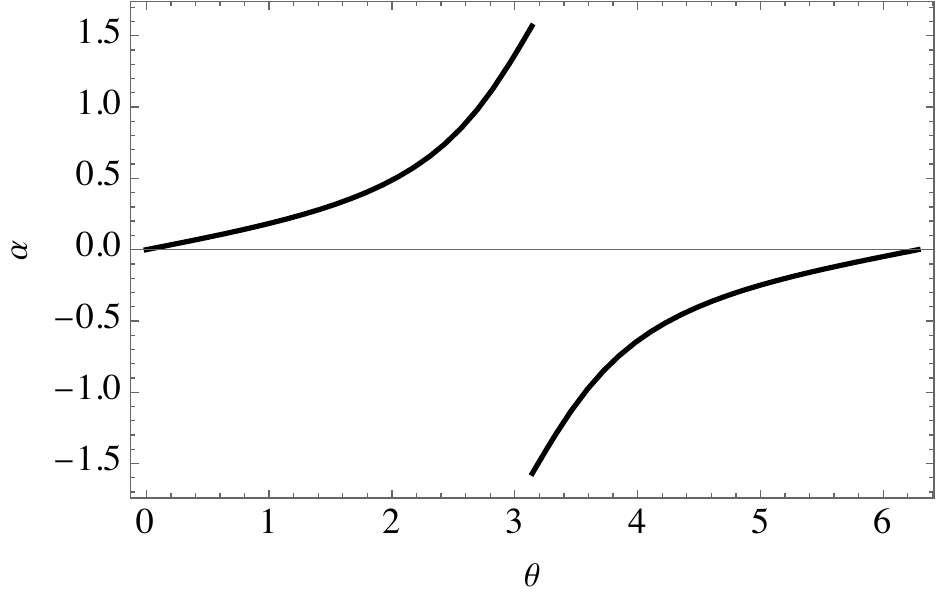}\,\,\,\,\,
\includegraphics[width=0.48\textwidth]{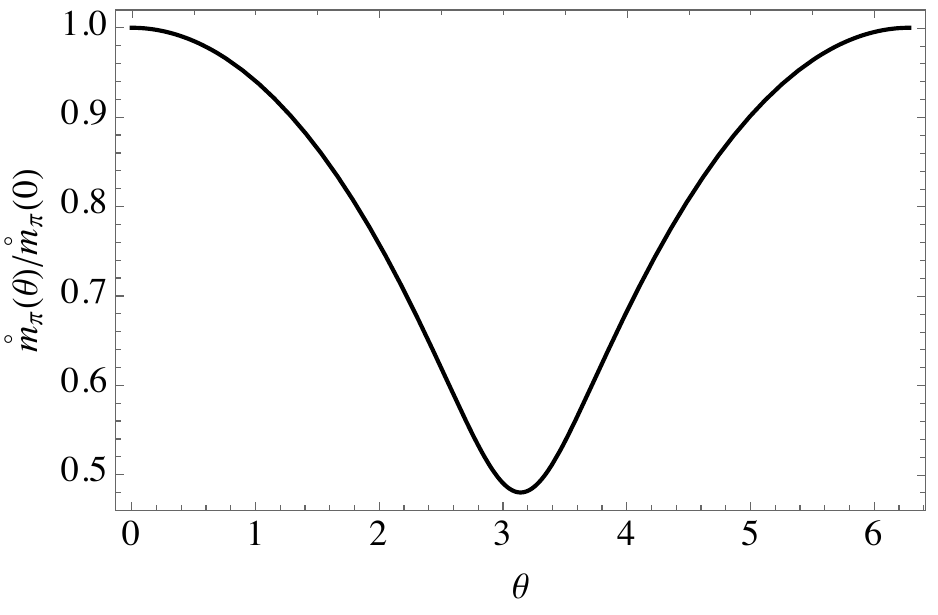}
\caption{(Left) Plot of the tree-level ground state value of $\alpha$ as a function of the vacuum angle $\theta$. (Right) Plot of tree-level pion mass as a function of $\theta$ (normalized by the corresponding $\theta=0$ value).} 
\label{fig:alpha-mass}     
\end{figure*}
These topological cumulants are non-vanishing as a consequence of the axial $U(1)$ anomaly~\cite{Crewther:1977ce,Witten:1979vv,Veneziano:1979ec,DiVecchia:1980yfw,SHIFMAN1980493} that prevents the vacuum angle from being rotated away. The experimental observation of the neutron dipole moment constrains  the angle to $\theta\lesssim 10^{-11}$~\cite{Afach:2015sja,kim2010axions}. Values of $\theta$ many orders of magnitude larger than the upper bound do not affect observables significantly~\cite{ubaldi2010effects,lee2020theta} -- in the absence of an adequate explanation for the zero value of the angle, $\theta$ is for all practical purposes zero. This is the strong $CP$ problem. 

The $U(1)$ axial anomaly, however, is better understood. Besides the color gauge symmetry, the $n$-flavor QCD Lagrangian in the chiral limit has an $U(n)_{L}\times U(n)_{R}$ flavor symmetry. The QCD vacuum possesses a non-zero chiral condensate, which breaks the vector subgroup $SU(n)_{A}$ leaving $SU(n)_{V}\times U(1)_{A}\times U(1)_{V}$ intact. However, the low-energy spectrum of QCD only possesses $n^{2}-1$ Goldstone bosons (pseudoscalars). Naively, if $U(1)_{A}$ were intact, one expects to observe parity partners (scalars) associated with the pions (3), kaons (4) and eta (1). However, no parity partners or even approximate parity partners are present in the spectrum. Additionally, if $U(1)_{A}$ were spontaneously broken, one expects to observe an isosinglet with an independent pion decay constant with a mass saturated at $\sqrt{3}m_{\pi}$. 

The breaking of $U(1)_{A}$ is rather strong (and expected to occur for instance through instantons~\cite{tHooft:1976rip, tHooft:1976snw,DelDebbio:2004ns}) -- the integration measure of the QCD partition function unlike the the QCD Lagrangian (in the massless limit) is not invariant under $U(1)_{A}$. Consequently the $\theta$-term of QCD
\begin{align}
\mathcal{L}_{\theta}&=-\frac{g^{2}\theta}{32\pi^{2}}\widetilde{G}G&
\mathcal{D}\overline{q}\,\mathcal{D}q&\rightarrow \exp\left[-i\int d^{4}x\,\frac{g^{2}\Theta n}{16\pi^{2}}\right]\mathcal{D}\overline{q}\,\mathcal{D}q\ ,
\end{align}
which is permitted by color gauge symmetry, cannot be rotated away through an axial $U(1)$ rotation of the quark field $q\rightarrow e^{-i\Theta \gamma_{5}}q$. Under the transformation, the measure rotates~\cite{Fujikawa:1979ay,LUSCHER1998342} (as shown above) with a choice of $\Theta=-\frac{\theta}{2n}$ eliminating the $\theta$-term at the expense of introducing an imaginary quark mass term in the QCD Lagrangian proportional to the quark mass, $m_{f}$ for each flavor $f$
\begin{align}
\mathcal{L}_{\rm mass}=-\sum_{f=1}^{n}m_{f}\overline{q}_{f}q_{f}\rightarrow -\sum_{f=1}^{n}m_{f}\overline{q}_{f}e^{-\theta\gamma_{5}}q_{f}\ .
\end{align}
For a chiral perturbation theory analysis of the cumulants~\cite{Adhikari:2021lbl,Adhikari:2021xra,adhikari2022topological}, one requires, in addition to the external magnetic field, an introduction of the $\theta$-term through a scalar source. 

The leading chiral perturbation theory Lagrangian 
\begin{align}
\mathcal{L}_{2}=\tfrac{1}{4}F^{2}\,{\rm Tr}\,[\,D_{\mu}\,\Sigma^{\dagger}D^{\mu}\,\Sigma\,]+\tfrac{1}{4}F^{2}\,{\rm Tr}\,[\,\chi\,\Sigma^{\dagger}+\chi^{\dagger}\,\Sigma\,]\ .
\end{align}
is parameterized in terms of $SU(n)$ fields, $\Sigma$, the scalar source $\chi=2B_{0}Me^{-i\theta/n}$ and the external magnetic field that enters through the covariant derivative, $D_{\mu}\Sigma=\partial\Sigma+i[Q,ieA_{\mu}^{\rm ext}]$. The presence of a $\theta$ vacuum rotates the vacuum that points in the (isospin) direction of the identity operator. For $n=2$, the rotation of the vacuum, characterized by $\alpha$, is understood from the isospin structure of the quark mass matrix, $M=\tfrac{1}{2}(m_{u}+m_{d})\,1+\tfrac{1}{2}(m_{u}-m_{d})\,\tau_{3}$, which favors the third direction in the presence of an isospin splitting
\begin{align}
\Sigma_{\alpha}&=\cos\alpha\,1+i\sin\alpha\,\hat{\phi}_{a}\tau_{a}&
\hat{\phi}_{a}\hat{\phi}_{a}&=1\ .
\end{align}
An obvious corollary is that in the strong isospin limit, the vacuum remains unrotated. Since the effect of the magnetic field on the free energy first enters at next-to-leading order, one requires the construction of the relevant Lagrangian and counterterms in terms of the correct field fluctuations. Since the rotation of the vacuum is axial, an appropriate rotation of the pion field fluctuations is required, $\Sigma=e^{i\frac{\alpha}{2}\tau_{3}}e^{i\frac{\phi_{a}\tau_{a}}{F}}e^{i\frac{\alpha}{2}\tau_{3}}$. A non-trivial check to the validity of this construction is the cancellation of divergences that arise in one-loop diagrams by the counterterms of chiral perturbation theory -- there are both magnetic field independent and dependent divergences. The latter leads to charge renormalization or equivalently a magnetic field renormalization that keeps $eB$ invariant. Additionally, the construction of the leading field-dependent free energy only requires the tree-level ground state even though the magnetic field alters the ground state orientation, a result that is evident upon noting the perturbative nature of the free energy, which allows for a perturbative evaluation of the ground state orientation. The resulting magnetic-field-dependent free energy has an analogous structure to that of the $\theta=0$ result
\begin{align}
\mathcal{F}_{B}(\theta)&=\frac{1}{2}B_{R}^{2}-\frac{1}{(4\pi)^{2}}\int_{0}^{\infty} \frac{ds}{s^{3}}e^{-\mathring{m}_{\pi}^{2}(\theta)s}\left[\frac{eBs}{\sinh eBs}-1+\frac{(eBs)^{2}}{6}\right]
\end{align}
\begin{align}
B_{R}&=Z_{B}B&
Z_{B}&=1+4e^{2}h^{r}_{2}+\frac{e^{2}}{6(4\pi)^{2}}\left(\log\frac{\Lambda^{2}}{\mathring{m}_{\pi}^{2}(\theta)}-1\right)
\end{align}
with the tree-level pion mass, $\mathring{m}_{\pi}(0)$ replaced by its $\theta$-vacuum value $\mathring{m}_{\pi}(\theta)$
\begin{align}
\mathring{m}^{2}_{\pi}(\theta)=\mathring{m}^{2}_{\pi}(0)\sqrt{1-\tfrac{4m_{u}m_{d}}{(m_{u}+m_{d})^{2}}\sin^{2}\tfrac{\theta}{2}}
\end{align}
that is shown in Fig.~\ref{fig:alpha-mass}. The mass decreases as a function of the vacuum angle reaching a minimum at $\theta=\pi$ and is symmetric about the angle. Also shown in Fig.~\ref{fig:alpha-mass} is the tree-level orientation of $\Sigma$ that is odd about $\theta=\pi$. The high-energy constant, $h_{2}^{r}(\Lambda)$, depends on the $\overline{\rm MS}$-bar scale $\Lambda$ and arises through field renormalization.  

\begin{figure*}[t!]
\centering
\includegraphics[width=0.48\textwidth]{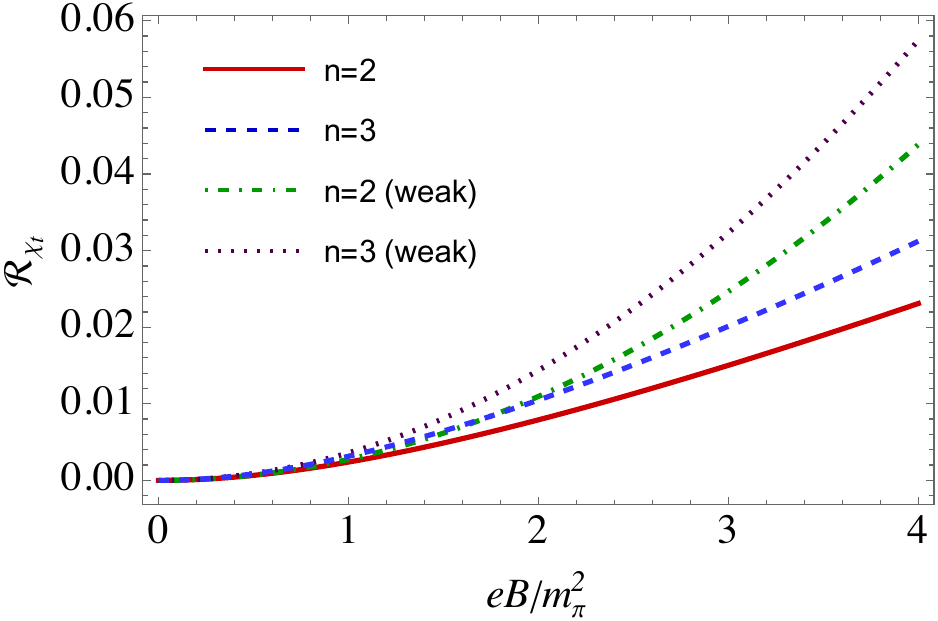}\,\,\,\,\,
\includegraphics[width=0.48\textwidth]{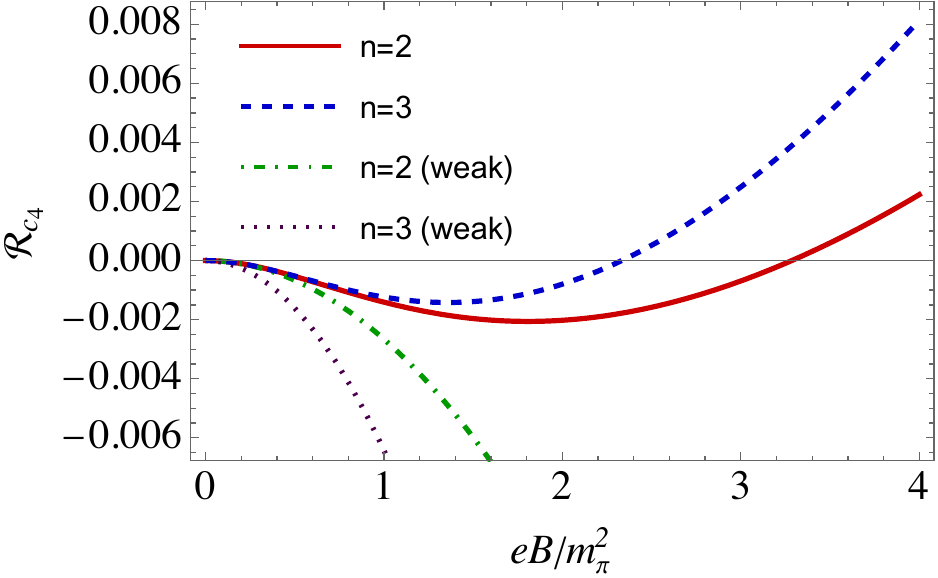}
\caption{Plot of the relative shifts of the topological susceptibility (left) and fourth cumulant (right) compared to tree-level values. The plots also include the small field approximations (up to quadratic order) to the relative shifts.} 
\label{fig:R-chit-c4}     
\end{figure*}

Since the pion mass squared appears in the negative exponential of the free energy integrand, the free energy has its largest value at $\theta=0$. The calculation of the impact of the magnetic field is now straightforward -- they are derivatives with respect to $\theta$ though one does need to take care to incorporate the $\theta$-dependent contribution of the renormalized magnetic field. The expressions for the shifts for the topological susceptibility and topological cumulants and their weak field behaviors are presented below in terms of the reduced mass $\overline{m}=\left(\frac{1}{m_{u}}+\frac{1}{m_{d}}\right)^{-1}$
\begin{align}
\chi_{t,B}&=-B_{0}\,\overline{m}\int_{0}^{\infty}\frac{ds}{(4\pi s)^{2}}e^{-m_{\pi}^{2}s}\left[\frac{eBs}{\sinh eBs}-1\right]
=B_{0}\,\overline{m}\left[\frac{1}{6(4\pi m_{\pi})^{2}}\right](eB)^{2}+\mathcal{O}\left\{(eB)^{4}\right\}\ ,\\
c_{4,B}&=B_{0}\,\overline{m}^{\,4}\left(\frac{1}{m_{u}^{3}}+\frac{1}{m_{d}^{3}}\right)\int_{0}^{\infty}\frac{ds}{(4\pi s)^{2}}e^{-m_{\pi}^{2}s}\left[\frac{eBs}{\sinh eBs}-1\right]
-3B_{0}^{2}\,\overline{m}^{2}\int_{0}^{\infty}\frac{ds}{(4\pi)^{2}s}e^{-m_{\pi}^{2}s}\left[\frac{eBs}{\sinh eBs}-1\right]\\
&=\left[-B_{0}\,\overline{m}^{\,4}\left(\frac{1}{m_{u}^{3}}+\frac{1}{m_{d}^{3}}\right)+\frac{3B_{0}^{2}\overline{m}^{\,2}}{m_{\pi}^{2}}\right]\frac{1}{6(4\pi m_{\pi})^{2}}(eB)^{2}+\mathcal{O}\{(eB)^{4}\}\ .
\end{align}
A plot of the cumulants and their weak field behavior is presented in Fig. \ref{fig:R-chit-c4} with the two-flavor approximations exhibiting a larger range of validity compared to three-flavor. We conclude this subsection by presenting low-energy theorems that relate the shift of the cumulants to quark condensates and susceptibilities,
\begin{align}
(n=2)\ \ \ \ \ \ \ \ \ \ \ \ \ \chi_{t,B}&=-\overline{m}\,\langle\, \bar{q}_{f}q_{f}\,\rangle_{B}&
c_{4,B}&=\overline{m}^{4}\sum_{q_{f}=u,d}\frac{\langle\,\overline{q}_{f}q_{f}\,\rangle_{B}}{m_{q_{f}}^{3}}+3\overline{m}^{2}\,\chi_{q_{f},B}\ .
\end{align}
The shift in the topological susceptibility is proportional to the shift of each quark condensate, $q_{f}=u,d$. The fourth cumulant is also proportional to the shift of the quark condensates and to the quark susceptibility, which measures the change in the free energy to second order changes in the quark mass.

Fascinatingly, these are manifestations of Ward-Takahashi identities that hold more generally in QCD and are constructed through functional derivatives and subsequent space-time integration of the axial rotated and unrotated versions of the QCD partition function. In particular, one can use functional derivatives to relate the topological susceptibility to the quark condensate and a space-time integral of a time-ordered two-point correlation of the $CP$-odd operator $\overline{q}M\gamma_{5}q$
\begin{align}
\nonumber
\chi_{t,B}&=-\frac{1}{n^{2}}\langle\,\overline{q}Mq\,\rangle_{B}
+\frac{i}{n^{2}}\int d^{4}x\,\left\langle\,\mathcal{T}\,\overline{q}(x)M\gamma_{5}q(x)\overline{q}(0)M\gamma_{5}q(0)\,\right\rangle_{B}\ ,
\end{align}
allowing for the determination of the latter, which vanishes in the isospin limit
\begin{align}
\nonumber
(n=2)\ \ \ \ \ \ \ \ \ 
i\int d^{4}x\, \langle \mathcal{T}\bar{q}(x)M\gamma_{5}q(x)\bar{q}(0)M\gamma_{5}q(0)\rangle_{B}
&=\frac{m_{d}-m_{u}}{m_{u}+m_{d}}\left[-m_{u}\langle\bar{u}u\rangle_{B}+m_{d}\langle\bar{d}d\rangle_{B}\right]\ .
\end{align}
While we forgo the discussion of the three-flavor low energy theorem, we refer the interested reader to Ref.~\cite{Adhikari:2021lbl}. The difference from the two-flavor case arises due to the interaction of charged kaon loops with the external field. The resulting low energy theorem for the topological susceptibility is simultaneously proportional to all three quark condensates though the expression for the fourth cumulant is rather more involved. Nevertheless, in the limit of infinite strange quark masses, which is equivalent to  integrating out heavy mesons (excluding the pions), the low energy theorems reduce to that of two-flavor chiral perturbation theory. 

\subsection{Finite volume thermodynamics}\label{sub:finite-volume-thermodynamics}
	
Lattice QCD calculations offer a more comprehensive avenue to investigate the impact of magnetic fields on topological cumulants~\cite{brandt2022qcd}. Surprisingly, until very recently, systematic characterization of finite volume effects on lattice thermodynamic observables and condensates in magnetic backgrounds did not exist~\cite{adhikari2023qcd}. In this section, we discuss the most important aspects of this recent development beginning with a discussion of magnetic periodic boundary conditions and the resulting finite volume Green's function that encodes finite volume QCD thermodynamics.
	
	Naturally, we work in \textit{Euclidean space} with $x_{\mu}=(x_{0},x_{1},x_{2},x_{3})$ and assume zero temperature or equivalently infinite extent in the Euclidean time direction. Utilizing the fully asymmetric gauge $A_{\mu}=(0,-Bx_{2},0,0)$, it is clear that the background gauge field is only periodic up to a gauge transformation~\cite{al2009discrete}
	\begin{align}
	A_{\mu}(x+L_{2}\hat{x}_{2})&=A_{\mu}(x)+\partial_{\mu}(\theta_{2}+\Lambda_{2})&
	\phi(x+L_{2}\hat{x}_{2})&=e^{-iQ(\theta_{2}+\Lambda_{2})}\phi(x)
	\end{align}
that is characterized by $\Lambda_{2}=-BL_{2}x_{1}$ and a twist angle $\theta_{2}$. Protecting gauge invariance under periodic transformations, therefore, requires an imposition of a boundary gauge transformation of the (positively charged) pion field, which we have denoted $\phi$ for notational simplicity. There are further such periodic gauge transformations induced by twist angles in the remaining Euclidean directions.  A consequence of magnetic periodicity is the existence of gauge invariant objects, beyond electric and magnetic fields, namely Wilson lines,
\begin{align}
\label{eq:WL}
W_{2}(x_{1})&=e^{-iQ(\theta_{2}-BL_{2}x_{1})}&
W_{1}(x_{2})&=e^{-iQ(\theta_{1}+BL_{1}x_{2})}\ .
\end{align}
that wrap around the first and second spatial directions, respectively.  

Uniqueness of the pion field upon periodic translations in the transverse directions leads to quantization of magnetic flux. More transparently, we consider the equalities
\begin{align}
\phi(x+L_{1}\hat{x}_{1}+L_{2}\hat{x}_{2})&=W_{2}(x_{1}+L_{1})\phi(x+L_{1}\hat{x}_{1})=e^{-iQ\theta_{1}}W_{2}(x_{1}+L_{1})\phi(x)\\
\phi(x+L_{2}\hat{x}_{2}+L_{1}\hat{x}_{1})&=e^{-iQ\theta_{1}}\phi(x_{2}+L_{2}\hat{x}_{2})=e^{-iQ\theta_{1}}W_{2}(x_{1})\phi(x)\ .
\end{align}
derived from imposing magnetic translations in different orders. Requiring the final results be identical leads to the condition that magnetic flux, $\Phi=eBL_{1}L_{2}$ is quantized in units of $2\pi N_{\Phi}$.

The study of thermodynamics is conducted through the construction of the relevant Green's function. The infinite volume, charged Green's function, $G^{\infty}_{+}(x',x)\equiv\langle\,\pi^{+}(x')\pi^{-}(x)\,\rangle$ in a magnetic background was constructed in the middle of the previous century~\cite{Schwinger:1951nm} and has been utilized extensively within chiral perturbation theory literature to construct the free energy, chiral condensate and the renormalized magnetization. For a finite volume analysis, we require its finite volume counterpart, $G_{+}(x'x)$ that satisfies the Green's function relation 
\begin{align}
(-D_{\mu}D_{\mu}+m^{2})G_{+}(x',x)=\prod_{\mu=0}^{3}\delta_{L_{\mu}}(x'_{\mu}-x_{\mu})
\end{align}
in addition to magnetic periodic boundary conditions. The $\delta$-functions are finite volume counterparts and possess compact support, i.e. they are non-vanishing for coincident points modulo periodicity. The resulting Green's function is related to its infinite volume counterpart through a sum over images~\cite{tiburzi2014neutron}
\begin{align}
\label{eq:GFV}
G_{+}(x',x)&=\sum_{\nu_{\mu}\in\mathbb{Z}}e^{iQ\theta_{1}\nu_{1}}[W^{\dagger}_{2}(x_{1})]^{\nu_{2}}G_{+}^{\infty}(x'+\nu_{\mu}L_{\mu},x)\ ,
\end{align}
that unlike the zero field counterpart is dressed by translation symmetry breaking inverse Wilson lines. The infinite volume Green's function, $G_{\infty}(x',x)\equiv e^{ieB\Delta x_{1}\overline{x}_{2}}g_{\infty}(\Delta x)$, contains a translational symmetry breaking phase and the following translationally symmetric contribution
\begin{align}
\label{eq:Ginf}
g_{\infty}(\Delta x)&=\frac{1}{(4\pi)^{2}}\int_{0}^{\infty}\frac{ds}{s^{2}}\frac{eBs}{\sinh{eBs}}e^{-m_{\pi}^{2}s}\exp\left(-\frac{1}{4s}\left[\frac{eBs}{\tanh eBs}\Delta x_{\perp}^{2}+\Delta x_{\parallel}^{2}\right]\right)\ 
\end{align}
with the full expression being coordinate independent in the coincident limit. We have used for notational compactness $\Delta x_{\mu}=x'_{\mu}-x_{\mu}$, $\overline{x}_{\mu}=\tfrac{1}{2}(x_{\mu}'+x_{\mu})$. 

\begin{figure*}[t!]
\centering
\includegraphics[width=0.31\textwidth]{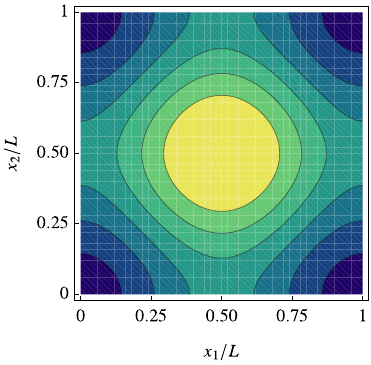}\,\,\,\,\,
\includegraphics[width=0.31\textwidth]{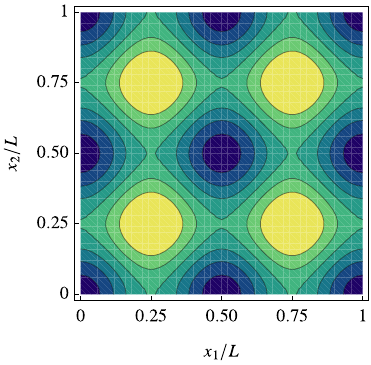}\,\,\,\,\,
\includegraphics[width=0.31\textwidth]{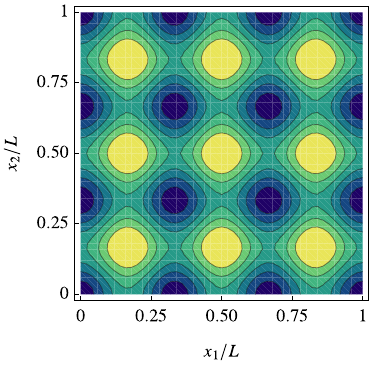}
\caption{Density plots of the local chiral condensate, $N_{\Phi}=1,2,3$ respectively, on the transverse plane (normalized by the size $L$).} 
\label{fig:cc}     
\end{figure*}

\begin{figure*}[t!]
\centering
\includegraphics[width=0.48\textwidth]{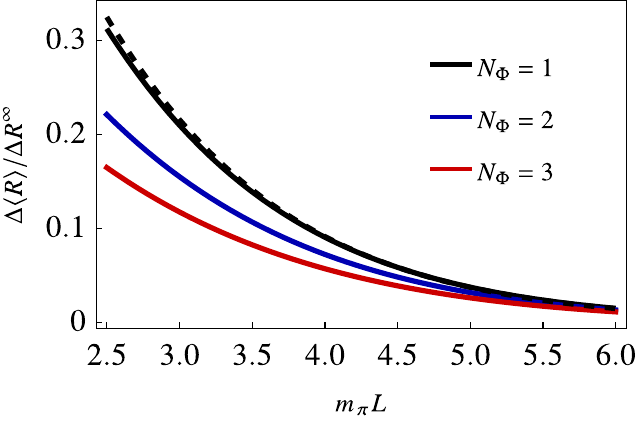}\,\,\,\,\,
\includegraphics[width=0.48\textwidth]{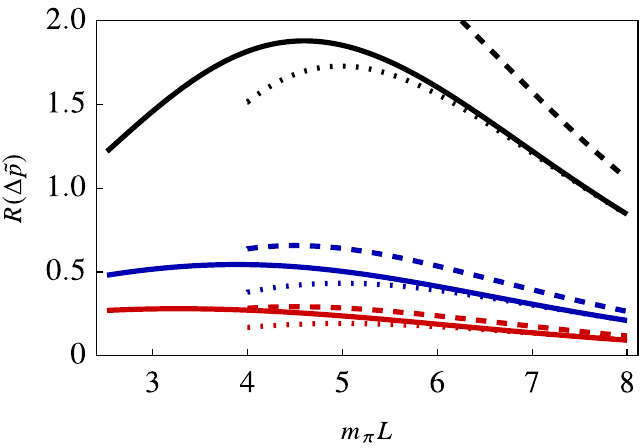}
\caption{(Left) Plot of the magnetic field dependence of the finite volume effect on the chiral condensate, Eq~(\ref{eq:DR-DRinf}), as function of $m_{\pi}L$. (Right) Plot of finite volume effect on the magnetic pressure anisotropy, Eq.~\ref{eq:RDp}, as function of $m_{\pi}L$. The exact results for presented in solid lines while the leading asymptotic behavior are shown in dashed lines. The asymptotic behavior that incorporates subleading corrections are shown in dotted lines. } 
\label{fig:Rcc-Rp}     
\end{figure*}

The periodic Green's function, however, does not share this property with translational symmetry broken down to $Z_{N_{\Phi}}$ in each of the transverse directions, a property encoded in the translational properties of the Wilson lines, which for integer $n$ satisfies
\begin{align}
W_{1}(x_{2}+\tfrac{n}{N_{\Phi}}L_{2})&=W_{1}(x_{2})&
W_{2}(x_{1}+\tfrac{n}{N_{\Phi}}L_{2})&=W_{2}(x_{1})\ .
\end{align} 
Discrete translation symmetry in the transverse directions is also evident in the (local) chiral condensate, defined in terms of the QCD partition function $Z$ and a scalar source $S(x)$, $\langle\,\bar{\psi}(x)\psi(x)\,\rangle=\left.-\frac{\delta\log Z}{\delta S(x)}\right|_{S(x)\rightarrow m_{q}}$. We construct the dimensionless ratio
\begin{align}
R(x_{\perp})&=\frac{\langle\,\overline{\psi}(x)\psi(x)\,\rangle-\langle\,\overline{\psi}\psi\,\rangle^{\infty}}{\langle\,\overline{\psi}\psi\,\rangle_{0}^{\infty}}
\end{align}
that characterizes deviations of the chiral condensate from its infinite volume counterpart, normalized by the tree level condensate. Chiral perturbation theory predicts a doubly periodic structure in the transverse directions 
\begin{align}
R(x_{\perp})&=-\int_{0}^{\infty}\frac{ds}{(4\pi sF_{\pi})^{2}}e^{-m_{\pi}^{2}s}\left[3e^{-\frac{L^{2}}{4s}}+\frac{eBs}{\sinh eBs}\left\{2e^{-\frac{L^{2}}{4s}}+2\cos(\tfrac{2\pi N_{\Phi}x_{1}}{L})+2\cos(\tfrac{2\pi N_{\Phi}x_{2}}{L})\right\}\right]\ ,
\end{align}
with leading finite volume corrections arising through transverse windings with magnitude one, $|\boldsymbol{\nu}|=1$. While twist angles have been set to zero, their presence add independent phases in the periodic functions that appear in the transverse directions. The first contribution arises from the neutral sector while the second contribution is due to charged pions -- the latter contains a spatially homogeneous piece that arises through windings in the longitudinal direction. The large volume limit is defined through $m_{\pi}L\rightarrow\infty$ and taken assuming fixed flux, $\frac{eBs}{\sinh eBs}\rightarrow 1$, with the leading contribution characterized by
\begin{align}
R(x_{\perp})
&=-\left[\frac{5}{2}+\cos\left(\frac{2\pi N_{\Phi}x_{1}}{L}\right)+\cos\left(\frac{2\pi N_{\Phi}x_{2}}{L}\right)\right]\times\frac{m_{\pi}^{2}}{F_{\pi}^{2}}\frac{e^{-m_{\pi}L}}{(2\pi m_{\pi}L)^{3/2}}\ .
\end{align}
In Fig.~\ref{fig:cc} we plot the local chiral condensate for the lowest three flux quanta, $N_{\Phi}=1,2,3$, which is periodic on the transverse plane. 
The periodic structure vanishes upon averaging due to projection to contributions arising from zero windings. The resulting average shift (normalized by the tree level condensate)
\begin{align}
\langle R\rangle-\langle R\rangle_{B=0}&=-\int_{0}^{\infty}\frac{ds}{(4\pi sF_{\pi})^{2}}e^{-m_{\pi}^{2}s}\left[\left(\frac{eBs}{\sinh eBs}-1\right)\left(\vartheta_{3}(0,e^{-\frac{-L_{3}^{2}}{4s}})-1\right)\right]\ .
\end{align}
characterizes the size of the finite volume effect on the shift of the chiral condensate induced by the magnetic field. Comparison with its infinite volume counterpart 
\begin{align}
R^{\infty}-R^{\infty}_{B=0}&=-\int_{0}^{\infty}\frac{ds}{(4\pi sF_{\pi})^{2}}e^{-m_{\pi}^{2}s}\left[\left(\frac{eBs}{\sinh eBs}-1\right)\right]\ ,
\end{align}
is performed through their ratio that is plotted in Fig.~\ref{fig:Rcc-Rp},
\begin{align}
\label{eq:DR-DRinf}
\frac{\Delta\langle R\rangle}{\Delta R^{\infty}}&=\frac{\langle R\rangle-\langle R\rangle_{B=0}}{R^{\infty}-R^{\infty}_{B=0}}=e^{-m_{\pi}L}\sqrt{2\pi m_{\pi}L}\left[1+\mathcal{O}\left(\frac{1}{m_{\pi}L}\right)+\mathcal{O}\left(\frac{N_{\Phi}^{2}}{(m_{\pi}L)^{2}}\right)+\mathcal{O}(e^{-m_{\pi}L})\right]\ .
\end{align}
The asymptotic structure is notably independent of flux quanta as is the leading power law correction with flux quanta dependence first appearing in the subleading correction. The asymptotic behavior most closely agrees with the lowest flux quanta result with agreement already excellent for box sizes with $m_{\pi}L=2.5$ which is in significant contrast to higher fluxes. For $m_{\pi}L=4$, the size of finite volume corrections is approximately $\gtrsim 10\%$ for the lowest flux while for higher fluxes, finite size effects are significantly smaller as the Landau wavefunctions fit more comfortably in the finite box.

In addition to finite volume corrections on the chiral condensate, it is also possible to consider such corrections on the matter contribution to the free energy, $\mathcal{F}_{B}$ and renormalized magnetization, which measures the response of the free energy to infinitesimal variations of the magnetic field, $\mathcal{M}_{r}=-\frac{\partial\mathcal{F}_{B}}{\partial B}$. The latter is studied in lattice QCD through its relation to pressure anisotropy~\cite{bali2013magnetic}. Since magnetic flux is quantized on the lattice, it is more natural and computationally cheaper to fix flux rather than the magnetic field when varying the size of the box
\begin{align}
p_{i}&=\left.-\frac{L_{i}}{V}\frac{\partial F_{B}}{\partial L_{i}}\right|_{B,L_{j\neq i}}\ \ ,\ \  
\widetilde{p}_{i}=\left.-\frac{L_{i}}{V}\frac{\partial F_{B}}{\partial L_{i}}\right|_{\Phi, L_{j\neq i}}\ .
\end{align}
Pressure characterizes the response of the free energy, $F_{B}=V\mathcal{F}_{B}$ to changes in the size of the system. The finite volume correction, $\mathcal{F}_{B}^{FV}$ to the infinite volume free energy, $\mathcal{F}_{B}^{\infty}$, that excludes the pure gauge, renormalized contribution,
\begin{align}
\mathcal{F}_{B}^{\infty}&=-\frac{1}{(4\pi)^{2}}\int_{0}^{\infty}\frac{ds}{s^{3}}\ e^{-m^{2}s}\left[\frac{eBs}{\sinh eBs}-1-\frac{1}{6}(eBs)^{2}\right]\\
\mathcal{F}_{B}^{FV}&=-\frac{1}{(4\pi)^{2}}\int_{0}^{\infty}\frac{ds}{s^{3}}\ e^{-m^{2}s}\left[\theta_{3}(\tfrac{L_{3}^{2}}{4s})-1\right]\left[\frac{eBs}{\sinh eBs}-1\right]\ ,
\end{align}
requires no further renormalization. Rather fascinatingly, finite volume effects from the transverse plane only appears through flux quantization -- transverse lengths do not appear explicitly in the finite volume correction. The longitudinal extent, however, appears in the finite volume free energy through a Jacobi theta function. The two pressures in the longitudinal direction depend on the longitudinal extent and are identical,
\begin{align}
\widetilde{p}_{3}&=p_{3}=-\mathcal{F}_{B}-L_{3}\frac{\partial \mathcal{F}_{B}}{\partial L_{3}}&
\widetilde{p}_{\perp}-p_{\perp}=-B\mathcal{M}_{r}\ .
\end{align}
However, the transverse pressures, $\widetilde{p}_{\perp}$, and $p_{\perp}$, in the two definitions differ due to an additional contribution proportional to renormalized magnetization that appears when the flux is fixed. The pressure anisotropy, $\Delta\widetilde{p}$, characterizes the pressure difference at fixed flux between the transverse and longitudinal directions
\begin{align}
\label{eq:pressure}
\Delta\widetilde{p}&=\widetilde{p}_{\perp}-\widetilde{p}_{3}&
\Delta\widetilde{p}^{\,\infty}&=\lim_{L_{3}\rightarrow\infty}\left(\,\widetilde{p}_{\perp}-\widetilde{p}_{3}\,\right)=-B\mathcal{M}_{r}^{\infty}\ .
\end{align}
We compare this pressure anisotropy with its infinite volume counterpart, $\Delta\widetilde{p}^{\,\infty}$, constructed by taking the infinite longitudinal extent limit while keeping $eB$ fixed. This is equivalent to comparing the finite volume magnetization with its infinite volume counterpart. For this comparison, we plot the quantity
\begin{align}
\label{eq:RDp}
R(\Delta\widetilde{p})&=\frac{\Delta \widetilde{p}-\Delta \widetilde{p}^{\,\infty}}{\Delta\widetilde{p}^{\,\infty}}=\frac{30}{7(2\pi)^{3/2}N_{\Phi}^{2}}(m_{\pi}L)^{9/2}e^{-m_{\pi}L}\left[1+\mathcal{O}\left(\frac{1}{m_{\pi}L}\right)+\mathcal{O}(e^{-m_{\pi}L})\right]\ .
\end{align}
and its large volume, asymptotic structure in Fig.~\ref{fig:Rcc-Rp}. The size of finite volume corrections are substantial, significantly more so than the chiral condensate even for the largest magnetic fields. Unlike the chiral condensate, the renormalized magnetization is small and scales cubically with the magnetic field for small fields. Additionally, the asymptotic value of $R(\Delta\widetilde{p})$ is reached slowly. The leading asymptotic behavior arises exclusively through the term proportional to the longitudinal derivative of the finite volume free energy, Eq.~(\ref{eq:pressure}). The leading relative correction is a power law (with coefficient $-13/8$) that arises through the finite volume contribution to $-B\mathcal{M}_{r}$.  Its inclusion significantly improves agreement with the exact result (shown in dotted lines).

Finite volume corrections to thermodynamic observables and condensates in a magnetic field are substantial since the effect of a magnetic field enters chiral perturbation theory at the same order as finite size effects. As such significantly large volumes are required to minimize the effect -- for the lowest quanta, the chiral condensate exhibits corrections of size $\sim10\%$ for $m_{\pi}L=4$ and an analysis of the neutral pion magnetic polarization, omitted in this review, presents corrections of a similar size, $\sim10\%$ for $m_{\pi}L=4$~\cite{adhikari2023qcd}. The corrections to the renormalized magnetization are very significant even for $m_{\pi}L=8$ due to their smallness. These results, while obviously relevant to lattice QCD calculations, may also be of theoretical and experimental interest in heavy ion collisions where finite size effects may be important for certain observables.

\section{
QED Fermions in a noisy magnetic field background: Perturbation theory and renormalized quasiparticles }\label{Munoz}

Inhomogeneous and rapidly fluctuating magnetic fields are present in important physical scenarios, particularly in early stages of heavy-ion collisions~\cite{Alam_2021,ayala2022anisotropic,Inghirami_2020,Ayala:2019jey,Ayala:2017vex} and the corresponding genesis of the quark-gluon plasma~\cite{Busza_2018,Hattori_2016,Hattori_2018,Buballa_2005}. Their effects can influence the physical properties of both charged as well as neutral particles, the latter due to the quantum mechanical fluctuations of the vacuum, leading to the creation of virtual charged fermion-antifermion pairs. 
Classical, static and uniform background magnetic fields have been extensively studied in the context of the gluon polarization tensor~\cite{Hattori_2013,Hattori_2016,Ayala_Pol_020,Hattori_2018} in relation to the emergence of the vacuum birefringence phenomena~\cite{Hattori_2013,Ayala_Pol_020}, and the propagation of fermions in such magnetized background~\cite{Ayala_Hernandez_2021}, as expressed by the self-energy, that leads to the definition of a magnetic mass and to an spectral broadening~\cite{Ayala_mass_021} due to Landau level mixing. Nevertheless, since spatio-temporal fluctuations with respect to a {\it{finite}} background magnetic field may indeed exist in the  different aforementioned physical scenarios~\cite{Inghirami_2020}, in the present work we shall study their effect on the renormalization of the fermion propagator itself in QED. 

As discussed in a recent work~\cite{Castano_Munoz_PhysRevD.107.096014}, a perturbative treatment of such fluctuations in the framework of the replica method~\cite{Parisi_Mezard_1991,Kardar_Parisi_1986} allows us to show that their effect leads to a renormalization of the charge $e\rightarrow z_3 e$ and an effective refraction index $v'/c = z^{-1}$. We further show~\cite{Castano_Munoz_PhysRevD.107.096014} that $z_3$ and $z$ depend not only on the energy scale, as usual, but also on the magnitude of the average magnetic field $|\mathbf{B}|$, as well as on the strength of its spatial fluctuations, that we define as $\Delta_B$.

\subsection{The model}

Let us start by assuming a system of QED fermions $\psi(x)$ immersed in a classical static magnetic field background, that nevertheless possesses random spatial fluctuations with respect to an otherwise constant mean value. For the gauge fields $A^{\mu}(x)$, we shall distinguish three physically different contributions~\cite{Castano_Munoz_PhysRevD.107.096014}
\begin{eqnarray}
A^{\mu}(x) \rightarrow A^{\mu}(x) + A^{\mu}_\text{BG}(x) + \delta A^{\mu}_\text{BG}(\mathbf{x}).
\label{eq_Atot}
\end{eqnarray}

Here, $A^{\mu}(x)$ represents the dynamical photonic quantum field, while BG stands for ``background'', representing the presence of a classical external field imposed by the experimental conditions. Moreover, for this BG contribution, we consider the effect of static (quenched) white noise spatial fluctuations $\delta A^{\mu}_\text{BG}(\mathbf{x})$ with respect to the mean value $A_\text{BG}^{\mu}(x)$, satisfying the statistical properties
\begin{eqnarray}
\langle \delta A^{j}_\text{BG}(\mathbf{x}) \delta A^{k}_\text{BG}(\mathbf{x}')\rangle &=&  \Delta_{B}\delta_{j,k}\delta^{3}(\mathbf{x}-\mathbf{x}'),\nonumber\\
\langle \delta A^{\mu}_\text{BG}(\mathbf{x})\rangle &=& 0.
\label{eq_Acorr}
\end{eqnarray}

These statistical properties are represented by a Gaussian functional distribution of the form
\begin{eqnarray}
dP\left[ \delta A^{\mu}_\text{BG} \right] = \mathcal{N} e^{-\int d^3x\,\frac{\left[\delta A_\text{BG}^{\mu}(\mathbf{x})\right]^2}{2 \Delta_B}}
\mathcal{D}\left[\delta A_\text{BG}^{\mu}(\mathbf{x})\right].
\label{eq_Astat}
\end{eqnarray}

In the context of heavy-ion collisions (HIC), we expect strong spatial fluctuations in the strong magnetic fields $\mathbf{B} = \nabla\times\mathbf{A}_{BG}$ generated within small spatial regions in the early stages of individual collisions, where the characteristic length scale is $L \sim \sqrt{\sigma}$, with $\sigma$ being the effective cross-section. In these collisions, the axial (z-direction) component of the magnetic field dominates, such that on average we have $\langle \mathbf{B} \rangle = \hat{e}_3\,B$. However, there are also smaller transverse components $\delta B_x$ and $\delta B_y$, such that the fluctuation of the field within the small collision region can be estimated to be on the order $(\delta B)^2 \sim (\delta B_x)^2 + (\delta B_y)^2$. Since many such collisions occur at different points in space, an approximate model for this physical scenario is provided by the spatial random noise Eq.~\eqref{eq_Acorr}. By dimensional analysis, the magnitude of $\Delta_B$ is of the order:
\begin{equation}
\Delta_B \sim \left(\delta B\right)^2\,L^{5} \sim \left(\delta B\right)^2\,\sigma^{5/2}.
\label{eq_DeltaB}
\end{equation}

As discussed in our recent work~\cite{Castano_Munoz_PhysRevD.107.096014}, this analysis leads to the following estimation
\begin{equation}
\Delta_B \sim \pi^{5/2}\left(\delta B\right)^2r_0^5N^{5/3}\left(\frac{N_\text{part}}{2N}\right)^{5/3}.
\end{equation}

In peripheral heavy-ion collisions, the magnetic field fluctuations along the transverse plane are approximately $|e\,\delta B| \sim m_\pi^2/4$, where $m_\pi$ is the pion mass \cite{PhysRevC.83.054911, castano2021effects}. For instance, in a Au+Au collision with $N=197$, and if $N_\text{part}/N=1/2$, we obtain $\Delta \equiv e^2\Delta_B \sim 2.6~\text{MeV}^{-1}$,
while for less central collisions with $N_\text{part}/N=1/8$ we have
$\Delta \sim 0.26~\text{ MeV}^{-1}$.

The Lagrangian for this model is a superposition of two contributions
\begin{eqnarray}
\mathcal{L} = \mathcal{L}_\text{FBG} + \mathcal{L}_\text{NBG},  
\end{eqnarray}
where the first one represents the system of Fermions (and photons) immersed in the deterministic background field (FBG)
\begin{eqnarray}
\mathcal{L}_\text{FBG} = \bar{\psi}\left(\ii\slashed{\partial} - e \slashed{A}_\text{BG} - e \slashed{A}   - m \right)\psi-\frac{1}{4}F_{\mu\nu}F^{\mu\nu},
\label{eq_LFBG}
\end{eqnarray}
with $F_{\mu\nu}=\partial_{\mu}A_{\nu}-\partial_{\nu}A_{\nu}$ the electromagnetic tensor for the quantum photon gauge fields. 
In contrast, the second term represents the interaction between the Fermions and the static classical noise (NBG), represented by the spatial fluctuations $\delta A^{\mu}_{BG}(x)$
\begin{eqnarray}
\mathcal{L}_\text{NBG} = \bar{\psi}\left( - e \delta\slashed{A}_\text{BG} \right)\psi.
\label{eq_LDBG}
\end{eqnarray}

The generating functional (in the absence of sources) for a given realization of the noisy  fields is given by
\begin{eqnarray}
Z[A] = \int \mathcal{D}[\bar{\psi},\psi]
e^{\ii\int d^4 x \left[ \mathcal{L}_\text{FBG} + \mathcal{L}_\text{NBG}  \right]}.
\end{eqnarray}

To study the physics of this system, we need to calculate the statistical average over the magnetic background noise $\delta A_\text{BG}^{\mu}$ of the $\overline{\ln Z}$. For this purpose, we apply the replica method, which is based on the following identity~\cite{Parisi_Mezard_1991}
\begin{eqnarray}
\overline{\ln Z[A]} = \lim_{n\rightarrow 0}\frac{\overline{Z^n[A]}-1}{n}.
\end{eqnarray}

Here, we defined the statistical average according to the Gaussian functional measure in Eq.~\eqref{eq_Astat}, and $Z^n$ is obtained by incorporating an additional ``replica" component for each of the Fermion fields, i.e. $\psi(x)\rightarrow \psi^{a}(x)$, for $1\le a \le n$. The ``replicated" Lagrangian has the same form as Eqs.~\eqref{eq_LFBG} and \eqref{eq_LDBG}, but with an additional sum over the replica components of the Fermion fields. Therefore, the averaging procedure leads to
\begin{eqnarray}
\overline{Z^n[A]} &=& \int 
\prod_{a=1}^{n}\mathcal{D}[\bar{\psi}^{a},\psi^{a}]
\int \mathcal{D}\left[\delta A_\text{BG}^{\mu}\right]e^{-\int d^3x\,\frac{\left[\delta A_\text{BG}^{\mu}(\mathbf{x})\right]^2}{2 \Delta_B}}\nonumber\\
&&\times e^{\ii\int d^4 x \sum_{a=1}^n \left( \mathcal{L}_\text{FBG}[\bar{\psi}^a,\psi^a] + \mathcal{L}_\text{NBG}[\bar{\psi}^a,\psi^a] \right)}\nonumber\\
&=& \int 
\prod_{a=1}^{n}\mathcal{D}[\bar{\psi}^{a},\psi^{a}] e^{\ii \bar{S}\left[\bar{\psi}^a,\psi^a;A \right]},
\label{eq_repl}
\end{eqnarray}
where in the last step the Gaussian integral over the background noise was explicitly calculated, leading to the definition of the effective averaged action for the replica system~\cite{Castano_Munoz_PhysRevD.107.096014}
\begin{eqnarray}
\bar{S}\left[\bar{\psi}^a,\psi^a;A \right]
&=& \int d^4 x \left(\sum_{a}\bar{\psi}^{a}\left(\ii\slashed\partial -  e \slashed{A}_\text{BG} - e \slashed{A} - m  \right)\psi^{a}-\frac{1}{4}F_{\mu\nu}F^{\mu\nu}\right)\nonumber\\
&+& \ii\frac{e^2\Delta_{B}}{2}\int d^4x\int d^4 y\sum_{a,b}\sum_{j=1}^{3}\bar{\psi}^{a}(x)\gamma^{j}\psi^{a}(x)\bar{\psi}^{b}(y)\gamma_{j}\psi^{b}(y)\delta^{3}(\mathbf{x}-\mathbf{y}).
\label{eq_Savg}
\end{eqnarray}

The result of the averaging procedure is an effective interacting theory, with an instantaneous local interaction proportional to the fluctuation amplitude $\Delta_B$ that characterizes the magnetic noise, as defined in Eq.~\eqref{eq_Acorr}. The ``free'' part of the action corresponds to Fermions in the average background classical field
$A_\text{BG}^{\mu}(x)$. We choose this background to represent a uniform, static magnetic field along the $z$-direction $\mathbf{B} = \hat{e}_3 B$, using the gauge~\cite{Dittrich_Reuter}
\begin{eqnarray}
A_\text{BG}^{\mu}(x) = \frac{1}{2}(0,-B x^2, B x^1,0).
\end{eqnarray}
Therefore, this allows us to use directly the Schwinger proper-time representation of the free-Fermion propagator dressed by the background field~\cite{Schwinger_1951,Dittrich_Reuter},
\begin{eqnarray}
S_F(x,x') = \Phi(x,x')\int\frac{d^4 p}{(2\pi)^4}
e^{-ip\cdot(x-x')}S_F(p),
\end{eqnarray}
Since the effective interaction is strictly local (i.e. proportional to $\delta^{(3)}(\mathbf{x}-\mathbf{y})$), the Schwinger phase is a global common factor in the corresponding Dyson equation for the dressed propagator. Therefore, we only need to consider the translational-invariant part of the Schwinger propagator 
\begin{eqnarray}
\left[S_\text{F}(k)\right]_{a,b}
=-\ii\delta_{a,b}\int_{0}^{\infty}\frac{d\tau}{\cos(\qB \tau)}
e^{\ii\tau\left(k_{\parallel}^2 - \mathbf{k}_{\perp}^2\frac{\tan(\qB\tau)}{\qB\tau}-m^2 + \ii\epsilon \right)}\left\{
\left[\cos(\qB\tau) + \ii\gamma^1\gamma^2\sin(\qB\tau)  \right](m + \slashed{k}_{\parallel})+\frac{\slashed{k}_{\perp}}{\cos(\qB \tau)}
\right\},
\label{eq_Sprop}
\end{eqnarray}
which is  diagonal in the replica space. Moreover, since it is an explicit function of the average magnetic field $\mathbf{B} = \nabla\times\mathbf{A}_\text{BG}$ (rather than a function of the background gauge field $A_\text{BG}^{\mu}$), it is also gauge-invariant. Here, as usual, we separated the parallel from the perpendicular directions with respect to the background external magnetic field by splitting the metric tensor as $g^{\mu\nu} = g_{\parallel}^{\mu\nu} + g_{\perp}^{\mu\nu}$, with
\begin{eqnarray}
g_{\parallel}^{\mu\nu} &=& \text{diag}(1,0,0,-1),\nonumber\\
g_{\perp}^{\mu\nu} &=& \text{diag}(0,-1,-1,0),
\end{eqnarray}
thus implying that for any 4-vector, such as the momentum $k^{\mu}$, we write $\slashed{k} = \slashed{k}_{\perp} + \slashed{k}_{\parallel}$,
and $k^2 = k_{\parallel}^2 - \mathbf{k}_{\perp}^2$. In particular, $k_{\parallel}^2 = k_0^2 - k_3^2$,
while $\mathbf{k}_{\perp}=(k^1,k^2)$ is the Euclidean 2-vector lying in the plane perpendicular to the field, such that its square-norm is  $\mathbf{k}_{\perp}^2 = k_1^2 + k_2^2$.

As we showed in our previous work~\cite{Castano_Munoz_PhysRevD.107.096014}, the Schwinger propagator can be expressed exactly as follows,
\begin{eqnarray}
\left[S_\text{F}(k) \right]_{a,b}=-\ii\delta_{a,b}\left[ 
\left( m + \slashed{k}_{\parallel} \right)\mathcal{A}_1
+  \ii\gamma^{1}\gamma^{2}\left( m + \slashed{k}_{\parallel} \right)  \mathcal{A}_2
+ \mathcal{A}_3 \slashed{k}_{\perp}
\right]\nonumber
\label{propSchwinger}
\end{eqnarray}
where we defined the scalar function
\begin{eqnarray}
\mathcal{A}_1(k,B) = \int_{0}^{\infty}d\tau e^{\ii\tau\left( k_{\parallel}^2 - m ^2 + \ii\epsilon\right) -\ii\frac{\mathbf{k}_{\perp}^2}{\qB}\tan(\qB \tau) },
\end{eqnarray}
that clearly reproduces the inverse scalar propagator (with Feynman prescription) in the zero-field limit
\begin{eqnarray}
\lim_{B\rightarrow 0}\mathcal{A}_1(k,B) = \frac{\ii}{k^2 - m^2 + \ii\epsilon}.
\end{eqnarray}
The other coefficients are nothing but momentum-derivatives of this basic scalar function, defined as follows
\begin{subequations}
\bea
\mathcal{A}_2(k,B)\equiv\int_0^\infty d\tau ~\tan(\qB\tau)e^{\ii\tau\left(k_\parallel^2-\tb{\tau}\mathbf{k}_\perp^2-m^2+\ii\epsilon\right)}
=\ii \qB\frac{\partial\mathcal{A}_1}{\partial(\mathbf{k}_{\perp}^2)},
\eea
\bea
\mathcal{A}_3(k,B)\equiv\int_0^\infty \frac{d\tau}{\cos^2(\qB\tau)}e^{\ii\tau\left(k_\parallel^2-\tb{\tau}\mathbf{k}_\perp^2-m^2+\ii\epsilon\right)}=\mathcal{A}_1+(\ii \qB)^2\frac{\partial^2\mathcal{A}_1}{\partial(\mathbf{k}_{\perp}^2)^2}.
\eea
\label{A2A3properties}
\end{subequations}

Moreover, with these definitions it is straightforward to calculate the inverse of the Schwinger propagator Eq.~\eqref{propSchwinger} as~\cite{Castano_Munoz_PhysRevD.107.096014}:
\bea
\hat{S}_\text{F}^{-1}(k)&=&\frac{\ii}{\mathcal{D}(k)}\left[\left(m - \slashed{k}_\parallel\right)\mathcal{A}_1- \ii\gamma^1\gamma^2\left(m - \slashed{k}_\parallel\right)\mathcal{A}_2-\mathcal{A}_3\slashed{k}_\perp\right],
\label{Sinverse}
\eea
where
\bea
\mathcal{D}(k)=\mathcal{A}_3^2\mathbf{k}_\perp^2-\left(\mathcal{A}_1^2-\mathcal{A}_2^2\right)\left(k_\parallel^2-m^2\right).
\label{Denprop}
\eea

Then, all the relevant expressions will be given in terms of $\mathcal{A}_1$.

\subsection{Perturbation theory: Self-energy and dressed propagator}
\begin{figure}[h!]
    \centering
    \includegraphics[scale=0.5]{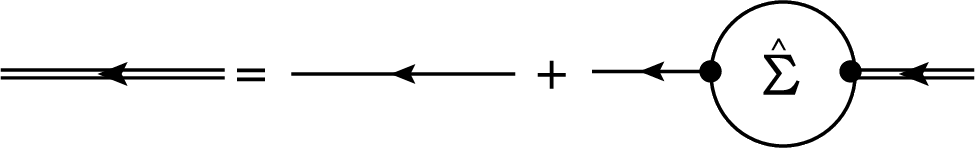}
    \caption{Dyson equation for the "dressed" propagator (double-line), in terms of the free propagator (single-line) and the self-energy $\Sigma$.}
    \label{fig:DiagramSelfEnergy2}
\end{figure}

\begin{figure}[h!]
    \centering
    \includegraphics[scale=0.45]{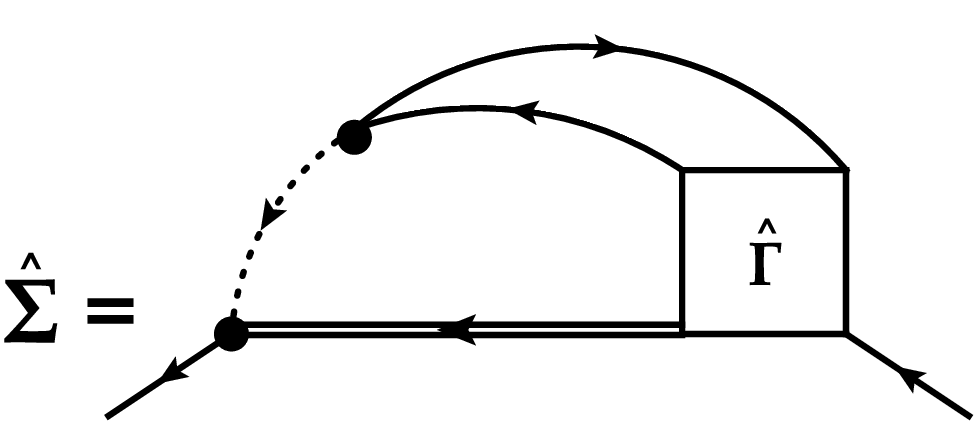}
    \caption{Skeleton diagram representing the self-energy for the effective interacting theory. The dashed line is the disorder-induced interaction $\Delta_B$, while the box $\hat{\Gamma}$ represents the 4-point vertex function.}
    \label{fig:DiagramSelfEnergy1}
\end{figure}

The effective action in Eq.~\eqref{eq_Savg} reveals a resulting fermion-fermion interaction arising from the average over the background magnetic noise. Starting from a free Fermion propagator, as defined by Eq.~\eqref{eq_Sprop}, we include the magnetic noise-induced interaction effects by ``dressing" the propagator with a self-energy, as shown diagrammatically in the Dyson equation depicted in Fig.~\ref{fig:DiagramSelfEnergy2}. We remark that for this theory, the skeleton diagram for the self-energy is represented in Fig.~\ref{fig:DiagramSelfEnergy1}. 

\begin{figure}
    \centering
    \includegraphics[scale=0.5]{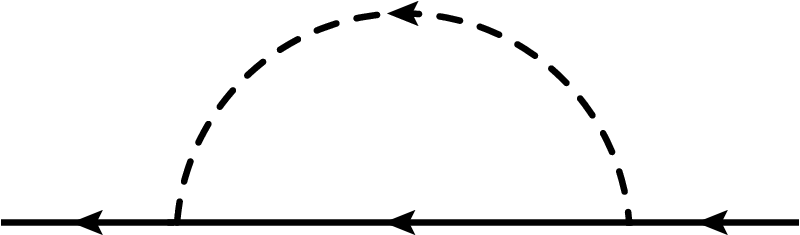}
    \caption{Self-energy diagram at first order in $\Delta = e^2\Delta_B$.}
    \label{fig:selfenergydiagram}
\end{figure}
At first-order in $\Delta \equiv e^2\Delta_B$, the self-energy depicted in the Feynman diagram in Fig.~\eqref{fig:selfenergydiagram} is given by the integral expression~\cite{Castano_Munoz_PhysRevD.107.096014}
\bea
\hat{\Sigma}_\Delta(q) &=&(\ii\Delta)\int\frac{d^3p}{(2\pi)^3}\gamma^j\hat{S}_F(p+q;p_0=0)\gamma_j\nonumber\\
&=& \frac{i(\ii\Delta)}{(2\pi)^3}\int d^3 p\left\{3\left(\gamma^0 q_0-m \right)\mathcal{A}_1(q_0,p_3;\mathbf{p}_{\perp})+ \ii\gamma^1\gamma^2\left(m - q_0\gamma^0 \right)(\ii \qB)\frac{\partial}{\partial \mathbf{p}_{\perp}^2}\mathcal{A}_1(q_0,p_3;\mathbf{p}_{\perp})
\right\}.\nonumber
\label{SigmaA1}
\eea

After some elementary transformations~\cite{Castano_Munoz_PhysRevD.107.096014}, the self-energy (valid at all orders in the background average magnetic field $B$) is
\bea
\hat{\Sigma}_\Delta (q) 
=\frac{\ii(\ii\Delta)}{(2\pi)^3}
\left[3\left(\gamma^0 q_0 - m\right)\widetilde{\mathcal{A}}_1(q_0)- \ii\gamma^1\gamma^2(\ii\pi \qB)\left(m - q_0\gamma^0 \right)\widetilde{\mathcal{A}}_2(q_0)
\right],
\label{SigmaGeneral}
\eea
where we have defined:
\bea
\widetilde{\mathcal{A}}_1(q_0)&\equiv&\int d^3 p\mathcal{A}_1(q_0,p_3;\mathbf{p}_{\perp}),\nn\\
\widetilde{\mathcal{A}}_2(q_0)&\equiv&\int_{-\infty}^{+\infty}dp_3\mathcal{A}_1(q_0,p_3;\mathbf{p}_{\perp} = 0).
\label{Atildes}
\eea

Inserting this self-energy into the Dyson equation, as depicted diagrammatically in Fig.~\ref{fig:DiagramSelfEnergy2}, we obtain the dressed inverse propagator at first-order in $\Delta$
\bea
\hat{S}_\Delta^{-1}(k)=\hat{S}_\text{F}^{-1}(k)-\hat{\Sigma}_\Delta,
\label{SF}
\eea
so that by using Eqs.~(\ref{Sinverse}) and~(\ref{SigmaGeneral}) we explicitly obtain~\cite{Castano_Munoz_PhysRevD.107.096014}
\begin{eqnarray}
S_{\Delta}^{-1}(q) &=& \frac{\ii z}{\mathcal{D}(q)}\left[ 
\left(m - q_0\gamma^{0}- z^{-1}q_3\gamma^3  \right)\mathcal{A}_1(q)
-z_3\left(\ii\gamma^1\gamma^2  \right)
\left(m - q_0\gamma^{0} - z^{-1}q_3\gamma^3  \right)\mathcal{A}_2(q)-\ii \mathcal{A}_3(q)z^{-1}\slashed{q}_{\perp}
\right]\nonumber\\
&=& \frac{\ii z}{\mathcal{D}(q)}\left[ 
\left(m - \tilde{\slashed{q}}_{\parallel}  \right)\mathcal{A}_1(q)
-z_3\left(\ii\gamma^1\gamma^2  \right)
\left(m - \tilde{\slashed{q}}_{\parallel}  \right)\mathcal{A}_2(q)-\ii \mathcal{A}_3(q)\tilde{\slashed{q}}_{\perp}\right],
\label{SDeltainverse}
\end{eqnarray}
where in the last line we defined the four-vector $\tilde{q}^{\mu} = (q^0,z^{-1}\mathbf{q})$ that incorporates the definition of the effective refraction index $v'/c= z^{-1}$ due to the random magnetic fluctuations. Here, $z$ and $z_3$ are the renormalization factors for the wave fermion field and the charge, respectively. While $z$ will emerge as a global factor in the dressed propagator, the factor $z_3$ will only be associated to the tensor structures involving the spin-magnetic field interaction $e\sigma_{\mu\nu}F^{\mu\nu}_\text{BG} = i\gamma_1\gamma_2 e B$. 
The explicit algebraic expressions, at arbitrary order in the background magnetic field, are~\cite{Castano_Munoz_PhysRevD.107.096014}
\begin{subequations}
 \bea
&&z= 1+\frac{3\ii\Delta }{(2\pi)^3}\frac{\widetilde{\mathcal{A}}_1(q_0)}{\mathcal{A}_1(q)}\mathcal{D}(q),
\label{z}
\eea
\bea
&&z_3=\frac{1-\frac{\ii\pi(\ii\Delta)(\qB) }{(2\pi)^3}\frac{\widetilde{\mathcal{A}}_2(q_0)}{\mathcal{A}_2(q)}\mathcal{D}(q)}{1+\frac{3\ii\Delta }{(2\pi)^3}\frac{\widetilde{\mathcal{A}}_1(q_0)}{\mathcal{A}_1(q)}\mathcal{D}(q)},
\label{z3}
\eea
 \bea
 m'=m,
 \eea
 and
 \begin{eqnarray}
 \frac{v'}{c} = z^{-1} = \left(1 + \frac{3\ii\Delta }{(2\pi)^3}\frac{\widetilde{\mathcal{A}}_1(q_0)}{\mathcal{A}_1(q)}\mathcal{D}(q)\right)^{-1}
 \end{eqnarray}
\end{subequations}
By comparing Eq.~\eqref{SDeltainverse} with Eq.~\eqref{Sinverse}, it is clear that they possess the same tensor structure. Therefore, by means of the elementary properties of the Dirac matrices, this expression can be readily inverted to obtain the ``magnetic noise-dressed" fermion propagator~\cite{Castano_Munoz_PhysRevD.107.096014}
\begin{eqnarray}
&&S_{\Delta}(q) = -\ii z^{-1}\frac{\mathcal{D}(q)}{\tilde{\mathcal{D}}(q)}\left[ 
\left( m + \tilde{\slashed{q}}_{\parallel} \right)\mathcal{A}_1(q)\right.\nonumber\\ 
&&\left.+ \ii z_3\gamma^1\gamma^2 \left( m + \tilde{\slashed{q}}_{\parallel} \right)
\mathcal{A}_2(q) + \mathcal{A}_3(q)\tilde{\slashed{q}}_{\perp}
\right],
\end{eqnarray}
where $\mathcal{D}(q)$ was defined in Eq.~\eqref{Denprop}, and
\begin{eqnarray}
\tilde{\mathcal{D}}(q) = \mathcal{A}_3^2 z^{-2}\mathbf{q}_{\perp}^2 - \left( \mathcal{A}_1^2 - \mathcal{A}_2^2 \right)\left( z^{-2}q_{\parallel}^2 - m^2 \right)
\end{eqnarray}

Let us now discuss the explicit magnetic field and magnetic noise dependence of the renormalized parameters defined in Eqs.~\eqref{z} and~\eqref{z3}, i.e $z$ and $z_3$. For this purpose, we shall distinguish three different regimes, corresponding to the very weak, the intermediate and the ultra-intense magnetic field, respectively.

\subsubsection{Very weak field $eB/m^2\ll 1$}
As shown in detail in~\cite{Castano_Munoz_PhysRevD.107.096014}, for very weak fields $eB/m^2\ll 1$ the function $\mathcal{A}_1(k,B)$ can be expanded in terms of the power series
\begin{eqnarray}
\mathcal{A}_1(k,B) = \frac{\ii}{\mathcal{D}_\parallel }\left(
1 + \sum_{j=1}^{\infty}\left( \frac{\ii \qB}{\mathcal{D}_\parallel }  \right)^j \mathcal{E}_{j}(x)\right),
\label{eq_A1_exp1}
\end{eqnarray}
where for notational simplicity, we defined the ``parallel" inverse scalar propagator
\begin{eqnarray}
\mathcal{D}_\parallel  = k_{\parallel}^2 - m^2 + \ii\epsilon,
\end{eqnarray}
and the dimensionless variable $x = \mathbf{k}_{\perp}^2/\qB$. We also defined the polynomials $\mathcal{E}_{j}(x)$, as those generated by the function $e^{-\ii x \tan v}$, i.e.
\begin{eqnarray}
\mathcal{E}_{j}(x) = \lim_{v\rightarrow 0}\frac{\partial^j}{\partial v^j}\left( e^{-\ii x \tan v} \right).
\end{eqnarray}
At the lowest order, and after subtracting the divergent vacuum contribution
from Eq.~\eqref{eq_A1_exp1}, we have~\cite{Castano_Munoz_PhysRevD.107.096014}
\begin{eqnarray}
\mathcal{A}_1(k,B) - \mathcal{A}_1(k,0) = \frac{-2\ii \left(e B\right)^2\mathbf{k}_{\perp}^2}{\left[k^2 - m^2 + \ii\epsilon \right]^4}+ O((eB)^4).\nonumber\\
\label{eq:A1low}
\end{eqnarray}

Therefore, using this weak field expansion of the propagator, we obtain~\cite{Castano_Munoz_PhysRevD.107.096014}
\bea
\widetilde{\mathcal{A}}_1(q_0)
=-\frac{\pi^2}{6}\frac{(\qB)^2}{(q_0^2-m^2)^{3/2}} + \mathcal{O}((eB)^4).
\label{eq:tildeA1Weak}
\eea
In addition, at order $\mathcal{O}((\qB)^2)$ we also obtain~\cite{Castano_Munoz_PhysRevD.107.096014}
\bea
\widetilde{\mathcal{A}}_2(q_0)
=\frac{\pi}{\sqrt{q_0^2-m^2}}+ \mathcal{O}((eB)^4).
\label{eq:A2tildeweak}
\eea

Therefore, from Eqs.~(\ref{z}),~(\ref{eq:A1low}), ~(\ref{eq:tildeA1Weak}), and~(\ref{eq:A2tildeweak}), we can directly evaluate the renormalization parameters to obtain
 \bea
z&=&1+\frac{3\ii\Delta }{(2\pi)^3}\frac{\widetilde{\mathcal{A}}_1(q_0)}{\mathcal{A}_1(q)}\mathcal{D}(q)\nn\\
&=& 1+ O((eB)^4),
\eea
and similarly from Eq.~\eqref{z3}
\bea
z_3 &=& 1 + O((eB)^4).
\eea
\subsubsection{Intermediate field}

For intermediate magnetic field intensities, we  calculate the integral $\mathcal{A}_1$
by means of an expansion in terms of Landau levels~\cite{Castano_Munoz_PhysRevD.107.096014}. For this purpose, let us consider the generating function of the Laguerre polynomials\cite{Gradshteyn2}
\begin{eqnarray}
e^{-\frac{x}{2}\frac{1-t}{1+t}} = (1 + t)e^{-x/2}\sum_{n=0}^{\infty}(-t)^nL_{n}^{0}(x),
\end{eqnarray}
since
\begin{eqnarray}
e^{-\ii x \tan v} &=& \exp\left[-x\left(1 - e^{-2 \ii v} \right)/\left(1 + e^{-2 \ii v} \right)\right]\\
&=& \left(1 + e^{-2 \ii v} \right) e^{-x}
\sum_{n=0}^{\infty}(-1)^n e^{-2 \ii n v} L_{n}^{0}(2 x)\nonumber
\label{eq_gf}
\end{eqnarray}

Therefore, we have (for $x = \mathbf{k}_{\perp}^2/\qB$) after evaluating the corresponding integrals~\cite{Castano_Munoz_PhysRevD.107.096014}
\begin{eqnarray}
\mathcal{A}_1(k) 
= \ii \frac{e^{-x}}{\mathcal{D}_\parallel }
\left[ 
1 + \sum_{n=1}^{\infty}\frac{(-1)^n\left[ 
L_{n}^{0}(2x) - L_{n-1}^{0}(2x)
\right]}{1 - 2 n\frac{\qB}{\mathcal{D}_\parallel }}
\right].\nonumber
\label{eq_A1_Land_main}
\end{eqnarray}

Inserting this expression into the definition of $\widetilde{\mathcal{A}}_1$ of Eq.~\eqref{Atildes}, we obtain
\bea
\widetilde{\mathcal{A}}_1(q_0)&=&\int d^3 p\mathcal{A}_1(q_0,p_3;\mathbf{p}_{\perp})\nn\\
&=& \mathcal{I}_1 + \sum_{n=1}^{\infty}(-1)^{n}\mathcal{I}_{2,n}
\label{eq_A1B}
\eea

Here, we defined~\cite{Castano_Munoz_PhysRevD.107.096014}
\bea
\mathcal{I}_1&=&\ii\int d^3 p\frac{e^{-\mathbf{p}_{\perp}^2/\qB}}{q_0^2-p_3^2-m^2+\ii\epsilon}\nn\\
&=&\frac{\pi^2\qB}{\sqrt{q_0^2-m^2+\ii\epsilon}},
\label{eq_I1}
\eea
and
\bea
\mathcal{I}_{2,n}&=&\ii \int d^3 p\, e^{-\mathbf{p}_{\perp}^2/\qB}\frac{
L_{n}\left(\frac{2\mathbf{p}_{\perp}^2}{\qB}\right)-L_{n-1}\left(\frac{2\mathbf{p}_{\perp}^2}{\qB}\right)}{q_0^2-p_3^2-m^2-2n\qB+\ii\epsilon}\nn\\
&=& 2\pi e B (-1)^n \int_{-\infty}^{\infty}\frac{dp_3}{q_0^2 - m^2 - p_3^2 - 2 n q B + \ii\epsilon}\nn\\
\label{eq_I2n}
\eea

\begin{figure}
    \centering
    \includegraphics[scale=0.6]{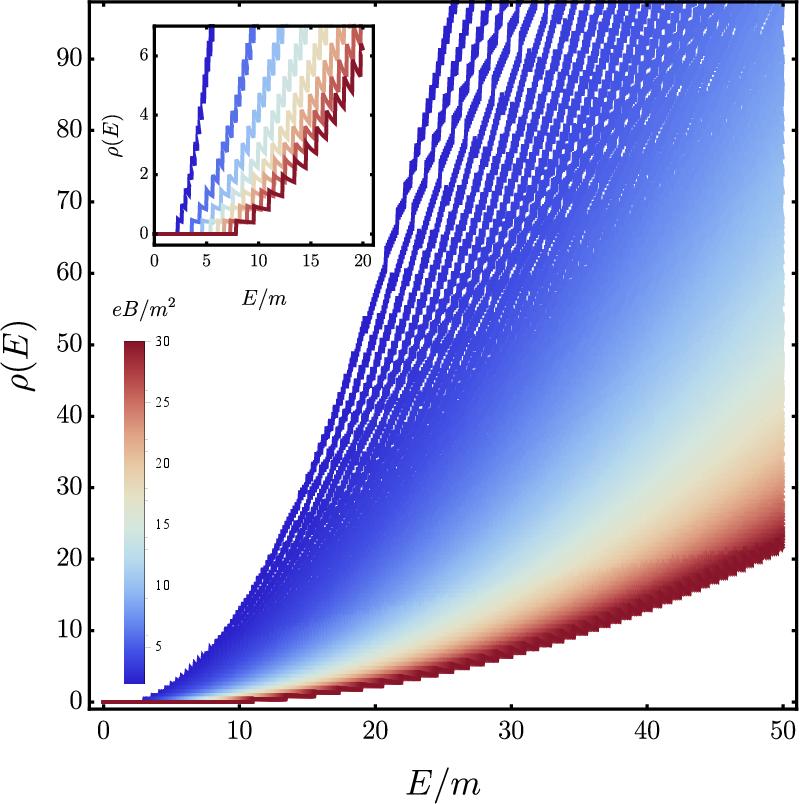}
    \caption{Spectral density for the Landau level spectrum $\rho(E)$, as a function of the dimensionless energy scale $E/m$, for different values of the average background magnetic field $eB/m^2$. The inset is shown in order to appreciate in detail the staircase pattern produced by the discrete Landau levels.}
    \label{fig:densityofstates}
\end{figure}

Inserting Eq.~\eqref{eq_I1} and Eq.~\eqref{eq_I2n} into Eq.~\eqref{eq_A1B}, we obtain (after shifting the index $n\rightarrow n+1$)
\bea
&&\tilde{\mathcal{A}}_{1}(q_0) = 
\frac{\pi^2(\qB)}{\sqrt{q_0^2-m^2+\ii\epsilon}}\\
&&+ 2\pi e B \int_{-\infty}^{+\infty}dp_3\sum_{n=0}^{\infty}\frac{1}{q_0^2 - m^2 - p_3^2 - 2(n+1)\qB + \ii\epsilon}\nn
\label{eq_A11}
\eea

Let us introduce the density of states for Landau levels
\bea
\rho(E) = \int_{-\infty}^{\infty}\frac{dp_3}{2\pi}\sum_{n=0}^{\infty}\delta\left(  E - E_{n}(p_3)\right),
\label{eq_DOS}
\eea
with the dispersion relation for the spectrum
\bea
E_{n}(p_3) = \sqrt{p_3^2 + m^2 + 2(n+1)\qB}.
\label{eq_spec}
\eea
As shown in detail in~\cite{Castano_Munoz_PhysRevD.107.096014}
\bea
\rho(E) &=& \Theta(E - \sqrt{m^2 + 2 e B})\frac{E}{\pi\sqrt{\qB}}\nn\\
&&\times\left[ 
\zeta\left(\frac{1}{2},\frac{E^2 - m^2 - 2 e B}{2 e B} - N_{max}(E)  \right)\right.\nn\\
&&\left.- \zeta\left(\frac{1}{2},\frac{E^2 - m^2 }{2 e B}  \right)
\right],
\label{eq:rho}
\eea
where we defined
\bea
N_{max}(E) = \lfloor \frac{E^2 - m^2}{2 e B}-1 \rfloor,
\eea
with $\lfloor x\rfloor$ the integer part of $x$, and $\zeta(n,z)$ is the Riemann zeta function. The spectral density Eq.~\eqref{eq:rho} is represented in Fig.~\ref{fig:densityofstates}, as a function of the dimensionless energy scale $E/m$. For large magnetic fields  $eB/m^2 \gg 1$, the spectral density displays a clear staircase pattern, where each step represents the contribution arising from a single Landau level $n = 0, 1,\ldots$. On the other hand, for weak magnetic fields $eB/m^2\ll 1$, the spectral density exhibits a denser, quasi-continuum behavior.

With these definitions, we obtain from Eq.~\eqref{eq_A11} the exact expression~\cite{Castano_Munoz_PhysRevD.107.096014}
\bea
\tilde{\mathcal{A}}_{1}(q_0) = 
\frac{\pi^2 \qB}{\sqrt{q_0^2-m^2+\ii\epsilon}}+ 4\pi^2 e B \int_{-\infty}^{+\infty}dE \frac{\rho(E)}{q_0^2 - E^2 + \ii\epsilon},\nn\\
\eea

\begin{figure}[h!]
    \centering
    \includegraphics[scale=0.6]{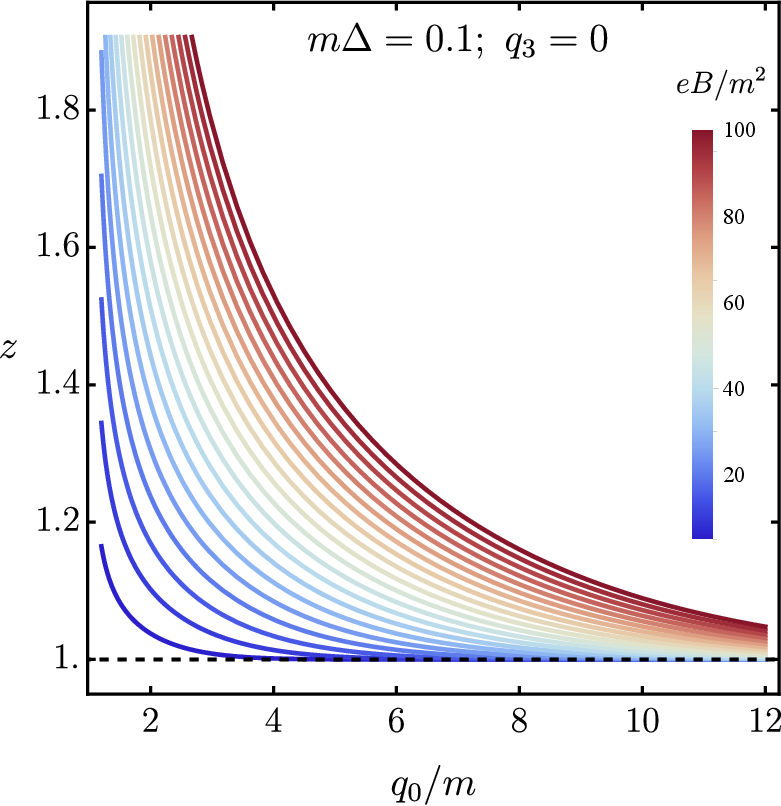}
    \caption{Wavefunction renormalization factor $z$ as a function of the dimensionless energy scale $q_0/m$. Here $q_3 = 0$.}
    \label{fig:zvsp0_strongfield}
\end{figure}

\begin{figure}[h!]
    \centering
    \includegraphics[scale=0.6]{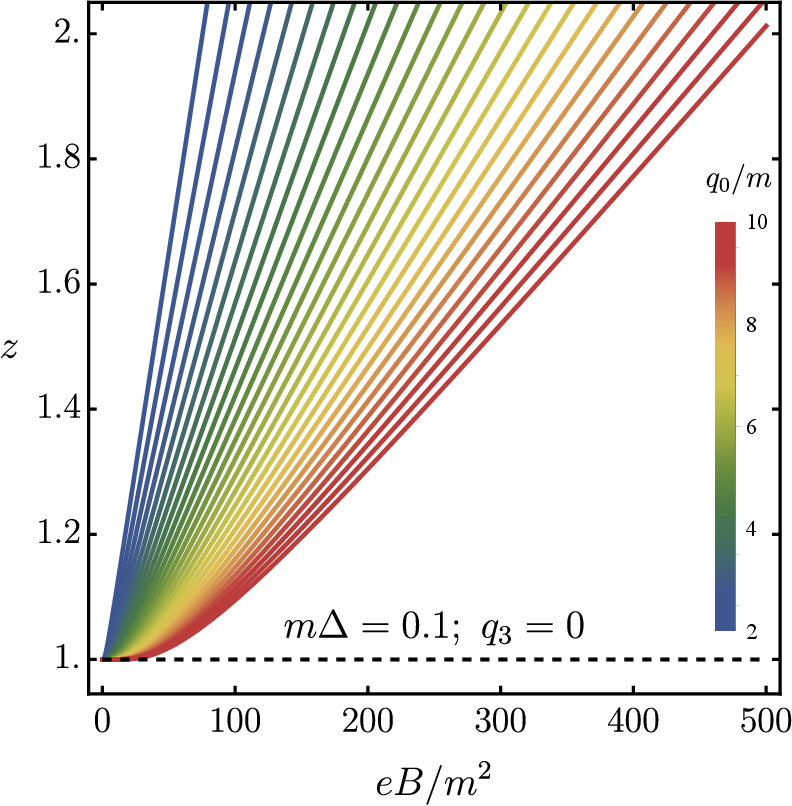}
    \caption{Wavefunction renormalization factor $z$ as a function of the dimensionless magnetic field scale $e B/m^2$. Here $q_3=0$, and $\qB/m^2\in[10,500]$}
    \label{fig:zvsB_strongfield}
\end{figure}

\begin{figure}[h!]
    \centering
    \includegraphics[scale=0.6]{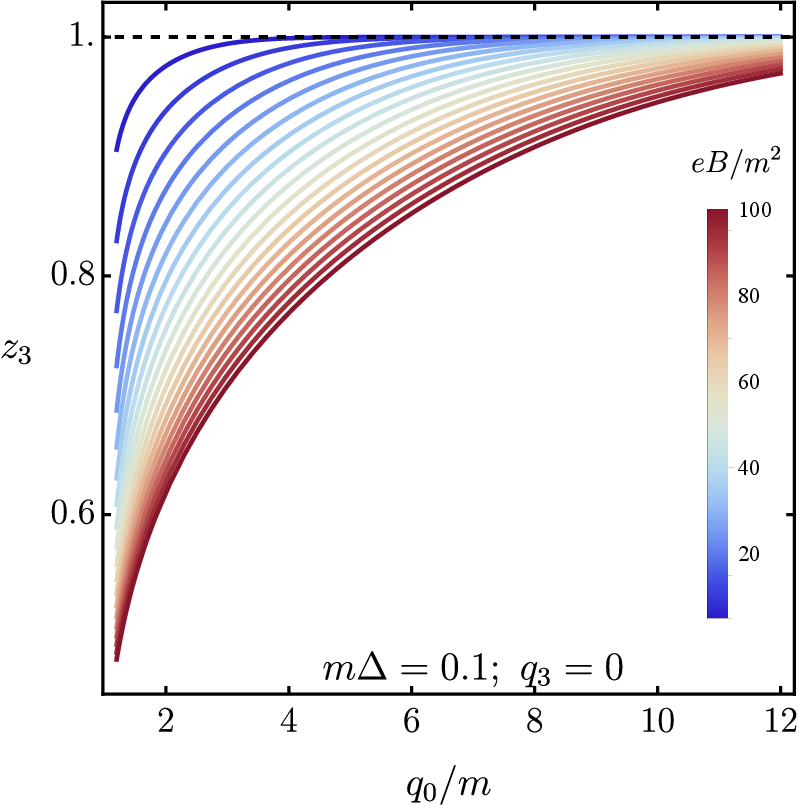}
    \caption{Charge renormalization factor $z_3$ as a function of the dimensionless energy scale $q_0/m$. Here $q_3=0$.}
    \label{fig:z3vsp0_strongfield}
\end{figure}

\begin{figure}[h!]
    \centering
    \includegraphics[scale=0.6]{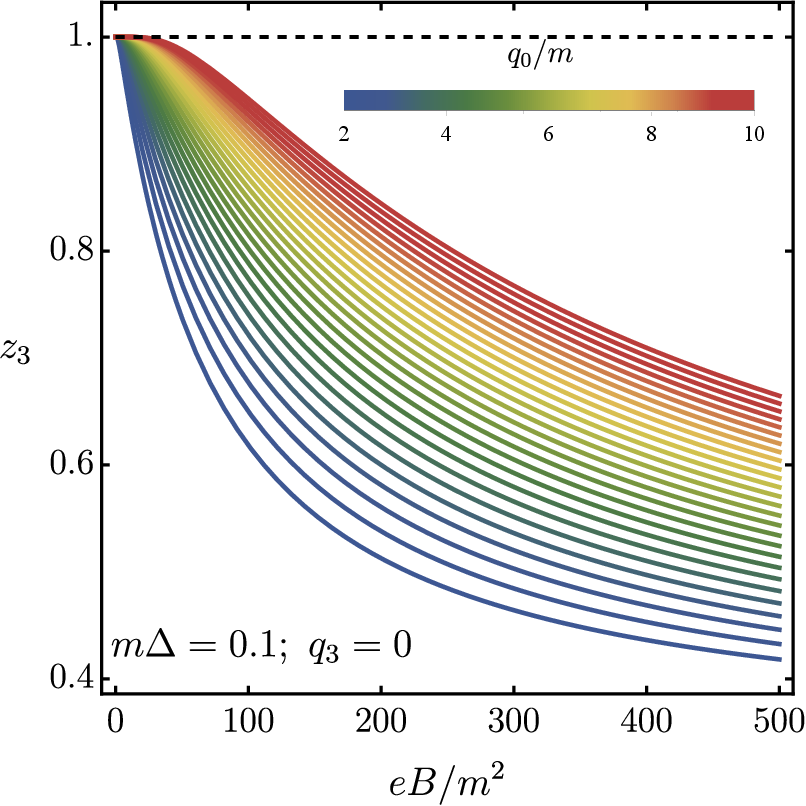}
    \caption{Charge renormalization factor $z_3$ as a function of the dimensionless magnetic field scale $e B/m^2$. Here $q_3=0$, and $\qB/m^2\in[10,500]$}
    \label{fig:z3vsB_strongfield}
\end{figure}

\begin{figure}
    \centering
    \includegraphics[scale=0.6]{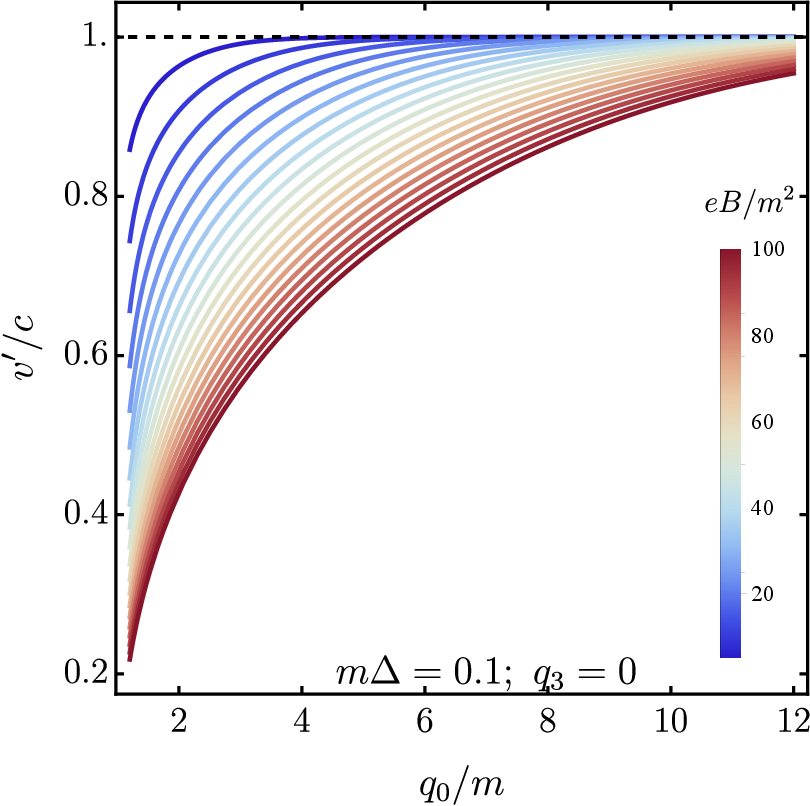}
    \caption{Effective refraction index $v'/c$ as a function of the dimensionless energy scale $q_0/m$. Here $q_3=0$.}
    \label{fig:vvsp0_strongfield}
\end{figure}

\begin{figure}
    \centering
    \includegraphics[scale=0.6]{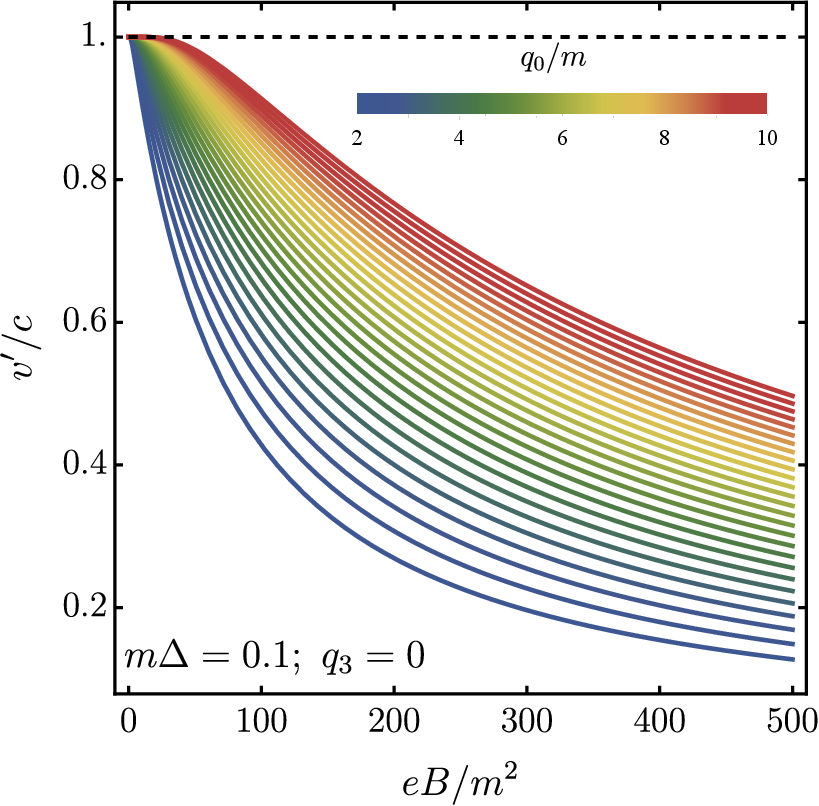}
    \caption{Effective refraction index $v'/c$ as a function of the dimensionless magnetic field scale $e B/m^2$. Here $q_3=0$, and $\qB/m^2\in[10,500]$}
    \label{fig:vvsB_strongfield}
\end{figure}

On the other hand, from Eq.~\eqref{Atildes} we have~\cite{Castano_Munoz_PhysRevD.107.096014}
\bea
\widetilde{\mathcal{A}}_2(q_0)&=&\int_{-\infty}^{\infty} dp_3 p\mathcal{A}_1(q_0,p_3;\mathbf{p}_{\perp}=0)\\
&=&\frac{\pi}{\sqrt{q_0^2-m^2+\ii\epsilon}}.\nn
\eea
In order to evaluate the formulas for $z$ in Eq.~\eqref{z} and $z_3$ in Eq.~\eqref{z3}, we obtain explicit expressions for the following coefficients (with $x = \mathbf{k}_{\perp}^2/(\qB)$)~\cite{Castano_Munoz_PhysRevD.107.096014}
\bea
\mathcal{A}_2 &=& \ii\qB\frac{\partial\mathcal{A}_1}{\partial\mathbf{k}_{\perp}^2} = \frac{e^{-x}}{\mathcal{D}_{\parallel}}\left[ 
1 + \sum_{n=1}^{\infty}\frac{(-1)^n\left(
L_{n}(2x) + 2 L_{n-1}^{(1)}(2x) - L_{n-1}(2x) - 2 \theta_{n-2}\cdot L_{n-2}^{(1)}(2x)
\right)}{1 - 2\frac{n e B}{\mathcal{D}_{\parallel}}}
\right]\\
\mathcal{A}_3 &=& \mathcal{A}_1 + (\ii eB)^2 \frac{\partial^2\mathcal{A}_1}{\partial(\mathbf{k}_{\perp}^2)^2}\nn\\
&=& \ii \frac{e^{-x}}{\mathcal{D}_{\parallel}} \sum_{n=1}^{\infty}\frac{(-1)^n\left(
L_{n}(2x) - L_{n-1}(2x) + 2 L_{n-1}^{(1)}(2x) + 4 \theta_{n-2}\cdot L_{n-2}^{(2)}(2x) - 2 L_{n-1}^{(1)}(2x)
- 4 \theta_{n-3}\cdot L_{n-3}^{(2)}(2x) 
\right)}{1 - 2\frac{n e B}{\mathcal{D}_{\parallel}}}\nn
\eea
where $\theta_{n - k}$ is the Heaviside step function
\begin{eqnarray}
\theta_{n-k} = \left\{\begin{array}{cc}1, & n \ge k\\0, & \text{otherwise}\end{array}\right.
\end{eqnarray}
Using these expressions, we evaluate
\be
\mathcal{D}(k) = \mathcal{A}_3^2\mathbf{k}_{\perp}^2 -
\left( \mathcal{A}_1^2 - \mathcal{A}_2^2 \right)(k_{\parallel}^2 - m^2),
\ee
and finally calculate $z$ and $z_3$ with Eq.~\eqref{z} and Eq.~\eqref{z3}, respectively. These results can be appreciated in Figs.\ref{fig:zvsp0_strongfield}--\ref{fig:vvsB_strongfield}, as a function of the energy scale $q_0/m$, as well as the magnitude of the average background magnetic field $eB/m^2$, respectively.
\subsubsection{Ultra-intense (LLL) field $eB/m^2\gg 1$}
In the ultra-intense magnetic field regime $eB/m^2\gg 1$,
we obtain the corresponding asymptotic expression for $\A_1(q)$ by considering only the lowest-Landau level (LLL) $n = 0$ in Eq.~\eqref{eq_A1_Land_main}. Therefore, we have~\cite{Castano_Munoz_PhysRevD.107.096014}
 \bea
    \A_1(q) \sim \ii\frac{e^{-\mathbf{q}_\perp^2/\qB}}{q_\parallel^2-m^2},
\eea
and the corresponding expressions for its derivatives are~\cite{Castano_Munoz_PhysRevD.107.096014}
\bea
    \A_2(q)=\frac{e^{-\mathbf{q}_\perp^2/\qB}}{q_\parallel^2-m^2},
\eea
and~\cite{Castano_Munoz_PhysRevD.107.096014}
\bea
    \A_3(q)=0.
\eea
Similarly, we also obtained~\cite{Castano_Munoz_PhysRevD.107.096014}
\bea
\mathcal{D}(q)=2\frac{e^{-2\mathbf{q}_\perp^2/\qB}}{q_\parallel^2-m^2}.
\eea
Finally, the integrals of $\A_1(q)$ are given, in this approximation, by the expressions~\cite{Castano_Munoz_PhysRevD.107.096014}
\begin{subequations}
 \bea
    \widetilde{\mathcal{A}}_1(q)=\frac{\pi^2 \qB}{\sqrt{q_0^2-m^2}},
\eea
\bea
\widetilde{\mathcal{A}}_2(q_0)=\frac{\pi}{\sqrt{q_0^2-m^2}}.
\eea
\end{subequations}

Applying these asymptotic results for the ultra-strong field regime, and substituting into the general definitions Eq.~\eqref{z} and Eq.~\eqref{z3}, we obtain explicit analytical expressions for the renormalization factors $z$ and $z_3$, respectively, as follows~\cite{Castano_Munoz_PhysRevD.107.096014}
\bea
z&=&1+\frac{3}{4}\frac{\Delta(\qB)e^{-\mathbf{q}_\perp^2/\qB}}{\pi\sqrt{q_0^2-m^2}},
\eea
and
\bea
z_3 &=& \left(
1 + \frac{\Delta(\qB)e^{-\mathbf{q}_\perp^2/\qB}}{4\pi\sqrt{q_0^2-m^2}}
\right)z^{-1}\nonumber\\
&=& \frac{1+\frac{\Delta(eB)e^{-\mathbf{q}_\perp^2/(eB)}}{4\pi\sqrt{q_0^2-m^2}}}{1+\frac{3}{4}\frac{\Delta(eB)e^{-\mathbf{q}_\perp^2/(eB)}}{\pi\sqrt{q_0^2-m^2}}}
\eea

Remarkably, while for very large magnetic fields $z \sim eB/m^2$ grows linearly, $z_3$ instead converges asymptotically to the constant, field-independent limit
\bea
\lim_{\frac{eB}{m^2}\rightarrow\infty}z_3 = \frac{1}{3}.
\label{eq:z3lim}
\eea


\subsection{Results}
In Figs.~\ref{fig:zvsp0_strongfield}--\ref{fig:z3vsB_strongfield}, we display the behavior of the renormalization factors $z$ and $z_3$ as a function of the energy scale $q_0/m$, and the average background magnetic field $eB/m^2$, respectively, in the whole range of both parameters. As shown in Fig.~\ref{fig:zvsp0_strongfield}, $z$ presents a monotonically decreasing behaviour as a function of the energy $q_0/m$, that asymptotically reaches the limit $z\rightarrow 1$ as $q_0/m\gg 1$, for all values of the average background magnetic field $eB/m^2$. The physical interpretation is that the quasi-particle renormalization due to the random magnetic field fluctuations tends to be negligible as the energy of the propagating fermions becomes very large, but in contrast it can be quite strong at low energy scales. This trend is also consistent with the effective refraction index $v'/c = z^{-1}$, as shown in Figs.~\ref{fig:vvsp0_strongfield},\ref{fig:vvsB_strongfield} for low energy scales, $v'/c < 1$, indicating a strong renormalization of the effective group velocity of the propagating quasi-particles due to the presence of the magnetic background fluctuations. In contrast, for larger energies the renormalization is weaker, thus recovering the asymptotic limit $v'/c\rightarrow 1$ as $q_0/m\gg 1$. Since low energy and momentum components in the Fourier representation of the propagator correspond to long-wavelength components in configuration space, our results indicate that such long-wavelength components are more sensitive to the spatial distribution of the magnetic fluctuations, and hence experience a higher degree of decoherence, thus reducing the corresponding group velocity. In contrast, the high-energy Fourier modes of the propagator, corresponding to short-wavelength components in the configuration space, are less sensitive to the presence of spatial fluctuations of the background magnetic field.

Concerning the charge renormalization factor $z_3$ displayed in Fig.\ref{fig:z3vsp0_strongfield}, it experiences a strong effect $z_3 < 1$ at low quasi-particle energies $q_0/m$, but in contrast this effect becomes negligible a large energy scales $q_0/m\gg 1$, since $z_3\rightarrow 1$ as an asymptotic limit. In physical terms, we interpret this as a charge screening due to the spatial magnetic fluctuations in the background, consistent with the aforementioned interpretation for the effective index of refraction as a function of energy. On the other hand, as displayed in Fig.~\ref{fig:z3vsB_strongfield}, $z_3$ decreases as a function of the average background magnetic field intensity, thus achieving an asymptotic limit $z_3\rightarrow 1/3$ as shown in Eq.~\eqref{eq:z3lim}.

\subsection{Conclusions}

In this section, we have studied the effects of quenched, white noise spatial fluctuations in an otherwise uniform background magnetic field, over the properties of the QED fermion propagator. This configuration is important in different physical scenarios, including heavy-ion collisions and the quark-gluon plasma, where spatial anisotropies of the background magnetic field may be present. Upon averaging over magnetic fluctuations using the replica formalism, we obtained an effective action in the replica fields, leading to an effective particle-particle interaction proportional to the strength $\Delta$ of the spatial auto-correlation function of the background noise. 
Our perturbative results show that, up to first order in $\Delta$, the propagator retains its form, thus representing renormalized quasi-particles with the same mass $m'=m$, but propagating in the medium with a magnetic field and noise-dependent index of refraction $v'/c = z^{-1}$, and effective charge $e' = z_3 e$, where $z$ and $z_3$ are renormalization factors. We showed that $z$ presents a monotonically decreasing behaviour as a function of the energy $q_0/m$, that reaches the asymptotic limit $z\rightarrow 1$ as $q_0/m\gg 1$, for all values of the average background magnetic field $eB/m^2$. In physical terms, this shows that the quasi-particle renormalization due to the random magnetic field fluctuations, while being significant at low energy scales, tends to be negligible as the energy of the propagating fermions becomes very large. This trend is also observed in the effective refraction index $v'/c = z^{-1}$, since at low energy scales $v'/c < 1$, indicating a strong renormalization of the effective group velocity of the propagating quasi-particles due to the presence of the magnetic background fluctuations. In contrast, for larger energy scales the effect becomes weaker, thus recovering the asymptotic limit $v'/c\rightarrow 1$ as $q_0/m\gg 1$.

On the other hand, our results show that the effective quasi-particle charge experiences a strong renormalization $z_3 < 1$ at low energies $q_0/m$, while the effect becomes negligible a large energy scales $q_0/m\gg 1$, since $z_3\rightarrow 1$ as an asymptotic limit. We interpret this as a charge screening due to the spatial magnetic fluctuations in the background. On the other hand, $z_3$ tends to decrease as a function of the average background magnetic field intensity, achieving an asymptotic limit $z_3\rightarrow 1/3$ as shown in Eq.~\eqref{eq:z3lim}. We remark that, in a semi-classical picture the relevant length-scale is the Landau radius $\ell_B = 1/\sqrt{eB}$, representing the typical size of the ``cyclotron radius" of the helycoidal trajectories that propagate along the magnetic field axis. Therefore, the stronger the magnetic field, the smaller the Landau radius, and hence the quasi-particle propagator is modulated towards higher momentum and energy components that, as previously discussed, are more sensitive to the magnetic noise renormalization effects, as is verified by the trend observed both in $z_3$ and in $v'/c$, that strongly decrease as the average magnetic field intensity increases $eB/m^2\gg 1$.

The analysis and results presented in this work only concern the study of the quasi-particle fermion propagator in the noisy magnetic field background. However, the effective model obtained via the replica method and its consequences can be extended towards the study of other physical quantities, such as the photon polarization tensor. 

\section{
QED Fermions in a noisy magnetic field background from the perspective of  the effective action approach }\label{Loewe}

\subsection{Introduction}
In this section, we discuss an extended analysis of the work carried out in Ref.~\cite{Replicas_1}. The present approach explores the noisy magnetic background scenario, this time from the perspective of a non-perturbative approach. The basic idea is that the gauge field $A_{mu}$ can be decomposed according to
\begin{eqnarray}
A^{\mu}(x) \rightarrow A^{\mu}(x) + A^{\mu}_\text{BG}(x) + \delta A^{\mu}_\text{BG}(x)
\end{eqnarray}
\noindent
where the first term on the right side in the previous equation represents the quantum photon field, whereas BG refers a classical background associated to the presence of an external magnetic field, which was assumed to be static and uniform. An additional novelty of the present analysis is the consideration of space-time fluctuations whereas in~\cite{Replicas_1} only spatial dependent fluctuations were taken into account. The last term in the previous equation, $\delta A_{BG}^\mu (x)$, encodes the white-noise fluctuations with respect to the mean value $A^{\mu}_\text{BG}(x)$ that satisfy
\begin{eqnarray}
\langle \delta A^{j}_\text{BG}(x) \delta A^{k}_\text{BG}(x')\rangle_{\Delta} &=&  \Delta_{B}\delta_{j,k}\delta^{(4)}(x-x'),\nonumber\\
\langle \delta A^{\mu}_\text{BG}(x)\rangle_{\Delta} &=& 0.
\end{eqnarray}
If we consider the case of relativistic heavy-ion collisions it is quite natural to incorpore  the presence of spatio-temporal fluctuations in the background field, corresponding to the most general case. In our previous discussion ~\cite{Replicas_1}, only spatial fluctuations as source of the noisy effects were considered. Here we extend such scenario. The source for such noisy gauge field will be a classical current $J^{\mu}_{cl}(x) + \delta J_{cl}^{\mu}(x)$. The noisy gauge fields are solutions of $\square (A_{BG}^{\mu} + \delta A_{BG}^{\mu}) = J_{cl}^{\mu}(x) + \delta J_{cl}^{\mu}(x)$, and the background noisy magnetic field $\mathbf{B}_{BG} + \delta \mathbf{B}_{BG} = \nabla\times(\mathbf{A}_{BG} + \delta\mathbf{A}_{BG} )$ satisfies the Gauss law $\nabla\cdot(\mathbf{B}_{BG} + \delta \mathbf{B}_{BG}) = 0$. These statistical properties are represented by a Gaussian functional distribution of the form
\begin{eqnarray}
dP\left[ \delta A^{\mu}_\text{BG} \right] = \mathcal{N} e^{-\int d^4x\,\frac{\left[\delta A_\text{BG}^{\mu}(x)\right]^2}{2 \Delta_B}}
\mathcal{D}\left[\delta A_\text{BG}^{\mu}(x)\right],
\end{eqnarray}
with the corresponding statistical average over background fluctuations defined by
\begin{eqnarray}
\langle \hat{O} \rangle_{\Delta} = \int dP\left[ \delta A^{\mu}_\text{BG} \right] \hat{O}\left[ \delta A^{\mu}_\text{BG}\right].
\label{eq_def_average}
\end{eqnarray}
\noindent
There is a number of important physical scenarios where the presence of strong magnetic fields determine the dynamics of relativistic particles. The quark-gluon plasma~\cite{Busza_2018,Hattori_2016,Hattori_2018,Buballa_2005} and heavy-ion collisions~\cite{Alam_2021,Inghirami_2020,Ayala:2019jey,Ayala:2017vex} are among them. For technical reasons, as we said, the theoretical analysis of such systems was simplified to the configuration of a constant (both static and uniform), such that the Schwinger proper time formalism can be applied~\cite{Schwinger_1951,Dittrich_Reuter,Dittrich_Gies}. However, in a more realistic description of such phenomena, the electromagnetic field may develop spatio-temporal patterns~\cite{Inghirami_2020,Alam_2021} that will then in principle modify the associated physical predictions. We discussed such possibility in our recent work \cite{Replicas_1}, in the frame of  a perturbative analysis to conclude that such magnetic noise effects may indeed be relevant. In this work,we emphasize tis point again, we revisit the problem from a non-perturbative point of view, in order to shed further light into such effects over a broader range of magnetic noise intensities. 
Following the analysis presented in our previous work~\cite{Replicas_1},
we shall consider a physical scenario where a classical and static magnetic field background, possessing local random fluctuations, modifies the quantum dynamics of a system of fermions. For this purpose,
we shall assume the standard QED theory involving fermionic fields $\psi(x)$, as well as gauge fields $A^{\mu}(x)$. In the latter, we shall distinguish three physically different contributions~\cite{Replicas_1}
\begin{eqnarray}
A^{\mu}(x) \rightarrow A^{\mu}(x) + A^{\mu}_\text{BG}(x) + \delta A^{\mu}_\text{BG}(x).
\end{eqnarray}
Here, $A^{\mu}(x)$ represents the dynamical photon quantum field, while BG stands for ``background", thus capturing the presence of a classical external field $\mathbf{B} = \langle\nabla\times\mathbf{A}_\text{BG}(x)\rangle_{\Delta}$, assumed to be static and uniform as imposed by the experimental conditions. Moreover, for this BG contribution, we consider the effect of white noise fluctuations $\delta A^{\mu}_\text{BG}(x)$ with respect to the mean value $A_\text{BG}^{\mu}(x)$, satisfying the statistical properties
\begin{eqnarray}
\langle \delta A^{j}_\text{BG}(x) \delta A^{k}_\text{BG}(x')\rangle_{\Delta} &=&  \Delta_{B}\delta_{j,k}\delta^{(4)}(x-x'),\nonumber\\
\langle \delta A^{\mu}_\text{BG}(x)\rangle_{\Delta} &=& 0.
\label{eq_Acorr2}
\end{eqnarray}
We remark that the sources of the noisy gauge field are, in the context of heavy-ion collisions, the anisotropic and incoherent distribution of ionic beams in the initial stages prior to thermalization, that are modeled here as a classical current $J^{\mu}_{cl}(x) + \delta J_{cl}^{\mu}(x)$. Therefore, our noisy gauge fields are solutions of $\square (A_{BG}^{\mu} + \delta A_{BG}^{\mu}) = J_{cl}^{\mu}(x) + \delta J_{cl}^{\mu}(x)$, and hence by construction the background noisy magnetic field $\mathbf{B}_{BG} + \delta \mathbf{B}_{BG} = \nabla\times(\mathbf{A}_{BG} + \delta\mathbf{A}_{BG} )$ satisfies the Gauss law $\nabla\cdot(\mathbf{B}_{BG} + \delta \mathbf{B}_{BG}) = 0$. 

We also remark that in contrast with our previous work~\cite{Replicas_1}, where we assumed only spatial fluctuations, here we shall assume spatio-temporal fluctuations in the background gauge fields.
These statistical properties are represented by a Gaussian functional distribution of the form
\begin{eqnarray}
dP\left[ \delta A^{\mu}_\text{BG} \right] = \mathcal{N} e^{-\int d^4x\,\frac{\left[\delta A_\text{BG}^{\mu}(x)\right]^2}{2 \Delta_B}}
\mathcal{D}\left[\delta A_\text{BG}^{\mu}(x)\right],
\label{eq_Astat2}
\end{eqnarray}
with the corresponding statistical average over background fluctuations defined by
\begin{eqnarray}
\langle \hat{O} \rangle_{\Delta} = \int dP\left[ \delta A^{\mu}_\text{BG} \right] \hat{O}\left[ \delta A^{\mu}_\text{BG}\right].
\label{eq_def_average2}
\end{eqnarray}
In heavy-ion collisions (HIC), strong magnetic fields $\mathbf{B} = \nabla\times\mathbf{A}_\text{BG}$ are generated locally within a small spatial region whose characteristic length scale is $L \sim \sqrt{\sigma}$, with $\sigma$ being the effective cross-section. In these collisions, the dominant component of the magnetic field is along the axial z-direction, such that on average we have $\langle \mathbf{B} \rangle = \hat{e}_3\,B$. There are also smaller transverse components $\delta B_x$ and $\delta B_y$ such that we can estimate the fluctuation of the field within the small collision region to be on the order of $(\delta B)^2 \sim (\delta B_x)^2 + (\delta B_y)^2$. Since many such collisions occur at different points in space and their time-span $\delta \tau \sim L'/c$, an approximate model for this physical scenario is provided by the magnetic random noise Eq.~\eqref{eq_Acorr2}. By dimensional analysis, the magnitude of $\Delta_B$ is of the order:
\begin{equation}
\Delta_B \sim \left(\delta B\right)^2\,L^{5} L' \sim \left(\delta B\right)^2\,\sigma^{5/2} L'.
\end{equation}

The effective cross-section for a nuclear collision can be estimated as the fraction $f$ of the area of perfectly central collisions between two nuclei, each with a radius of $r_A$
\begin{equation}
\sigma = f\pi r_A^2.
\end{equation}

Here, $f$ represents the fraction of the geometrical cross-section $\sigma_\text{geom}$, which is defined as the area of the circle with a radius of $r_1 + r_2 = 2R$ in a maximum peripheral collision, and the cross-section $\sigma_b$ for a peripheral collision with impact parameter $b$ \cite{bartke2008introduction, castano2021effects}:
\begin{equation}
f = \frac{\sigma_b}{\sigma_\text{geom}} = \left(\frac{N_\text{part}}{2N}\right)^{2/3},
\end{equation}
where $\sigma_b$ describes an effective nucleus of radius $b$. The nuclear radius is always written as $r_A = r_0N^{1/3}$, where $N$ is the number of nucleons per ion and $r_0 \sim 10^{-3}\,\text{MeV}^{-1}$. Here, $N_\text{part}$ is the number of participants corresponding to the effective nucleus.

Therefore, under these considerations, we have
\begin{equation}
\Delta_B \sim \pi^{5/2}\left(\delta B\right)^2r_0^5N^{5/3}\left(\frac{N_\text{part}}{2N}\right)^{5/3}L'.
\end{equation}

In peripheral heavy-ion collisions, the magnetic field fluctuations along the transverse plane are approximately $|e\,\delta B| \sim m_\pi^2/4$, where $m_\pi$ is the pion mass \cite{PhysRevC.83.054911, castano2021effects}. For an Au+Au collision with $N=197$, and if $N_\text{part}/N=1/2$, we obtain (for $L' \sim r_A$)
\begin{equation}
e^2\Delta_B \sim 1.6\times 10^{-6}\,\text{MeV}^{-2},
\end{equation}
or for less central collisions with $N_\text{part}/N=1/8$:
\begin{equation}
e^2\Delta_B \sim 1.6\times10^{-7}\,\text{ MeV}^{-2}.
\end{equation}
In general, for many technical details we will refer in what follows to Ref.~\cite{Castano-Yepes:2023brq}, the original article were we presented our analysis.
As we shall later show, the effects of magnetic noise are controlled by the dimensionless parameter $\Delta = e^2\Delta_B m_f^2$, where $m_f$ is the fermion mass. If one considers the mass of the proton, then the relevant dimensionless scale would be
\begin{eqnarray}
\Delta_{\text{proton}} \sim 0.16 - 1.6,
\end{eqnarray}
 whereas if one considers the mass of the constituent quark species

\begin{eqnarray}
\Delta_{\text{quark}} \sim 0.018 - 0.18.
\end{eqnarray}
These considerations allow us to write the Lagrangian for this model as a superposition of two terms
\begin{eqnarray}
\mathcal{L} = \mathcal{L}_\text{FBG} + \mathcal{L}_\text{NBG}.
\label{eq_LAG}
\end{eqnarray}
The first term represents the system of fermions (and photons) immersed in the deterministic background field (FBG)
\begin{eqnarray}
\mathcal{L}_\text{FBG} = \bar{\psi}\left(\ii\slashed{\partial} - e \slashed{A}_\text{BG} - e \slashed{A}   - m_f \right)\psi-\frac{1}{4}F_{\mu\nu}F^{\mu\nu},
\label{eq_LFBG2}
\end{eqnarray}
where $F_{\mu\nu} = \partial_{\mu} A_{\nu}- \partial_{\nu} A_{\mu}$ is the strength tensor for the dynamical quantum gauge fields (photons). The second term in the lagrangian Eq.~\eqref{eq_LAG} represents the interaction between the Fermions and the classical noise (NBG)
\begin{eqnarray}
\mathcal{L}_\text{NBG} = \bar{\psi}\left( - e \delta\slashed{A}_\text{BG} \right)\psi.
\label{eq_LDBG2}
\end{eqnarray}

The generating functional (in the absence of sources) for a given realization of the noisy  fields is given by
\begin{eqnarray}
Z[A] = \int \mathcal{D}[\bar{\psi},\psi]
e^{\ii\int d^4 x \left[ \mathcal{L}_\text{FBG} + \mathcal{L}_\text{NBG}  \right]}.
\end{eqnarray}
We need the generator of connected correlation functions to extract the physics of this system. However, as the presence of disorder is modeled by a statistical ensemble of different realizations of the magnetic background noise $\delta A_\text{BG}^{\mu}(x)$, we need to calculate the disorder-averaged generator of connected correlation functions $\langle \ln Z \rangle_{\Delta}$. For this purpose, we apply the replica method, which is based on the following identity~\cite{Parisi_Mezard_1991}
\begin{eqnarray}
\langle \ln Z[A]\rangle_{\Delta} = \lim_{n\rightarrow 0}\frac{\langle Z^n[A]\rangle_{\Delta}-1}{n}.
\label{eq_ln}
\end{eqnarray}
Here, we defined the statistical average according to the Gaussian functional measure of Eq.~\eqref{eq_Astat2} as in Eq.~\eqref{eq_def_average2}, and $Z^n$ is obtained by incorporating an additional ``replica" component for each of the Fermion fields, i.e. $\psi(x)\rightarrow \psi^{a}(x)$, for $1\le a \le n$. The ``replicated" Lagrangian has the same form as Eqs.~\eqref{eq_LFBG2} and~\eqref{eq_LDBG2}, but with an additional sum over the replica components of the Fermion fields. Therefore, the averaging procedure leads to
\begin{eqnarray}
\langle Z^n[A] \rangle_{\Delta} &=& \int 
\prod_{a=1}^{n}\mathcal{D}[\bar{\psi}^{a},\psi^{a}]
\int \mathcal{D}\left[\delta A_\text{BG}^{\mu}\right]e^{-\int d^4x\,\frac{\left[\delta A_\text{BG}^{\mu}(x)\right]^2}{2 \Delta_B}}\nonumber\\
&&\times e^{\ii\int d^4 x \sum_{a=1}^n \left( \mathcal{L}_\text{FBG}[\bar{\psi}^a,\psi^a] + \mathcal{L}_{DBG}[\bar{\psi}^a,\psi^a] \right)}\nonumber\\
&=& \int 
\prod_{a=1}^{n}\mathcal{D}[\bar{\psi}^{a},\psi^{a}] e^{\ii \bar{S}\left[\bar{\psi}^a,\psi^a;A \right]},
\label{eq_repl2}
\end{eqnarray}
where in the last step we explicitly performed the Gaussian integral over the background noise, leading to the definition of the effective averaged action for the replica system
\begin{eqnarray}
\ii\,\bar{S}\left[\bar{\psi}^a,\psi^a;A \right]
&=& \ii\,\int d^4 x \left(\sum_{a}\bar{\psi}^{a}\left(\ii\slashed\partial -  e \slashed{A}_\text{BG} - e \slashed{A} - m_f  \right)\psi^{a}-\frac{1}{4}F_{\mu\nu}F^{\mu\nu}\right)\nonumber\\
&-& \frac{e^2\Delta_{B}}{2}\int d^4x\int d^4 y\sum_{a,b}\sum_{j=1}^{3}\bar{\psi}^{a}(x)\gamma^{j}\psi^{a}(x)\bar{\psi}^{b}(y)\gamma_{j}\psi^{b}(y).
\end{eqnarray}
Clearly, we end up with an effective interacting theory between vector currents corresponding to different replicas, with a coupling constant proportional to the fluctuation amplitude $\Delta_B$ that characterizes the magnetic noise, as defined in Eq.~\eqref{eq_Acorr2}. 

The ``free'' part of the action corresponds to Fermions in the average background classical field
$A_\text{BG}^{\mu}(x)$. We choose this background to represent a uniform, static magnetic field along the $z$-direction $\mathbf{B} = \hat{e}_3 B$, using the gauge~\cite{Dittrich_Reuter}
\begin{eqnarray}
A_\text{BG}^{\mu}(x) = \frac{1}{2}(0,-B x^2, B x^1,0).
\label{eq_BGauge}
\end{eqnarray}

In addition, as we shall focus on the analysis for the fermion propagator, we shall not consider the photons in this scenario $A^{\mu}(x) = 0$, and hence its corresponding strength tensor $F_{\mu\nu} = 0$.

\subsection{Auxiliary boson fields}
Let us now introduce a Hubbard-Stratonovich transformation via a set of complex boson fields $Q_{j}(x)$, by means of the Gaussian integral identity
\begin{eqnarray}
&&e^{ -\frac{e^2\Delta_B}{2}\int d^4 x\int d^4 y \sum_{a,b}^{n}\sum_{j=1}^{3}\bar{\psi}^{a}(x)\gamma^{j}\psi^{a}(x)\bar{\psi}^{b}(y)\gamma_{j}\psi^{b}(y) }\nonumber\\
&&= \mathcal{N} \left[\prod_{j=1}^{3}\int\mathcal{D}Q_j (x)\mathcal{D}Q_j^{*} (x) \right]
 e^{-\frac{2}{\Delta_{B}} \int d^4 x \left|Q_{j}(x)\right|^2 + \ii e \int d^4 x Q_{j}(x)\sum_{a=1}^n \bar{\psi}^{a}(x)\gamma^{j}\psi^{a}(x)- \ii e \int d^4 x Q_{j}^{*}(x)\sum_{a=1}^n \bar{\psi}^{a}(x)\gamma^{j}\psi^{a}(x)}\nonumber\\ 
\end{eqnarray}
With this transformation into the averaged, replicated $n^{th}$ power of the generating functional Eq.~\eqref{eq_repl2}, we obtain the equivalent form

\begin{eqnarray}
\langle Z^n[A_\text{BG}]\rangle_{\Delta} &=& \mathcal{N} \left[\prod_{j=1}^{3}\int\mathcal{D}Q_j (x)\mathcal{D}Q_j^{*} (x)\right]
e^{-\frac{2}{\Delta_{B}} \int d^4 x\, \left|Q_{j}(x)\right|^2  }\nonumber\\
&&\times
\left[\prod_{a=1}^{n}\int\mathcal{D}[\bar{\psi}^{a},\psi^{a}]\right] e^{\ii\int d^4 x \left\{\sum_{a=1}^{n}\bar{\psi}^{a}(x)\left(\ii\slashed\partial -  e \slashed{A}_\text{BG} - m_f  - e \gamma^{j}(Q_j - Q_j^*)\right)\psi^{a}(x)\right\}  }\nonumber\\
&=& \mathcal{N} \left[\prod_{j=1}^{3}\int\mathcal{D}Q_j (x)\mathcal{D}Q_j^{*} (x) \right]
e^{-\frac{2}{\Delta_{B}} \int d^4 x\, \left|Q_{j}(x)\right|^2  }
\left[\det\left( \ii\slashed\partial -  e \slashed{A}_\text{BG} - m_f   - e \gamma^{j}\left( Q_j - Q_j^* \right) \right)  \right]^{n}\nonumber\\
&=& \mathcal{N} \left[\prod_{j=1}^{3}\int\mathcal{D}Q_j (x)\mathcal{D}Q_j^{*} (x) \right]
e^{-\frac{2}{\Delta_{B}} \int d^4 x\, \left|Q_{j}(x)\right|^2  + n{\rm{Tr}}\ln\left[ \ii\slashed\partial -  e \slashed{A}_\text{BG} - m_f  - e \gamma^{j}\left( Q_j - Q_j^* \right)\right]  }
\label{eq_barZn}
\end{eqnarray}

Based on this identity, and combined with Eq.~\eqref{eq_ln}, we obtain the effective action
\begin{eqnarray}
&&\ii\,S_\text{eff}[A_\text{BG}] - \ii\,S_{0}[A_\text{BG}]  = \langle \ln Z[A_\text{BG}]\rangle_{\Delta} - \ln Z_0[A_\text{BG}]\nonumber\\
&=&  \mathcal{N} \left[\prod_{j=1}^{3}\int\mathcal{D}Q_j (x)\mathcal{D}Q_j^{*} (x)  \right]
e^{-\frac{2}{\Delta_{B}} \int d^4 x\, \left|Q_{j}(x)\right|^2} \lim_{n\rightarrow 0}\frac{1}{n}\left[e^{n{\rm{Tr}}\ln\left[ \ii\slashed\partial -  e \slashed{A}_\text{BG}  - m_f - e\gamma^{j}(Q_j-Q_j^*) \right]  }-1\right]- \ln Z_0[A_\text{BG}]\nonumber\\
&=& \langle {\rm{Tr}} \ln\left( \ii\slashed\partial -  e \slashed{A}_\text{BG} - m_f  - e \gamma^{j}(Q_j - Q_j^*) \right) \rangle_{\Delta} - {\rm{Tr}} \ln\left( \ii\slashed\partial -  e \slashed{A}_\text{BG} - m_f  \right),
\label{eq_effact}
\end{eqnarray}
where $\langle (\cdot) \rangle_{\Delta}$ represents the average over the Gaussian functional measure of the complex fields $Q_{j}(x)$. 
Let us define the inverse fermion propagator, including the classical background field, as follows
\begin{eqnarray}
S_\text{F}^{-1}(x-y) = \left( \ii\slashed{\partial} - e \slashed{A}_\text{BG} - m_f \right)_{x}\delta^{(4)}(x-y)
\label{eq_Schwingernonoise}
\end{eqnarray}

Therefore, for the effective action we have
\begin{eqnarray}
\ii\,S_\text{eff}[A_\text{BG}] &=& \langle {\rm{Tr}}\ln\left( \ii\slashed{\partial} - e\slashed{A}_\text{BG} - m_f - e \gamma^{j}(Q_j - Q_j^*)  \right) \rangle_{\Delta}\nonumber\\
&=& \langle {\rm{Tr}}\ln\left(  S_\text{F}^{-1} - e \gamma^{j}(Q_j - Q_j^*) \right) \rangle_{\Delta}\nonumber\\
&=& {\rm{Tr}}\ln S_\text{F}^{-1}\nonumber\\ 
&+& 
\langle {\rm{Tr}}\ln\left( \mathbf{1} - e S_\text{F} \gamma^{j}(Q_j - Q_j^*)   \right) \rangle_{\Delta}.
\end{eqnarray}

By noticing that $\ii\,S_{0}[A_\text{BG}] = {\rm{Tr}}\ln S_\text{F}^{-1}$, we have
\begin{eqnarray}
\ii\,S_\text{eff}[A_\text{BG}] &-& \ii\,S_{0}[A_\text{BG}]\nonumber\\ &=& \langle {\rm{Tr}}\ln\left( \mathbf{1} - e S_\text{F} \gamma^{j}(Q_j - Q_j^*)  \right) \rangle_{\Delta}\nonumber\\
\label{eq_deltaS}
\end{eqnarray}
where the right hand side contains all the effects of the noise.
\subsection{Saddle-point approximation (Mean-field)}
The result in Eq.~\eqref{eq_deltaS} can be expressed in the explicit functional integral form
\begin{eqnarray}
\ii\,S_\text{eff}[A_\text{BG}] - \ii\,S_{0}[A_\text{BG}] = \mathcal{N} \left[\prod_{j=1}^{3}\int\mathcal{D}Q_j (x)\mathcal{D}Q_j^{*} (x)  \right]
e^{-\frac{2}{\Delta_{B}} \int d^4 x\, \left|Q_{j}(x)\right|^2 + \ln\left[{\rm{Tr}}\ln\left( \mathbf{1} - e S_\text{F} \gamma^{j}(Q_j - Q_j^*)  \right)\right]}.
\label{eq_effS_saddle}
\end{eqnarray}

In order to study the effects of the background noise, we shall adopt a mean-field approximation, by searching for the saddle-point of the exponent in Eq.~\eqref{eq_barZn}

\begin{eqnarray}
\frac{\delta}{\delta Q_j(x)}\left\{-\frac{2}{\Delta_B}\int d^4 y Q_{l}(y)Q_{l}^{*}(y) + \ln\left[{\rm{Tr}}\ln\left[ \mathbf{1}- e S_\text{F} \gamma^{l}(Q_l - Q_l^*)  \right]\right]\right\} &=& 0,\nonumber\\
\frac{\delta}{\delta Q_j^{*}(x)}\left\{-\frac{2}{\Delta_B}\int d^4 y Q_{l}(y)Q_{l}^{*}(y)+ \ln\left[{\rm{Tr}}\ln\left[ \mathbf{1}- e S_\text{F} \gamma^{l}(Q_l - Q_l^*)  \right]\right]\right\} &=& 0.
\label{eq_variational}
\end{eqnarray}

This condition leads to the equation (assuming homogeneous solutions of the form $Q_j(x)\equiv Q_j$)
\bea
Q_j^{*} &=& -e\frac{\Delta_B}{2} \frac{{\rm{Tr}}\left[S_\text{F}\gamma^{j}\left( \mathbf{1}- e S_\text{F} \gamma^{l}(Q_l - Q_l^*)   \right)^{-1}\right]}{{\rm{Tr}}\ln\left[ \mathbf{1}- e S_\text{F} \gamma^{l}(Q_l - Q_l^*)  \right]}
\nonumber\\
\eea
From the second equation in Eq.~\eqref{eq_variational}, we obtain the additional condition
\bea
Q_j^{*} = - Q_j.
\eea
We can combine both equations, by noticing that
\bea
Q_j + Q_j^{*} &=& 0\nonumber\\
q_j \equiv Q_j - Q_j^{*} &=& 2\, \ii\, \Im Q_j\ne 0,
\label{eq_qj}
\eea
where the second line implies that $q_j$ is a pure imaginary number, and it satisfies the non-linear equation
\bea
q_j &=& e\Delta_B \frac{{\rm{Tr}}\left[S_\text{F}\gamma^{j}\left( \mathbf{1}- e S_\text{F} \gamma^{l}q_l   \right)^{-1}\right]}{{\rm{Tr}}\ln\left[ \mathbf{1}- e S_\text{F} \gamma^{l}q_l  \right]}.
\label{eq_MFexact}
\eea

The numerator of this equation can be expanded as an infinite geometric series as follows
\begin{eqnarray}
&&{\rm{Tr}}\left[S_\text{F}\gamma^{j}\left( \mathbf{1}- e S_\text{F} \gamma^{l}q_l   \right)^{-1}\right] = \sum_{l=0}^{\infty}e^{l}{\rm{Tr}} \left[ S_\text{F}\gamma^{j}\left(  S_\text{F}\gamma^{k}q_k\right)^{l} \right]\nonumber\\
&&=  \sum_{l=1}^{\infty}e^{l}\left[\Pi_{\alpha=1}^{l}q_{k_{\alpha}} \right]{\rm{Tr}} \left[ S_\text{F}\gamma^{j}  S_\text{F}\gamma^{k_{1}}S_\text{F}\gamma^{k_{2}}\ldots S_\text{F}\gamma^{k_{l}}  \right].
\end{eqnarray}

On the other hand, the denominator can also be expanded by means of the Taylor series for the natural logarithm, as follows
\bea
&&{\rm{Tr}}\ln\left[ \mathbf{1}- e S_\text{F} \gamma^{l}q_l  \right]
= \sum_{l=1}^{\infty}\frac{e^l}{l}{\rm{Tr}}\left[ \left(S_F \slashed{q}\right)^{l} \right]\nn\\
&=& \sum_{l=1}^{\infty}\frac{e^l}{l}\left[\Pi_{\alpha=1}^{l}q_{k_{\alpha}} \right]{\rm{Tr}}\left[   S_\text{F}\gamma^{k_{1}}S_\text{F}\gamma^{k_{2}}\ldots S_\text{F}\gamma^{k_{l}} \right].
\eea
\begin{figure*}
    \centering
    \includegraphics[width=1\textwidth]{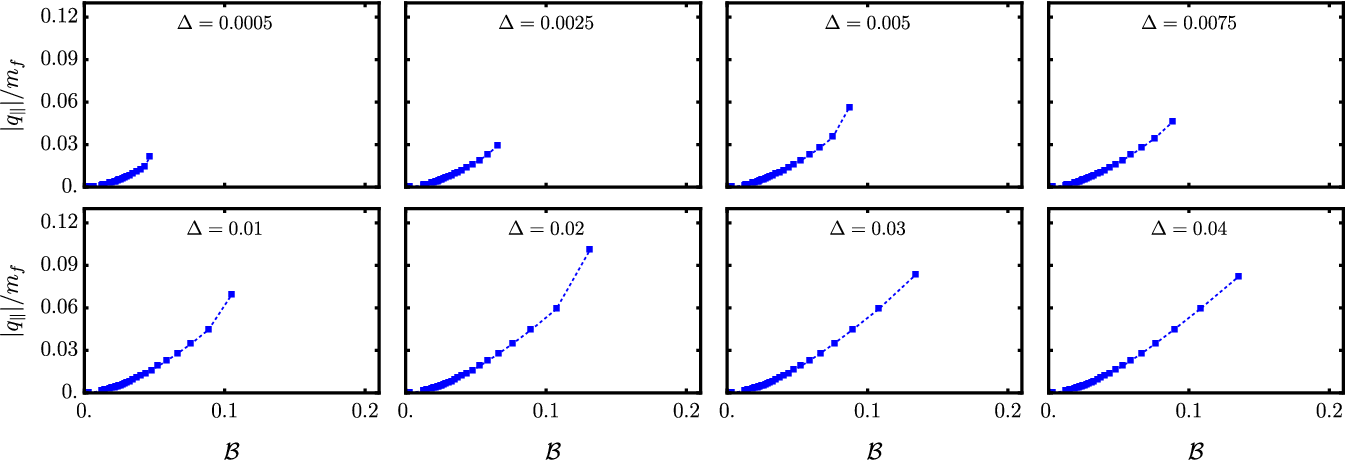}
    \caption{Non-trivial solutions of Eq.~\eqref{eq:det1} for the Case 1 as a function of $\mathcal{B}$ for $\Delta\in[0.1,0.04]$. The dashed line represents the smooth envelope connecting the discrete non-trivial solutions.}
    \label{fig:q3VseB1}
\end{figure*}

From the exact expression Eq.~\eqref{eq_MFexact}, we can extract the leading contribution by retaining terms up to third order in the $q_j$ coefficients in the numerator, while retaining up to second order in the denominator
\begin{eqnarray}
q_j = e\Delta_B \frac{e\mathcal{M}^{jl}q_l + e^3\mathcal{M}^{jlmn}q_l q_m q_n}{\frac{e^2}{2}\mathcal{M}^{mn}q_m q_n},
\label{eq_system}
\end{eqnarray}
where we defined the matrix coefficients
\begin{eqnarray}
\mathcal{M}^{jl} &=& {\rm{Tr}} \left[ S_\text{F}\gamma^{j}S_\text{F}\gamma^{l}\right]\nn\\
&=& \int \frac{d^4 k}{(2\pi)^4} {\rm{tr}} \left[ S_\text{F}(k)\gamma^{j}S_\text{F}(-k)\gamma^{l}\right],
\label{m2}
\end{eqnarray}
and
\begin{eqnarray}
&&\mathcal{M}^{jlmn} = {\rm{Tr}} \left[ S_\text{F}\gamma^{j}S_\text{F}\gamma^{l}S_\text{F}\gamma^{m}S_\text{F}\gamma^{n}\right]
\label{m4}
\\
&=& \int \frac{d^4 k}{(2\pi)^4} {\rm{tr}}\left[ S_\text{F}(k)\gamma^{j}S_\text{F}(k)\gamma^{l}S_\text{F}(k)\gamma^{m}S_\text{F}(k)\gamma^{n}\right].\nn
\end{eqnarray}

Here, ${\rm{tr}}[ \cdot ]$ stands for trace over the space of Dirac matrices. The Schwinger propagator in Fourier space is defined by Eq.~\eqref{eq_Sprop} and, more importantly for calculation purposes, by its alternative form Eq.~\eqref{propSchwinger}.

We can analyze the possible solutions to Eq.~\eqref{eq_system}, by casting it into the form of a quasi-linear system
\begin{eqnarray}
\left( \Delta_B \mathbf{\mathcal{M}} + \tilde{\mathbf{\mathcal{M}}}[\mathbf{q}] \right)\mathbf{q} = 0,
\label{eq_linear}
\end{eqnarray}
where we defined
\begin{eqnarray}
\left[\tilde{\mathbf{\mathcal{M}}}[\mathbf{q}]\right]^{jl} \equiv \left( -\frac{1}{2}\delta^{jl}\mathcal{M}^{mn} + e^2\Delta_B \mathcal{M}^{jlmn} \right)q_m q_n.
\label{eq_Mtilde_gen}
\end{eqnarray}

There is always a trivial solution $\mathbf{q} = 0$ to Eq.~\eqref{eq_linear}. However, nontrivial solutions $\mathbf{q}$ may exist provided that the (nonlinear) matrix coefficient is singular, i.e.
\begin{eqnarray}
\det \left( \Delta_B \mathbf{\mathcal{M}} + \tilde{\mathbf{\mathcal{M}}}[\mathbf{q}] \right) = 0.
\label{eq_secular}
\end{eqnarray}

In order to analyze this second condition, we need to evaluate the matrix coefficients explicitly. For this purpose, we first discuss the mathematical representation of the Schwinger propagator in the next section.

\begin{figure*}
    \centering
    \includegraphics[width=1\textwidth]{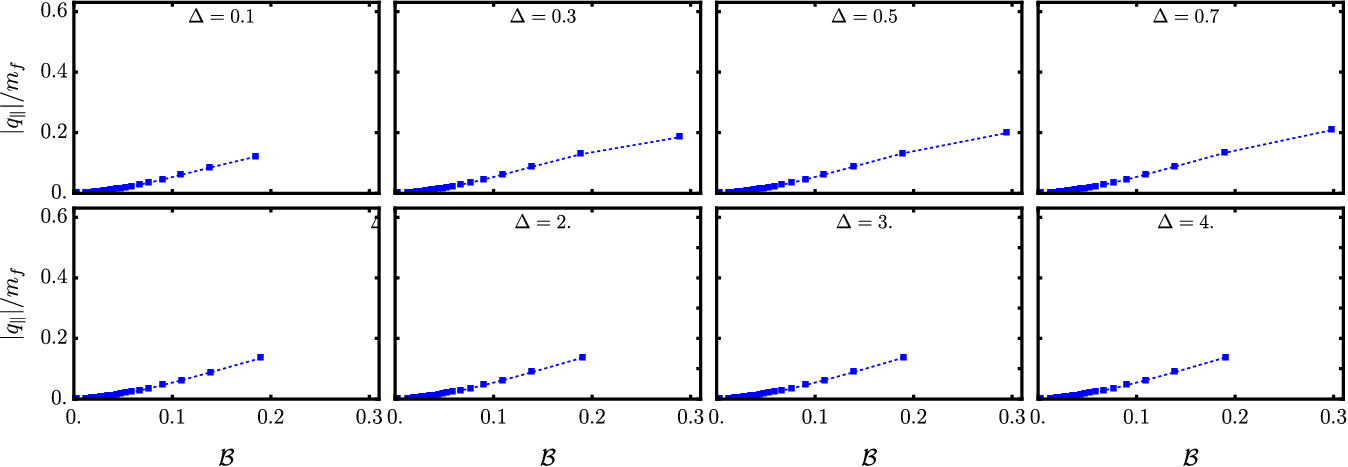}
    \caption{Non-trivial solutions for $q_3$ of Eq.~\eqref{eq:det1} as a function of $eB$ for $\Delta_B\in[0.1,4]$.}
    \label{fig:q3VseB2}
\end{figure*}

\subsection{The Schwinger propagator}

As discussed in the previous section, the matrix coefficients depend on traces and integrals involving products of the fermion propagator immersed in the constant background magnetic field.
Therefore, this allows us to use directly the Schwinger proper-time representation of the free-Fermion propagator dressed by the background field, whose direction is chosen along the $z$-axis, $\mathbf{B} = \hat{e}_3\,B$, as follows~\cite{Schwinger_1951,Dittrich_Reuter}
\begin{eqnarray}
\left[S_\text{F}(k)\right]_{a,b}
&&= -\ii\delta_{a,b}\int_{0}^{\infty}\frac{d\tau}{\cos(\qB \tau)}
e^{\ii\tau\left(k_{\parallel}^2 - \mathbf{k}_{\perp}^2\frac{\tan(\qB\tau)}{\qB\tau}-m^2_f + \ii\epsilon \right)}\nonumber\\
&&\times\left\{
\left[\cos(\qB\tau) + \gamma^1\gamma^2\sin(\qB\tau)  \right](m_f + \slashed{k}_{\parallel})\right.\nonumber\\
&&\left.+\frac{\slashed{k}_{\perp}}{\cos(\qB \tau)}
\right\},
\end{eqnarray}
which is clearly diagonal in replica space. Here, as usual, we separated the parallel from the perpendicular directions with respect to the background external magnetic field by splitting the metric tensor as $g^{\mu\nu} = g_{\parallel}^{\mu\nu} + g_{\perp}^{\mu\nu}$, with
\begin{eqnarray}
g_{\parallel}^{\mu\nu} &=& \text{diag}(1,0,0,-1),\nonumber\\
g_{\perp}^{\mu\nu} &=& \text{diag}(0,-1,-1,0),
\end{eqnarray}
thus implying that for any 4-vector, such as the momentum $k^{\mu}$, we write
\begin{eqnarray}
\slashed{k} = \slashed{k}_{\perp} + \slashed{k}_{\parallel},
\end{eqnarray}
and
\begin{eqnarray}
k^2 = k_{\parallel}^2 - \mathbf{k}_{\perp}^2,
\end{eqnarray}
respectively. In particular, we have $k_{\parallel}^2 = k_0^2 - k_3^2$,
while $\mathbf{k}_{\perp}=(k^1,k^2)$ is the Euclidean 2-vector lying in the plane perpendicular to the field, such that its square-norm is  $\mathbf{k}_{\perp}^2 = k_1^2 + k_2^2$.
The Schwinger propagator can alternatively be expressed as~\cite{Replicas_1}
\begin{eqnarray}
&&\left[S_\text{F}(k) \right]_{a,b} = -\ii\delta_{a,b}
\left[
\left( m_f + \slashed{k} \right)\mathcal{A}_{1}\right.\nn\\
&&\left.
+ (\ii \qB) \gamma^{1}\gamma^{2}\left( m_f + \slashed{k}_{\parallel} \right)\frac{\partial\mathcal{A}_1}{\partial \mathbf{k}_{\perp}^2} + \left(\ii \qB \right)^2\slashed{k}_{\perp}\frac{\partial^2\mathcal{A}_1}{\partial (\mathbf{k}_{\perp}^2)^2}
\right]\nonumber\\
&&=-\ii\delta_{a,b}\left[ 
\left( m_f + \slashed{k}_{\parallel} \right)\mathcal{A}_1
+  \gamma^{1}\gamma^{2}\left( m_f + \slashed{k}_{\parallel} \right)  \mathcal{A}_2
+ \mathcal{A}_3 \slashed{k}_{\perp}
\right]\nonumber\\
\end{eqnarray}

Here, we defined the function
\begin{eqnarray}
\mathcal{A}_1(k,B) = \int_{0}^{\infty}d\tau e^{\ii\tau\left( k_{\parallel}^2 - m_f^2 + \ii\epsilon\right) -\ii\frac{\mathbf{k}_{\perp}^2}{\qB}\tan(\qB \tau) },
\label{eq:A1}
\end{eqnarray}
that clearly reproduces the scalar propagator (with Feynman prescription) in the zero-field limit
\begin{eqnarray}
\lim_{B\rightarrow 0}\mathcal{A}_1(k,B) = \frac{\ii}{k^2 - m_f^2 + \ii\epsilon}, 
\end{eqnarray}
and its derivatives
\begin{subequations}
\bea
\mathcal{A}_2(k,B)&\equiv&\int_0^\infty d\tau ~\tan(\qB\tau)e^{\ii\tau\left(k_\parallel^2-\tb{\tau}\mathbf{k}_\perp^2-m^2_f+\ii\epsilon\right)}\nn\\
&=&\ii \qB\frac{\partial\mathcal{A}_1}{\partial(\mathbf{k}_{\perp}^2)},
\eea
\bea
\mathcal{A}_3(k,B)&\equiv&\int_0^\infty \frac{d\tau}{\cos^2(\qB\tau)}e^{\ii\tau\left(k_\parallel^2-\tb{\tau}\mathbf{k}_\perp^2-m_f^2+\ii\epsilon\right)}\nn\\
&=&\mathcal{A}_1+(\ii \qB)^2\frac{\partial^2\mathcal{A}_1}{\partial(\mathbf{k}_{\perp}^2)^2}.
\eea
%
\end{subequations}

As we showed in detail in our previous work~\cite{Replicas_1}, an exact representation for the function \eqref{eq:A1} is given by
\begin{eqnarray}
\mathcal{A}_1 (k) &=& \frac{\ii\,e^{-\frac{k_{\perp}^2}{eB}}}{2 e B} e^{-\frac{\ii\pi}{2 e B}\left( k_{\parallel}^2 - m_f^2 + \ii\epsilon  \right)}\Gamma\left(-\frac{k_{\parallel}^2 - m_f^2 + \ii\epsilon}{2 e B}  \right)\nonumber\\ &&\times U\left( -\frac{k_{\parallel}^2 - m_f^2 + \ii\epsilon}{2 e B},0,\frac{2k_{\perp}^2}{e B} \right),
\label{eq:A1_hyper}
\end{eqnarray}
where $\Gamma(z)$ is the Gamma function, while $U(a,b,z)$ represents the Tricomi's Confluent Hypergeometric function.

%

\subsection{Results and Discussion}
\begin{figure*}
    \centering
    \includegraphics[scale=1.2]{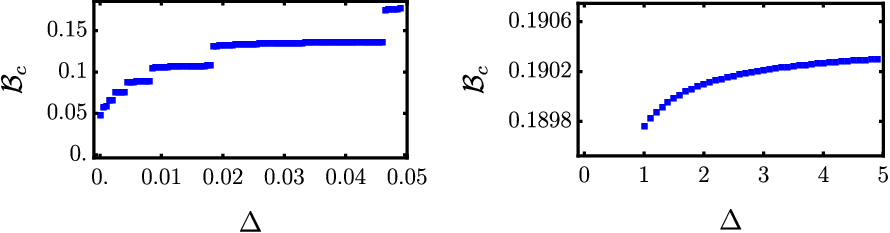}
    \caption{Critical magnetic field $\mathcal{B}_c$ of Fig.~\ref{fig:q3VseB2} as a function of $\Delta$, in two different ranges of such parameter.}
    \label{fig:Bc_vs_Delta_high_1}
\end{figure*}

For the analysis of our numerical results, it is convenient to define the following dimensionless groups,
\bea
\Delta &=& e^2 \Delta_B m_f^2,\nn\\
\mathcal{B} &=& \frac{e B}{m_f^2},
\eea
respectively. 
The matrices $\mathcal{M}^{ij}$ and $\mathcal{M}^{ijkl}$, as defined by Eq.~\eqref{m2} and Eq.~\eqref{m4} respectively, are calculated by tracing over the Dirac matrix space. All technical details concerning the evaluation of traces as well as the resulting momentum integrals are explained in appendices A and B in~\cite{Castano-Yepes:2023brq}. The interested reader is referred to those Appendices for further mathematical details.

In order to analyze the existence and features of non-trivial solutions for the order parameter $\boldsymbol{q} = (q_1,q_2,q_3)$,
we solve the secular equation Eq.~\eqref{eq_secular}, by assuming two different symmetry conditions, according to the directions orthogonal ($\perp$) and parallel ($\parallel$) to the magnetic field, respectively.

\subsection{Case 1: $q_3^2 \equiv q_{\parallel}^2$, with $q_1 = q_2 = 0$}

If we impose the condition $q_1=q_2=0$ into Eq.~\eqref{eq_Mtilde_gen}, we have
\begin{eqnarray}
\left[\tilde{\mathcal{M}}[q_{\parallel}]\right]^{jl}
&=&\left( -\frac{1}{2}\delta^{jl}\mathcal{M}^{33} + e^2\Delta_B \mathcal{M}^{jl33} \right)q_{\parallel}^2\nonumber\\
&\equiv& \mathcal{C}^{jl}_{\parallel}q_{\parallel}^2.
\label{eq_case1}
\end{eqnarray}
Therefore, substituting this reduced expression into the secular equation Eq.~\eqref{eq_secular}, we obtain
\bea
{\rm{det}}\left( \Delta_B \mathcal{M} + q_{\parallel}^2 \mathcal{C}_{\parallel}   \right) = 0.
\label{eq:det1}
\eea
Furthermore, using elementary matrix properties and given that the matrix $\mathcal{M}$ is non-singular, we can manipulate the expression above as follows
\bea
&&{\rm{det}}\left( \Delta_B \mathcal{M} + q_{\parallel}^2 \mathcal{C}_{\parallel}   \right) =\\\nonumber 
&&{\rm{det}}\left( \Delta_B \mathcal{M} \right) \cdot {\rm{det}}\left( \mathbf{1}_3 + q_{\parallel}^2 \Delta_B^{-1} \mathcal{M}^{-1} \mathcal{C}_{\parallel}   \right) = 0.
\eea
Given that $\mathcal{M}$ is non-singular, the above expression implies the secular condition
\bea
{\rm{det}}\left( \mathbf{1}_3 + q_{\parallel}^2 \Delta_B^{-1} \mathcal{M}^{-1} \mathcal{C}_{\parallel}   \right) = 0.
\label{eq_sec_case1}
\eea
\begin{figure}
    \centering
    \includegraphics[width=0.8\textwidth]{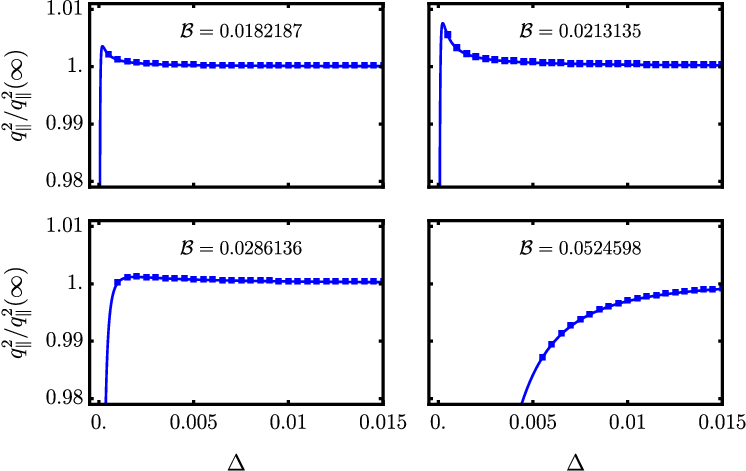}
    \caption{Discrete solutions for the order parameter (normalized by its asymptotic limit) as a function of $\Delta$, for fixed $\mathcal{B}$ (filled squares). The continuous line represents the envelope function defined by Eq.\eqref{eq:Sigmoid_parallel} before the conditions of Eq.\eqref{eq_qj} have been applied.}
    \label{fig:q3VsDelta}
\end{figure}
Our analysis at this point is consistent, up to second order powers in the components of the order parameter. Therefore, applying the elementary identity ${\rm{det}}\left(\mathbf{1} + \epsilon X  \right) = 1 + \epsilon {\rm{tr}}X + O(\epsilon^2)$, we expand Eq.~\eqref{eq_sec_case1} to obtain
\bea
q_{\parallel}^2 &=& -\frac{\Delta_B}{{\rm{tr}}\left( \mathcal{M}^{-1} \mathcal{C}_{\parallel}  \right)}\nonumber\\ 
&=& -\frac{\Delta_B}{-\frac{1}{2}\mathcal{M}^{33}{\rm{tr}}\left( \mathcal{M}^{-1}\right) + e^2\Delta_B\rm{tr}\left( \mathcal{M}^{-1} \mathcal{\tilde{M}}^{(33)}  \right)}\nonumber\\
&=& \frac{\chi_{\parallel}\Delta_B}{1 + \mathcal{K}_{\parallel}\Delta_B},
\label{eq:Sigmoid_parallel}
\eea
where we defined the parameters
\bea
\chi_{\parallel} &=& \frac{2}{\mathcal{M}^{33}\,{\rm{tr}}\left( \mathcal{M}^{-1}\right)}\nonumber\\
\mathcal{K}_{\parallel} &=& \frac{-2e^2\rm{tr}\left( \mathcal{M}^{-1} \mathcal{\tilde{M}}^{(33)}  \right)}{\mathcal{M}^{33}\,{\rm{tr}}\left( \mathcal{M}^{-1}\right)},
\eea
as well as the reduced matrix
\bea
\left[\tilde{\mathcal{M}}^{(33)}\right]^{jl} \equiv \mathcal{M}^{jl33}.
\eea

Figure~\ref{fig:q3VseB1} illustrates the non-trivial solutions of Eq.~\eqref{eq:det1} for Case 1, as a function of the external magnetic field and various values of $\Delta$. It can be observed that there exists a region where the discrete solutions exhibit a monotonically increasing pattern with a smooth envelope, abruptly terminated at a point beyond which only the trivial solution $q_\parallel=0$ exists. We refer to this point as the {\it critical magnetic field} $\mathcal{B}_c$. A similar scenario arises when the magnitude of $\Delta$ is 
increased, as demonstrated in Fig.~\ref{fig:q3VseB2}. Furthermore, it is worth noting that for higher values of $\Delta$, the solutions become approximately identical, and the $\mathcal{B}_c$ converges toward a specific limit as is shown in Fig.~\ref{fig:Bc_vs_Delta_high_1}.


In line with the results presented above, it is worth emphasizing that a specific combination of parameters $(\Delta,\mathcal{B})$ plays a pivotal role in giving rise to a discrete set of non-trivial solutions characterized by $q_\parallel\neq 0$, or in the context of Eq.~\eqref{eq_qj}, resulting in purely imaginary values for $q_j$. This behavior is depicted in Figure~\ref{fig:q3VsDelta}, offering valuable insights into the system's dynamics. Figure~\ref{fig:q3VsDelta} illustrates a spectrum of these parameters, revealing that for fixed values of $\mathcal{B}$ not all the values of $\Delta$ produce non-trivial solutions (those are discrete). We have also shown in Figure~\ref{fig:q3VsDelta}, by a dashed line, the smooth envelope of those discrete points.
Nevertheless, for higher values of $\Delta$ the solutions become a quasi-continuum and saturate to the value $q_\parallel^2(\infty)$ defined as
\bea
q_\parallel^2(\infty)=\lim_{\Delta\to\infty}q_\parallel^2=\frac{\chi_\parallel}{\mathcal{K}_\parallel},
\eea
where $q_\parallel^2$ is given in Eq.~\eqref{eq:Sigmoid_parallel}.
In the opposite limit, for very small values of $\Delta$, Eq.~\eqref{eq:Sigmoid_parallel} shows that the order parameter follows an approximately linear trend with a slope defined by
\bea
\chi_{\parallel} = \lim_{\Delta_B\rightarrow 0}\frac{\partial}{\partial \Delta_B} q_{\parallel}^2.
\eea
\begin{figure*}
    \centering
    \includegraphics[width=1\textwidth]{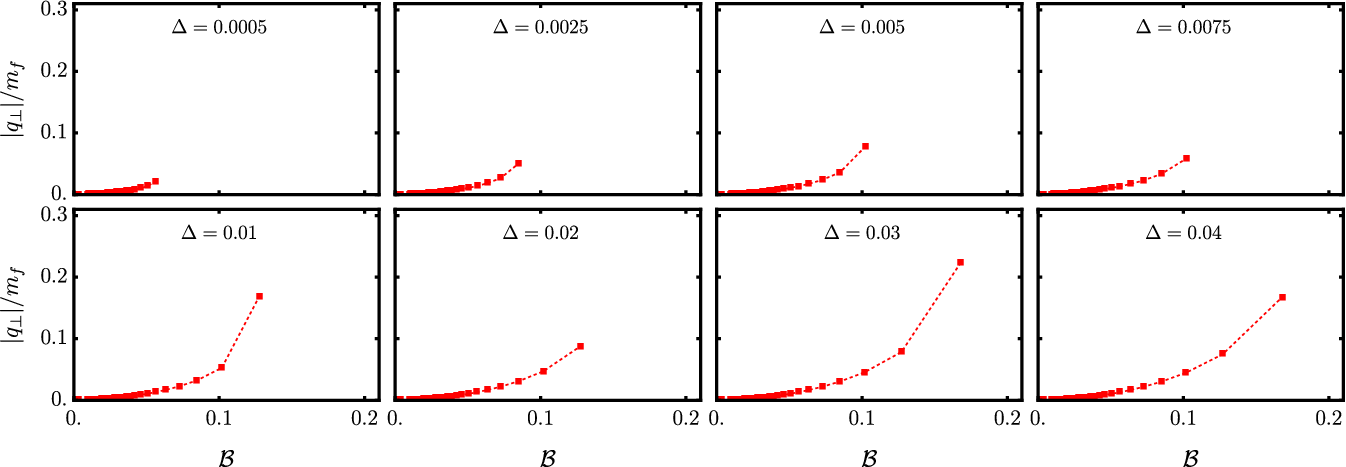}
    \caption{Non-trivial solutions of Eq.~\eqref{eq:det1} for the Case 2 as a function of $\mathcal{B}$ for $\Delta\in[0.1,0.04]$. The dashed line is the smooth envelope connecting those discrete non-trivial solutions.}
    \label{fig:qpVseB1}
\end{figure*}

\subsection{Case 2: $q_1^2 = q_2^2 \equiv q_{\perp}^2$, with $q_3 =0$ }

In this case, the matrix Eq.~\eqref{eq_Mtilde_gen} reduces to
\bea
&&\left[\tilde{\mathcal{M}}[q_{\perp}]\right]^{jl} = \left(-\frac{1}{2}\delta^{jl}\left( \mathcal{M}^{11} + \mathcal{M}^{22} + \mathcal{M}^{12} + \mathcal{M}^{21} \right)\right.\nonumber\\
&&\left.+ e^2 \Delta_{B}
\left( \mathcal{M}^{jl11} + \mathcal{M}^{jl22} + \mathcal{M}^{jl12} + \mathcal{M}^{jl21}     \right)
\right) q_{\perp}^2\nonumber\\
&&\equiv \mathcal{C}_{\perp}^{jl} q_{\perp}^2.
\eea
For this case, the secular Eq.~\eqref{eq_secular} reduces to
\bea
{\rm{det}}\left( \Delta_B \mathcal{M} + q_{\perp}^2 \mathcal{C}_{\perp}   \right) = 0,
\eea
and repeating the same procedure as described in Case 1, we obtain
\bea
{\rm{det}}\left( \mathbf{1}_3 + q_{\perp}^2 \Delta_B^{-1} \mathcal{M}^{-1} \mathcal{C}_{\perp}   \right) = 0.
\label{eq_sec_case2}
\eea

Finally, just as in Case 1, we retain only second order powers of $q_{\perp}$ in Eq.~\eqref{eq_sec_case2} to arrive at the explicit algebraic solution
\bea
q_{\perp}^2 = -\frac{\Delta_B}{{\rm{tr}}\left( \mathcal{M}^{-1} \mathcal{C}_{\perp}  \right)}= \frac{\chi_{\perp}\Delta_B}{1 + \mathcal{K}_{\perp}\Delta_B},
\label{eq:Sigmoid_perp}
\eea
where we defined the parameters
\bea
\chi_{\perp} &=& \frac{2}{\left(\sum_{m,n=1,2}\mathcal{M}^{mn}\right)\,{\rm{tr}}\left( \mathcal{M}^{-1}\right)},\nonumber\\
\mathcal{K}_{\perp} &=& \frac{-2e^2\sum_{m,n=1,2}{\rm{tr}}\left( \mathcal{M}^{-1} \mathcal{\tilde{M}}^{(mn)}  \right)}{\left(\sum_{m,n=1,2}\mathcal{M}^{mn}\right)\,{\rm{tr}}\left( \mathcal{M}^{-1}\right)},
\eea
and the reduced matrices
\bea
\left[\mathcal{\tilde{M}}^{(mn)}\right]^{jl}
= \mathcal{\tilde{M}}^{jlmn}.
\eea
As already discussed in Eq.~\eqref{eq_qj}, the order parameter $q_j$ (for $j=1,2,3$) is a pure imaginary number. 
\begin{figure}
    \centering
    \includegraphics[scale=1.2]{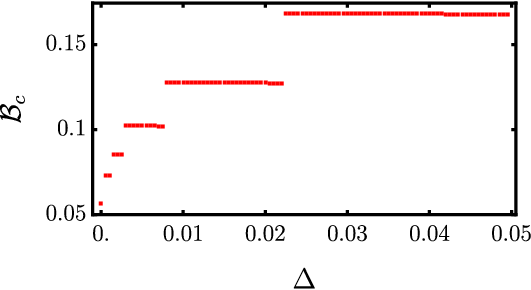}
    \caption{Critical magnetic field $\mathcal{B}_c$ of Fig.~\ref{fig:qpVseB1} as a function of $\Delta$ for the Case 2.}
    \label{fig:Bc_vs_Delta_high_1_qp}
\end{figure}

\begin{figure}
    \centering
    \includegraphics[width=0.8\textwidth]{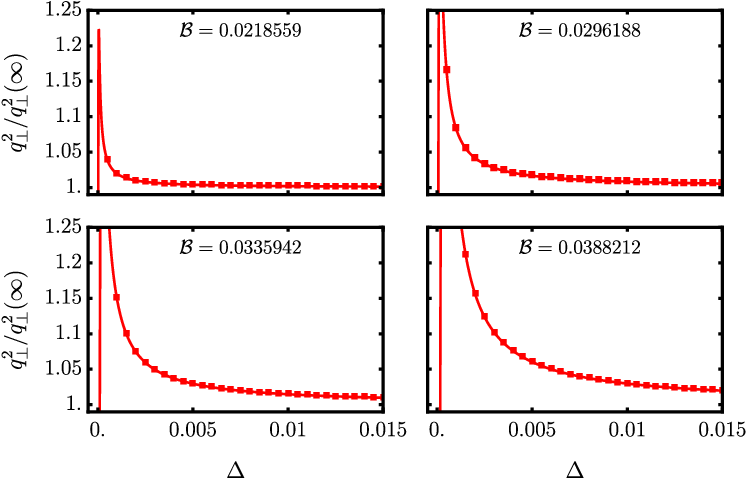}
    \caption{Discrete solutions for the order parameter (normalized by its asymptotic limit) as a function of $\Delta$, for fixed $\mathcal{B}$ (filled squares). The continuous line represents the envelope function defined by Eq.\eqref{eq:Sigmoid_perp} before the conditions of Eq.\eqref{eq_qj} have been applied.}
    \label{fig:qperpVsDelta}
\end{figure}

Figure \ref{fig:qpVseB1} illustrates the non-trivial solutions for $q_\perp$ in Case 2, considering various values of $\Delta$. Similar to Case 1, these solutions are discrete and highly dependent on the magnetic field and noise parameter. Notably, the critical magnetic field in this scenario is lower compared to Case 1. Furthermore, Fig.~\ref{fig:Bc_vs_Delta_high_1_qp} demonstrates the behavior of the critical magnetic field $\mathcal{B}_c$, which exhibits a distinct functional form from that shown in Fig.~\ref{fig:Bc_vs_Delta_high_1}. The relationship between $q_\perp^2$ and $\Delta$, with an envelope given by Eq.~\eqref{eq:Sigmoid_perp}, is depicted in Fig.~\ref{fig:qperpVsDelta}, where discrete non-trivial solutions at particular values of $\Delta$ are permissible for a constant magnetic field. Here, we can also identify the value at which the sigmoidal Eq.~\eqref{eq:Sigmoid_perp} saturates as a function of the noise $\Delta_B$,
\bea
q_\perp^2(\infty)=\lim_{\Delta\to\infty}q_\perp^2=\frac{\chi_\perp}{\mathcal{K}_\perp}.
\eea
In the opposite limit, for very small values of $\Delta$, Eq.~\eqref{eq:Sigmoid_perp} shows that the order parameter follows an approximately linear trend with a slope defined by

\bea
\chi_{\perp} = \lim_{\Delta_B\rightarrow 0}\frac{\partial}{\partial \Delta_B} q_{\perp}^2.
\eea
Consequently, we can infer that the underlying physics in both cases is closely related, and the mechanism leading to the emergence of a non-trivial value of the order parameter maintains a consistent nature and interpretation, regardless of the specific values of $(q_1, q_2, q_3)$.


\subsection{Physical interpretation of the order parameter $Q_j$}
The physical interpretation of the order parameter components can be obtained from the functional representation Eq.~\eqref{eq_barZn}, {\it{before}} we integrate out the fermions, to obtain, via saddle-point, the mean expectation value

\begin{eqnarray}
Q_j = \ii e \Delta_B \langle \langle \bar{\psi}\gamma_j \psi\rangle\rangle_{\Delta} = - Q_j^*,
\end{eqnarray}
where here the double bracket stands for the statistical average over the classical noise in the background field, as well as the quantum expectation value of the corresponding observable, which clearly is a component of the vector current for the fermions, $j = 1,2,3$.

It is important to remark the physical effect that this order parameter, at the mean-field level, produces in the disorder-averaged fermion propagator, that is given by 
\begin{equation}
\ii S_{\text{F},\Delta}^{-1} (x - y) = \left( \ii\slashed{\partial} - e \slashed{A}_\text{BG} - m_f - e \slashed{q} \right)_x\delta^{(4)}(x-y), 
\end{equation}
where we defined $\slashed{q} = \gamma^{j} q_{j}$, for $q_j = Q_{j} - Q_{j}^{*}$ the order parameter. Clearly then, the differential equation for the propagator in the presence of the magnetic noise is given by
\begin{eqnarray}
\left( \ii\slashed{\partial} - e\slashed{A}_\text{BG} - m_f - e \slashed{q}  \right)_x S_{\text{F},\Delta}(x-y) = \ii \delta^{(4)}(x - y).
\label{eq_SDelta}
\end{eqnarray}

It is straightforward to verify that, if $S_\text{F} (x-y)$ represents the Schwinger propagator in the presence of the average background magnetic field, then the function
\begin{eqnarray}
S_{\text{F},\Delta}(x-y) = e^{- \ii e q\cdot(x-y)} S_\text{F}(x-y)
\label{eq_disphase}
\end{eqnarray}
is a solution to Eq.~\eqref{eq_SDelta}. Indeed, by direct substitution, we have
\begin{eqnarray}
\left( \ii\slashed{\partial} - e\slashed{A}_\text{BG} - m_f - e \slashed{q}  \right)_x S_{\text{F},\Delta}(x-y)
&=& \left( \ii\slashed{\partial} - e\slashed{A}_\text{BG} - m_f - e \slashed{q}  \right)_x \left[ e^{- \ii e q\cdot(x-y)} S_F(x-y) \right]\nonumber\\
&=& e^{- \ii e q\cdot(x-y)} \left( \ii\slashed{\partial}  - e\slashed{A}_\text{BG} - m_f   \right)_x S_\text{F}(x-y) = \ii \delta^{(4)}(x - y),
\end{eqnarray}
where in the last step we applied the definition Eq.~\eqref{eq_Schwingernonoise} for the Schwinger propagator in the absence of noise.
Therefore, we conclude that at the level of the disorder-averaged propagator, Eq.~\eqref{eq_disphase} introduces an exponential damping effect given that the order parameter $q_j$ is a pure imaginary number.

\subsection{Summary}

In this section, we considered a system of QED fermions submitted to an external, classical magnetic field. In particular, we studied the effects of white noise in this magnetic field with respect to an average uniform value $\langle\mathbf{B}\rangle_{\Delta} = \hat{\mathbf{e}}_3 B$, as a function of the standard deviation $\Delta_B$, over the fermion propagator. As discussed in the introduction to this section, this represents a statistical model for the actual scenario in heavy-ion collisions, where strong magnetic fields emerge for very short times within small spatial regions, whose size is of the order of the scattering cross-section. Since several such collisions occur at different points in space, the physical situation can be represented by a statistical ensemble, for different realizations of the magnetic field fluctuations, which are then described as a random variable.

We analyzed our model by applying the replica formalism, that led us to an effective action in terms of auxiliary boson fields. A mean field analysis of the corresponding effective action reveals that the magnetic noise effects can be captured by an order parameter, whose physical interpretation is the statistical ensemble average of the expectation value of the fermion vector current components. Therefore, non-trivial solutions where this order parameter acquires a non-zero value break the U(1) gauge symmetry in the system, as a consequence of the statistical noise in the background magnetic field. An interesting feature of such non-trivial solutions is that they exist only for certain discrete values of the average background magnetic field. Such discrete values can be identified to be in correspondence with the quantized Landau levels associated to the average background field. This feature is then consistent with the interpretation of the order parameter as the ensemble average of the fermion current. In addition, for a fixed value of the disorder strength characterized by $\Delta_B$, we find an upper critical value of the average background magnetic field $\mathcal{B}_c$, beyond  which the non-trivial solutions cease to exist in favour of a vanishing order parameter. This region of parameter space is then characterized by a dominance of the average background field over noise, whose effect then becomes negligible. In contrast, in the limit of very strong magnetic noise $\Delta_B\rightarrow\infty$, we observe that the order parameter asymptotically saturates to a constant finite value $q_{\parallel,\perp}^2(\infty)$ that depends on the field, but is independent of $\Delta_B$, as can be clearly seen from Eq.~\eqref{eq:Sigmoid_parallel} and Eq.~\eqref{eq:Sigmoid_perp}, respectively.

Remarkably, in the context of the fermion propagator, we showed that the order parameter, which is strictly imaginary, represents a finite screening length that leads to weak localization effects. Our present analysis is restricted to the fundamental level of the fermion propagator, but its consequences could manifest themselves in physical observables, such as effective collision rates for certain processes.

\section{Photon mass generation from magnetic fluctuations}\label{Castano}

\subsection{Introduction}
The importance of the role of magnetic fields in processes related to high-energy physics has been garnering attention in recent years. In particular, there is growing theoretical evidence indicating that during the early stages of a heavy-ion collision (HIC), very intense magnetic fields are created, which must be included in current calculations and data analysis~\cite{David_2020,PhysRevC.91.064904}. Moreover, these magnetic fields are expected to have a direct impact on the final observables. Therefore, it is crucial to understand their role and the modifications of the observables that arise from magnetic-related contributions. For instance, the emission of prompt photons during these early stages has been studied, revealing that the inclusion of external magnetic fields in the participant's region introduces an additional source of photons carrying a characteristic elliptic flow~\cite{Ayala:2017vex,Ayala2020,PhysRevC.106.064905,JIA2023138239}. This phenomenon arises from the breaking of spatial symmetry due to the presence of a privileged direction marked by the magnetic field. 

On the other hand, the creation of intense magnetic fields in HICs coincides with the Color Glass Condensate (CGC) stage, where the medium is purely described by a high-density gluon occupation~\cite{LAPPI2006200,MCLERRAN201471, Harland-Lang2015, PhysRevC.106.034904}. Indeed, the prompt photon production by gluon fusion or splitting processes is enabled by the presence of magnetic fields which are otherwise not allowed. Therefore, it is important to ask: How can the magnetic fields modify the dispersive relations of gauge fields, such as gluons and photons? In an effort to answer that question, the one-loop polarization tensor for photons and gluons in the presence of an external magnetic field has been computed, both for strong and weak limits~\cite{Ayala2021}. Those works showed that the presence of constant magnetic fields doesn't modify the dispersion relation of such particles, i.e., they don't generate an associated magnetic mass. The latter is achieved by computing the poles of the inverse propagator at the limit of vanishing four momenta, which is crucial for preserving the $U(1)$ symmetry and ensuring the fulfillment of the Ward-Takahashi identity~\cite{PhysRevD.101.036016}.

Clearly, assuming the existence of a constant magnetic field in HICs is not very realistic. In fact, simulations of collisions for various centralities, energies, and species indicate that although the magnetic field in the collision center is predominantly oriented in one spatial direction, there are fluctuations in other directions~\cite{castano2021effects}. In some cases, these fluctuations are of the order of $1/4$ of the highest value of the magnetic field. Following this scenario, in previous works, we proposed to study the impact of classical background magnetic fields, which possess stochastic fluctuations, on the properties of a QED medium~\cite{PhysRevD.107.096014, PhysRevD.108.116013}. By employing the so-called {\it replica trick}~\cite{mezard1991replica}, we derived an effective interaction term for QED fermions in the presence of such a noisy magnetic field. This approximation leads, at the perturbative level, to a renormalization of the fermion propagator, which now describes quasi-particles propagating in a dispersive medium. The results from both perspectives revealed that magnetic fluctuations break the $U(1)$ symmetry, and thus, the dispersion relations for gluons and photons may be modified.

In this section, by computing the one-loop polarization tensor in the limit of vanishing four momenta, we show that gauge fields may acquire a magnetic mass due to fluctuations in the external magnetic field. The structure of the paper is as follows: in Sec.~\ref{sec:The one-loop polarization tensor}, we propose the expression for the one-loop polarization tensor that incorporates white-noise stochastic fluctuations. In Sec.~\ref{sec:The one-loop polarization tensor in the limit}, we summarize the results in the limit $p^\mu\to0$, where the expressions are analytically computed. In Sec.~\ref{sec:Mass of gauge fields by magnetic fluctuations}, we present the analysis to determine the magnetic masses for the gauge fields. Finally, in Sec.~\ref{sec:Summary and conclusions}, we summarize and present our conclusions. This manuscript is based on the published version that can be found in Ref.~\cite{PhysRevD.109.056007}.

\subsection{The one-loop polarization tensor}\label{sec:The one-loop polarization tensor}
\begin{figure}[h!]
    \centering
    \includegraphics[scale=0.27]{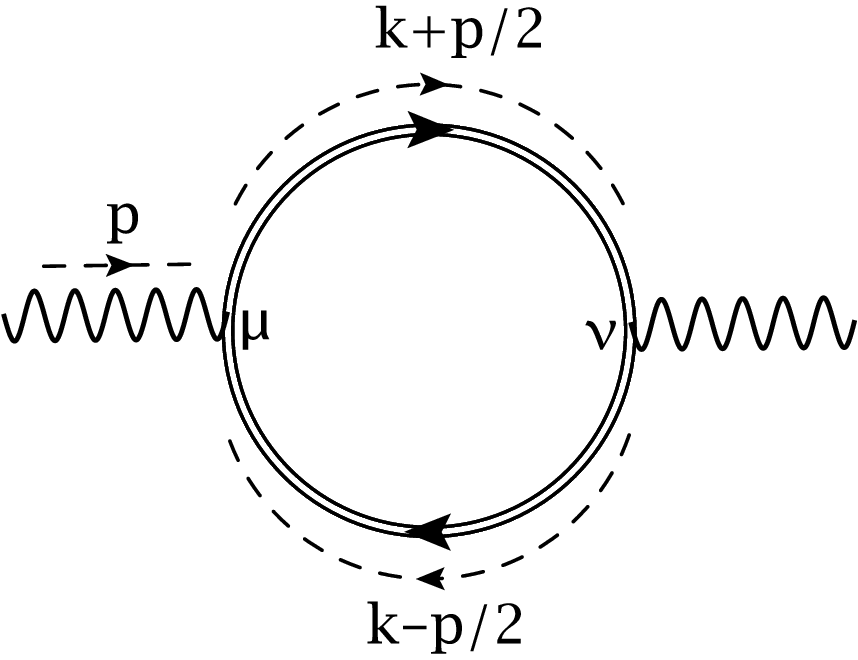}\hspace{0.2cm}\includegraphics[scale=0.27]{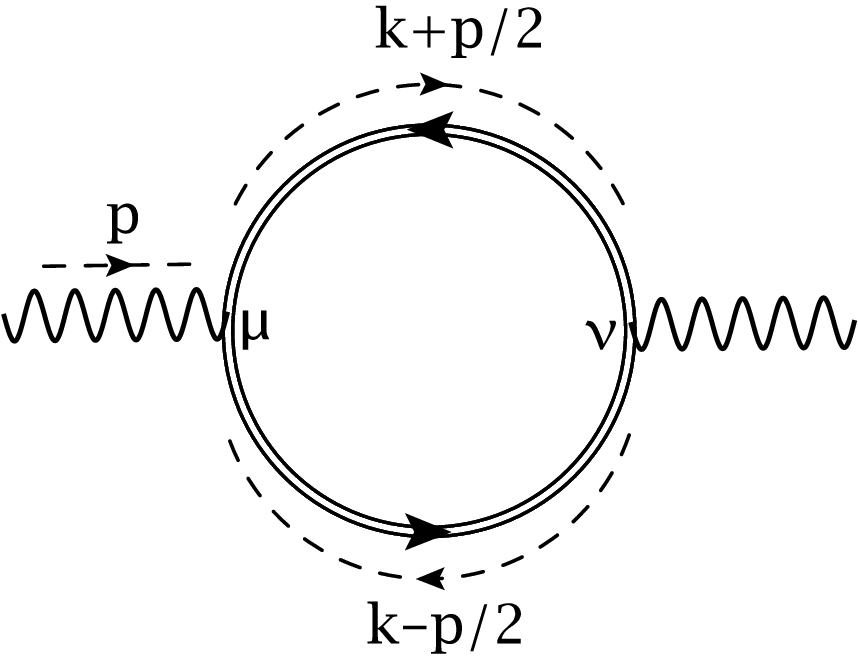}
    \caption{Feynman diagrams contributing to the one-loop photon polarization tensor depict arrows in the propagators indicating the direction of charge flow, while dashed arrows represent momentum flux.}
    \label{fig:polarization}
\end{figure}

To proceed with our analysis, we express the one-loop contribution to the photon/gluon polarization tensor, as illustrated in Fig.~\ref{fig:polarization}, as follows
{\small
\bea
\ii\Pi_\Delta^{\mu\nu}&=&-\frac{1}{2}\int\frac{d^4k}{(2\pi)^4}\text{Tr}\Big\{\ii q_f\gn\ii S_\Delta^{(-)}\left(k\right)\ii q_f\gm\ii S_\Delta^{(-)}(k-p)\Big\}-\frac{1}{2}\int\frac{d^4k}{(2\pi)^4}\text{Tr}\Big\{\ii q_f\gn\ii S_\Delta^{(+)}(-k+p)\ii q_f\gm\ii S_\Delta^{(+)}(-k)\Big\},
\label{eq:PloTenDef}
\eea
}
where $q_f$ represents the electric charge of the fermion within the loop, accounting for both particle and antiparticle contributions.

To perform an analytical computation of~Eq.\eqref{eq:PloTenDef}, we will utilize the renormalized fermion propagator in the presence of static (quenched) white noise spatial fluctuations, focusing particularly on the regime of a strong external magnetic field~\cite{PhysRevD.107.096014,PhysRevD.108.116013}. Specifically, the effective interaction between fermions arises from averaging over the background magnetic noise. We account for the effects of magnetic noise-induced interaction by modifying the Schwinger propagator with a self-energy, as illustrated diagrammatically in the Dyson equation shown in Fig.~\ref{fig:DiagramSelfEnergy2}. It is noteworthy that in this theory, the skeleton diagram for the self-energy is depicted in Fig.~\ref{fig:DiagramSelfEnergy1}, and the dressed propagator is determined by

\bea
\ii S_\Delta^{(\pm)}(p)&=&C(p)\left(m+\gamma_0 p^0+\frac{\gamma_3 p^3}{z(p)}\right)\mathcal{P}^{(\pm)}(p),\nn\\
\label{eq:Prop_Delta}
\eea
where $m$ is the fermion mass and
\begin{subequations}
\bea
C(p)&=&\ii\frac{z(p)\,e^{-\mathbf{p}_\perp^2/|q_f B|}}{p_\parallel^2-z^2(p)m^2},
\eea
\bea
z(p)=1+\frac{3}{4}\frac{\Delta|q_f B|e^{-\mathbf{p}_\perp^2/|q_f B|}}{\pi\sqrt{p_0^2-m^2}}.
\eea
\bea
z_3(p)=\frac{1}{z(p)}\left(1+\frac{\Delta|q_f B|e^{-\mathbf{p}_\perp^2/|q_f B|}}{4\pi\sqrt{p_0^2-m^2}}\right),
\eea

\bea
\mathcal{P}^{(\pm)}(p)=\frac{1}{2}\left[1\pm\sgn(q_f B)\ii z_3(p)\gamma^1\gamma^2\right].
\eea
\end{subequations}

As outlined in Ref.~\cite{PhysRevD.107.096014}, the computation of the fermion propagator self-energy was carried out to order $\Delta \equiv q_f^2 \Delta_B$. Thus, to preserve consistency with this approximation level, we expand Eq.~\eqref{eq:Prop_Delta} as follows
\bea
\ii S_\Delta^{(\pm)}(p)&=&\ii S_0^{(\pm)}(p)+\ii\Delta\left(\frac{\qfB}{2\pi}\right)\Big[\Theta_1(p)(\slashed{p}_\parallel+m)\mathcal{O}^{(\pm)}-\Theta_2(p)\gamma^3\mathcal{O}^{(\pm)}\pm\Theta_3(p)\ii\gamma^1\gamma^2(\slashed{p}_\parallel+m)\Big]+\text{O}(\Delta^2),
\eea
where
\bea
\ii S_0^{(\pm)}(p)&=&2\ii\frac{e^{-\pt^2/|q_f B|}}{\ppar^2-m^2}(\slashed{p}_\parallel+m)\mathcal{O}^{(\pm)}
\eea
is the fermion propagator in the presence of an intense magnetic field, the spin-projection operator is given by
\bea
\mathcal{O}^{(\pm)}&=&\frac{1}{2}\left[1\pm\sgn(q_f B)\ii\gamma^1\gamma^2\right],
\eea
and we defined the functions
\begin{subequations}
\bea
\Theta_1(p)&\equiv&\frac{3(\ppar^2+m^2)e^{-2\pt^2/|q_f B|}}{(\ppar^2-m^2)^2\sqrt{p_0^2-m^2}} ,
\eea
\bea
\Theta_2(p)&\equiv&\frac{3p_3e^{-2\pt^2/|q_f B|}}{(\ppar^2-m^2)\sqrt{p_0^2-m^2}},
\eea
\bea
\Theta_3(p)&\equiv&\frac{e^{-2\pt^2/|q_f B|}}{(\ppar^2-m^2)\sqrt{p_0^2-m^2}}.
\eea
\end{subequations}

Here, we distinguished between the parallel ($\parallel$) and perpendicular ($\perp$) Minkowski subspaces, delineated by their respective orientations relative to the background external magnetic field, by decomposing the metric tensor as
\begin{subequations}
    \bea
g^{\mu\nu}=\gp^{\mu\nu}+\gt^{\mu\nu},
\eea
where
\bea
\gp^{\mu\nu}&=&\text{diag}(1,0,0,-1)\nn\\
\gt^{\mu\nu}&=&\text{diag}(0,-1,-1,0).
\eea
\end{subequations}

The latter implies that for any four-vector
\begin{subequations}
    \bea
    p^\mu&=&\ppar^\mu+p_\perp^\mu,
\eea
we get
\bea
p^2=\ppar^2-\pt^2,
\eea
with
\bea
\ppar^2&=&p_0^2-p_3^2\nn\\
\pt^2&=&p_1^2+p_2^2.
\eea
\end{subequations}

\subsection{The one-loop polarization tensor in the limit $p^\mu\to0$}\label{sec:The one-loop polarization tensor in the limit}
From Eq.~\eqref{eq:PloTenDef}, the one-loop polarization tensor takes the form
\bea
\ii\Pi_\Delta^{\mu\nu}&=&\ii\Pi_0^{\mu\nu}+\ii\frac{q_f^2\qfB\Delta}{4\pi}\sum_{i=1}^{3}T_i^{\mu\nu},
\eea
where $\ii\Pi_0^{\mu\nu}$ is the one-loop polarization tensor in the strong field limit and in the absence of fluctuations~\cite{PhysRevD.83.111501}:
\bea
\ii\Pi_0^{\mu\nu}=\frac{\ii q_f^2\qfB}{4\pi^2}e^{-\pt^2/2\qfB}\ppar^2\mathcal{I}_0(\ppar^2)\left(\gp^{\mu\nu}-\frac{\ppar^\mu\ppar^\nu}{\ppar^2}\right),
\eea
with
\bea
\mathcal{I}_0(\ppar^2)=\int_0^1dx\frac{x(1-x)}{x(1-x)\ppar^2-m^2}.
\eea

Hence, in the limit $p^\mu\to0$, the tensor is given by
\bea
\lim_{p^\mu\to0}\ii\Pi_\Delta^{\mu\nu}=\ii\frac{q_f^2\qfB\Delta}{4\pi}\sum_{i=1}^{3}\lim_{p^\mu\to0}T_i^{\mu\nu},
\eea
where
\begin{subequations}
\bea
\lim_{p^\mu\to0}T_1^{\mu\nu}= -\frac{\ii \qfB}{32\pi m}\gp^{\mu\nu},
\eea
\bea
\lim_{p^\mu\to0}T_2^{\mu\nu}= -\frac{\ii \qfB}{2\pi m}\left(\gp^{\mu\nu}+2\delta_3^\mu\delta_3^\nu\right),
\eea 
and
\bea
\lim_{p^\mu\to0}T_3^{\mu\nu}=\frac{\ii\qfB}{3\pi m}\left[\gt^{\mu\nu}-\frac{1}{4}\gp^{\mu\nu}\right].
\eea
\label{T1T2T3_p0}
\end{subequations}

Further details of the calculation can be found in Ref.~\cite{PhysRevD.109.056007}.

\subsection{Mass of gauge fields by magnetic fluctuations }\label{sec:Mass of gauge fields by magnetic fluctuations}
To understand how magnetic field fluctuations might influence the potential generation of mass in the gauge fields, we identify the poles of the propagator. Utilizing the Dyson equation, the inverse of the propagator is given by
\begin{equation}
\left[D^{\mu\nu}(p)\right]^{-1} = \left[D^{\mu\nu}_0(p)\right]^{-1} - \ii\Pi^{\mu\nu}(p),
\end{equation}
where the ``free" photon propagator, in the Feynman gauge is
\begin{equation}
D^{\mu\nu}_0(p) = \frac{-\ii\,g^{\mu\nu}}{(p^2 + \ii\epsilon)}
\end{equation}

Adopting the method detailed in Ref.~\cite{lebellac}, the poles corresponding to the dynamic mass appear as the coefficients of $g^{\mu\nu}_{\parallel}$ and $g^{\mu\nu}_{\perp}$, respectively, when evaluating the limits $p_0\to0$ and $\mathbf{p}\to0$ in $\ii\Pi^{\mu\nu}(p)$. So then, from Eqs.~\eqref{T1T2T3_p0}:
\bea
\left[D^{\mu\nu}(p)\right]^{-1}=\ii g^{\mu\nu}_{\parallel}\left( p^{2} + \ii M_{\parallel}^2 + \ii\epsilon \right)+ \ii g^{\mu\nu}_{\perp}\left( p^{2} - \ii M_{\perp}^2 + \ii\epsilon \right) + 3 \ii M_{\perp}^2\,\ii \delta_3^\mu\delta_3^\nu+ \ldots,
\eea
where we defined the magnetic effective masses in both parallel and transverse projections by the coefficients
\bea
M_\parallel^2&\equiv&\frac{59\alpha_\text{em}\mathcal{B}^2\widetilde{\Delta}}{96\pi}m^2,\nn\\
M_\perp^2&\equiv&\frac{\alpha_\text{em}\mathcal{B}^2\widetilde{\Delta}}{3\pi}m^2.
\eea

We emphasize the physical interpretation of these effective masses by comparing them with the poles of the photon propagator along each polarization direction, given by
\be
p^2 + \ii M_{\perp,\parallel}^2 = p_0^2 - \mathbf{p}^2 + \ii M_{\perp,\parallel}^2
\ee
which indicates an effective photon dispersion relation for each polarization direction as
\be
\omega_{\perp,\parallel}(\mathbf{p}) = \sqrt{\mathbf{p}^2 - \ii M_{\perp,\parallel}^2 },
\ee
where the corresponding effective mass is complex
\be
m_{\perp,\parallel} \equiv \left(-\ii M_{\perp,\parallel}^2 \right)^{1/2} = \frac{1 - \ii}{\sqrt{2}}M_{\perp,\parallel}.
\ee

The real part signifies the typical damping effect, while the imaginary part introduces an oscillatory component. This characterization aligns with our understanding of random magnetic fluctuations creating an effective dispersive medium for both fermions and photons (or gluons) equally.

\subsection{Summary}\label{sec:Summary and conclusions}
In this section, we focused on the one-loop contribution to the gluon and photon polarization tensor when these gauge fields are in a medium subjected to a strong, external, and classical magnetic field with white-noise stochastic fluctuations. These fluctuations are incorporated into the fermion propagators by correcting the {\it free} fermion propagators using the so-called {\it replica trick}, which averages the fluctuations over the QED Lagrangian. This approach implies the existence of an effective model with fermion-fermion interactions, which is perturbatively expanded to find the propagator at order $\Delta$. By explicitly computing the tensor, we showed that it doesn't satisfy the Ward-Takahashi identity, indicating a break in the $U(1)$-QED symmetry. This finding is supported by previous works, using both perturbative and effective Lagrangian methods~\cite{PhysRevD.107.096014,PhysRevD.108.116013}.

The lack of $U(1)$ symmetry suggests a possible generation of magnetic mass for the considered gauge fields. In fact, by following the standard process to find the poles of the photon's propagator, we analytically computed the factor associated with these magnetic masses. The results show that the generated masses possess imaginary parts, indicating that in a fluctuating magnetic medium, a refractive index may appear. Importantly, the gauge field mass differs in each direction with respect to the magnetic field, also indicating birefringent effects due to the violation of Lorentz symmetry.

\subsection{Conclusions and outlook}
\label{Concl}

In this Part we have discussed recent theoretical developments whose aim is to elucidate properties of strongly interacting matter under the influence of magnetic fields. We have tackled the problem from the perspective of hadron as well as of fundamental QCD degrees of freedom. From the hadron perspective, special attention has been devoted to the magnetic driven change of meson masses, interactions as well as the relation of the topological cumulants with the chiral condensate in a magnetic background for infinite as well as finite volume. It is found that light meson masses are influenced by the mixing between
pseudoscalar, vector and axial vector mesons induced by the magnetic field. Arguments to explore the quark-gluon structure using relatively low magnetic fields have been presented by discussing the calculation of meson form factors in meson exchange processes. It was also shown that finite volume corrections to thermodynamic observables and condensates in a magnetic background are substantial since magnetic effects enter into the chiral perturbation theory description at the same order than finite size effects.
 
From the point of view of fundamental QCD degrees of freedom, the properties of hot and dense strongly interacting matter in a very strong magnetic field have also been explored using pQCD. It is found that medium loop corrections are
negligible, as compared to the free term, for very large magnetic fields and physical choices of the renormalization running
scale. Also, possible effects of the presence of a strong magnetic field at pre-equilibrium during a heavy-ion reaction have been studied in terms of the photon production channel. It is shown that both the yield and $v_2$ get enhanced allowing for an extra contribution that could explain the slight differences found when applying state-of-the-art descriptions to photon production in a relativistic heavy-ion collision environment. Finally, the consequences of a non-uniform, but instead fluctuating field, such as the one that would be more likely to be produced in physical situations, has been also explored. The presence of a noisy background modifies the particle propagators and shows up as an energy-dependent effective index of refraction causing screening for charged fermions as well as the generation of a photon/gluon mass.

The review discussed in this Part is clearly non-exhaustive, but it provides a picture of some of the state-of-the-art efforts to explore the strong interaction using as a probe a magnetic field. Important aspects of these studies such as the properties of screening masses and their relation to IMC~\cite{Ayala:2023llp,Fayazbakhsh:2012vr, Fayazbakhsh:2013cha,Ding:2020hxw, Ding:2022tqn, Sheng:2020hge}, the enhancement of particle emission induced by a magnetic background~\cite{Jaber-Urquiza:2023sal}, effects on the Mott transition~\cite{Avancini:2018svs,Mao:2019avr}, among others, have not been explored but certainly constitute avenues that promise to produce a clearer picture of the properties of strongly interacting matter probed by magnetic fields.

	\newpage
	\graphicspath{{./Figures_CME_Lat/}}

\newcommand{\chapabstract}[1]{
    \begin{quote}
        \small
        \rule{16.75cm}{1pt}\\
        #1
        \vskip-4mm
        \rule{16.75cm}{1pt}
\end{quote}}

	\part{Anomalous transport and non-perturbative phenomena on and off the lattice}\label{III}
		

	
	
	
	\section{Introduction}
	Background electromagnetic fields source a variety of fascinating phenomena in non-Abelian quantum field theories like Quantum Chromodynamics (QCD) or the electroweak theory. For the strong interactions, all characteristic non-perturbative features of the QCD medium, like chiral symmetry breaking, confinement or topology, are affected by background fields in a substantial and non-trivial manner. Topological effects in particular, when combined with background electromagnetic fields or vorticity, can lead to novel transport phenomena. Due to the prominent role played by the axial anomaly for them, these are referred to as anomalous transport phenomena and are also found to arise in various condensed matter physics systems. Similarly intricate non-perturbative effects are generated by background fields in the electroweak theory.
	
	In this Part, we discuss recent developments pertaining to the impact of background magnetic fields on non-perturbative phenomena in the realm of QCD and the electroweak theory. A special emphasis is put on the study of anomalous transport phenomena in QCD as well as on low-dimensional QED, relevant for certain condensed matter systems.
	Most of these effects are deeply non-perturbative and require either first-principles lattice field theory simulations or effective approaches like hydrodynamics.
	
	We begin the discussion with lattice QCD simulations of anomalous transport phenomena. In Sec.~\ref{sec:Marko}, the leading-order conductivities are determined on the lattice for the chiral magnetic effect (CME) and the chiral separation effect (CSE) in equilibrium. The CSE coefficient is observed to be strongly suppressed in the chiral symmetry broken region and to approach its perturbative massless value as the temperature grows. In turn, the in-equilibrium CME coefficient is found to vanish for all temperatures. The chiral separation effect is investigated for nonzero chemical potentials in two-color QCD in Sec.~\ref{sec:Buividovich} in the presence of heavy flavors. The latter are found to enhance the CSE conductivity and, simultaneously, also lead to a reduction of the electric conductivity. This type of behavior is discussed in relation to the Kondo effect.

	Next, novel anomalous transport phenomena are studied in a hydrodynamic approach in Sec.~\ref{sec:Kaminski}, highlighting their potential phenomenological applications for various physical systems. 
	Hydrodynamics is based on symmetries, while the quark-gluon plasma generated in heavy-ion collisions, neutron stars and astrophysical plasma is subject to symmetry-breaking, e.g., due to the chiral anomaly of QCD and strong external magnetic fields, which break parity and isotropy, respectively. Thus, in Sec.~\ref{sec:Kaminski}, the construction of the proper hydrodynamic description of such systems is reviewed taking these broken symmetries into account, which gives rise to novel hydrodynamic transport effects. 
	One well-known hydrodynamic transport effect arising from the breaking of parity in a strong magnetic field is the CME. Sec.~\ref{sec:Kaminski} discusses this effect far away from equilibrium in a holographic model, focusing on the dependence of the time-evolution of the CME currents on initial conditions.
	
	Moving from the QCD setting to non-relativistic condensed matter systems, analogous anomalous transport phenomena can be found. The discovery of Dirac and Weyl materials, in which the charge carriers obey a relativistic-like equation of motion, allowed for close analogies between high energy and condensed matter phenomena. While materials of this kind in 3-space dimensions have been successfully explored, presenting an Abelian version of the CME, in Sec.~\ref{sec:Mizher} the yet poorly explored (2+1)-dimensional case is discussed. The motivation for considering these materials	is presented and new analogies that include these systems are proposed. As a specific example, a new Hall effect, whose construction was inspired by the symmetry patterns of the CME, is discussed in detail for planar Weyl materials.

	Besides anomalous transport effects, background magnetic fields affect further non-perturbative features of quantum field theories. In Sec.~\ref{sec:Kotov}, the QCD phase diagram is investigated at nonzero baryon density in the presence of magnetic fields. By determining the fluctuations of conserved charged in this magnetized and dense environment, the equation of state of the system is constructed. Magnetic fields are expected to affect the phase structure not just in QCD, but also in the electroweak theory. Sec.~\ref{sec:Chernodub} reports about the latest developments in the study of a superconducting phase emerging for magnetic fields comparable to the electroweak scale. This novel phase is found to involve inhomogeneous condensates of both $W$ and $Z$ bosons with a vortex structure that has features of both fluids and solids. The transition between the normal and superconducting phases is found to be an analytic crossover, which is followed, at a higher field, by another crossover to a symmetric phase with restored electroweak symmetry.

	\section{Anomalous transport coefficients from lattice QCD}
	\label{sec:Marko}
	%

\newcommand{\CME}{C_{\text{CME}}}
\newcommand{\CSE}{C_{\text{CSE}}}
\newcommand{\B}{\mathcal{B}}
\newcommand{\Q}{\mathcal{Q}}
\renewcommand{\Tr}{\textmd{Tr}}

	\subsection{Introduction}
	Anomalous transport phenomena provide a very interesting testing ground for the properties of the QCD vacuum. The interplay between electromagnetic fields and vorticities with quantum anomalies yields a wide variety of these transport effects, whose properties have been subject of intense study in the last decades.
	
	The most celebrated of these phenomena is the chiral magnetic effect (CME) \cite{Fukushima:2008xe}, the generation of a vector current in the presence of a magnetic field and chiral imbalance. Its possible existence is not only interesting from a purely theoretical point of view, but it is also currently subject to an intense experimental effort. It was detected in condensed matter systems \cite{Li:2014bha} and it is actively searched for in heavy-ion collisions \cite{STAR:2013ksd,STAR:2014uiw,STAR:2021mii}.
	
	Another example of these effects is the chiral separation effect (CSE) \cite{Son:2004tq,Metlitski:2005pr}, which corresponds to the existence of an axial current in dense and magnetized matter. Although often regarded as a ``dual'' of CME, the equilibrium properties of these two phenomena greatly differ, as it will be discussed in this work. Together, the CME and the CSE are expected to form the chiral magnetic wave (CMW) \cite{Kharzeev:2010gd}, a collective excitation that would arise in these extreme environments.  
	
	The main objective of this work is to review the analytic calculation of the CME/CSE conductivities (precisely defined below) for non-interacting fermions, highlighting the importance of proper UV regularization, as well as in comparison to lattice calculation results. We also present the results for the conductivities in the presence of interactions using full lattice QCD simulations. A detailed discussion of all these results can be found in \cite{Brandt:2024wlw,Brandt:2023wgf}. For a recent review on lattice investigations of anomalous transport phenomena and, in general, on effects of background electromagnetic fields in QCD, see~\cite{Endrodi:2024cqn}.
	
	\subsection{Conductivities}
	Let us consider a system in the presence of a constant background magnetic field $B$, which without the loss of generality we choose to point in the third spatial direction. We can define the charged vector and axial currents
\begin{equation}
	J_{\nu}^\Q=\sum_f \dfrac{q_f}{e} \,\frac{T}{V}\int \dd^4x \, \bar\psi_f(x) \gamma_\nu \psi_f(x)\,, \quad 	J_{\nu5}^\Q=\sum_f \dfrac{q_f}{e} \,\frac{T}{V}\int \dd^4x \, \bar\psi_f(x) \gamma_\nu \gamma_5  \psi_f(x)\,,
	\label{eq:qveccurdef}
\end{equation}
where $f=u,d,s,\ldots$ labels the quark flavors and $q_f$ are the corresponding electric charges.
Analogously, we can consider the baryon vector and  axial currents
\begin{equation}
	J_{\nu}^\B=\frac{1}{3}\sum_f \,\frac{T}{V}\int \dd^4x \, \bar\psi_f(x) \gamma_\nu \psi_f(x)\,, \quad 	J_{\nu5}^\B=\dfrac{1}{3}\sum_f \,\frac{T}{V}\int \dd^4x \, \bar\psi_f(x) \gamma_\nu \gamma_5  \psi_f(x)\,.
	\label{eq:bveccurdef}
\end{equation}
To parameterize the zeroth component of these currents (the density and the chiral density), we use chemical potentials, which can either couple to the electric charge $\mu^\Q$, $\mu_5^\Q$ or to the baryon number $\mu^\B$, $\mu_5^\B$. In the following we consider the physically motivated case of a charged vector current and a baryon chiral chemical potential for the CME, as well as a charged axial current and a baryon chemical potential for the CSE.\footnote{For non-interacting fermions, there is no ambiguity in the choice of any possible combination. However, in full QCD different combinations change the result in the case of CSE \cite{Brandt:2023wgf}, but not for CME \cite{Brandt:2024wlw}. In the different setups there is an overall factor $C_{\rm dof}$ which we use to rescale all our results (see \cite{Brandt:2023wgf} for a more detailed discussion).} For clarity, we drop these indices in the rest of the manuscript.

For studying both effects, we will consider the first coefficient in the Taylor expansion of the conductivities, i.e. the linear response to weak $B$ and $\mu_5/\mu$
\begin{align}
	\label{eq:ccme}
	&\langle J_{3} \rangle=\CME\, \mu_5 \, e B + \mathcal{O}(\mu_5^3, B^3)\,,\\
	 &\langle J_{35} \rangle=\CSE\, \mu \, e B + \mathcal{O}(\mu^3, B^3)\,.
	 \label{eq:ccse}
\end{align}
We will refer to $\CME/\CSE$ interchangeably as conductivity coefficients or simply conductivities. We can study these coefficients via derivatives of the currents with respect to the appropriate chemical potential 
\begin{align}
	\label{eq:j3der}
	&\eval{\pdv{\langle J_{3} \rangle}{\mu_5}}_{\mu_5=0}= C_{\text{CME}} \, e B\,,\\[1em]
	&\eval{\pdv{\langle J_{35} \rangle}{\mu}}_{\mu=0}= C_{\text{CSE}} \, e B\,.
	\label{eq:j35der}
\end{align}
We note that the expectation values on the left-hand side are to be evaluated at zero $\mu/\mu_5$ but finite magnetic field $B$. This enables a lattice QCD study of $\CSE$, since simulations at finite chemical potential suffer from a sign problem. Using the analogous approach when studying CME is motivated by the possibility of using the same ensembles and that although simulations at $\mu_5\neq 0$ are possible, an implementation with staggered fermions is technically challenging \cite{Brandt:2024wlw}.

To understand the precise nature of these effects in equilibrium, we present in the next section the study of $\CME$ and $\CSE$ for a system of free fermions.   
	
\subsection{Non-interacting fermions}
The seminal paper \cite{Fukushima:2008xe} established CME as a non-dissipative, stationary effect, whose conductivity is fixed by the axial anomaly and hence topologically protected. It was further argued that it would survive in an equilibrium setup, with the same conductivity as the real-time effect. However in thermal equilibrium, the focus of our current investigation, Bloch's theorem forbids the global flow of conserved currents and so the $\CME$ must vanish \cite{Yamamoto:2015fxa}\footnote{Notice that Bloch's theorem allows local currents to flow, which have been discussed very recently in~\cite{Brandt:2024fpc}.}. The role of UV regularization in the vanishing result has been widely raised in the CME literature, see \cite{Hou:2011ze,Buividovich:2013hza,Zubkov:2016tcp,Horvath:2019dvl,Banerjee:2021vvn} and references therein. We will now also illustrate it by the calculation of $\CME$ for a system of free fermions. In addition, we present an analogous calculation of $\CSE$ where we reproduce the known results for this conductivity \cite{Son:2004tq,Metlitski:2005pr}.
\subsubsection{Analytic results}
We consider one flavor of a colorless, non-interacting fermion of charge $q$ and mass $m$, in the presence of a constant and homogeneous magnetic field $B$. We start our discussion with the calculation of $\CME$ at $T=0$, which accounts to the calculation of the two-point function
\begin{align}
	\left.\frac{\partial\langle J_{3}\rangle}{\partial\mu_5}\right|_{\mu_5=0}
	=  
	\frac{T}{V}\int \dd^4x \int \dd^4y \,\langle \bar\psi(x)\gamma_3\psi(x)\bar\psi(y)\gamma_0\gamma_5\psi(y)\rangle\,.
	\label{eq:cme1}
\end{align}
We regularize the UV behavior using the Pauli-Villars method (PV) in a textbook manner \cite{Itzykson:1980rh}. Unlike a momentum cutoff, this regulator respects gauge invariance in QED, which turns out to be a very important ingredient in the study of the CME. We introduce three new regulator fields with coefficients $c_s$ and masses $m_s$, reserving $s=0$ for the physical field (with physical mass $m$). The parameters are then 
\begin{align}
	c_0&=c_1=1\,,\quad c_2=c_3=-1\,,\\
	m_0^2&=m^2\,,\quad m_1^2=m^2+2\Lambda^2\,,\quad m_2^2=m_3^2=m^2+\Lambda^2\,,
\end{align}
with $\Lambda\to\infty$ to be taken at the end of the calculation. 

Using Wick's theorem we can rewrite the right hand side of Eq.~\eqref{eq:cme1} to obtain
\begin{align}
	\label{eq:cme_PV_start}
	\CME \, qB = \frac{iT}{V}\sum_{s=0}^{3} c_s\int \dd^4 x\int \dd^4y \Tr\left[\gamma_3 S_s(x,y) \gamma_0 S_s(y,x)\right]\,,
\end{align}
where the PV fields are already taken into account under the sum over $s$, and $S_s$ is the fermion propagator for the field $s$ in a homogeneous magnetic background. The latter reads~\cite{Shovkovy:2012zn}
\begin{align}
	S_s(x,y) = \Phi(x,y)\int\frac{\dd^4p}{(2\pi)^4}{\rm \,e\,}^{-ip(x-y)}\widetilde S_s(p)\,.
\end{align}
Here 
\begin{align}
	\Phi(x,y) = \exp\left[iqB(x_1+y_1)(x_2-y_2)/2\right]\,,
\end{align}
is the Schwinger phase, and
\begin{align}
	\widetilde S_s(p) = \int_0^\infty \dd z {\rm \,e\,}^{iz(p_0^2 - m_s^2-p_3^2) - i\frac{p_1^2+p_2^2}{|qB|}\tan{(z|qB|)}} \left[\slashed{p}+m_s+(p_1\gamma_2-p_2\gamma_1)\tan(zqB)\right]
	\left[1-\gamma_1\gamma_2\tan(zqB)\right]\,.
\end{align}	

Carrying out the traces we obtain
\begin{align}
	\CME \,qB=-4\sum_{s=0}^{3} c_s &\int\frac{\dd^4p}{(2\pi)^4}\int_0^\infty \dd z_1\, \dd z_2 {\rm \,e\,}^{i(z_1+z_2)(p_0^2 - m_s^2-p_3^2)}\left(m_s^2-p_0^2-p_3^2\right) \nonumber\\
	&{\rm \,e\,}^{- i\frac{p_1^2+p_2^2}{qB}[\tan{(z_1qB)}+\tan{(z_2qB)}]}\left[\tan(z_1qB)+\tan(z_2qB)\right]\,.
\end{align}
Now we Wick rotate the momentum components $p_0 = ip_4$ and the Schwinger parameters $iz_1 = Z_1\,,iz_2 = Z_2$. The momenta integrals are simple Gaussians, and evaluating them factorizes the magnetic field dependence, uncovering an explicitly linear behavior in $B$. The formula is simplified to
\begin{align}
	\CME =\sum_{s=0}^3c_s \dfrac{m_s^2}{4\pi^2} \int^\infty_0 \dd Z_1 \int^\infty_0 \dd Z_2\, \dfrac{e^{-m_s^2 (Z_1+Z_2)}}{Z_1+Z_2}\,.
\end{align}
Performing the integrals, we arrive at
\begin{align}
	\label{eq:cmecancel}
	\CME =\dfrac{1}{4\pi^2} \sum_{s=0}^3c_s =0\,.
\end{align}
In this expression, the relevance of the regulator is clear. The PV fields cancel the contribution from the physical field, yielding a vanishing $\CME$ in equilibrium.

It is straightforward to generalize this calculation to finite temperatures $T$, replacing the frequency component of the integral by a sum over fermionic Matsubara frequencies
\begin{align}
	\int \frac{\dd p_0}{2\pi} \to iT\sum_{n=-\infty}^\infty\,,
\end{align}
while also replacing $p_0\to i \omega_n= i 2\pi T(n+1/2)$ in $\widetilde S_s(p)$. This does not change the result for $\CME$, showing that the current vanishes for temperature independently.

An analogous calculation for $\CSE$ leads, however, to a non-vanishing result. We start from
\begin{align}
	\left.\frac{\partial\langle J_{35}\rangle}{\partial\mu}\right|_{\mu=0}
	=  
	\frac{T}{V}\int \dd^4x \int \dd^4y \,\langle \bar\psi(x)\gamma_0\psi(x)\bar\psi(y)\gamma_3\gamma_5\psi(y)\rangle\,.
	\label{eq:cse1}
\end{align}
Now
\begin{align}
	C_{\rm CSE} \,qB=4i\sum_{s=0}^{3} c_s T\sum_{n}&\int\frac{\dd^3p}{(2\pi)^3}\int_0^\infty \dd z_1\, \dd z_2 {\rm \,e\,}^{i(z_1+z_2)(-\omega_n^2 - m_s^2-p_3^2)}\left(m_s^2-\omega_n^2+p_3^2\right) \nonumber\\
	&{\rm \,e\,}^{- i\frac{p_1^2+p_2^2}{qB}[\tan{(z_1qB)}+\tan{(z_2qB)}]}\left[\tan(z_1qB)+\tan(z_2qB)\right]\,.
\end{align}
The $p_{1,2}$ integrals can be carried out and the expression simplifies to
\begin{align}
	C_{\rm CSE} =-\frac{1}{2\pi^2}\sum_{s=0}^{3} c_s T\sum_{n}&\int_{-\infty}^\infty \dd p_3\int_0^\infty \dd z_1\, \dd z_2 {\rm \,e\,}^{i(z_1+z_2)(-\omega_n^2 - m_s^2-p_3^2)}\left(m_s^2-\omega_n^2+p_3^2\right)\,,
\end{align}
allowing the evaluation of the $z_1$ and $z_2$ integrals as well. We find
\begin{align}
	C_{\rm CSE} =-\frac{1}{2\pi^2}\sum_{s=0}^{3} c_s T \sum_{n=-\infty}^\infty \int_{-\infty}^\infty \dd p_3\, \frac{m_s^2-\omega_n^2+p_3^2}{\left(\omega_n^2+p_3^2+m_s^2\right)^2}\,.
\end{align}

\begin{figure}
	\centering
	\includegraphics{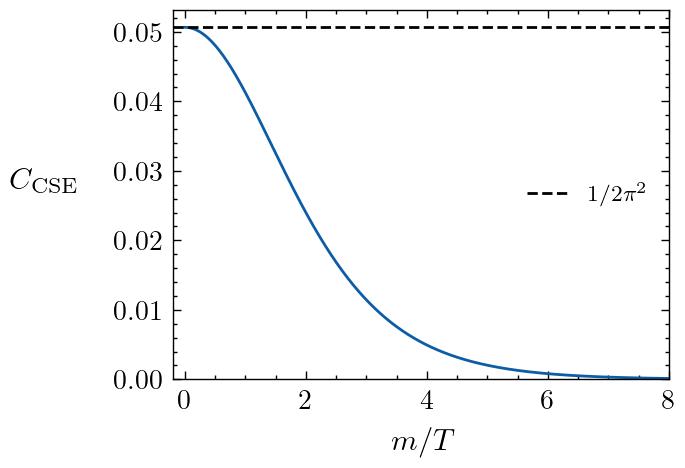}
	\caption{Leading coefficient in the CSE conductivity for non-interacting fermions as a function of $m/T$.}
	\label{fig:cseanaly}
\end{figure}

The Matsubara sum then evaluates to
\begin{align}
	C_{\rm CSE} =-\frac{1}{2\pi^2}\sum_{s=0}^{3} c_s \int_{-\infty}^\infty \dd p_3\, \frac{\dd n_F(E_p^{(s)})}{\dd E_p^{(s)}}\,,
\end{align}
where $n_F(x)=({\rm e}^{x/T}+1)^{-1}$ is the Fermi-Dirac distribution and $E_p^{(s)}=\sqrt{p_3^2+m_s^2}$. The derivative of the Fermi-Dirac distribution vanishes for infinite masses, therefore only the $s=0$, physical term remains from the sum over PV fields,
\begin{align}
	C_{\rm CSE} =-\frac{1}{\pi^2} \int_{0}^\infty \dd p_3\, \frac{\dd n_F(E_p)}{\dd E_p}=
	\frac{1}{2\pi^2} \int_{0}^{\infty} \dd p_3 \, \left[ 1+\cosh( \sqrt{p_3^2+(m/T)^2} ) \right]^{-1}\,.
\end{align}
We plot the result of the integral in Fig.~\ref{fig:cseanaly}, which shows that $\CSE$ is a non-trivial function of $m/T$. Note that Bloch's theorem does not forbid the existence of CSE, since the axial current is not conserved due to the finite mass and the chiral anomaly.

\subsubsection{Lattice results}
With these analytical results at hand, we now turn to the lattice formulation. First of all, we check the results in the non-interacting case to benchmark our lattice setup. For a detailed discussion on the lattice observables and the definition of the currents, see~\cite{Brandt:2024wlw,Brandt:2023wgf}. For free fermions we use two distinct discretizations: Wilson and staggered quarks. In both formulations it is possible to define vector and axial currents which fulfill the Ward identities (see~\cite{Sharatchandra:1981si} for staggered and~\cite{Karsten:1980wd} for Wilson), which are point-split, meaning that they connect fermions at nearest neighbor sites.

For comparison, in the Wilson formulation it is also possible to define a local vector current, involving fermion fields at the same point, which however is not conserved on the lattice.

Nevertheless, it is commonly used in the study of hadron properties, since it has the correct quantum numbers. It was also used to study CME with Wilson fermions in QCD and in the quenched approximation \cite{Yamamoto:2011gk,Yamamoto:2011ks}, obtaining $\CME\neq0$. It is therefore important to study the effect of using the non-conserved current, in order to clarify its role in the non-vanishing CME result.

\begin{figure}[ht]
	\centering
	\subfloat{{\includegraphics[width=8cm]{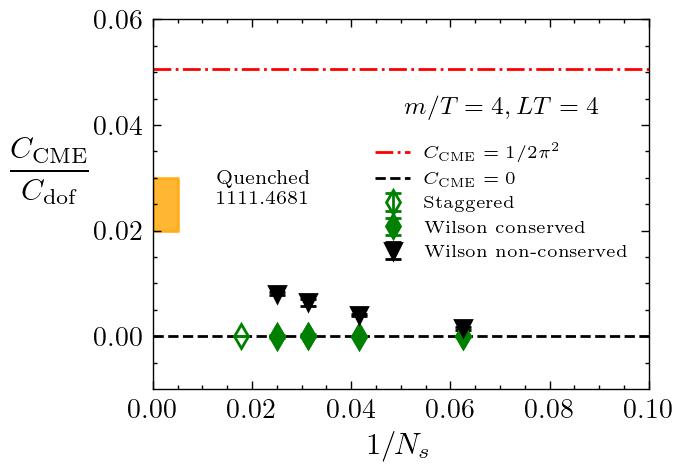} }}%
	\qquad
	\subfloat{{\includegraphics[width=8cm]{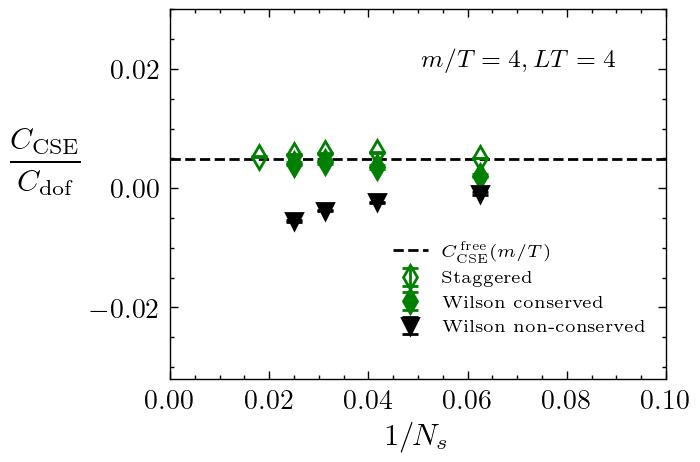} }}%
	\caption{Lattice spacing dependence of the CME (left panel) and the CSE (right panel) coefficients for non-interacting fermions. Besides the correct, conserved vector currents, we also use a local, non-conserved current for Wilson quarks -- this gives an unphysical result for both effects. A comparison to the quenched study~\cite{Yamamoto:2011ks}, which used a non-conserved vector current for the CME, is also included.}%
	\label{fig:free}%
\end{figure}

On the lattice, the coefficients are obtained by calculating the expectation values \eqref{eq:j3der} and \eqref{eq:j35der} at different values of $eB$, and then performing a linear fit. The slope of the fit (rescaled by the appropriate factor accounting for the charges of different flavors) corresponds to the desired conductivities. In Fig.~\ref{fig:free}, we show the results for $\CME$ and $\CSE$ for free fermions with a mass of $m/T=4$. The calculation is performed on a four-dimensional lattice with $N_s$ spatial points and $N_t$ temporal points, with $a$ the lattice spacing. The spatial volume is $V=L^3=(aN_s)^3$ and the temperature is $T=(aN_t)^{-1}$. 

On the left panel of Fig.~\ref{fig:free}, we can see the results for the CME. Both Wilson and staggered fermions, when using a conserved vector current, yield a vanishing $\CME$. This is in agreement with the arguments presented earlier, since the Euclidean lattice formulation is inherently in equilibrium and uses a gauge invariant regularization. However, when a local vector current is used with Wilson fermions, the continuum limit deviates from zero, which is an indication that the $\CME\neq0$ in \cite{Yamamoto:2011ks} is very likely linked to the use of a non-conserved vector current. In particular our free results suggestively approach the continuum limit estimation of \cite{Yamamoto:2011ks} in the quenched theory.

It is important to emphasize at this point that the absence of the CME is not due to the lack of chirality in the system. It can be shown, both analytically and on the lattice, that the axial susceptibility, which corresponds to the linear response of the chiral density to a weak $\mu_5$, is not vanishing \cite{Brandt:2024wlw}.

The right panel of Fig.~\ref{fig:free} shows the result for the CSE. Similarly to the CME, when the appropriate currents are used, the results in the continuum limit using either discretization agree with eq.~\eqref{fig:cseanaly} for $m/T=4$, while the non-conserved Wilson current shows a deviation. The same behavior was found for other values of $m/T$.

From these results we learn the importance of using the proper discretization of the current on the lattice, as well as a positive cross-check of our setup. This enables us to proceed to the full QCD -- this is what we discuss next.

\subsection{Results}
\subsubsection{Quenched theory}
We now present the results on how interactions affect the CME and the CSE. We start with an intermediate step: the quenched theory. This setup is not only less computationally challenging, but also enables a direct comparison with existing lattice studies of these effects in the literature.

We perform simulations with Wilson and staggered fermions as valence quarks, while the configurations are generated using the plaquette action (these configurations were already used in~\cite{Bali:2017ian,Bali:2018sey}). The pion mass is tuned to be $M_\pi\approx415$ MeV in the staggered case, while for Wilson fermions we use $M_\pi\approx710 $ MeV. The latter is to best compare to the CME study \cite{Yamamoto:2011ks}. Besides zero temperature simulations we do a temperature scan in the range from 260 MeV to 520 MeV.

In our setup, $\CME$ was found to be zero in the whole range of temperatures, see \cite{Brandt:2024wlw} for details. This is a first confirmation that interactions do not affect the arguments presented before, and will be further reinforced by the results in dynamical QCD.

\begin{figure}[ht]
	\centering
	\includegraphics[width=9cm]{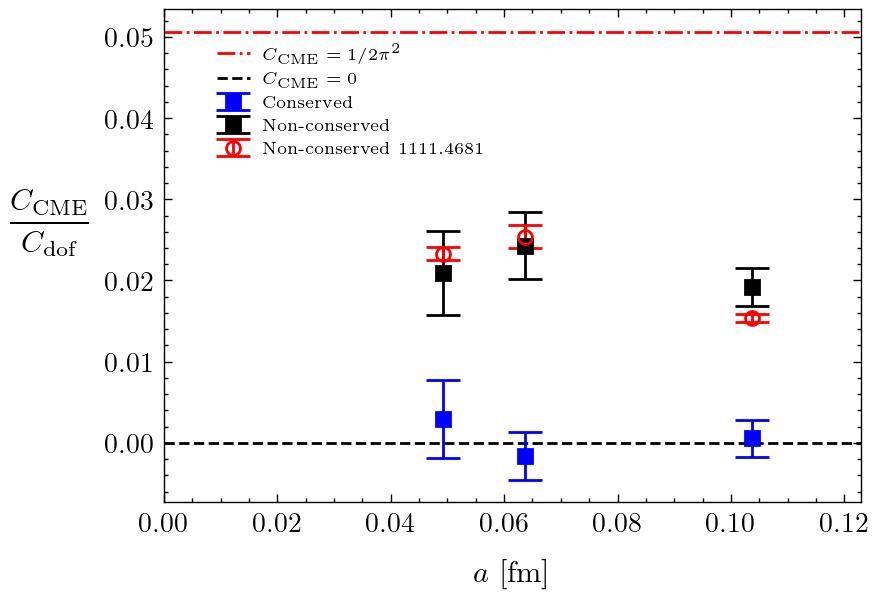}
	\caption{Lattice spacing dependence of the CME coefficient in quenched QCD, using conserved (blue) and non-conserved (black) vector currents, including a comparison to~\cite{Yamamoto:2011ks}.}
	\label{fig:compare}
\end{figure}

However, there is another important aspect that can be studied. As already mentioned, we can directly compare our setup with the one used in \cite{Yamamoto:2011ks}, where the vector current was directly measured at finite $B$ and $\mu_5$. We show the comparison in Fig.~\ref{fig:compare}. As it can be seen, when a non-conserved vector current is used in our setup, we reproduce the results from \cite{Yamamoto:2011ks}, deviating from zero. However, when a conserved current is used, $\CME$ vanishes. We conclude that the $\CME\neq0$ result is just an artifact of using a local vector current, and it shows the importance of using the appropriate discretization of currents to study anomalous transport effects on the lattice.

For the quenched study of the CSE, there is a subtlety related to the effect of non-trivial Polyakov loop sectors, see~\cite{Brandt:2023wgf} for details. Once this is taken into account, the conductivity was found to be suppressed at low temperatures and approach the value corresponding to the free massless fermion value for high $T$. The latter behavior is to be expected due to asymptotic freedom, and is in qualitative agreement with a quenched study with overlap fermions with massless valence quarks~\cite{Puhr:2016kzp}.

\subsubsection{Full QCD}
In this last subsection we summarize our results for full dynamical QCD using $2+1$ flavors of stout-improved staggered fermions with physical quark masses. The conductivities are calculated for a wide range of temperatures, from $T\approx0$ MeV to 400 MeV.  

\begin{figure}[ht]
	\centering
	\includegraphics[width=9cm]{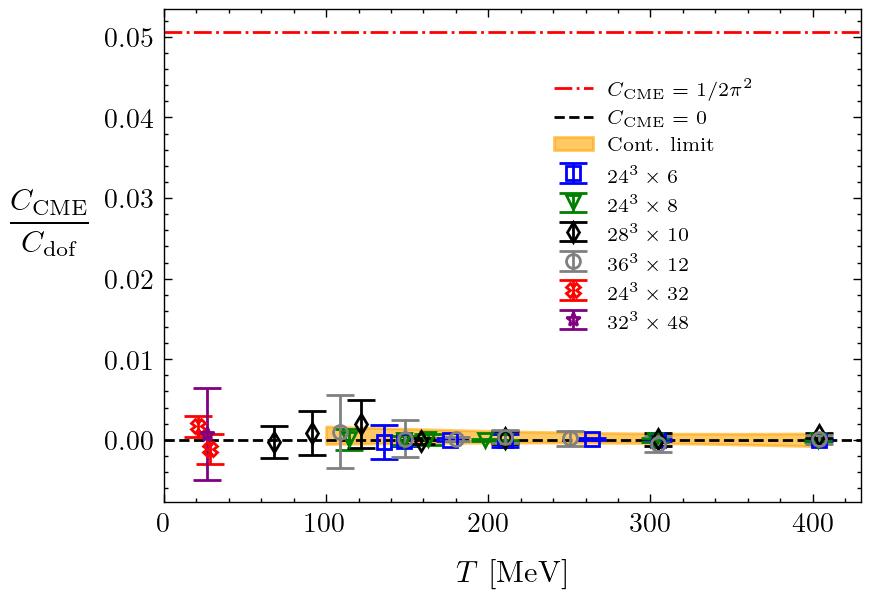}
	\includegraphics[width=9cm]{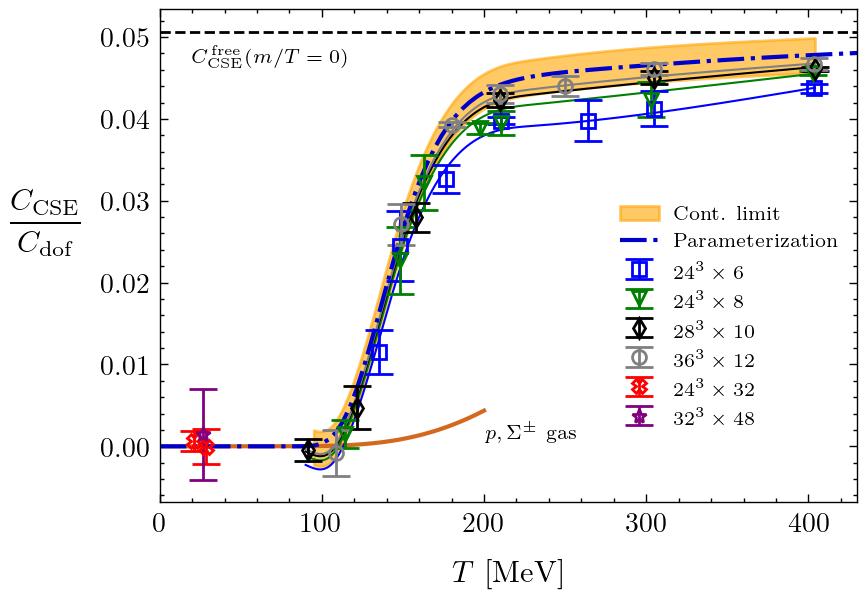}
	\caption{Left panel: continuum extrapolation of the CME coefficient in full QCD~\cite{Brandt:2024wlw}. Right panel: continuum extrapolation of the CSE coefficient in full QCD, with a comparison to a model involving a free gas of $p$ and $\Sigma^\pm$ hadrons at low temperature~\cite{Brandt:2023wgf}.}
	\label{fig:fullcme}
\end{figure}

In the left panel of Fig.~\ref{fig:fullcme} we present the results for the CME. The coefficient vanishes within errors for all temperatures, which confirms the arguments presented throughout this work about the absence of the CME in thermal equilibrium, generalizing it to QCD. Analogously as in the case of free fermions, it can be shown that the nonzero value of $\CME$ is not due to the absence of chirality -- the axial susceptibility in full QCD is also nonzero. In particular, this observable shows a gradual increase around the crossover temperature $T_c$ towards the free massless value for high $T$~\cite{Brandt:2024wlw}.

Finally, we present the results for $\CSE$ in the right panel of Fig.~\ref{fig:fullcme}. Unlike $\CME$, it shows a highly non-trivial structure in equilibrium. $\CSE$ is highly suppressed at low $T$, with a sharp increase around the chiral transition temperature, approaching the analytical value for free fermions $\CSE(m/T=0)=1/2\pi^2$ at high temperatures \cite{Brandt:2023wgf}. The suppression at low $T$ is in agreement with a two-color QCD study \cite{Buividovich:2020gnl}, and it can be further cross-checked with a hadron resonance gas type calculation. The results are also in qualitative agreement with the recent results obtained using holographic QCD~\cite{Gallegos:2024qxo}.
   
\subsection{Summary and outlook}
In this section, we have studied the equilibrium aspects of two anomalous transport phenomena, the chiral magnetic effect and the chiral separation effect. We have shown an analytic calculation of the conductivity for these effects, pointing out the importance of UV regularization. While the CSE shows a non-trivial behavior as a function of $m/T$ even in thermal equilibrium, the CME may only be present out-of-equilibrium. We cross-checked these results with a lattice evaluation with Wilson and staggered fermions, which showed complete agreement with the analytical results if the appropriate discretization of currents was used. The latter turns out to be the most crucial aspect of the simulation.

We then discussed interacting lattice QCD results, in the quenched approximation and in the full dynamical theory. In the quenched theory for the CME, we were able to directly compare to the simulations of~\cite{Yamamoto:2011ks}, where a non-zero result for $\CME$ was obtained, and show that the reason behind this is the use of a non-conserved vector current on the lattice. If a conserved one is used, the CME vanishes within errors for all temperatures. In turn, the CSE conductivity was found to be suppressed at low $T$ and approach the free massless fermion limit of $\CSE$ at high $T$. 

These results from the quenched theory are confirmed in full dynamical QCD with staggered fermions tuned to the physical point, where the CME was found to vanish for a wide range of temperatures. Regarding the CSE, its conductivity vanishes for temperatures lower than the crossover temperature, with a sharp increase over it, finally approaching the non-interacting massless fermion limit. Although directly detecting $\CSE$ in experiments is very challenging since it amounts to an axial current, the structure of the temperature dependence of the conductivity could play a role in the detection of the chiral magnetic wave.

An important open question is the out-of-equilibrium behavior of $\CME$ in QCD~\cite{Kharzeev:2009pj,Banerjee:2022snd}. Lattice QCD simulations do not have direct access to real-time properties, since only Euclidean correlators can be calculated. However, it is possible to learn about these properties by inferring the spectral function from the Euclidean correlator~\cite{Buividovich:2024bmu} through spectral reconstruction. This is a mathematically ill-posed problem and therefore a complicated and delicate procedure, where small uncertainties in the original correlator may impact the final result substantially. However, there is a wide variety of tools that alleviate the difficulties of this problem. As an outlook, we plan to use spectral reconstruction methods to calculate the out-of-equilibrium $\CME$ in QCD, which promises a deeper understanding of the effect of interactions for the CME.   
	

	\section{Enhancement of Chiral Separation Effect due to heavy quark flavours: a signature of QCD Kondo effect?}
	\label{sec:Buividovich}
	
	\newcommand{\lr}[1]{ \left( #1 \right) }
\newcommand{\lrs}[1]{ \left[ #1 \right] }
\newcommand{\lrc}[1]{ \left\{ #1 \right\} }
\renewcommand{\vev}[1]{ \langle \, #1 \, \rangle }

\renewcommand{\Tr}{ {\rm Tr} \, }
\renewcommand{\tr}{ {\rm Tr} \, }

\renewcommand{\ket}[1]{ \, | #1 \rangle }
\renewcommand{\bra}[1]{ \langle #1 | \, }

\newcommand{\expa}[1]{ \exp{\left( #1 \right)} }
\renewcommand{\abs}[1]{\left| #1 \right|}

\subsection{Introduction}

Chiral Separation Effect (CSE) is one of the macroscopic transport phenomena generated by microscopic quantum anomalies - violation of classical symmetries at the quantum level. CSE originates in Adler-Bell-Jackiw axial anomaly and leads to the generation of a macroscopic axial current $j^A$ in the direction of an external magnetic field $B$ \cite{Vilenkin:1980fu,Son:2004tq,Metlitski:2005pr}:
\begin{eqnarray}
\label{eq:CSE_basic}
 j^A_k = \bar{q} \, \gamma_5 \, \gamma_k \, q = \sigma_{CSE}\lr{\mu, T} \, B_k ,
\end{eqnarray}
where $k = \lrc{x, y, z}$ enumerates spatial dimensions, $\bar{q}$, $q$ are fermion (quark) fields, $\gamma_5$ and $\gamma_k$ are the Dirac $\gamma$-matrices, $\mu$ is the fermion chemical potential and $T$ is the temperature. For free non-interacting fermions with unit electric charge, the corresponding transport coefficient $\sigma_{CSE}\lr{\mu, T}$ takes a particularly simple form
\begin{eqnarray}
\label{eq:sigma_CSE_free}
 \sigma_{CSE}\lr{\mu}  = \frac{\mu}{2 \pi^2} ,
\end{eqnarray}
where the coefficient $1/\lr{2 \pi^2}$ is often identified with the universal coefficient of the $U\lr{1}_A$ axial anomaly.

In combination with other anomalous transport phenomena, CSE can lead to observable signatures in heavy-ion collision experiments, such as electric quadrupole moment of quark-gluon plasma \cite{Burnier:2011bf}, and novel hydrodynamical excitations such as the Chiral Magnetic Wave \cite{Kharzeev:2010gd}.

As anomalous transport phenomena became the subject of intense theoretical research in the past decades, it was quickly realized that CSE can get both perturbative \cite{Gorbar:2013upa} and non-perturbative \cite{Newman:2005as,Avdoshkin:2017cqp} corrections in Quantum Chromodynamics (QCD) and Quantum Electrodynamics (QED) where interactions between fermions are mediated by gauge bosons. In view of potential experimental detection, in particular, in the data obtained during the RHIC isobar run \cite{STAR:2021mii}, it becomes important to quantify the strength of possible corrections to CSE in QCD. Currently, the only way to quantify such corrections entirely from first principles is provided by lattice gauge theory simulations. 

CSE is particularly suitable for lattice gauge theory simulations, as it exists in a state of thermal equilibrium and involves only well-defined lattice observables. The Chiral Vortical Effect (CVE) is another anomalous transport phenomenon which exists in a thermal equilibrium state. However, lattice signatures of CVE involve expectation values of the energy-momentum tensor \cite{Braguta:2013loa,Braguta:2014gea}, which is one of the most challenging quantities to define and measure on the lattice. In contrast, the Chiral Magnetic Effect (CME) is known to vanish in thermal equilibrium state \cite{Buividovich:2013hza}, which makes it conceptually difficult to simulate CME on the lattice.

A particular challenge for studying CSE in first-principle lattice QCD simulations is the infamous fermionic sign problem at finite baryon density $\mu$, which is necessary for the CSE to exist. For this reason, first lattice studies of CSE \cite{Puhr:2016kzp} were performed in quenched $SU\lr{3}$ lattice gauge theory with overlap valence quarks, where no statistically significant deviations from free quark result were found on both sides of the confinement-deconfinement transition.

The next step is to consider CSE in one of the QCD-like gauge theories where the fermionic sign problem is absent due to some special symmetry, like $SU\lr{2}$ gauge theory with $N_f=2$ light dynamical quarks. However, background magnetic field breaks the time-reversal invariance which ensures the positive-definiteness of path integral weight in $SU\lr{2}$ gauge theory. This does not allow to study CSE in simulations with finite background magnetic field, as in \cite{Puhr:2016kzp}. A way to resolve this difficulty is to use the linear response approximation and expand the expectation value of electric current in powers of magnetic field. While this expansion does not work for uniform magnetic fields by virtue of total flux quantization in a finite volume, one can consider magnetic fields with nonzero wavelength $k$ and zero total flux. This leads to the following Kubo formula for the CSE coefficient $\sigma_{CSE}\lr{\mu, T}$ \cite{Landsteiner:2012kd}:
\begin{eqnarray}
\label{eq:CSE_Kubo}
 \sigma_{CSE}\lr{\mu, T, k_z} 
 = 
 \vev{ j^A_x\lr{k_z} j^V_y\lr{-k_z} }/ k_z . 
\end{eqnarray}
The expectation value in this formula is taken with respect to the thermal equilibrium state without background magnetic field. $j^A_x\lr{k_z}$ and $j^V_y\lr{k_z}$ are the Fourier transforms of the local axial and vector current densities $j^A_k\lr{\vec{r}} = \bar{q}\lr{\vec{r}} \gamma_5 \gamma_k q\lr{\vec{r}}$ and $j^V_k\lr{\vec{r}} = \bar{q}\lr{\vec{r}} \gamma_k q\lr{\vec{r}}$ with momentum that is directed along the $z$ axis (that is, the other components $k_x$ and $k_y$ are equal to zero). The relevant limit for hydrodynamical description of CSE is $k_z \rightarrow 0$.

The Kubo formula (\ref{eq:CSE_Kubo}) was used in \cite{Buividovich:2020gnl} to study CSE in finite-density $SU(2)$ LGT with dynamical quarks. In this work, a suppression of the CSE response was found at low temperatures and densities. Phenomenologically, this suppression can be described by the formula $\sigma_{CSE}\lr{\mu, T} \sim \rho_V\lr{\mu, T}$, where $\rho_V\lr{\mu, T}$ is the electric charge density induced by finite chemical potential $\mu$ \cite{Avdoshkin:2017cqp}.

Subsequent effort to measure CSE in full lattice QCD simulations with $SU\lr{3}$ used linear expansion in the chemical potential $\mu$ in fixed background magnetic field $B$ to circumvent the fermionic sign problem \cite{Velasco:2022gaw,Brandt:2023wgf}, confirming the suppression of CSE at low temperatures.

\subsection{CSE and the QCD Kondo effect}

The suppression of anomalous transport phenomena in low-temperature QCD makes perfect sense, as non-Abelian gauge fields tend to confine chiral fermions, thus preventing them from participating in collective transport responses. This common-sense expectation makes it particularly exciting to find physical mechanisms that could actually \emph{enhance} anomalous transport responses. 

In this section we demonstrate one such mechanism related to the presence of heavy quarks. As demonstrated in \cite{Yasui:2013xr,Hattori:2015hka}, the amplitude of scattering of light quarks off heavy quarks of mass $m_Q$ is enhanced as $\log\lr{m_Q}$ in the vicinity of the Fermi surface of light quarks. This phenomenon is analogous to Kondo effect in condensed matter physics, and has been termed ``the QCD Kondo effect''. The enhancement of heavy-light quark interactions near the Fermi surface leads to the formation of the Kondo condensate $\vev{\bar{Q} \, q}$, where $\bar{Q}$ is the heavy quark field and $q$ stands for light quarks \cite{Yasui:2016svc}. At the mean-field level, this effect can be described by an additional term in the Lagrangian \cite{Suenaga:2020oeu}:
\begin{eqnarray}
\label{eq:Kondo_mean_field}
 \mathcal{L} 
 \rightarrow 
 \mathcal{L} 
 +
 \Delta \lr{\bar{q} Q + \bar{q} \gamma_{\mu} \frac{k_{\mu}}{|k|} Q } .
\end{eqnarray}
This additional coupling between heavy and light quarks induces a divergence of the density of states of the light quarks $q$ in the vicinity of the Fermi surface \cite{Suenaga:2020oeu}. The dominant contribution to the axial-vector current-current correlator (\ref{eq:CSE_Kubo}) comes from momenta in the vicinity of the Fermi surface (see Fig.~1 in \cite{Buividovich:2013hza}), thus an enhanced density of states in this region due to the Kondo effect is likely to lead to an enhancement of the Chiral Separation Effect. This is precisely what was found in the mean-field calculation in \cite{Suenaga:2020oeu}.

\subsection{Lattice setup: $SU\lr{2}$ lattice gauge theory with $N_f = 2 + 1$ light and heavy quark flavours}

A beautiful mean-field prediction for the enhancement of the CSE response due to the presence of heavy quarks clearly calls for a first-principle check in lattice gauge theory simulations. Since the full lattice QCD with $SU\lr{3}$ gauge group suffers from the fermionic sign problem at finite chemical potential $\mu$, in this Contribution we study the enhancement of CSE in $SU\lr{2}$ lattice gauge theory with $N_f = 2 + 1$ light and heavy quark flavours.

In our study we use gauge field configurations generated using the standard HMC algorithm with $N_f=2$ mass-degenerate rooted staggered fermions (with bare mass $m_q^{stag} = 0.005$), one extra flavour of heavy rooted staggered fermion, and a tree-level improved Symanzik gauge action. We use ensembles with heavy fermion masses equal to $a \,m_Q = 1$, $a \, m_Q = 0.5$ and $a \, m_Q = 0.2$, with numbers of gauge configurations ranging from $200$ to $1000$. We use lattices with spatial size $L_s = 24$ and variable temporal sizes ranging between $L_t = 10$ and $L_t = 22$. We work at fixed gauge coupling $\beta = 1.7$ and hence at fixed lattice spacing (which we do not determine), changing the temperature by changing $L_t$. The chiral crossover occurs at $L_t \approx 16$. Overall, our lattice setup is similar to the one used in the works \cite{Buividovich:2020dks,Buividovich:2020gnl}.

To ensure the absence of fermionic sign problem because of the extra heavy flavour, chemical potential is only introduced for light flavours. This is justified because, physically, only the existence of the Fermi surface for light quarks is crucial for the Kondo effect. Furthermore, because of large quark masses, heavy flavours are much less sensitive to chemical potential.

For valence quarks, we use Wilson-Dirac fermions with HYP-smeared gauge links \cite{Hasenfratz:2001hp}. The main reason to use Wilson-Dirac valence quarks is that the axial current operator entering (\ref{eq:CSE_Kubo}) is easy to implement. We use the conventional point-split definition of $j^A$, introduced in \cite{Bochicchio:1985xa}, as well as the conserved lattice vector current operator.

The Wilson-Dirac bare mass for light flavours is tuned to $m_{WD} = -0.21$ to achieve the same pion mass $m_{\pi} \approx 0.16$ as for staggered quarks that are used as sea quarks. The $\rho$-meson mass on the same ensemble is approximately $m_{\rho} \approx 0.4$, thus our pion mass is still quite large. On the other hand, large $m_{\pi}$ ensures the smallness of finite-volume artifacts on our moderately large lattices. 

In order to work in the QCD-like regime of the $SU\lr{2}$ lattice gauge theory, we work at a small value of the chemical potential $a \, \mu = 0.05 < a \, m_{\pi}/2$. The reason to use such a small value is that at $\mu > m_{\pi}/2$ the theory enters the diquark condensation regime, where physics is quite different. 

\begin{figure}[h!tpb]
    \centering
    \subfloat[$L_t = 14$]{\includegraphics[width=0.49\textwidth]{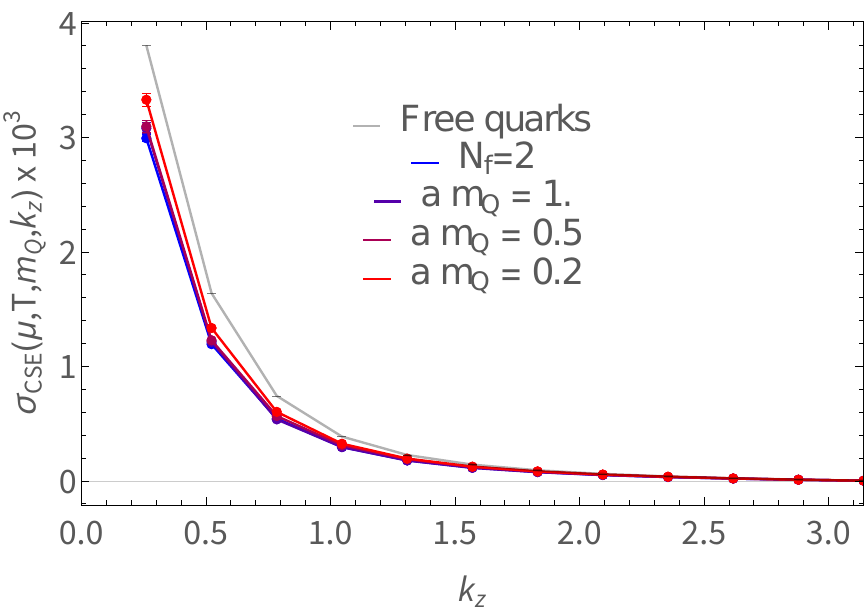}}
    \subfloat[$L_t = 22$]{\includegraphics[width=0.49\textwidth]{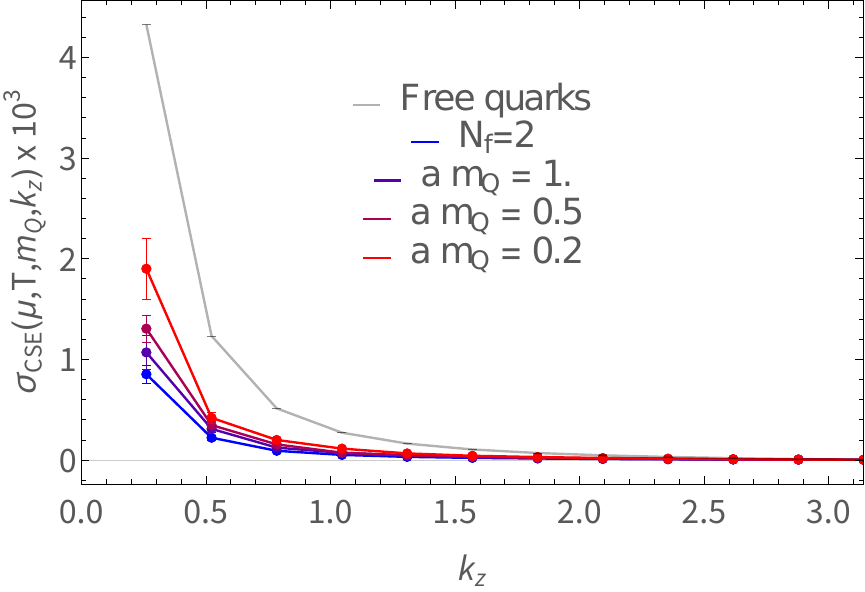}}\\
    \caption{The CSE transport coefficient $\sigma_{CSE}\lr{\mu, T, k_z}$ as a function of lattice momentum $k_z$ at $a \mu = 0.05$ and at two different temperatures: $T > T_c$ on the left ($L_t = 14$) and $T < T_c$ on the right ($L_t = 22$).}
    \label{fig:CSE}
\end{figure}

\subsection{CSE enhancement in $N_f = 2 + 1$ $SU\lr{2}$ gauge theory}

As in the work \cite{Buividovich:2020gnl}, we use the axial-vector current-current correlator (\ref{eq:CSE_Kubo}) at finite momentum $k_z$ to extract the momentum-dependent CSE coefficients $\sigma_{CSE}\lr{\mu, T, m_Q, k_z}$. It is important to stress that we only consider the contribution of light quarks into this correlator, in order to isolate the trivial enhancement of CSE due to increased number of flavours from the nontrivial physics that might be associated with the Kondo effect.

The lattice results for $\sigma_{CSE}\lr{\mu, T, m_Q, k_z}$ are shown on Fig.~\ref{fig:CSE} for different values of the heavy quark mass. We use two values of temperature above and below the chiral crossover temperature which corresponds to temporal lattice size $L_t \approx 1\lr{a T_c} \approx 16$. For comparison, we also present the free-quark result obtained on the same lattice and the same fermion action (with bare Wilson-Dirac mass $m_{WD} = 0.005$), as well as numerical results for $N_f = 2$ light flavours taken from \cite{Buividovich:2020gnl}. The latter results formally correspond to the limit $m_Q \rightarrow +\infty$, where heavy quarks completely decouple.

At the temperature $1/\lr{a T} = L_t = 14$ (left plot on Fig.~\ref{fig:CSE}), the presence of heavy quarks results in just a minor modification of the CSE transport coefficient, although the trend towards enhancement is already visible. More or less all lattice results are reasonably close to the corresponding free quark result. 

\begin{figure}
    \centering
     \subfloat[Chiral condensate of light quarks as a function of temporal lattice size (inverse temperature) for different values of heavy quark mass.]{\includegraphics[width=0.49\textwidth]{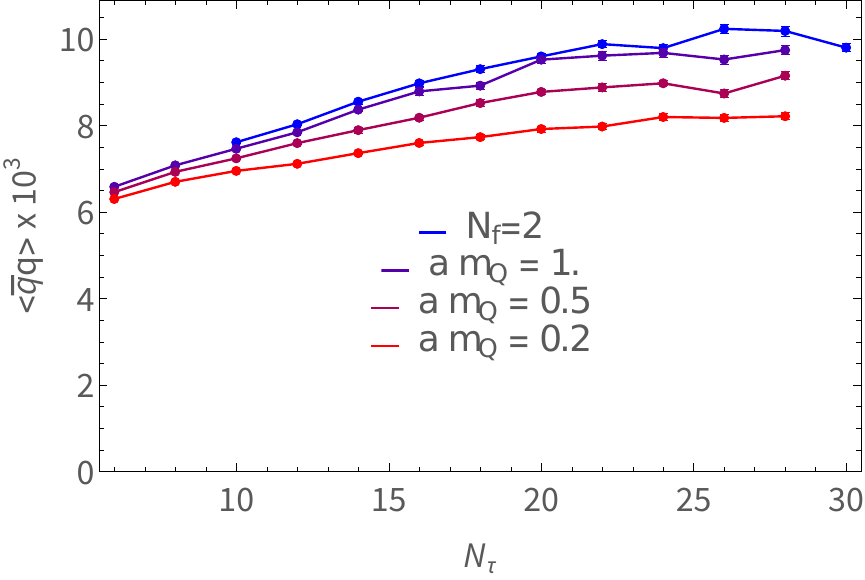}\label{fig:qbarq}}
    \subfloat[DC limit of electric conductivity in $N_f=2+1$ $SU\lr{2}$ lattice gauge theory as a function of temporal lattice size (inverse temperature) for different values of heavy quark mass.]{\includegraphics[width=0.49\textwidth]{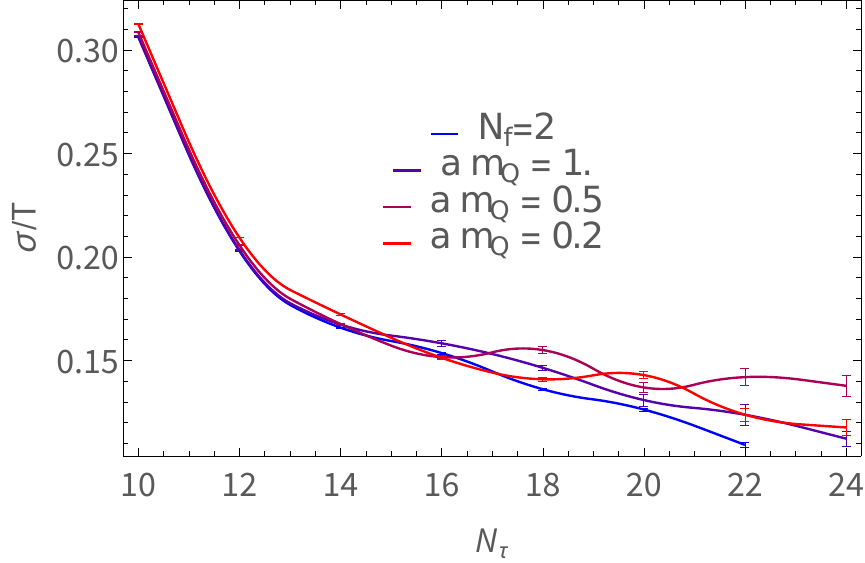}\label{fig:conductivity}}
    \caption{Heavy quark mass dependence of electric conductivity and chiral condensate.}
    \label{fig:conductivity_qbarq}
\end{figure}

On the other hand, at the lower temperature  $1/\lr{a T} = L_t = 22$ (left plot on Fig.~\ref{fig:CSE}), CSE enhancement becomes obvious, and the addition of heavy quark with bare mass $a \, m_Q = 0.2$ changes the low-momentum limit of $\sigma_{CSE}\lr{\mu, T, m_Q, k_z}$ by almost a factor of two in comparison with the result for two light flavours (blue line/data points). However, we never observe an enhancement that would go beyond the corresponding free-fermion result.

One possible explanation for CSE enhancement is that adding more quark flavours drives the theory away from the confinining regime with spontaneous chiral symmetry breaking. We check to what extent is the chiral symmetry breaking for light quarks because of the presence of heavy quarks, on Fig.~\ref{fig:qbarq} we also present lattice results for the chiral condensate of light quarks, $\vec{\bar{q}q}$, at different heavy quark masses and different temperatures. We indeed see that the chiral condensate decreases as the heavy quark mass goes down and they influence the dynamics more and more. However, this effect is not very large. For example, for $1/\lr{a T} = L_t = 22$ and $m_Q = 0.2$, the chiral condensate is only around 25\% smaller than for the two-flavour case ($m_Q \rightarrow + \infty$). This has to be compared with a factor of two increase in the CSE transport coefficient $\sigma_{CSE}$. It seems therefore that the relatively mild tendency towards restoration of spontaneously broken chiral symmetry with more quark flavours cannot explain much more significant increase of the CSE response. 

A paradigmatic signature of the Kondo effect in condensed matter physics is a non-monotonous dependence of the electric DC conductivity $\sigma$ on temperature , expressed by the famous formula \cite{10.1143/PTP.32.37}
\begin{eqnarray}
\label{eq:Kondo_paradigmatic}
 \frac{1}{\sigma\lr{T}} = \frac{1}{\sigma_0} + a \, T^2 + c_m \log\lr{T^{-1}} + b \, T^5 .
\end{eqnarray}
While the $b \, T^5$ term is related to electron-phonon interactions, the first three terms are universal and from the Fermi liquid behavior of fermion degrees of freedom. This formula predicts a maximum of electric conductivity (equivalently, a minimum of resistivity) at some intermediate temperature. Similar results were also obtained in the context of QCD Kondo effect using renormalization group techniques \cite{Yasui:2017bey}. 

In order to check whether we can observe such a non-monotonous dependence of conductivity on temperature, we also calculated electric conductivity for our ensemble of gauge configurations. To this end we calculated Euclidean correlators of vector currents (again, considering only the light quark contributions) and using the Backus-Gilbert algorithm to extract conductivity, as in \cite{Buividovich:2020dks}.  

Our results for the DC limit of electric conductivity are shown on Fig.~\ref{fig:conductivity}. We only observe a monotonous decrease of $\sigma\lr{T}$ towards lower temperatures, without any statistically significant dependence on the heavy quark mass. 

\subsection{Discussion and conclusions}

Our lattice results qualitatively confirm the enhancement of the CSE response in the low-temperature regime in the presence of heavy quarks, first predicted in the mean-field approximation in \cite{Suenaga:2020oeu}. However, a complementary study of the temperature dependence of electric conductivity, which is a ``smoking gun'' for Kondo physics, suggests that we are not in the Kondo regime yet. In particular, the value of the chemical potential $a \, \mu = 0.05$ used in our study may be too small for the formation of a well-defined Fermi surface. This is definitely true in the free quark case, where only the few lowest discrete lattice momenta $k_i = \frac{2 \pi m_i}{L_s}$ would contribute to the would-be Kondo effect. However, the fact that we do observe a significant enhancement of CSE in the presence of heavy quarks suggests that the mechanism of this enhancement may be even more general than the Kondo effect as such.

Two ways to reach the true Kondo regime for $SU\lr{2}$ lattice gauge theory would be: 
\begin{itemize}
 \item To use very large spatial lattice sizes $L_s$, where there would be sufficiently many discrete lattice momenta within the Fermi surface $\epsilon\lr{k} < \mu$. If we want to stay in the QCD-like phase of $SU\lr{2}$ lattice gauge theory with $\mu < m_{\pi}/2$, $L_s$ should be of the order of few hundred sites, which is unfeasible with currently available computational resources.
\item To use very large values of $\mu \gg m_{\pi}/2$. Even though such large values are in the diquark condensation phase, Fermi surface can still be formed \cite{Hands:2006ve,Bornyakov:2017txe}. There are good reasons to expect that this weak-coupling, low-temperature, high-density regime should be similar in both $SU\lr{2}$ gauge theory and in full QCD. This direction seems more promising for future exploration.
\end{itemize}

Another important feature of the Kondo effect is that formally the Kondo coupling is enhanced towards heavier fermions $\log\lr{m_Q}$ \cite{10.1143/PTP.32.37}. However, in lattice QCD simulations of the grand canonical ensemble the density of heavy fermions is suppressed as $e^{-\mu m_Q}$, which is a much stronger factor. Qualitatively, this explains why in our case the CSE response increases towards smaller $m_Q$. A cleaner way to study Kondo physics as such would be simulations of the canonical ensemble with fixed density of heavy quarks, rather than at fixed chemical potential \cite{Li:2011ee}. In the limit of $m_Q \rightarrow + \infty$, heavy quarks can also be introduced as insertions of Polyakov loops in lattice observables. We leave these refinements for future work.

Experimentally, the enhancement of CSE in the presence of heavy quarks can be used to construct refined experimental probes, where signatures of CSE or the closely related Chiral Magnetic Wave/electric quadrupole moment \cite{Burnier:2011bf} would be correlated with the abundance of heavy mesons or baryons (e.g. charmonium/bottomonium).

	\section{Novel hydrodynamic transport coefficients in extreme magnetic fields \& the CME far from equilibrium}
	\label{sec:Kaminski}

	%
	%
	%
	%
	%

\subsection{Introduction}\label{sub:hydroHoloIntro}
%
Relativistic hydrodynamics is a state-of-the art framework for describing plasmas generated in heavy-ion collisions~\cite{Romatschke:2017ejr}, astrophysical plasmas~\cite{Shore:1992xv,Janka:2012wk}, and neutron star mergers~\cite{Faber:2012rw,Baiotti:2016qnr}.   
Hydrodynamics is based on the symmetries of the system to be described, 
in particular, standard hydrodynamics assumes parity-invariance and spatial isotropy. 
However, it is experimentally established~\cite{Blackie:1960,Bernstein:2011bx} that the chiral anomaly of QCD,  responsible for the fast pion decay, breaks the parity invariance~\cite{Adler:1969gk,Bell:1969ts}. 
In addition, strong magnetic fields, $B$, of leading order in the hydrodynamic derivative expansion (yet small compared to the temperature scale $T$, {i.e.}~$B\ll T^2$) break spatial isotropy. 
%
Consequently, the hydrodynamic description of QCD-plasma subject to strong magnetic fields needs to be constructed under these restricted symmetry constraints (anisotropy and parity-violation).

\begin{figure}[]
\centering
%
\includegraphics[width=0.23\textwidth,angle=270]{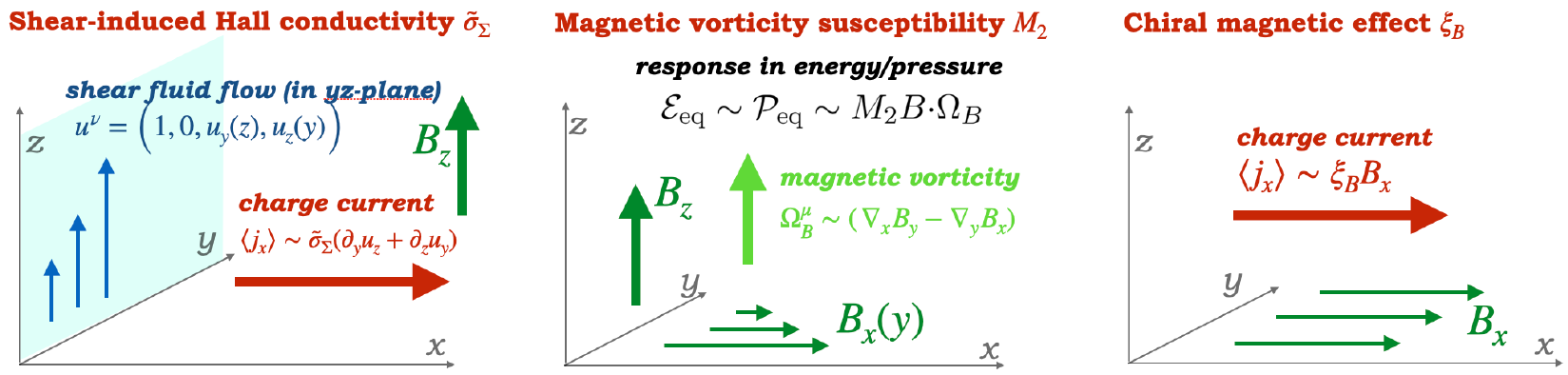} 
%
%
\caption{\label{fig:parityViolatingTransport} 
Parity-violating transport effects arising in strong magnetic fields: the shear-induced Hall current is a novel effect requiring strong magnetic fields, just like the novel energy/pressure shift induced by a strong vortical magnetic field in equilibrium. The chiral magnetic effect (CME) already appears at weak and persists at strong magnetic fields. All three depicted transport effects require a nonzero chemical potential in addition to parity-violation and strong magnetic field. 
There are 27 transport coefficients with Kubo formulae derived in~\cite{Ammon:2020rvg}. 
   } 
\end{figure}

As a major result, novel transport effects arise~\cite{Kovtun:2016lfw,Hernandez:2017mch,Kovtun:2018dvd,Ammon:2020rvg}. 
Among those novel coefficients there is the non-dissipative \emph{shear-induced Hall conductivity}~\cite{Ammon:2020rvg}. This coefficient measures the response of a charge current (for example, in $x$-direction) to a shear in the fluid velocity perpendicular to it (for example, in the $(y,z)$-plane), see Fig.~\ref{fig:parityViolatingTransport}. 
There are also novel susceptibilities in equilibrium, for example, the \emph{perpendicular magnetic vorticity susceptibility}~\cite{Ammon:2020rvg}, which measures the shift in the equilibrium energy density and pressure when the plasma is subject to a vortical magnetic field, see Fig.~\ref{fig:parityViolatingTransport}. 
How these novel effects are derived and interpreted is outlined in section~\ref{sub:hydroTransport}, see~\cite{Ammon:2020rvg} for a detailed interpretation. 

Another major result due to the anisotropy is that known transport coefficients such as the shear viscosity take distinct values along the plane containing the magnetic field (parallel) versus the plane perpendicular to it~\cite{Critelli:2014kra,Ammon:2020rvg}. At large magnetic fields, the parallel shear viscosity divided by the entropy density can drop below the holographic \emph{KSS value}~\cite{Kovtun:2004de} and can even vanish~\cite{Critelli:2014kra,Ammon:2020rvg}. 
Remarkably, transport coefficients known from 2+1-dimensional systems, such as Hall viscosities, emerge due to anisotropy~\cite{Ammon:2020rvg}. These altered shear and Hall viscosities can affect the evolution of elliptic flow and thus may have phenomenological consequences, since analyses like~\cite{Bernhard:2019bmu} need to be refined.  

One transport process arising from parity-violation through the chiral anomaly already at weak magnetic fields, i.e.~at sub-leading order in the hydrodynamics gradient expansion, is the chiral magnetic effect (CME)~\cite{Vilenkin:1978is,Vilenkin:1979ui,Kharzeev:2004ey,Son:2009tf}. This effect at strong magnetic fields appears at leading order in the derivative expansion. Thus, the CME at strong magnetic fields becomes a potential equilibrium phenomenon, as explained in the main text. The CME predicts an electric charge current in response to an external magnetic field in presence of an axial charge imbalance~\cite{Kharzeev:2004ey,Son:2009tf}, see Fig.~\ref{fig:parityViolatingTransport}. 
The CME was experimentally verified in Weyl semi-metals, see for example~\cite{Li:2014bha}, while results from heavy-ion experiments are being debated~\cite{STAR:2021mii}. 
Two major insights into the CME stemming from holographic plasma are (i) the rigorous time evolution of charged plasma states in magnetic fields~\cite{Fuini:2015hba,Ammon:2016fru,Cartwright:2020qov,Ghosh:2021naw,Cartwright:2021maz,Grieninger:2023myf}, see section~\ref{sub:holoCME}, and (ii) the dynamical generation of axial charge imbalance~\cite{Grieninger:2023wuq}. 
%
\subsection{Novel transport coefficients in extreme magnetic fields from chiral hydrodynamics~\cite{Ammon:2020rvg}}\label{sub:hydroTransport}
Hydrodynamic constitutive relations can be derived either from a generating functional~\cite{Jensen:2012jh,Banerjee:2012iz,Grozdanov:2013dba,Kovtun:2016lfw,Haehl:2015pja,Crossley:2015evo,Jensen:2018hse,Liu:2018kfw,Jain:2023obu,Mullins:2023ott} or constructed directly~\cite{LL6,Kovtun:2012rj}. For the chiral hydrodynamic constitutive relations, a hybrid approach turns out to be most efficient, deriving their equilibrium part from a generating functional and constructing their non-equilibrium part directly. 
\paragraph{Equilibrium generating functional \& constitutive relations.}
Considering hydrodynamics as an effective field theory, one notes that the hydrodynamic framework is constrained by the existence of an equilibrium generating functional. For systems with finite correlation lengths that are in equilibrium, and with sources that vary on length scales much larger than the correlation length, the generating functional $W_s = -i \ln Z[g,A]$ is a local functional of the Killing vector $V$ and the sources~\cite{Jensen:2012jh,Banerjee:2012iz}. The generating functional can then be expanded in a derivative expansion. The traditional zero derivative terms are the temperature $T$, fluid velocity $u^\mu$ and chemical potential $\mu$, which can all be writen in terms of the Killing vector $V$ and the external sources $A$ and $g$. In addition, for systems in strong magnetic fields $B^\mu = \frac{1}{2} \epsilon^{\mu\nu\rho\sigma} u_\nu F_{\rho\sigma}$, one can take $B^\mu$ and $B^2=B^\mu B^\nu g_{\mu\nu}$ to be order zero in derivatives~\cite{Kovtun:2016lfw,Hernandez:2017mch,Ammon:2020rvg}. In this case, the generating functional contains corrections at first order in the derivative expansion, $W_s \supset \int d^4x \sqrt{-g} M_n(T,\mu,B^2) s_n$ where
$M_n$ are thermodynamic susceptibilities and $s_n$ are scalars containing only one derivative and that do not vanish in equilibrium such as $B^\mu \partial_\mu \left( \frac{B^2}{T^4}\right)$ and $\epsilon^{\mu\nu\rho\sigma} F_{\mu\nu} F_{\rho\sigma}$, see table 1 of~\cite{Ammon:2020rvg}. 

The equilibrium constitutive relations which relate the one point functions of the stress energy tensor $T^{\mu\nu}$ and the current $J^\mu$ to the hydrodynamic variables can be derived by varying the equilibrium generating functional with respect to the sources, and they depend on the susceptibilities. For example,
\begin{equation}\label{eq:M2}
    \mathrm{pressure}\sim T^{\mu\nu} (g_{\mu\nu}+u_\mu u_\nu) \sim M_2 \epsilon_{\mu\nu\rho\sigma}B_\mu u_\nu \partial_\rho B_\sigma\,,  \quad \mathrm{energy}\sim T^{\mu\nu}u_\mu u_\nu \sim M_2 \epsilon_{\mu\nu\rho\sigma}B_\mu u_\nu \partial_\rho B_\sigma\,,
\end{equation}
indicates that there is an equilibrium response, a shift in the energy and pressure, to a non-uniform magnetic field. 
\paragraph{The chiral anomaly.} For systems with a chiral anomaly, the generating functional is not gauge invariant, 
$
    \delta_\alpha W_{cons} = \frac{C}{24} \int d^4x \sqrt{g} \alpha \epsilon^{\mu\nu\rho\sigma}F_{\mu\nu} F_{\rho\sigma}
$,
where $C$ is the anomaly coefficient.\footnote{For simplicity, a microscopic theory with one single anomalous $U(1)$-symmetry is considered here.} The consistent current found by varying the generating functional $W_{cons}$ with respect to the sources is not gauge invariant, so it is useful to work with a covariant current $J^\mu_{cov} = J^\mu_{cons} - \frac{1}{6} C \epsilon^{\mu\nu\rho\sigma} A_\nu F_{\rho \sigma}$. Expanding the generating functional order by order in derivatives, and varying it with respect to the sources we find the equilibrium constitutive relations for fluids with a chiral anomaly
\begin{equation}\label{eq:const-anom}
    T_A^{\mu\nu} = T_s^{\mu\nu} + \xi_T u^{(\mu} \Omega^{\nu)} + \xi_{TB} u^{(\mu} B^{\nu)} \,,  \quad  J_{cov}^{\mu} = J_s^{\mu} + \xi\, \Omega^{\mu} + \xi_{B} B^{\mu}\,.
\end{equation}
In the above, $T^{\mu\nu}_s$ and $J^\mu_s$ are the equilibrium stress tensor and current for a system without an anomaly, and 
\begin{equation}\label{eq:xis}
    \xi = \xi_{TB} = \frac{C}{2}\mu^2 + c_1 T^2 \,, \quad \xi_B = C\mu\,, \quad \xi_T = \frac{C}{3}\mu^3 + 2 c_1 T^2\mu\,,
\end{equation}
are the chiral transport coefficients. The chiral magnetic conductivity $\xi_B$ parametrizes the response of the current to a magnetic field (the CME, see Fig.~\ref{fig:parityViolatingTransport}), the chiral vortical heat conductivity $\xi_T$ the response of the heat current to the fluid vorticity, and the chiral vortical conductivity $\xi=\xi_{TB}$ the response of the current to the fluid vorticity, and of the heat current to a magnetic field. The coefficient $c_1$ is related to the mixed gauge-gravitational anomaly~\cite{Jensen:2012kj}, and there is an additional coefficient $c_2$~\cite{Ammon:2020rvg} (it vanishes for CPT invariant systems), which is set to zero in this section.

\paragraph{Equilibrium Kubo formulae.} Varying the equilibrium constitutive relations twice with respect to the sources we find the static Kubo formulae for thermodynamic coefficients. For example, for $\mathbf{k} = k_z \mathbf{\hat{z}}$, and a magnetic field $\mathbf{B} = B_0 \mathbf{\hat{z}}$ 
\begin{equation}\label{eq:Kubo-stat}
    \langle T^{xz}(\mathbf{k}) T^{yz} (-\mathbf{k}) \rangle = -2 i B^2_0 M_2 k_z  \,, \quad \langle J^x(\mathbf{k}) J^y(-\mathbf{k})\rangle \sim -i \xi_B k_z \,.
\end{equation}
Note that Eq.~\eqref{eq:Kubo-stat} together with Eq.~\eqref{eq:xis} implies that the chiral magnetic effect can manifest itself on the level of two-point functions in an inhomogenous equilibrium state in presence of a strong magnetic field.\footnote{This is because the factor of momentum $k_z$ (after a Fourier transform) stems from a spatial derivative acting on the chemical potential $\mu$. Thus, in momentum space one obtains $\langle J^x(k_z) J^y(-k_z)\rangle \sim -i C \mu k_z$, if an equilibrium state with inhomogenous $\mu(\mathbf{x})$ is subject to a magnetic field of zeroth order in the derivative expansion. 
Note, however, the restrictions imposed by the generalized Bloch theorem~\cite{Yamamoto:2015fxa}. 
In addition, we note that lattice computations involving massive fermions report not to find any CME in equilibrium states~\cite{Brandt:2024wlw}, see this topic reviewed in~\cite{Endrodi:2024cqn}. 
} 

\paragraph{Out-of-equilibrium constitutive relations.} The non-equilibrium terms in the constitutive relations can be constrained by symmetry arguments.\footnote{It is possible to formulate a non-equilibrium generating functional which sources the non-equilibrium constitutive relations~\cite{Jensen:2012jh,Banerjee:2012iz,Grozdanov:2013dba,Kovtun:2016lfw,Haehl:2015pja,Crossley:2015evo,Jensen:2018hse,Liu:2018kfw,Jain:2023obu,Mullins:2023ott}. However, doing so significantly increases the complexity of the formalism so we approach the non-equilibrium terms with the standard phenomenological approach~\cite{LL6,Kovtun:2012rj} for simplicity.} For a relativistic fluid in strong magnetic fields, there are 14 independent hydrodynamic transport coefficients, 10 of which are dissipative and 4 are non-dissipative. An example of a non-dissipative transport coefficient is the novel \emph{shear-induced Hall conductivity} $\tilde{\sigma}_\Sigma$ (a dissipative one is the novel \emph{shear-induced conductivity} $\sigma_\Sigma$),\footnote{The shear-induced conductivity is labelled $c_8$ and the shear-induced Hall conductivity is labelled $c_{10}$ in~\cite{Ammon:2020rvg}.}
\begin{equation}
    J^\mu \sim \sigma_\Sigma \Sigma^\mu + \tilde{\sigma}_\Sigma \tilde{\Sigma}^\mu\,,  \quad \Sigma^{\mu} = \left(\Delta^{\mu\nu} - B^\mu B^\nu/B^2 \right) \sigma_{\nu\rho}B^\rho/\sqrt{B^2}\,, \quad \tilde{\Sigma}^\mu = \epsilon^{\mu\nu\rho\sigma} u_\nu B_\rho \Sigma_\sigma /\sqrt{B^2}.
\end{equation}
where $\sigma^{\mu\nu}= \Delta^{\mu\alpha} \Delta^{\nu\beta} \left(\nabla_\alpha u_\beta + \nabla_\beta u_\alpha - \frac{2}{3} g_{\alpha\beta} \nabla{\cdot} u \right)$ is the fluid shear tensor and $\Delta^{\mu\nu} = g^{\mu\nu} + u^\mu u^\nu$. 

\paragraph{Hydrodynamic Kubo formulae.} Finding the Kubo formulae for the hydrodynamic transport coefficients requires solving the linearized hydrodynamic equations. For example, for a magnetic field $\mathbf{B} = B_0 \mathbf{\hat{z}}$, one finds~\cite{Ammon:2020rvg}
\begin{equation}\label{eq:Kubo}
\begin{aligned}
    &\lim_{\omega\to 0}\frac{1}{\omega}{\rm Im} \, G_{T^{tx}T^{xz}}(\omega,\mathbf{k}=0) \sim {\sigma}_\Sigma \,, \quad \lim_{\omega\to 0}\frac{1}{\omega}{\rm Im}\, G_{T^{tx} T^{yz}}(\omega,\mathbf{k}=0)    \sim\tilde{\sigma}_\Sigma \,.
\end{aligned}    
\end{equation}

\paragraph{Eigenmodes.} With the hydrodynamic equations and the constitutive relations, one can study the eigenmodes of small perturbations around equilibrium~\cite{Ammon:2020rvg}. For example, for momenta $\mathbf{k}|| \mathbf{B}$, we find two ``sound'' waves and the chiral magnetic wave
\begin{equation}\label{eq:eigenmodes}
    \omega = v_{\pm} k -  \frac{i \Gamma_{\pm}}{2} k^2\,, \quad \omega = v_0 k - i D_{||} k^2\,.
\end{equation}
The velocities $v_0$ and $v_\pm$ and damping coefficients $\Gamma_\pm$ and $D_{||}$ depend on the transport coefficients in the constitutive relations. Remarkably, the chiral magnetic wave only propagates when there is an anomaly, \textit{i.e.,} $v_0 \propto C$, and the ``sound'' waves propagate at different speeds depending on the direction they travel in, \textit{i.e.,} $|v_+-v_-| \propto C$.

\begin{figure}[]
\centering
\includegraphics[width=0.32\textwidth,angle=0]{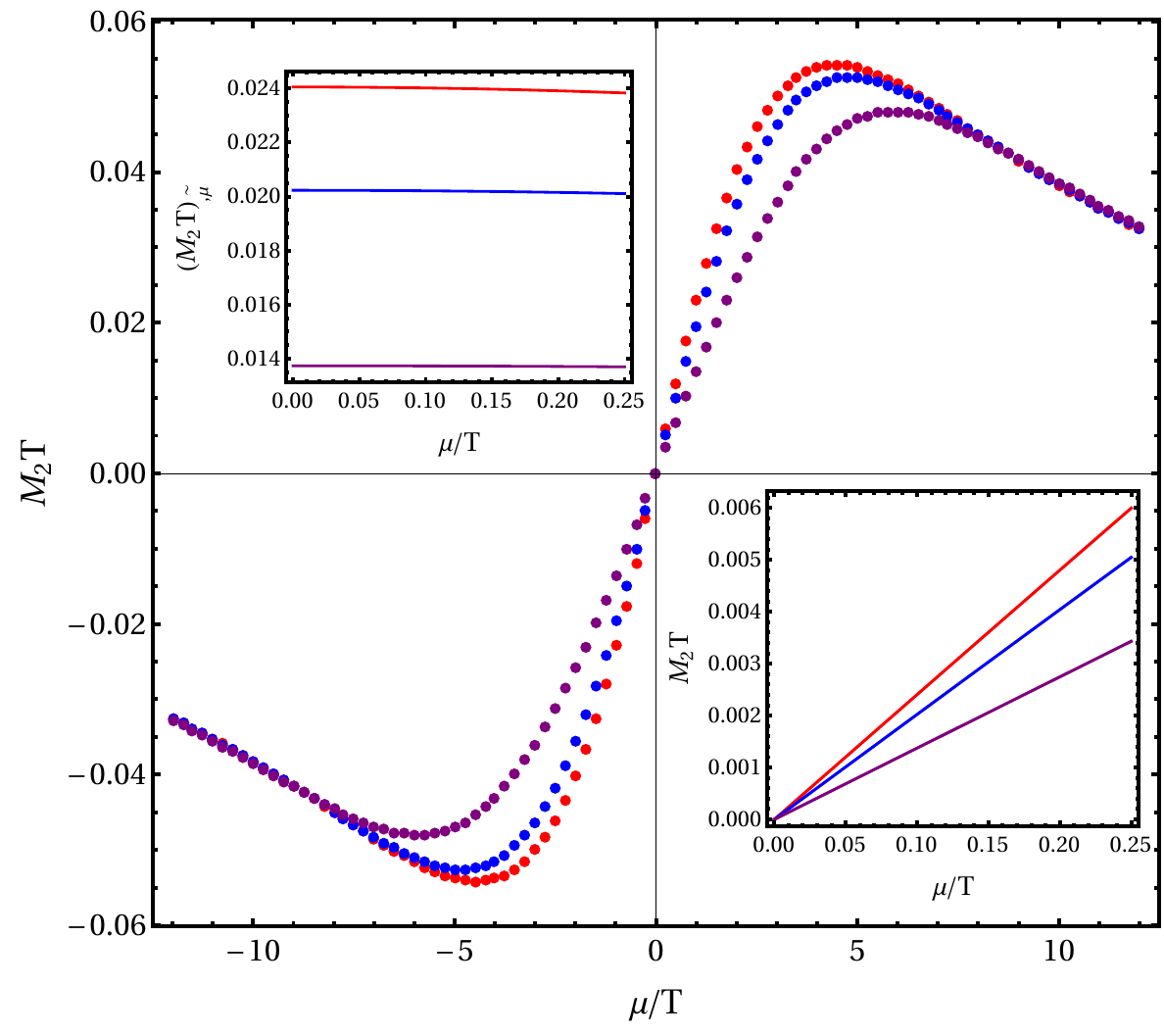} \hfill
\includegraphics[width=0.3\textwidth,angle=0]{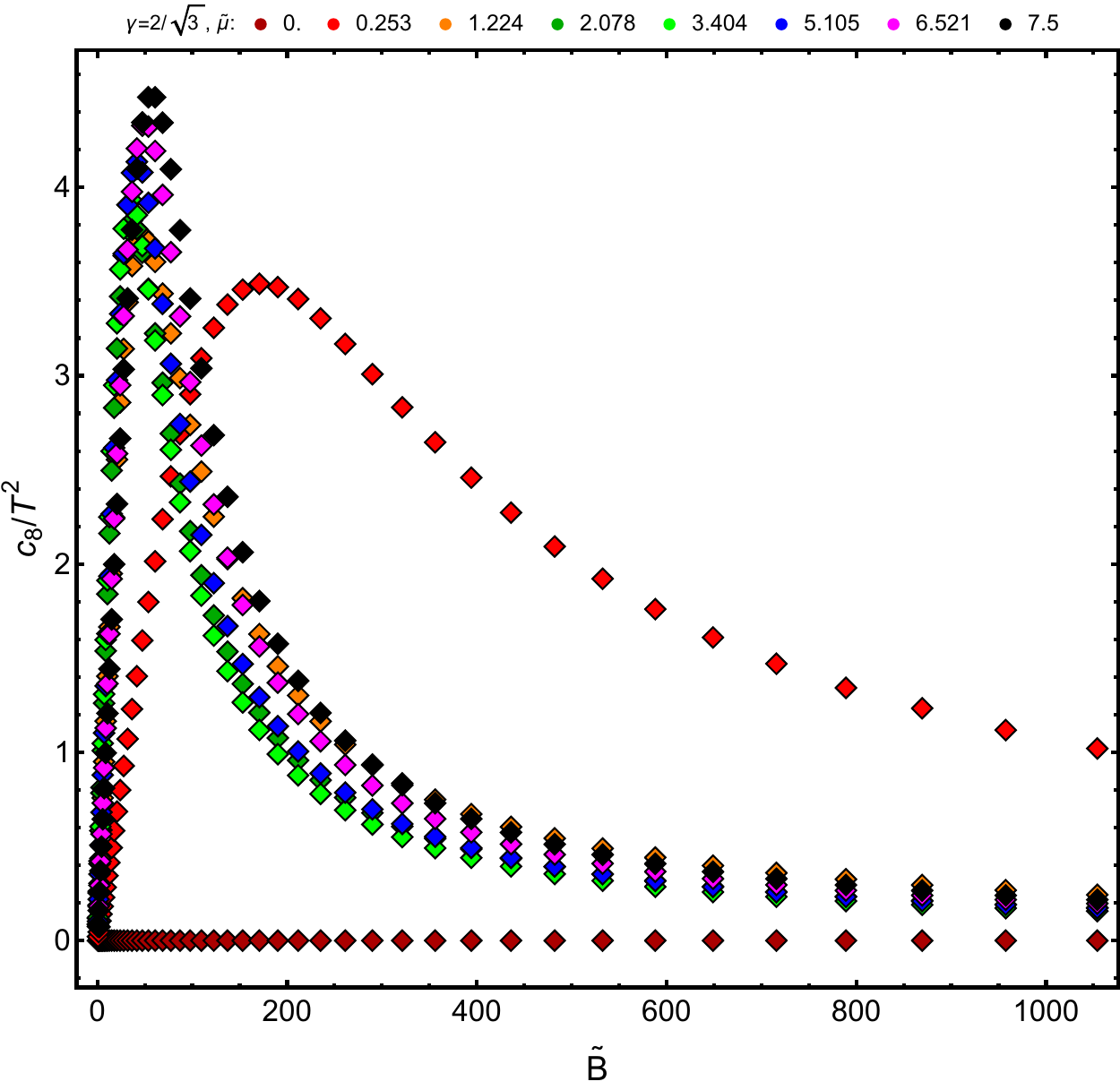} \hfill
\includegraphics[width=0.3\textwidth,angle=0]{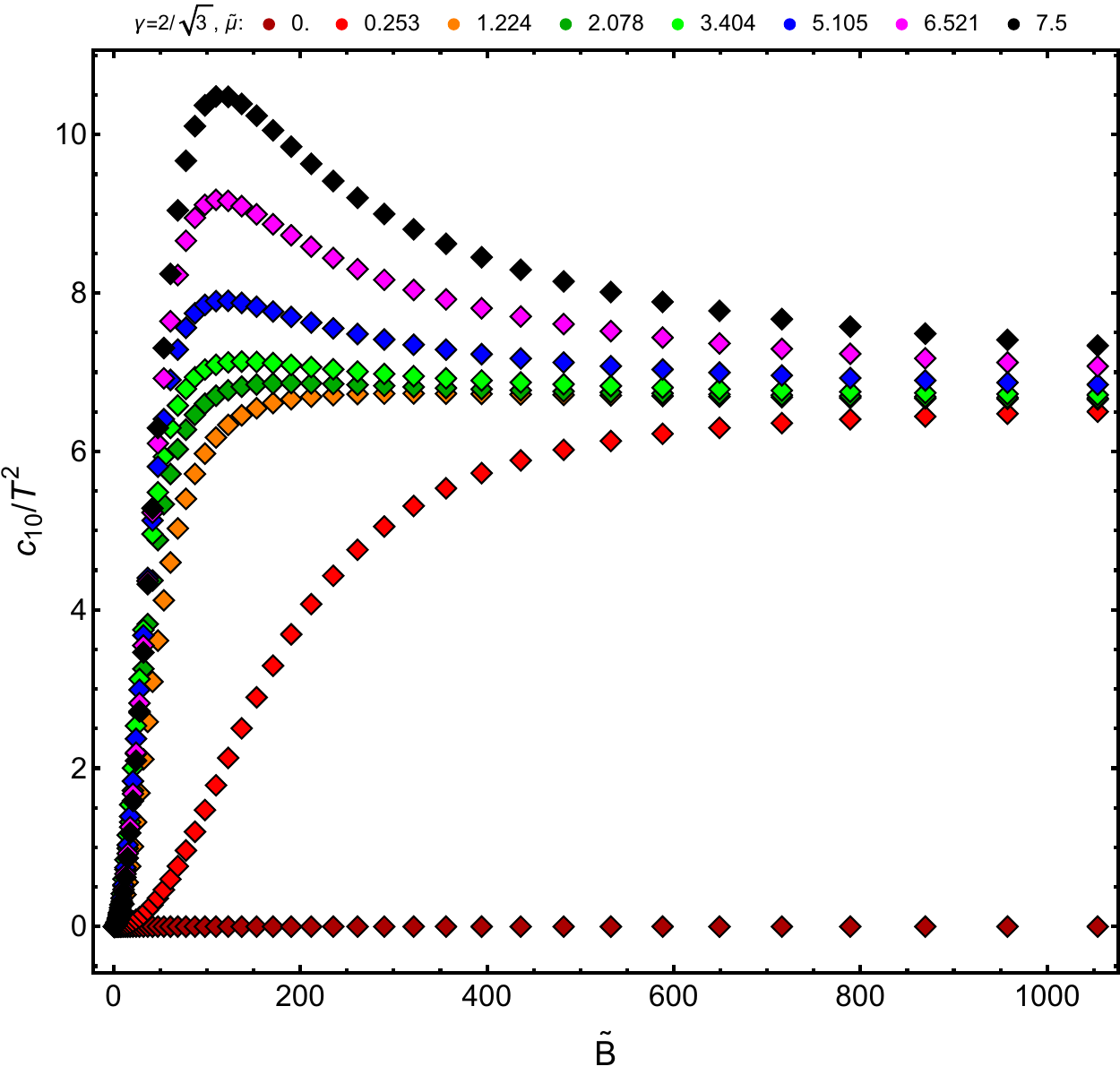} 
\caption{\label{fig:transport}%
Transport coefficients in the presence of the chiral anomaly (parametrized by $\gamma = 2/\sqrt{3}\sim C$).
Without chiral anomaly or for vanishing chemical potential, all three transport coefficients are zero. Left: The dimensionless thermodynamic susceptibility $M_2 T$ as a function of the axial chemical potential $\mu/T$. Different curves correspond to magnetic fields 
$\tilde{B}=B/T^2 =
\{0.05, 12.5, 30\}$ (red, blue, purple). 
Middle: The dimensionless shear-induced conductivity 
${\sigma}_\Sigma/T^3$. 
Right:  The dimensionless shear-induced Hall conductivity 
$\tilde{\sigma}_\Sigma/T^2$.
} 
\end{figure}

\paragraph{Physics of transport coefficients.} The hydrodynamic framework described above uncovers 14 non-equilibrium transport coefficients (5 more are related through Onsager relations), 4 chiral transport coefficients and 5 thermodynamic transport coefficients~\cite{Ammon:2020rvg}. An example of a thermodynamic transport coefficient is the magnetic vortical susceptibility $M_2$ appearing in Eq.~\eqref{eq:M2}. It parametrizes the response of the energy and pressure to a non-uniform magnetic field, see Fig.~\ref{fig:parityViolatingTransport}. It also paramterizes the shear response to a non-zero Poynting vector $\mathbf{B}\times \mathbf{E}$. For example, for $\mathbf{B} = B_0 \mathbf{\hat{z}}$ and $\mathbf{E}=E_0 \mathbf{\hat{y}}$, there is a shear term $T^{xz} \sim \frac{\partial M_2}{\partial\mu} B_0^2 E_0$. The chiral conductivities are completely fixed in terms of the anomaly coefficients $C$ and $c_1$, see Eq.~\eqref{eq:xis}. They parametrize the response of the charge current and heat current to a magnetic field and fluid vorticity (see Eq.~\eqref{eq:const-anom}), and are essential for propagation of the chiral magnetic wave in~\eqref{eq:eigenmodes}. Thermodynamic and chiral transport coefficient are equilibrium quantities, and they can be computed or experimentally observed in equilibrium configurations, see Eq.~\eqref{eq:Kubo-stat}. The shear induced conductivities ${\sigma}_\Sigma$ and $\tilde{\sigma}_\Sigma$ are examples of non-equilibrium transport coefficients. They measure the current response to a shear in the fluid velocity. For example, for a system with shear in the ($y,z$)-plane subject to a magnetic field $B = B_0 \hat{z}$, the current responds as 
$J^x \sim {\tilde\sigma}_\Sigma \left(\partial_y u_z + \partial_z u_y \right)$, see Fig.~\ref{fig:parityViolatingTransport} and 
$J^y \sim {\sigma}_\Sigma \left(\partial_y u_z + \partial_z u_y \right)$. Note that the $\tilde{\sigma}_\Sigma$ response is orthogonal to the perturbation, and constitutes a Hall-like response of the current to the shear, which turns out to be dissipationless. Non-equilibrium transport coefficients require time dependent perturbations to compute or measure experimentally, see for example the hydrodynamic Kubo formulae in Eq.~\eqref{eq:Kubo}. For a detailed discussion of all the transport coefficients, see section 2.5 of~\cite{Ammon:2020rvg}.\footnote{A relation between the conformal anomaly and the Nernst effect was first observed in~\cite{Chernodub:2017jcp} and shown generally using hydrodynamics in~\cite{Ammon:2020rvg}.} 
\paragraph{A holographic model displaying chiral hydrodynamic behavior.} As an explicit example that the above transport coefficients are non-zero for a microscopic theory with a chiral anomaly, a simple holographic model was used in~\cite{Ammon:2020rvg} to compute these coefficients for a charged plasma in quantum field theory with a chiral anomaly at strong coupling and in presence of a (strong) magnetic field. The model in question is a simplified version of the action~\eqref{eq:actionSt} with the vector field $V$ (and $F$), the St\"uckelberg field $\theta$ and the gauge field mass $m_s$ all set to zero. The dual theory corresponds to a sector of ${\cal N}=4$ Super-Yang-Mills (SYM) with gauge group $SU(N_c)$ at large $N_c$ and large 't Hooft coupling $\lambda$.\footnote{The regime of applicability of hydrodynamics in strong magnetic fields for these systems has recently been considered in~\cite{Cartwright:2024rus}.}  
The transport coefficients can be computed by first computing holographic two point functions numerically, and using the Kubo formulae~\eqref{eq:Kubo-stat} and \eqref{eq:Kubo}. See Fig.~\ref{fig:transport} for the plots of $M_2$, ${\sigma}_\Sigma$ and $\tilde{\sigma}_\Sigma$ in this model.\footnote{Note the different conventions for the Chern-Simons factor $\gamma$ compared to section~\ref{sub:holoCME}.}

\subsection{The chiral magnetic effect far from equilibrium~\cite{Cartwright:2021maz,Grieninger:2023myf}}
\label{sub:holoCME}
\noindent\textbf{Holographic CME model.} The holographic model used in~\cite{Cartwright:2021maz,Grieninger:2023myf}, as far as we know originally due to \cite{Jimenez-Alba:2014iia}, is given by
\begin{equation}\label{eq:actionSt}
     \!S\!=\!\dfrac{1}{2 \kappa_5^2}\! \int\limits_{\mathcal{M}}\!\! \dd^5x \sqrt{-g}\left[R+\dfrac{12}{L^2}\!-\!\dfrac{1}{4}(F^2\!+\!F_{(5)}^2)\! -\! \dfrac{m_s^2}{2}(A_{\mu}-\partial_{\mu}\theta)^2\!
    +\! \dfrac{\alpha}{3} \epsilon^{\mu\nu\rho\sigma\tau}\! (A_\mu\!-\!\partial_{\mu}\theta)\left(3 F_{\nu\rho}F_{\sigma\tau} + F_{\nu\rho}^{(5)}F_{\sigma\tau}^{(5)}\right)\right] + S_\text{GHY}+S_\text{ct},
\end{equation}
\noindent
where $S_\text{GHY}$ is the Gibbons-Hawking-York boundary term to make the variational problem well defined, $L$ is the anti-de Sitter (AdS) radius, $\kappa_5^2$ is the Newton constant, $\alpha$ the Chern-Simons coupling and $m_s$ the mass of the gauge field. The Levi-Civita tensor is defined as $\epsilon^{\mu \nu \rho \sigma \tau}=\epsilon(\mu \nu \rho \sigma \tau)/\sqrt{-g}\,$. The St\"uckelberg field is denoted as $\theta$ whereas the field strengths are defined as $F=dV$ and $F_{(5)}=dA\,$. $\theta(x^{\mu})$ couples the operator $\text{Tr}\{G\wedge \tilde{G}\}$ ($G$ is the gluon field strength) thus playing the role of the $\theta$ angle~\cite{Gursoy:2014ela}. Note that the gluon field strength does not appear explicitly but is mediated through $\theta$. Moreover, the axial gauge field couples to $\theta$ through the mass term and hence the dual axial current is non-conserved due to the non-Abelian anomaly. Finally, the axial gauge field and the vector gauge field are coupled through the Chern-Simons term accounting for the Abelian anomaly. \\

\noindent\textbf{CME in holographic Bjorken flow.} 
In both~\cite{Cartwright:2021maz, Grieninger:2023myf} time dependent solutions for the metric, the two gauge fields and the St\"uckelberg field were obtained working with generalized Eddington-Finkelstein coordinates and a metric ansatz consistent with having the dual plasma expanding homogeneously and uniformly in a direction perpendicular to the magnetic field. 
This produces a vector magnetic field whose magnitude is given by $B^1=\frac{B}{ \tau}$ as seen by a co-moving observer at the boundary. 
In addition, the gauge field ansatz allows for choosing nonzero axial charge density and for a resulting vector current (CME current) along the magnetic field.

Solving the Einstein-Maxwell equations of motion arising from~\eqref{eq:actionSt}, one obtains  
the expectation value of the energy-momentum tensor, the vector and axial currents via the holographic correspondence~\cite{Cartwright:2021maz,Grieninger:2023myf} 
\begin{equation}  \kappa_5^2\braket{T_{ab}}=\text{diag}\left(\epsilon(\tau), P_1(\tau), P_2(\tau), P_3(\tau)\right)\, , \quad 2\kappa_5^2\braket{J^a_5}=(Q_5(\tau),0,0,0)\, , \quad 2\kappa_5^2\braket{J^a}=(0,J_{CME}(\tau),0,0) \, .
\end{equation}
From the Ward identity $\partial_a\braket{T^{ab}}=0$, the evolution equation for the energy density follows 
\begin{equation}
 \partial_\tau \epsilon(\tau)-\frac{P_1(\tau)}{\tau }-\frac{P_2(\tau)}{\tau }-\frac{B^2}{8 \tau^3 } +\frac{2 \epsilon (\tau )}{\tau }=0 \, .\label{eq:EQ_for_energy}
\end{equation}
When $B/\tau^3$ can be ignored and $P_1\sim P_2\sim\epsilon(\tau) /3$, this reduces to $\tau\partial_\tau \epsilon(\tau)+4/3 \epsilon(\tau)=0$ which is the evolution equation describing a conformal fluid undergoing Bjorken expansion~\cite{Bjorken:1982qr}. This asymptotic behavior of the dual fluid is highly useful for making instructive choices of the initial data, and parameters, so that the results of the calculation can be compared, as closely as possible, to collider experiments which generate the QGP (and potentially produce chiral magnetic phenomena). 
Holographic time evolution~\cite{Chesler:2008hg,Chesler:2013lia} requires an initial choice of the spatial metric components and the field $V$ on the initial time slice as well as an initial choice of: energy density, axial charge density and magnetic field~\cite{Cartwright:2021maz,Grieninger:2023myf}. In addition, one must fix the gravitational coupling $\kappa_5$ and Chern-Simons coupling $\alpha$.
Fixing the initial energy density, axial charge density {etc.}~is typically done by ensuring that the evolution of these flow through values associated with those used in hydrodynamic modeling of the QGP. For instance, for the energy density, initial data is chosen by appealing to the asymptotic behavior of Eq.~\eqref{eq:EQ_for_energy}, while for the Chern-Simons coupling $\alpha$, one matches the value of the anomaly coefficient of three-flavor QCD. 
For further information and different matching choices, we refer the interested reader to the detailed discussions given in~\cite{Cartwright:2021maz,Grieninger:2023myf}. 
\begin{figure}[htbp]
\centering
    \includegraphics[width=0.48\textwidth]{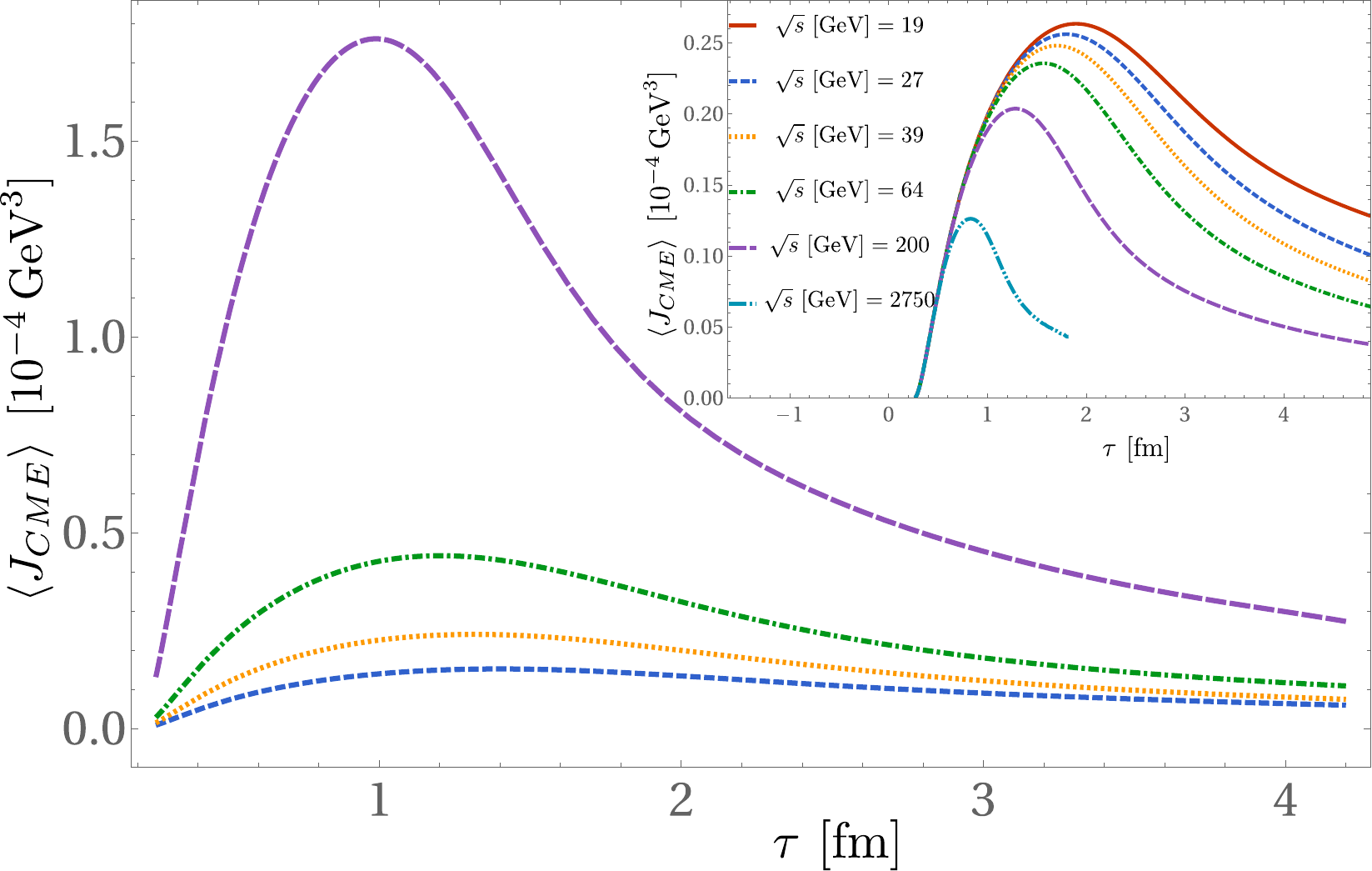} \hfill
    \includegraphics[width=0.48\textwidth]{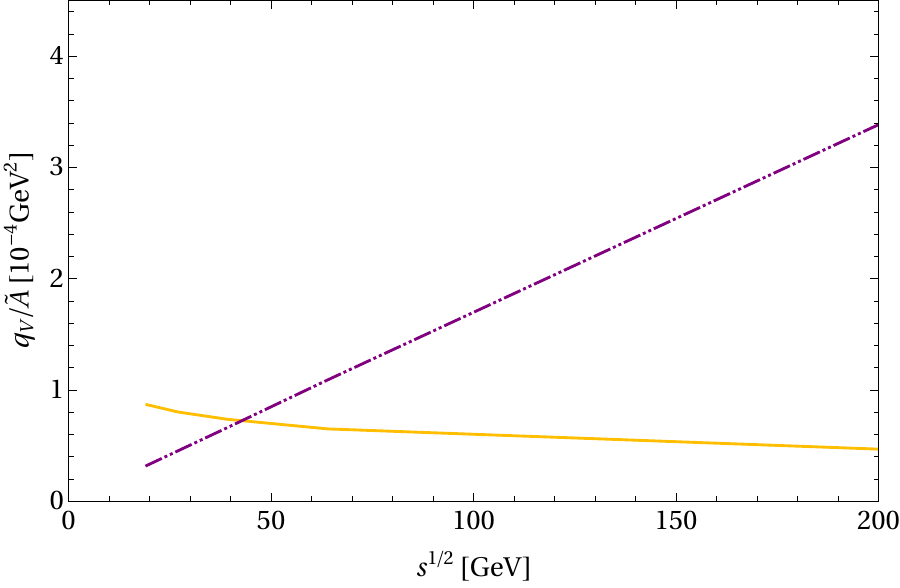} 
   \caption{
   \textit{Left:} The holographic CME current $\vev{J_{CME}}$ along the magnetic field is displayed for different beam energies (initial energy densities). The inset image displays the same, but for initial data which holds fixed the choice of the initial axial charge density $\vev{J^0_{(5)}}= 0.00032 \,\si{GeV}^3$, and the initial magnetic field $e B\approx m_\pi^2$. \textit{Right:} The time integrated CME current, yielding the accumulated charge as a result of the CME current. The purple dashed line corresponds to the main figure on the left, the yellow line corresponds to the inset. 
   \label{fig:vec_current}} 
\end{figure}
%

\noindent\textbf{Physical questions.} With a holographic setup capable of mimicking basic aspects of a heavy-ion collision (HIC), the goals of~\cite{Cartwright:2021maz,Grieninger:2023myf} are stated as: ``Given reasonable assumptions on the initial conditions, what is the magnitude and evolution of the CME current during a holographic HIC? How does the collision energy alter the CME current? Does enough CME signal build up if one incorporates the fact that chiral charge is generated and dissipates dynamically via topological processes?'' \\

\noindent\textbf{Holographic CME results.} 
The results of these calculations are best separated by $m_s=0$~\cite{Cartwright:2021maz} and $m_s\neq 0$~\cite{Grieninger:2023myf}. Beginning with $m_s=0$, a sample of the CME currents obtained from the numerical solutions is displayed in the left image of Fig.~\ref{fig:vec_current}. Both the image and inset image show the CME current generated during the evolution initially grow but quickly begin to decrease as the axial charge is diluted by the expansion of the medium.  The inset image shows the result of choosing the same initial value for the magnetic field and the same initial axial charge density for each choice of initial energy density. 
The main image displays the CME current generated when allowing for the initial values of the axial charge density and magnetic field to depend on the initial energy density. 

To better characterize the size of the CME response, consider the time integrated response, yielding the charge accumulated per area $q_V/\tilde{A}$, displayed as a function of beam energy in the right image of Fig.~\ref{fig:vec_current}. The yellow line corresponds to CME current in the inset image of the left figure in Fig.~\ref{fig:vec_current} (yellow line), 
the purple dashed line is the result for the left main image of Fig.~\ref{fig:vec_current} (purple dashed line). 
Interestingly these two curves in Fig.~\ref{fig:vec_current} (right), have opposite behavior as a function of beam energy, $s^{1/2}$. The more realistic choice of allowing for the energy dependence of the axial charge and magnetic field (purple curve) results in a larger CME signal at larger beam energies. 
\begin{figure*}
    \centering
  \includegraphics[scale=0.42]{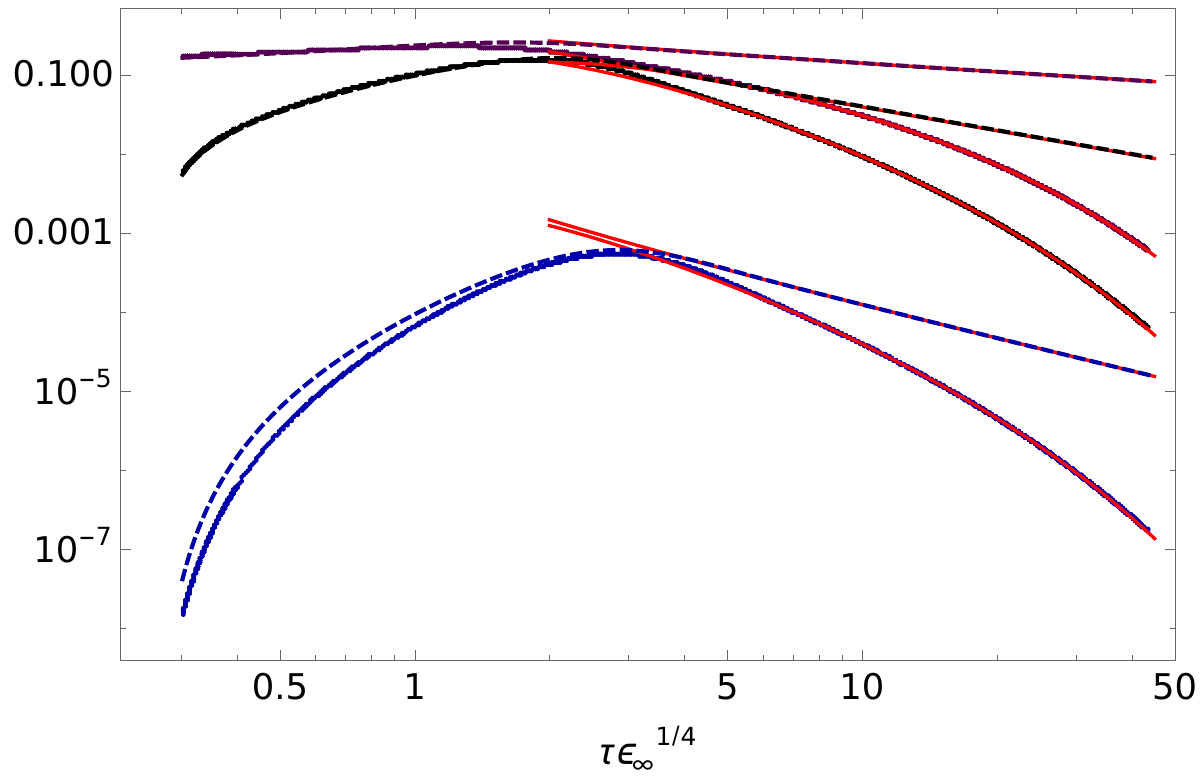}\hfill\includegraphics[scale=0.46]{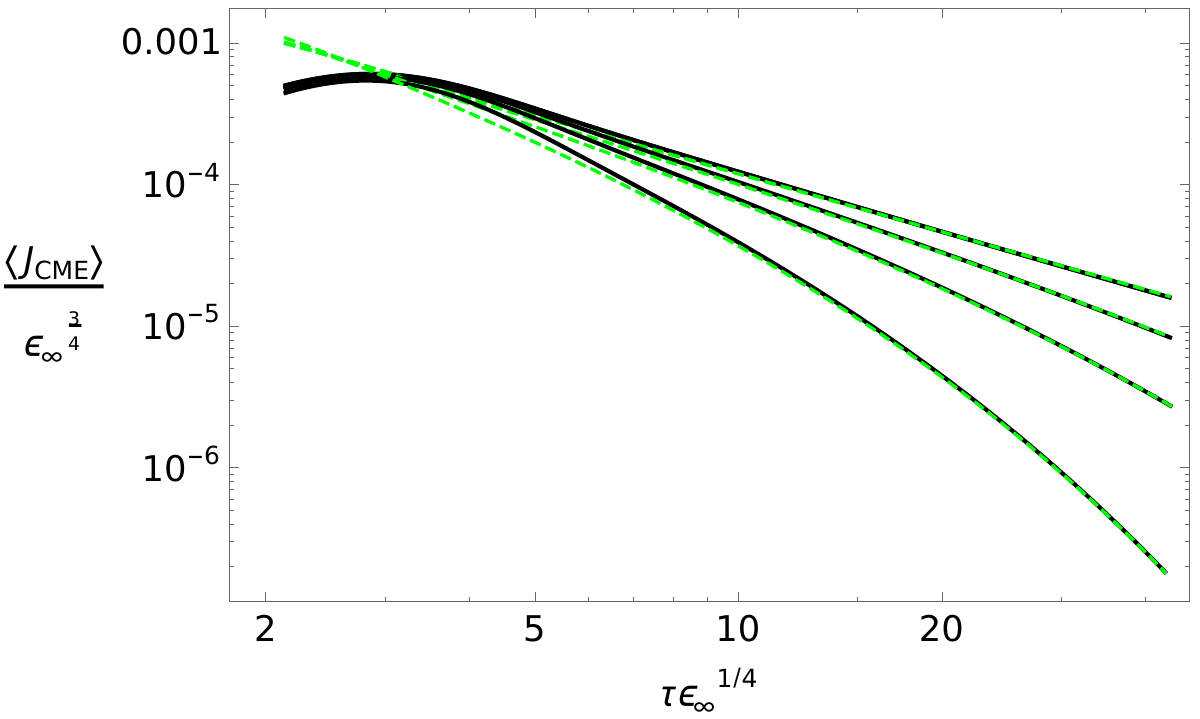}
   \caption{\textit{Left:} $\langle J_\text{CME}\rangle/\epsilon_\infty^{3/4}$ (blue), $n_5/e/\epsilon_\infty^{3/4+\Delta/4}$ (black) and $A_v(1)$ (purple). The dashed lines correspond to $\Delta=1.25\cdot 10^{-7}$ ($m_s\approx 0)$ and the solid lines to $\Delta=0.3$ ($m_s\not= 0)$. The red lines are fits outlined in the main text. \textit{Right:} Double logarithmic plot showing the late time behavior of the chiral magnetic current for different values of $\Delta$ (black lines). The green dashed lines are formula \eqref{eq:horizonexpression} for the corresponding values. The plots are for the $\sqrt{s}=200\,\si{\GeV}$ initial conditions (taken from \cite{Grieninger:2023myf}).  \label{fig:exp4}}
\end{figure*}
%
In the case of $m_s\not=0$, the model incorporates axial charge generation and relaxation by means of the non-Abelian anomaly thus incorporating the effect of topological dynamics. This has the advantage that axial charge does not have to be artificially introduced but rather is generated dynamically. 
The fact that the axial current is non-conserved for $m_s\not =0$ is reflected in the fact that it acquires an anomalous dimension (like the singlet $U(1)_A$ current in QCD)  $\text{dim($ \langle J_5\rangle$)}=3+\Delta$, where $\Delta \equiv -1+\sqrt{1+m_s^2}$.
In the expanding non-Abelian plasma, the Chern-Simons diffusion rate and hence the axial charge relaxation rate are time dependent due to the falloff of the magnetic field with time ($B\sim 1/\tau$). For small values (or for $\alpha=0$) of the Abelian anomaly, the axial charge relaxation rate increases for increasing the strength of the magnetic field. If the Abelian anomaly is strong enough, the opposite is true and the axial charge relaxation rate decreases~\cite{Grieninger:2023wuq}. The specific value of the Abelian anomaly considered in~\cite{Grieninger:2023myf}, lies in the latter regime and hence axial charge relaxation is accelerated towards later times. The modified relaxation dynamics also impacts the chiral magnetic current which relies on the axial charge and hence is decaying faster. Overall, we find that for $m_s\not =0$, the chiral magnetic current decays faster and reaches lower peak values compared to the left figure in Fig.~\ref{fig:vec_current} i.e. when $m_s=0$. The falloff follows $\langle J_\text{CME}\rangle/\epsilon_\infty^{3/4}\sim A e^{B\tau+C/\tau}\,\tau^D$ where the exponents $B,C,D$ are monotonically increasing functions of $m_s$, see left side in Fig.~\ref{fig:exp4}. Note that for $m_s=0$, the late time decay follows $\langle J_\text{CME}\rangle\sim 1/\tau^{4/3}$. For details and values see \cite{Grieninger:2023wuq}. Remarkably, the late time behavior of the chiral magnetic current may be captured in terms of the horizon value of the temporal component of the axial gauge field $A_v(\tau,1)$, the magnetic field, the strength of the Abelian anomaly $\alpha$ and the strength of the non-Abelian anomaly $m_s$
\begin{equation}\label{eq:horizonexpression}
    \langle J_\text{CME}\rangle=\frac{\alpha}{3(1-\Delta)}A_v(\tau,1)\,B(\tau),
\end{equation}
where $A_v(\tau,1)$ mimicks the role of an axial chemical potential in our system out-of-equilibrium with explicitly broken $U(1)_A$ symmetry. The excellent agreement can be seen in the right side of Fig.~\ref{fig:exp4}. Note the structural similarity of~\eqref{eq:horizonexpression} to the hydrodynamic formula $J_\mathrm{CME}\sim \xi_B \, B \sim C \, \mu \, B $, see Eq.~\eqref{eq:const-anom} and Eq.~\eqref{eq:xis}. 

\subsection{Discussion and Conclusions}\label{sub:holoHydroDiscussion}
%
\noindent\textbf{Chiral hydrodynamics in strong magnetic fields.} 
The first line of major progress reviewed in Sec.~\ref{sub:hydroTransport} stems from developing hydrodynamics with magnetic fields as an effective field theory:  
\emph{In strong magnetic fields, a large number of novel transport coefficients and susceptibilities can arise}~\cite{Ammon:2020rvg,Kovtun:2016lfw,Hernandez:2017mch}, see also Fig.~\ref{fig:parityViolatingTransport}. 
The relevance of these novel terms in QCD plasma has to be estimated. If they turn out to be relevant, they will influence for example the time evolution of elliptic flow (since one of the shear viscosities divided by entropy density can vanish), the time evolution of the magnetic field itself, and the detectability of the chiral magnetic effect. Consequently, relevant effects should be included in hydrodynamic codes used for the analysis of heavy-ion collision data or for the simulation of neutron star evolution, especially for magnetars. 
Already at weak magnetic fields, the chiral magnetic effect is sufficient~\cite{Kaminski:2014jda} to account for the observed~\cite{Chatterjee:2005mj} neutron star kicks accelerating proto neutron stars to approximately 1,000 km per second. In strong magnetic field hydrodynamics, this kick mechanism may be amplified. 
In addition to what was discussed here, there are several noteworthy transport coefficients, such as novel conductivities, novel Hall conductivities, novel Hall viscosities, various bulk viscosities, the Nernst effect, and novel susceptibilities, see section 2.5 of~\cite{Ammon:2020rvg} for a detailed interpretation.

\noindent\textbf{Relations to lattice QCD.} 
Utilizing the chiral hydrodynamics constructed in~\cite{Ammon:2020rvg}, it may be possible to compute the eight thermodynamic coefficients, including novel susceptibilities such as $M_2$, in lattice QCD with magnetic fields~\cite{Bali:2011qj,Bali:2012zg,Bali:2014kia}. 
Comparison of lattice QCD data in various strong magnetic fields suggests that the ratio of transverse to longitudinal pressure as function of $T^2/B$ matches that of $\mathcal{N}=4$ SYM from fairly small to large values of $T^2/B$~\cite{Endrodi:2018ikq}. This motivates that holographic models provide realistic estimates for QCD in strong magnetic fields, especially in the limit $T^2\gg B$ considered in this contribution. Holographic results can be compared to lattice results, see for example~\cite{Endrodi:2018ikq,Buividovich:2016ulp}.

\noindent\textbf{Chiral magnetohydrodynamics.} 
The hydrodynamic section~\ref{sub:hydroTransport} considered external (non-dynamical) magnetic fields. An extension of the presented framework to include dynamical magnetic fields into magnetohydrodynamics (MHD) was demonstrated for non-chiral hydrodynamics~\cite{Hernandez:2017mch,Grozdanov:2016tdf}. A subset of the possible terms of chiral MHD was derived in~\cite{Landry:2022nog}. A full derivation of all possible terms in chiral MHD is an open problem, but straightforward with the methods outlined in this contribution. 
Correspondingly, on the gravity side of holographic duals, the standard boundary conditions can be modified in order to obtain dynamical gauge fields in the boundary gauge theory~\cite{Marolf:2006nd,Grozdanov:2017kyl,Ahn:2022azl}. This technique allows to holographically model the interaction of dynamical electric and magnetic fields with charged plasma, which should be applied to model the generation and evolution of electric and magnetic fields in HICs. This will help estimate the CME in HICs, which is one goal of the \emph{AdS4CME Collaboration}~\cite{AdS4CME}.    

\noindent\textbf{Extending chiral hydrodynamics to vector and axial symmetry.} 
The chiral hydrodynamic framework discussed in section~\ref{sub:hydroTransport} only involves one single $U(1)$, which is broken by an anomaly~\cite{Ammon:2020rvg}. This suffices to tell what type of transport effects can appear in either a conserved vector or an anomalous axial current. In order to match QCD more closely, in future work, two U(1)-symmetries, one global axial and one vector gauge symmetry, should be considered.  
Similarly, the non-Abelian anomaly should be included in the construction of chiral hydrodynamics.

\noindent\textbf{Holographic models of the CME generation.} 
The second line of major progress reviewed in Sec.~\ref{sub:holoCME} is based on the holographic modeling of dynamically evolving plasma in search for the CME~\cite{Cartwright:2021maz,Grieninger:2023myf}:  
\emph{Collectively, the time evolution of all the considered holographic CME currents very strongly depends on the choice of initial conditions and parameter choices, see Fig.~\ref{fig:vec_current} and~\ref{fig:exp4}.}  
%
One model~\cite{Cartwright:2021maz} states that the energy dependence of the charge accumulated due to the CME current hinges on the initial conditions of the Bjorken-like flow, compare the yellow solid curve to the dashed purple curve in Fig.~\ref{fig:vec_current}. 
In the latter case, the accumulated charge due to the CME is increasing monotonically with the collision energy $s^{1/2}$, indicating that LHC would have a better chance of seeing the CME than RHIC. 
In an extension of~\cite{Cartwright:2021maz}, axial charge is dynamically generated~\cite{Grieninger:2023myf}.\footnote{ 
A holographic model recently predicted the dynamically generated spatial distribution of axial charge and its time evolution~\cite{Grieninger:2023wuq}. The spatial extent of correlations of electric currents which are responsible for the experimental manifestations of the chiral magnetic effect in heavy-ion collisions are quite large ($\sim 1\  {\rm fm}$), grow with time, and behave consistent with sphaleron dynamics. Such results should be compared to lattice QCD for the sphaleron rate in three flavor QCD~\cite{Bonanno:2023thi}. 
Axial charge profiles dynamically generated from holographic models~\cite{Grieninger:2023wuq} should be used as input for hydrodynamic codes modeling HICs.} This leads to a faster decay of the CME current and to lower peak values compared to~\cite{Cartwright:2021maz}. A fully dynamical study including dynamical magnetic field and dynamically generated axial charges is a future goal for holographic models~\cite{Grozdanov:2017kyl}.
%

\noindent\textbf{Main conclusions.} 
In conclusion, the novel hydrodynamic effects arising in extreme magnetic fields, see Sec.~\ref{sub:hydroTransport} have the potential to heavily impact various disciplines, ranging from heavy-ion collisions and neutron star physics, to the description of astrophysical plasma. 
Holographic models including magnetic fields, serve as discovery tools for new physics, such as the hydrodynamic chiral vortical effect~\cite{Erdmenger:2008rm,Banerjee:2008th,Son:2009tf}, and as testing grounds for proposed hydrodynamic effects~\cite{Ammon:2020rvg}. They can even be pushed to investigate plasma dynamics far from equilibrium, as illustrated in the CME case study~\cite{Ammon:2016fru,Cartwright:2020qov,Ghosh:2021naw,Cartwright:2021maz,Grieninger:2023myf}, see Sec.~\ref{sub:holoCME}.


	\section{Anomalous transport in low dimensions}
	\label{sec:Mizher}

	\newcommand{\ba}{\begin{eqnarray}}
\newcommand{\ea}{\end{eqnarray}}

\subsection{Introduction}

The chiral magnetic effect (CME) is a remarkable mechanism based on anomalous properties of the quark-gluon plasma (QGP), predicted almost two decades ago \cite{Fukushima:2008xe,Kharzeev:2007jp}. Due to its potential impact in different areas of physics, an intense activity - on both theoretical and experimental fronts - has been carried out aiming to achieve a robust observation of it. The proposal developed by Kharzeev and collaborators beautifully describes the microscopical theory behind it, connecting abstract concepts like the topological properties of the configurations of non-Abelian gauge fields to tangible quantities that may be manifest in the final state of the particles that reach the detectors in large accelerators colliding heavy-ions. 

As it is well-known in the community, the mechanism predicts the generation of an electric current driven by the chiral anomaly associated to a magnetic field. After hadronization of the quarks and gluons, this would result in a difference in the number of positively charged pions and negatively charged pions in opposite hemispheres. Unfortunately, it is very hard to have control over the two elements that are fundamental for the emergence of the CME in the QGP. On one hand, the topological configurations of the gauge fields are confirmed by lattice QCD, but its rate, size of domains, etc, in the actual plasma  are extremely hard to predict, being possible that the signal is so small that it will be completely washed in experiments. On the other hand, only recently the first experimental evidence of the huge magnetic field generated in non-central collisions predicted more than a decade ago was achieved. The most reliable results involve ultra-peripheric collisions, which is not a suitable environment for the CME \cite{Brandenburg:2021lnj}. More dramatic than that, it is not clear if the field lasts long enough to really affect the system. This is under investigation and a consensus has not been reached. 

Besides that the fundamental ingredients of the CME are far from being under control, the dynamics of the system involving large thermal fluctuations is very likely to interfere in the signal. All this factors have represented obstacles hard to overcome during the past decade in the run for obtaining an unambiguous signal of the CME. The community is currently tuned to new analysis on isobar collisions \cite{Kharzeev:2022hqz}, hoping to catch the evidence of this important phenomenon.

Nevertheless, the intense work accomplished during the last couple of decades has build a solid and sophisticated theoretical framework based mainly on quantum field theory to describe the CME and other transport phenomena in the QGP. A few years ago the theoretical knowledge developed in the context of the QGP gained an alternative application. It has been shown that in certain materials the interaction of the charge carriers with the underlying lattice produces quasi-particles whose dynamics is described by Dirac equation, mimicking the behavior of relativistic particles \cite{Novoselov2005}. This behavior was predicted for honeycomb lattices before the experimental discovery of Dirac materials and it was one of the first properties to be tested in graphene after its synthesis. 

This connection with relativistic fermions allowed for the search of phenomena predicted in the context of high energy physics, in Dirac/Weyl material. This is specially fruitful on what concerns transport phenomena, including the CME and other effects proposed to be observed in the QGP. A few years ago, an Abelian version of the CME effect was observed in ZrTe$_5$ \cite{Li:2014bha}. The successful setup produces the chiral anomaly by turning on colinear electric and magnetic fields, responsible for generating the necessary chiral imbalance. More details on how the chiral anomaly operates in these systems can be found in \cite{Li:2016vlc}. The core of the observation is on the quadratic dependence of the electric current on the magnetic field. This happens because the magnetic field plays a double role, inducing the chiral chemical potential and aligning the spins. Therefore, the raise of chiral imbalance in this case is controlled since the fields are external and their intensity can be tuned. This represents an important advantage on what concerns feasibility, in contrast to the CME in the QGP. On the other hand, in materials the particles suffer scattering and the chirality is not preserved, existing a relaxation time that dictates the typical scale for which the effect fades. After the first observation other groups have reported similar results in other materials \cite{PhysRevLett.53.2449,Li2015,PhysRevX.5.031023,PhysRevB.93.115414}.

All the materials where macroscopic manifestation of the chiral anomaly was observed were in (3+1) dimensions. This is expected since it is not possible to define chirality in odd dimensions. In this work we propose to extend the analogy, exploring materials in (2+1)D, but considering mechanisms based on the parity anomaly rather than the chiral anomaly. Considering a symmetry pattern totally analogous to the CME, we were able to predict a novel anomalous Hall effect in planar materials. From this point we followed two avenues. The first was searching for lattice configurations where the symmetry pattern arises naturally. The second, was checking if the symmetry breaking could be generated by interaction with external fields, in a similar fashion as the CME in  ZrTe$_5$.

\subsection{Honeycomb lattices}

\begin{figure}
\begin{center}
\includegraphics[width=5cm]{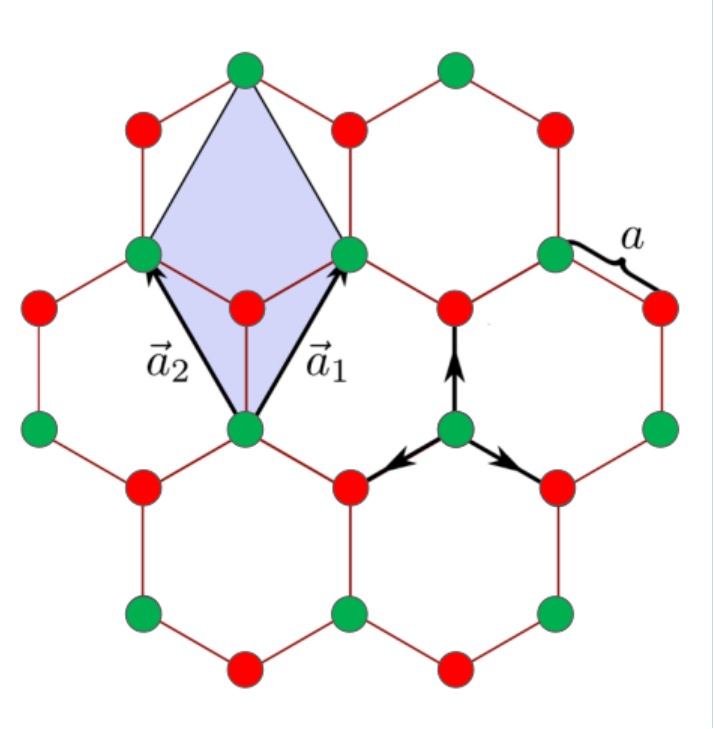}
\caption{Honeycomb lattice represented as two triangular sublattices.}
\label{fig:honeycomb}
\end{center}
\end{figure}

We aim to explore planar materials containing relativistic-like fermions. We work with honeycomb lattices since in this kind of structure it is very clear how the relativistic behavior appears. Although the pristine honeycomb does not satisfied the required conditions, we start by considering its corresponding Lagrangian and on top of that we add the necessary interventions.

A pure honeycomb lattice, like in graphene, can be represented as in Fig.\ref{fig:honeycomb}. It is usual to describe it using two triangular sublattices, represented in Fig.\ref{fig:honeycomb} in different colors. When calculating the band structure one can check that there are two inequivalent points where the energy vanishes. Those are called Dirac points. Expanding the Hamiltonian around these points, one obtains a linear dispersion relation, what is a clear signature of a relativistic-like behavior. It is also usual to consider sublattice and Dirac points as degrees of freedom of the fermionic system. If one represents the fermions in 4-component spinors, in addition to the linear dispersion relation another feature typical from relativistic systems emerges: the equation of motion is written in terms of Dirac matrices and taking the continuous limit the Lagrangian that comes out is totally analogous to the Lagrangian of massless Quantum Electrodynamics in (2+1)D (QED$_3$), \cite{Gusynin:2007ix,katsnelson},
\begin{eqnarray}
\mathcal{L}&=&\sum_s \bar{\psi}_s\left(i\gamma^0\hbar D_t +i\hbar v_F\gamma^xD_x + i\hbar v_F \gamma^y D_y \right)\psi_s.
\label{eq:lagrangian}
\end{eqnarray}
Here the index $s$ labels the (electron) spin and the covariant derivative $D_\alpha=\partial_\alpha-(ie/\hbar c) A_\alpha$. $A_\alpha$ is the gauge field associated to the electromagnetic interaction and the fields $\psi_s\equiv \psi_s(t,{\bf r}) $ denote four-component spinors that account for both Dirac point (also called valley) and sublattice (also called pseudospin) index. Note that based on the structure of Dirac matrices, the valley index is analogous to chirality quantum number in systems in (3+1)D.

\begin{figure}
\begin{center}
\includegraphics[width=4cm]{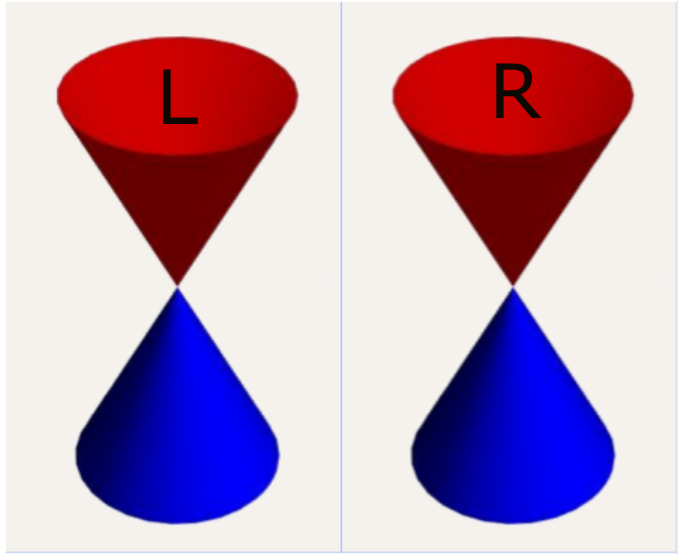}
\caption{
\label{fig:bands}
Blue cones represent valence bands and red cones represent conduction band. Honeycomb lattice generates gapless band structure, where the bands touch in one point. From the six points of vanishing energy, only two are non-equivalent. The Dirac structure that comes out naturally in the Lagrangian allows us to build an analogy between the two pairs of cones and the chirality degree of freedom. }
\end{center}
\end{figure}

To make the valley degree of freedom more transparent we rewrite our Lagrangian using the pseudo-chiral projections. We decompose the fermion field as
\begin{equation}
\psi=\psi_k + \psi_{k'}=\frac{1}{2}\left(1+ \gamma^5 \right)\psi + \frac{1}{2}\left(1- \gamma^5 \right)\psi.
\end{equation}
Doing this we can write the Lagrangian in terms of each valley separately,
\begin{eqnarray}
\mathcal{L}&=& \sum_{s,\chi=k,k'}\bar{\psi}_{\chi,s}[i\gamma^0\hbar\partial_t +i\hbar v_F\gamma^xD_x + i\hbar v_F \gamma^y D_y]\psi_{\chi,s},
\label{eq:chirallagrangian}
\end{eqnarray}
Up to this point, this description is applicable to honeycomb lattice whose lattice points are all identical, as it happens in graphene. The band structure around the Dirac points can be schematically represented in Fig.\ref{fig:bands}(a). There is no gap between the valence and conduction band, and they touch in one point. The two species, analogous to the chiralities, can be represented at this stage by two symmetric pairs of cones. 

\begin{figure}
\begin{minipage}{.45\textwidth}
\begin{center}
\includegraphics[width=4cm]{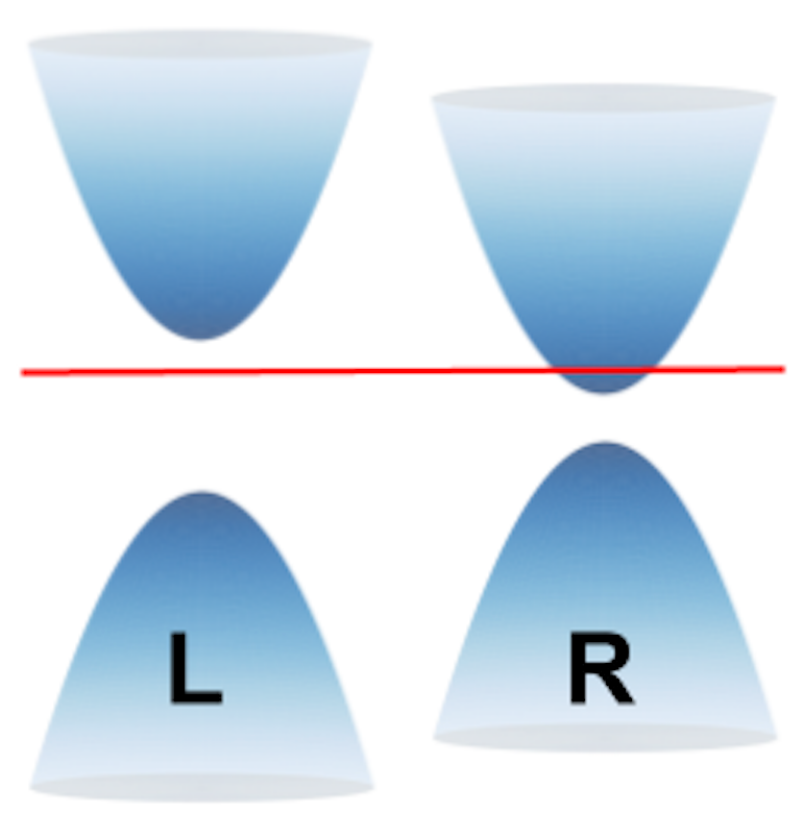}\\
(a)
\end{center}
\end{minipage}
\hspace{0.5cm}
\begin{minipage}{.45\textwidth}
\begin{center}
\includegraphics[width=4cm]{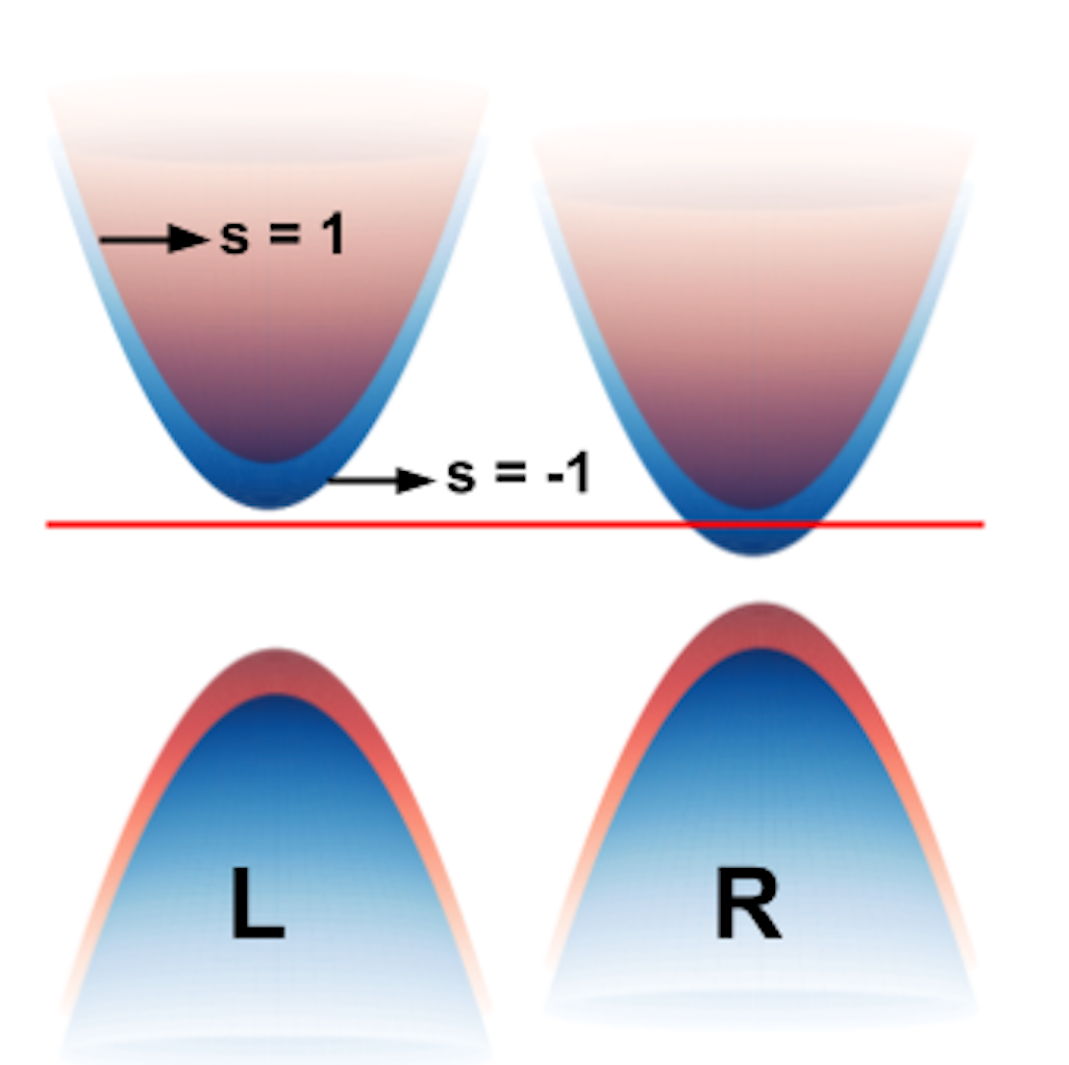}\\
(b)
\end{center}
\end{minipage}
\caption{The presence of a mass in the Lagrangian signals the opening of a gap between the valence and conducting bands. We propose a configuration where the gaps opened in each pair of cones is different, mimicking the imbalance of chirality in the CME. This situation is represented in the left panel of the figure, where we also added a chemical potential indicated by the red line, controlling the Fermi level. On the right, besides opening a gap, we also lift the spin degeneracy in order to avoid cancellation of the current of particles with opposite spin. }
\label{fig:asymmetric_cones}
\end{figure}

Following a close analogy with the CME, we want now to induce a specific symmetry breaking that will play the role of the imbalance in chirality responsible for the CME. We achieve this, exploring the connection of the valley degree of freedom and chirality of the quarks in Quantum Chromodynamics (QCD), described above. In order to mimic the imbalance in chirality responsible for inducing the CME in the QGP, we need to break the symmetry between the two components in Eq.(\ref{eq:chirallagrangian}). The way we choose to achieve this is by adding masses to the Lagrangian, where the magnitude of the masses are different for each projection in Eq.~(\ref{eq:lagrangian}). The resulting Lagrangian is \cite{Mizher:2018dtf}
\begin{eqnarray}
\mathcal{L}&=& \sum_{s,\chi=k,k'}\bar{\psi}_{\chi,s}[i\gamma^0\hbar\partial_t +\mu\gamma^0+i\hbar v_F\gamma^xD_x + i\hbar v_F \gamma^y D_y+m_{s,\chi}\gamma^z]\psi_{\chi,s},
\label{eq:chirallagrangian_proj}
\end{eqnarray}
where we also included the chemical potential, $\mu$. Without loss of generality, we may assume $m_km_k'>0$. This is the final Lagrangian for the fermions we employ in our procedure. In Fig.\ref{fig:asymmetric_cones} we represent the Dirac cones after gap opening, considering different gaps for each specie of fermion. The red line represents the chemical potential, strategically placed in a way that only one pair of cones are contributing to the current. This point will be clarified in a moments. 

\vspace{0.8cm}

\vspace{0.8cm}

\subsection{Reduced-QED}

In our analysis we aim to explore reactions of the system to external stimuli of electromagnetic fields. On doing this we must be careful since the electrons in the systems of interest are constrained to the plane while the electromagnetic fields are not. Neither the Quantum Electrodynamics in (3+1)D (QED$_4$) nor Quantum Electrodynamics in (2+1)D (QED$_3$) are suitable to describe it and we need to use a framework that consider systems with different dimensionality interacting. We adopt the theory known as reduced-QED (RQED) \cite{PhysRevD.64.105028} or pseudo-QED (PQED) \cite{MARINO1993551} as our framework. This is a scale invariant theory \cite{Dudal:2018pta}, where the gauge fields living in (3+1)D suffer a dimensional reduction in order to obtain their effective action on the plane where the fermions are constrained. For a comprehensive review on this theory, including formal aspects and applications, see \cite{Olivares:2021svj}.  

After the dimensional reduction, fermions in (2+1)D are added and a new effective theory completely defined in (2+1)D is obtained. The action is given by
\be
S_{eff}=-\frac{e^2}{8}\int d^3x d^3x' j^\mu (x)\frac{1}{\sqrt{-\square_E}} j^\mu (x').
\label{eq:rqed_action}
\ee
Here, since the fermions are constrained to the plane, the third space component of the current, $j^3 = 0$. It can be easily verified that the action in Eq.(\ref{eq:rqed_action}) can be obtained from the Lagrangian
\be 
\mathcal{L}_{RQED} =-\frac{1}{4}F^{\mu\nu} \left[\frac{2}{\sqrt{\square}}\right]F_{\mu\nu}-e j^\mu A_\mu+\mathcal{L}_M+\mathcal{L}_{GF},
\label{eq:Lagrangian}
\ee 
where $\mathcal{L}_M$ is the fermion Lagrangian given by Eq.(\ref{eq:chirallagrangian_proj}) and $\mathcal{L}_{GF}$ is a gauge fixing term. This is the Pseudo or Reduced-QED Lagrangian. Note that the main consequence of this modified action compared to the usual QED$_4$ is that the dependence of the photon propagator on momentum will appear as $1/q$ rather than the inverse of the square dependence obtained in QED$_4$. This softens the IR behavior of the theory.

\subsection{Anomalous Hall effect}

After defining the Lagrangian, the next step is to investigate the transport coefficients. We are particularly interested in the perpendicular conductivity, and work under the assumption that the system responds linearly to external fields, 
\be
\langle \vec j \rangle=\sigma_{xy} E \vec e_x,
\ee
where $j$ is an electric current also in-plane but perpendicular to the applied $\vec E$. The transverse current is usually associated to Hall effect and we expect it to be topologically protected, that is the motivation of our interest. Of course one could also calculate the direct current $j_{xx}$, but this would give us the usual Ohm law. Also, we are interested in the DC current, for which $\omega\rightarrow 0$, nevertheless we will use the trick to consider a time dependent field and vector potential and take the limit until the end,
\be\label{b}
\vec{E}=E e^{ - i\omega t} \vec{e}_y\,,\qquad \vec{a}=-\frac{iE}{\omega}e^{-i\omega t } \vec{e}_y.
\ee
To calculate the conductivity we apply Kubo formula of linear response theory. We follow the classical derivation described in \cite{Tong:2016kpv} to obtain
\begin{eqnarray}
\langle{\vec j(t)}\rangle= \frac{i}{\hbar}  \bra{0} \int_{-\infty}^{t} d\tau [\Delta H(\tau),\vec j(t)]\ket{0}.
\end{eqnarray}
getting the Kubo relation for the DC anomalous conductivity,
\be\label{sigma}
 \sigma_{xy} = \lim_{\omega\to0}\frac{1}{\hbar\omega}\left\{ \int_{0}^{\infty} d\tau e^{i\omega\tau}  \bra{0} [j_y(0),j_x(\tau)]\ket{0}\right\}.
\ee
The detailed development of the calculation of the conductivity can be found in \cite{Dudal:2021ret}. The current-current correlator is closely connected to the polarization tensor and can be written in terms of the latter as
\be
\sigma = \lim_{\omega\to0}\frac{1}{\hbar \omega} \Pi^{yx}. 
\ee
Kubo formula is particularly interesting in this context, since it was proved that the 1-loop contribution for the photon self energy is exact in this case and higher order contributions are identically zero \cite{Dudal:2018mms}, which means that Coleman-Hill theorem applies to PQED. Finally, calculating the polarization tensor \cite{Dudal:2021ret} the result for the conductivity is given by
\be
\sigma =-\lim_{\omega\to0} \frac{1}{\hbar\omega} \frac{m}{|m|}\frac{e^2}{4\pi} \theta(m^2-\mu^2)\varepsilon_{yx0}\omega= \frac{e^2}{4\pi\hbar}\frac{m}{|m|}\theta(m^2-\mu^2)
\ee
for each one of the two-component spinor. Note that the result is independent of the magnitud of the mass and depends only on its signal. The theta function controls the contribution from each valley, in a way that for a fixed gap, one can turn on and off the conductivity by adjusting the chemical potential. Summing up the contributions from the two valleys,
\be\label{naam}
\sigma_{xy} = \frac{e^2}{4\pi\hbar}\left(\frac{m_k}{|m_k|}\theta(m_k^2-\mu^2)-\frac{m_{k'}}{|m_{k'}|}\theta(m_{k'}^2-\mu^2)\right).
\ee
The expression above shows that if the gaps for both pair of cones is the same, the conductivity and consequently the current $j_{xy}$ vanishes. Conversely, if we are able to tune the chemical potential such that $m_k>0$, $m_{k'}>0$ and $m_{k'}^2<\mu^2< m_{k}^2$, only one specie of electrons, coming from one pair of cones, will contribute to the current and the cancellation does not occur. In addition, we verify that considering the spin as an additional degree of freedom, may also cause a cancellation. The symmetries involved in a spin flip (particularly how it reacts to time reversal symmetry) are translated in the Lagrangian as a flip of the mass signal. Therefore, if non-polarized electrons are playing the game, the conductivity vanishes again and no transverse current is generated by applying an electric field. We thus need to lift the spin degeneracy and adjust the chemical potential between the conduction bands of associated to each spin type, as showed in Fig.\ref{fig:asymmetric_cones}. 

To obtain this result, we assumed that asymmetric gaps could be produced without caring about how to generate it. Now it is time to investigate how this could be produced in real setups. We followed two avenues. The first one was looking for materials whose structure naturally yields to the desired symmetry pattern. We have mapped features and symmetries and from that we have proposed a few candidates to harbor the effect. Details and candidates can be found in \cite{Dudal:2021ret}. The other possibility we explored, was to check if interaction with external fields could lead to dynamical mass generation. This investigation is described in the next subsection.

\subsection{External fields generating anomalous transport}

As we discussed in the last couple of subsections, a gap opening in the band structure is translated in the continuous limit as a mass term for the fermions in the Lagrangian. As a possibility to obtain the mass structure required to activate the Hall effect described in the last subsection, we investigated if interaction with external fields can drive dynamical mass generation \cite{Olivares:2020eko}.  

In (3+1)D parallel electric and magnetic fields are the responsible for the chiral anomaly and flip of the chirality of some of the electrons \cite{Li:2014bha}. In our analogy, we seek for mechanisms that can break the symmetry between the pair of cones that are connected by parity transformation. We choose to verify if the same configuration can generate this asymmetry, since this combination of fields is odd under parity transformations. In terms of the gauge fields, the presence of parallel electric and magnetic fields can be encoded in a Chern-Simons term, 
\begin{equation}
\mathcal{L}_{CS}=\frac{\kappa}{2}\varepsilon^{\mu\nu\rho}A_\mu \partial_\nu A_\rho - A_\mu J^\mu.
\label{CSLagrangian}
\end{equation}
Following the line we described previously, we apply PQED in order to deal with the difference in dimensionality of fermion and gauge fields, and add the Chern-Simons term to obtain the total Lagrangian
\begin{eqnarray} 
		\mathcal{L}_{RQED}^ {CS}&=&-\frac{1}{4}F^{\mu\nu}\frac{2}{(-\Box)^{1/2}}F_{\mu\nu}+\bar{\psi}(i\gamma^\mu\partial_\mu+e\gamma^\mu A_\mu\\ \nonumber
		&+& m_e + \tau m_o)\psi
		+\frac{1}{2\zeta}(\partial\cdot A)^2 + \frac{-i\theta}{4}\varepsilon^{\mu\nu\rho}A_\mu \partial_\nu A_\rho,
		\label{massive_Lagrangian}
	\end{eqnarray}
Note that in this Lagrangian we included a specific combination of mass terms that may open a gap in the system. Specifically, we allow for the general case where an ordinary Dirac mass $m_e\bar{\psi}\psi$ can emerge. On top that, there is another mass term that potentially may be generated and is particularly interesting to us, known as Haldane mass $m_o\bar{\psi}\tau\psi$, with $\tau=[\gamma_3,\gamma_5]/2$. This term does not break chiral symmetry, but rather violates parity (and time-reversal) and because the CS is parity violating we expect a relation between them. This choice is justified because working with chiral projectors, it yields to different magnitude of mass for each chirality.

 As stated before, we are interested in dynamical mass generation driven by quantum effects. For this reason, we adopt the Schwinger-Dyson formalism to verify if the interaction with the external fields generates a mass term non-perturbatively. In this approach the bare propagators are corrected by the self-energy where all the propagators involved receive corrections, generating an infinite tower of coupled equations to be solved. In practice it is necessary to truncate the system of equations at a certain level. The corresponding equations for the two-point functions are
	\begin{eqnarray} 
		S^{-1}(p)&=&S_0^{-1}(p)-\Xi(p),\\
		\Delta^{-1}_{\mu\nu}(p)&=& \Delta^{-1}_{0\mu\nu}(p)-\Pi_{\mu\nu}(p)
		\label{ec1}
	\end{eqnarray}
The bare photon propagator derived from the Lagrangian Eq.~(\ref{massive_Lagrangian}) can be written as \cite{Dudal:2018mms}:
\begin{eqnarray}\label{photon_prop_tree}\nonumber
\hat \Delta_{\mu\nu}(\vec{q})&=&\frac{1}{2q}\frac{1}{(1+\theta^2)}\left(\delta_{\mu\nu}
-\frac{q_\mu q_\nu}{q^2}\right)\\&&+\frac{\zeta}{q^2}\frac{q_\mu q_\nu}{q^2}-\frac{1}{2q^2}\frac{\theta}{(1+\theta^2)}\epsilon_{\mu\nu\rho}q^{\rho}
 \,.
 \label{photon_prop}
\end{eqnarray}
where the influence of the Chern-Simons term is quantified by the parameter $\theta$. Note that in the limit $\theta \rightarrow 0$, the photon propagator recover the usual form within PQED. To account for non-perturbative corrections, we assume a general form for the fermion propagator:
\begin{equation}
    S_F^{-1}(p)=A_e(p)\slashed{p} + A_o(p)\tau \slashed{p} - B_e(p) -B_o(p) \tau,
\end{equation}
and for the sake of simplicity, we adopt the raibow-ladder approximation, where a bare photon-fermion vertex is considered. A study of more a sophisticated vertex applied to PQED can be found in \cite{Albino:2022efn}. Given that, we start from the gap equation 
	\begin{equation}
		S^{-1}_\pm(p)=S^{-1}_{0\pm}(p)+4\pi\alpha\int\frac{d^3k}{(2\pi)^3}\gamma^\mu S_{\pm}(k)\gamma^\nu \Delta_{\mu\nu}(q),
		\label{projected_prop}
	\end{equation}
where
	\begin{eqnarray}
		S^{-1}_\pm(p)&=&(\slashed{p}+M_\pm(p))\chi_{\pm}\nonumber\\
		S_0^{-1}(p)&=&\slashed{p}\chi_{\pm}\nonumber\\
		S_{\pm}(k)&=&-\frac{\slashed{k}+M_{\pm}(k)}{k^2+M^2_{\pm}(k)}\chi_{\pm}.
			\end{eqnarray}
Here $M_+=m_e+m_o$, $M_-=m_e-m_o$ and $\Delta_{\mu\nu}$ is the photon propagator given by Eq.~(\ref{photon_prop}). 
The details of this calculation can be followed in \cite{Olivares:2020eko} and the solution within our approach is given by:
			 \begin{eqnarray}
			 	M_\pm(p)&=&p^{-\frac{1}{2}\sqrt{1-\frac{\alpha}{\alpha_c}}-\frac{1}{2}} \left(c_2 p^{\sqrt{1-\frac{\alpha}{\alpha_c}}}+c_1\right)\nonumber\\
			 	&+&f(\theta, p)\nonumber\\
			 	&=& \Lambda e^{-\pi/\sqrt{\alpha/\alpha_c-1}}+f(\theta,p),
			 	\label{SD_solution}
			 	 \end{eqnarray}
where 
\begin{equation}
\alpha_c=\frac{\pi}{8}(1+\theta^2)\label{crit}
\end{equation}
and
\begin{eqnarray} \nonumber
f(\theta, p)&=&\Bigg[\pi  \left(\theta ^2+1\right) \left(\mp \frac{2\alpha  \theta }{\theta ^2+1}\right) \bigg(\kappa \left(\alpha +3 \pi  \left(\theta ^2+1\right)\right)\\ \nonumber
&&\left(\Lambda ^3+14 p^3\right)-54 p^4 \left(\alpha + \pi  \left(\theta ^2+1\right)\right)\bigg)\Bigg]\\
&\times&\frac{1}{18 \kappa p^2 \left(\alpha + \pi  \left(\theta ^2+1\right)\right) \left(\alpha +3 \pi  \left(\theta ^2+1\right)\right)}.
\label{f}
\end{eqnarray}

\begin{figure}
\begin{minipage}{.45\textwidth}
\begin{center}
\includegraphics[width=6cm]{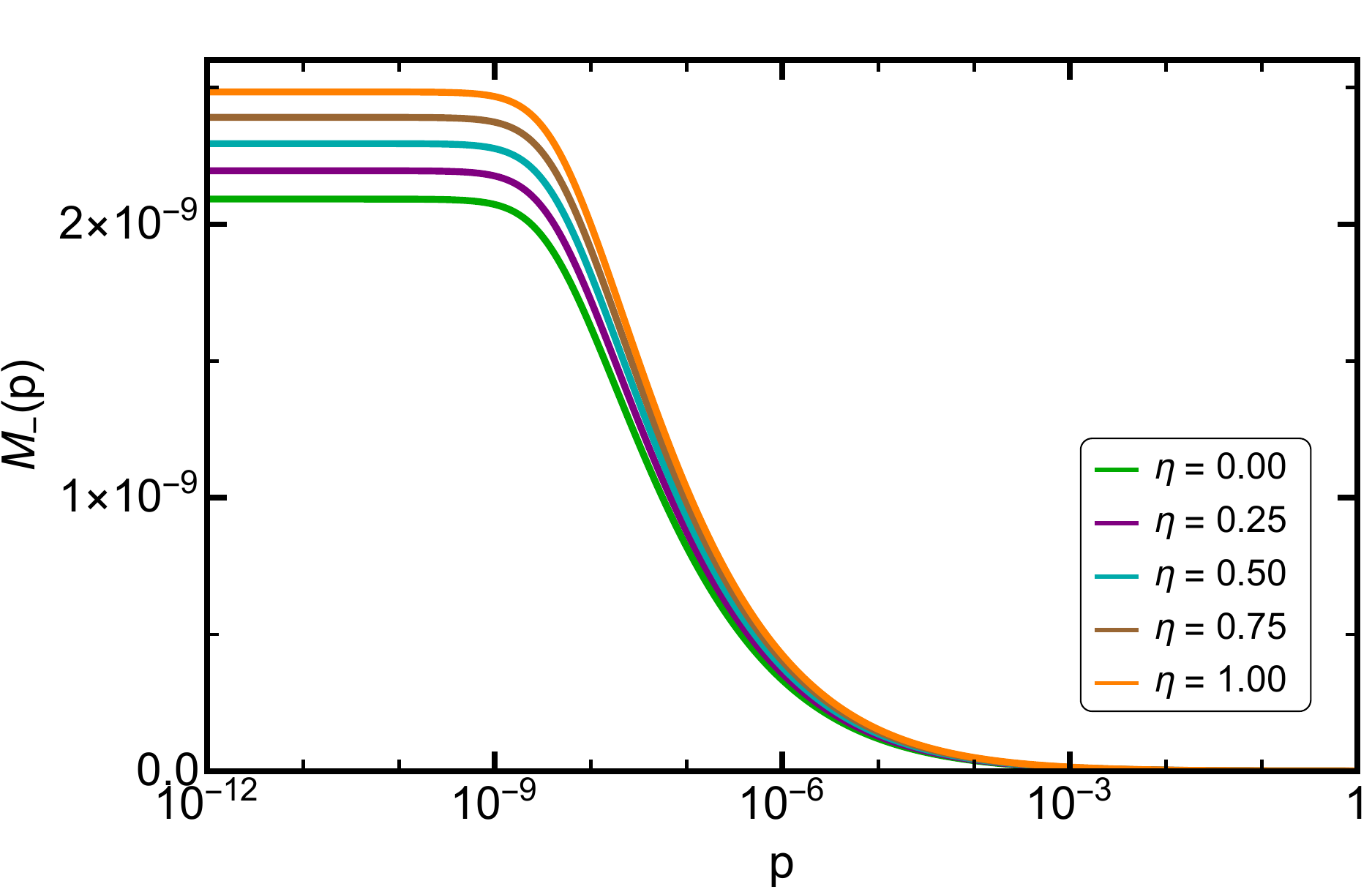}
\end{center}
\end{minipage}
\hspace{0.5cm}
\begin{minipage}{.45\textwidth}
\begin{center}
\includegraphics[width=6cm]{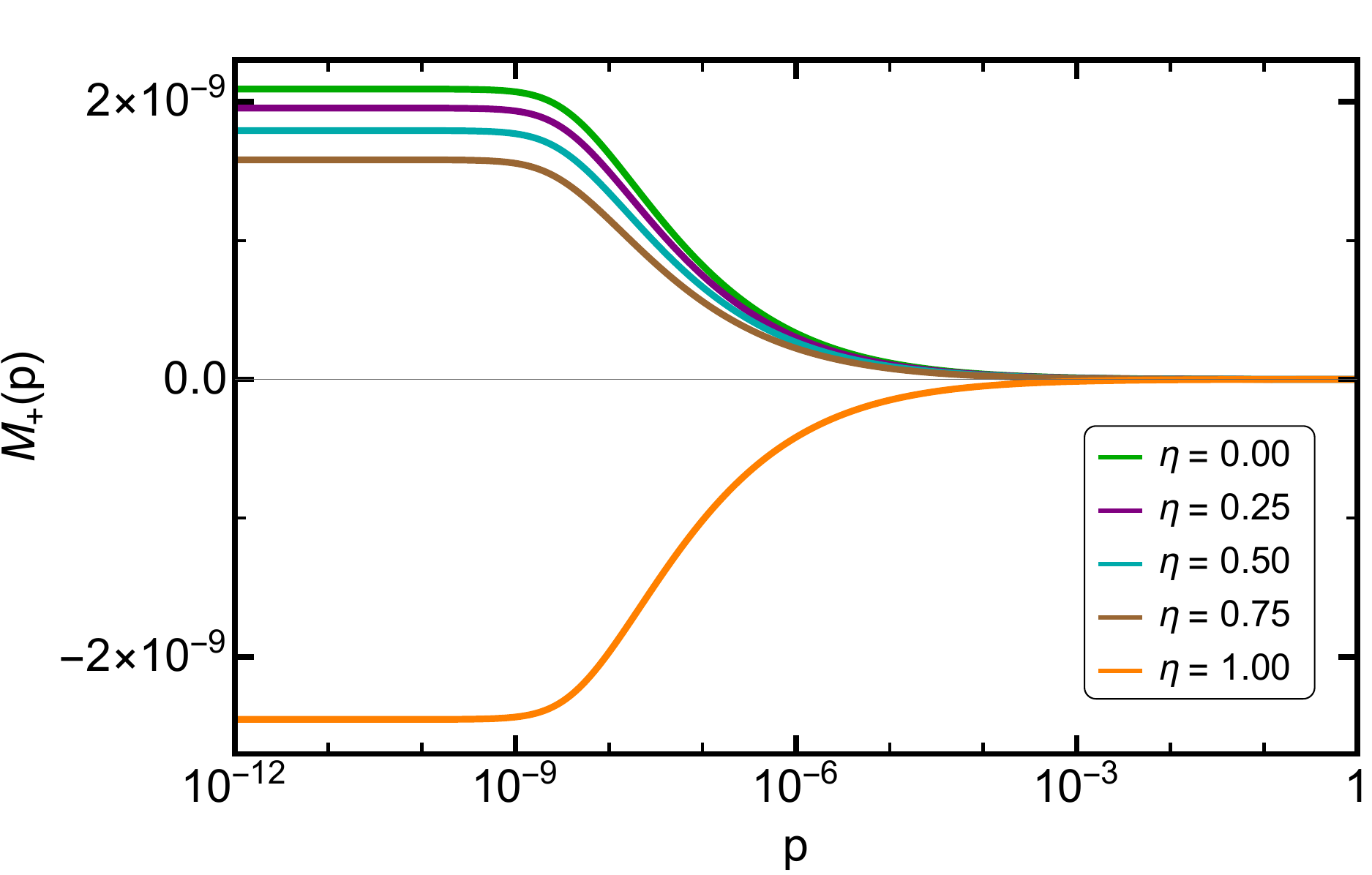}
\end{center}
\end{minipage}
\caption{M$_+$ and M$_-$ extracted from the gap equation as a function of external momentum for different values of the parameter $\theta$. Here $\theta=\eta \theta_c$ with $0\leq\eta\leq1$ and fixed value of the coupling $\alpha=1.07 \alpha_c$}
\label{fig:Mp_Mm}
\end{figure}

\begin{figure}
\begin{center}
\includegraphics[width=6cm]{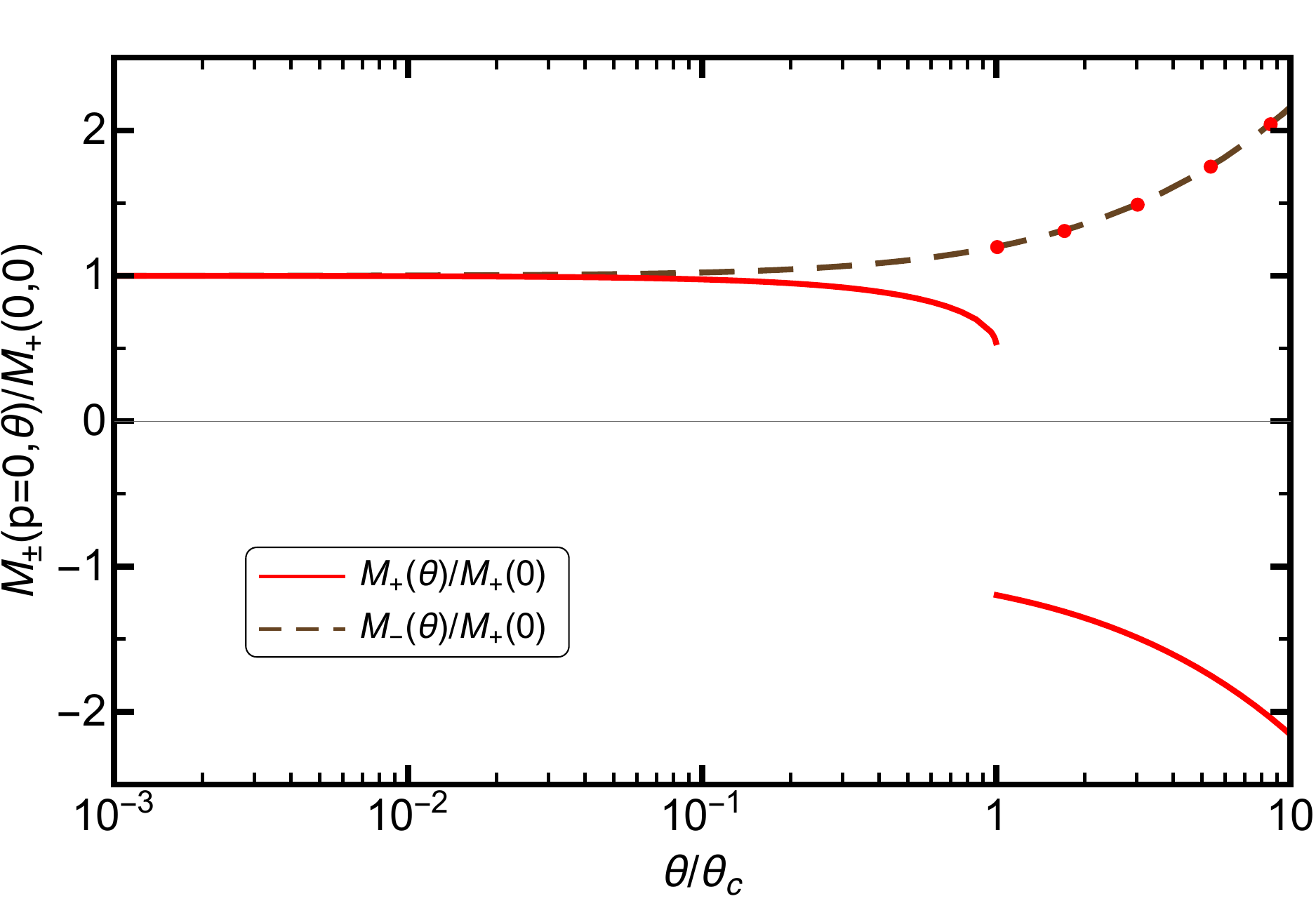}
\caption{$M_\pm$ re-scaled by their $\theta=0$ value. The full line corresponds to $M_+$ and dashed line $M_-$. The red dots are the absolute value of $M_+$ above the critical $\theta$.}
\label{fig:mass}
\end{center}
\end{figure}

The first term in Eq.~(\ref{SD_solution}) presents the so called Miransky scaling and corresponds to the solution in the absence of the Chern-Simons term. As usual, one can notice from Eq.~(\ref{SD_solution}) that there is a critical value of the coupling above which the solutions exists. It means that provided the coupling has a suitable value, both the Dirac and Haldane masses can be generated by interaction with external field in the configuration considered here. The solutions of Eq.~(\ref{SD_solution}) are represented in Fig.\ref{fig:Mp_Mm} as a function of external momentum for different values of the parameter $\theta$. In Fig.\ref{fig:mass} the values of $M_+$ and $M_-$ are represented in the static limit. We can clearly see that from a certain critical value of $\theta$ the masses associated to different chiralities assume different values, indicating non-vanishing value of Dirac and Haldane mass. 

\subsection{Summary}

Here we summarize results connecting the chiral magnetic effect and analogues in condensed matter in low dimensions. Although the CME itself is forbidden in odd dimensions, we show that a close analogy is possible exploring the parity anomaly rather than the chiral anomaly to produce a conserved current. We explore the well-known fact that the valley degree of freedom in honeycomb lattices corresponds to chirality on what concerns their group theory and propose to open different gaps in the two pair of Dirac cones as a way to mimic the imbalance of chirality that is a fundamental element in the CME in the quark-gluon plasma. The result is a topologically protected current that resembles a valley-Hall effect. We then proceeded to investigate how to physically generate the gaps we need and follow two lines. The first is to perform ab initio simulations of electronic structure to identify candidates that may intrinsically harbor the effect. The second track we follow is that the gaps may be open by interaction with external fields. To investigate non-perturbative effects, we calculated the Schwinger-Dyson equations for the system and showed that for a given critical coupling, two types of mass terms emerge in the Lagrangian. These combination is precisely what is needed to achieve the gaps required. Along all the procedure we adopt the framework of pseudo-QED, that accounts for gauge fields in (3+1)D interacting with fermions constrained to move in planar space. 

	\section{Magnetic fields, nonzero densities and QCD on the lattice}
	\label{sec:Kotov}
	
	\subsection{Introduction}

Basic ingredients for chiral effects are in general some external field and density of some charge. The most renowned example, chiral magnetic effect \cite{Fukushima:2008xe}, appears in systems with external magnetic field and chiral density. External magnetic field and nonzero baryon density are expected to lead to an axial current, it is an essense of the Chiral Separation Effect (CSE)\cite{Son:2004tq,Metlitski:2005pr}.

It is also expected that large magnetic fields and densities can be created in peripheral heavy ion collisions \cite{Kharzeev:2007jp,Skokov:2009qp}, possibly leading to the emergence of chiral effects. For this reason in the current section we would like to discuss in general the properties of QCD in external magnetic field with nonzero baryon density with a particular attention to the bulk properties, including the phase diagram and the Equation State.

QCD thermodynamics is described by several conserved charges, including baryon number $B$, electric charge $Q$ and strangeness $S$, which in grand canonical description are coupled to the corresponding chemical potentials $\mu_B$, $\mu_Q$, $\mu_S$. In the lattice study one typically works with chemical potentials $\mu_u$, $\mu_d$, $\mu_s$ coupled to the density of $u$, $d$ and $s$ quarks. Formulae connecting the chemical potentials in two different bases read:
\begin{eqnarray}
    \mu_u=&\mu_B/3+2\mu_Q/3\quad&\mu_B=\mu_u+2\mu_d\nonumber\\
    \mu_d=&\mu_B/3-\mu_Q/3\quad&\mu_Q=\mu_u-\mu_d\\
    \mu_s=&\mu_B/3-\mu_Q/3-\mu_S\quad&\mu_S=\mu_d-\mu_s\nonumber
\end{eqnarray}

In general one could study the properties of QCD as a function of all three chemical potentials $\mu_B$, $\mu_Q$, $\mu_S$. However, in this section we mainly discuss and present the results for the case $\mu_s=0,\mu_u=\mu_d=\mu_B/3=\mu$, which approximately corresponds to the zero strangeness $\langle S \rangle\approx0$ and might approximate the conditions realized in heavy ion collisions. More detailed tuning of the chemical potentials to meet the conditions created in heavy ion collisions (given by zero strangeness $S=0$ and electric-to-baryon charge ratio $Q/B\approx0.4$) can be found in \cite{Guenther:2017hnx} (without magnetic field) and in \cite{MarquesValois:2023ehu} (with magnetic field).

\subsection{Phase diagram}

First we would like to summarize our knowledge about the QCD phase diagram as a function of three variables: temperature, external magnetic field and density.

At zero magnetic field and vanishing baryon density the phase with the broken chiral symmetry at low temperatures and the chirally restored phase are known to be separated by a wide crossover \cite{Aoki:2006we}.

The phase diagram of QCD in the plane temperature - magnetic field was studied extensively on the lattice \cite{Bali:2011qj,DElia:2010abb,Ilgenfritz:2013ara,Ding:2020inp,Endrodi:2015oba,DElia:2021yvk}. It is currently known that nonzero magnetic field decreases the pseudo-critical temperature of the chiral phase transition - the phenomenon is called inverse magnetic catalysis effect. The transition becomes sharper at larger magnetic field, possibly becoming a first-order phase transition at very large magnetic fields $eB\sim5-10$~GeV$^{2}$ \cite{Endrodi:2015oba,DElia:2021yvk}. 

The phase diagram of QCD in the plane baryon density (or chemical potential) - temperature has very important phenomenological meaning and is under active study by multiple groups (for recent reviews see, e.g., \cite{Guenther:2022wcr,Aarts:2023vsf}). The main difficulty for lattice studies of the phase diagram with nonzero baryon density is the sign problem \cite{Muroya:2003qs}. Standard approaches to overcome this problem include analytical continuation from imaginary $\mu_B$ or Taylor expansion around $\mu_B=0$. Both these methods are based on the fact that the chiral phase transition at zero and small values of $\mu_B$ is a crossover,
thus for not very large values of the baryon chemical potential physical observables are analytic functions of the system parameters in general and in particular of the $\mu_B$. In this case one can obtain results for QCD at nonzero $\mu_B$ either by performing simulations at imaginary $\mu_B$, which is free from the sign problem, and then analytically continue to real values of $\mu_B$, or by calculating the Taylor coefficients of various quantities at $\mu_B=0$ and then performing a numerical summation of the Taylor series. For this reason the precise data on the phase diagram of QCD at nonzero baryon density are only known for small and moderate values of the chemical potential $\mu_B/T\leq2.5-3$ \cite{Borsanyi:2022qlh,Bollweg:2022fqq}.

Much less is known about the phase diagram of QCD as a function of all three parameters: temperature, nonzero baryon density and magnetic field. Lattice QCD studies of this problem were carried out in \cite{Braguta:2019yci}. To overcome the sign problem, in this paper the authors performed the simulations at the imaginary values of the baryon chemical potential and analytically continued the results to the real values of $\mu_B$. This approach is justified since the phase transition in QCD is a crossover for small and moderate values of magnetic field and chemical potential.

In the paper \cite{Braguta:2019yci} two different phase transitions were discussed: the chiral, which was determined from the chiral observables: the chiral condensate and the chiral susceptibility, and the deconfinement phase transition, which was determined using the single quark entropy \cite{Weber:2016fgn}. Since the finite temperature phase transition in QCD is a crossover rather then a genuine phase transition, different observables can follow different behaviour and predict slightly different transition temperature and transition width.

\begin{figure}
    \centering
\includegraphics[width=15cm]{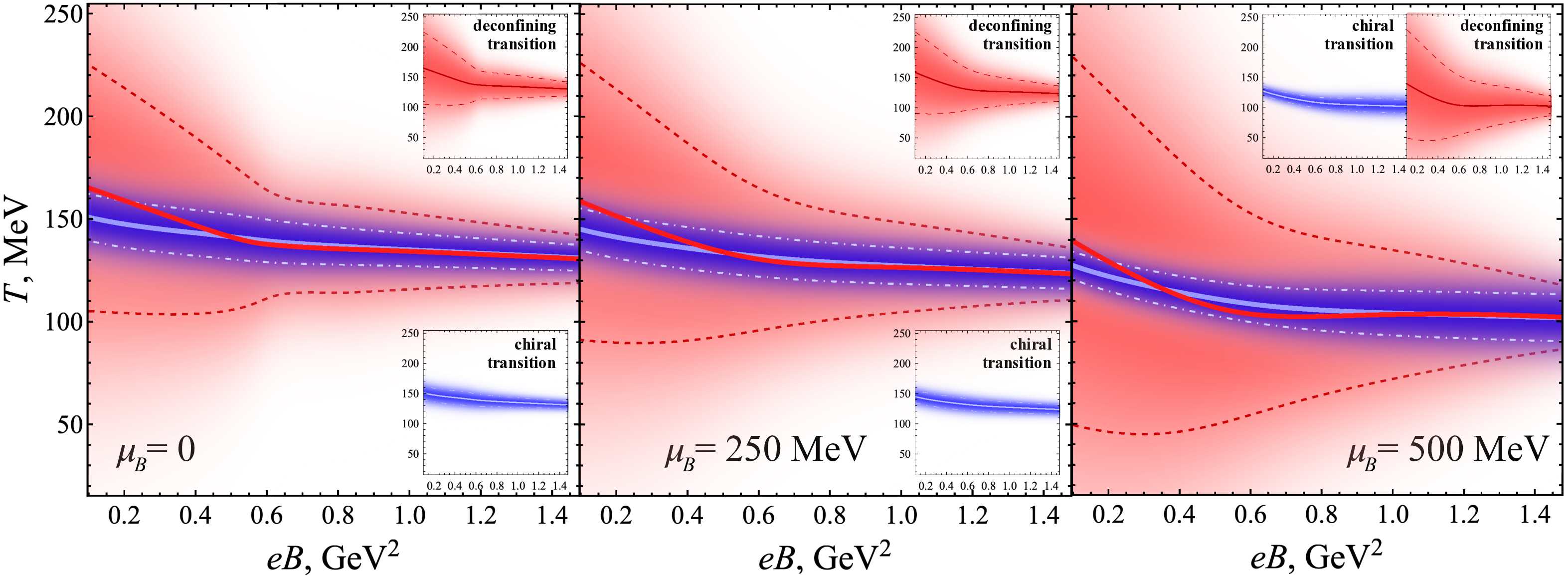}
\caption{(From \cite{Braguta:2019yci}). The tentative phase diagram in the plane temperature-magnetic field $T-eB$ for several values of the baryon chemical potential $\mu_B=0, 250$ an $500$ MeV.}
\label{fig:phasediagram}
\end{figure}

The final sketch of the phase diagram from \cite{Braguta:2019yci} is presented in Fig.\ref{fig:phasediagram}. In this Figure the phase digram of QCD in the plane $T$-$eB$ is shown for several values of the baryon chemical potential $\mu_B=0$, $\mu_B=250$ \text{MeV} and $\mu_B=500$ \text{MeV}, the data for the latter two values of the magnetic field were obtained using the method of analytical continuation. We would like to stress, that the method of analytical continuation might be questionable for the largest studied value of $\mu_B=500$ \text{MeV}, however we present it here in order to enhance the effect of the corresponding parameters on the phase transition. We can learn from this plot that both magnetic field and baryon density decrease the pseudo-critical temperature of the chiral and deconfining phase transition and when both these parameters are turned on their effect is just combined. Note, however, that the width of the phase transition has a nontrivial dependence on the parameters - with small magnetic field it decreases with $\mu_B$ and at large magnetic fields it increases with $\mu_B$. For the parameters under study no clear signatures of the first-order phase transition were observed.

\subsection{Equation of State}
\label{subsec:eos}

Equation of State of plays a pivotal role in studies of the quark-gluon plasma, providing an important input for hydrodynamical simulations. It describes how thermodynamic quantities, including pressure $p$, energy density $\epsilon$ and others depend on the parameters of the system. In this Section we discuss, how magnetic field changes the QCD Equation of State.

The basic quantity for the Equation of State is pressure $p$, which is given by the logarithm of the partition function $Z$:
\begin{equation}
    p=\frac{T}{V}\ln Z(\mu_B,\mu_S,\mu_Q,T,V,eB) ,
\end{equation}
here $T$ is the system temperature and $V$ is its volume. Since the partition function $Z$ and its logarithm cannot be determined in lattice calculations, the standard way which is used for the determination of the pressure $p$ is a calculation of its derivatives. For example, baryon density is given by the derivative with respect to the baryon chemical potential: $n_B=\frac{\partial p}{\partial \mu_B}$ and can be directly calculated on the lattice. Given that simulations at nonzero $\mu_B$ are hampered by the sign problem, one typically studies the Taylor expansion of the pressure $p$ in terms of the chemical potential $\mu_B$:

\begin{equation}
    \frac{p}{T^4}=c_0+c_2\left(\frac{\mu_B}{T}\right)^2+c_4\left(\frac{\mu_B}{T}\right)^4+c_6\left(\frac{\mu_B}{T}\right)^6+O\left(\left(\frac{\mu_B}{T}\right)^8\right).
\label{CME_eq:pressure}
\end{equation}

It naturally leads to the following Taylor expansion of $n_B$ in terms of $\mu_B$: 
\begin{equation}
    \frac{n_B}{T^3}=2c_2\frac{\mu_B}{T}+4c_4\left(\frac{\mu_B}{T}\right)^3+6c_6\left(\frac{\mu_B}{T}\right)^5+O\left(\left(\mu_B/T\right)^7\right)
\end{equation}

Without external density $\mu_B=0$ the Equation of State is given by the coefficients $c_0$ and its dependence on the temperature and magnetic field. Coefficient $c_0$ was extensively studied in \cite{Levkova:2013qda,Bonati:2013vba,Bali:2014kia}, where it was found that magnetic field leads to a large additional contribution to the thermodynamic observables at magnetic fields $eB\sim 0.5$ GeV$^2$.

\begin{figure}
    \centering
    \includegraphics[width=5cm]{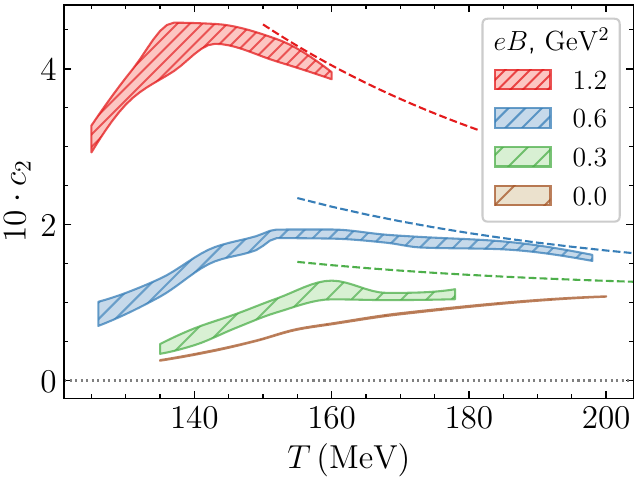}
    \includegraphics[width=5cm]{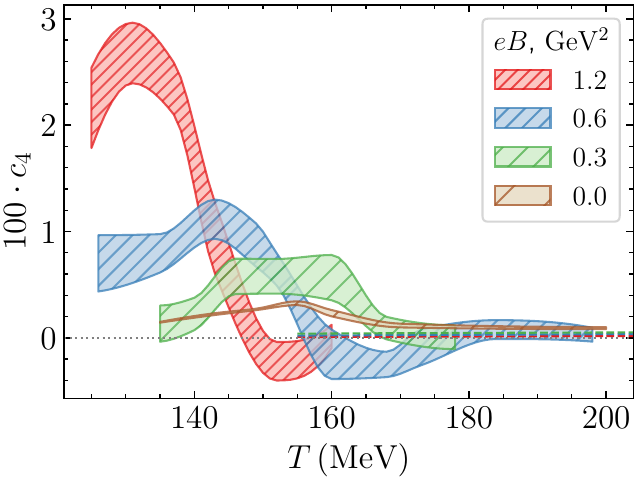}
    \includegraphics[width=5cm]{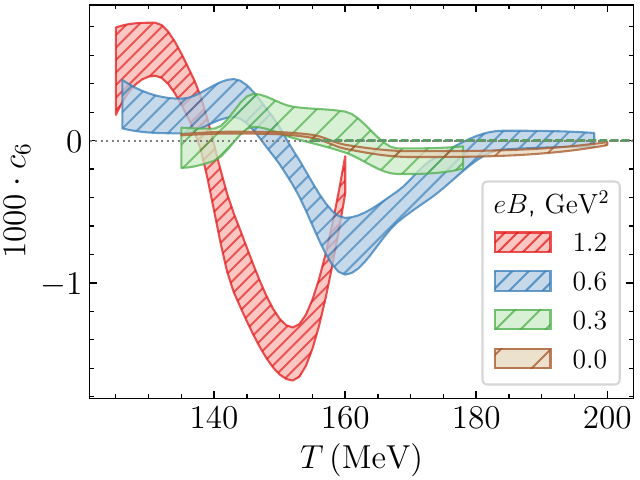}
    \caption{(From \cite{Astrakhantsev:2024mat}). The dependence of the coefficients $c_2$, $c_4$, $c_6$ on the temperature for several values of magnetic field. Data for $eB=0$ were calculated based on results of \cite{DElia:2016jqh}, data for nonzero magnetic field are from \cite{Astrakhantsev:2024mat}. Dashed lines correspond to the ideal gas approximation.
    }
    \label{fig:c2c4c6}
\end{figure}

\begin{figure}
    \centering
    \includegraphics[width=5cm]{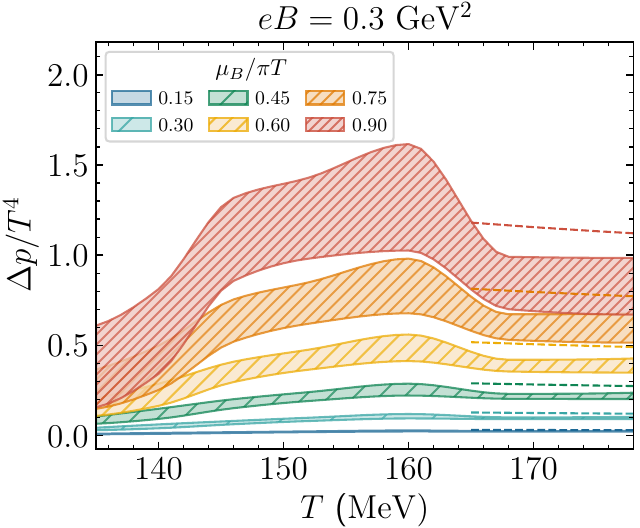}
    \includegraphics[width=5cm]{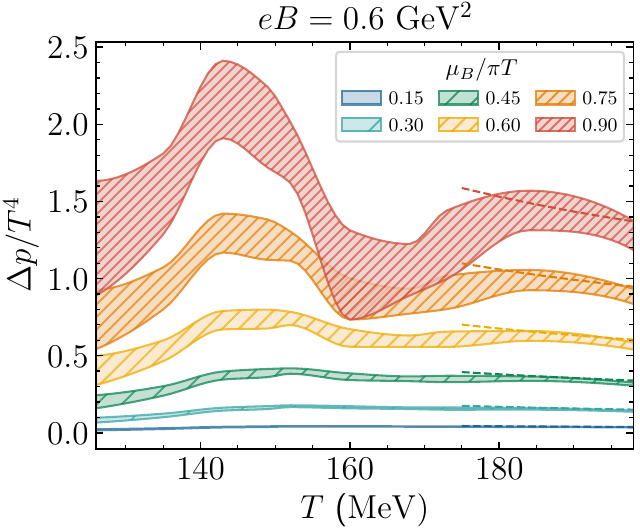}
    \includegraphics[width=5cm]{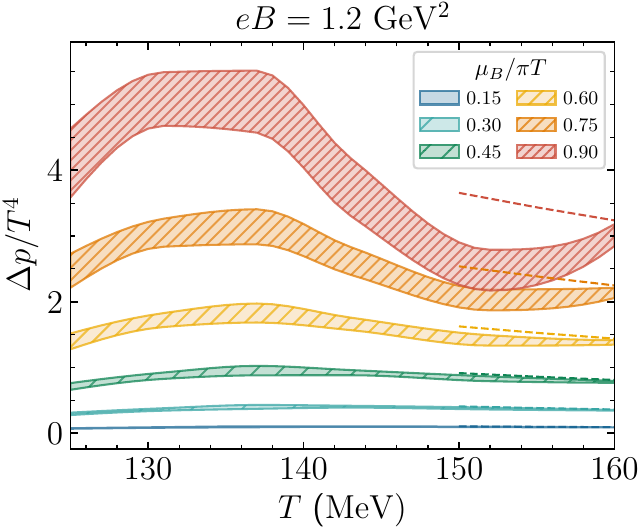}
    \caption{(From \cite{Astrakhantsev:2024mat}). Excess of the pressure $\Delta p=p(\mu_B)-p(\mu_B=0)$ due to nonzero baryon density $\mu_B$ for three different values of magnetic field $eB=0.3$ GeV$^2$, $eB=0.6$ GeV$^2$ and $eB=1.2$ GeV$^2$. Dashed lines correspond to the ideal gas approximation.
    }
    \label{fig:pressure}
\end{figure}

The behaviour of the coefficients $c_2$, $c_4$ and $c_6$ and their dependence on the magnetic field was studied in \cite{Astrakhantsev:2024mat}. In Fig.~\ref{fig:c2c4c6} we present the dependence of these coefficients on the temperature for several values of magnetic field $eB=0.0$ (from \cite{DElia:2016jqh}), $eB=0.3$, $eB=0.6$, $eB=1.2$ GeV$^2$. The main conclusion is that the coefficients $c_2$, $c_4$ and $c_6$ and, consequently, the pressure $p$ are significantly enhanced by the external magnetic field. At large magnetic fields their temperature dependence exhibits a peak structure. One can also expect that at high temperatures the system is described by an (almost) ideal gas approximation. Its predictions are also presented in Fig.~\ref{fig:c2c4c6} by dashed lines, which seem to agree with the lattice results at high temperatures above the chiral phase transition. Finally, in Fig.~\ref{fig:pressure} we present an additional contribution to the pressure (\ref{CME_eq:pressure}) coming from the baryon density $\Delta p=p(\mu_B)-p(0)$ for several values of magnetic field $eB=0.3$, $0.6$ and $1.2$ GeV$^2$. From Fig.~\ref{fig:c2c4c6} and ~\ref{fig:pressure} we can see that nonzero magnetic field not only significantly increases the pressure, but also modifies its dependence on the temperature and the baryon chemical potential $\mu_B$.

In Ref.~\cite{MarquesValois:2023ehu} the first results  for the QCD Equation of State beyond the approximation $\mu_s=0$, $\mu_u=\mu_d$ are presented. The authors tuned the chemical potentials to maintain the conditions $n_S=0$, $n_Q/n_B=0.4$, which correspond to the initial conditions in heavy ion collision. The authors also determine the leading order coefficient $c_2$ for these conditions, which is also significantly enhanced by the magnetic field.

\subsection{Fluctuations of the conserved charges}

Fluctuations of the conserved charges: baryon density (B), electric charge (Q) and strangeness (S) are closely related to the QCD phase diagram and its Equation of State, probing different degrees of freedom. Moreover, these quantities (more precisely, their combinations) can be easily accessible experimentally, building a bridge between lattice calculations and heavy ion collision experiments. 

Coefficients that describe the fluctuations, are defined as:

\begin{equation}
\chi^{BQS}_{ijk}=\left.\frac{\partial^{i+j+k}(p/T^4)}{ \partial^i(\mu_B/T)\partial^j(\mu_Q/T)\partial^k(\mu_S/T)}\right|_{\mu_B=\mu_Q=\mu_S=0}.
\end{equation}

Due to the sign problem in QCD, on the lattice the Equation of State is computed as the Taylor expansion of the pressure $p$ in $\mu_B$, $\mu_S$ and $\mu_Q$, thus naturally producing the values for the coefficients $\chi^{BQS}_{ijk}$. Without magnetic field the behaviour of the fluctuations $\chi^{BQS}_{ijk}$ was extensively studied on the lattice \cite{Borsanyi:2022qlh,Bollweg:2022fqq}.

Behaviour of the fluctuations up to the second order in QCD with the external magnetic field was studied in \cite{Ding:2021cwv,Ding:2023bft,MarquesValois:2023ehu}. In \cite{Ding:2021cwv} quadratic fluctuations and correlations of the conserved charges were calculated on the lattice at heavier than physical pion mass $m_{\pi}\approx 220$~MeV. It was found that magnetic field significantly changes the properties of the system. First, values of the coefficients $\chi^{BQS}_{ijk}$ notably grow with increasing magnetic field. Second, the temperature dependence of these coefficients at high magnetic fields starts to be non-monotonous, exhibiting a peak around the pseudo-critical temperature. Finally, the position of an inflection point or a peak in the temperature dependence of $\chi^{BQS}_{ijk}$ is shifted towards lower temperatures, indicating that the change in the degrees of freedom occurs at lower temperatures. 
In \cite{MarquesValois:2023ehu} the quadratic fluctuations and the conditions corresponding to the strange neutrality were studied for nonzero values of magnetic field and at the physical value of the pion mass. Results presented in \cite{MarquesValois:2023ehu} seem to exhibit the same qualitative behaviour as in \cite{Ding:2021cwv}. Note that the results for the Equation of State of QCD in magnetic field discussed in Sec.~\ref{subsec:eos} also show the same trend. In \cite{Ding:2023bft} the authors studied specific combinations of observables: the correlation between the baryon charge and the electric charge $\chi^{BQ}_{11}$ and the ratio of the chemical potentials $\mu_Q/\mu_B$ (tuned to the conditions corresponding to the heavy ion collision experiments). Based on their results which indicate that these quantities significantly grow at sufficiently large magnetic field, it was proposed that one can use these combination $\chi^{BQ}_{11}$ and $\mu_Q/\mu_B$ to estimate the values of the magnetic field produced in real heavy-ion collision experiments (the so-called magnetometer).

\subsection{Summary}

Bulk properties of QCD under such pairs of external parameters as density-temperature and magnetic field-temperature were extensively studied in lattice simulations. However, much less is known about the properties of QCD in the three-dimensional space temperature-density-magnetic field. In this section we reviewed current status of the lattice studies of this system: phase diagram, Equation of State and related to it conserved charge fluctuations. Our knowledge is limited due to the sign problem for QCD at nonzero density, for this reason, the studies are restricted mainly to the Taylor expansion of the corresponding quantities and small values of the baryon chemical potential $\mu_B/T<2.5-3$. However, we see that magnetic fields and baryon density significantly modify the properties of QCD, the critical temperature and possibly the nature of the phase transition. The results of the presented studies also suggest some observables that can quantify the value of the magnetic field in experiments (magnetometer).

Presented studies do not go to larger values of magnetic field, in particular, they even do not come close to the CEP, which is predicted to exist at very large values of magnetic field. Possible existence of this CEP and its connection to the chiral CEP in QCD in the $\mu-T$ plane could be a subject of possible future studies. Among other possible directions one could consider studies of the phase diagram and other properties of QCD with other fields and other parameters, including the chiral chemical potential $\mu_5$, electric field $eE$, non-homogeneous fields and others.

	\section{Superconductivity of vacuum in an extreme magnetic field: evidence from the Electroweak sector of the Standard model and implications for QCD}
	\label{sec:Chernodub}

\subsection{Introduction and Motivation}

\subsubsection{Strong magnetic fields in Nature}

An intense magnetic field can drastically affect various properties of physical systems and can even create new states of matter. For example, the strongest steady fields in Earth laboratories reach values of about a few tens of Tesla (T), which is enough to observe the quantum Hall effect in graphene at room temperature~\cite{Novoselov2007room}. In plasma physics, a magnetic field of a similar scale is designed to hold the thermonuclear plasma of temperature $1.5 \times 10^{8}\,\mathrm{K}$ at ITER Tokamak~\cite{ITER}. The strained graphene sheets may generate synthetic magnetic fields up to enormous values of about 300\,T~\cite{Levy2010strain}. These are, perhaps, the strongest steady fields that we can reach in Earth conditions. 

More powerful, persistent magnetic fields can, of course, exist in astrophysical environments. The fields with the strengths of the order $10^{8 \dots 11}\,\mathrm{T}$ are supposed to be present at the surfaces of highly magnetized neutron stars, called magnetars~\cite{Olausen2014mcgill}. The magnetar SGR 1806-20 possesses the strongest recorded value of about $10^{11}\,{\rm T}$ which has been confirmed to from investigation of bright flares of short-duration $\gamma$-ray bursts of this star~\cite{Terasawa2005}. In the cores of the magnetars, one expects to encounter even more powerful steady fields up to $10^{14}\, \mathrm{T}$~\cite{Lai1991cold, Cardall_2001}. 

According to theoretical estimates, much stronger but transient magnetic fields of the order of $10^{14 \dots 16}\,\mathrm{T}$ appear for a short time of $10^{- (23 \dots 24)}\,{\rm s}$, in quark-gluon plasma which is routinely created in non-central relativistic heavy-ion collisions at RHIC and LHC~\cite{Skokov:2009qp, Deng:2012pc}. The field of $10^{14}\,\mathrm{T}$ has been recently confirmed by the STAR experiment at the Relativistic Heavy Ion Collider \cite{Abdulhamid2024}. 

The cosmological electroweak phase transition might have developed a very strong magnetic field of the electroweak strength, $10^{20}\,\mathrm{T}$, which could have influenced the subsequent evolution of the Universe~\cite{Vachaspati:1991nm, Grasso:2000wj}. While this field bears the word ``weak'' in its name, it is among the strongest fields in the Standard Model of particles. Similarly, immense fields can possibly exist even in the modern Universe in the vicinity of the magnetized black holes~\cite{Maldacena:2020skw, Ghosh:2020tdu}. 

The presence of a background magnetic field can alter the phase diagram of fundamental field theories, impacting the transitions between various states of matter. This phenomenon bears significant relevance to the early Universe, as, during the cooling stages of its evolutionary stages, the Universe underwent a sequence of high-temperature transitions. Notably, the electroweak symmetry breaking occurred at temperature $T^{\mathrm{EW}} = 159.5(1.5) \,\mathrm{GeV} \simeq 2\times 10^{15}\,\mathrm{K}$ via a smooth crossover transition~\cite{DOnofrio:2015gop}. It has been immediately followed by the color confinement (hadronization) transition, that was accompanied by the chiral symmetry restoration at $T^{\mathrm{QCD}} = 156.5(1.5) \,\mathrm{MeV} \simeq 2\times 10^{12}\,\mathrm{K}$~\cite{HotQCD:2018pds}. The confining and chiral transitions, described by quantum chromodynamics (QCD), also have a crossover nature: they overlap and appear at the same temperature range. While the electroweak crossover does not experience a qualitatively noticeable modification at moderate (in the electroweak scale) $10^{20}\,\mathrm{T}$ magnetic fields~\cite{Kajantie:1998rz}, the magnetic-field background of the strength of a few GeV squared (corresponding to $10^{17}\,\mathrm{T}$) can generate a new end-point thus turning the finite-temperature QCD crossover into a first-order phase transition~\cite{DElia:2021yvk}. Surprisingly, the magnetic field background of sufficiently powerful strength can also influence the properties of the vacuum itself, which, by definition, contains no matter et al. Below, we will discuss the magnetic properties of a cold, empty vacuum devoid of matter.


\subsubsection{Characteristic scales of magnetic field in particle physics}

The impact of a magnetic field on a particular phenomenon is set by comparison of its strength with the relevant mass or length scales involved. For example, for electromagnetic interactions described by quantum electrodynamics, the relevant intensity of the magnetic field is set by the Schwinger limit,\footnote{We use the units $\hbar = c = 1$.}
\begin{align}
    B^{\mathrm{QED}} = m_e^2/e \simeq 4 \times 10^9\,\mathrm{T}\,,    
\end{align}
determined by the electron mass $m_e$. At this strength -- which is already bypassed by the fields near the surface of magnetars~\cite{Olausen2014mcgill} -- the vacuum acquires optical birefringence properties~\cite{Adler:1971wn} and can act as a ``magnetic lens'' which is able to distort and magnify images~\cite{Shaviv1999magnetic} similarly to the galactic-scale gravitational lens. 

Strong interactions, described by quantum chromodynamics, become sensitive to the presence of the magnetic field when it reaches the strength of the hadronic mass scale,
\begin{align}
    B^{\mathrm{QCD}} \sim m_p^2/e \sim 10^{16}\,\mathrm{T}\,,
    \label{eq_B_QCD}
\end{align}
where $m_p$ is the proton mass. Such fields lead to the effect of magnetic catalysis~\cite{Klevansky:1989vi, Klimenko:1991he, Shovkovy:2012zn}, which implies, in particular, a persistent enhancement of the chiral symmetry breaking in QCD vacuum as the external magnetic field strengthens. 

Electroweak interactions are influenced by much stronger magnetic fields,
\begin{align}
    B^{\mathrm{EW}}_W = m_W^2/e \simeq 1.1 \times 10^{20}\,\mbox{T}\,,
\label{eq_eBc1}
\end{align}
determined by the $W$-boson mass $m_W \simeq 80.4\, {\rm GeV}$, which sets a typical mass scale in the electroweak sector. More than thirty years ago, it was suggested that such fields generate instability of vacuum, which proceeds via an onset of the condensation of the $W$ bosons. The presence of the instability can be deduced from the classical equations of motion of the electroweak model~\cite{Skalozub1978, Skalozub:1986gw, Ambjorn:1988fx, Ambjorn:1988tm, Ambjorn:1988gb, Ambjorn:1989bd}. Much later, it became clear that the transition into this exotic electroweak vacuum state marks a radical change in the transport properties of the vacuum, which acquires anisotropically superconducting~\cite{Chernodub:2010qx} and superfluid~\cite{Chernodub:2012fi} properties. 

For completeness, we also mention that the electroweak vacuum should possess the second phase transition at an even higher magnetic field:
\begin{align}
B^{\mathrm{EW}}_H = m_H^2/e \simeq 2.7 \times 10^{20}\,\mbox{T}\,,
\label{eq_eBc2}
\end{align}
which is determined by the Higgs mass $m_H = 125.1\,\mathrm{GeV}$. Above this field, $B_{c2}$, the electroweak symmetry is expected to be restored~\cite{Salam:1974xe,Linde:1975gx,Ambjorn:1989bd}. 

\subsubsection{Vacuum instability in the electroweak sector and the puzzle in QCD}

The $W$ boson plays a crucial role in the instability of the electroweak vacuum because this particle has three important properties: it is an (i) electrically charged (ii) vector particle that (iii) possesses a large gyromagnetic ratio (the Land\'e factor), $g=2$. As a result, the $W$ mesons strongly couple to the magnetic field via their large magnetic moments. The importance of these properties can be understood from the energy spectrum of a free particle of a mass $m$ placed in the background of an external steady uniform magnetic field $B$:
\begin{align}
    \varepsilon_{n,s_z}^2(p_z) = p_z^2+(2 n - g s_z + 1) e B + m^2\,.
\label{eq_energy_levels}
\end{align}
Here $n$ is a nonnegative integer number that labels the Landau levels, $s_z = 0, \pm 1$ is the spin projection on the axis of the field $z$, and $p_z$ is the particle momentum along the $z$ axis. We assume that $eB>0$.

With an increase of the magnetic field, the mass of the $W$ polarization branch~\eqref{eq_energy_levels} with the quantum numbers $n=0$, $s_z = +1$, $p_z = 0$ decreases, $M^2(B) = m^2 - e B$, and becomes a purely imaginary quantity if the magnetic field exceeds the critical value $B_c = m^2/e$. Notice that this effect arises due to the big gyromagnetic ratio $g = 2$. For $W$ bosons, the critical field is given by Eq.~\eqref{eq_eBc1}, above which the vacuum experiences a tachyonic instability associated with a second-order phase transition~\cite{Skalozub:1986gw, Ambjorn:1988fx, Ambjorn:1988tm, Ambjorn:1988gb, Ambjorn:1989bd}. The vacuum turns into an inhomogeneous vortex-dominated phase which will be discussed in more detail below. In realistic quantum field theories, the $W$ bosons interact with other fields, so that this mechanism should be further scrutinized to ascertain its robustness in the presence of interactions. 

One can reasonably wonder whether analogous phenomena can exist in other physical systems that also host electrically charged vector particles. A similar mechanism has indeed been suggested to operate at the level of QCD that possesses electrically charged mesonic bound states with vector quantum numbers, the $\rho$-meson excitations~\cite{Chernodub:2010qx}. In this case, the $\rho$-meson mass $m_\rho \simeq 770\, {\rm MeV}$ determines the free-limit value of the critical field, $B_{c} = m_\rho^2/e$, that appears to be of the order of the characteristic field in QCD, $B_{c} \sim B^{\rm QCD}$~\eqref{eq_B_QCD}. If this mechanism works, then the vacuum may also acquire electromagnetic superconducting properties mediated by the condensation of $\rho^+$ mesons. 

The existence of this intriguing superconducting state of the QCD vacuum is supported by low-energy models~\cite{Chernodub:2011mc} and holographic approaches~\cite{Bu:2012mq, Callebaut:2011ab}. At the same time, it is questioned by symmetry arguments~\cite{Hidaka:2012mz} and effective field theory calculations~\cite{Andreichikov:2013zba}. Furthermore, this exotic phase avoids a direct first-principle detection in numerical simulations~\cite{Hidaka:2012mz, Bali:2017ian}, presumably because of numerical difficulties associated with the inhomogeneity of the charged condensate and the smooth crossover nature of the suspected superconducting transition~\cite{Chernodub:2012zx, Chernodub:2013uja, Li:2013aa}. Moreover, the first-principle lattice simulations of QCD do not display any signature of a tachyonic instability in a zero-temperature QCD vacuum exposed to high magnetic fields~\cite{Hidaka:2012mz, Luschevskaya:2014lga, Taya:2014nha, Luschevskaya:2015bea, Bali:2017ian, Luschevskaya:2024iic}. In other words, the masses of vector mesons do not vanish at the range of the relevant fields~\eqref{eq_B_QCD} in the apparent contradiction to the classical analysis within effective models. While these results would be sufficient to exclude the presence of the superconducting phase in QCD vacuum at strong fields, we show below that the same questions can also be asked at the level of the electroweak sector there the tachyonic instability associated with a second-order phase transition, was also predicted by theoretical analysis~\cite{Ambjorn:1988fx, Ambjorn:1988tm, Ambjorn:1988gb, Ambjorn:1989bd}. Remarkably, despite the absence of any tachyonic instability, the electroweak vacuum still has a superconducting electroweak phase~\cite{Chernodub:2022ywg}. This observation gives hope for a search for a similar phase in QCD, addressed in more detail in this paper. 

The key argument in favor of the existence of the new QCD phase, which is consistent with available numerical results, is that the transition from the usual hadron phase to the superconducting phase at the field strengths of the order of the QCD scale~\eqref{eq_B_QCD} may proceed via a smooth crossover, which prevents any instability. While this statement sounds questionable in view of the simple analysis of the free-particle spectrum~\eqref{eq_energy_levels}, the first-principle numerical simulations show that no instability occurs in the electroweak model, where the vacuum enters a similar $W$-condensed state via a magnetic-field-induced crossover. In other words, in the absence of a second-order phase transition, the instability should never appear in the electroweak model. Still, the magnetic field produces the superconducting phase in the electroweak model. 

Thus, a QCD transition to the new phase can also occur at a nonvanishing $\rho$-meson mass in the absence of any thermodynamic singularity~\cite{Chernodub:2013uja}. The latter scenario has a hypothetical character since it requires confirmation from a first-principle simulation. In this respect, the electroweak model provides us with a compelling playground, backed by the similarity of the superconducting mechanisms in both QCD and electroweak systems.

Notice that even in the electroweak model, where the condensation of the $W$ bosons has robustly been predicted at the classical level, the existence of the magnetic-field-induced phase was questioned in the theoretical literature before the numerical first-principle results~\cite{Chernodub:2022ywg} became available. At the classical level, the formation of the periodic vortex lattice in the background magnetic field has been established not only analytically but also numerically~\cite{Ho:2020ltr}. However, this classical-level scenario, together with the arguments based on loop computations~\cite{Nielsen:1978rm}, has been questioned in Ref.~\cite{Skalozub:2014epa} where it was shown that quantum corrections could add a radiative term to the classical $W$ mass in such a way that the mass does not vanish at the critical field $B_c = B^{\rm EW}_{W}$. Accordingly, it was concluded that no thermodynamic instability should appear in the electroweak model. Moreover, earlier numerical simulations of the electroweak model in the background magnetic field did not reveal the presence of the vortex-dominated phase around the finite-temperature electroweak crossover~\cite{Kajantie:1998rz}, which could be explained by a destructive role of strong thermal fluctuations. Therefore, the very existence of the superconducting phase in an electroweak model is a rather nontrivial fact. Below, we review the first-principle numerical results obtained in the electroweak model which demonstrate the existence  ~\cite{Chernodub:2022ywg}.

Before closing this section, it is worth mentioning that the mechanisms that drive the vacuum superconductivity at QCD~\cite{Chernodub:2010qx} and electroweak~\cite{Chernodub:2010qx, Chernodub:2012fi} scales are very similar to reentrant superconductivity that was suggested to occur in very clean superconducting materials in high---in the condensed matter standards---magnetic fields~\cite{Rasolt1992}. This phenomenon is suggested to be catalyzed by magnetic-field-induced condensation of Cooper pairs of electrons in a $p$-wave bound state. The $p$-wave Cooper pair is a condensed-matter counterpart of the $\rho^+$ meson, the latter being an electrically charged bound state of a quark and an anti-quark. The strength of the magnetic field required for this phenomenon to occur is suggested to be in the range of a few dozen Teslas.~\cite{Rasolt1992}

\subsection{Magnetic-field induced vortex phase in the electroweak model}

We consider the bosonic sector of the Electroweak model with the Lagrangian
\begin{align}
    {\mathcal L}_{\rm EW} = -\frac{1}{2} \tr (W_{\mu \nu} W^{\mu \nu}) - \frac{1}{4} Y_{\mu \nu} Y^{\mu \nu} + (D_\mu \phi)^\dagger (D^\mu \phi) - \lambda \Bigl(\phi^\dagger \phi - \frac{v^2}{2} \Bigr)^2\,,
\label{eq_LEW}
\end{align}
where $W_{\mu \nu}^a = \partial_\mu W_\nu^a - \partial_\nu W_\mu^a + i g \varepsilon^{abc} W_\mu^b W_\nu^c$ and $Y_{\mu \nu} = \partial_\mu Y_\nu - \partial_\nu Y_\mu$ are the field strengths of the SU(2) gauge field $W_\mu^a$ and $U(1)_Y$ hypercharge gauge field $Y_\mu$, respectively. These fields interact with the complex scalar Higgs doublet $\phi \equiv (\phi_1,\phi_2)^T$ via the covariant derivative, $D_\mu = \partial_\mu + \frac{i}{2} g W_\mu^a \sigma^a + \frac{i}{2} g' Y_\mu$, where $\sigma^a$ ($a=1,2,3$) are the Pauli matrices. The ratio of the $U(1)$ and $SU(2)$ gauge couplings, $g'/g = \tan \theta_W$, defines the electroweak mixing (Weinberg) angle $\theta_W$ with
$\sin^2 \theta_W \equiv 1 - m_W^2/m_Z^2 = 0.22290(30)$~\cite{CODATA2018}. The dimensionless coupling $\lambda$ controls the self-interaction of the Higgs field, and the only dimensionful parameter $v$ gives, at low temperature, the vacuum expectation value to the Higgs field, $\langle\phi\rangle \neq 0$, breaking electroweak symmetry down to its electromagnetic subgroup, $SU(2)_W\times U(1)_Y \to U(1)_{\mathrm{e.m.}}$. We do not consider fermions as they do not play any qualitatively important role in the phenomena that we are interested in. 

The particle content of the model is simple. In the $T = 0$ broken phase, the Higgs field acquires the mass $m_H = \sqrt{2 \lambda} v$. The model also possesses the massless photon, $A_\mu = W_\mu^3 \sin \theta_W + Y_\mu \cos \theta_W$, and three massive gauge bosons, which include the electrically (off-diagonal) charged $W$ bosons $W^\pm_\mu = W^1_\mu \pm i W^2_\mu$, and the neutral (diagonal) $Z$ boson, $Z_\mu = W_\mu^3 \cos \theta_W - Y_\mu \sin \theta_W$, with the masses $m_W = g v/2$ and $m_Z = m_W/\cos \theta_W$, respectively.

The electroweak model~\eqref{eq_LEW} in the background of the magnetic field has been extensively studied in Ref.~\cite{Chernodub:2022ywg} using first-principle numerical simulations. In simulations, we used the hypermagnetic field $B_Y$ associated with the hypergauge field $Y_\mu$ which, in the broken phase, corresponds to the usual magnetic field $B$ via the relation $g' B_Y = eB$ and plays an independent role in the symmetry-restored phase, in which the magnetic field cannot be defined. The results of Ref.~\cite{Chernodub:2022ywg} reveal the existence of three phases separated by two (pseudo)critical magnetic fields:
\begin{align}
    e B_{c1} = 0.68(5) m_W^2\,,
    \qquad\
    e B_{c2} = 0.99(2)m_H^2\,.
\end{align}
The first critical field $B_{c1} \simeq 0.7 B_W^{\rm EW}$ appears to be about 30\% weaker than the value~\eqref{eq_eBc1} predicted by the classical theoretical analysis that does not take into account quantum fluctuations. However, the second critical field $B_{c2} \simeq B_H^{\rm EW}$ agrees precisely with the theoretical value~\eqref{eq_eBc2} within the accuracy of 1\%.

\begin{figure}[!htb]
\includegraphics[width=0.95\textwidth,clip=true]{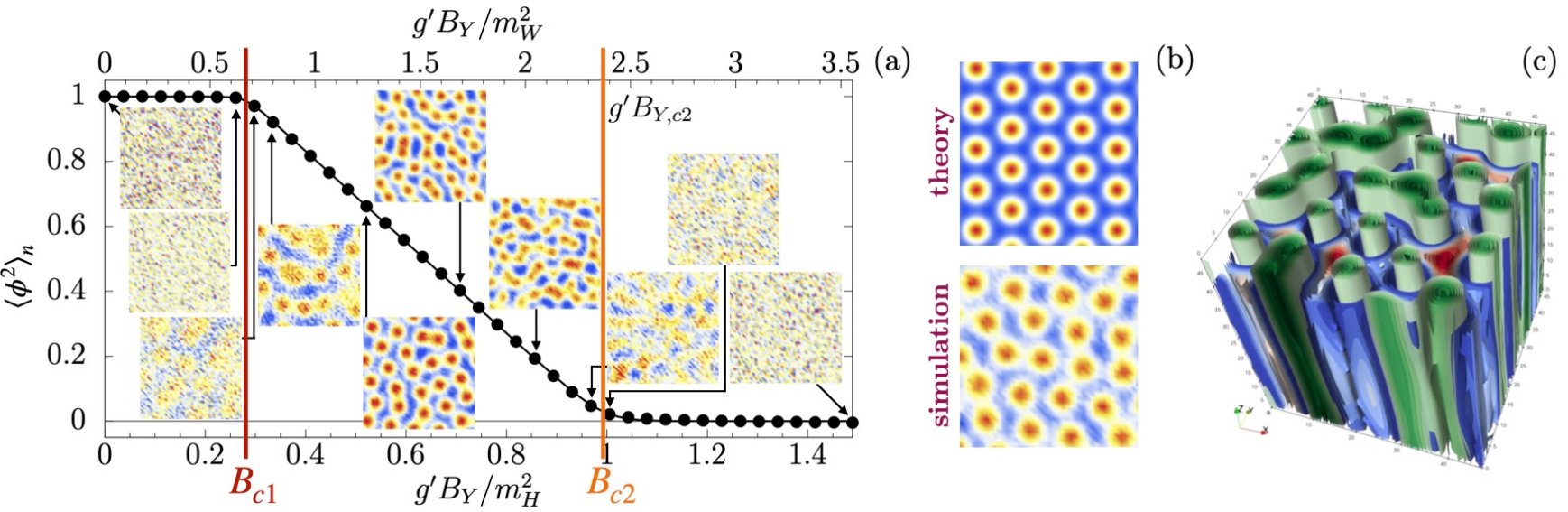}
\caption{(a) A normalized value of Higgs condensate squared $\langle\phi^2\rangle_n$. The insets represent the density plots of the $Z_{12}$ fluxes in the cross-sections normal to the magnetic field axis obtained in numerical simulations at the values of the magnetic field shown by the arrows. (b) An illustration of a hexagonal crystalline order of the vortices predicted by the theory (top) and the result obtained in the numerical simulations (bottom); (c) A typical 3d configuration in the superconducting phase ($B_{c1} < B < B_{c2}$) in the magnetic field background $e B = 1.1 m_W^2$. The blue and red (green) surfaces show equipotential surfaces of the $W$ (Higgs) condensate in complementary regions. Figure adapted from Ref.~\cite{Chernodub:2022ywg}.}
\label{CME_fig1}
\end{figure}

The existence of these phases can be readily seen in Fig.~\ref{CME_fig1}(a), where a normalized expectation value of the Higgs condensate is shown against the magnetic field. One can distinguish (i) the low-field phase at $B < B_{c1}$, where the Higgs expectation value is independent of the strength of the magnetic field; (ii) the intermediate phase at $B_{c1} < B < B_{c2}$, where the field gradually diminishes; and (iii) the symmetry restored phase at $B > B_{c2}$. Thus, the phase structure of the electroweak model in a strong magnetic field is indeed nontrivial, and it agrees rather well with the predictions of the theory. However, what is the nature of the intermediate phase and the bordering phase transitions?

Theoretically, the intermediate phase should be described by a classical inhomogeneous $W$-condensate solution. It can best be imagined as a crystal made of parallel vortex-like structures that share a geometric similarity with the lattice of Abrikosov vortices of a conventional type-II superconductor. For realistically heavy Higgs masses, $m_H > m_Z$, the vortices in the $W$ condensate arrange themselves into a hexagonal lattice~\cite{Skalozub:1986gw, MacDowell:1991fw, Tornkvist:1992kh}. It is worth also noticing that the similarity between the magnetized vortex phase of the electroweak vacuum and a type-II superconductor is geometric rather than physical because the properties of the two systems differ significantly. For example, the $W$ vortices expel the magnetic field from their cores~\cite{Ambjorn:1988fx, Ambjorn:1992ca} while the Abrikosov vortices carry the magnetic field lines inside their cores~\cite{LL9}. In addition, the vacuum in this exotic state exhibits superconductivity (associated with the condensation of the $W$ bosons~\cite{Chernodub:2010qx}) and superfluidity (supported by the condensation of the $Z$ bosons~\cite{Chernodub:2012fi}) only along the direction of the magnetic field, being an insulator in the transverse plane~\cite{Chernodub:2010qx, Chernodub:2011gs}. On the contrary, an ordinary superconductor supports dissipationless transport in all spatial directions. 

The numerical results on the inhomogeneities of the vacuum phases are presented in the insets of Fig.~\ref{CME_fig1}(a) corresponding to different values of the background magnetic field. The snapshots of the cross-sections of typical lattice configurations do indeed reveal the presence of flux-like vortices in the intermediate phase. Contrary to a crystalline hexagonal structure suggested by the theory, the vortices form an amorphous medium, which can be characterized as a disordered solid or a liquid but not a crystal. We attribute the difference between the numerical results and the theoretical predictions on the structure of the inhomogeneous state to the quantum fluctuations that were not taken into account by the semiclassical estimations. A qualitative comparison between the vortex arrangement in the cross-section 2d plane proposed by the theory and found in the simulations is given in Fig.~\ref{CME_fig1}(b), while the 3d visualization of a typical field configuration found in the simulations is presented in Fig.~\ref{CME_fig1}(c). Our simulations also reveal that at a higher magnetic field, the solid state of the vacuum melts down (at zero temperature) and disappears: the vacuum enters the homogeneous symmetry-restored state at $B > B_{c2}$. This observation agrees with earlier theoretical estimations~\cite{Olesen:1991df, VanDoorsselaere:2012zb}. 


What is also interesting---and this property should definitely be crucial in an analysis of the corresponding phase in QCD---is that both numerically observed transitions are smooth crossovers. As it is demonstrated in Ref.~\cite{Chernodub:2022ywg}, the susceptibility of the Higgs field does not possess a local maximum across either $B_{c1}$ or $B_{c2}$ as it could be expected in a case of a phase transition of first or second order, or in a case of a strong crossover. Therefore, these transitions are not marked with thermodynamic singularities. As a consequence, no instability may occur. 

\subsection{Magnetic-field induced vortex phase in QCD}

We expect that a similar scenario can also be realized in QCD, where the role of the $W$ bosons is played by the charged $\rho$ mesons. The numerical simulations of the electroweak model reveal the absence of any tachyonic instability at the critical transition field, while the vortex phase does exist at the intermediate values of the magnetic field. Likewise, the tachyonic instability has not been found in QCD as well. However, do vortex-like structures appear in QCD in the background magnetic field? This question has been addressed in Ref.~\cite{Braguta:2012fol} in the quenched approximation in the simplest, two-color version of the theory. In this section, we highlight certain findings of that paper. 

The superconducting ground state should be characterized by the presence of the mixed quark-antiquark condensates, composed of $u$ and $d$ quarks, $\langle \bar u \gamma_1 d\rangle = \rho$ and $\langle \bar u \gamma_2 d\rangle =  - i \rho$, that carry the quantum numbers of the electrically charged $\rho$ mesons. Effective models predict that these condensates should have an inhomogeneous vortex-like structure in the plane normal to the axis of the magnetic field. This phenomenon can be interpreted as the $\rho$-meson condensation. 

To probe the emergence of these inhomogeneous states, Ref.~\cite{Braguta:2012fol} employed lattice Monte-Carlo simulations of $SU(2)$ Yang-Mills lattice gauge theory. The quark fields were introduced by the overlap lattice Dirac operator $\mathcal{D}$ with exact chiral symmetry \cite{Neuberger:1997fp}. The background magnetic field has been introduced via the twisted spatial boundary conditions~\cite{Al-Hashimi:2008quu}. The quarks, treated in the quenched approximation, were subjected to the background magnetic field in the range $eB = (0 \dots 2.14)\, \mbox{GeV}^2$, which includes the critical field~\eqref{eq_B_QCD}.

Due to the oscillating phase of the complex $\rho$-meson condensate and the motion of vortices, the calculation of the local value of this condensate appears to be a challenging task. Therefore, the inhomogeneous properties of the vacuum were accessed via the following  $\rho$-meson correlator:
\begin{align}
    \phi(x) \equiv \phi(x;A,B) = \vev{\rho^\dagger(0) \rho(x)}_{A,B} \equiv  \tr\lr{\frac{1}{\mathcal{D}_u(A,B) + m} \, \gamma_{\mu} \, \frac{1}{\mathcal{D}_d(A,B) + m} \, \gamma_{\nu}}\,,
    \label{eq_phi_lat}
\end{align}
computed in the background of both the non-Abelian gauge field $A$ and the Abelian magnetic field $B$. The point $x=0$ has been taken at the origin of the lattice. 

The potential inhomogeneous properties of the $\rho$-meson condensation can be probed in the spirit of Ref.~\cite{Bali:1994de}, where the existence of a confining chromoelectric flux tube has been revealed in lattice Yang-Mills theory. The flux tube has been introduced with the help of a rectangular Wilson loop ${\cal W}$ as a source of heavy quark-antiquark pair and then studied the local energy density operator ${\mathcal E} = {\mathcal E}(x)$ as a probe of the flux tube formed due to the presence of the pair. 

Working in analogy to Ref.~\cite{Bali:1994de}, one can construct the normalized scalar energy of the $\rho$-meson field ${\mathcal E}(x)$, the normalized electric (super)current density $j_\mu(x)$ generated by the $\rho$-meson field, and the local vortex density $\upsilon(x)$, respectively:
\begin{align}
    {\mathcal E}(x) & = \frac{|D_\mu \phi(x)|^2}{|\phi(x)|^2}\,, \qquad D_\mu = \partial_\mu - i e A_\mu\,, 
\label{eq_energy}\\
j_\mu(x) & = \frac{\phi^*(x) {\overrightarrow D}_\mu \phi(x) - \phi^*(x) {\overleftarrow D}_\mu \phi(x)}{2 i |\phi(x)|^2}\,, 
\label{eq_current}\\
\upsilon(x) & = {\mathrm{sing}} \, {\mathrm{arg}} \, \phi(x) \equiv \frac{\epsilon^{ab}}{2\pi} \frac{\partial}{\partial x_a} \frac{\partial}{\partial x_b} \arg \phi(x)\,, \qquad a,b=1,2\,.
\label{eq_vortex}
\end{align}

\begin{figure}[!htb]
\includegraphics[width=0.85\textwidth,clip=true]{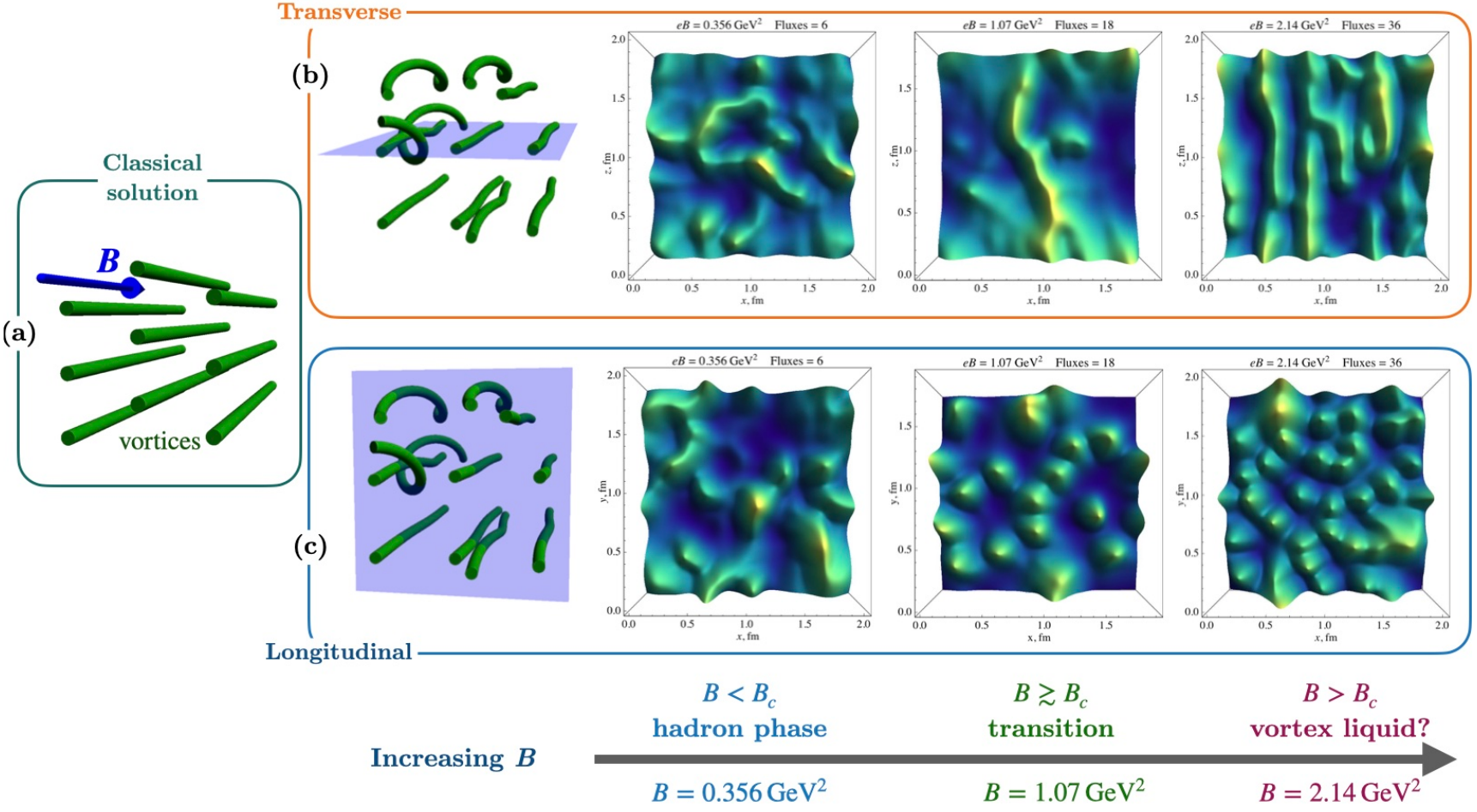}
\caption{(a) An illustration of the classical vortex solution in the superconducting QCD vacuum. The vortices arrange themselves parallel to the background magnetic field in a hexagonal structure; (b) the longitudinal and (c) transverse cross-sections of typical gauge-field configurations in quenched SU(2) QCD as revealed by the energy correlator~\eqref{eq_energy}, preceded by an illustration of the relevant cross-section for a partially disordered, realistic configuration. Figure adapted from Ref.~\cite{Braguta:2012fol}.}
\label{fig2}
\end{figure}

Our experience with the electroweak model tells us that the ideal semiclassical hexagonal lattice of $\rho$ vortices, shown in Fig.~\ref{fig2}(a), should get perturbed by quantum fluctuations, forming a disordered solid or a fluid. The fluctuations impact vortex patterns in the cross-sections. For example, a transverse $(x,y)$ cross-section of the volume may contain some linear segments of the vortices instead of a set of localized points in an ideal vortex lattice. In addition, the linear vortex instructions may disappear from one longitudinal $(x,z)$ plane as the vortices may propagate to another slice. The orientations of the cross-sections are illustrated in the first plots of the panels in Figs.~\ref{fig2}(b) and \ref{fig2}(c), respectively.

Within the core of a physical $\rho$-vortex, the energy density is elevated relative to the surrounding regions. Consequently, if physical $\rho$-vortices emerge within a sufficiently strong magnetic field, one would expect to observe localized point-like energy density concentrations in the transverse $(x,y)$ plane, as indeed seen in Fig.~\ref{fig2}(b). In the picture, the magnetic field is oriented along the $z$-axis. We can also expect the formation of long linear structures in the longitudinal $(x,z)$ plane, which is also the case according to Fig.~\ref{fig2}(c). In these figures, typical examples of the distribution of the energy density in the transverse $(x,y)$ and longitudinal $(x,z)$ planes are shown for weak ($eB = 0.356\,\mbox{GeV}^2$), moderate ($eB = 1.07\,\mbox{GeV}^2$) and high ($eB = 2.14\,\mbox{GeV}^2$) magnetic fields.

In accordance with our qualitative expectations, at a low magnetic field, the vortex lattice is not formed. At the moderate magnetic field, the appearance of a coherent vortex structure is seen. However, the vortices are not rigidly ordered in the transverse plane, and they are not quite parallel to the magnetic field. At a higher magnetic field, the physical picture shown in Fig.~\ref{fig2} becomes visually consistent with the presence of a melted lattice (liquid) of the $\rho$ vortices. This picture of the vortex formation in QCD~\cite{Braguta:2012fol} strongly resembles the formation of the superconducting $W$ vortices in electroweak model~\cite{Chernodub:2022ywg}.

One can also numerically observe~\cite{Braguta:2012fol} that the lumps in the energy density~\eqref{eq_energy} are surrounded by circulating electric currents~\eqref{eq_current} while the position of the energy lumps correspond to the locations of the vortex cores as revealed by the vorticity density~\eqref{eq_vortex}. These expected features correspond to the properties of the $\rho$ vortices obtained from the analytical solutions in Ref.~\cite{Chernodub:2010qx}. Thus, the results of quenched two-color QCD strongly suggest that similar superconducting vortex patterns may also emerge in realistic QCD with dynamical quarks.

\subsection{Conclusions}

First-principle numerical simulations of the electroweak model in a strong magnetic field reveal the presence of an inhomogeneous phase made of vortices parallel to the magnetic field~\cite{Chernodub:2022ywg}. Contrary to various theoretical estimations, the transition to this phase proceeds via a very smooth crossover, which does not show any signature of a tachyonic instability, claimed to be responsible for the emergence of this new phase. Still, the inhomogeneous phase does exist, and most of its features qualitatively coincide with the predictions coming from the semiclassical analysis~\cite{Skalozub:1986gw, Ambjorn:1988fx, Ambjorn:1988tm, Ambjorn:1988gb, Ambjorn:1989bd}. A notable difference between the theory and simulations appears to be in the geometrical structure of this phase, which, contrary to the hexagonal crystalline order, resides in an exotic disordered solid/liquid state. The observed condensation of the $W$ and $Z$ fields implies that the vacuum in this phase should also be an unusual, strongly anisotropic superconductor and, simultaneously, an anisotropic superfluid~\cite{Chernodub:2010qx, Chernodub:2012fi}. Such anisotropic superconducting properties also imply that the vacuum becomes a hyperbolic metamaterial that behaves as a diffractionless 'perfect lens.'~\cite{Smolyaninov:2011wc, Smolyaninov:2012yh, Smolyaninov:2013ma}. 

The results obtained in the electroweak model---where the existence of the superconducting phase can be revealed analytically at a classical level---suggest that a similar situation can occur in QCD: the superconducting inhomogeneous phase may be reached in strong magnetic fields in the absence of a tachyonic instability and with a perfectly analytical behavior of masses of the $\rho$ mesons. In the absence of the Higgs field in QCD, which has a global expectation value in the vortex phase in the electroweak model, the existence of the inhomogeneous phase of QCD is difficult to reveal. The observables that could show the existence of the vortex structure should perhaps be related to local correlators of the $\rho$-meson fields such as energy~\eqref{eq_energy}, electric current~\eqref{eq_current}, and vorticity~\eqref{eq_vortex} that were shown to exhibit robust vortex-like features in quenched QCD in strong magnetic fields~\cite{Braguta:2012fol}.

    \section{Outlook}
    
    The chiral magnetic effect in QCD is by now understood as an inherently  out-of-equilibrium phenomenon. The main challenge for the theory remains to find reliable ways to estimate the size of the effect including interactions. Comparison to heavy-ion experiments can be carried out through hydrodynamic simulations, which need the CME conductivity as an input. Via linear response theory, out-of-equilibrium transport coefficients can be determined using lattice Monte-Carlo simulations of equilibrium QCD. Nevertheless, purely non-equilibrium properties are hidden in the spectral function, in constrast to the Euclidean correlators directly available on the lattice. As a desirable extension of current lattice simulations, an analytic continuation (spectral reconstruction) must be performed to obtain the out-of-equilibrium CME conductivity. The calculation of this conductivity from first-principle lattice QCD simulations could, in combination with other effective theory approaches to strongly interacting matter phenomenology, provide valuable information towards the goal of detecting the CME in heavy-ion collision experiments.
    
    Lattice simulations turned out to be very efficient in studying  equilibrium properties of QCD, such as the QCD phase diagram or its equation of state. The introduction of an external magnetic field, from one side, leads to plenty of new phenomena, including inverse magnetic catalysis, possible critical endpoints, enhancement of fluctuations and others. From another side, it makes the parameter space of the system very large and complicates lattices studies, especially if one takes the continuum limit. Despite recent progress in understanding QCD bulk properties in the temperature-magnetic field-baryon density space, there are still multiple open issues, which could be studied further. These include, e.g., the existence of the critical endpoint at large magnetic fields and its connection to the chiral critical endpoint in QCD in the $\mu-T$ plane, or the existence of new non-trivial phases. A more detailed study of QCD properties in a larger parameter space, including strangeness or electric chemical potential, as well as the tuning of the parameters to the conditions relevant for the heavy ion collision experiments would be beneficial. One could also possibly extend existing studies and results to various other parameters, including the chiral chemical potential $\mu_5$, electric field $eE$, non-homogeneous fields and others.
    
    Novel chiral hydrodynamic effects arising in extreme magnetic fields, see Sec.~\ref{sec:Kaminski}, have the potential to heavily impact various disciplines. 
    Developing chiral hydrodynamics as an effective field theory has lead to the conclusion that in strong magnetic fields, a large number of novel transport coefficients and susceptibilities can arise, see also Fig.~\ref{fig:parityViolatingTransport}. 
    The relevance of these novel transport effects for QCD plasma, neutron star mergers, and astrophysical plasma has to be estimated. If they turn out to be relevant, they should be included in hydrodynamic codes used for the analysis of heavy-ion collision data or for the simulation of neutron star evolution, especially for magnetars. 
    Such effects can, for example, account for the observed neutron star kicks. 
    A subset of the novel transport effects are susceptibilities obeying time-independent Kubo relations, and can thus be computed in equilibrium from lattice QCD. 
    For the example of the \emph{magnetic vorticity susceptibility}, $M_2$, Eq.~\eqref{eq:Kubo-stat} and Fig.~\ref{fig:parityViolatingTransport} illustrate what would need to be calculated for such susceptibilities: In a homogenous magnetic field pointing in $z$-direction, one needs a $y$-dependent magnetic field in $x$-direction in order to have non-vanishing \emph{magnetic vorticity}. In this state, the $\langle T^{xz}(+k) T^yz(-k)\rangle$ correlator needs to be calculated. No time-dependence is needed here, only spatial inhomogeneity, making $M_2$ and its peers ($M_1$, $M_3$, $M_4$, and $M_5$) 
    a well defined target for future lattice computations. 
    A complete extension of the chiral hydrodynamics framework from Sec.~\ref{sec:Kaminski} into chiral magnetohydrodynamics, including dynamical magnetic fields obeying Maxwell's equations  
    is an open problem. 
    In addition, the framework should be extended to include axial $U(1)$ \emph{and} vector $U(1)$ symmetries simultaneously. 
    Taking chiral transport far away from equilibrium, holographic models show that the time evolution of the CME currents very strongly depends on the choice of initial conditions and parameter choices. 
    A fully dynamical study including dynamical magnetic field and dynamically generated axial charges is a future goal for holographic models. 
    In general, holographic models including magnetic fields, serve as discovery tools for new physics and as testing grounds for novel hydrodynamic effects. They can even be pushed to investigate plasma dynamics far from equilibrium, as illustrated in the CME case study. 

    
Another facet of the aforementioned transport phenomena with great potential to be explored is their condensed matter analogues. The table-top setups used in this kind of study are enormously simpler than high energy experiments and have attracted attention of distinct communities, providing technological applications, insight on fundamental physics and hopefully will provide some insight on their non-Abelian cousins. So far, a few groups have claimed to have observed the CME in (3+1)D materials. However, it is still hard to state firmly that the samples really belong to the Weyl type. ARPES (Angle-resolved Photoemission Spectroscopy) technique has given support on this issue but at this point it is very useful to build analogues in materials whose charge carriers can be more easily identified as relativistic-like. Reminding that the CME is one example of a plethora of transport phenomena, anomalous or not, that may take place in the quark-gluon plasma, increasing our understanding and confidence on the nature of these materials and having other kinds of structures that may harbor analogues will allow for the search of other transport mechanisms such as chiral separation effect, chiral magnetic wave, etc.

Finally it is worth to remark that by constructing these type of analogues is a rich procedure to export benefits from high energy to condensed matter, since several theoretical aspects of the theory of anomalies and transport in general have been well developed and understood by particle physicists. On the other hand, it would be breakthrough if one could formulate how to explore tabletop setups in order to have more concrete insight about their hard to access high energy counterparts. 



The relevance of the strong magnetic fields of the electroweak scale to the physics of the present-day Universe poses an intriguing question. Such fields with the the typical electroweak magnitudes of $10^{20}$~T may still exist in the atmospheres of the extreme magnetically charged black holes. These magnetized Reissner-Nordstr\" om black holes should have zero temperature, implying that they do not radiate Hawking radiation and are, potentially, stable. In-falling matter will, however, heat the atmospheres of the black holes and lead to thermal fluctuations of the electroweak superconducting vortices. Being positioned above the event horizon of such a black hole, the vortices should emit high-energy rays that, after a gravitational well red-shift, will appear, for a distant observer, as radiation in a lower-energy electromagnetic spectrum. If the population of such black holes is statistically significant, their radiation can potentially have an observable imprint on the cosmic background.

Our idea poses the challenging question of finding properties of the superconducting electroweak phase subjected to the strong gravitational field of a black hole. To this end, the lattice gauge theory methods adapted to curved space-times can provide us with insight into the realistic structure of this vortex-populated phase and give us vital knowledge for detecting magnetic black holes using astrophysical observations.


	\newpage
	\graphicspath{{./Figures_Eff_Models/}}

%
%
\part{Effective Models}
\label{Eff_Mod}

\section{Introduction}
\label{EFF_intro}

Strongly interacting matter is one of the many fascinating physical systems being intensively investigated in the last decade.
Nuclear matter, quark matter and even cold atoms may be considered strongly interacting matter, though in different scales.
Most of the problems in dealing with such systems arise from the (infinite) degrees of freedom and from the fact that the
couplings are strong (non-perturbative). Without any extreme conditions, that may come from higher temperatures, higher
densities, strong magnetic fields or any other environmental condition, strong interacting matter is already a complex system
due to its many-body nature. This means that the description of such system has to undergo some approximation, even in ab initio
frameworks. Many different approaches have been used and most of what we currently know about strongly interacting matter
is the result of a great effort during decades in order to understand systems that are supported by strong interactions.

Many effective models have been developed to describe few-nucleon systems, nuclear matter, nuclei and QCD matter.
Although they were designed to describe some specific properties, they provided a great knowledge on strong interactions
that would be extremely difficult to acquire directly from the fundamental theory with first-principle calculations. In this Part, several effective approaches aimed to describe strongly interacting magnetized matter
will be discussed.

 %
 %

 \section{Modified NJL models}

Among the many effective models that have been developed to study strongly interacting
matter under extreme conditions, there is one approach that rely on the NJL model \cite{Nambu:1961tp} for
a basic strong interaction, in general, two and three-flavor quark matter have been considered. The unknown details are represented by a thermo-magnetic
coupling $G(eB,~T)$ which encodes the temperature and magnetic field effects.
The first
attempts in this direction were made one decade ago and considered an analytical {\it ansatz}
for the thermo-magnetic coupling \cite{Farias:2014eca,Ferreira:2014kpa}. Later, in 2017, an improvement of
this approach was developed by numerically building $G(eB,~T)$ with the constrain that
the average of $u$ and $d$ condensates obtained with lattice QCD are reproduced by the
effective model for given values of $T$ and $eB$ \cite{Farias:2016gmy}. The beauty of this
approach is that it reproduces the inverse magnetic catalysis observed in lattice QCD
calculations in a very simple way, a feature that the standard NJL model fails do describe.

\begin{figure}[!htb]
    \begin{center}
        \includegraphics[scale=0.9]{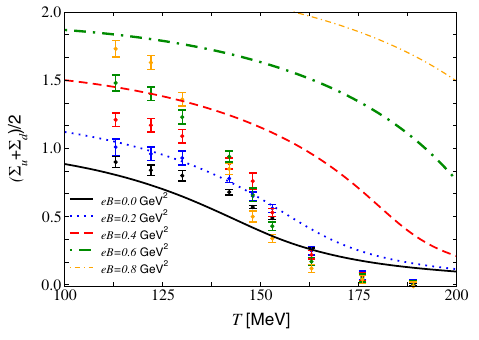}
        \hspace{1cm}
        \includegraphics[scale=0.9]{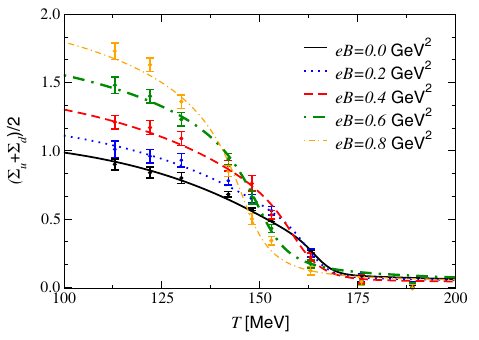}
        \caption{Average $u$ and $d$ quark condensates as a function of the temperature
        for some values of the magnetic field, for fixed coupling $G$ (left panel) and
        thermo-magnetic coupling $G(eB,T)$ (right panel).}
        \label{fig1-ssa}
    \end{center}
\end{figure}

In Fig. \ref{fig1-ssa}, we show how the average $u$ and $d$ quark condensates obtained
from lattice QCD simulations can be reproduced in the NJL model by introducing a
thermo-magnetic coupling, $G(eB,T)$ , which is constructed by a fit to the lattice
QCD results. Note that the standard NJL model gives a very poor description of the
{\it ab-initio} results.

\begin{figure}[!htb]
    \begin{center}
        \includegraphics[scale=0.9]{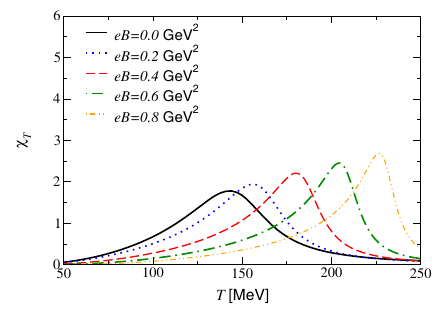}
        \hspace{1cm}
        \includegraphics[scale=0.9]{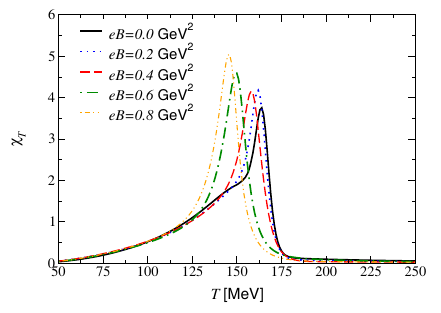}
        \caption{Thermal susceptibilities as a function of the temperature
        for some values of the magnetic field, for fixed coupling $G$ (left panel) and
        thermo-magnetic coupling $G(eB,T)$ (right panel).}
        \label{fig2-ssa}
    \end{center}
\end{figure}

\begin{figure}[!htb]
    \begin{center}
        \includegraphics[scale=0.9]{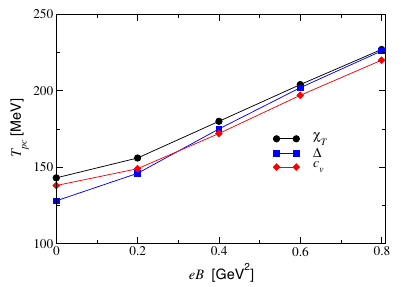}
        \hspace{1cm}
        \includegraphics[scale=0.9]{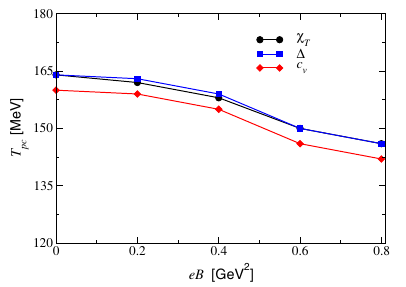}
        \caption{Pseudo-critical temperature as a function of the magnetic field
        in three strategies, for a fixed coupling $G$ (left panel) and for the
        thermo-magnetic coupling $G(eB,T)$ (right panel).}
        \label{fig3-ssa}
    \end{center}
\end{figure}

With the thermo-magnetic coupling, the inverse magnetic catalysis is obtained with
the thermo-magnetic coupling, as can be observed in the thermal susceptibilities displayed
in Fig. \ref{fig2-ssa} and in the pseudo-critical temperature for the chiral transition
shown in Fig. \ref{fig3-ssa}.

\begin{figure}[!htb]
    \begin{center}
        \includegraphics[scale=0.9]{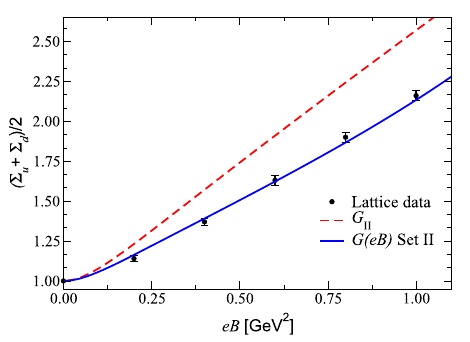}
        \hspace{1cm}
        \includegraphics[scale=0.9]{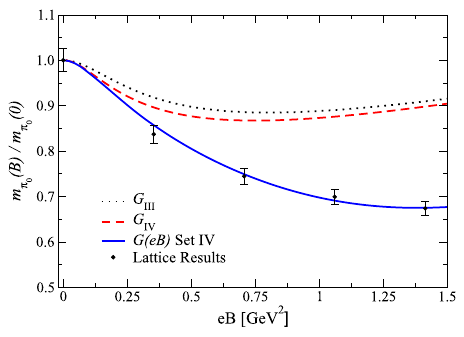}
        \caption{Left panel: average $u$ and $d$ quark fields as a function of the magnetic field, for fixed coupling $G$ (dashed red line) and
        thermo-magnetic coupling $G(eB,T)$ (solid blue line), compared to lattice QCD
        (black points). Right panel: Pion mass as a function of the magnetic field, for fixed coupling $G$ (dashed red line and dotted black line) and
        thermo-magnetic coupling $G(eB,T)$ (solid blue line), compared to lattice QCD
        (black points)}
        \label{fig4-ssa}
    \end{center}
\end{figure}

Another interesting result is obtained, at zero temperature, by using a $B$-dependent
coupling which is also built by performing a fit procedure with lattice QCD results
\cite{Avancini:2016fgq}. This can be observed in Fig. \ref{fig4-ssa} where we display
the condensates and the pion mass as functions of the magnetic field at $T=0$.
The B-dependent coupling is constructed to reproduce the average condensates, and the pion
mass is a prediction based on the new coupling.
In ref.\cite{Avancini:2016fgq}, we emphasize the importance of a proper regularization for the excellent agreement between the pion pole mass calculation and LQCD results.  This issue is a key point for any calculation using effective models under extreme electromagnetic fields, although, many authors still do not use correct regularization techniques causing, in general, unphysical results for the observables. A common signal of improper regularizations is unphysical oscillations that are artifacts of a wrong calculation \cite{Avancini:2019wed}.

\begin{figure}[!htb]
    \begin{center}
        \includegraphics[scale=0.38]{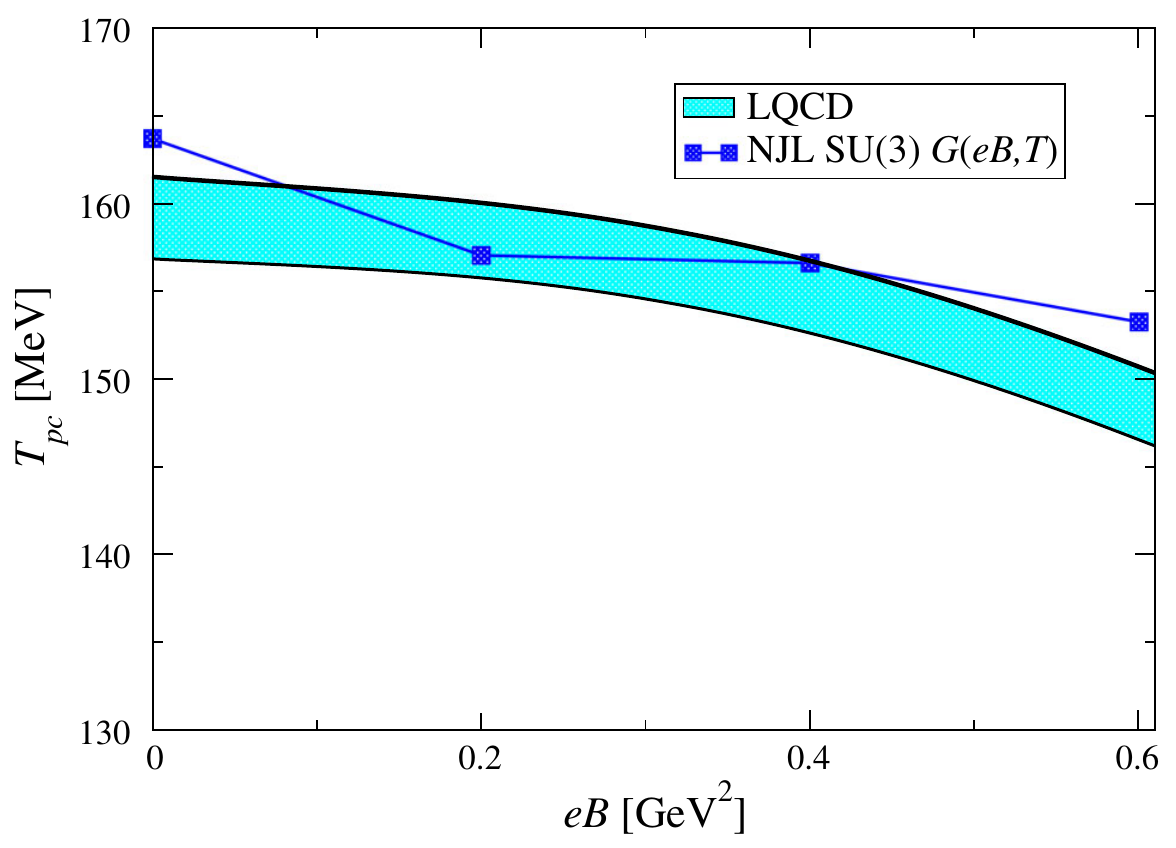}
        \hspace{1cm}
        \includegraphics[scale=0.9]{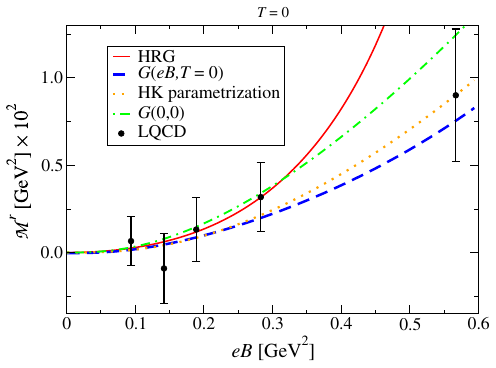}
        \caption{Left panel: pseudo-critical temperature as a function of the magnetic field
        with the thermo-magnetic coupling $G(eB,T)$. Right panel: renormalized magnetization
        at zero temperature, as a function of the magnetic field, in different calculations.}
        \label{fig5-ssa}
    \end{center}
\end{figure}

\begin{figure}[!htb]
    \begin{center}
        \includegraphics[scale=0.9]{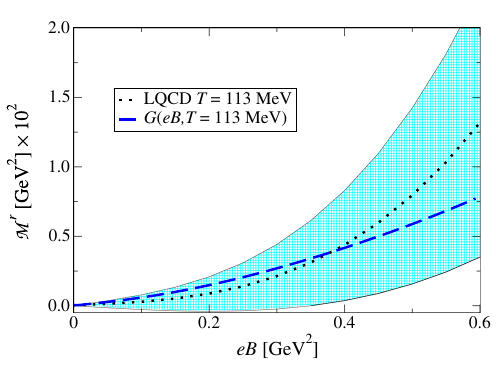}
        \hspace{1cm}
        \includegraphics[scale=0.9]{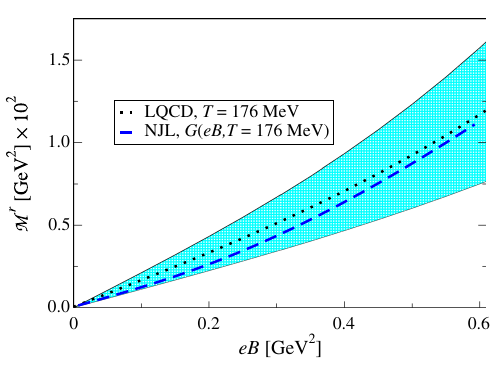}
        \caption{Renormalized magnetization as a function of the magnetic field, compared to
        lattice QCD simulations, below the chiral transition (left panel) and above the
        chiral transition (right panel).}
        \label{fig6-ssa}
    \end{center}
\end{figure}

\begin{figure}[!htb]
    \begin{center}
        \includegraphics[scale=0.9]{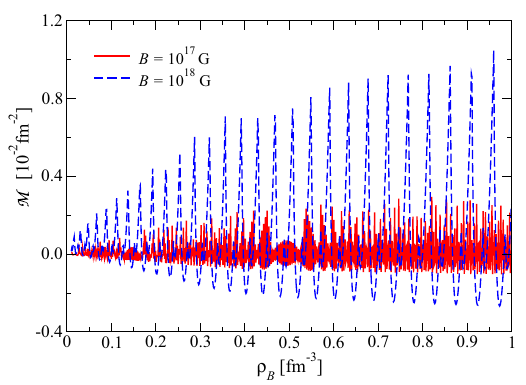}
        \hspace{1cm}
        \includegraphics[scale=0.9]{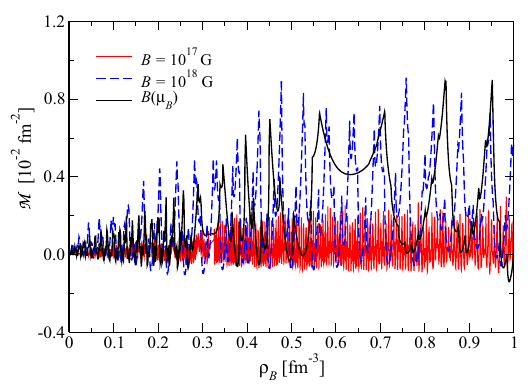}
        \caption{Magnetization as a function of the barionic density for two values of the
        magnetic field.}
        \label{fig7-ssa}
    \end{center}
\end{figure}

The effects produced by the thermo-magnetic coupling also affects
the magnetization of the hadronic matter \cite{Tavares:2021fik}.
A numerical evaluation is summarized in Figs. \ref{fig5-ssa} and
\ref{fig6-ssa}, where the pseudo-critical temperature and the
renormalized magnetization, obtained in our modified NJL model,
are compared to lattice QCD calculations. In another study \cite{Avancini:2017gck}, we computed the magnetization in the case of
zero temperature and finite density. Figure \ref{fig7-ssa} displays
the magnetization of cold hadronic matter for barionic densities between
zero and $1~{\rm fm}^{-3}$, considering two values of the magnetic field.
In the presence of strong magnetic fields, the rotational symmetry breaking due to the magnetic field, makes the parallel and perpendicular pressure related to the direction of the external field to be different, and the size of this difference is determined by the magnetization.  Our results in \cite{Avancini:2017gck} show that for typical magnetic fields found in magnetars, i.e., a special class of neutron stars with strong surface magnetic fields, the parallel and perpendicular pressures are almost identical.

\begin{figure}[!htb]
    \begin{center}
        \includegraphics[scale=2.0]{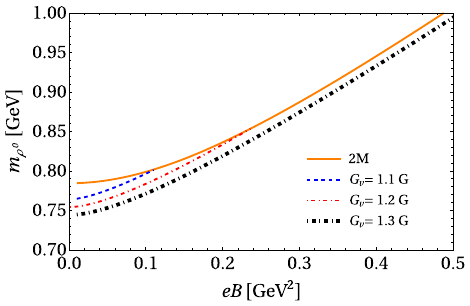}
        \caption{Neutral $\rho$ meson mass as a function of the magnetic field for different values of the vector interaction, compared to the effective quark mass times two.}
        \label{fig8-ssa}
    \end{center}
\end{figure}

As a last sample of a recent application of the NJL model, we show the
calculation of the neutral $\rho$ meson pole mass in a full random phase
approximation, considering vector interactions \cite{Avancini:2022qcp}.
In Fig. \ref{fig8-ssa}, we show the $\rho^0$ mass as a function of the
magnetic field background for three different values of the vector interaction.

	\clearpage
%
%
\section{Background magnetic field helps fixing model parameters}\label{sub:eff_mod_para_fix_cai}

\subsection{Motivation}\label{sssec:mot_cai}
In this section of the review, we will briefly focus on how the known properties of a QCD medium in the presence of an external magnetic field such as magnetic catalysis (MC) and inverse magnetic catalysis (IMC) can be utilised to constrain the parameter space of the model. In turn, the constraints can be tested by predicting some observables. It is also interesting to observe how these medium properties can be used to understand better the working principles of such effective models.

	After being first observed in lattice QCD, the IMC effect has been observed in various effective models, including the NJL model (local version), the LSM, and the QM model by tweaking the models, particularly by introducing a magnetic field and a temperature-dependent coupling constant~\cite{Farias:2014eca,Ferreira:2014kpa,Farias:2016gmy}. This obviously requires the addition of further parameters to the models. On the other hand, the non-local NJL model can capture the IMC effect through its very construction and does not require the introduction of any additional parameters.

	We realise that the running of the coupling constant is the crucial property that is responsible for producing the IMC effect. In the non-local NJL model, the coupling constant becomes momentum-dependent through the form factor. On the other hand, in local versions, the coupling constant becomes a running one only after it is made dependent on temperature and magnetic field.

	This particular feature of the non-local NJL model gives us an opportunity to check whether such a model is capable of reproducing other known observables predicted by lattice QCD, even without introducing any new external parameters. In the process, it hands us the scope to constrain the model parameters which are otherwise completely unknown or not so precisely known.

	One such parameter is the strength of the 't Hooft determinant term in a 2-flavour NJL model~\cite{Buballa:2003qv}. Its value is governed by a dimensionless parameter, $c$. There exists literature that provides us with some possible ranges of it~\cite{Frank:2003ve,Boomsma:2009yk}. However, no efforts had been made to directly determine its value until recently, when a 2-flavour non-local NJL model in the presence of a magnetic field was utilized to fit its value~\cite{Ali:2020jsy}.

	Once its value is estimated, it can be further tested by calculating observables such as topological susceptibility $(\chi_t)$ at zero magnetic field and predicting its $(\chi_t)$ values for non-zero magnetic field~\cite{Ali:2020jsy}. This value of $c$ is further utilised in a study that investigates the intertwined effects of the 't Hooft determinant term and a difference in light quark masses on the QCD corrections to the pion mass difference~\cite{Ali:2021zsh}.

	For this part of the review, we will first provide the required formalism in short and then briefly describe the process of fitting the parameter $c$ while describing some associated results. We, then, present the results obtained with that fitted $c$ value for different observables

	\subsection{Determining the strength of the 't Hooft determinant term}
	 \label{sssec:mod_n_para_fit_cai}

	 The version of the Lagrangian that we utilise here is a 2-flavour non-local NJl model written in the Euclidean space as,
	 \begin{equation}
        {\cal L}_{\rm NJL}= {\bar{\psi}}\left(-i\slashed\partial+\hat{m}\right)\psi-G_1\left\{j_a(x)j_a(x)+\tilde{j}_a(x)\tilde{j}_a(x)\right\} -G_2\left\{j_a(x)j_a(x)-\tilde{j}_a(x)\tilde{j}_a(x)\right\}\;,
        \label{eq:njl_cai}
     \end{equation}
	 where the currents are given by
	 \begin{equation}
        j_{a}(x)/\tilde{j}_{a}(x)=\int d^4z\ {\cal H}(z) \bar{\psi}\left(x+\frac{z}{2}\right)\Gamma_{a}/\tilde{\Gamma}_{a}\,\psi\left(x-\frac{z}{2}\right).
        \label{eq:currents_cai}
     \end{equation}
     The $\Gamma$'s are defined as $\Gamma=(\Gamma_0,\vec{\Gamma}_{j})=(\mathbb{I},i\gamma_5\vec{\tau}^j)$ and $\tilde{\Gamma}=(\tilde{\Gamma}_0,\tilde{\vec{\Gamma}}_{{j}})=(i\gamma_5,\vec{\tau}^j)$ with $\vec{\tau}=(\tau^1,\tau^2,\tau^3)$ representing the Pauli matrices. The current quark mass matrix, $\hat{m}=m\times\mathbb{I}$, with $m_u=m_d=m$. ${\cal H}(z)$ is the form factor, which if taken to be a $\delta$-function reduces the Lagrangian in Eq.~\ref{eq:njl_cai} into a local one. It is important to note that the second part of the interaction term breaks the $U(1)_A$ term. Thus, putting $G_2=0$ makes the Lagrangian symmetric under $SU(2)_V\times SU(2)_A\times U(1)_V\times U(1)_A$ in the chiral limit.

     In the absence of isospin symmetry breaking (equal non-zero current quark mass does not break it), the symmetry allows only $\langle\bar\psi\psi\rangle$ condensate that depends on the $G_1+G_2$ combination. On the other hand, if there is an isospin symmetry-breaking agent such as isospin chemical potential $(\mu_I)$ and/or magnetic field $(eB)$ there is an explicit breaking of $SU(2)_V$ symmetry. In that circumstance, one can have another condensate such as $\langle\bar\psi\tau_3\psi\rangle$ that couples to the fermions with coupling strength $G_1-G_2$. This motivates to parameterise the coupling constants $G_1=(1-c)G_0/2$ and $G_2=cG_0/2$ in terms of a single coupling constant, $G_0$ and the parameter, $c$. Its value $(c)$ determines the strength of the 't Hooft determinant term, and $c=1/2$ reduces the Lagrangian in Eq.~\ref{eq:njl_cai} to that of the usual NJL model.

     It is well-known how to obtain the free energy starting from the mean-field version of the Lagrangian given in Eq.~\ref{eq:njl_cai}. Eventually, to perform numerical calculation a few more things need to be known. One of them is the form of ${\cal H}(z)$, the form factor. The Fourier transform of which is denoted as $g(p)$ and can only be a function of $p^2$ due to the Lorentz invariance: $g(p)=e^{-p^2/\Lambda^2}.$ The other important things are the parameters in the model such as $m$, $G_0$ and $\Lambda$. It should be mentioned that $\Lambda$ does not serve as a cut-off in the same way as in the usual local model; instead, it sets the scale of the theory.

     These parameters are fixed by fitting known QCD observables such as pion mass $(m_\pi)$, pion decay constant $(F_\pi)$ and the chiral condensate $(\langle\bar\psi\psi\rangle)$. Occasionally, these observables are chosen from different studies. However, in the pursuit of determining the $c$ parameter~\cite{Ali:2020jsy} these are taken from a single lattice QCD study~\cite{Fukaya:2007pn} to maintain consistency. One needs to be careful while fitting such observables obtained in lattice QCD calculation, for which generally the scale is set at $2$ GeV. On the other hand, the scale of the effective model is roughly $1$ GeV. In order to express the lattice QCD results at a desired scale of $1$ GeV one can utilise the perturbative renormalisation group (RG) running~\cite{Giusti:1998wy}.

     After utilising perturbative RG running the condensate value becomes $\langle\bar{\psi}\psi\rangle^{1/3}|_{\mu=1\,{\rm GeV}}=224.8(3.7)\,{\rm MeV}$. On the other hand, $F_\pi$ being a scale-independent quantity remains the same, $F_\pi=87.3(5.6)\,{\rm MeV}$. The pion mass is taken to be $135\,{\rm MeV}$. To explore the whole observable space while fitting the model parameters one needs to include the errors in both in $\langle\bar{\psi}\psi\rangle$ and $F_\pi$. This makes a total of nine different sets of observables to fit. Out of which, considering the four corner sets along with the central one will be enough to understand all possible physical scenarios arising from different combinations.

     Out of those five observable sets, the most important three are quoted here along with the fitted parameter values in Table~\ref{tab:para_cai}. The rest of the sets can be found in Ref.~\cite{Ali:2020jsy}. The letters C, H and L stand for the central, highest and lowest values of the observables, respectively. For each parameter set, the first letter corresponds to the value of the condensate and the second to that of the pion decay constant. With the fitted parameters, one can now proceed with the calculation. So far the whole discussion is in a vacuum. To introduce the magnetic field the non-local currents in Eq.~\ref{eq:njl_cai} need to be modified as
     \begin{equation}
        j_{a}(x)/\tilde{j}_{a}(x)=\int d^4z\ {\cal H}(z) \bar{\psi}\left(x+\frac{z}{2}\right)W^\dagger\left(x+\frac{z}{2},x\right)\Gamma_{a}/\tilde{\Gamma}_{a}W\left(x,x-\frac{z}{2}\right)\,\psi\left(x-\frac{z}{2}\right),
        \label{eq:currents_eB_cai}
     \end{equation}
     where, $W(s,t)={\rm P}\,{\rm exp}\left[-i{\hat Q\int_s^t dr_\mu A_\mu(r)}\right]$ with $r$ running over arbitrary path and connecting $s$ and $t$. $A_\mu(x)$ represents the background magnetic field and ${\hat Q}={\rm diag} (q_u,q_d)$ with $q_u$ and $q_d$ being the electric charge of $u$ and $d$ quark, respectively.

     It is to be noted here that the very definition of the currents in Eq.~\ref{eq:currents_eB_cai} helps produce the IMC effect in a non-local NJL model. To get to the non-zero temperature Matsubara formalism is used, which connects the Euclidean time component to the temperature.

\begin{figure}[!htb]
    \begin{center}
        \includegraphics[scale=0.9]{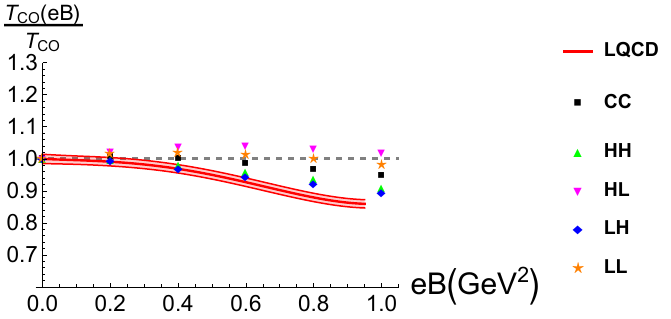}
        \caption{QCD phase diagram in the $T-eB$ plane for different parameter sets and also in lattice QCD.}
        \label{fig:QCD_pd_cai}
    \end{center}
    \end{figure}

    \begin{table}[!htbp]
	\centering
	\begin{tabular}{|m{2.8cm} | m{2.8cm} | m{1.8cm} | m{1.6cm} || m{1.8cm} |m{1.6cm}| m{1.8cm} | m{1.6cm}| }
		\hline
		& $\langle\bar{\psi_f}\psi_f\rangle^{1/3}$(MeV) & $m_\pi$(MeV) & $F_\pi$(MeV)& $F_{\pi,0}$(MeV)& $m$(MeV) & $G_0({\rm GeV^{-2}})$ & $\Lambda$(MeV) \\
		\hline
		Parameter Set CC  & 224.8    & 135  & 87.3   & 84.25 & 5.87  & 43.34   & 697.22   \\
		\hline
		Parameter Set HH  & 228.6  & 135  & 92.9   & 90.63 & 6.31  & 57.15   & 660.46   \\
		\hline
		Parameter Set LH  & 221.1    & 135  & 92.9   & 91.00 & 6.94  & 80.26   & 605.05   \\
		\hline
    \end{tabular}
    \caption{Central and two other sets along the fitted parameters of the model given on the right of the
	double bars.}
	\label{tab:para_cai}
    \end{table}
    In Fig.~\ref{fig:QCD_pd_cai}, the QCD phase diagram (PD) in the $T-eB$ plane is shown. The band represents the lattice QCD data~\cite{Bali:2011qj}. The plot displays five data sets~\cite{Ali:2020jsy} including the three given in Table~\ref{tab:para_cai} that reproduces the decreasing behaviour of the crossover temperature $(T_{\rm CO})$. The IMC effect is also captured for these three sets. It is to be mentioned that although decreasing $T_{\rm CO}$ and the IMC effect accompany each other, the simultaneous occurrence of the two is not guaranteed~\cite{DElia:2018xwo}. The $T_{\rm CO}$ is calculated from the inflection point of the condensate average for each value of $eB$ following the lattice QCD given definition of the condensate~\cite{Bali:2012zg} as.
    \begin{equation}
        \Sigma_{B,T}^f=\frac{2m}{{\cal N}^4}\left[\langle\bar\psi_f\psi_f\rangle^{\rm reg}_{B,T}-\langle\bar\psi_f\psi_f\rangle^{\rm reg}_{0,0}\right]+1,
        \label{eq:sig_scaled_lat_cai}
    \end{equation}
    where ${\cal N}=(m_{\pi}F_{\pi,0})^{1/2}$ with $F_{\pi,0}$ being the pion decay constant in the chiral limit.
    From the three sets which capture the decreasing behaviour of $T_{\rm CO}$, the sets LH and HH do it better and are comparable. Thus, these are the two parameter sets for which the fitted $c$-values are further considered.

    \begin{minipage}{\textwidth}
      \begin{minipage}[b]{0.49\textwidth}
        \centering
        \includegraphics[scale=0.6]{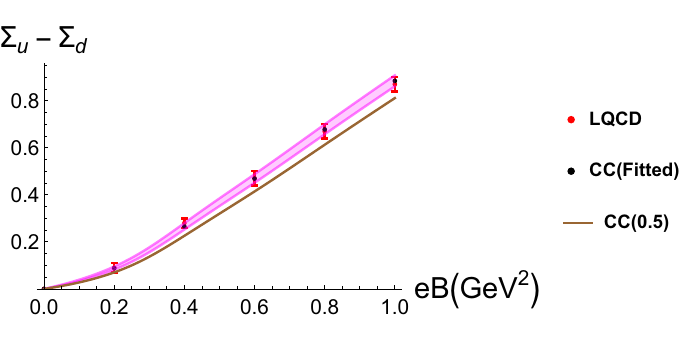}
        \captionof{figure}{The condensate difference fitted in $c$.}
        \label{fig:c_fit_cai}
      \end{minipage}
    \hfill
      \begin{minipage}[b]{0.49\textwidth}
        \centering
        \begin{tabular}{|m{2.8cm} | m{2.2cm} | m{2cm} |  }
		\hline
		                 &   c        &  $\chi^2$ per DoF  \\
		\hline
		Parameter Set HH  &  $0.044 \pm 0.079$   &   $ 0.149 $\\
		\hline
		Parameter Set LH  &  $0.149 \pm 0.103$   &   $ 0.634 $  \\
		\hline
        \end{tabular}
        \captionof{table}{$\chi^2$ fitting of condensate difference in $c$.}
        \label{tab:fit_c_values_cai}
      \end{minipage}
    \end{minipage}\\

    In Fig.~\ref{fig:c_fit_cai}, a specimen of the fitting of the light quark condensate difference in $c$ is shown for the set CC. This is done for zero temperature and the result for $c=1/2$ is also shown. Similar figures can be obtained for other parameter sets. Two such fitted values for HH and LH sets are given in Table~\ref{tab:fit_c_values_cai}.

    To narrow down further on the best possible parameter set and decide on the $c$-value the $\eta^*$ phenomenology is evoked, $\eta^*$ is nothing but the mass of the fluctuations in the isoscalar pseudoscalar chanel~\cite{Dmitrasinovic:1996fi}. However, it cannot be directly used to fit for $c$ as there is no physical particle representing the $\eta^*$. But physical consideration can help constrain $c$. Considering it to be an admixture of $\eta$ and $\eta'$, the mass of $\eta^*$ to be a few times $\pi^0$. Imposing the physically motivated constrain that $M_{\eta^*}>400\,{\rm MeV}$ it is possible to find a lower bound on $c$~\cite{Ali:2020jsy}. It is found to be $c>0.12$. With this constraint, the only viable parameter set is LH and the corresponding fitted value of $c=0.149^{+0.103}_{-0.029}$.

    \begin{figure}[!htb]
    \begin{center}
        \includegraphics[scale=0.58]{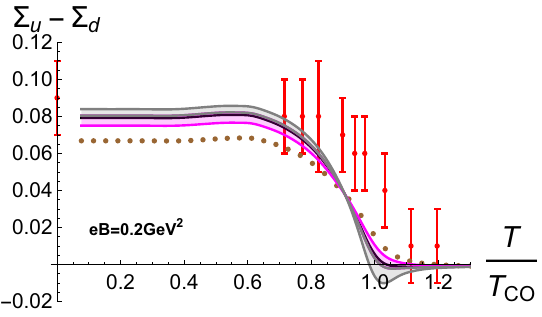}
        \includegraphics[scale=0.58]{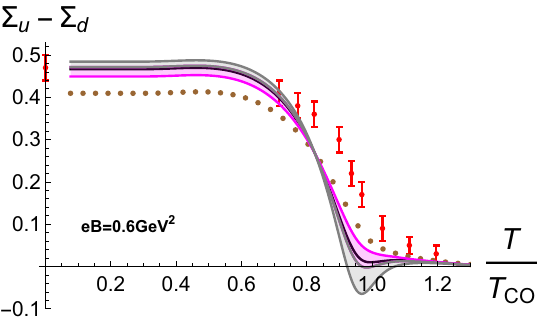}
        \includegraphics[scale=0.58]{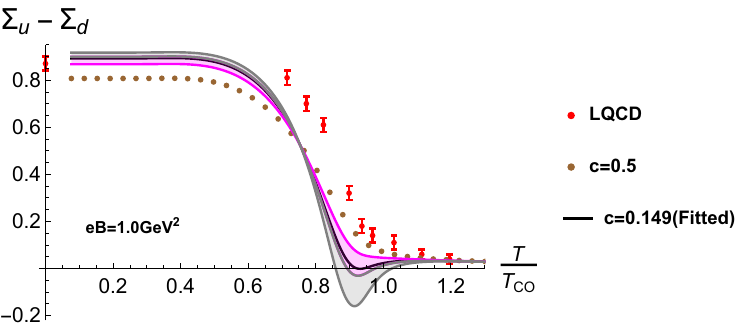}
        \caption{The condensate difference for the fitted value of $c$ in the LH parameter set.}
        \label{fig:cond_diff_cai}
    \end{center}
    \end{figure}
    With the fitted value of $c$ at zero $T$, the lattice-defined condensate difference at non-zero $T$ for different values of $eB$ is obtained as a prediction of the model. This is shown in Fig.~\ref{fig:cond_diff_cai} along with the lattice QCD result. The gray band in the figure gets excluded with a consideration of $M_{\eta^*}>400\,{\rm MeV}$ resulting in a better match with the lattice QCD result.

    \subsection{Topological susceptibility}
    \label{sssec:top_sus_cai}

\begin{figure}[!htb]
    \begin{center}
        \includegraphics[scale=0.65]{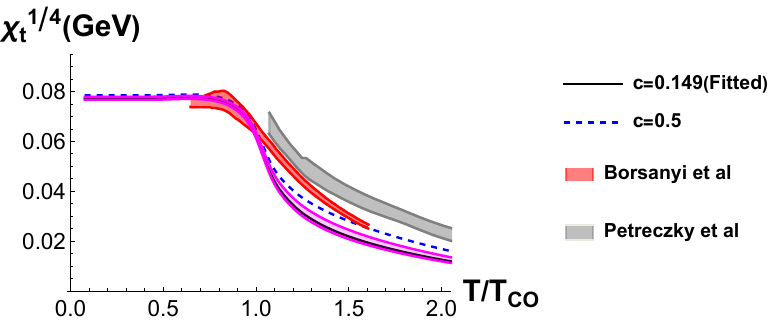}
        \includegraphics[scale=0.65]{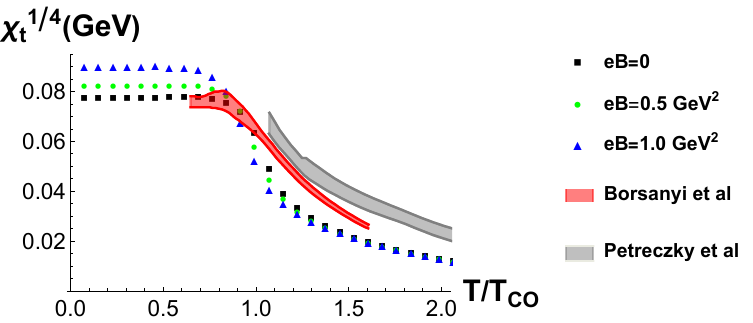}
        \caption{$\chi_t$ for zero $eB$ (left panel) and non-zero $eB$ (right panel). The red and the gray bands are lattice results.}
        \label{fig:top_sus_cai}
    \end{center}
    \end{figure}

    The CP-violating topological term that can be added to the QCD Lagrangian is written as~\cite{weinberg_1996},
    \begin{equation}
    \Delta {\cal{L}} = \epsilon^{\mu\nu\sigma\lambda}\frac{\theta}{64 \pi^2}{\cal F}^a_{\mu\nu}{\cal F}^a_{\sigma\lambda} \;,
    \label{eq:top_lag_cai}
    \end{equation}
    where $\theta$ can be related to the axion field $(a)$ as $\theta=a/f_a$, with $f_a$ being axion decay constant. This term can be removed by making a suitable $U(1)_A$ transformation on the fermionic fields. Under such a transformation the first interaction term in Eq.~\ref{eq:njl_cai} remains invariant but the second one picks up phase as
    \begin{align}
        \,G_2\left[\mathrm{cos}\theta\left\{j_a(x)j_a(x)-\tilde{j}_a(x)\tilde{j}_a(x)\right\}+2\mathrm{sin}\theta\left\{j_{0}(x)\tilde{j}_{0}(x)-j_{i}(x)\tilde{j}_{i}(x)\right\}\right],
        \label{eq:lag_axion_cai}
    \end{align}
    with $i$ running from $1$ to $3$. With this modification the free energy $(\Omega)$ is calculated and from there the topological susceptibility $(\chi_t)$,
    \begin{align}
        \chi_t=\frac{d^2\Omega}{d \theta^2}\bigg |_{a=0}.
        \label{eq:top_sus_cai}
    \end{align}
    In the left panel of Fig~\ref{fig:top_sus_cai}, we show the sensitivity of $\chi_t$ to $c$ for zero $eB$. We have both the fitted $c$-value and $c=1/2$, the standard NJL model results. For both values of $c$, $\chi_t$ below $T_{\rm CO}$ is the same and is within the lattice result given by the red band~\cite{Borsanyi:2016ksw}. With increasing temperature, they become different and fall at a faster rate than the lattice results~\cite{Borsanyi:2016ksw,Petreczky:2016vrs}. However, at non-zero $eB$ (in the right panel), the model predicts an enhancement in $\chi_t$ below $T_{\rm CO}$, which could be tested in future lattice QCD simulations.

    \subsection{QCD corrections to the pion mass difference}
    \label{sssec:pion_mass_diff_cai}
\begin{figure}
    \begin{center}
        \includegraphics[scale=0.4]{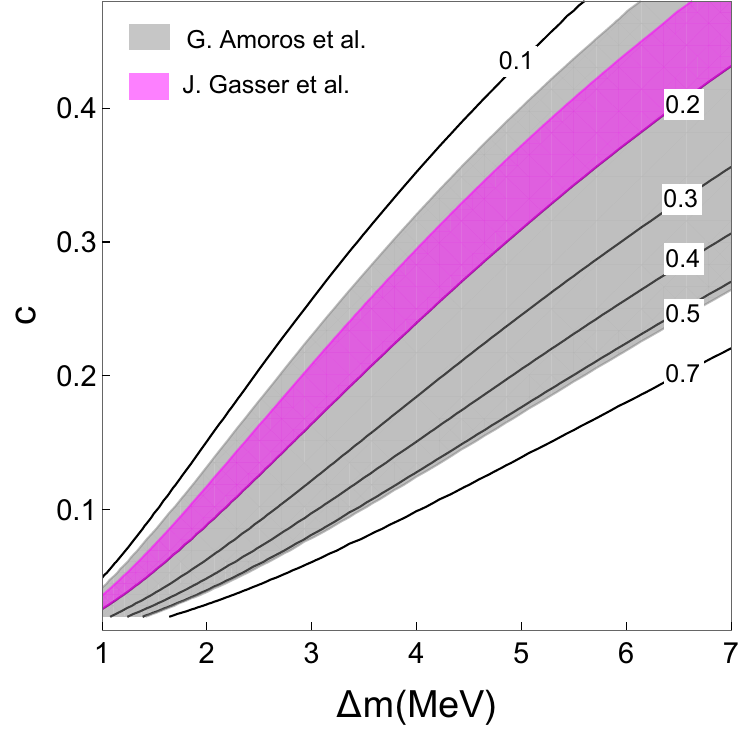}
        \caption{Contour plot of $(\Delta M_\pi)$ in $\Delta m$ and $c$. The bands are from $\chi$PT calculation (please see text).}
	\label{fig:contour_cai}
    \end{center}
    \end{figure}

    We know that the charged and neutral pion mass difference $(\Delta M_\pi)$ can arise because of both the QED and QCD contributions. The current quark mass difference $(\Delta m)$ is responsible for the QCD corrections. In a 2-flavour non-local NJL model (Eq.~\ref{eq:njl_cai}) this QCD contribution has been calculated~\cite{Ali:2021zsh} and it is found that 't Hooft determinant parameter, $c$ plays an important role when intertwined with $\Delta m$. This is shown using a contour plot in Fig.~\ref{fig:contour_cai}, which has been drawn for different $\Delta M_\pi$-values. The magenta and grey bands are from $\chi$PT calculation~\cite{Gasser:1984gg} and~\cite{Amoros:2001cp}, respectively.

\begin{figure}
    \begin{center}
        \includegraphics[scale=0.72]{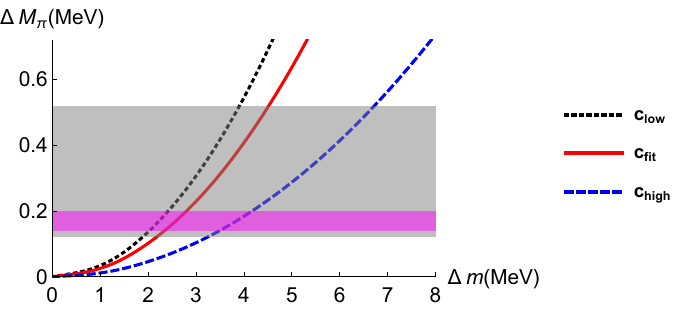}
        \includegraphics[scale=0.6]{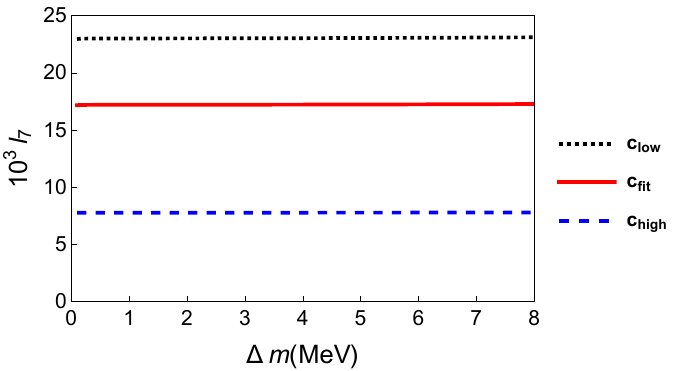}
        \caption{Left panel: $\Delta M_\pi$ as a function of $\Delta m$ for the fitted $c$ value from Ref.~\cite{Ali:2020jsy}. The bands are obtained using $\chi$PT calculations. Right panel: the $\Delta m$ dependence of the low energy constant $l_7$. For details, please check the main text.}
        \label{fig:Mass_diff_cfit_cai}
    \end{center}
    \end{figure}

    In the left panel of Fig.~\ref{fig:Mass_diff_cfit_cai}, $\Delta M_\pi$ is plotted as a function of $\Delta m$ with the fitted value of $c=0.149^{+0.103}_{-0.029}$. Using that fitted $c$-value and different $\Delta M_\pi$ values from Refs.~\cite{Gasser:1984gg,Amoros:2001cp}, one can determine the allowed range of $\Delta m$ in the model as
    \begin{align}
    \begin{split}
        \Delta m &=2.59\left(_{-0.24}^{+0.22}\right)\left(_{-0.35}^{+1.26}\right) {\rm{MeV\; for}}\; \Delta M_\pi\; \text{from Ref.~\cite{Gasser:1984gg}} {\rm{,\;and}}\\
        \Delta m &=3.55\left(_{-1.37}^{+0.97}\right)\left(_{-0.48}^{+1.73}\right) {\rm{MeV\; for}} \Delta M_\pi\; \text{from Ref.~\cite{Amoros:2001cp}}\;.
    \end{split}
    \end{align}
    In turn, these $\Delta m$-values enable us to extract the chiral perturbation theory low-energy constant, $l_7$, given in the right panel of Fig.~\ref{fig:Mass_diff_cfit_cai}. It is also observed that the value of $l_7$ increases with increasing $c$. The estimated $l_7$ value using the 2-flavour non-local NJL model is $l_7=17.2^{+5.8}_{-9.4}\times 10^{-3}$. Such an estimation is within the range provided by other studies~\cite{Gasser:1983yg,GrillidiCortona:2015jxo,Frezzotti:2021ahg}.
%
%
%
%
%

\newpage
\section{van der Waals-type models}\
\label{Lourenco}
\vspace{-1.5cm}
\subsection{Overview}

Another class of effective models for description of strongly interacting matter, more specifically protons and neutrons as degrees of freedom at different regimes of temperature and density, is based on classical extension of ideal gases thermodynamics, namely, the van der Waals~(vdW) model~\cite{book-greiner_1995,book-landau_1975}, i.e., the nucleon is considered as a finite size hard-core object~(not a particle anymore). In the context of the nuclear matter system, the exclude volume~(EV) effect is responsible for the repulsive part of the nuclear interaction, and the intermediate range inter-particle attraction is represented by the pressure reduction. However, this phenomenology is not enough to completely describe nucleonic systems for which quantum statistical effects are clearly non-negligible (the classical Maxwell-Boltzmann fluid is not suitable), and even relativistic effects can also be important in certain domains. A complete treatment in which these effects are also taken into account was performed for the first time in Ref.~\cite{PRC91-064314_2015}. The energy density at $T=0$ for symmetric nuclear matter~(SNM) in this approach reads
\begin{eqnarray}
\mathcal{E}(\rho)=\frac{2(1-b\rho)}{\pi^2}\int_0^{k_F^*}dk\,k^2(k^2+M^2)^{1/2} - a\rho^2,
\label{deev}
\end{eqnarray}
where $M$ is the nucleon rest mass, $\rho$ is the total density, $k_F^*=(3\pi^2\rho^*/2)^{\frac{1}{3}}$, $\rho^* = \rho/(1-b\rho)$, and $b$ is the EV parameter. The two free parameters ($a$, and $b$) are found by imposing to the model a bound state at the saturation density $\rho_0$ ($\approx 0.15$~fm$^{-3}$) with the correct value for the binding energy ($\approx -16$~MeV~\cite{ARNS21-93_1971}). A similar implementation is addressed in Ref.~\cite{APPB47-1943_2016}. The extension of this model to finite temperature regime allows the study of first-order liquid-gas phase transition in nuclear matter at moderate temperatures, and the thermodynamical analyses at sufficiently high temperatures. In this case, the Fermi-Dirac distribution is not a step function as in the zero temperature regime. As an example of $T>0$ calculations, we cite the critical temperature found in Ref.~\cite{PRC91-064314_2015}, $T_c\simeq 19.7$~MeV, close to some experimental estimates for this quantity~\cite{PRL89-212701_2002,PRC67-011601_2003,PRC87-054622_2013}. Furthermore, attractive and repulsive nucleon-nucleon interactions of the nucleonic vdW model were included in the hadron resonance gas model in Ref.~\cite{PRL118-182301_2017}. The second-order susceptibilities of some conserved charges calculated from this approach are shown to be compatible with lattice data for a large range of~$T$. An improvement to the aforementioned description was proposed in Ref.~\cite{PRC96-015206_2017}, where the generalized version of the model takes into account possible density dependence of the constants $a$ and $b$. The new energy density expression is constructed by making $b\rightarrow\mathcal{B}(\rho)$ and $a\rightarrow\mathcal{A}(\rho)$ in Eq.~\eqref{deev}, in the notation of Refs.~\cite{ApJ882-67_2019,JPG47-035101_2020}. A proper choice for the function $\mathcal{A}(\rho)$ selects the type of real gas model to be used in the description of~SNM. Redlich-Kwong-Soave~\cite{CR44-233_1949,CES27-1197_1972}, Peng-Robinson~\cite{IECF15-59_1976}, and Clausius equations of state are possible options, see Ref.~\cite{PRC96-015206_2017}. The function $\mathcal{B}(\rho)$ accounts for the method used to implement the excluded volume to the system, such as the one based on the Carnahan-Starling~(CS) model~\cite{IECF15-59_1976}. The traditional EV procedure is recovered by taking $\mathcal{B}(\rho)=b$. This particular choice, along with $\mathcal{A}(\rho)=a$, identifies the previous nucleonic vdW model.

A direct challenge for these models is to make them more capable of describing as many strongly interacting matter properties as possible. In this direction, we address the reader to Ref.~\cite{PRC96-045202_2017} for a generalized multicomponent system vdW model, useful to treat asymmetric nuclear matter. In a different way, the authors of Refs.~\cite{ApJ882-67_2019,JPG47-035101_2020} proposed a new form for the $\mathcal{A}(\rho)$ function, namely,
\begin{equation}
\mathcal{A}(\rho)= \frac{a}{(1+b\rho)^n},
\label{arho}
\end{equation}
and a new term in the energy density, proportional to the squared difference between protons and neutrons densities inspired by the popular relativistic mean-field~(RMF) models~\cite{PR464-113_2008,PRC90-055203_2014,PRC109-055801_2024}. It mimics the exchange of the $\rho$ meson and, consequently, generates the inequality of the numbers of protons and neutrons in the system. The four parameters of this model, named as density-dependent vdW~(\mbox{DD-vdW}) model, are determined in order to reproduce the values of saturation density, binding energy, incompressibility, and symmetry energy (the last three quantities evaluated at $\rho=\rho_0$ and $y_p=0.5$, with $y_p$ being the proton fraction of the system).

\subsection{Causality}

\begin{figure}[!htb]
\hspace{1.5cm}
\includegraphics[scale=0.29]{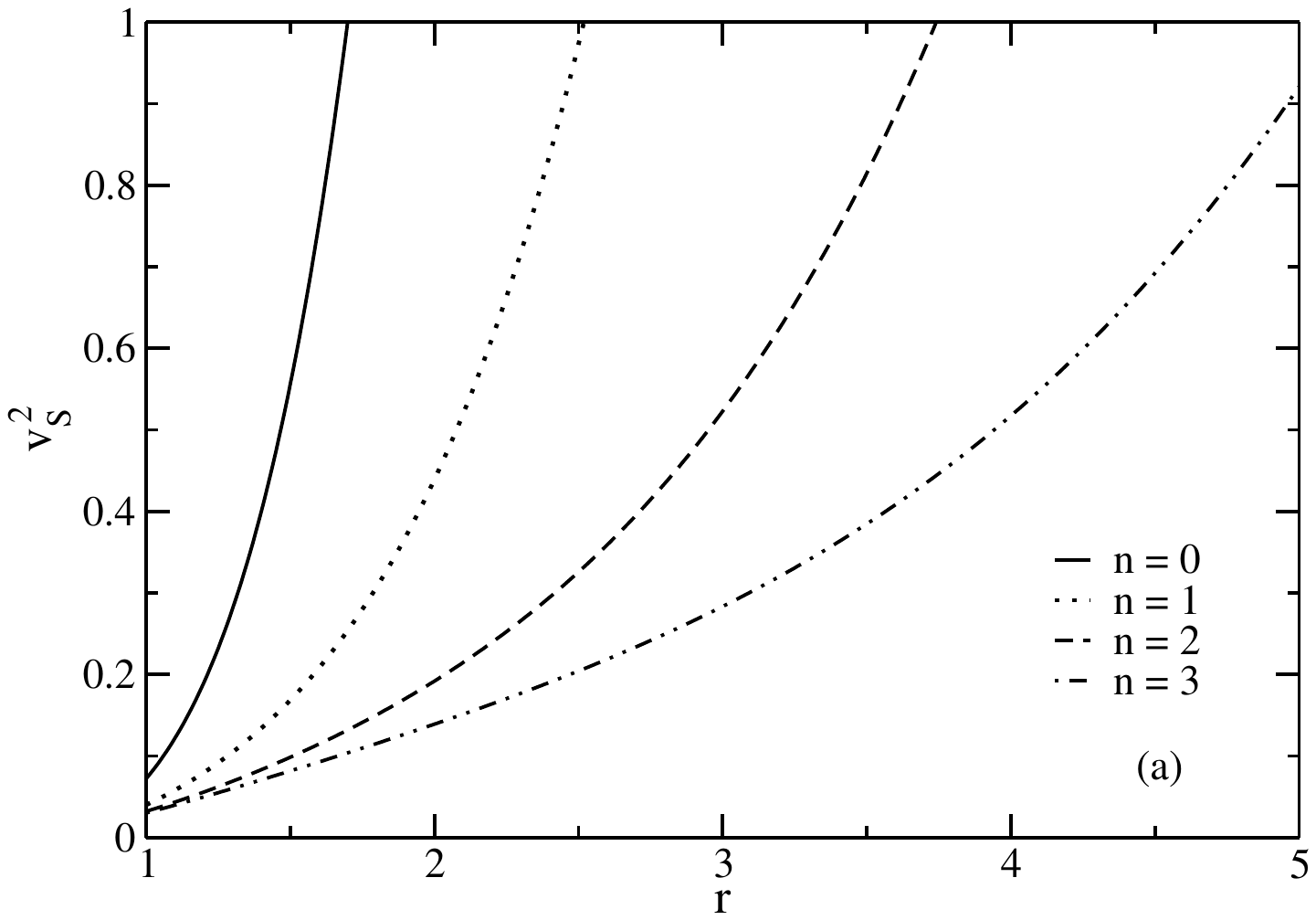}
\hspace{-1.2cm}
\includegraphics[scale=0.29]{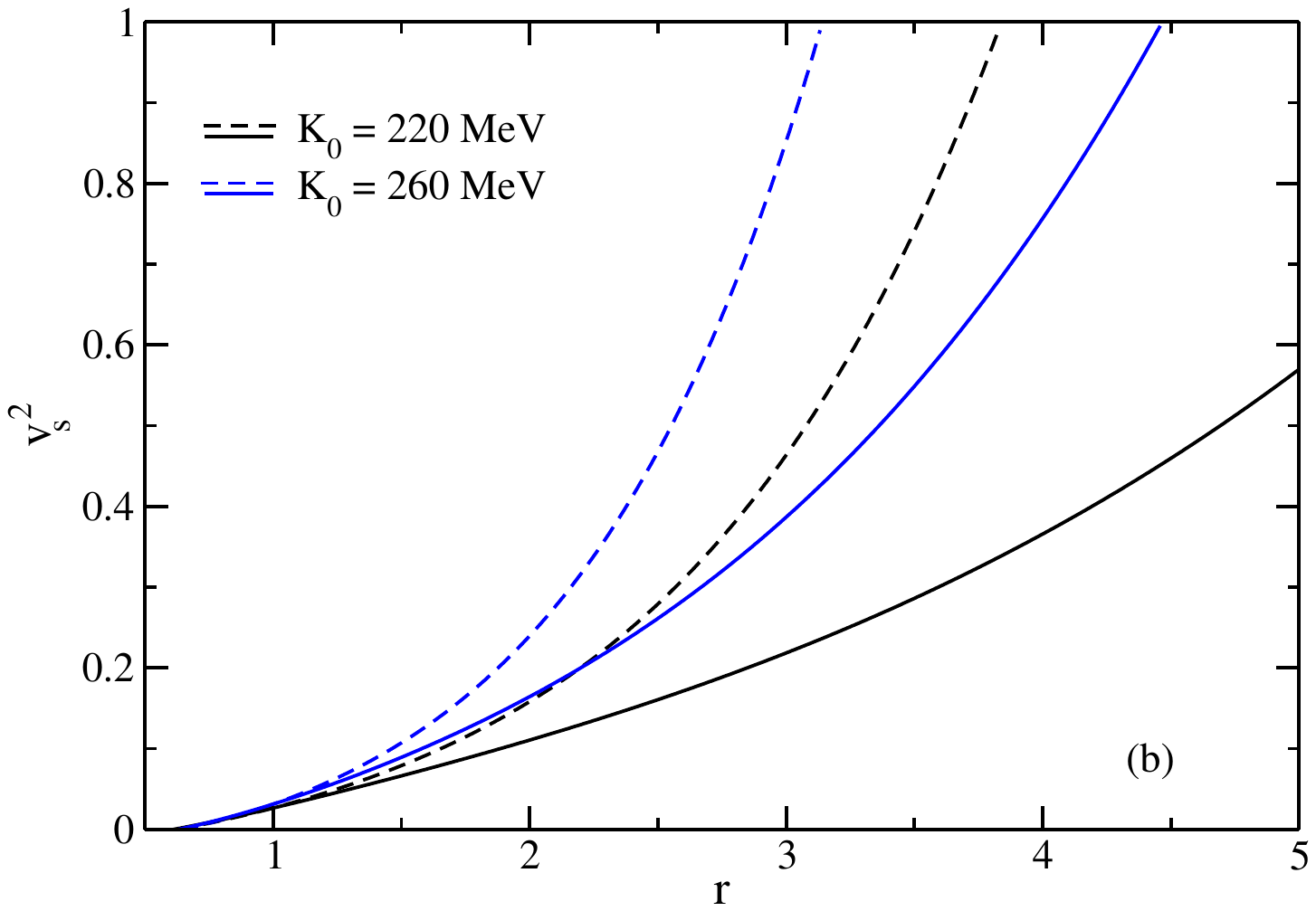}
\vspace{-0.8cm}
\caption{Squared sound velocity as a function of the ratio $r=\rho/\rho_0$ for (a) \mbox{DD-vdW} model, (b) CCS model with (full lines) and without (dashed lines) SRC included, for incompressibility values of $K_0=220$~MeV and $K_0=260$~MeV. Results for SNM at $T=0$ in both panels.}
\label{figvs}
\end{figure}

A typical drawback of EV models is the broken causality at high-density regimes for nuclear matter at $T=0$ generating, as a consequence, superluminal equations of state. The Lorentz contraction of the hard-sphere nucleons should be taken into account in order to correct this issue, and then, it could be interpreted as a nucleon with the effective volume being a decreasing density-dependent function. An attempt to use the CS mechanism for this aim is still limited since the range of densities in which the model is causal is still not enough for studying, for instance, compact stars (a common dense matter system in which different hadronic models can be probed). However, the use of Eq.~\eqref{arho} in addition to the CS procedure in the \mbox{DD-vdW} model is shown to extend the region of causality, as depicted in Fig.~\ref{figvs}{\color{blue}a}.
\vspace{-1.0cm}
Another way of circumventing this problem is to implement the Lorentz contraction in the rest frame of the medium in a proper way, as done in Ref.~\cite{NPA807-251_2008} (effect presented in two finite size nucleons). One more option to overcome the difficulty in dense matter, with a nice application in neutron stars, was proposed in Ref.~\cite{ApJ871-157_2019} with the use of the induced surface tension, generated by the repulsion between the nucleons at the surface and the outer nucleons. The increase of the causality range, in comparison with the simplest version of the vdW model, allows the generation of equations of state capable of reproducing massive stars, as well as configurations compatible with SNM constraints, such as the flow constraint~\cite{Sci298-1592_2002}.

\subsection{Short-range correlations}

One more very important feature observed in nuclear systems is the phenomenon of short-range correlations~(SRC)~\cite{Sci346-614_2014,Nat560-617_2018,Nat566-354_2019,Nat578-540_2020,Sci320-1476_2008}, recently included into vdW-like models in Ref.~\cite{MNRAS523-4859_2023}. SRC was probed in some experiments, such as those performed at the Thomas Jefferson National Accelerator Facility~\cite{Sci320-1476_2008}, where it was observed that collisions of very energetic incident particles in target nuclei induce the removal of correlated pairs of nucleons, mostly neutron-proton pairs, with large relative momentum due to the tensor force in the neutron-proton isosinglet channel. As a consequence, the single-nucleon momentum distribution ($T=0$ case) is modified in comparison with the usual step function. The asymmetric matter version of this quantity can be written as~\cite{PRC92-011601_2015}
\begin{eqnarray}
n_{p,n}(k) = \left\{
\begin{array}{ll}
\Delta_{p,n}, & 0<k<k_F^{p,n},
\\ \\
C_{p,n}\dfrac{(k_F^{p,n})^4}{k^4}, & k_F^{p,n}<k<\phi_{p,n} k_F^{p,n},
\end{array}
\right.
\label{htm}
\end{eqnarray}
with $\Delta_{p,n}=1 - 3C_{p,n}(1-1/\phi_{p,n})$, $C_p=C_0[1 - C_1(1-2y_p)]$, $C_n=C_0[1 + C_1(1-2y_p)]$, $\phi_p=\phi_0[1 - \phi_1(1-2y_p)]$ and $\phi_n=\phi_0[1 + \phi_1(1-2y_p)]$. Here the normalization condition was used, i.e., the momentum integral of $n_{p,n}(k)$ is imposed to be equal to $\rho_{p,n}=(k_F^{p,n})^3/3\pi^2$, with $k_F^{p,n}$ being the Fermi momentum of protons/neutrons. The constants $C_0$, $C_1$, $\phi_0$, and $\phi_1$ are determined from SRC experiments, as pointed out in Ref.~\cite{PPNP99-29_2018}. It is worth noticing the new structure in $n_{p,n}(k)$ induced by SRC: the depletion under the nucleon Fermi momentum, and the so-called high momentum tail~(HMT) above~$k_F^{p,n}$. The shape of the HMT, proportional to $1/k^4$, is a property also observed in other quantum systems, such as two-component cold fermionic atoms~\cite{PPNP99-29_2018}.  Data from nucleon knockout reactions induced by high-energy particles can be used to determine the HMT fraction in SNM and in pure neutron matter, around $28\%$ and $1.5\%$ respectively~\cite{Sci346-614_2014,PRC92-045205_2015,PRC91-025803_2015}.
\vspace{-1.7cm}
\begin{figure}[!htb]
\hspace{1.5cm}
\includegraphics[scale=0.29]{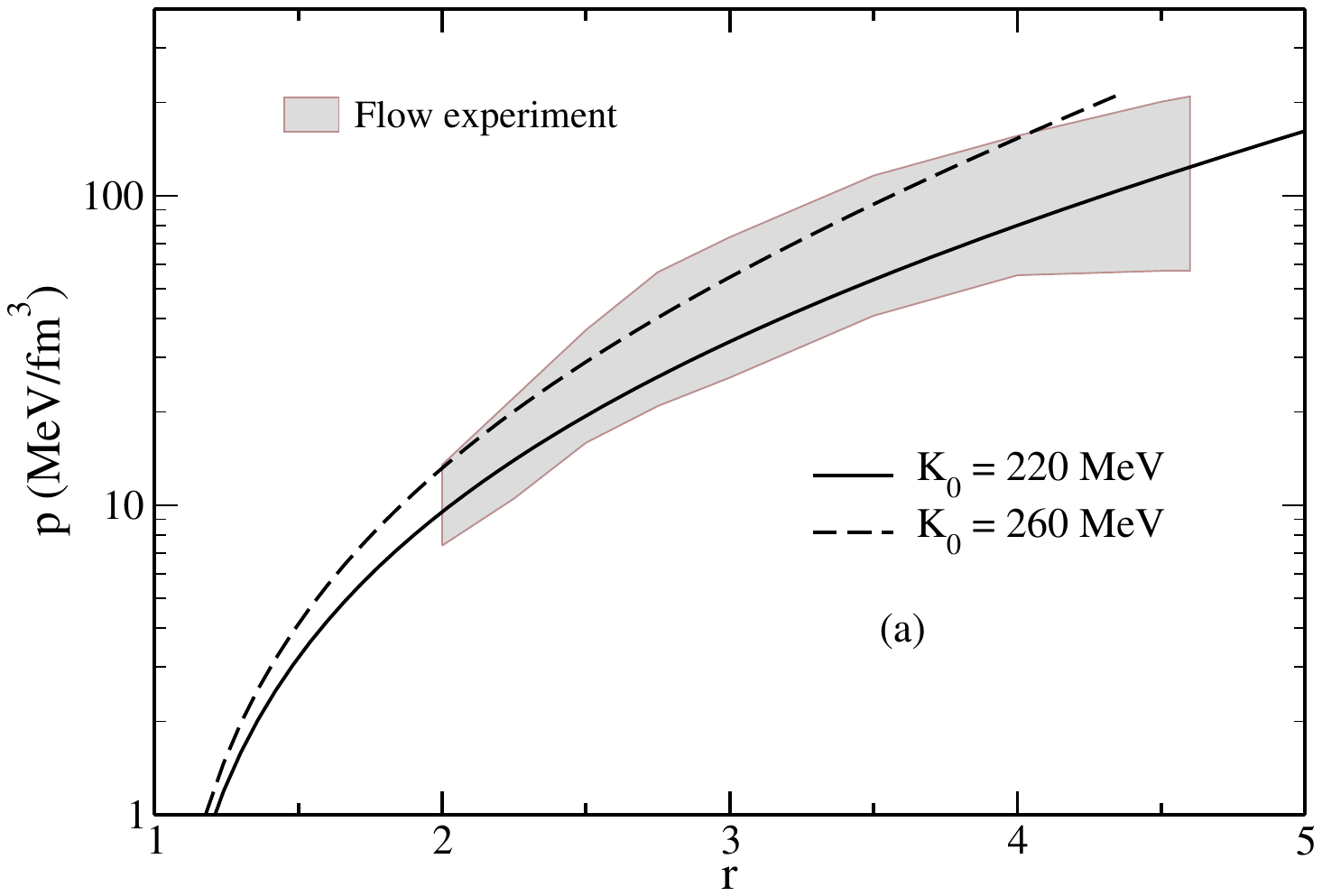}
\hspace{-1.2cm}
\includegraphics[scale=0.29]{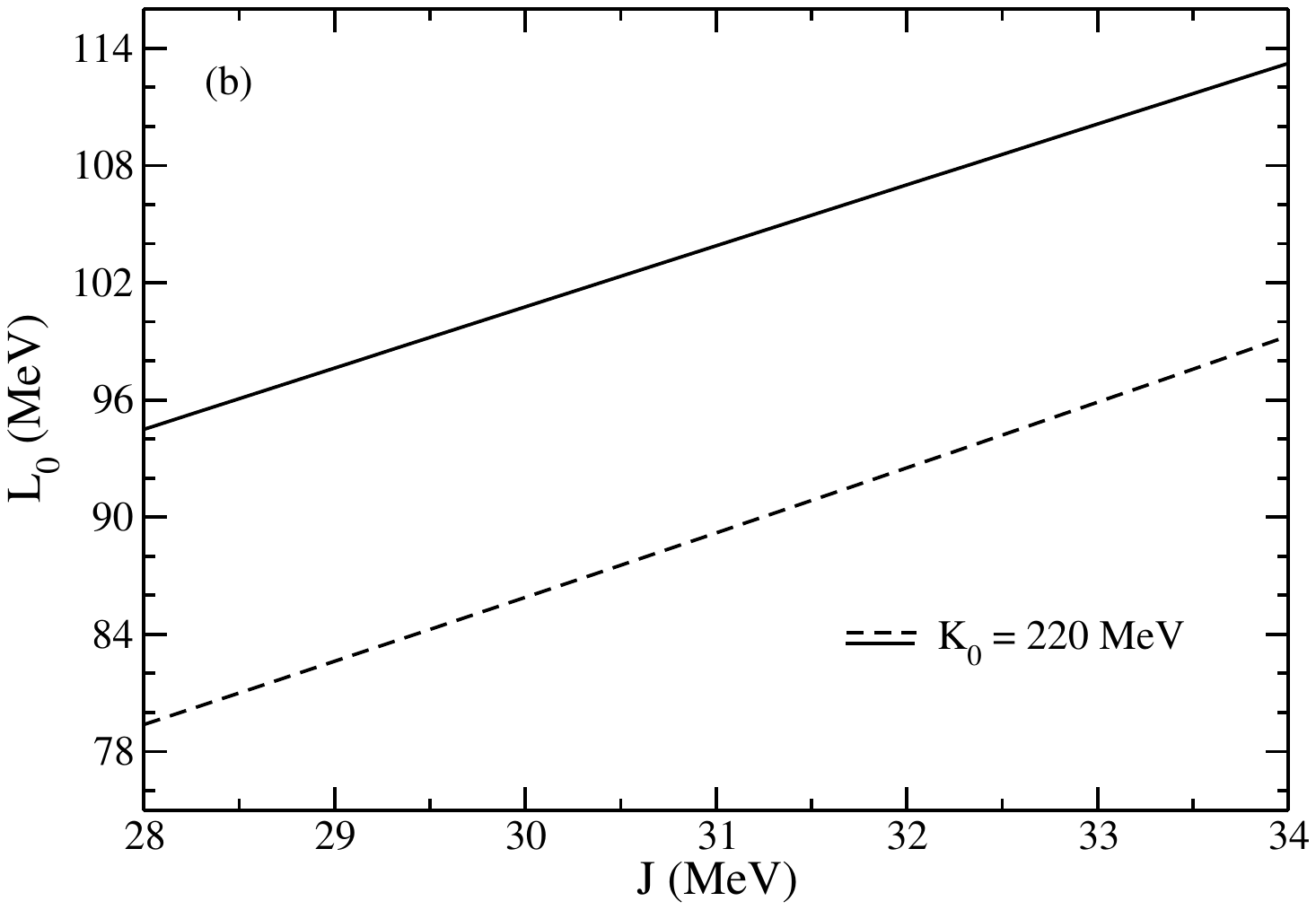}
\vspace{-0.6cm}
\caption{CCS model for SNM at $T=0$: (a) pressure as a function of ratio $r=\rho/\rho_0$ for $K_0=220$~MeV and $K_0=260$~MeV, model with SRC. Flow constraint extracted from Ref.~\cite{Sci298-1592_2002}. (b) The slope of symmetry energy ($L_0$) as a function of symmetry energy ($J$), both quantities at the saturation density. Model for $K_0=220$~MeV with (full lines) and without (dashed lines) SRC included.}
\label{figpressjl}
\end{figure}

An interesting application of SRC in dense fermionic system was analyzed in Ref.~\cite{PRC93-014619_2016}, where it is shown that one of the effects of this particular phenomenology is the capability of producing more massive neutron stars in comparison with the ones obtained from equations of state without SRC included. Furthermore, the average nucleon kinetic energy extracted from the RMF model used in Ref.~\cite{PRC93-014619_2016} is shown to be compatible with electron-nucleus scattering data, feature not observed if SRC are not implemented in the model. The same is valid for the kinetic symmetry energy of the SRC induced model, i.e, this quantity is in agreement with results provided by microscopic many-body theories~\cite{EL97-22001_2012,PRC89-044303_2014,EPJA50-13_2014}. Later on, further studies with \mbox{RMF-SRC} models presented: (i) approximate analytical expressions for the nucleon chemical potentials~\cite{PRC101-065202_2020}, as well as the impact of SRC in the constraints established by the LIGO/Virgo Collaboration concerning the tidal deformabilities of compact stars, (ii) study on neutron star matter admixed with dark matter~\cite{MNRAS517-4265_2022,PRD105-023008_2022,PRD106-043010_2022}, and (iii) influence of SRC on the pasta phase~\cite{EPJA59-209_2023}. As already mentioned, SRC were also applied to vdW-like models~\cite{MNRAS523-4859_2023}, more specifically, in parametrizations of the real gas Clausius model with CS method of excluded volume adopted and the density-dependent function given by $\mathcal{A}(\rho)=a/(1+c\rho)$, model named as \mbox{CCS-SRC}. In this case, it is verified that SRC move the break of causality to higher densities in comparison with the model without this phenomenology implemented, as displayed in Fig.~\ref{figvs}{\color{blue}b}. At this dense matter region, it is also observed that \mbox{CCS-SRC} model completely satisfies the flow constraint~\cite{MNRAS523-4859_2023}, see Fig.~\ref{figpressjl}{\color{blue}a}. It is also worth noticing the enhancement in the symmetry energy slope caused by SRC in the \mbox{CCS-SRC} model, as we can see in Fig.~\ref{figpressjl}{\color{blue}b}. It is verified as well, that the values obtained for this quantity are in agreement with the range of $L_0=(106\pm 37)$~MeV~\cite{PRL126-172503_2021}, compatible with the updated results provided by PREX-2 collaboration concerning the neutron skin thickness of $^{208}\rm Pb$~\citep{PRL126-172502_2021}. However, other studies are pointing out smaller ranges for $L_0$, see for instance Ref.~\cite{PRL127-232501_2021}, in which the range $L_0=(54\pm 8)$~MeV was obtained from theoretical uncertainties of the parity-violating asymmetry in $^{208}\rm Pb$. Another different range, namely, $L_0=(37-66)$~MeV, was found in Ref.~\cite{NatPhys18-1196_2022} from ab initio calculations for the $^{208}\rm Pb$ nucleus. Analysis from the CREX Collaboration~\citep{PRL129-042501_2022}, performed in Ref.~\cite{PAR6-30_2023}, determined the range of $L_0=(-5\pm 40)$~MeV. Combined studies of results from PREX-2 and CREX through Bayesian inference in Ref.~\cite{PRC108-024317_2023}, and through a covariance analysis in Ref.~\citep{PRC107-055801_2023}, found the ranges of found $L_0=15.3^{+46.8}_{-41.5 }$~MeV and $L_0=(82.32\pm22.93)$~MeV, respectively.

\subsection{More improvements and new challenges}

It is expected that at a high-density regime, the strongly interacting matter at zero temperature undergoes a hadron-quark phase transition. Such a phenomenon can be described by means of two different models representing the respective sectors of the system at this extreme condition. Nevertheless, this is not the unique procedure used for this description, see for instance, the concept of ``quarkyonic matter'', presented in Ref.~\cite{NPA796-83_2007}, and later on discussed in Ref.~\cite{PRL122-122701_2019} in terms of a momentum space shell structure. In this particular phase of matter, deconfined quarks are surrounded by confined baryons, and a hadron-quark mixture is verified. The dynamical mechanism governing this phenomenon is the following: at a low-density regime, baryons degrees of freedom are responsible for the maximum momentum shell width (radius of the sphere in the momentum space). On the other hand, as density increases it is observed the emergence of a quark Fermi sea and, consequently, the shell width is decreased since at this stage quark matter is energetically more favorable than the baryonic one. An inner ``quark'' sphere appears, and the width is measured from the outer baryon surface to the inner quark one. Another picture of the hadron-quark mixture was studied in Ref.~\cite{PLB841-137942_2023}, where the opposite mechanism is proposed: confined baryons are surrounded by a shell of deconfined quarks, i.e., the quark surface is now the outer one, in contrast to the baryonic inner surface. The system described by this configuration is called ``baryquark matter''~\cite{PLB841-137942_2023}. Both, quarkyonic and baryquark matter properties were recently incorporated into vdW-like models in Ref.~\cite{PRC108-045202_2023} with some interesting features observed, such as the value of the transition density (transition from normal matter to the quarkyonic/baryquark one) as being around $(1.5-2.0)\rho_0$; the need of introducing a regulator to avoid the singular behavior in the sound velocity at the onset of quark appearance; the baryquark matter being energetically favored; among other findings.

A natural improvement yet to come in this particular approach for describing strongly interacting matter at extreme conditions is the development of unified vdW-like models capable of simultaneously describing all aforementioned phenomenology shown in this section and in previous ones, as well as account for other current challenges such as~(i) implementation of strangeness by means of inclusion of hyperons~\cite{SA4-187_1960} with the aim of solving the so-called ``hyperon-puzzle''; (ii) addition of magnetic field in the equations of state in order to enable the models to describe magnetars~\cite{SSR191-315_2015} as well as the magnetic field configuration presented in rapidly rotating pulsars; (iii) full agreement with the new observational astrophysical data coming from the advent of the multi-messenger astronomy, such as gravitational wave observations related to the binary neutron star merger (event called GW170817~\cite{PRL119-161101_2017}); and (iv) compatibility with observations provided by the Neutron Star Interior Composition Explorer (NICER) X-ray timing telescope, coupled to the International Space Station in 2017, concerning the neutron star radius~\cite{AJL887-L21_2019,AJL887-L24_2019,AJL918-L27_2021,AJL918-L28_2021}.

 \newpage

 \section{Supersymmetric Quantum Mechanics and the Dynamics of Dirac Matter Under Extermal Electromagnetic Fields}\label{Raya}

Supersymmetry or SUSY is a theoretical framework in particle physics that proposes a symmetry between fermions and bosons in extensions to the Standard Model. The first mathematical construction dates back to the Wess-Zumino model~\cite{Wess:201649}. Supersymmetry gained traction as it offered elegant solutions to several theoretical challenges (see, for example, Ref.~\cite{MartinBook}), such as the hierarchy problem, Higgs stabilization, grand unification of fundamental interactions and even provided a candidate for dark matter. Despite the lack of direct experimental confirmation of its existence to date, supersymmetry remains a pivotal and highly studied component of modern theoretical physics, promising profound insights into the fundamental structure of the universe.

Supersymmetry in quantum mechanics (SUSY-QM) is a powerful framework that offers significant advantages and insights in the study of quantum systems, making it a valuable tool in theoretical and mathematical physics~\cite{COOPER1988}. One of the primary uses of SUSY-QM is its ability to provide exact solutions to certain quantum mechanical problems that are otherwise difficult to solve. By introducing a symmetry between bosonic and fermionic states, SUSY-QM allows for the construction of partner Hamiltonians whose spectra are related, simplifying the analysis of complex systems. SUSY-QM is instrumental in understanding the algebraic structures underlying quantum mechanics. It offers a natural context for exploring the properties of shape-invariant potentials, which are central to solving the Schrödinger equation for a variety of potentials~\cite{COOPER1988}. This has practical applications in fields such as molecular physics, statistical mechanics, optics and condensed matter physics, just to mention some examples.

In discussing the main ideas of SUSY-QM, one often starts from two Schr{\"o}dinger-like Hamiltonians $H^{\pm}$ which are intertwined by means of an operational relation
\begin{equation} \label{1}
    H^{+}L^{-} = L^{-}H^{-}.
\end{equation}
This is know as intertwining relation, where $L^{-} = d/dx + w(x)$ is the intertwining operator, and $w(x)$ the superpotential function. The corresponding potentials $V^{\pm}(x)$ can be written in terms of the superpotential as
\begin{equation} \label{2}
    V^{\pm}(x) - \epsilon= w^{2}(x) \pm \frac{dw(x)}{dx},
\end{equation}
where $\epsilon$ is a constant called factorization energy. Thus, the supersymmetric partner Hamiltonians $H^{\pm}$ turn out to be factorized as
\begin{equation} \label{3}
    H^{+} -\epsilon = L^{-}L^{+}, \quad H^{-} - \epsilon = L^{+}L^{-},
\end{equation}
being $L^{+} = \left(L^{-}\right)^{\dagger} = - d/dx + w(x)$.
All these elements satisfied the supersymmetric algebra, which is defined with the following commutation rules:
\begin{equation}
    [H_{SS},Q^{\pm}] = 0,\quad \{Q^{+},Q^{-}\} = H_{SS}.
\end{equation}
Often, the supersymmetric Hamiltonian $H_{SS}$ and the superchargers are written as $2\times 2$ matrices of the form
\begin{equation}
    Q^{+} = \begin{pmatrix}
        0 & L^{+} \\
        0 & 0
    \end{pmatrix},\quad
    Q^{-} = \begin{pmatrix}
        0 & 0 \\
        L^{-} & 0
    \end{pmatrix},\quad
    H_{SS} = \begin{pmatrix}
        H^{-}-\epsilon & 0 \\
        0 & H^{+}-\epsilon
    \end{pmatrix}.
\end{equation}
The primary use of the supersymmetric transformations defined in Eq. \eqref{1} is to construct a supersymmetric partner Hamiltonian $H^{+}$ from an initial Hamiltonian $H^{-}$ whose eigenfunctions $\psi^{-}_{n}(x)$ and eigenvalues $E_{n}$ are known. This process also involves determining the eigenfunctions $\psi^{+}_{n}(x)$ of $H^{+}$ from those of $H^{-}$. It is worth noting that a consequence of the supersymmetric algebra is that the supersymmetric partner Hamiltonians are either isospectral or their spectra differ only by one energy level (the factorization energy level), depending on the square-integrability conditions of the respective eigenfunctions. The supersymmetric algorithm to build a supersymmetric partner Hamiltonian is the following: Given a seed solution $u(x)$ satisfying $H^{-}u(x) = \epsilon u(x)$, the superpotential function can be obtained as
\begin{equation}
    w(x) = - \frac{du(x)/dx}{u(x)},
\end{equation}
thus, using Eq.\eqref{2}, we can calculate the supersymmetric partner potential $V^{+}(x)$. Moreover, from Eqs. \eqref{1} and \eqref{3}, assuming the eigenfunctions $\psi^{-}_{n} (x)$ are normalized, we arrive at
\begin{equation}
    \psi_{n}^{\pm}(x) = \frac{L^{\mp}\psi^{\mp}_{n}(x)}{\sqrt{E_{n}-\epsilon}}.
\end{equation}
Finally, the eigenfunction $\psi^{+}_{\epsilon}$ of $H^{+}$ corresponding to the factorization energy $\epsilon$ is given by
\begin{equation}
    \psi_{\epsilon}^{+}(x) = \frac{1}{u(x)},
\end{equation}
except for a normalization factor. This algorithm describes a first-order supersymmetric transformation since $L^{\pm}$ are first-order differential operators. However, the supersymmetric algorithm can be generalized if the intertwining operators are taken as second-order differential ones. In this process, two seed solutions and their respective factorization energies are necessary to construct the supersymmetric partner Hamiltonian but the intertwining relation in Eq. \eqref{1} remains. In general, we can carry out a $k$-th order supersymmetric transformation. For a more complete discussion on this issue, see Refs. \cite{Cooper2001,Gangopadhyaya2010,Junker2019,Fernandez2004,fernandez2019} and references within.


General $k$-th order SUSY-QM is a hard nut to crack. Yet, second-order SUSY-QM has been proven to be a very useful tool. 
In a second-order SUSY transformation, the Hamiltonians $H^{\pm}$ are intertwined by a second-order operator $L_{2}^{-}$, which can be written as
\begin{equation}
    L^{-}_{2} = \frac{d^{2}}{dx^{2}} + \eta(x)\frac{d}{dx} + \gamma(x),
\end{equation}
where $\eta(x)$ and $\gamma(x)$ are functions to be determined. The corresponding intertwining relation is
\begin{equation}
    H^{+}L^{-}_{2} = L^{-}_{2}H^{-}.
\end{equation}
The SUSY partner potentials $V^{\pm}(x)$ and the function $\gamma(x)$ can be written in terms of the function $\eta(x)$ as follows:
\begin{equation}
    \begin{aligned}
        &V^{+}(x) = V^{-}(x) + 2\frac{d\eta(x)}{dx},\quad \gamma(x) = \frac{\eta^{2}(x)}{2} - \frac{d\eta(x)/dx}{2} -V^{-}(x) +\frac{\epsilon_1+\epsilon_{2}}{2}, \\
        &V^{-}(x) = \frac{d^{2}\eta(x)/dx^{2}}{2\eta(x)} - \left(\frac{d\eta(x)/dx}{2\eta(x)}\right)^{2} - \frac{d\eta(x)}{dx} + \frac{\eta^{2}(x)}{4} + \left(\frac{\epsilon_1+\epsilon_2}{2}\right) + \left(\frac{\epsilon_1 - \epsilon_{2}}{2\eta(x)}\right)^{2}.
    \end{aligned}
\end{equation}
This time, there are two factorization energies $\epsilon_{1}$ and $\epsilon_{2}$ associated with two seed solutions $u_{1}(x)$ and $u_{2}(x)$, which are solutions of the Hamiltonian $H^{-}$. In terms of the seed solutions, the function $\eta(x)$ is given by
\begin{equation}
    \eta(x) = - \frac{dW(u_1(x),u_2(x))/dx}{W(u_1(x),u_2(x))},
\end{equation}
being $W(f,g)=fg'-f'g$ the Wronskian. Furthermore, the eigenfunctions $\psi^{\pm}_{n}(x)$ for the Hamiltonians $H^{\pm}$ are related by the following equalities:
\begin{equation}
    \psi^{\pm}_{n}(x) = \frac{L^{\mp}_{2}\psi^{\mp}_{n}(x)}{\sqrt{(E_{n}-\epsilon_1)(E_n-\epsilon_{2})}},
\end{equation}
while the functions $\psi^{+}_{\epsilon_j}$ turn out to be
\begin{equation}
    \psi_{\epsilon_1}^{+}=\frac{u_2(x)}{W(u_1(x),u_2(x))} ,\quad \psi_{\epsilon_2}^{+} = \frac{u_1(x)}{W(u_1(x),u_2(x))}.
\end{equation}
Finally, the products of the intertwining operators $L^{\pm}_{2}$ are second-degree polynomials on $H^{\pm}$, i.e.,
\begin{equation}
    L^{-}_{2}L^{+}_{2}= (H^{+}-\epsilon_1)(H^{+}-\epsilon_2 ),\quad  L^{+}_{2}L^{-}_{2}= (H^{-}-\epsilon_1)(H^{-}-\epsilon_2).
\end{equation}

First and second order SUSY-QM have been widely implemented in exploring the dynamics of non-relativistic charged particles in external static electromagnetic fields. This is valuable for exploring the dynamics of Dirac particles under similar circumstances, as we show below.

\subsection{SUSY and the Dirac equation}

The supersymmetric structure of the Dirac equation reveals a deep connection between SUSY and relativistic quantum mechanics. The Dirac equation, which describes the behavior of fermions such as leptons and quarks, can be reinterpreted within a supersymmetric framework, highlighting symmetries between different quantum states.
%
%
The Dirac equation in natural units (\( \hbar = c = 1 \)) is given by:
\begin{equation}
(i \gamma^\mu \partial_\mu - m) \psi = 0,
\end{equation}
where \( \gamma^\mu \) are the Dirac matrices, \( \partial_\mu \) is the four-gradient, \( m \) is the mass of the particle, and \( \psi \) is the Dirac spinor.
%
%
%
%
To reveal the SUSY structure in the Dirac equation, we consider the stationary form of the Dirac equation:
\begin{equation}
H \psi = E \psi,
\end{equation}
with
\begin{equation}
H = \vec\alpha \cdot \vec p + \beta m .
\end{equation}
Here, \( \alpha_i=\gamma^0\gamma^i \) and \( \beta=\gamma^0, \) with \( \vec p \) the momentum operator.

In a suitable representation, the Hamiltonian \( H \) can be decomposed into a block form that resembles a pair of supersymmetric partner Hamiltonians:
\begin{equation}
H = \begin{pmatrix}
0 & A \\
A^\dagger & 0
\end{pmatrix} \;,
\end{equation}
where \( A \) and \( A^\dagger \) are operators that act on the components of the spinor \( \psi \).
%
%
The supercharges \( Q \) and \( Q^\dagger \) can be defined as:
\begin{equation}
Q = \begin{pmatrix}
0 & 0 \\
A & 0
\end{pmatrix}, \quad Q^\dagger = \begin{pmatrix}
0 & A^\dagger \\
0 & 0
\end{pmatrix}\,. \end{equation}
These supercharges satisfy the SUSY algebra:
\begin{equation}
\{Q, Q^\dagger\} = H, \quad Q^2 = (Q^\dagger)^2 = 0\;.
\end{equation}
%
%
In this framework, the operators \( A \) and \( A^\dagger \) correspond to the creation and annihilation operators in the context of the Dirac equation. They map solutions of the Dirac equation to one another, revealing the underlying supersymmetric structure.

The supersymmetric structure shows that for every solution of the Dirac equation with positive energy, there is a corresponding solution with negative energy, forming a supersymmetric pair. This pairing reflects the symmetry between particles and antiparticles.

%
%

\subsection{Bound states in external electromagnetic fields}

The influence of external magnetic fields into the Dirac equation can be straightforwardly included through the minimal substitution replacement $\partial_\mu \to D_\mu=\partial_\mu +ieA_\mu$, where $A_\mu$ is the vector potential giving raise to the external fields. Bound states for Dirac fermions have been widely explored for static fields. In the case of magnetic fields, the uniform magnetic field configuration has lead to a lot of interest. For large magnetic fields compared to the rest mass of the Dirac particle, it is known that such configuration gives rise to the relativistic Landau levels, which are of the utmost importance in many branches of physics~\cite{MIRANSKY20151}. Going beyond this configuration, adding the influences of static electric fields, either parallel or perpendicular to the magnetic field, and the plane wave configuration are among the handful of examples that can be exactly solvable by standard technique~\cite{Schwinger}.

SUSY-QM provides the means to analytically find the bound states in static magnetic fields with varying profiles, such as an exponentially decaying field, a Pöschl-Teller-like well, hyperbolic well, a trigonometric singular well, a singular field, a hyperbolic singular field~\cite{COOPER1988,Kuru2009,Concha_2018}. Moreover, second-order SUSY-QM gives modifications of these profiles~\cite{Fernandez2020,Fernandez2021}. Even, it is possible obtain complex fields~\cite{Fernandez2022}. Many features of the SUSY structure of 2D Dirac fermions in uniform magnetic fields were discussed in~\cite{Hernandez-Ortiz_2012}.

The effect of combined electric and magnetic fields has been valuable to explore transport properties of Dirac matter. For instance,
under the influence of the parallel electric and magnetic fields, a chiral magnetic current is induced in ZrTe\(_5\)~\cite{Li2016}. This current is directly proportional to the applied magnetic field strength and the chiral chemical potential difference induced by the electric field.
A negative longitudinal magnetoresistance is observed, which is a signature of the chiral anomaly. This effect manifests as a decrease in electrical resistance with increasing magnetic field strength when the fields are aligned. This is the signature of the Chiral Magnetic Effect, which has been observed in several Dirac-Weyls semimetals as predicted first in the context of heavy ion collisions~\cite{Kharzeev:2007jp,Fukushima,KHARZEEV2014133}.

Recently, a configuration in which static and non-uniform parallel electric and magnetic fields was considered \cite{PerezPedraza2024}. Such configuration is applied to a 3D Dirac material and SUSY-QM is used to solve the eigenvalue problem, finding bound states for hyperbolic and trigonometric profiles. It is worth mentioning that the zero-energy bound state has an associated probability current due to the chiral symmetry conservation, which is perpendicular to the fields but in the same plane, thus defining a chiral planar Hall effect (PHE). This phenomenon has been also addressed in Weyl semimetals with perturbations, such as strains, in other words, the Weyl semimetals are in the presence of electromagnetic pseudofields, and an in-plane transverse voltage appears in the material \cite{Alberto2023}. Furthermore, in the particular case of bilayer graphene, a non-linear PHE was studied, as a consequence of an orbital effect from the in-plane magnetic field on the electrons without spin-orbit coupling \cite{Langari2022}.

\subsection{Green Functions}

In many physical situations, finding the wave functions for electronic states it is equally useful as to know the corresponding propagator. However, because the asymptotic {\em in-} and {\em out-}states in the background of a classical electromagnetic field do not correspond to plane waves, representing the two-point function is cumbersome, making it nearly impossible to express the propagator in a closed form except in a few specific cases, such as for uniform electric or magnetic fields (either parallel or perpendicular) and plane wave electromagnetic fields~\cite{Schwinger}. To address this, alternative representations have been developed. Among these, the Schwinger method~\cite{Schwinger} and the spectral representation, the Ritus method~\cite{RITUS1972,Ritus1974,Ritus1978} which was pedagogically discussed in~\cite{Murguia2010}, provide ways to express the propagator in a closed form.

We focus our attention to the expansion of  the propagator in the basis of Ritus functions which correspond to the eigenfunctions of the operator $(\gamma\cdot\Pi)^2$ where $\Pi_\mu=p_\mu +eA_\mu$ is the canonical momentum operator that includes the effect of the external magnetic field through minimal coupling (with  $A_\mu$ denoting the corresponding vector potential and $e$ is the elementary charge) and $\gamma^\mu$ denote the Dirac matrices.
Considering a general situation in which a static magnetic field of general spatial profile points perpendicularly to the plane of motion of Dirac fermions,  working in a Landau-like gauge, we introduce an electromagnetic potential  $A^{\mu}=(0,0, \mathcal{W}_{0}(x),0)$, where $\mathcal{W}_{0}(x)$ is a scalar function such that $\mathcal{W}_{0}'(x)=\partial_x \mathcal{W}_{0}(x)$ defines the profile of the field.
In these circumstances, the fermion propagator cannot be diagonalized on a plane wave  basis because, as mentioned before, the asymptotic states of these fermions in a background field do not correspond to the kinematical momentum eigenfunctions. Motivated by this observation, we notice that  the Green function for Dirac particles, $G(z,z')$, satisfies
\begin{equation}
((\gamma \cdot \Pi) -m)G(z,z')=\delta^{(3)}(z -z'), \label{Green function}
\end{equation}
with $z^\mu=(t,x,y,z)$, $\gamma^\mu$ denoting the Dirac matrices  and $\Pi_\mu$ is the canonical momentum. We omit the Lorentz index in the vectors to keep a shorthand notation when necessary. Since $G(z,z')$ commutes with $(\gamma \cdot \Pi)^2$, we expand the propagator on the basis of the eigenfuctions of the later, namely,
\begin{equation}
(\gamma \cdot \Pi)^2\mathbb{E}_p(z)=p^2\mathbb{E}_p(z), \label{Ritus}
\end{equation}
where the eigenvalue $p^2$ can be any real number corresponding, as we shortly will see, to the magnitude squared of the vector $p^\mu$ (or simply $p$ to avoid cumbersome notation) that labels the functions $\mathbb{E}_p(z)$. We refer to the functions $\mathbb{E}_p(z)$ as the Ritus eigenfunctions~\cite{RITUS1972,Ritus1974,Ritus1978}.
It can be directly verified that these functions $\mathbb{E}_p(z)$ fulfill the closure and completeness relations
\begin{subequations}
\begin{align}\label{orthogonality}
\int \mathrm{d}^4z\, \Bar{\mathbb{E}}_{p'}(z)\mathbb{E}_p(z)&=\mathbb{I}\,\widehat{\delta}(p-p'), \\
\int \mathrm{d}^4p\, \mathbb{E}_p(z')\Bar{\mathbb{E}}_p(z)&=\mathbb{I}\,\widehat{\delta}(z-z'),
\end{align}
\end{subequations}
with $\Bar{\mathbb{E}}_p(z)=\gamma^0\mathbb{E}^{*}_{p}(z)\gamma^0$ and $\mathbb{I}$ is the unit matrix. The hat over the $\delta$ function refers to the situation when quantum numbers induced by the external fields should be considered discrete labels, whereas those that remain unafected are continuous variables.

In order to construct the Ritus eigenfunctions,  we notice that the operator
\begin{equation}
(\gamma \cdot \Pi)^2=\gamma^{\mu}\gamma^{\nu}\Pi_{\mu}\Pi_{\nu}=\Pi^2 + \frac{e}{2}\sigma^{\mu\nu}F_{\mu\nu},
\end{equation}
where $F_{\mu\nu}=\partial_\mu A_\nu - \partial_\nu A_\mu$ is the electromagnetic field strength tensor and $\sigma^{\mu\nu}=i[\gamma^\mu, \gamma^\nu]/2$. For a static magnetic field pointing perpendicularly to the plane, the only non-vanishing components of these tensors are
\begin{equation}
F_{12}=-F_{21}=\mathcal{W}_{0}'(x), \quad \sigma^{12}=\Sigma_{3}.
\end{equation}
Then, the eigenvalue Eq.~(\ref{Ritus}) becomes
\begin{equation}
(\Pi^{2}+e\Sigma_{3}\mathcal{W}_{0}'(x))\mathbb{E}_p(z)=p^2\mathbb{E}_p(z),
\end{equation}
from where we observe that the Ritus eigenfunctions are actually block matrices, depending upon the dimensionality of the representation of the Dirac matrices $\gamma^\mu$. 
Notice that the Ritus functions are labeled by a subscript $p$, which is the shorthand notation of the vector $p^\mu=(p_0,p_2,p_3,k)$ is a vector that contains the eigenvalues of the operators $i\partial_t$, $-i\partial_y$, $\partial_z$ and $\mathcal{H}_{\sigma}$, respectively, and whose norm squared corresponds to the eigenvalue in Eq.~(\ref{Ritus}). That is, the components of the vector $p^\mu$ are the numbers such that
\begin{equation}
i\partial_t\mathbb{E}_p(z)=p_0\mathbb{E}_p(z), \quad i\partial_y\mathbb{E}_p(z)=-p_2\mathbb{E}_p(z), \quad
i\partial_z\mathbb{E}_p(z)=p_3\mathbb{E}_p(z), \quad
\mathcal{H}_{\sigma}\mathbb{E}_p(z)=k\mathbb{E}_p(z),
\end{equation}
with $\mathcal{H}_{\sigma}=-(\gamma \cdot \Pi)^2 + \Pi^{2}_{0}+\Pi^2_3$. These eigenvalues allow us to write the scalar functions as
\begin{equation}
E_{p,\sigma}(x)=e^{-i(p_0t-p_2y-p_3z)}F_{k,p_2,\sigma}(x), \label{funciones de Ritus}
\end{equation}
where $\sigma=\pm1$ are the eigenvalues of $\sigma_3$ and the functions $F_{k,p_2,\sigma}(x)$ satisfy
\begin{equation}
[-\partial_{x}^{2} + (p_2 + e\mathcal{W}_{0}(x))^2 - e\sigma \mathcal{W}_{0}'(x) ]F_{k,p_2,\sigma}=(k+p_3^2)F_{k,p_2,\sigma}, \label{Pauli Hamiltonian}
\end{equation}
which corresponds to a Pauli equation for a particle with mass $m=1/2$ and gyromagnetic factor $g=2$. This equation possesses a supersymmetric structure as we will briefly discuss below. Thus, $F_{k,p_2,\sigma}(x)$ are the solutions of the  equations in~(\ref{Pauli Hamiltonian}) associated to each of the supersymmetric-partner potentials
\begin{equation}
V_0^{\sigma}(x)=(p_2 + e\mathcal{W}_{0}(x))^2 -e\sigma \mathcal{W}_{0}'(x).
\end{equation}
From now on, we fix the value $\sigma=1$. Then, we have the required ingredients to construct the Ritus eigenfunctions from a first-order supersymmetric formalism.

Therefore, we can use these functions to diagonalize the fermion propagator  $S(z,z')$ in momentum space in the same way plane waves are used to define the Fourier transform,
\begin{equation}
S(z,z')=\int {\rm d}^4p\,{\rm d^4p'}\,\mathbb{E}_p(z)\ S_F(p,p')\ \Bar{\mathbb{E}}_{p'}(z')\;.
\end{equation}
Inserting this Green's functions in Eq. (\ref{Green function}), using the property \cite{RITUS1972}
\begin{equation}
(\gamma \cdot \Pi)\mathbb{E}_p(z)=\mathbb{E}_p(z)(\gamma \cdot \bar{p})\;,
\end{equation}
where $\bar{p}$ is the shorthand notation to define the three-momentum vector  $\bar{p}^\mu=(p_0, 0, \sqrt{k},p_3)$ that satisfies $\bar{p}^{2}=p^2=p_0^{2}+p_3^2 -k$ \cite{Murguia2010} and the properties (\ref{orthogonality}), the propagator in momentum space takes the form
\begin{equation}
S_{F}(p)=\frac{1}{\gamma\cdot\Bar{p}-m},
\end{equation}
similar to the free-particle propagator, but the momentum $\bar{p}$, which carries the quantum numbers induced on the dynamics of Dirac fermions by in the presence of the external field. In the configuration space, we write the propagator as
\begin{align}
S(z,z')&=\int {\rm d}^{4}p\ {\rm d}^{4}p'\ \mathbb{E}_{p}(z)\left[\frac{1}{\gamma\cdot\Bar{p}-m}\right]\Bar{\mathbb{E}}_{p'}(z')\nonumber\\
&=\int {\rm d}^{4}p\ {\rm d}^{4}p'\mathbb{E}_{p}(z)\left[\frac{\gamma\cdot\Bar{p}+m}{p^{2}-m^{2}}\right]\Bar{\mathbb{E}}_{p'}(z').
\end{align}
From this expression we can find the value of the  vacuum current density
\begin{equation}
j^\mu=Tr[\gamma^\mu S(z,z)].
\end{equation}
The propagator has been widely explored for constant magnetic fields in the context of the phenomenon of magnetic catalysis (see, for example, \cite{Leung96,Leung97,FERRER2000287,Ferrer98,Raya2008,Scoccola2024}). Extensions to exponentially decaying magnetic fields were considered in~\cite{Sadooghi:2012wi} exploring the local electric current correlation function. In low dimensions, this propagator has been discussed for uniform and exponentially decaying fields~\cite{Murguia2010,Raya2010}. It has also been discussed for second-order SUSY-generated magnetic fields generated from these profiles as seed fields~\cite{Concha-Sanchez_2022}. Furthermore, the same Ritus basis has been used to derive the Foldy-Wouthuysen transformation in external magnetic fields, rendering it to a free from with the quantum numbers induced by the field~\cite{Murguia:2010zz}.

 %


\noindent
\section{Quark condensate nonequilibrium dynamics under a strong magnetic field} 
\label{Gastao}

\vspace{0.25cm} 
In this section, we address the non-equilibrium dynamics of the quark condensate under a strong magnetic field. Time evolution is a common feature of strongly interacting matter in settings such as the early universe, collapsing stars, and heavy-ion collisions. A complete understanding of the condensate dynamics in these settings requires solving a complex nonequilibrium quantum many-body problem in QCD. Although solving such a problem is beyond our present capabilities, we can nevertheless gain insight into it by focusing on some general aspects that transcend features peculiar to a specific setting. Within such a perspective, we~employ a Langevin field equation to describe the dynamics~\cite{Krein:2021sco}. Langevin field equations are widely used in field theory treatments of dynamical phase transitions~\cite{Goldenfeld:1992qy,Onuki:2002}; they are either motivated phenomenologically or derived from a microscopic model for the system of interest{\textemdash}for an extensive list of references on these approaches, see Ref.~\cite{Krein:2021sco}. In this communication, we discuss a recent derivation~\cite{Krein:2021sco} of a Langevin field equation from the linear sigma model~\cite{GellMann:1960np} with quarks (LSMq) coupled to an external magnetic field. We~also present new results~\cite{AFrazon:2024}, not contemplated in the original publication~\cite{Krein:2021sco}, regarding the long-time evolution of condensate and its thermalization.   

The Langevin equation we discuss here was derived in Ref.~\cite{Krein:2021sco} using the Schwinger-Keldysh closed-time path (CTP) effective action~\cite{Schwinger:1960qe,Keldysh:1964ud} and the Feynman-Vernon influence functional (IF)~\cite{Feynman:1963fq} formalisms of nonequilibrium quantum field theory~\cite{Calzetta:2008iqa,Bellac:2011kqa}. More specifically, Ref.~\cite{Krein:2021sco} built on the semiclassical approach developed in Ref.~\cite{Nahrgang:2011mg} to include magnetic-field effects. The derived Langevin equation contains magnetic field-dependent damping and noise kernels that reflect the condensate's interactions with a locally equilibrated magnetized quark medium. These kernels determine time scales associated with the chiral crossover dynamics in the presence of a magnetic field. 

The explicit numerical results presented here relate to a quench scenario of dynamical phase transitions, using parameters relevant to heavy-ion collisions. In this scenario, the thermodynamic state of the system changes much more rapidly than any variation in an external parameter or field. A classic example is a temperature quench in a spin system, where a sudden drop in temperature drives the system irreversibly from a spin-disordered phase into a spin-ordered phase. Although our results focus on parameters pertinent to heavy-ion collisions, the derived Langevin equation is also applicable to other contexts, such as the early universe and collapsing stars.

\subsection{Langevin equation from the LSMq}

We outline the derivation in Ref.~\cite{Krein:2021sco} of the Langevin field equation for the condensate. Due to the lack of space, we cannot go into much detail, but the reader is directed to that reference for an in-depth discussion. To make the discussion selfcontained and set the notation, we write the LSMq Lagrangian density used to derive the Langevin equation describing the quark condensate dynamics: 
\begin{equation}
{\mathcal L} = \bar{q}[i \slashed{\partial} 
- g(\sigma + i\gamma_5 {\bm \tau} \cdot {\bm \pi} ) ] q 
+ \frac{1}{2}\left[ \partial_{\mu}\sigma\partial^{\mu}\sigma 
+ \partial_{\mu}{\bm \pi}\cdot\partial^{\mu}{\bm \pi}\right] -  \frac{\lambda}{4} (\sigma^2 + {\bm \pi}^2 -v^2)^2 - h_{q}\sigma - U_0,
\label{U}
\end{equation}
where $q = (u,d)^T$ is a fermion isodoublet representing the light $u$ and $d$ quark Dirac fiels, and ${\bm \pi}$ and $\sigma$ are respectively a pseudoscalar-isotriplet and scalar-isoscalar fields that Yukawa-couple to the quarks with strength~$g$. Finally, $U_0$ is an arbitrary constant that sets to zero of the classical ground-state energy. We use the metric signature $g^{\mu\nu} = (1, -1, -1, -1)$ and the Bjorken-Drell~\cite{Bjorken:1965zz} conventions for the Dirac $\gamma^{\mu}$ matrices, for which $\{\gamma^\mu, \gamma^\nu\} = 2 g^{\mu\nu}$.  One can fit the parameters of the model to chiral physics observables{\textemdash}a fit at the classical level, for example, sets the parameters as: $h_q = f_\pi m^2_\pi$, $v^2 = f^2_\pi - m^2_\pi/\lambda$, $m^2_\sigma = 2  \lambda f^2_\pi + m^2_\pi$, and $m_q = g \langle \sigma \rangle$.  Here $f_\pi$ and $m_\pi$ are the pion weak decay constant and mass, $m_\sigma$ the $\sigma$-meson mass, and $m_q$ the constituent-quark mass. The external magnetic field is introduced by the minimal substitution, in that $\partial_\mu \rightarrow D_\mu = \partial_\mu + i q A_\mu$, where $A_\mu$ the electromagnetic vector field and $q$ is the quark charge. In this study pions are neglected; therefore, the $\sigma \leftrightarrow \bar{q}q$ processes are the only source of dissipation, as we discuss at the end of this section. The parameters of the model were set so that in the mean-field approximation and $B=0$ it leads to a crossover transition; the values are~\cite{Nahrgang:2011mg} $g = 3.3$ and $\lambda = 20$. These lead to $m_\sigma = 604$~MeV and $m_q =290$~MeV. The crossover temperature is $T_{\rm pc} \simeq 150$~MeV, close to the lattice QCD value~\cite{Borsanyi:2020fev}, $T_{\rm pc} = 160$~MeV. Although the LSMq in the mean-field approximation does not reproduce lattice QCD results close to the chiral transition when $B\neq 0$, it nevertheless provides a great deal of insight in many aspects of magnetic field effects in hadron physics. In addition, going beyond the mean-field approximation by e.g. including  it can be improved to self-coupling and boson-fermion coupling .  

\begin{figure}[h]
\includegraphics[width=0.9\linewidth]{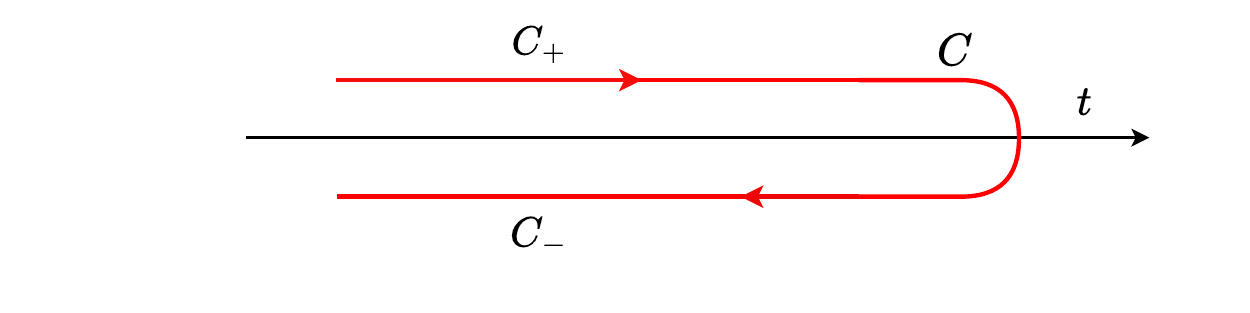} 
\vspace{-0.5cm}
\caption{Schwinger-Keldish contour.}
\label{fig1}
\end{figure}

The starting point for the derivation of the Langevin field equation for~$\sigma$ is the semi-classical effective action:
\begin{equation}
\Gamma[\sigma,S] = \Gamma_{\rm cl}[\sigma] + i \, {\rm Tr} \ln S - i \, {\rm Tr} \left(i\slashed{D}
- m_0\right)S + \Gamma_2[\sigma, S] \,,
\label{S-semi}
\end{equation}
where $\Gamma_2[\sigma,S]$ contains the 2PI diagrams, for which it was taken the fermion tadpole:
\begin{equation} 
\Gamma_2[\sigma,S] = g \int_{\cal C} d^4x \, 
{\rm tr} \left[ S^{++}(x,x) \sigma^+(x) 
+ S^{--}(x,x) \sigma^-(x) \right] \,,
\end{equation}
where $\sigma^\pm$ mean fields and the quark propagators $S_{++}$, $S_{--}$, and $S_{\pm}$ are defined on the Schwinger-Keldysh contour~$\mathcal{C}$ shown in Fig.~\ref{fig1}, and  $tr$ means trace over color, flavor, and Dirac indices.
The two fields $\sigma^{\pm}(x)$ are not independent; they couple through the CTP boundary condition $\sigma^{+}(T,\bm{x}) = \sigma^-(T,\bm{x})$ where $T \rightarrow + \infty$ (borders of the CTP contour) so that  there is a single mean field $\sigma(x)$ given by the equation of motion motion~\cite{Calzetta:2008iqa}:
\begin{equation}
\frac{\delta \Gamma[\sigma,S]}{\delta \sigma^+(x)}{\Big |}_{\sigma^+=\sigma^-=\sigma}
= \frac{\delta \Gamma[\sigma,S]}{\delta \sigma^-(x)}{\Big |}_{\sigma^+=\sigma^-= \sigma} = 0\,.
\end{equation}
Next, quarks are integrated out by using their equation of motion obtained by varying the action w.r.t. to the components of $S$:
\begin{equation}
\dfrac{\delta \Gamma[\sigma,S]}{\delta S^{ab}(x,y)} = 0\,,
\end{equation}
which leads to
\begin{equation}
\left(i\slashed{D} - g \sigma_0(x) \right) S^{ab} (x,y) 
- \int_{\cal C} d^4z \, \dfrac{\delta \Gamma_2[\sigma,S]}{\delta S^{ac}(x,z)} 
\, S^{cb}(z,y) 
=  i \delta^{ab} \delta^{(4)}(x-y) \;.
\end{equation}

The above equations of motion are very difficult to solve, even numerically. A~tractable approximate scheme to solve those equations is to ``expand around the local equilibrium'', in a sense similar to the hydrodynamic description of many-particle systems. That is, the full $\sigma$ and $S$ are written as a sum of terms with the leading contribution being the corresponding equilibrium quantities $\sigma_0$ and $S_{\rm thm}$ at fixed values of temperature~$T$ and magnetic field. Specifically:
\begin{align}
\sigma^a(x) &=  \sigma^a_0(x) + \delta\sigma^a(x)\,,
\label{sig-exp} \\[0.25true cm]
S^{ab}(x,y) &= S^{ab}_{\rm thm}(x,y) + \delta S^{ab}(x,y) 
+ \delta^2 S^{ab}(x,y) + \cdots\;, \label{S-exp}
\end{align}
in which the leading terms are solutions of
\begin{align} 
&\frac{\delta \Gamma_{\rm cl}}{\delta \sigma^a_0(x)} = - g {\rm Tr}S^{aa}(x,x)\,, 
\label{sig0-eq}
\\[0.3true cm]
&\left[ i\slashed{D} - m_0 - g \, \sigma_0(x) \right] S^{ab}_{\rm thm}(x,y) 
= - i \delta^{ab} \delta^{(4)}(x-y)\;,
\label{Sth-eq}
\end{align}
and the terms $\delta S^{ab}(x,y)$ and $\delta^2 S^{ab}(x,y)$ are constructed iteratively~\cite{Nahrgang:2011mg,Krein:2021sco}. Equation~\eqref{Sth-eq} was solved in Ref.~\cite{Krein:2021sco} for a constant magnetic field and using the lowest Landau-level (LLL) approximation, which is valid only for very strong magnetic fields. The equation is easily solved in momentum space. Taking the magnetic field pointing in the $z$-direction, one can define transverse and longitudinal momentum components ${\bm p}^2_\perp = p^2_x + p^2_y$, $p_{\parallel}^2 = p_{0}^2 -p_{z}^2$, in terms of which the individual $S^{ab}_{thm}$ components of the thermomagnetic quark propagator are given by~\cite{Krein:2021sco}: 
\begin{align}
S^{++}_{thm}(p) &= e^{-{{\bm p}^2_\perp}/{\vert q_f B\vert}}  \, A(p) 
\left[\frac{i}{p_{\parallel}^2 - m^2_q + i\epsilon} 
- 2\pi n_F(p_0) \delta(p^2_\parallel - m^2_q) \right],
\label{S++thm} \\[0.2true cm]
S^{+-}_{thm}(p) &= e^{-{{\bm p}^2_\perp}/{\vert q_f B\vert}}  \, A(p) 
2\pi \delta(p^2_\parallel - m^2_q) \left[\theta(-p_0) - n_F(p_0)\right] ,
\label{S+-thm} \\[0.2true cm]
S^{-+}_{thm}(p) &= e^{-{{\bm p}^2_\perp}/{\vert q_f B\vert}}  \, A(p) 
2\pi \delta(p^2_\parallel - m^2_q) \left[\theta(p_0) - n_F(p_0)\right] ,
\label{S-+thm} \\[0.2true cm]
S^{--}_{thm}(p) &= e^{-{{\bm p}^2_\perp}/{\vert q_f B\vert}}  \, A(p) 
\left[\frac{-i}{p_{\parallel}^2 - m^2_q - i\epsilon} 
- 2\pi n_F(p_0) \delta(p^2_\parallel - m^2_q) \right],
\label{S--thm}
\end{align} 
where  $A(p) = (\slashed{p}_{\parallel} + m_q) \left[1 + i \gamma^1 \gamma^2 {\rm sign}(qB) \right]$ with  
$\slashed{p}_{\parallel} = \gamma^0 p_0 - \gamma^3 p_z$ and $m_q = g \sigma_0$, $q_u = 2 e/3$, $q_d = - e/3$, where $e = 1/\sqrt{137}$, and $n_F(p_0)$ is the Fermi-Dirac distribution:
\begin{equation} 
n_F(p_0) = \frac{1}{e^{|p_0|}/T + 1}.
\label{FD}
\end{equation} 
The asymmetry between the dependencies ${\bm p}_\perp$ and $p_z$ in these propagator components imposes an asymmetry in the Langevin field equation, as we shall discuss in the following. 

The traditional method of varying the action $\Gamma[\sigma,S]$ w.r.t. $\sigma^{\pm}$ to obtain an e.o.m. is not well defined because the action is complex. The Feynman-Vernon trick is used to identify the imaginary part with a noise source coupling linearly to the field such that a real action is obtained and the variation w.r.t. $\sigma$ leads to a Langevin equation for~$\sigma$. The resulting Langevin equation features dissipation and noise kernels with memory. The presence of memory in those kernels obstructs an analytical treatment, and further simplification is necessary to deal with such effects analytically. A common way to proceed is to use the so-called~{\em linear harmonic approximation}, in that the memory dynamics is effectively taken into account by soft-mode (long wavelengths) harmonic oscillations around a mean field\textemdash{for details, see Refs.~\cite{Nahrgang:2011mg,Rischke:1998qy,Xu:1999aq,Krein:2021sco}}. This leads to the following Langevin equation in momentum-space~\cite{Krein:2021sco}:
\begin{equation} 
\frac{\partial^2\sigma(t,{\bm p})}{\partial t^2} 
+ {\bm p}^2 \, \sigma(t,{\bm p}) + \eta_\sigma(\bm{p}_\perp) \,
\frac{\partial \sigma(t,{\bm p})}{\partial t} 
+ F_\sigma(t,{\bm p}) 
= \xi_\sigma(t,{\bm p}), 
\label{EoM-mom}
\end{equation} 
where $\eta_\sigma({\bm p}_\perp)$ is the transverse-momentum-dependent dissipation coefficient
\begin{equation} 
\eta_\sigma({\bm p}_\perp) = g^2 \frac{N_c}{4\pi} \left[1 - 2n_F(E_\sigma({\bm p}_\perp)/2T)\right] 
\frac{1}{E^2_\sigma({\bm p}_\perp)}
\sqrt{E^2_\sigma({\bm p}_\perp) - 4m^2_q} 
\sum_{f=u,d} |q_f B|  \; e^{-{\bm p}^2_\perp/2|q_fB|} ,
\label{eta-final}
\end{equation}
with $E_\sigma(\bm{p}_\perp) = (\bm{p}^2_\perp + m^ 2_\sigma)^ {1/2}$. The noise field $\xi_\sigma(t, {\bm p})$ has zero mean, $\langle \xi_\sigma(t, {\bm p})\rangle_\xi= 0$, and its correlationis also controlled by~$\eta_\sigma(t,\bm{p}_\perp)$, namely:
\begin{equation} 
\langle \xi_\sigma(t, {\bm p})  \xi_\sigma(t, {\bm p}) \rangle_\xi = 
(2\pi)^3 \delta({\bm p} + {\bm p}')  \, \delta(t-t') \, N_\sigma({\bm p}_\perp),
\label{noise-cf} 
\end{equation} 
where 
\begin{equation} 
N_\sigma({\bm p}_\perp)  = \eta({\bm p}_\perp) \, E_\sigma({\bm p}_\perp) \, \coth (E_\sigma({\bm p}_\perp)/2T) ,
\label{noise-final}
\end{equation} 
a feature reflecting the fluctuation-dissipation theorem. Finally, $F_\sigma(t,{\bm p})$ is given by
\begin{equation} 
F_\sigma(t,{\bm p}) = \int d^3x \, e^{-i {\bm p} \cdot {\bm x}} \, \left[ \lambda \, \sigma(t,\bm{x}) \left(\sigma^2(t,\bm{x}) - v^2\right) + g\,\rho_s(\sigma_0) 
+ h_q \right] ,
\label{def-f}
\end{equation} 
with the scalar density $\rho_s(\sigma_0)$ given by the sum $\rho_s(\sigma_0) = \rho^{BT}_s(\sigma_0) + \rho^B_s(\sigma_0)$, where $\rho^{BT}_s(\sigma_0)$ depends on~$B$ and~$T$
and $\rho^B_s(\sigma_0)$ depends only on~$B$:
\begin{align} 
\rho^{BT}_s(\sigma_0) &= - \frac{N_c}{\pi^2} \, m_q \left(|q_u B|+|q_d B|\right)  
\int^\infty_0 dp_z \, \frac{n_F(E_q(p_z))}{E_q(p_z)}, \label{rhoBT} \\
%
\rho^B_s(\sigma_0) &= - \frac{N_c}{2\pi^2} \, m_q\, \sum_{f=u,d} |q_fB| \left[\ln\Gamma(x_f) 
- \frac{1}{2} \ln 2\pi  + x_f - \frac{1}{2}(2x_f -1)\ln x_f\right],
\label{rhoB}
\end{align}  
where $N_c =3$ is the number of colors, $n_F(E_q)$ is the Fermi-Dirac distribution given in Eq.~\eqref{FD} with $E_q = \sqrt{p^2_z + m^2_q}$, $x_f = m^2_q/(2|q_fB|)$, and $\Gamma(x)$ the Euler gamma function. Equation~\eqref{EoM-mom} displays the $({\bm p}_\perp, p_z)$ asymmetry imposed by the magnetic field, as mentioned above. More specifically, the asymmetry comes through the ${\bm p}_\perp$ dependence of the dissipation coefficient{\textemdash}Ref.~\cite{Nahrgang:2011mg} deals with zero magnetic fields and obtains a momentum-independent~$\eta_\sigma$, namely:
\begin{equation}
\eta_\sigma(eB=0) =  g^2 \frac{2N_c}{4\pi} \left[1 - 2n_F(m_\sigma/2T)\right] 
\frac{1}{m^2_\sigma}
\left(m^2_\sigma - 4m^2_q\right)^{3/2} .
\label{eta-eB0}
\end{equation}

Equation~\eqref{EoM-mom} describes the nonequilibrium dynamics of the condensate, the field $\sigma$. Compared to the traditional mean field equation of motion for~$\sigma$, Eq.~\eqref{EoM-mom} is modified by two important physical effects: dissipation and fluctuation, the first is described by the first-order time derivative and the second by the noise field~$\xi$. Both effects are driven by the dissipation coefficient~$\eta_\sigma$, which is sourced by the $\sigma \leftrightarrow \bar{q}q$ processes; from Eq. \eqref{eta-final}, one sees that if $E_\sigma < 2 m_q$, $\eta_\sigma$~vanishes. The magnetic field affects both the sigma mass and the quark mass, and the temperature dependence of $\eta_\sigma$ might be very different from that of $B=0$, as we discuss shortly. If pions were included, they would also contribute to dissipation through the $\sigma \leftrightarrow \pi \pi$ processes. 

\subsection{Explicit results} 

Before presenting explicit numerical solutions of the Langevin equation~\eqref{EoM-mom} and analyzing the effect of the magnetic field, it is instructive to recall well-known qualitative facts about the solutions at short and long times. We consider a~high-to-low temperature quench, such that at $t=0$ the temperature $T \gg T_{\rm pc}$ suddenly drops to a value $T < T_{\rm pc}$. Right after the temperature quench, the field $\sigma$ is still small and one can neglect the term $\sigma^3$ in $F_\sigma$ and analytically solve \eqref{EoM-mom} for $\sigma(t,{\bm p})$ of the field, since the equation becomes linear in~$\sigma(t,{\bm p})$. The solution for the field average $\langle \sigma'(t,\bm{p})\rangle_\xi \equiv \overline{\sigma}(t,\bm{p})$ is of the form:
\begin{equation}
\overline{\sigma}(t,\bm{p}){\big|}_{t \simeq  0} =  A_-({\bm p}) \, e^{\lambda_+({\bm p}) \, t/2 } 
+ A_-({\bm p}) \, e^{\lambda_-({\bm p}) \, t/2 }, \hspace{0.5cm} 
\lambda_\pm(\bm{p}) = - \eta_\sigma  
\pm \sqrt{\eta^2_\sigma + 4 (\lambda v^2 - \bm{p}^2)} ,
\label{short}
\end{equation}
where $A_\pm(\bm{p})$ are determined by initial conditions and are irrelevant for the following discussion. This result shows the well-known short-time ``explosion'' of an order parameter after a temperature quench~\cite{Fraga:2004hp,Koide:2006vf}. The explosion, i.e., the short-time exponential growth, is driven by long-wavelength fluctuations, i.e., fluctuation modes with $\bm{p}^2 < \lambda v^2$, for which $\omega_+ > 0$, with the zero mode $\bm{p} = 0$ giving the most important contribution. However, short-wavelength fluctuations, for which $\bm{p}^2 > \lambda v^2$, decay exponentially with time (for $\bm{p}^2 > \eta^2_\sigma/4 + \lambda v^2$ they are damped oscillations). The time scales involved in this short-time behavior depend on the values of $\eta_\sigma(\bm{p}_\perp)$ and, of course, on $\lambda v^2$, which is related to the $\sigma$ mass, as we discuss shortly. As time increases, the linear equation cannot be used and one needs to solve the full, nonlinear equation. The initial explosion of the explosion is stopped by the term $\sigma^3$ in $F_\sigma$. For long times, the field eventually thermalizes, that is, it reaches an equilibrium value. Denoting $\sigma'(t,\bm{p})$ the fluctuations around the equilibrium value $\sigma_{\rm eq}$, the Langevin equation for $\sigma'(t,\bm{x})$ can, again, be linearized, and the resulting equation for $\sigma'(t,\bm{x})$ is in the form of a stochastically driven damped harmonic oscillator. The average of $\sigma'(t,\bm{x})$ is given by the homogeneous solution of this equation, namely: 
%
%
%
\begin{equation}
\overline{\sigma}'(t,\bm{p}) _{t \rightarrow  \infty} =  e^{- \eta_\sigma t/2 } \left[ A_-({\bm p}) \, e^{i \omega({\bm p}) \, t } 
+ A_-({\bm p}) \, e^{-i \omega({\bm p}) \, t }  \right], \hspace{0.5cm} 
\omega(\bm{p}) = \sqrt{\bm{p}^{2} + \lambda \left(3 \sigma^2_{\rm eq} - v^2\right)
- \eta^2_\sigma/4 }.
\label{long}
\end{equation}
For the parameter values used in this work, the frequencies $\omega(\bm{p})$ are real for any value of $\bm{p}^2$, a characteristic that implies that the long-time behavior of the field is in the form of damped oscillations, as should be, since the system has to reach equilibrium for a nonzero value of~$\eta_\sigma$. 

Next, we present numerical results of the full Langevin equation and make contact with the short- and long-time behaviors just discussed. We first discuss the impact of a strong magnetic field on the dissipation coefficient~$\eta_\sigma$, which is a key parameter in the time evolution of $\sigma$, as seen above. We limit our discussion to the contribution of the zero mode $\bm{p}_\perp =0$ to $\eta_\sigma$, since it is the most important mode in the dynamics of~$\sigma$, as seen above. Therefore, from Eq.~\eqref{eta-final}, one sees that the value of $\eta_\sigma$ is nonzero only when $m_\sigma > 2 m_q$. For $eB = 0$, the study in~\cite{Nahrgang:2011mg} has shown that $\eta_\sigma$ vanishes below the pseudocritical temperature $T_{\rm pc} = 150$~ MeV. For $eB \neq 0$, the situation changes substantially~\cite{Krein:2021sco}: $m_\sigma$, $m_q$ and the transition temperature increase (magnetic catalysis)~\cite{Ferrari:2012qta}, $m_\sigma$ and can be greater than $2 m_q$ and so $\eta_\sigma$ can be nonzero even below the new critical temperature. In Table~\ref{tab:eta} we show selected results for $\eta_\sigma$ from~\cite{Krein:2021sco} for two values of temperature and $eB$~{\textemdash}we recall that we are restricted to use relatively high values of $eB$ due to the use of the LLL approximation. We have chosen the temperature values $T = 50$~MeV and $T = 100$~MeV to emphasize a quench scenario for the simulation results we present shortly. The results shown in the table reveal that, for a given temperature, $\eta_\sigma$ increases with~$eB$. This result is driven partially by the linear dependence on $eB$ shown in Eq.~\eqref{eta-final}, which is a direct consequence of the dimensional reduction brought about by the magnetic field: the dimensional reduction replaces the $(m^2_\sigma - 4 m^2_q)^{3/2}$ in the zero field expression for $\eta_\sigma$ in Eq.~\eqref{eta-eB0} by $(m^2_\sigma - 4 m^2_q)^{1/2} \, eB$. On the other hand, for a given value of $eB$, $\eta_\sigma$ decreases with $T$, because both $m_\sigma$ and $m_q$ decrease with~$T$.

\begin{table} 
\centering
\caption{Values of the zero-mode dissipation coefficient, $\eta_\sigma(\bm{p}_\perp=0)$. }
\label{tab:eta}
\begin{tabular}{ccccccccc} 
    \hline
    {$eB/m^2_\pi$ } & { } & $T$ [MeV] & {} & $m_\sigma$ [MeV] & { } & $m_q$ [MeV] & { } & $\eta_\sigma$ [fm$^{-1}$] \\ 
    \hline
      & { } &   50  & { } & 601 & { }  & 307 & { }   &  0  \\[-0.35true cm] 
  0   & { } & {}    & { } & { } & { }  & { } & { }   & { }  \\[-0.35true cm]
      & { } & 100   & { } & 448 & { }  & 271 & { }   &  0  \\ 
     \hline
%
%
%
%
      & { } &  50  & { } & 726 & { }  & 342 & { }   & 1.12  \\[-0.35true cm] 
 10   & { } & {}   & { } & { } & { }  & { } & { }   & { }   \\[-0.35true cm]
      & { } & 100  & { } & 690 & { }  & 339 & { }   & 0.62  \\
    \hline
      & { } &  50  & { } & 810 & { }  & 370 & { }   & 1.89  \\[-0.35true cm] 
 15   & { } & {}   & { } & { } & { }  & { } & { }   & { }   \\[-0.35true cm]
      & { } & 100  & { } & 775 & { }  & 367 & { }   & 1.49  \\
    \hline
\end{tabular}
\end{table}

\begin{figure}[h]
\includegraphics[width=0.5\linewidth]{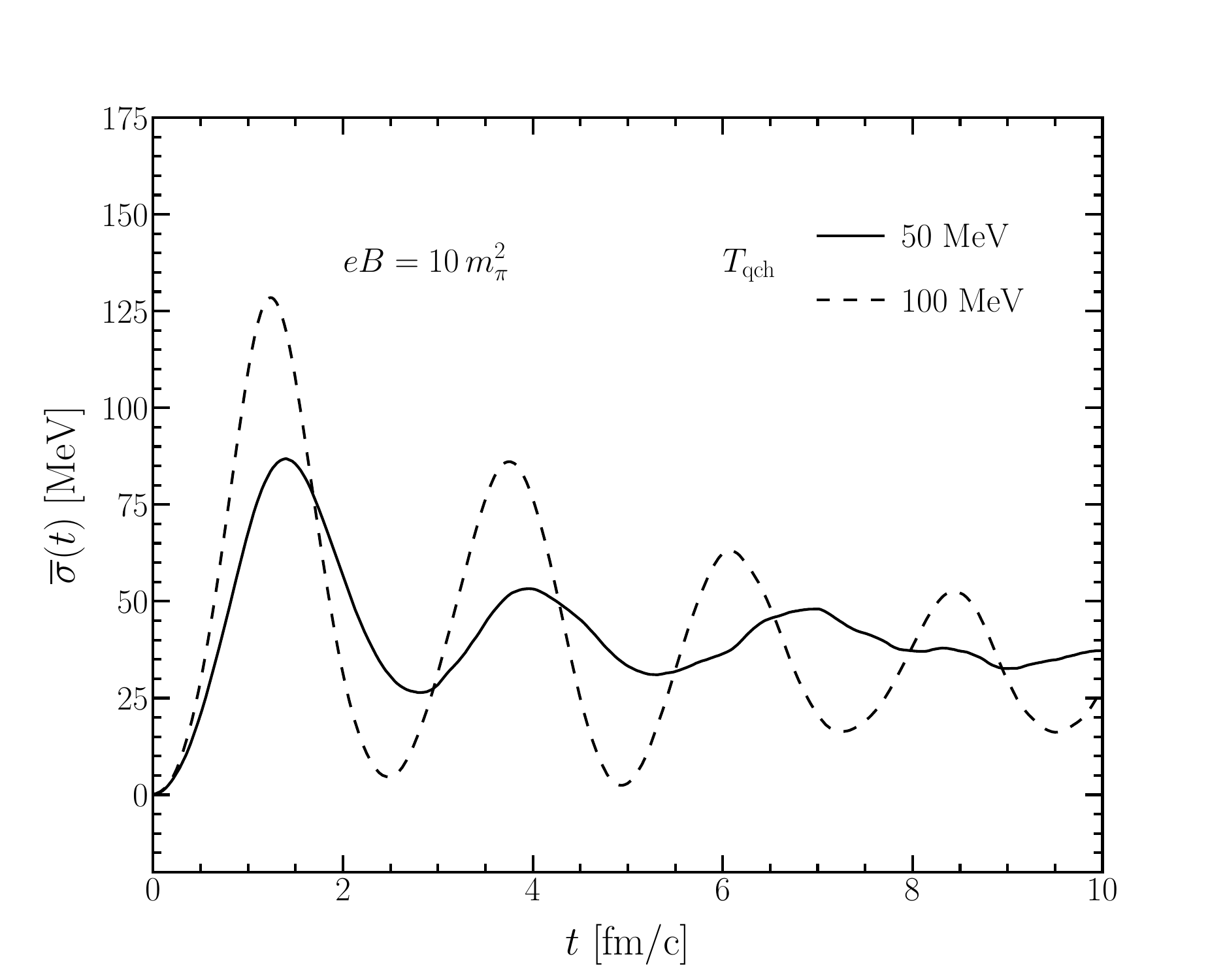} 
\hspace{-1.0cm}
\includegraphics[width=0.5\linewidth]{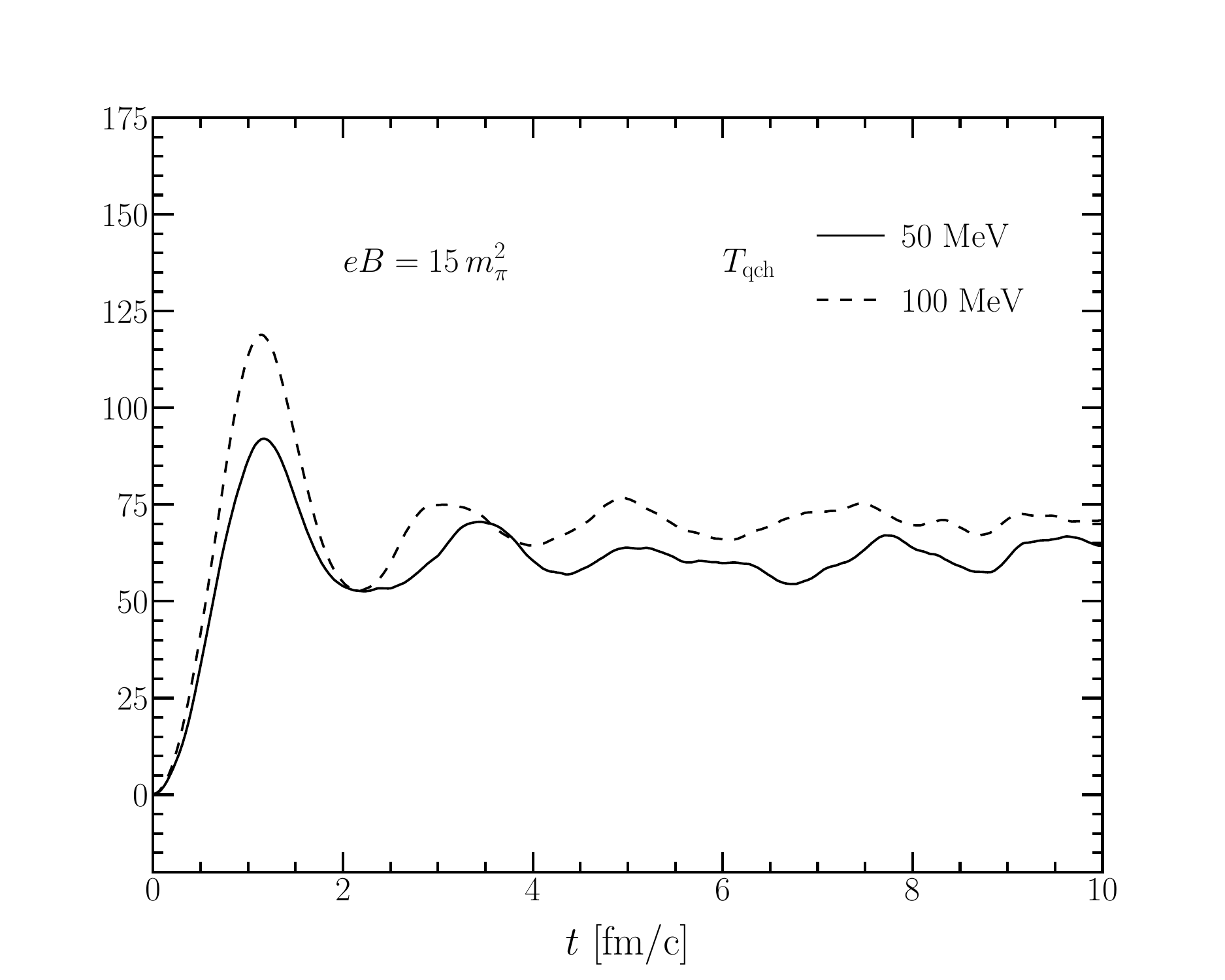} 
\vspace{-0.5cm}
\caption{Time dependence of average of the quark condensate corresponding to the temperature and $eB$ values in Table~\ref{tab:eta}.  }
\label{GK_fig1}
\end{figure}

We numerically solved the Langevin equation in coordinate space. We solved the equation in the $(x,y)$ plane, perpendicular to the magnetic field; in doing so, we took the field independent of the $z$~coordinate. We used a leapfrog algorithm to evolve the field in discrete time and a finite differences to discretize the spatial derivatives~\cite{AFrazon:2024,farias2007nonequilibrium,Cassol-Seewald:2007oak}. Figure~\ref{GK_fig1} displays the results for the spatial average of $\sigma$, that is, the average of $\sigma$ over the lattice, with the lattice extent being~$10$~fm. The results displayed are for two temperature quenches, $T_{\rm qch} = 50$~MeV and $T_{\rm qch} = 100$~MeV, and two values of the magnetic field, $eB = 50$~ MeV and $eB = 100$~MeV, the same as those in Table~\ref{tab:eta}. 

Both the short-time fast growth and the long-time damped oscillations discussed above are clearly visible in both panels of the figure. Short-time growth occurs within a time interval of about of~$1$~fm/c after the quench in all cases. The growth is controlled by $\lambda_+(0) =  \sqrt{4\lambda v^2 + \eta^2_\sigma} - \eta_\sigma > 0 $, which grows with~$T$ because $\eta_\sigma$ decreases with~$T$ for the parameter values used in this work. Thermalization of the field occurs within a time interval of $10$~fm/c for both temperature quenches and $eB = 15 \, m^2_\pi$, whereas for $eB = 10 \, m^2_\pi$, only for $T_{\rm qch} = 50$~MeV it termalized within that time interval. We recall from Eq.~\eqref{long} that the thermalization of the field is determined mainly by the value of $\eta_\sigma$ which, for $eB = 10 \, m^2_\pi$ for $T_{\rm ech} = 50$~MeV is the smallest among all the values considered in Table~\ref{tab:eta}.

%
\subsection{Conclusions and perspectives}
\label{sec:conclusions}

We addressed the non-equilibrium dynamics of the quark condensate under a strong magnetic field within the framework developed in Ref.~\cite{Krein:2021sco}. That framework bases the dynamics on a mean-field Langevin equation derived from the linear sigma model using the Schwinger-Keldysh closed-time path effective action and the Feynman-Vernon influence functional formalisms of non-equilibrium quantum field theory. We summarized the main steps involved in the derivation of the Langevin field equation and presented new results regarding the long-time evolution of the condensate and discussed its thermalization, results that were not contemplated in the original publication~\cite{Krein:2021sco}. Although the results presented have omitted important physical effects, they served the purpose of getting insight and developing an intuition on what to expect in a more complete calculation. Among other effects, a more complete calculation must include those due to pions and take into account the time dependence of the magnetic field. In addition, to address phenomenological consequences of the time evolution of the quark condensate in a heavy-ion setting, one needs to address the expansion and magnetohydrodynamics of the medium. 
We close by remarking that the framework developed in Ref.~\cite{Krein:2021sco} can be adapted to study magnetic field effects on the QCD phase transition in the early universe and in the interior of magnetized 
compact stars (magnetars). 

    \newpage \graphicspath{{./Figures_dense_magnetized_matter_phase_diagram/}}

 \part{Dense Magnetized Matter and Phase Diagram}
\label{densemagnetizedmatter}
	\section{Introduction}
 \label{DENSE_intro}
 
Quantum chromodynamics (QCD) encompasses multiple phases characterized by different degrees of freedom~\cite{Gross:2022hyw}. Quarks and gluons, though fundamental to the theory, are not directly observable experimentally; instead, we observe their bound states, namely mesons and baryons. It is well known that phenomena at low energies are primarily governed by the dynamics of the lightest mesons. Considerable effort has been devoted to exploring QCD under extreme conditions such as very high temperatures and densities~\cite{Fukushima:2010bq}. A significant challenge lies in understanding the physics governing quark-gluon plasma (QGP), a novel state of matter observed experimentally as a thermally equilibrated, deconfined phase of nuclear matter. Numerous experiments involving heavy-ion collisions (HIC) are actively investigating this unique phase of QCD matter to provide insights into constructing an accurate QCD phase diagram.
 
From a theoretical perspective, substantial efforts have been directed towards elucidating the QCD phase diagram. However, our understanding remains limited primarily due to the necessity of performing QCD calculations in the nonperturbative regime, which is currently impractical. Challenges in the ab initio lattice QCD approach, particularly in regions of moderately high densities due to the sign problem, further complicate this endeavor~\cite{Nagata:2021ugx,Splittorff:2007ck}. Consequently, much of our current understanding of the QCD phase diagram derives from effective models, which offer predictions for phenomena inaccessible through lattice techniques.
 
Recent years have seen a notable focus on the generation of strong magnetic fields during noncentral heavy-ion collisions, reaching strengths up to $10^{20}$ G. These fields, emerging immediately after the collision, can significantly impact the phases of QCD, often exceeding or matching the $\Lambda_{QCD}$ scale. The study of strong magnetic fields in systems described by QCD has seen significant growth in various fields, including magnetars, non-central heavy-ion collisions, neutron star mergers, and early universe physics. Recent reviews provide comprehensive insights into the phase structure and transitions of QCD matter under strong magnetic fields~\cite{Andersen:2014xxa,Miransky:2015ava,Bandyopadhyay:2020zte,Hattori:2023egw,Endrodi:2024cqn}. The number of works investigating quark/nuclear matter under intense magnetic fields has expanded greatly. Studies using lattice QCD simulations have explored the effects of strong magnetic fields on hadron spectra and scenarios of chiral symmetry restoration, yielding nontrivial results such as inverse magnetic catalysis~\cite{Bali:2012zg} at high temperatures and unexpected behaviors of neutral mesons~\cite{Ding:2022tqn}.

In this Part, we delineate recent theoretical advancements that enhance the continuous exploration of the magnetic field effects on the QCD phase diagram and dense magnetized matter. In section~\ref{sec2} a new mechanism for accelerating proto-neutron stars involving the chiral separation effect is proposed, which induces an axial vector current in dense environments. This study focuses on neutrino scattering interactions with the background axial vector current of electrons. It has become clear that the presence of anisotropy, whether in the magnetic field or density within momentum space, is essential for generating a nonzero recoil effect. This phenomenon, termed "chiral anisotropy conversion", requires conditions such as a strong magnetic field and moderate anisotropy. Under these conditions, the results suggest that chiral anisotropy conversion can lead to velocities of the order of typical pulsar kicks, comparable to velocities observed in nature.

In Section~\ref{sec3}, we examine the impact of incorporating ring diagrams into the four-quark interaction on the quark gap equation within a chiral model. This in-medium coupling is shown to decrease the chiral transition temperature in certain chiral models and has the potential to induce inverse magnetic catalysis at finite temperatures. Furthermore, the influence of confining forces on this phenomenon will also be analyzed.

In Sec.~\ref{sec4} we apply finite energy sum rules (FESR) in dense nuclear matter and in the presence of a constant and uniform external magnetic field to obtain information on different hadronic and QCD parameters.
Nucleon coupling currents and hadronic continuum thresholds can be obtained from the nucleon-nucleon correlator. The hadronic threshold signals confinement or deconfinement, so it is the main parameter provided in this formalism.
The objective of this work is the study of the nuclear axial coupling constant, which can also be obtained through the correlation of nucleons and axial-vector current.
This scenario of dense, magnetized matter emulates the interior of magnetars, so it may give an idea of what to expect there. The direct signal is the neutrino emission through the Urca process, which is a function of nuclear axial coupling. As a result, it is confirmed that dense matter tends to deconfine, while the magnetic field promotes confinement. It is also confirmed that the axial coupling at nuclear saturation density is $\sim 1$. Apparently, in this order of density values, the magnetic field has no important relevance, but it does for the most dilute or extremely dense nuclear matter.

In Sec.~\ref{sec5} we investigate the momentum diffusion coefficients of heavy quarks in a hot, magnetized medium, examining a wide spectrum of external magnetic field strengths. Our methodology involves systematically integrating the magnetic field's effects by using effective gluon and quark propagators designed for such a medium. To achieve gauge independence and derive analytical form factors applicable to all Landau levels, we utilize the hard thermal loop approach to the effective gluon propagator. By applying this resummed gluon propagator along with the generalized Schwinger quark propagator, we are able to analytically determine both longitudinal and transverse momentum diffusion coefficients for charm and bottom quarks, going beyond the static approximation.

In Sec.~\ref{sec6} the exploration of the magnetized QCD phase diagram is of considerable interest due to its potential to replicate fundamental properties of matter found in magnetars, the primordial universe, and peripheral ultrarelativistic heavy-ion collisions. Effective theories and models are commonly employed in the non-perturbative QCD regime because of their simplicity, which facilitates the investigation of conditions involving high temperatures, densities, and magnetic fields. Additionally, novel environments and effects could provide connections between effective models and lattice QCD. In this context, the quark anomalous magnetic moment (AMM) presents an opportunity to probe new phenomena within the magnetized QCD phase diagram. Current literature, using the Nambu–Jona-Lasinio model, predicts notable effects such as potential first-order phase transitions induced by strong magnetic fields, inverse magnetic catalysis at both zero and finite temperatures, and various non-physical oscillations in quark condensates. These phenomena are often associated with regularization procedures that intertwine magnetic field and vacuum contributions within the thermodynamic potential. This study will elucidate why these effects result from regularization issues and how the vacuum magnetic regularization (VMR) scheme can effectively address these challenges.
 \newpage
\section{Axial-vector current and implications to the phase diagram and the pulsar kick}\label{sec2}


\subsection{Induced vector and axial-vector currents}

It is a widely accepted idea that the magnetic field coupled with the vector (axial vector) field induces the axial vector (vector) current, which is concisely summarized as
$j_{\mathrm{A}}^\mu \propto \varepsilon^{\mu\nu\rho\sigma} (\partial_\nu \phi_{\mathrm{V}}) F_{\rho\sigma}$
and
$j_{\mathrm{V}}^\mu \propto \varepsilon^{\mu\nu\rho\sigma} (\partial_\nu \phi_{\mathrm{A}}) F_{\rho\sigma}$.
The best known examples include the chiral magnetic effect (CME) and the chiral separation effect (CSE)~\cite{Fukushima:2018grm,Kharzeev:2015znc}, that is,
\begin{equation}
  \boldsymbol{j}_{\mathrm{V}} = \frac{\mu_5  e\boldsymbol{B}}{2\pi^2}\,,\qquad 
  \boldsymbol{j}_{\mathrm{A}} = \frac{\mu e\boldsymbol{B}}{2\pi^2}
\end{equation}
for single Dirac fermion with the electric charge $e$.  Here, $\mu$ is the chemical potential coupled to the fermion number (if the fermion represents a single-flavor quark, $\mu$ is the quark chemical potential), while $\mu_5$ is the chiral chemical potential.  We note that a finite mass should modify $\boldsymbol{j}_{\mathrm{A}}$, while the CME current $\boldsymbol{j}_{\mathrm{V}}$ is robust with respect to infrared scales such as the mass and the temperature.

Usually, these vector expectation values are considered as realization of topological transport.  Of course, this interpretation is completely legitimate.  Nevertheless, it is strange that almost nobody had taken account of $\boldsymbol{j}_{\mathrm{V}}$ and $\boldsymbol{j}_{\mathrm{A}}$ for the phase diagram research of magnetized quark matter.  Actually, it is a common approach to postulate an effective interaction of four fermions, that is,
\begin{equation}
  \mathcal{L}_{\mathrm{int}} = -\sum_i G_i \bar{\psi} \Gamma_i  \psi\, \bar{\psi}\Gamma_i \psi\,.
\end{equation}
The interaction bases should contain the scalar, the pseudo-scalar, the vector, the axial-vector, and the spin (anti-symmetric tensor) channels.  Therefore, the mean-field interaction from $\Gamma=\gamma^\mu$ (and $\gamma^\mu\gamma^5$) has a contribution proportional to the squared quantity of $\boldsymbol{j}_{\mathrm{V}}$ (and $\boldsymbol{j}_{\mathrm{A}}$, respectively).  Let us specifically consider the CME current only.  Then, the mean-field interaction reads~\cite{Fukushima:2010zza}:
\begin{equation}
  \mathcal{L}_{\mathrm{int MF}} = -G_{\mathrm{V}} \boldsymbol{j}_{\mathrm{V}} \cdot \boldsymbol{j}_{\mathrm{V}} + 2G_{\mathrm{V}} \boldsymbol{j}_{\mathrm{V}}\, \bar{\psi}\boldsymbol{\gamma}\psi\,. 
\end{equation}
The first term is the mean-field energy and the second term is regarded as the gauge coupling.  That is, in view of the Dirac kinetic term, $\bar{\psi} i \gamma_\mu \partial^\mu \psi$, the above second term can be incorporated in the replacement of $\partial^\mu \to D^\mu=\partial^\mu - iA^\mu$ with $\boldsymbol{A} = -2G_{\mathrm{V}}\boldsymbol{j}_{\mathrm{V}}$.

Now, it is straightforward to write down the grand potential.  For the concrete setup, let us take the magnetic field direction along the $z$ axis.  The grand potential should be
\begin{equation}
  \Omega_{j_{\mathrm{V}}}/V
  = G_{\mathrm{V}} (j_{\mathrm{V}}^z)^2 + \Omega_A/V
  = G_{\mathrm{V}} (j_{\mathrm{V}}^z)^2 - \frac{|eB|}{2\pi}
  \sum_{s, n} \alpha_{s,n} \int \frac{dp^z}{2\pi} \omega_s(p)\,,
\end{equation}
where the dispersion relation is~\cite{Fukushima:2008xe}
\begin{equation}
  \omega_s^2 = m^2 + \Bigl[ |\boldsymbol{p}| + \mathrm{sgn}(p^z) s \mu_5 \Bigr]^2\,,
  \qquad
  |\boldsymbol{p}|^2 = (p^z + A^z)^2 + 2|eB|n
\end{equation}
with the Landau level $n$.  The spin degeneracy factor is $\alpha_{s,n}=1$ for $n\neq0$ and $\alpha_{s,0}=\delta_{s,+}$ for $s=\pm$.  In the ordinary case of equilibrated quark matter, the chiral chemical potential $\mu_5$ should be vanishing, and the phase diagram is not modified unless $\mu_5\neq 0$ is introduced by hand.

Still, it is an interesting question how $G_{\mathrm{V}}$ may have impact to $\boldsymbol{j}_{\mathrm{V}}$.  The mean-field, $\boldsymbol{j}_{\mathrm{V}}$, should be determined from the self-consistency condition, i.e.,
\begin{equation}
  \frac{\partial (\Omega_{j_{\mathrm{V}}}/V)}{\partial \boldsymbol{j}_{\mathrm{V}}} = 0\,.
\end{equation}
We shall treat the induced gauge field as a perturbation, and then the perturbative expansion gives the following expression:
\begin{equation}
  j_{\mathrm{V}}^z \simeq \frac{\partial(\Omega_A/V)}{\partial A^z}\biggr|_{A^z=0}
  + \frac{\partial^2 (\Omega_A/V)}{\partial A^{z2}} \biggr|_{A^z=0}
  A^z = 
  \frac{|eB|}{2\pi^2} \mathrm{sgn}(eB)\mu_5
  -2G_{\mathrm{V}} \chi_{\mathrm{V}}^{(0)} j_\mathrm{V}^z \,.
\end{equation}
Here, the correction from the vector interaction is proportional to the susceptibility defined by
\begin{equation}
  \chi_{\mathrm{V}}^{(0)} = \frac{\partial^2 (\Omega_A/V)}{\partial A^{z2}}\biggr|_{A^z=0}
  = \frac{|eB|}{2\pi^2} \biggl( 1 + \frac{2\Lambda^2}{3|eB|} \biggr)\,.
\end{equation}
It should be noted that the first term appears from the Lowest Landau Level contribution (that is the derivative of the CME current) and the second term from the higher Landau levels has scheme dependence with the ultraviolet cutoff $\Lambda$.  In the above expression, $\Lambda$ is the four-vector cutoff, that is, $\Lambda^2 = 2|eB|n + \Lambda_n^2$ with the longitudinal momentum cutoff $\Lambda_n$. The numerical value of the coefficient in front of $\Lambda^2$ may differ for other regularization such as the proper-time regularization.  In fact, $\chi_{\mathrm{V}}^{(0)}$ is nothing but the mass of the gauge field $A^z$, and it must be zero if $|eB|\to 0$.  Therefore, the term $\propto \Lambda^2$ should be an artifact from the explicit breaking of gauge symmetry due to naive momentum cutoff.  This implies that the properly renormalized value of $\chi_{\mathrm{V}}$ should be $\chi_{\mathrm{V}}=|eB|/2\pi^2$.  Then, finally, we can conclude that the self-consistent CME current should take the following form:
\begin{equation}
  \boldsymbol{j}_{\mathrm{V}} = \frac{1}{1+2G_{\mathrm{V}} \chi_{\mathrm{V}}}
  \frac{\mu_5 e\boldsymbol{B}}{2\pi^2}\,.
\end{equation}
This very simple exercise tells us a stimulating lesson.  If we impose a condition that the CME formula should hold regardless of the interaction, one possible interpretation of the above result is that the input variables receive in-medium modification.  Because $|eB|$ is renormalization invariant, it is unlikely that $\boldsymbol{B}$ is screened.  Thus, we see a renormalization of the chiral chemical potential as
\begin{equation}
  \mu_5 ~\to~
  \frac{\mu_5}{1+2G_{\mathrm{V}} \chi_{\mathrm{V}}}
  = \frac{\mu_5}{1+G_{\mathrm{V}} |e\boldsymbol{B}|/\pi^2}\,.
\end{equation}
We should be aware that $\mu_5$ can be medium modified just like $\mu$ shifted by the vector interaction.  It is noteworthy that the effective $\mu_5$ is significantly suppressed if $G_{\mathrm{V}}$ is positive and large.
This is qualitatively understandable.  Suppose that the magnetic field is sufficiently strong to justify the dimensional reduction along the longitudinal dynamics.  Then, in such a (1+1)-dimensional system, the spatial component of the vector current is identified as the temporal component of the axial-vector current.  A finite $\mu_5$ leads to a nonzero chirality $n_5$, so that the axial-vector interaction gives rise to renormalization of $\mu_5$.

A more interesting question, which is potentially relevant to the phase diagram research, is the effect of the axial-vector current from the CSE at finite density and magnetic field.  In the same way as the previous discussions, we can consider the mean-field interaction as
\begin{equation}
  \mathcal{L}_{\mathrm{int ML}} = -G_{\mathrm{A}} \boldsymbol{j}_{\mathrm{A}}
  \cdot \boldsymbol{j}_{\mathrm{A}} + 2G_{\mathrm{A}} \boldsymbol{j}_{\mathrm{A}}
  \bar{\psi}\boldsymbol{\gamma}\gamma_5 \psi\,.
\end{equation}
This extra term causes the axial gauge field,
$-\boldsymbol{A}_5\cdot \bar{\psi} \boldsymbol{\gamma}\gamma_5\psi$, with
$\boldsymbol{A}_5 = -2G_{\mathrm{A}} \boldsymbol{j}_{\mathrm{A}}$.  It is a nontrivial observation that this induced axial gauge field results in a finite density as
\begin{equation}
  j_{\mathrm{ind}}^0 = -\frac{1}{2\pi^2}
  \boldsymbol{A}_5\cdot e\boldsymbol{B}\,.
\end{equation}
Then, a natural extension we may want to ask is whether an axial-vector current is further induced or not.  The short answer is yes, but the final expression depends on the cutoff and renormalization scheme again.  We can parametrize it with an unknown coefficient $\alpha$.
Then, the same argument as before leads to
\begin{equation}
  j_{\mathrm{A}}^z \simeq
  \frac{|eB|}{2\pi^2} \mu - 2G_{\mathrm{A}} \chi_{\mathrm{A}}^{(0)} j_{\mathrm{A}}^z
\end{equation}
with the susceptibility given by
\begin{equation}
  \chi_{\mathrm{A}}^{(0)} = \frac{\partial^2(\Omega_{A_5}/V)}{\partial A_5^{z2}}
  \biggr|_{A_5^z=0}\,,\qquad
  \chi_{\mathrm{A}} = \alpha \frac{|eB|}{2\pi^2}\,.
\end{equation}
Here again, the susceptibility has the divergence from the Landau sum, that is, the naive expression involves $\alpha = 1 + 2\sum_{n>0}$ with $n$ the nonzero Landau level, which could be somehow renormalized.
In such a way, after all, the in-medium CSE current is modified into the following screened form:
\begin{equation}
  \boldsymbol{j}_{\mathrm{A}} = \frac{1}{1+2G_{\mathrm{A}} \chi_{\mathrm{A}}}
  \frac{\mu e\boldsymbol{B}}{2\pi^2}\,.
\end{equation}
We have seen that these induced currents may be affected by the interaction effects in extreme environments.  So far, there are not many theoretical works to study the in-medium modification effects, but definitely, the question deserves further investigations in the future.  Now, does it make sense to think of an opposite possibility that the medium properties are influenced by these currents?

\subsection{Inhomogeneous condensation and the axial-vector interaction}

Let us deepen our thinking about the physical meaning of the axial-vector interaction.  The low-energy effective theory in general would accommodate the axial-vector interaction like
\begin{equation}
  \mathcal{L}_{\mathrm{int}} = -G_{\mathrm{A}} \bar{\psi}\gamma_\mu \gamma_5\psi\; \bar{\psi}\gamma^\mu \gamma_5\psi\,.
\end{equation}
Using the Dirac representation, it is easy to take the nonrelativistic limit;
$\psi = (\varphi,\, \boldsymbol{\sigma}\cdot\boldsymbol{p}/(E+m) \varphi)^{\mathsf{T}}$ in which the lower component is negligible in the nonrelativistic treatment.  In this approximation, we immediately see,
$\bar{\psi}\boldsymbol{\gamma}\gamma_5\psi \simeq \varphi^\ast \boldsymbol{\sigma}\varphi$.  Therefore, the above interaction,
$\mathcal{L}_{\mathrm{int}}\sim G_{\mathrm{A}} \varphi^\ast \boldsymbol{\sigma}\varphi\cdot \varphi^\ast\boldsymbol{\sigma}\varphi$, represents the spin-spin interaction in the scattering of heavy fermions.

Such interaction vertices are also known in nuclear physics since long time ago.  In the conventional notation, the short-range Landau-Migdal interaction is parametrized as
\begin{equation}
  f_{NN} + g_{NN} \boldsymbol{\sigma}_1 \cdot \boldsymbol{\sigma}_2
  + f'_{NN} \boldsymbol{\tau}_1 \cdot \boldsymbol{\tau}_2
  + \biggl(\frac{f_{\pi}^2}{m_\pi^2}\biggr) g'_{NN} (\boldsymbol{\sigma}_1 \cdot \boldsymbol{\sigma}_2)
  (\boldsymbol{\tau}_1 \cdot \boldsymbol{\tau}_2)\,,
\end{equation}
where $\boldsymbol{\sigma}_i$ and $\boldsymbol{\tau}_i$ are the spin and the isospin operators, respectively.  For the $p$-wave pion condensation, the most important term is the last one with $g'_{NN}$ together with the counterparts with $\Delta$, i.e., $g'_{N\Delta}$ and $g'_{\Delta\Delta}$.  Although $\Delta$ is a spin-isospin 3/2 baryon, the generalization is straightforward with the replacement of $\boldsymbol{\sigma}_i$ and $\boldsymbol{\tau}_i$ by the transition spin-isospin operators.  The task for the investigation of the pion condensation is the computation of the pion dispersion relation $D_\pi^{-1}(\omega,\boldsymbol{k})$ with increasing baryon density.  The propagator inverse involves the pion self-energies (which are dominated by the $p$-wave $\pi N$ interaction) as well as the diagrams with the Landau-Migdal interaction.  Then, one may find $D_\pi^{-1}(\omega=0, k=k_c)=0$ at a threshold density $\rho=\rho_c$ depending on the values of $g'_{NN}$, $g'_{N\Delta}$, and $g'_{\Delta\Delta}$.  Experimentally, $g'_{NN}$ and $g'_{N\Delta}$ are constrained, while $g'_{\Delta\Delta}$ is undetermined; see Ref.~\cite{Tatsumi:2003fa} for a review.  Many nuclear theorists are not in favor of the pion condensation based on the analysis with the universality ansatz, $g'_{NN}\sim g'_{N\Delta}\sim g'_{\Delta\Delta}=0.6$-$0.8$, but the case is not closed yet.  These calculations and interpretations are usually presented in the nonrelativistic language, but the full relativistic translation can be easily done with the covariant form of the axial-vector current interaction.  It should be worth approaching the possibility of the pion condensation from the higher density where the relativistic formulation is more appropriate.

Surprisingly, what is discussed in the context of quark matter has remarkable similarity.  In a sense, if quark-hadron phases are continuously connected, quark matter may well have a dual of the pion condensed phase.  Let us simplify the story by assuming an effective (1+1)-dimensional system.  The dimensional reduction could occur in the strong magnetic field limit or even near the Fermi surface at very high density.  Then, in the chiral limit, the chemical potential can be absorbed in the fermion basis rotation as
\begin{equation}
  \bar{\psi}\bigl[ (\partial_4 - \mu)\gamma^4 + \partial_z \gamma^z \bigr]\psi
  = \bar{\psi}' (\partial_4 \gamma^4 + \partial_z \gamma^z ) \psi'\,,
\end{equation}
where
\begin{equation}
  \psi = e^{\mu\gamma^z \gamma^4 z} \psi' \,,\qquad
  \bar{\psi} = \bar{\psi}' e^{\mu\gamma^z \gamma^4 z}\,.
\end{equation}
At first glance, the chemical potential is just eliminated from the theory, and physical observables seem to be independent of $\mu$.  If this is true, the derivative of the grand potential with respect to $\mu$ is naturally zero, meaning that the density is zero even though the chemical potential was originally finite.  Is it really possible to delete all the fermions by the basis rotation only?  This plausible conclusion is drawn from careless momentum shift implied by the basis rotation.  We may indeed split the fermion into two components according to the eigenvalues of $\gamma^z \gamma^4$.  Because of the anti-commutative nature, $\gamma^z \gamma^4$ is an anti-Hermitian matrix, and its eigenvalues are $\pm i$.  Thus, we can introduce two eigenstates; $\gamma^z\gamma^4 \psi_+ = +i \psi_+$ and $\gamma^z\gamma^4 \psi_- = -i\psi_-$.  The basis rotation takes a simple form of $e^{\pm i\mu z}$, which results in the momentum shift by $\pm \mu$.

We can express the grand potential in the form of the phase space integration of the zero-point oscillation energy.  In the chiral limit, the energy dispersion relation is simply $\varepsilon(p)=|p|$.  If the original theory has a mass term, $\bar{\psi}\psi$, it is not invariant under the basis rotation.  Nevertheless, we may still consider a situation that some interaction yields a finite mass in terms of $\psi'$.  If this is the case, the dispersion relation is $\varepsilon(p)=\sqrt{p^2 + m^2}$.  Then, $\psi_\pm$ bases are mixed and the grand potential is written as
\begin{equation}
  \Omega(\mu)/V = -\int_{-\Lambda+\mu}^{\Lambda-\mu} \frac{dp}{2\pi}\,
  \frac{\varepsilon(p)}{2} - \int_{-\Lambda-\mu}^{\Lambda+\mu}
  \frac{dp}{2\pi}\, \frac{\varepsilon(p)}{2}\,,
\end{equation}
Interestingly, for $\Lambda \gg \mu$, we can extract the finite-density correction as
\begin{equation}
  \Omega(\mu)/V = \Omega(\mu=0)/V - \frac{\mu^2}{2\pi}\,,
\end{equation}
from which we derive the density as $n_{\mathrm{rotated}}=-\partial(\Omega/V)/\partial\mu = \mu/\pi$.
We note that the density is mass independent even though the energy dispersion relation explicitly contains a finite mass.  This mass independence is quite important.  Let us presume a system of (1+1)-dimensional fermion without any basis rotation at all.  Then, with a dynamically generated mass, $M$, the density should be $n_{\mathrm{non-rotated}}=p_{\mathrm{F}}/\pi\, \theta(\mu-M)$ with $p_{\mathrm{F}}=\sqrt{\mu^2 - M^2}$.  It is obvious that $n_{\mathrm{rotated}} > n_{\mathrm{non-rotated}}$ for $M\neq 0$, and so the free energy with the basis rotation should be smaller than the non-rotated case.  This energy consideration explains why the basis rotation to eliminate the $\mu$ term from the Lagrangian is more favored.  After the change of the basis, a condensate $\langle\bar{\psi}' \psi'\rangle$ is formed, which generates a dynamical mass but does not suppress the density due to the anomalous nature as we have elaborated above.  Suppose that we have a homogeneous condensate, $\Sigma=\langle\bar{\psi}'\psi'\rangle\neq 0$, this state has a helical condensate once expressed in terms of the original basis.  That is, it is easy to see~\cite{Kojo:2009ha}:
\begin{equation}
  \langle\bar{\psi}\psi\rangle = \Sigma \cos(2\mu z)\,,\qquad
  \langle\bar{\psi}\gamma^z\gamma^4\psi\rangle = -\Sigma \sin(2\mu z)\,,
\end{equation}
which is conventionally called the chiral spirals.  In this simple (1+1)-dimensional system, $\gamma^z\gamma^4$ could be identified as a counterpart of $\gamma_5$, and then, the above expresses an alternative helical condensate with $\sigma$ and $\eta$ mesons.  Interestingly enough, such a structure is analogous to what is called the Alternating-Layer-Spin (ALS) structure of neutral pion condensate~\cite{Takatsuka:1978ku}, which is characterized by $\pi^0 \sim \cos(k_c z)$.

This type of helical condensate should naturally be affected by the Landau-Migdal-type interaction, specifically, $g$ for this example.  Of course, not in the (1+1)-dimensional case but in the (3+1)-dimensional case, one can generalize the helical condensate with $\sigma$ and $\pi^0$ with the isospin matrix inserted in the condensate, which is called the dual chiral density wave state~\cite{Tatsumi:2004dx}.  For such a system, a counterpart of $g'$ should be taken into account and the axial-vector iso-vector interaction like
\begin{equation}
  \mathcal{L}_{\mathrm{int}} = -G_{\mathrm{A}}' \, \bar{\psi}\gamma_\mu\gamma_5 \boldsymbol{\tau} \psi \cdot \bar{\psi}\gamma^\mu \gamma_5 \boldsymbol{\tau}\psi\,.
\end{equation}
would be necessary.  In the neutron star environment, isospin symmetry is broken and it is expected that the iso-nonsinglet axial-vector should arise.  For example, the component coupled to $\tau_3$ is nothing but the difference between the axial-vector current of $u$ quarks and that of $d$ quarks.  It is a quite nontrivial question whether this axial-vector iso-vector interaction in neutron star matter (nearly pure neutron matter) would favor or disfavor the helical condensate structures in quark matter similar to the ALS in nuclear matter.

\begin{figure}
  \centering 
  \includegraphics[width=0.5\textwidth]{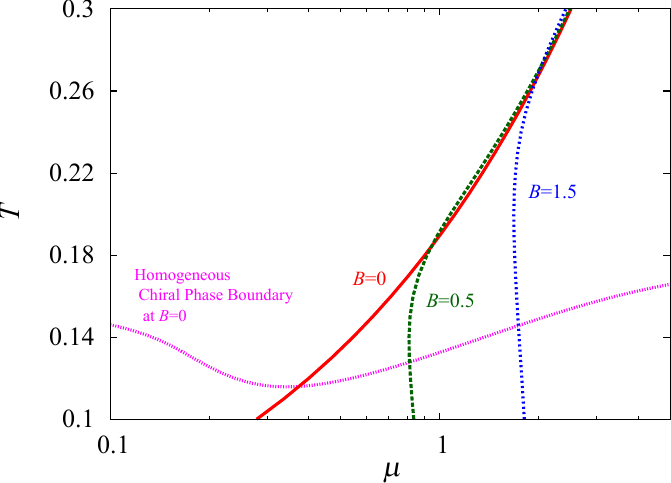}
  \caption{Phase boundaries calculated in the holographic QCD model (Sakai-Sugimoto model) with the dimensionless temperature $T$ and $\mu$.  The solid curve represents the onset of the spatially modulated phase at $B=0$, which is pushed up toward higher density with increasing $B$.  Figure is adapted from Ref.~\cite{Fukushima:2013zga}.}
  \label{fig:SSM}
\end{figure}

A systematic investigation is desirable in the future, but for the moment, we introduce a suggestive result from the holographic QCD model as shown in Fig.~\ref{fig:SSM}; see Ref.~\cite{Fukushima:2013zga} for details.  Here, the solid curve represents the onset of the spatially modulated phase at $B=0$.  In the region below this solid curve and above the dotted curve with the label, ``Homogeneous Chiral Phase Boundary'', only the inhomogeneous condensate is formed.  According to the previous arguments, stronger $B$ would render the system closer to the (1+1)-dimensional situation, and the chiral spirals should be preferred.  However, surprisingly, the onset curve is pushed up toward higher density with increasing $B$ as seen in Fig.~\ref{fig:SSM}.  One feasible account for this observation is that the background field of the axial-vector current enhances the effect of the axial-vector interactions that destroy the inhomogeneous condensates.  Although the holographic QCD model is a powerful approach including nonperturbative interactions, it is difficult to diagnose the microscopic mechanism behind the resulting answer.  To clarify the role played by the axial-vector interaction with strong magnetic field at finite density, model studies would provide us with useful insights.

\subsection{Axial-vector background scattering for the pulsar kick}

Finally, let us take one more example to demonstrate the implication of the mean-field background of the axial-vector current with magnetic field at finite density.  The proto-neutron star has strong $B$ and the neutrino density is so high that the mean-free-path of the neutrinos can be less than $1$m.  In cores of the proto-neutron star, the baryon density is high, and the neutrinos cannot escape outside, which defines the neutrino sphere from where the neutrinos can be emitted.  Inside of the neutrino sphere, scatterings makes the neutrinos thermally equilibrated with the electrons.

Since the electrons are charged particles, in response to the external magnetic field, they immediately develop the axial-vector current following the CSE formula.  There is no direct coupling between the magnetic field and the neutrino (except for the magnetic moment), but in the hydrodynamic regime, the axial-vector background may contain the neutrino component as well as the electrons.  If this is true, the CSE formula gives an estimate for the neutrino transport in the proto-neutron star along the magnetic direction.  Because the neutrino emission has a preferred direction reflecting the presence of the background current, the recoil accelerates the proto-neutron star.  In this way, the pulsar-kick mechanism can have a novel origin from the CSE current~\cite{Kaminski:2014jda,Yamamoto:2021hjs}.

\begin{figure}
  \centering 
  \includegraphics[width=0.25\textwidth]{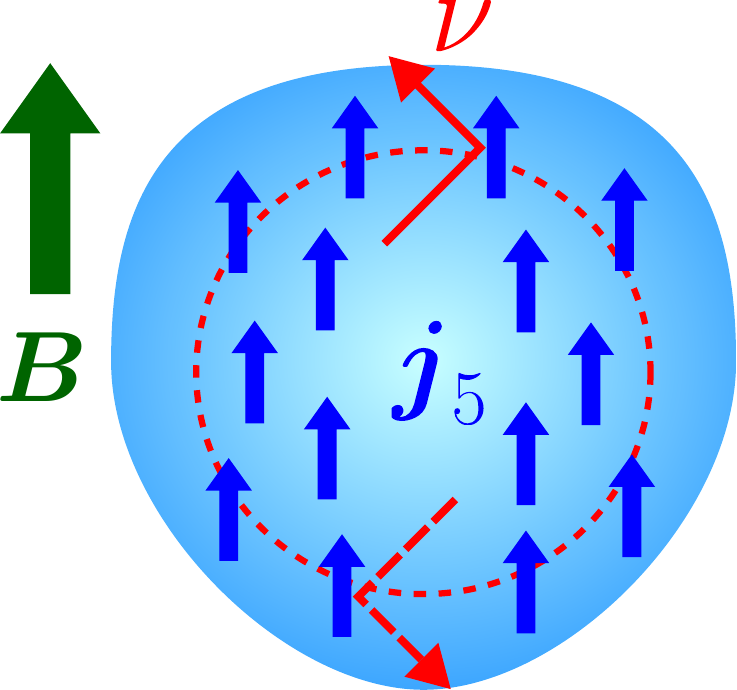}
  \caption{Neutrino scattering processes in the presence of the background field of the axial-vector current.  The dotted region schematically represents the neutrino sphere.  Figure is adapted from Ref.~\cite{Fukushima:2024cpg}.}
  \label{fig:kick}
\end{figure}

This scenario has a hypothesis of the hydrodynamic regime, however.  Even though the neutrinos are trapped in the neutrino sphere due to scatterings, it is not so conceivable that scatterings bend the neutrino motion coherently along the magnetic direction.  It is thus a challenging question what microscopic processes can realize the macroscopic transport of the neutrinos along the axial-vector current of the electrons.  This can be studied in a simplified treatment of the physical system in which the axial-vector current of the electrons is given as a background field~\cite{Fukushima:2024cpg}.  The weak interaction describes the interaction between the electrons and the neutrinos, and we introduce a mean-field approximation into those electron-neutrino vertices.  More specifically, the mean-field effective interaction is found as
\begin{equation}
  \mathcal{L}_{\mathrm{eff}} = \frac{3G_F}{2\sqrt{2}} j^\mu_{\mathrm{A},e}
  \, \bar{\nu}_e \gamma_\mu (1-\gamma_5) \nu_e\,.
\end{equation}
Using these vertices, we can consider the scattering processes such as
$\nu_e + j_{\mathrm{A},e} \to \nu_e$ and $\bar{\nu}_e + j_{\mathrm{A},e} \to \bar{\nu}_e$.
Now, the background field has a preferred direction proportional to $\boldsymbol{B}$, and it is a natural anticipation that the scatterings parallel to the current and anti-parallel to the current may exhibit different amplitudes.  In other words, the weak interaction explicitly breaks the parity symmetry.  Because the magnetic field is externally applied, the neutrino emission and absorption must have anisotropy.

Concrete calculations conclude that there is no difference between the upward scattering and downward scattering as schematically sketched in Fig.~\ref{fig:kick}.  This is understandable from the weak interaction.  For the anisotropic emission of the neutrino, the spin flip of the nucleus is necessary~\cite{Lai:1998sz}.  However, the background field of the axial-vector current cannot achieve the spin flipping and thus there is no anisotropic scattering.  Now, we learn a lesson;  the scattering between the electron axial-vector current and the neutrino is not sufficient to justify the hydrodynamic regime with the neutrino axial-vector current.  One should then deal with the scattering processes of the electron and the neutrino under the magnetic field.  Such a setup itself constitutes an alternative scenario for the pulsar-kick mechanism.

Although no kick is derived from the scattering between the electron axial-vector current and the neutrino, the presence of the current at least tends to focus the scattering directions and some beaming effects are expected.  The problem is that the neutrinos are emitted equally in both directions parallel and anti-parallel to the magnetic field.  Since the total energy of emitted neutrinos is huge, only a small imbalance would suffice to accelerate the pulsar.  If we see the supernovae simulations, the distributions of the magnetic field and the electron density profile, which both contribute to the axial-vector current, are not isotropic at all.  Therefore, it should be an acceptable assumption that the background field of the axial-vector current has spatial anisotropy.  Then, we can make a statement that the scattering on top of the spatially distorted axial-vector background is an efficient process to convert the spatial anisotropy to the acceleration (i.e., the chiral anisotropy conversion)~\cite{Fukushima:2024cpg}.

It should really be a challenging attempt to model the proto-neutron star evolution including the full dynamics of the neutrino and the background field effect of the electron axial-vector current.  It is always a subtle question how much effects can survive outside of the neutrino sphere where the density is small.  So far, the implication of the CSE current has been discussed based on the order of magnitude estimates.  To go beyond the order of magnitude estimates, the numerical simulations should be refined in the future studies.

 	\newpage
	\section{Driving chiral phase transition with ring diagram
}
 \label{sec3}


\subsection{Introduction}

Understanding the properties of QCD matter under extreme conditions, such as those created in ultra-relativistic heavy-ion collisions or in the cores of neutron stars, demands a reliable description of chiral symmetry restoration in a medium of partially deconfined quarks and gluons. Effective models are a robust exploratory tool for studying such dynamical systems. One advantage of these models is their ability to include or suppress specific interactions or diagrams, allowing for isolated examination of their effects. Additionally, they provide insights into the values of phenomenological parameters and their connections to the properties of underlying constituents. This approach is valuable for interpreting results from first-principle methods such as lattice QCD (LQCD).

Magnetic field provides an additional handle to probe the QCD phase diagram.  While most effective chiral models can capture the magnetic catalysis in vacuum, i.e. chiral condensate increases in magnitude at $T=0$ on increasing $B$, they predict the opposite trend from LQCD calculations~\cite{Bali:2011qj,Bruckmann:2013oba} on the magnetic field dependence of the chiral transition temperature~\cite{Boomsma:2009yk,Lo:2020ptj}.

There is yet another problem: The standard NJL model lacks a description of confinement.  A common remedy is to couple quarks to the Polyakov loop, resulting in the Polyakov-NJL (PNJL) model~\cite{pnjl,Fukushima:2017csk}. However, this class of models tends to overestimate the pseudo-critical temperature, predicting $T_{\rm ch} \approx 220$ MeV instead of the LQCD result of $T_{\rm ch} \approx 156.5$ MeV. Currently there is no satisfactory solution to this problem, and one has to resort to an {\it ad hoc} rescaling of the deconfinement transition temperature $T_d$ in the gluon potential from from its physical value of $T_d \approx 270$ MeV to about $T_d \approx 200$ MeV~\cite{shift1,shift2,mediumG1}.  Such an adjustment is far from ideal and indicates missing interactions in the original model.

In Ref.~\cite{Lo:2021buz}, we proposed a natural resolution to the aforementioned problems by including the polarization (ring) diagram. This approach drives the screening of the four-quark interaction, which lowers the $T_{\rm ch}$ at $B=0$ and leads to inverse magnetic catalysis at $B \neq 0$. A key merit of this method is that it introduces no new parameters and leaves vacuum properties unaffected, as the ring diagram vanishes in vacuum. In field-theoretical terms, this represents an improved truncation scheme beyond the standard mean-field approximation, specifically beyond the self-consistent Hartree-Fock method.

\subsection{Chiral Quark Model}

We review an effective chiral quark model motivated by the Coulomb Gauge
QCD~\cite{Quandt:2018bbu,conf,Lo:2021buz}. 
The Lagrangian density reads:
\begin{equation}
    \begin{split}
    \mathcal{L}(x) &= \bar{\psi}(x) \, (i \slashed{\partial}_x - m) \, \psi(x) \\
    & \quad - \frac{1}{2} \,
    \int d^4 y \,  \rho^a(x) \, V^{ab}(x, y) \, \rho^b(y)
    \end{split}
    \label{eq:model}
\end{equation}
where $m$ is the current quark mass, $\rho^a(x) = \bar{\psi}(x) \gamma^0 T^a
\psi(x)$ is the color quark current, and
$T^a$ is a generator of the $SU(N_c)$ symmetry group, with $a = 1, 2, \ldots, N_c^2-1$. 
For the class of models where the interaction potential $V$ is
instantaneous and color-diagonal, i.e.,

\begin{equation}
    V^{ab}(x, y) \rightarrow \delta^{ab} \times \delta(x^0-y^0) \,
    V(\vec{x}-\vec{y}),
\end{equation}
the gap equation for the dynamical quarks has been 
derived~\cite{conf}. It can be summarized as follows:

\begin{equation}
        S^{-1}(p) = \slashed{p} - m - \Sigma(p)
        \label{eq:sinv}
\end{equation}
where
\begin{equation}
    \Sigma(p) = C_F \, \int \frac{d^4 q}{(2 \pi)^4} \, V(\vec{p}-\vec{q}) \, i \, \gamma^0 S(q) \gamma^0.
        \label{eq:sfquark}
\end{equation}
The constant $C_F = \frac{N_c^2-1}{2 N_c}$ is introduced via the quadratic Casimir operator. 
A particularly transparent case is a contact interaction

\begin{equation}
V(\vec{p}-\vec{q}) \rightarrow V_0.
\end{equation}
The solution to the gap equation~\eqref{eq:sinv} becomes

\begin{equation}
    \begin{split}
        M &= m + {\rm Tr} \,(\Sigma) / {\rm Tr} \, (I) \\
        &= m + C_F \, V_0 \, \int \frac{d^3 q}{(2 \pi)^3} \,
        \frac{M}{2 E} \\
        &\quad \times (1 - 2 \, N_{\rm th}(E)),
    \end{split}
    \label{eq:cgap}
\end{equation}
where $E = \sqrt{q^2 + M^2}$ and $N_{\rm th}(E) = ({e^{\beta E}+1})^{-1} $ 
is the Fermi-Dirac distribution.
Equation~\eqref{eq:cgap} is of the same form as the familiar result for quark mass
($N_f$ flavors) in the model of Nambu-Jona and Lasinio
(NJL)~\cite{Klevansky1992}. 

The in-medium dressing of the interaction potential $V_0$ 
by the polarization tensor $\Pi_{00}$ is implemented via~\cite{conf}

\begin{equation}
    \tilde{V_0}^{-1} = {V_0}^{-1} - \frac{1}{2} \, N_f \, \Pi_{00}
    \label{eq:dress}
\end{equation}
where the polarization is given by
\begin{equation}
    \label{eq:pi00}
    \begin{split}
        \Pi_{00}(p^0, \vec{p}) &= \frac{1}{\beta} \, \sumint {\rm Tr} \left( \gamma^0
        S(q) \gamma^0 S(q+p) \right).
    \end{split}
\end{equation}
Here $\sumint$ denotes a Matsubara sum over the fermionic frequencies
($\omega_n = (2n+1) \, \pi/\beta$), and an integral over the momenta $d^3 q$.
Equation~\eqref{eq:dress} describes the screening of the gluon propagator by the Debye mass. The factor of $\frac{1}{2}$ in Eq.~\eqref{eq:dress} comes from the color structure, i.e. ${\rm Tr} \, T^a T^b = \frac{1}{2} \, \delta^{ab}$, and is essential to reproduce the known result of the perturbative Debye mass for QCD, rather than for QED.

\begin{figure}
	\resizebox{0.8\textwidth}{!}{%
	\includegraphics{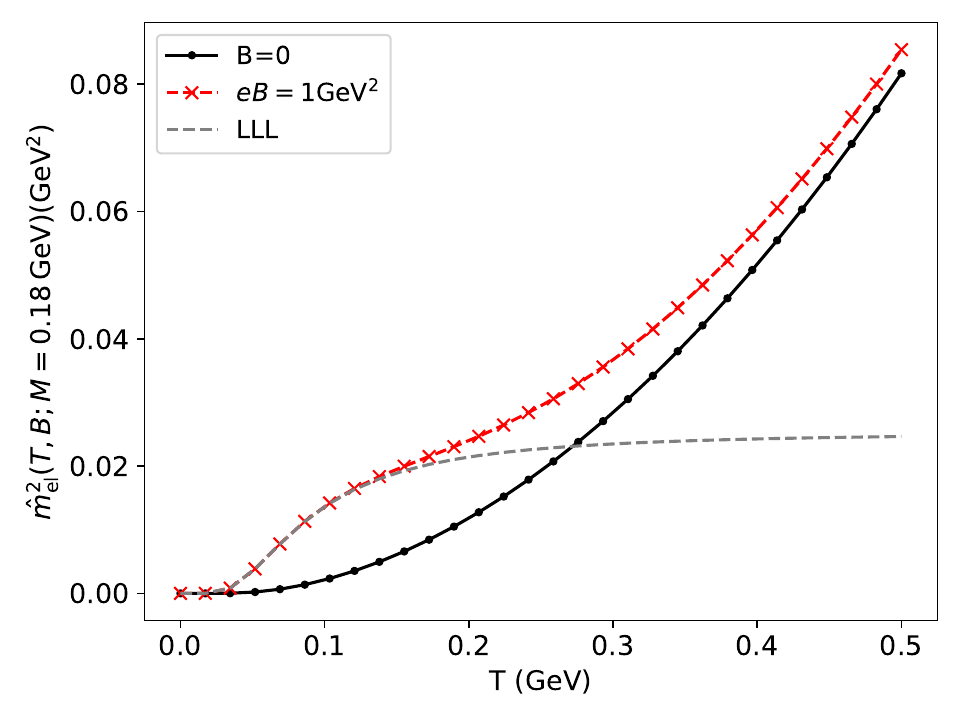}
	}
        \caption{The electric mass (Eq.~\eqref{eq:electric_mass}) for the 2-flavor case at fixed quark mass $M=0.18$ GeV, 
        as a function of temperature.}
\label{DENSE_fig1}
\end{figure}

For simplicity we only consider the screening by the static limit of $\Pi_{00}$. 
The vacuum part vanishes exactly. The matter part is related to the Debye (electric) mass via

\begin{equation}
    \begin{split}
    \hat{m}_{el}^2 &= -\frac{1}{2} \, N_f \,  \Pi_{00}(p^0=0, \vec{p} \rightarrow \vec{0}) \\
        &= \frac{1}{2} \, N_f \times  \int \frac{d^3 q}{(2 \pi)^3} \, 4 \beta N_{\rm th} (1-N_{\rm th}).
    \end{split}
    \label{eq:electric_mass}
\end{equation}
In Fig.~\ref{DENSE_fig1} we demonstrate a numerical calculation of the Debye mass Eq.~\eqref{eq:electric_mass} at fixed value of $M=0.18$ GeV (a typical value near the transition), and $N_f=2$ flavors. The results at finite $B$ are also shown. 

The following limits help interpret the results:

1. At $B=0$ and $M \rightarrow 0$ (or large $T$):
    \begin{equation}
        \hat{m}_{el}^2 \approx \frac{1}{2} \, N_f \times \frac{T^2}{3};
        \label{eq:lim1}
    \end{equation}

2. At $B \neq 0$, considering the contribution from the lowest Landau level (LLL):
    \begin{equation}
        \hat{m}_{el}^2 \approx \frac{1}{2} \,  \frac{\vert e_f \vert B}{4 \pi} \, \int
            \, \frac{d q_z}{2 \pi} \, \frac{4 \beta e^{\beta \sqrt{q_z^2+M^2}}}{(e^{\beta \sqrt{q_z^2+M^2}}+1)^2 },
    \end{equation}
    which for massless quarks reduces to:
    \begin{equation}
        \hat{m}_{el}^2 \rightarrow \frac{1}{2} \,  \frac{\vert e_f \vert B}{2 \pi^2}.
        \label{eq:LLL}
    \end{equation}

Observe how the exact numerical result follows the LLL at low temperature and eventually reaches the limit in Eq.~\eqref{eq:lim1} at high temperature. Note that the limit in Eq.~\eqref{eq:LLL} is not realized.

\begin{figure*}
	\resizebox{\textwidth}{!}{
	\includegraphics{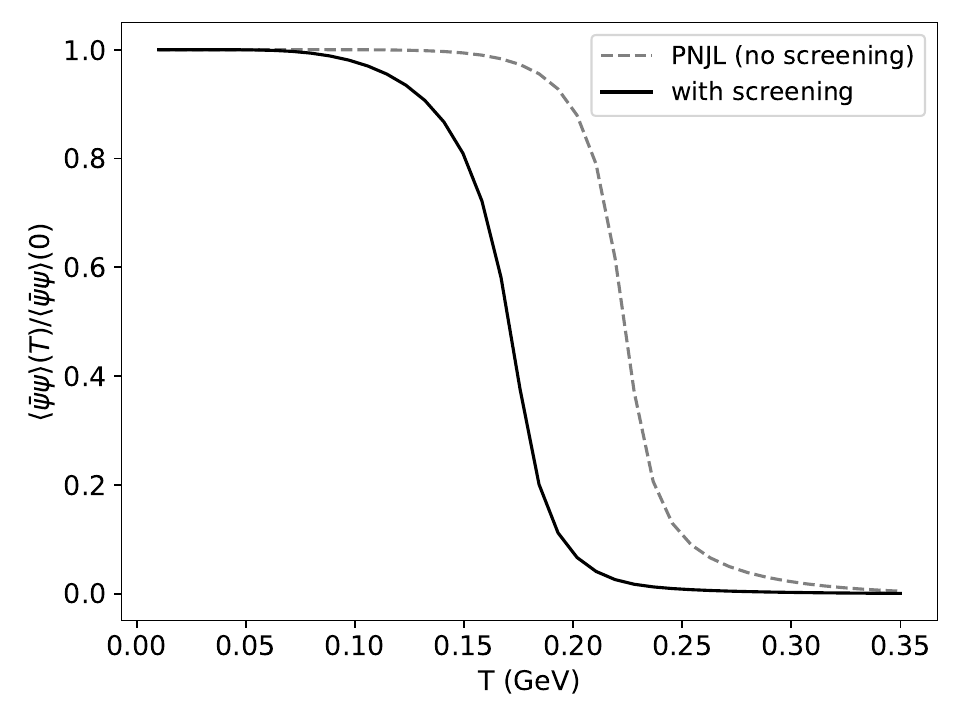}
	\includegraphics{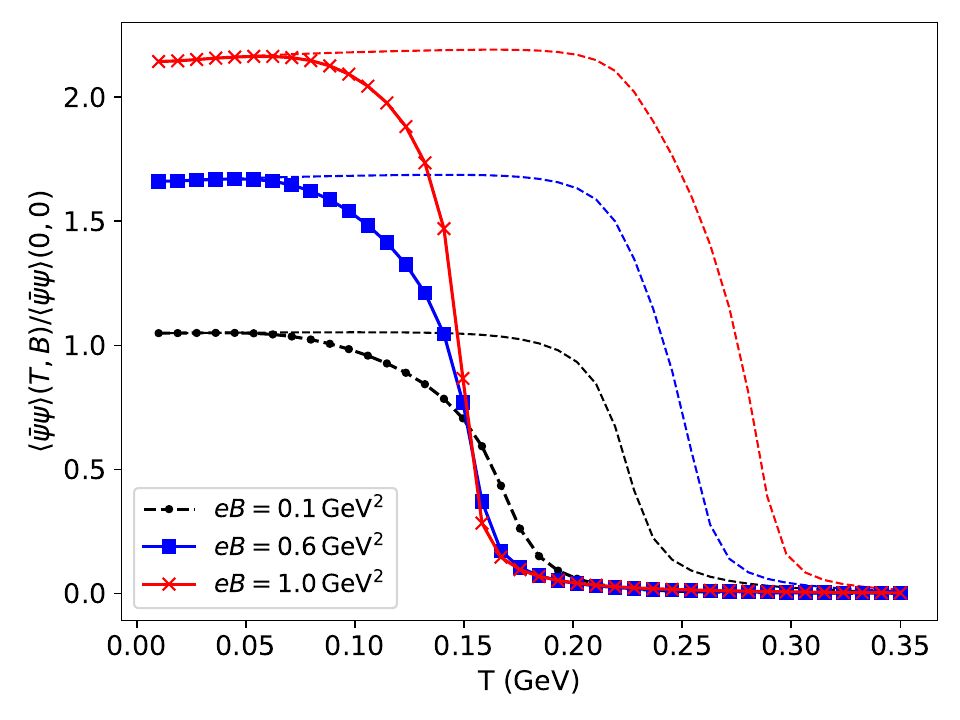}
	}
        \caption{The chiral condensate (with current quark mass contribution
        subtracted), normalized to the vacuum value, versus the temperature, at zero (left) and finite (right) magnetic field. 
        Dashed lines represent results obtained from a PNJL model without 
        screening effect. 
        The model with dressed coupling is capable of producing the inverse magnetic catalysis at finite temperature.}
\label{fig2pml}
\end{figure*}

Following the dressing of the four-quark coupling in Eq.~\eqref{eq:dress}, it is intuitive to see the effect of the ring diagram: the coupling is substantially weakened at high temperatures, allowing a transition at lower $T_{\rm ch}$. Additionally, at a fixed temperature, the coupling is weakened by the magnetic field, driving inverse magnetic catalysis. The actual prediction by the model involves a complex interplay between chiral symmetry and confinement. A typical result is shown in Fig.~\ref{fig2pml}.
See Ref.~\cite{Lo:2021buz} for details~\footnote{For completeness we give the
model setup here: we use $ \tilde{V}_0(T; M, \ell) \approx
\tilde{V}_0(T; \langle M \rangle, \ell)$, and fix $\langle M \rangle = 0.136$ GeV.
The set of model parameters is given by: 
$\Lambda = 1.076$ GeV, $G_{\rm NJL} \, \Lambda^2 = 4.232$, $\mathcal{R}_{4D}(q)
= e^{-q^8/\Lambda^8}$, 
and a current quark mass $m = 5$ MeV. 
In the vacuum, this gives: $f_\pi = 92.9$ MeV, $m_\pi = 137.8$ MeV, and
$\langle \bar{\psi} \psi \rangle = -(250 \, {\rm MeV})^3$ (per flavor).}.

\subsection{Towards Dense Matter}

To advance our understanding of chiral quarks at finite baryon density, it is essential to incorporate additional physical effects absent in the current model. Chief among these is restoring the momentum dependence of various quark dressing functions, such as the constituent mass function and wavefunction renormalization, through improved modeling of the confining potential~\cite{conf}. This would enhance our understanding of deconfinement beyond the statistical confinement picture employed~\cite{eric_cgauge,Lo:2020ptj}. Additionally, a lattice computation of the gluon Debye mass at low temperature with a finite magnetic field could validate the relevance of ring diagram screening to inverse magnetic catalysis.


\newpage
	\section{Nucleon axial coupling constant under magnetars conditions}\label{sec4}


\subsection{Introduction}
The study of dense nuclear matter and high-energy processes focuses the attention of researchers mainly on the study of quantum chromodynamics (QCD) phase transition in relativistic heavy ion collisions and the interior of neutron stars.
One of the particular problems for studying QCD under extreme baryon density is that lattice QCD simulations are not applicable under this regime, due to the well-known sign problem \cite{Nagata:2021ugx}.
This scenario provides the need to deal with effective models to describe the properties in the QCD medium, in particular at high baryon chemical potential, or baryon density.
The particular case of compact stars completely escapes the possibilities of lattice simulation, so the use of effective hadronic or QCD models with non-perturbative effects becomes the only way at the moment.
The possibility of extracting information from the hadronic sectors as well as from the quark degrees of freedom makes the QCD sum rules a versatile tool, since it can connect any hadronic model at intermediate energies to the general QCD description with non-perturbative effects parameterized by the operator product expansion (OPE).

Here we show the implementation of the finite energy sum rules (FESR) at high baryon density under the presence of an external magnetic field, considering nucleon correlators.
The calculation of the axial coupling constant will be presented in detail.

\bigskip

\begin{figure}
\begin{minipage}{0.3\textwidth}
\includegraphics[scale=0.7]{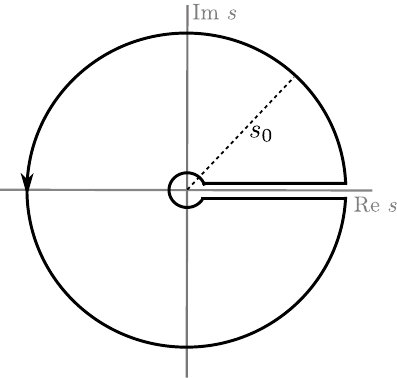}
\end{minipage}
\begin{minipage}{0.7\textwidth}
\qquad\qquad\includegraphics{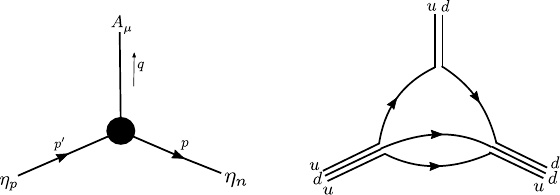}
\end{minipage}
\caption{ {\it Pac-man} contour used in FESR ({\bf left}).
Proton, neutron and axial-vector currents in the hadronic sector ({\bf center}) and in the  QCD sector  ({\bf right})}
\label{fig:contour-diagrams}
\end{figure}

There are different formalisms of addition rules, all of them with their own advantages and disadvantages.
Here FESR will be presented as a practical tool for the determination of various in-medium hadronic and QCD parameters based on quark-hadron duality.
Among the different types of sum rules in the literature, FESR establishes a clear criterion for the separation between the hadronic and QCD sectors.
Considering a form factor $\Pi(s)$ generated by current correlation, with $s=q^2$,
the FESR are obtained by integrating the form factor along the well-known contour {\it pac-man}, as shown in Fig.\,\ref{fig:contour-diagrams} (left).
The hadronic resonances lie on the positive real axis, while the QCD sector lies on the circle.
The radius of this circle, $s_0$, is the hadronic continuum threshold, denoted as the energy where the resonances begin to overlap.
At finite temperature, $s_0$ acts as an order parameter for the deconfinement phase transition.

From Cauchy's theorem, quark-hadron duality relates both sectors, i.e.
\begin{equation}
    \int_0^{s_0}\frac{ds}{\pi} s^N \text{Im}\Pi^\mathrm{\tiny had}(s)
   =-\oint_{s_0}\frac{ds}{2\pi i}s^N\Pi^\mathrm{\tiny QCD}(s) \,,
   \end{equation}
where the factor $s^N$ is an analytic kernel.
To relate these two sectors, non-perturbative effects are included, parameterized in terms of the operator product expansion (OPE) leading to
$
    \Pi(s) = C_0(s) + \sum_{n>0}C_n(s)\langle O_n\rangle/s^n,
$
where the first term corresponds to perturbative QCD, and the second term is the sum of operators corresponding to combinations of quark and gluon background fields.

\subsection{In-medium nucleon-nucleon correlator}

In order to describe nucleon currents we use Ioffe's interpolating function \cite{Ioffe:1981kw} for protons
\begin{equation}
    \eta _{p} (x) = \varepsilon ^{abc}[(u^a)^T (x)\; C\gamma _{\mu}u^b (x)]\gamma ^{\mu} \gamma _5 \,d^c (x), \label{eta}
\end{equation}
where $C$ is the charge conjugation matrix given by $C = i \gamma_0\gamma_2$.
The neutron current is the same but exchanging the quark flavors, i.e. $u\leftrightarrow d$.
In the hadronic sector $\eta (x)$ is defined  by their matrix elements as
$
\langle 0 | \eta _{N} (x) |N(p, s) \rangle = \lambda _{N} u_N(p,s)e^{ip\cdot x}, \label{Lambda}
$
being $\lambda _{N}$ the current-nucleon coupling, a phenomenological parameter,  and with $u_N$ the nucleon spinor.
The interpolating function can then  be expressed in terms of the asymptotic nucleon field as $\eta_N(x) =\lambda_N\Psi_N(x)$ which turns to be convenient if we want to calculate in-medium correlators.
The two-point function is
   $\Pi(q) = i \int d^4 x  \; e^{iq x}  \langle 0| T\,\eta (x) \bar{\eta } (0)| 0\rangle.  $

The procedure now consists of treating the interpolating function in terms of nucleon fields and calculating the time-ordered correlators by considering fermion propagators in a magnetic background.
Also he same procedure will be used in the QCD sector with quarks.
Here we will expand the propagator in power series of the magnetic field $B$, getting a general expression for the Green function $G(x,x') = e^{\Phi(x,x')}S(x-x')$, where $S$ is the local part of the propagator and $\Phi$ is the Schwinger phase.


The information of the presence of a uniform and constant magnetic field is provided by the propagators.
In the QCD sector, the correlators can be expanded in powers of $B$, which after contour integration becomes a series in powers of $(eB/s_0)^n$.
Considering that $s_0 > 1$\, GeV$^2$, the magnetic field can reach easily physical values for systems like relativistic heavy-ion collisions or magnetars.

The general expression for the local part of the fermion propagator in momentum space is ~\cite{Mukherjee:2018ebw}.
\begin{align}
S(p)= &\;\frac{i(\slashed{p}+m)}{p^2-m^2+i\epsilon}
-(qB)\frac{i\sigma_{12}(\slashed{p}_\parallel+m)}{(p^2-m^2+i\epsilon)^2}
+2i(qB)^2\frac{(\slashed{p}_\parallel+m)\left[p_\perp^2-\slashed{p}_\perp(\slashed{p}_\parallel-m)\right]}{(p^2-m^2+i\epsilon)^4}
-(\kappa B)\frac{i(\slashed{p}+m)\sigma_{12}(\slashed{p}+m)}{(p^2-m^2+i\epsilon)^2}
 +\dots,
\label{eq_SB}
\end{align}
where $\kappa$ is the anomalous magnetic moment of the fermion.
In the hadronic sector, it is customary to introduce a perpendicular fermion velocity denoting the symmetry breaking by the background magnetic field, so in this case $p=(p^0,v_\perp\boldsymbol{p}_\perp,p^3)$.
This velocity is related with the perpendicular screening mass as $m_\perp = m/v_\perp$.

In vacuum, the nucleon correlator can be decomposed as $\Pi(p)=\slashed{p}\Pi_1(p^2) +\Pi_2(p^2)$.
However, the magnetic field will introduce new structures which has to be combined with the electromagnetic tensor $F^{\mu\nu}$.
The most general decomposition of a correlator is given in germs of the Clifford algebra structures, scalar, pseudoscalar, vector, axial-vector, and tensor
\begin{equation}
\Pi=\Pi_S + i\gamma_5 \Pi_P + \gamma_\mu\Pi_V^\mu + \gamma_\mu \gamma_5 \Pi_A^\mu + \sigma_{\mu\nu}\Pi_T^{\mu\nu}.
\label{eq_Clifford}
\end{equation}
$\Pi_P=0$ unless topological anomalies be present.
So, the most general combinations of the vector, tensor and axial-vector components must involve $p_\mu$, $g_{\mu\nu}$, $\epsilon_{\mu\nu\alpha\beta}$ and $F_{\mu\nu}$, considering  that the correlator must be parity even:
\begin{equation}
    \Pi_V^\mu  =  p_\parallel^\mu\,\Pi_V^\parallel+p^\mu_\perp\ \Pi_V^\perp+\tilde p_\perp^\mu \tilde\Pi_V^\perp,
    \qquad
     \Pi_A^\mu  =  \tilde p_\parallel^{\mu}  \Pi_A,
        \qquad
\Pi_T^{\mu\nu} = \epsilon_\perp^{\mu\nu}\Pi_T^\perp
    + (p_\parallel^\mu p_\perp^\nu - p_\parallel^\nu p_\perp^\mu)\Pi_T^{\parallel\perp}
    + (p_\parallel^\mu \tilde p_\perp^\nu -  p_\parallel^\nu \tilde p_\perp^\mu)\tilde\Pi_T^{\parallel\perp},
        \label{eq_PiT}
\end{equation}
where $\tilde p_\perp^\mu\equiv \epsilon_\perp^{\mu\alpha}p_\alpha$,  $\tilde p_\parallel^\mu\equiv \epsilon_\parallel^{\mu\alpha}p_\alpha$, with the anti-symmetric tensors defined as $\epsilon_\perp^{\mu\nu} \equiv \epsilon^{0\mu\nu 3}$ and $\epsilon_\parallel^{\mu\nu}  = \epsilon^{\mu 12 \nu}$.

One may notice that we now have eight independent structures instead of two.
This abundance of information is compensated by the appearance of new condensates, where for example the gluon condensate will now be divided into chromoelectric and chromomagnetic condensate, along or perpendicular to the external magnetic field.

Each independent form factor can now be integrated along the FESR contour in the frame. $p_\perp=0$ with $p_\perp^2=s$.
FESR has the advantage of skipping the form factor calculation, because the contour integration can be done before the integration of the internal loop momentum.
This simplifies the calculations considerably \cite{Dominguez:2018njv,Dominguez:2020sdf}.

In \cite{Dominguez:2020sdf} the structures $\Pi_S$, $\Pi_V^\parallel$, $\Pi_V^\perp$ and $\Pi_T^\perp$ were used to calculate the proton and neutron current couplings$\lambda_p$, $\lambda_n$, the hadronic thresholds $s_p$, $s_n$, the mass parameters $m_n$, $m_p$, the perpendicular velocities $v^p_\perp$, $v_\perp^n$ and the spin polarization of the quark condensates $\langle \bar u\sigma_{12} u\rangle$ and $\langle \bar d\sigma_{12} d\rangle$.

\bigskip
QCD sum rules for baryons in dense baryonic medium has been strongly studied since the first set of papers \cite{Cohen:1991js,Cohen:1991nk,Furnstahl:1992pi,Jin:1993up,Jin:1994bh}.
The dense medium will break Lorentz symmetry and hence temporal and spatial coordinates will split. This is often parameterized by the introduction of the fourthvector $u=(1,0,0,0)$ denotes as the fourth-velocity in the thermal bath rest frame.

The structure of correlators now changes and new condensates appear.
All  correlator structures can be separated into  {\it even}  and  {\it odd}   contributions in terms of $p_0$ as
$
\Pi(p_0,\bs{p}^2) =
\Pi^e(p_0^2,\bs{p}^2)+p_0\Pi^o(p_0^2,\bs{p}^2)$.
In this case, the contour integration will be done in the frame $\boldsymbol{p}=0$ with $s=p_0^2$.
In the case of the effective nucleon propagator, assuming a generalized momentum-expanded self-energy, the breaking of the Lorentz symmetry will lead to some correction of the baryon chemical potential, as well as to the appearance of a velocity in the third component of the momentum.
This can be summarized in Eq\,\eqref{eq_SB} by setting $p=(\,p^0+\Delta\mu\, ,\,v_\perp \boldsymbol{p}_\perp \, , \, v_\parallel p^3\,)$ .

\subsection{Axial-vector coupling}

To obtain the nucleon axial coupling we need the following current correlator
\begin{equation}
 \Pi_\mu(x,y,z)=-\langle 0|\,{\cal T}\,\eta_p(x)A_\mu(y)\,\bar\eta_n(z)\,|0\rangle,
\end{equation}
where in the QCD sector $A_\mu(y)= \bar d(y)\,\gamma_\mu\gamma_5\, u(y)$
and in the hadronic sector it can be expressed in terms of the elements of the axial current matrix involving nucleons
\begin{align}
 \langle p',s'|A_\mu(y)|p,s\rangle = \bar u_p^{s'}(p')\,T_\mu(q)\,u_n^s(p) \,e^{iq\cdot y},
\end{align}
with $q=p'-p$ and with $T$ is the most general form is
\begin{equation}
 T_\mu(q) =
 G_A(t)\gamma_\mu\gamma_5 +G_P(t)\gamma_5 \frac{q_\mu}{2  m_N}
 +G_T(t)\sigma_{\mu\nu}\gamma_5  \frac{q_\nu}{2 m_N},
 \label{T_mu}
\end{equation}
 with $t=q^2$  and  $ m_N$ the vacuum nucleon mass.
 The nucleon Axial coupling constant is defined as $g_A\equiv G_A(0)$.
We can write the axial current related with nucleons in terms of the nucleon fields
\begin{align}
A_\mu(y) = \int d^4\xi\, \bar\psi_p(\xi)\tilde T_\mu(\xi-y)\psi_n(\xi) ,
\end{align}
where $\tilde T_\mu(x)$ is the inverse Fourier transformation of $T_\mu(q)$ in configuration space.
So now, we can explicitly deal with magnetic field dependent propagators.
The hadronic correlator is expressed diagrammatically in in Fig.\,\ref{fig:contour-diagrams} (center).
The Fourier transformation of the correlator is
\begin{equation}
 \Pi_\mu(p,p')=\int d^4y\, d^4z\, e^{-i(q\cdot y+p\cdot z)}\,\Pi_\mu(0,y,z).\label{Eq.Pi-momentum}
\end{equation}
which in the hadronic sector is, in vacuum case
\begin{equation}
 \Pi^\text{\tiny had}_\mu(p,p')=\lambda_n\lambda_p\frac{ (\slashed{p}+m_n)T_\mu(q)(\slashed{p}'+m_p)}{(p^2-m_n^2)(p'^2-m_p^2)}.
\end{equation}

Now we want to isolate $G_A(t)$ from the other form factors.
Tracing the correlator multiplied by a gamma matrix will produce the desired selection in the following way:
\begin{equation}
 \mathrm{tr}\,[\Pi_\mu(p,p')\,\gamma_\nu]=-4i\epsilon_{\mu\nu\alpha\beta}p^\alpha p'^\beta \Pi(s,s',t),
\end{equation}
with $s=p^2$, $s'=p'^2$.
$\Pi$ in the hadronic sector is
\begin{equation}
 \Pi^\text{\tiny had}(s,s',t)=\lambda_n\lambda_p\frac{G_A(t)+G_T(t)(m_n-m_p)/ m_N}{(s-m_n^2)(s'-m_p^2)}.
 \label{Pi_had}
\end{equation}
If we neglect the term $(m_n-m_p)/m_N$, this procedure isolates $G_A$.

The current correlator in the QCD sector is described diagrammatically in Fig\,\ref{fig:contour-diagrams} (right).
The same procedure applied to the QCD sector at $t=0$ produces at leading order
\begin{equation}
 \Pi^\text{\tiny pQCD}(s,s',0)=
 \frac{s^2\ln(-s/\mu^2)-s'^2\ln(-s'/\mu^2)}{(2\pi)^4\,(s'-s)}\\
 +\text{regular terms},
 \label{PI-pQCD}
\end{equation}
where $\mu$ is the $\overline{\text{MS}}$ scale.
Regular terms means terms without discontinuities on the real axes, nor singularities which would vanish when integrating in the FESR contour.
Now we have two variables so a double FESR \cite{Dominguez:1999ka,Villavicencio:2022gbr,Dominguez:2023bjb} must be used for each current channel, in this case the momentum associated to nucleon currents.
\begin{equation}
 \int_0^{s_p}\frac{ds'}{\pi} \,\mathrm{Im}_{s'}\!\!\int_0^{s_n}\frac{ds}{\pi}\,\mathrm{Im}_s\Pi^\text{\tiny had} (s,s',t)\\
 =\oint_{s_p}\frac{ds'}{2\pi i}\oint_{s_n}\frac{ds}{2\pi i}\,\Pi^\text{\tiny QCD}(s,s',t),
 \label{eq.FESR_had=QCD}
\end{equation}
where  $\mathrm{Im}_s f(s)\equiv\lim_{\epsilon\to 0}\mathrm{Im}f(s+i\epsilon)$.
Applying the double FESR to the proton-axial-neutron currents correlator projected to $G_A$ in the frame $t=0$ we get
\begin{equation}
 g_A\lambda_n\lambda_p
=\frac{1}{48\pi^4}\left[s_n^3\,\theta(s_p-s_n)
+s_p^3\,\theta(s_n-s_p)\right],
\label{eq.gA(p,n)}
\end{equation}
where $s_p$ and $s_n$ are the proton and neutron current hadronic thresholds.
The values of the current-couplings and the hadronic thresholds are obtained through the nucleon-nucleon correlator FESR \cite{Dominguez:2020sdf}.
In vacuum, both nucleon thresholds are equal so
$g_A= s_0^3/48\pi^4\lambda_N^2$.
In \cite{Villavicencio:2022gbr}, this value was calculated in vacuum by fitting the experimental value of $g_A$ and then searching for the allowed values of the quark condensate and the gluon condensate.
When medium effects are taken into account, it is expected that the the axial-vector contribution to the form factor splits into different components, according to the symmetry breaking induced by external magnetic field and thermal or dense bath
\begin{equation}
    G_A\gamma_\mu\to G_A^{(0)}\gamma_0+G_A^{(\perp)}\gamma^\perp_\mu+G_A^{(3)}\gamma_3+\tilde G_A F_{\mu\nu}\gamma^\nu.
\end{equation}
The case in the presence of magnetic field only was obtained in \cite{Villavicencio:2022gbr}, while the effects of both, baryon density and magnetic field was obtained in \cite{Dominguez:2023bjb}.
There was considered only the leading contribution in Eq.\,(\ref{eq.gA(p,n)}), medium effects will be present only through the nucleon-current couplings and nucleon hadronic thresholds.

\subsection{Results}

For the combined magnetic field and baryon density problem, we select only six FESR from the neutron-neutron and proton-proton correlators, in order to obtain the current-couplings $\lambda_n$, $\lambda_p$, the hadronic thresholds $s_n$, $s_p$, and the condensates $\langle \bar u\sigma_{12} u\rangle$ and $\langle \bar d\sigma_{12} d\rangle$.
The inputs considered for the in-medium condensates depend mainly on the baryon density, except for the quark condensate (see details in \cite{Dominguez:2023bjb}).
The first thing to be noticed is that, for all the range of values considered both for baryon density and magnetic field, $s_p > s_n$. This mean that the axial-vector coupling constant at finite baryon density and magnetic field is then
\begin{equation}
    g_A=\frac{1}{48\pi^4}\frac{s_n^3}{\lambda_n\lambda_p}.
\end{equation}

\begin{figure}
\begin{minipage}{.5\textwidth}
\includegraphics[scale=0.40]{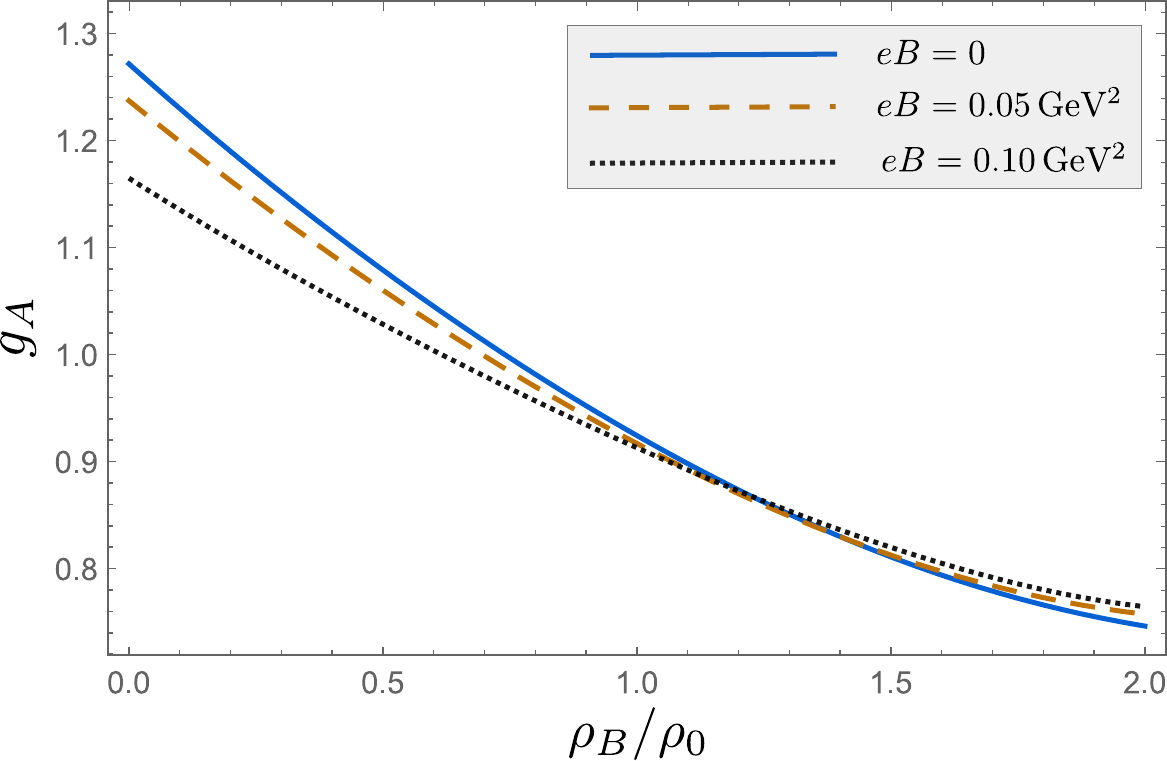}
\end{minipage}%
\begin{minipage}{.5\textwidth}
\includegraphics[scale=0.47]{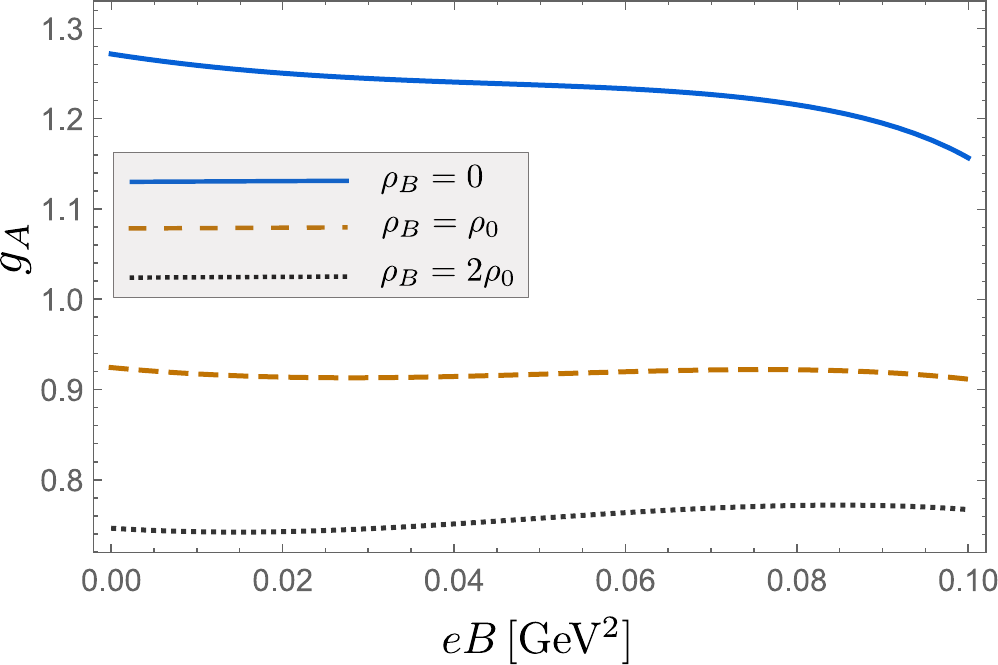}
\end{minipage}
\caption{In-medium nucleon axial coupling constant as a function of baryon density for different values of the magnetic field ({\bf left}), and as a function of the magnetic field for different values of the baryon density ({\bf right}).}
\label{fig:plot_gA}
\end{figure}

The medium evolution of the nucleon axial-vector coupling constant can be seen in Fig.\,\ref{fig:plot_gA} as function of the baryon density and  as function of the magnetic  field strength.
The axial-vector coupling constant decreases both with the baryon density and with the magnetic field.
It is worth to notice that at $B=0$ and at the nuclear saturation density $\rho_B=\rho_0 = 0.16 \text{ fm}^{-3}$ we get $g_A^*\equiv g_A(\rho_0)   \approx 0.92$, in agreement with other approaches which get the result $g_A^* \lesssim 1$  \cite{Rho:1974cx,Wilkinson:1974huj,Brown:1978zz,Park:1997vv,Carter:2001kw,Lu:2001mf,Suhonen:2017krv,Bass:2020bkl,Rho:2021zwm,Rho:2023vow}.

As can be seen in Fig.\,\ref{fig:plot_gA}, for $\rho_B\sim \rho_0$ there is no significant variation with the magnetic field, except for lower values of the baryonic density.
For $\rho \gtrsim 1.25 \rho_0$, $g_A$ appears to increase with the magnetic field and there is a flip in the behavior of the curves.
This behavior can be enhanced for higher values of baryon density and magnetic field, for example, near the core of a magnetar.
At present, these are the only results in the literature considering these two effects.
We hope that similar studies performed with other models and techniques can reproduce these results.

\newpage
	\section{Dynamics of Heavy quarks in a magnetised medium}
 \label{sec5}


\subsection{Introduction}

Heavy quarks have been extensively studied within the heavy-ion community as a prominent signature for characterising the properties of hot and dense quark matter. Considering the non-centrality of the HIC experiments and the subsequent generation of strong magnetic fields, studies related to heavy quarks have recently been extended to magnetised medium. Unlike most of the previous heavy quark (HQ) dynamical studies, in the present study we go beyond those limiting scenarios to tackle the most general case of arbitrary valued external magnetic fields.

The HQ momentum diffusion coefficients are crucial theoretical quantities needed to describe the evolution of heavy quarks and significantly impact the phenomenology of heavy quarks, influencing theoretical predictions for relevant experimental observables~\cite{Rapp:2018qla}. We use the widely adopted Langevin equations which assumes that heavy quarks receive random "kicks" from the thermal partons in the surrounding medium. In absence of the magnetic field and within the nonrelativistic static limit of HQ (i.e. $p\approx 0$, $M\gg T$, $p$ and $M$ being the HQ momentum and mass respectively) there is no anisotropy imposed on the system resulting in a single diffusion coefficient $\kappa$. Several studies have evaluated this $\kappa$ using various techniques~\cite{Caron-Huot:2007rwy,Madni:2022bea,Altenkort:2023oms}. Moving beyond the static limit involves attributing a finite velocity, where $\gamma v \lesssim 1$ (i.e. $p=\gamma M v \lesssim M$) to the heavy quark (HQ). This introduces an anisotropy in the system due to the HQ's motion in a preferred direction. Consequently, the momentum diffusion coefficient $\kappa$ splits into longitudinal and transverse components, denoted as $\kappa_L$ and, $\kappa_T$, which also has been extensively studied~\cite{Braaten:1991jj,Braaten:1991we,Thoma:1990fm,Moore:2004tg,Beraudo:2009pe}. 
 
The presence of an external magnetic field introduces additional anisotropy into the system, raising intriguing questions about how to incorporate the scale $eB$. Most current studies have adopted the Lowest-Landau-Level (LLL) approximation~\cite{Fukushima:2015wck,Singh:2020fsj,Bandyopadhyay:2021zlm}, or the weak magnetic field approximation\cite{Dey:2023lco}. The validity of these approximations depends on the strength and time-dependence of the magnetic field generated in non-central heavy-ion collisions (HICs). However, extending calculations to arbitrary external magnetic fields eliminates these constraints, freeing us from limitations imposed by the scale of $eB$.

To compute HQ momentum diffusion coefficients, which crucially depend on the HQ scattering rate, in the presence of arbitrary magnetic fields, it is necessary to generalise the effective gluon propagator for a hot magnetised medium beyond the LLL approximation. This involves using hard thermal loop (HTL) approximations to calculate the form factors of the effective generalised gluon propagator across all Landau levels. Recent approaches have also incorporated the magnetic field effect differently in calculations of HQ potential and HQ energy loss, where the medium's influence is assumed to be encapsulated solely through the medium-dependent Debye mass~\cite{Ghosh:2022sxi,Nilima:2022tmz,Jamal:2023ncn}. Within the static limit, comparing results from these methods provides insights into the limitations of using the simplified approximation of a medium-modified Debye mass.

\subsection{Formalism}
We examine a dynamic heavy quark (HQ) system (denoted as ( $P\equiv (M, p) \equiv (E, v)$)) interacting with an anisotropic magnetic field $\vec{B} = B \hat{z}$, deriving the complete expressions for longitudinal and transverse momentum diffusion coefficients for charm and bottom quarks. Despite the heavy quark mass $M\gg \sqrt{eB},T$ being the dominant scale, we do not impose additional scale hierarchies relative to $eB$ and $T$, unlike the LLL or weak field approximations. Following Refs.~\cite{Fukushima:2015wck,Bandyopadhyay:2021zlm}, we utilize the HTL approximation with the condition $\alpha_s eB \ll T^2$  (where $\alpha_s$ denotes the QCD running coupling), allowing us to neglect soft self-energy corrections of quarks and gluons during scattering rate calculations.

When the HQ moves in the presence of an external magnetic field, focusing on two distinct scenarios is beneficial: 
$\vec{v} \shortparallel \vec{B}$ and $\vec{v} \perp \vec{B}$. In the former case $\vec{v} \shortparallel \vec{B}$, two distinct diffusion coefficients $\kappa_L$ and $\kappa_T$ arise, which are linked to the HQ scattering rate as
 \begin{align}
\kappa_T (p) = \frac{1}{2}\int d^3q\frac{d~\Gamma(v)}{d^3q}q_\perp^2,~~ \kappa_L (p) = \int d^3q\frac{d~\Gamma(v)}{d^3q}q_z^2.
\label{coeffs_case1}
\end{align}
On the contrary, $\vec{v} \perp \vec{B}$ case generates three different diffusion coefficients $\kappa_j$'s ($j=\{x,y,z\}\equiv\{1,2,3\}$), i.e.
\begin{align}
\kappa_j (p) = \int d^3q\frac{d~\Gamma(v)}{d^3q}q_j^2.
\label{coeffs_case2}
\end{align}

It's clear that when we consider the static limit ($\vec{v}\rightarrow 0$) in a magnetized medium, only one anisotropy remains, dictated by the direction of $\vec{B}$, which causes the scenario $\vec{v} \perp \vec{B}$ to vanish.

From Eqs. \ref{coeffs_case1} and \ref{coeffs_case2}, it's evident that to calculate the momentum diffusion coefficients, we first need to determine the scattering rate or interaction rate of $2\leftrightarrow 2$ scatterings between the light quark/gluon and the heavy quark (i.e. $qH\leftrightarrow qH$ and $gH\leftrightarrow gH$). To evaluate these scattering rates, we employ an effective method initially introduced by Weldon~\cite{Weldon:1983jn}. In this approach, we express the $t$-channel $2\leftrightarrow 2$ scatterings involving the heavy quark as the cut/imaginary parts of the heavy quark self-energy, which can be formulated as
\begin{align}
\Gamma(P) = -\frac{1}{2E}~\frac{1}{1+e^{-E/T}}~\rm{Tr}\left[(\slashed{P}+M)~\rm{Im} ~\Sigma(p_0+i\epsilon,\vec{p})\right],
\end{align}
where $\Sigma(P)$'s represent the effective HQ self energy with an HTL resummed effective gluon propagator which duly incorporates the soft contributions. The detailed calculation of $\Gamma(P)$ and hence $\kappa_i$'s has been done in Ref.~\cite{Bandyopadhyay:2023hiv}.

\subsection{Results}

\begin{figure*}[h]
\begin{center}
\includegraphics[scale=0.45]{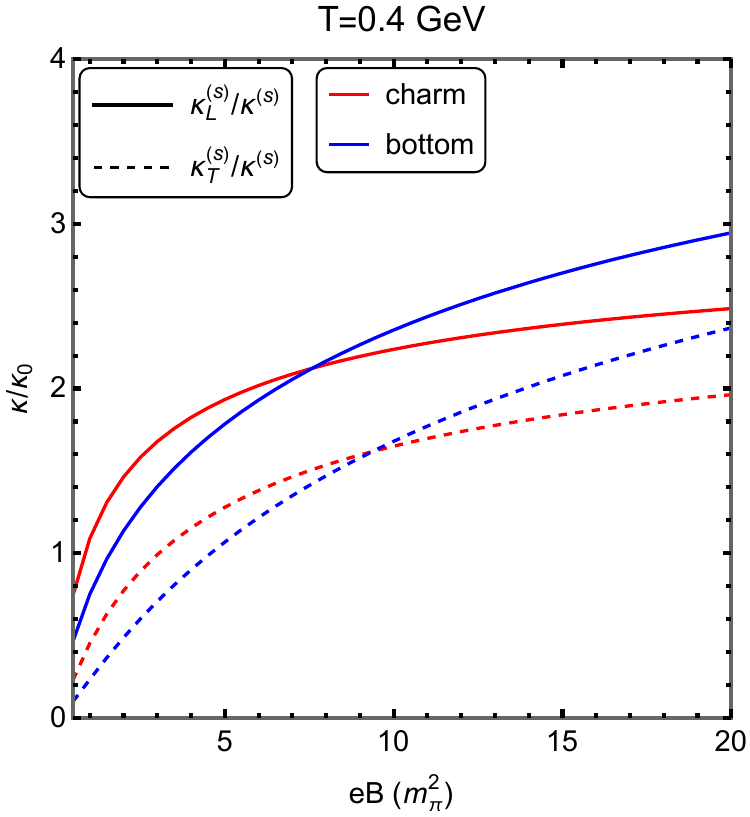}
\hspace{1cm}
\includegraphics[scale=0.45]{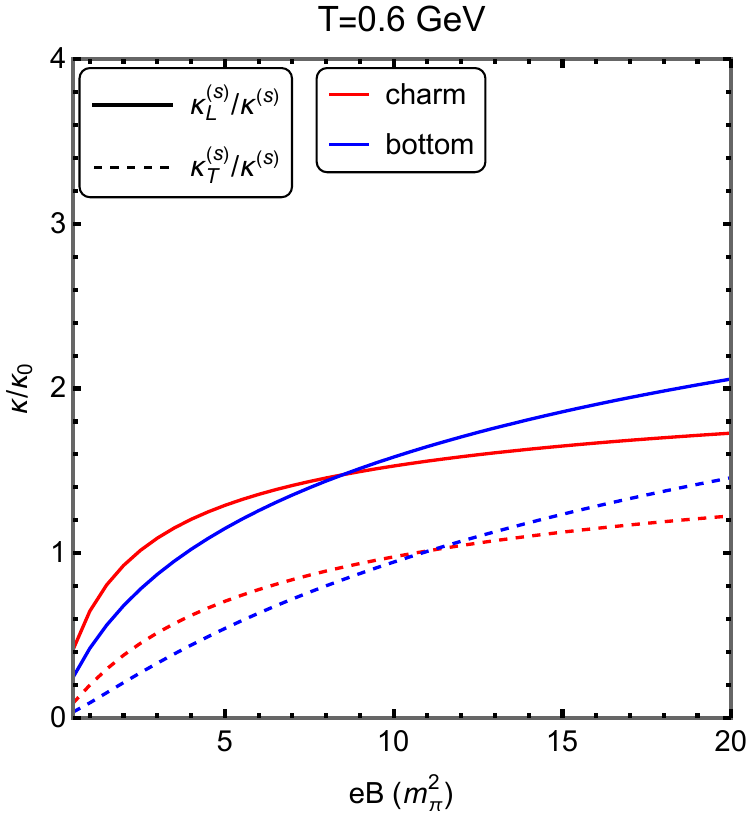}
\caption{The variation of the scaled momentum diffusion coefficients with respect to the magnetic field strength $eB$ is illustrated for longitudinal (solid lines) and transverse (dashed lines) diffusion in the static limit, considering both charm (red curves) and bottom (blue curves) quarks. This comparison is presented for two distinct temperatures: $T=0.4$ GeV (left panel) and $T=0.6$ GeV (right panel). The masses of charm and bottom quarks, denoted as $M$, are specified in the accompanying text.} 
\label{kappa_ratio_static_wzB}
\end{center}
\end{figure*}

We start with our most general static limit results for magnetized medium in Fig.~\ref{kappa_ratio_static_wzB}, by varying the ratio $\kappa_{L/T}^{(s)}/\kappa^{(s)}$ with respect to the external magnetic field. $\kappa^{(s)}$ is the zero magnetic field value of the single isotropic momentum diffusion coefficient. We opted for two relatively higher temperatures, specifically $T=0.4$ and $T=0.6$ GeV, to maintain consistency with the HTL approximation used throughout our calculations. Observations from the plots reveal that at lower values of $eB$, both the longitudinal (solid curves) and transverse (dashed curves) momentum diffusion coefficients exhibit a more pronounced rate of increase compared to higher $eB$ values. This effect is particularly prominent for charm quarks (red curves), leading to a crossover with the bottom quark (blue) curves. Throughout the magnetic field range depicted in Fig.~\ref{kappa_ratio_static_wzB}, the values of $\kappa_L$ consistently surpass those of $\kappa_T$ for both static charm and bottom quarks. Finally, as temperature increases, the overall ratio decreases, as expected due to the competing influences of $eB$ and $T$ scales.

\begin{figure*}[h]
\begin{center}
\includegraphics[scale=0.45]{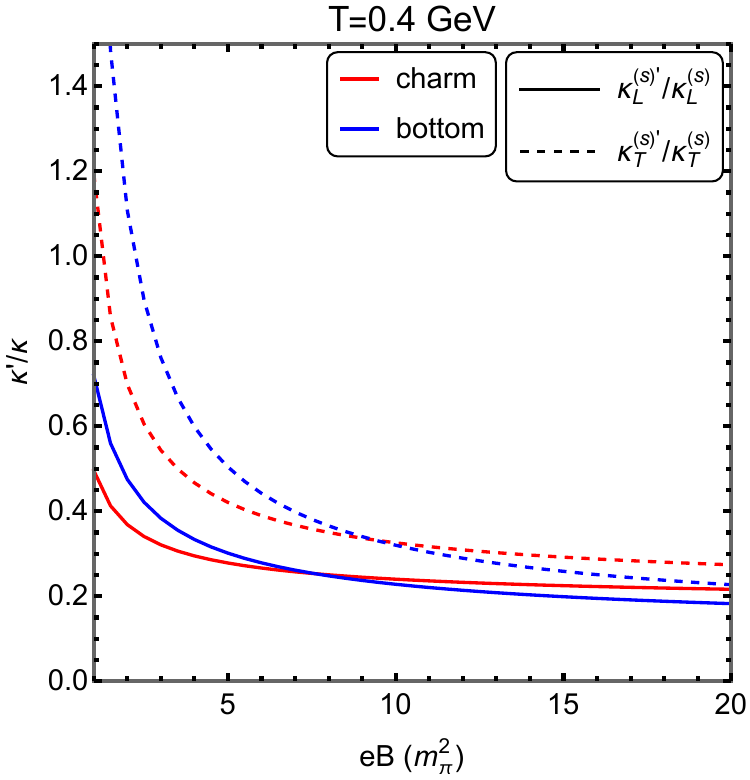}
\hspace{1cm}
\includegraphics[scale=0.45]{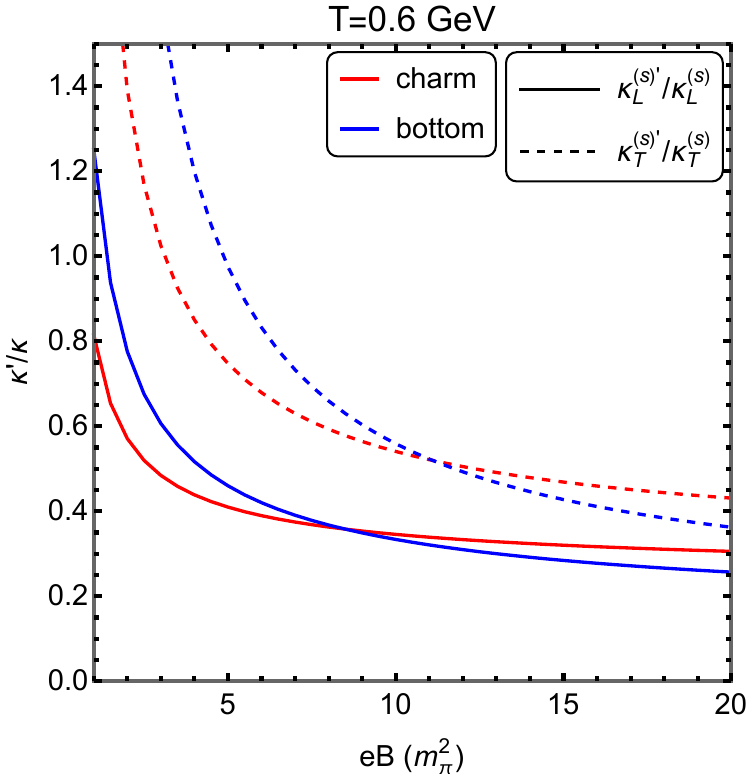}
\caption{The plots depict the variation of the ratio between the Debye mass approximated results ($\kappa'$) and the exact results ($\kappa$) with respect to the magnetic field strength $eB$. This comparison is shown for both longitudinal (solid lines) and transverse (dashed lines) momentum diffusion coefficients within the static limit, considering charm (red curves) and bottom (blue curves) quarks. The analysis is conducted at two distinct temperatures: $T=0.4$ GeV (left panel) and $T=0.6$ GeV (right panel).} 
\label{kappa_ratio_static}
\end{center}
\end{figure*}

In the present study, we also explore an alternative method to incorporate the influence of magnetic fields on heavy quark (HQ) scattering rates and momentum diffusion coefficients. Here, all medium effects are encapsulated through the medium-modified Debye screening mass. In Fig.~\ref{kappa_ratio_static}, we present comparisons between results obtained from the exact calculation ($\kappa$) and those from the Debye mass approximation ($\kappa'$) which illustrates the variation of the ratio $\kappa'/\kappa$ with respect to the magnetic field strength for two distinct temperatures, $T=0.4$ GeV and $T=0.6$ GeV. For both longitudinal and transverse components of charm and bottom quarks, the general trend is similar: the Debye mass approximated results consistently underestimate the exact results for larger values of $eB$ and overestimate them for smaller values of $eB$. These discrepancies are more pronounced for bottom quarks due to their heavier mass ($M_b = 4.18$ GeV) compared to charm quarks ($M_c = 1.27$ GeV). 

Additionally, it is noticeable that the ratio of transverse components ($\kappa_T'/\kappa_T$) is consistently larger than that of the longitudinal components ($\kappa_L'/\kappa_L$) across the entire range of $eB$ considered. This observation aligns with the findings in Fig.~\ref{kappa_ratio_static_wzB}. The dominance of $\kappa_L$ over $\kappa_T$ can be attributed to the predominant gluonic contributions in the $t$-channel scatterings studied in this context.

\begin{figure*}
\begin{center}
\includegraphics[scale=0.45]{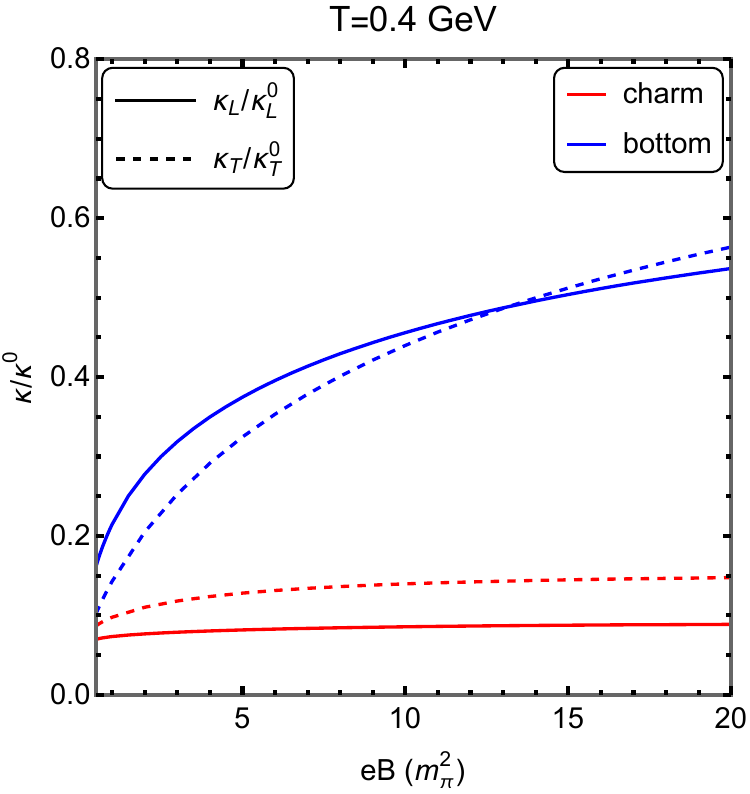}
\hspace{1cm}
\includegraphics[scale=0.45]{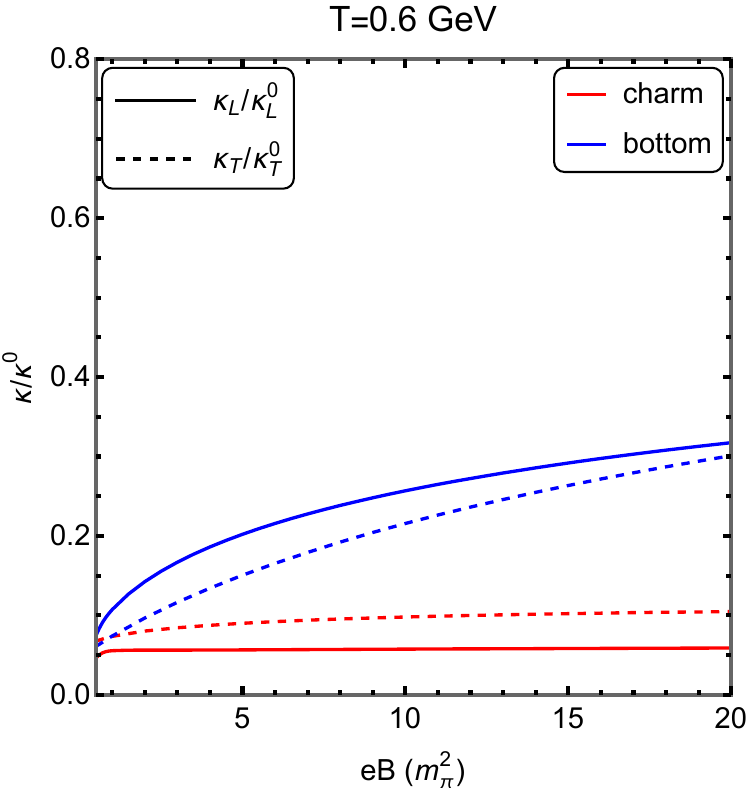}
\caption{$\vec{v}\shortparallel \vec{B}$ case : The longitudinal (solid curves) and transverse (dashed curves) momentum diffusion coefficients are depicted for charm (red curves) and bottom (blue curves) quarks as functions of the external magnetic field strength. These results are shown for two distinct temperatures: $T=0.4$ GeV (left panel) and $T=0.6$ GeV (right panel). The magnetized momentum diffusion coefficients are normalized relative to their values at $eB=0$. } 
\label{kappaveB_case1}
\end{center}
\end{figure*}

\begin{figure*}
\begin{center}
\includegraphics[scale=0.45]{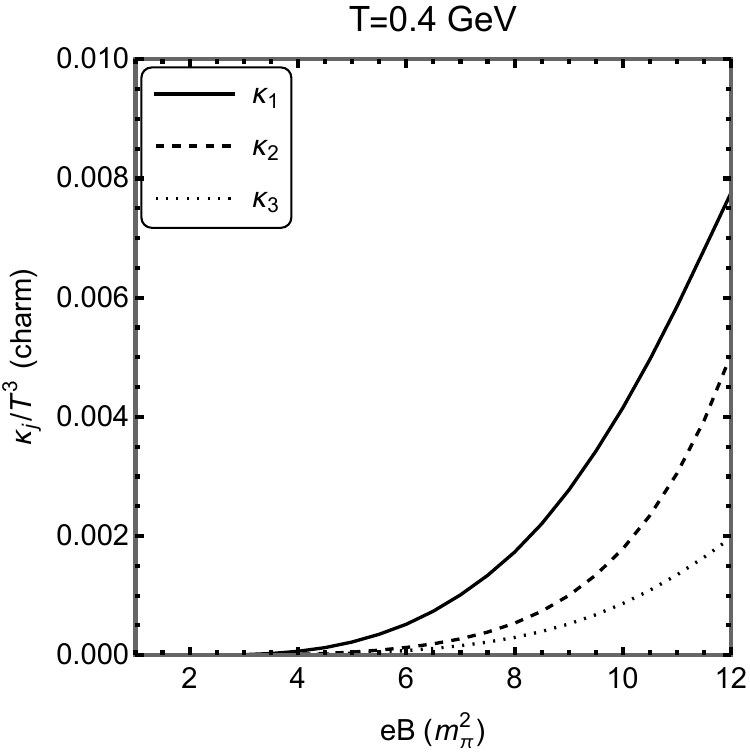}
\hspace{1cm}
\includegraphics[scale=0.45]{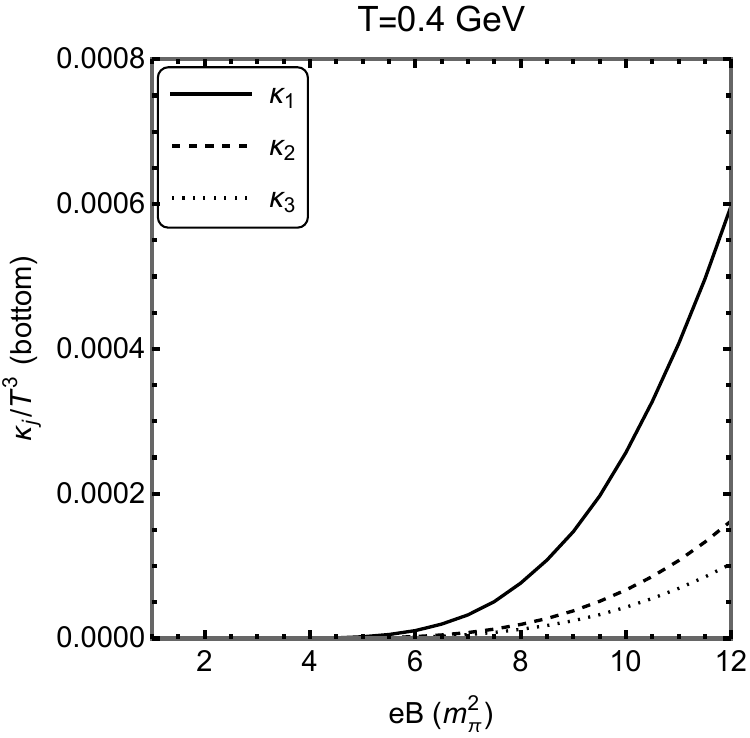}
\caption{$\vec{v}\perp \vec{B}$ case : The variation of the transverse components $\kappa_1$ (solid curves), $\kappa_2$ (dashed curves), and the longitudinal component $\kappa_3$ (dotted curves) of the momentum diffusion coefficient is illustrated for charm quarks (left panel) and bottom quarks (right panel) as functions of the external magnetic field. These results are presented at a fixed temperature of $T=0.4$ GeV. The magnetized momentum diffusion coefficients are normalized relative to $T^3$. } 
\label{kappaveB_case2}
\end{center}
\end{figure*}

Next we present our estimated results for longitudinal and transverse momentum diffusion coefficients beyond the static limit, while maintaining confinement within the small energy transfer regime and the scale hierarchy $M \ge p \gg T$. We fix the HQ momentum at $p = 1$ GeV and consider temperatures $T = 0.4$ GeV and $T = 0.6$ GeV, consistent with the applied HTL approximation.

Firstly, in the $\vec{v} \parallel \vec{B}$ case, we examine the variation of longitudinal and transverse momentum diffusion coefficients with external magnetic field in Fig.\ref{kappaveB_case1} for the aforementioned fixed temperatures. The coefficients are scaled relative to their respective values at $eB=0$. Similar to the static limit, we observe that for higher values of $eB$, the rate of change of $\kappa_{L/T}$ becomes relatively flat, especially for charm quarks (red curves). Interestingly, $\kappa_T$ (dashed curves) dominates over $\kappa_L$ (solid curves) throughout the range of $eB$ considered for charm quarks. Conversely, for bottom quarks (blue curves), $\kappa_L$ is larger than $\kappa_T$ until a higher threshold of $eB$, beyond which a crossover occurs. These crossovers reflect the interplay among the three competing scales $M$, $T$, and $eB$. Bottom quarks, with their higher mass, require a correspondingly higher $eB$ to exhibit similar behaviors in $\kappa_{L/T}$ as observed for charm quarks. The right panel of Fig.~\ref{kappaveB_case1} demonstrates this trend more clearly, where the absence of an immediate crossover indicates that bottom quarks would require an even higher $eB$ for similar behavior.

Figure~\ref{kappaveB_case2} illustrates the results for the perpendicular case, $\vec{v} \perp \vec{B}$. At $eB=0$, no such configuration exists, thus we scale the values of $\kappa_j$ with $T^3$ here. Transverse components $\kappa_1$ (solid curves) and $\kappa_2$ (dashed curves) dominate over the longitudinal component $\kappa_3$ (dotted curves) for both charm and bottom quarks. The specific choice of $p_\perp$ ensures $\kappa_1$ is larger than $\kappa_2$. Unlike previous cases, no saturation behavior is observed for any of the momentum diffusion coefficients at higher $eB$ values; instead, their rates of change visibly increase with increasing $eB$. Additionally, the values of momentum diffusion coefficients for bottom quarks (right panel) are approximately an order of magnitude lower compared to those for charm quarks (left panel).

Combining the results from both cases, we observe that for lower $eB$ values, the effect of the external magnetic field on various momentum diffusion coefficients is more pronounced when the HQ moves parallel to the magnetic field direction with higher momentum. This effect tends to saturate as $eB$ increases. Conversely, when the HQ moves perpendicularly to the external magnetic field, the effect of magnetic field on the momentum diffusion coefficients increases continuously with higher $eB$ values.

\subsection{Summary}

In this section, we have discussed the momentum diffusion coefficients of heavy quarks (HQ) within a magnetized medium. We have explored both general cases: (a) arbitrary values of the external magnetic field and (b) beyond the static limit by considering a finite HQ velocity. By comparing our exact results with an alternative Debye mass approximated procedure within the static HQ limit, we have clearly identified the limitations of the alternate approach. We emphasize the necessity of incorporating the full structure of gluon two-point correlation functions in a hot and magnetized medium. Across both static and non-static scenarios (particularly when the HQ moves parallel to the magnetic field), we observe a consistent pattern in the dependence of momentum diffusion coefficients on $eB$: rapid increase at lower $eB$ values followed by saturation, especially pronounced for charm quarks. Conversely, when the HQ moves perpendicularly to the magnetic field, we observe increasing changes in momentum diffusion coefficients with higher $eB$ values. Within the static limit, soft gluon scatterings dominate the longitudinal diffusion coefficient over the transverse counterpart, reflecting the dynamics of $t$-channel processes at leading order of strong coupling. However, beyond the static limit, we observe an opposite trend. The interplay among scales such as $M$, $p$, $eB$, and $T$ is clearly manifested in our findings. This study opens avenues for future research directions, such as investigating the in-medium evolution of HQ and its implications on experimental observables like directed and elliptic flow of open heavy flavor mesons.

	\newpage

	\section{The role of quark anomalous magnetic moment in magnetized quark matter}
 \label{sec6}



\subsection{Introduction}

The description of the magnetized QCD (Quantum Chromodynamics) phase diagram has remarkable developments in the past few years, with intensive efforts from both lattice QCD (LQCD) and effective model communities. In LQCD at zero density, it is well known the magnetic catalysis (MC), which is the increasing of the chiral condensate as function of the magnetic field \cite{Miransky:2015ava}. However, close to the pseudocritical temperature it is observed the inverse magnetic catalysis (IMC) \cite{Bali:2012zg}, which is the non-monotonic behavior of the quark condensate as a function of the magnetic field. Both effects are very important in order to describe the basic physics of the strongly interacting quark matter, but, unfortunately, just the MC effect is well described by effective models in the mean field approximation \cite{Bandyopadhyay:2020zte}.  These results have motivated several possibilities in effective approaches, as one can see in the references therein Ref. \cite{Andersen:2014xxa}. Particularly, in the case of the Nambu--Jona-Lasinio model \cite{Nambu:1961tp,Nambu:1961fr} one can explore beyond mean field approximations \cite{Mao:2016fha}, fittings of the coupling \cite{Bandyopadhyay:2019pml,Andersen:2014xxa,Farias:2014eca,Endrodi:2019whh,Farias:2016gmy} and, also including the anomalous magnetic moment (AMM) of the quarks  \cite{Singh:1985sg, Bicudo:1998qb, Fayazbakhsh:2014mca, Chaudhuri:2019lbw, Chaudhuri:2020lga, Ghosh:2020xwp, Xu:2020yag,Mei:2020jzn, Aguirre:2020tiy, Aguirre:2021ljk, Chaudhuri:2021skc, Ghosh:2021dlo, Wen:2021mgm, Farias:2021fci, Wang:2021dcy, Chaudhuri:2021lui, Lin:2022ied, Kawaguchi:2022dbq, Mao:2022dqn, Chaudhuri:2022oru, He:2022inw, Qiu:2023kwv, Islam:2023zyo, Qiu:2023ezo, Tavares:2023oln, Chaudhuri:2023djv, He:2024gnh, Mondal:2024eaw}.

In the recent literature, the quark AMM has been used as a new set of parameters in the NJL model, included in the Lagrangian by the Schwinger \textit{ansatz}. These parameters are obtained in order to fix the magnetic moment of proton and neutron \cite{Fayazbakhsh:2014mca}. Then, they can be used to explore several properties of the magnetized QCD phase diagram. Some of the modern predictions are claiming that, at low temperatures (zero temperature, in some cases), it is possible to obtain IMC and first-order phase transitions \cite{Fayazbakhsh:2014mca, Chaudhuri:2019lbw, Xu:2020yag, Ghosh:2020xwp, Chaudhuri:2020lga, Wen:2021mgm, Kawaguchi:2022dbq}. However, it is well established that these effects are not predicted by LQCD \cite{Bali:2012zg,DElia:2021yvk}. Also, all the phase transitions obtained in the region of interest, i.e., $eB< 1.0$ GeV$^2$, relevant to the heavy ion collisions, indicate crossover transitions \cite{Andersen:2021lnk}, with the possibility of first-order phase transitions at very strong magnetic fields, i.e., $eB\sim 8$ GeV$^2$ \cite{DElia:2021yvk}.  In this way, recent works \cite{Farias:2021fci,Tavares:2023oln} have been calling the attention that the use of inappropriate  regularization procedures can be the source of nonphysical phenomena. At zero AMM case, the methods called MFIR (magnetic field independent regularization) \cite{Avancini:2019wed} are claimed to eliminate nonphysical oscillations in quark condensate as a function of the magnetic field and also in several other quantities \cite{Avancini:2020xqe}. The main idea of these methods is to separate the contribution of the magnetic field from the natural divergences of the model, which must be regularized. In Ref. \cite{Farias:2021fci}, the authors adapt the Heisenberg-Weisskopf effective lagrangian, used in the context of the electrons with anomalous magnetic moment \cite{Dittrich:1977ee}, for quarks in the NJL model. Then the vacuum magnetic regularization (VMR) scheme \cite{Avancini:2020xqe}, which is a step-forward of the MFIR scheme, is applied. The results indicate very different behavior of the effective quark masses as function of the magnetic field, with no indication of first-order phase transition in the region $eB\leq0.3$ GeV$^2$ and just a very small window of IMC when considers sizable values of the quark AMM parameters. 

The solution for the difference between the model results is then detailed explained in Ref. \cite{Tavares:2023oln}, where one observes that the non-MFIR procedures hidden mass-dependent terms in the ultraviolet limit of the model which are inducing first-order phase transitions. These transitions can be obtained by both high enough magnetic fields or quark AMM values in the model, when it is considered as a free control parameter. If one uses the full lagrangian with the inclusion of the Schwinger \textit{ansatz}, it is possible to show that in the limit of $eB/M^2_0 \ll 1$, one can reproduce the Heisenberg-Weisskopf effective lagrangian, changing the electron mass for $M_0$, the effective quark mass in the vacuum. Then, the mass-dependent terms can be avoided in the UV limit, eliminating the possible first-order phase transitions.

In this section we revisit the main ideas explored in the recent literature, some of the basic results and how and what can still be done. The work has the following structure: In Section \ref{secII} we present the basic analytical structure of the NJL model in a constant magnetic field with quark AMM. In Section \ref{secIII} we define the two main regularization procedures we want to study. The discussion of the basic results are given in Section \ref{secIV} and the conclusions in Section \ref{secV}.

\subsection{Nambu--Jona-Lasinio SU(2) with quark AMM in a constant magnetic field}\label{secII}
The lagrangian density of the Nambu--Jona-Lasinio (NJL) model with two quark flavors possessing
anomalous magnetic moments (AMM) in an external electromagnetic field environment is given by
\cite{Fayazbakhsh:2014mca}
\begin{align}
\mathcal{L}&=\overline{\psi}\left(i \slashed D -  \hat{m}+\frac{1}{2}\hat{a}\sigma_{\mu\nu}F^{\mu\nu}\right)\psi
+G\left[(\overline{\psi}\psi)^{2}+(\overline{\psi}i\gamma_{5}\vec{\tau}\psi)^{2}\right],
\label{DENSE_lagrangian}
\end{align}
where $\psi$ is the quark fermion field $\psi=(\psi_u \quad \psi_d)^T$, the bare quark mass  matrix is $\hat{m}=\ $diag($m_u$, $m_d$), which is in the isospin limit $m_u = m_d \equiv m$, where $\hat{m}= m \mathds{1}$. The AMM matrix is $\hat{a}= \ $diag($a_u$, $a_d$). The phenomenological Pauli term $\frac{1}{2}\hat{a}\sigma_{\mu\nu}F^{\mu\nu}$ couples each quark  AMM flavor with the electromagnetic tensor. At one-loop approximation level, we have the following definitions 
\begin{align}
a_f= q_f \alpha_f \mu_B, \; \alpha_f=\frac{\alpha_e q_f^2}{2\pi}, \;  \alpha_e=\frac{1}{137} \;.
\label{amm_def}
\end{align} 
The electromagnetic gauge field is $A^\mu$ and the electromagnetic tensor is $F^{\mu\nu} = \partial^\mu A^\nu - \partial^\nu A^\mu$, Here, we adopt $Q$ as the diagonal quark charge matrix\footnote{We are using the Gaussian natural units, i.e., $1\,{\rm GeV}^2= 1.44 \times 10^{19} \, {\rm G}$ and $e=1/\sqrt{137}$.} $Q= \ $diag($q_u = 2 /3$, $q_d = -1/3$). The covariant derivative is given by $D^\mu =(\partial^{\mu} + i e Q A^{\mu})$ and $\slashed{D} = \gamma_\mu D^\mu$. The coupling constant is given by $G$. 

In a mean field approximation, the lagrangian becomes
\begin{align}
 \mathcal{L}=\overline{\psi}\left(i\slashed{D}-M+\frac{1}{2}\hat{a}\sigma_{\mu\nu}F^{\mu\nu}\right)\psi - \frac{(M-m)^2}{4G},
\end{align}
where the effective quark mass is defined as
\begin{align}
 M=m-2G \left \langle \overline{\psi}\psi \right \rangle, \label{gap}
\end{align}
where $\left \langle \overline{\psi}\psi \right \rangle$ is the chiral condensate. 
Working with constant AMM values, we can define a Bohr magneton as $\mu_B=e/2M$. As we are interested in constant values of quark AMM, in our case we have adopted  $\mu_B=e/2M_0$, where $M_0$ is the vacuum effective quark mass.
Choosing the Landau gauge, $A_{\mu}=\delta_{\mu 2}x_{1}B$, in order to satisfy 
 $\nabla \times \vec{A}=\vec{B}=B{\hat{e_{3}}}$ and $\nabla \cdot \vec{A}=0$, we
 can work with a constant magnetic field in the z-direction.

\subsubsection{Thermodynamic potential with quark AMM}
\label{potential}

The general structure of the thermodynamic potential at zero temperature with the quark AMM in a constant magnetic field environment is given by \cite{Farias:2021fci, Tavares:2023oln,Fayazbakhsh:2014mca}

\begin{eqnarray}
 \Omega = \frac{(M-m)^2}{4G}+\Omega^{\text{mag}}(M,B),
 \label{omega}
\end{eqnarray}

\noindent where $\Omega^{\text{mag}}$ is written as the magnetic field term, given by

{\small
\begin{eqnarray}
 \Omega^{\text{mag}}(M,B) = -N_c\sum_{f=u,d}|B_f|\sum_{n=0}^{\infty}\sum_{s=\pm 1}\int_{-\infty}^{\infty}\frac{dp_3}{4\pi^2}E_{n,s}^f \;,\label{omega1}
\end{eqnarray} }

\noindent in which $N_c=3$ is the number of colors; $B_f \equiv q_feB$, $s=\pm 1$ is the spin index, $n$ are the Landau levels and the quark energy dispersion relation is defined as

\begin{eqnarray}
 E_{n,s}^f=\sqrt{p_3^2+(M_{n,s}^f-sa_fB)^2}
\end{eqnarray}

\noindent where we define $M_{n,s}^f=\sqrt{|B_f|(2n+1-s_fs)+M^2}$ and $s_f=\text{sign}(q_f)$ as the charge sign function. 

Since Eq.(\ref{omega1}) is divergent, one need to adopt a regularization procedure. 

\subsection{Regularization}\label{secIII}

There are several choices used in the literature, which can be divided in two central regularization procedures, the VMR (vacuum magnetic regularization) \cite{Avancini:2020xqe}, which is a MFIR-type (magnetic field independent regularization) \cite{Avancini:2019wed} technique, and non-MFIR procedures, mostly based in Form-Factor functions \cite{Avancini:2019wed}. Recent works are using the VMR/MFIR regularization \cite{Wen:2021mgm, Farias:2021fci, He:2024gnh, Qiu:2023kwv, Tavares:2023oln} and Form-Factor procedures \cite{Fayazbakhsh:2014mca, Xu:2020yag, Lin:2022ied, Kawaguchi:2022dbq, Islam:2023zyo, Ghosh:2021dlo, Chaudhuri:2021lui, Chaudhuri:2022oru, Qiu:2023ezo, Wang:2021dcy} to deal with the AMM of the quark in the NJL model. Here we will discuss some of the basic properties of the most common regularization methods applied with the inclusion of quark AMM.

\subsubsection{Form-Factor regularization}

The idea of form-factor (FF) regularization resides in use some function to regularize the ultraviolet divergent integrals. We can multiply the vacuum contribution in Eq.(\ref{omega1}) by a FF the function used in Ref. (\cite{Fayazbakhsh:2014mca})

\begin{eqnarray}
 f_{n,s}^{f}(\Lambda,eB)=\frac{1}{1+\exp\left(\frac{((p_z^2+|q_feB|(2n+1-ss_f))^{1/2}-\Lambda)}{A}\right)}
\end{eqnarray}

\noindent where $\Lambda$ is the cutoff of the model and $A$ is a new parameter of regularization. We will call this type of FF function as FF$_1$.

The second FF function we will use is the FF$_2$, from Ref. \cite{Kawaguchi:2022dbq,Xu:2020yag}
\begin{eqnarray}
 f_{n,s}^{f}(\Lambda,eB)=\frac{\Lambda^{N}}{\Lambda^{N}+\left(p_z^2+|q_feB|(2n+1-ss_f)\right)^{N/2}}
\end{eqnarray}

\noindent where $N$ is a new regularization parameter. These two regulators should be sufficient to explore the main effects that we are interested.
However, there are a variety of different regularization prescriptions that are non-MFIR procedures with different FF functions that we will discuss briefly later.

\subsubsection{VMR regularization}

The magnetic field contribution of the thermodynamic potential Eq.(\ref{omega1}) at zero temperature in the VMR regularization is written in the Schwinger proper-time method as \cite{Tavares:2023oln}

\begin{eqnarray}
 \Omega^{\text{mag}}(M,B)
 &=& \frac{N_c}{8\pi^2}\sum_{f=u,d}\int_0^{\infty}\frac{d\tau}{\tau^3}e^{-\tau M^2}F_f(\tau), \label{omega2}
\end{eqnarray}

\noindent where the function $F_f(\tau)$ is given by

\begin{eqnarray}
F_f(\tau) 
 &=& e^{-\tau(a_fB)^2}\tau|B_f|\left[s_f\sinh(\tau2a_fBM) + F^{(2)}_f(\tau)\right],\label{Dk0}
\end{eqnarray}

\noindent and the function $F^{(2)}(\tau)$ is 

\begin{eqnarray}
   F^{(2)}_f(\tau) = \sum_{k=0}^{\infty}\frac{(\tau 2 a_f B M)^{2k}}{(2k)!}
   \sum_{n=0}^{k} \binom{k}{n}(-1)^{n}\left(\frac{|B_f|}{M^2}\right)^{n}
   \frac{d^n}{d(\tau |B_f|)^n}\coth(|B_f|\tau) .\label{F2}
\end{eqnarray}

Now, we have to apply the subtraction of divergences in the thermodynamic potential, Eq. (\ref{omega2}).
The integral diverges when $\tau \to 0$.
When considering the Taylor expansion of the function $F_f(\tau)$ around $\tau=0$ up to the order $\mathcal{O}(\tau^2)$, we obtain

\begin{eqnarray}
 F^0_f(\tau)&=&1+(a_fB)^2\tau+R_f(B_f,M)\tau^2 +\mathcal{O}(\tau^3),\; \tau\ll1, \label{Ftau0}
\end{eqnarray}

\noindent in the last expression, the coefficient of $\tau^2$ is given by

\begin{eqnarray}
 R_f(B_f,M)&=&\frac{|B_f|^2}{3}-\frac{(a_fB)^4}{6}+2(a_fB)^2M^2+s_f2|B_f|(a_fB)M.
\end{eqnarray}

Then the prescription of VMR can be applied as

\begin{eqnarray}
\Omega^{\text{mag}}(B,M) &=& [\Omega^{\text{mag}}(B,M) - \Omega^{VD}(B,M)] + \Omega^{VD}(B,M)\nonumber\\
 &\rightarrow& \Omega^{\text{mag}}_R(B,M) + \Omega^{VM}(B,M),\label{vmr}
\end{eqnarray}

\noindent where $\Omega^{\text{mag}}_R = \Omega^{\text{mag}}(B,M) - \Omega^{VD}(B,M)$ is the magnetic part of the regularized thermodynamic potential, $\Omega^{VD}(B,M)$ is the vacuum-divergent contribution and $\Omega^{VM}(B,M)$ is the vacuum-magnetic contribution, which must be regularized.

It is possible to show that, in the limit of $eB/M_0^2\ll1$, we can expand the $F_f(\tau)$ function in the thermodynamical potential in order to obtain

\begin{eqnarray}
 F_f(\tau)&=& e^{-\tau(a_fB)^2}\tau|B_f|\left[\frac{\cosh((\alpha_f+1)|B_f|\tau)}{\sinh(|B_f|\tau|)}\right]
 \label{Ftau2}.
\end{eqnarray}


The expansion of $F_f^0(\tau)$ is then, given by

\begin{eqnarray}
 F_f^0(\tau)=1+\frac{(B_f\tau)^2}{6}\left(3c_f^2-1\right)+\mathcal{O}(\tau^3).
 \label{Ftau0_MI}
\end{eqnarray}

\noindent which has no dependence on the effective quark masses.
The thermodynamical potential can be written as

\begin{eqnarray}
 \Omega= \frac{N_c}{8\pi^2}\sum_f\int_0^{\infty}\frac{d\tau}{\tau^3}e^{-\tau \mathcal{K}_{0,f}^2}\left[\tau|B_f|\frac{\cosh(c_f|B_f|\tau)}{\sinh(|B_f|\tau|)}\right],
\end{eqnarray}

\noindent in the last expressions we have defined $\mathcal{K}_{0,f}=\sqrt{M^2+(a_fB)^2}$  and $c_f = a_f+1$. This result is the same obtained in \cite{Farias:2021fci} where the one-loop Schwinger-Weisskopf lagrangian in \cite{Dittrich:1977ee} is adapted to the NJL model. The regularization of this expression is given by applying the prescription Eq.(\ref{vmr}) 

\subsection{Recent Achievements}\label{secIV}

The vast literature focuses in the case of magnetized Nambu--Jona-Lasinio model in the mean field approximation with constant values of quark AMM, based in the ideas developed in Ref. \cite{Fayazbakhsh:2014mca}. Then, the main difference between most works concerns in the regularization procedures. Several different properties of the magnetized QCD phase diagram have been explored, as the thermodynamics \cite{Chaudhuri:2022oru,Wen:2021mgm,He:2022inw,He:2024gnh}, transport coefficients \cite{Qiu:2023ezo} meson masses \cite{Xu:2020yag,Lin:2022ied,Chaudhuri:2019lbw,Chaudhuri:2020lga,Aguirre:2021ljk} and dilepton production \cite{Ghosh:2020xwp}.

Recently the study with quark AMM has been extended to non-constant values. In Ref. \cite{Kawaguchi:2022dbq}, three cases were considered, in which $\kappa_{u,d}=\text{constant}$, $\kappa_{u,d}\propto \sigma$ and $\kappa_{u,d}\propto \sigma^2$, where $\sigma$ is the chiral condensate, that depends on the magnetic field and temperature in the context of the FF$_2$. The results with $\kappa_{u,d}=\text{constant}$ reproduces well the forthcoming discussion.

Besides the FF regularization, there are other regularizations that can be applied in the group of non-MFIR procedures, as is the case of Pauli-Villars (PV) regularization. In Ref. \cite{Avancini:2020xqe}, the zero AMM case is explored and, in fact, PV regularization shows very similar result for the average quark condensates as a function of the magnetic field for both MFIR and non-MFIR groups. One of the clear advantages of using PV regularization in this case is to avoiding the nonphysical oscillations previously discussed. The non-zero AMM case with PV regularization in a non-MFIR procedure has been in the context of chiral and deconfinement phase transitions \cite{Mao:2022dqn,Mei:2020jzn} . In the case of the magnetic field dependent three momentum cutoff, some of the results are qualitative the same when compared to the Form-Factor regularization, as the decreasing of the pseudocritical temperature as a function of the magnetic field and oscillations in the effective quark masses \cite{Chaudhuri:2019lbw,Ghosh:2020xwp}.

It is possible to study the behavior of the quark AMM as a function of temperatures and magnetic fields in the NJL model as derived in Ref. \cite{Ghosh:2021dlo}, by evaluating the photon-quark-antiquark vertex function, which is out of scope of the present work. However, the study focuses in the use of non-MFIR procedures, and future developments with MFIR/VMR schemes could be useful in order to provide comparisons between different methods.

\subsubsection{Form-Factor procedures}

We can summarized the basic set of parametrizations used in FF$_1$ and FF$_2$ as

\begin{center}
\begin{table}[h]
\centering
\tabcolsep=0pt
\begin{tabular*}{30pc}{@{\extracolsep\fill}lcccc@{\extracolsep\fill}}
Regularization & $\Lambda$[MeV]  & $m_0$[MeV]  & G$\Lambda^2$  & Extra Parameters \\
\hline
FF$_1$ & 664.30  & 5.000  & 2.440  & A=0.05$\Lambda$   \\
FF$_2$ & 681.38  & 4.552  & 1.860  & N=10.00   \\
\end{tabular*}
\item Table 1: In FF$_1$ we follow Ref. (\cite{Fayazbakhsh:2014mca}). In FF$_2$ we follow Ref. (\cite{Xu:2020yag}).
\end{table}
\end{center}

The set of parameters can change depending on the vacuum values of quark condensate $\langle \bar{\psi}\psi\rangle$, the pion mass $m_{\pi}$ and pion decay constant $f_{\pi}$ \cite{Avancini:2020xqe}. However, the general qualitative behavior is preserved.


The quark AMM for each flavor is determined to match the experimental values of the magnetic moments of the proton and neutron~ \cite{Fayazbakhsh:2014mca}. Then, one can obtain two different sets of parameters, which depends on $k_f$, defined in Eq. (\ref{amm_def}) by $a_f=q_f\alpha_f\mu_B\rightarrow a_f=q_f ek_f$ . For the set 1, we can consider the higher values of AMM as
\begin{eqnarray}
 &&k_u^{[1]}=0.29016\hspace{0.4em} \text{GeV}^{-1},\quad k_d^{[1]}=0.35986\hspace{0.4em} \text{GeV}^{-1},\nonumber\\
 &&\alpha_u^{[1]}=0.242, \quad\quad\quad\quad\quad \alpha_d^{[1]} = 0.304,\label{set1}
\end{eqnarray}

\noindent while the second set 2 is given by
\begin{eqnarray}
&& k_u^{[2]}=0.00995\hspace{0.4em} \text{GeV}^{-1},\quad k_d^{[2]}=0.07975\hspace{0.4em} \text{GeV}^{-1},\nonumber\\
&& \alpha_u^{[2]}=0.006, \quad\quad\quad\quad\quad \alpha_d^{[2]} = 0.056.\label{set2}
\end{eqnarray}

\begin{figure}[ht]
\centering
\includegraphics[width=0.5\textwidth]
{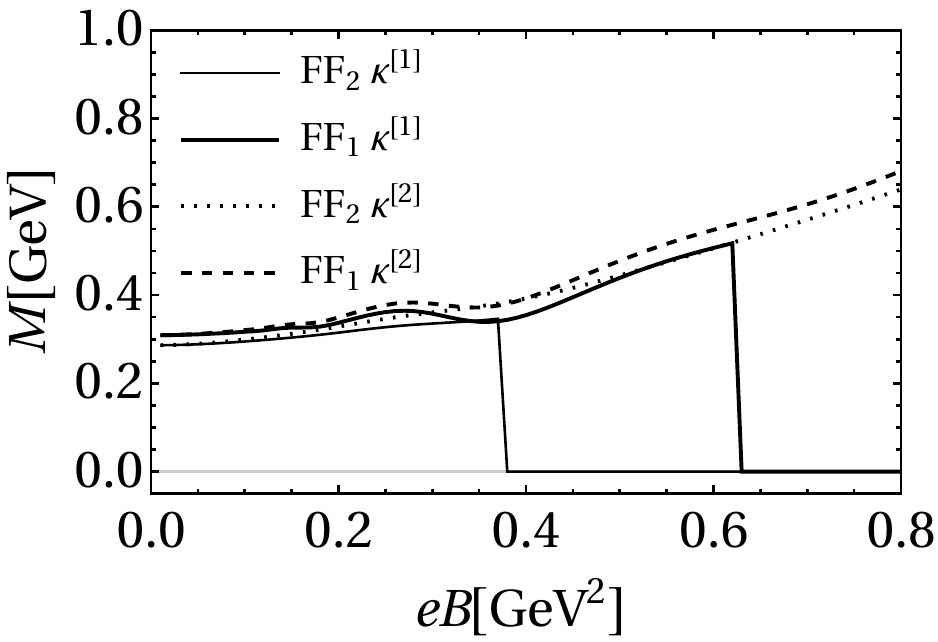}
\caption{Effective quark masses as a function of the magnetic field for the two sets of quark AMM with the form-factor regularizations FF$_1$ and FF$_2$.}
\label{mass}
\end{figure}

The set 2 of quark AMM values induces almost no difference when compared to the zero AMM case for the effective quark masses as a function of the magnetic field in which the magnetic catalysis effect is observed \cite{Tavares:2024myk}. The main ideas behind the application of FF regularization with quark AMM are represented in Figure \ref{mass}. The first aspect we want to evidence is the nonphysical oscillations observed in the effective quark masses, which are present due the non-MFIR regularizations, already observed in the zero AMM case \cite{Avancini:2019wed}. The second aspect concerns to phase transitions. For the set 1, the FF$_1$ regularization induces a possible first-order phase transitions close to $eB\sim 0.6$ GeV$^2$, while the FF$_2$ the transition occurs when $eB\sim 0.4$ GeV$^2$ \cite{Fayazbakhsh:2014mca,Kawaguchi:2022dbq}. Therefore, the basic aspects with quark AMM in the two-flavor NJL model are in qualitative agreement between the two form-factor regularizations explored in this work. Nevertheless, first-order phase transitions at low temperatures are predicted to happen at very strong magnetic fields, e.g., $eB_c\sim 8$ GeV$^2$ \cite{DElia:2021yvk}, which is far beyond the validity range of the model, since $eB_c\gg \Lambda^2$. In this way, we believe that the evaluations with FF regularization could not be allowing to reproduce well-known aspects of the magnetized QCD phase diagram and further developments are necessary in order to improve the applications with FF and quark AMM.

\subsubsection{MFIR/VMR procedures}

In the MFIR/VMR prescription, some seminal results have been obtained in Refs. \cite{Farias:2021fci,Tavares:2023oln}, in which the quark AMM effect has less prominent results when compared to the FF case. We adopt two sets of parameters for the VMR scheme with 3-D sharp cutoff in Ref. \cite{Farias:2021fci} and Proper-Time Ref. {\cite{Tavares:2023oln}, given in Table 2.

\begin{center}
\begin{table}[H]
\centering
\tabcolsep=0pt
\begin{tabular*}{25pc}{@{\extracolsep\fill}lccc@{\extracolsep\fill}}
Regularization & $\Lambda$[MeV]  & G$\Lambda^2$  & $m_0$[MeV]   \\
\hline
VMR$_1$ & 591.60  & 2.404  & 5.723    \\
VMR$_2$ & 886.62  & 4.001  & 7.383     \\
\end{tabular*}
\item Table 2: In VMR$_1$ we follow Ref. (\cite{Farias:2021fci}). In VMR$_2$ we follow Ref. (\cite{Tavares:2023oln}).
\end{table}
\end{center}

The first interesting result is that we can observe no oscillations in the quark condensate as a function of the magnetic field for zero temperatures when considering both sets of AMM parameters as observed in the left plot of Figure \ref{DENSE_fig2}. This is a signature of MFIR procedures as explored in detail in Ref. \cite{Avancini:2019wed}. 
  One can verify that there is no first-order phase transitions in the region of magnetic fields explored \cite{Farias:2021fci}. 
In what concerns to inverse magnetic catalysis, we can look to the right plot of Figure \ref{DENSE_fig2}. For the set 2, there is almost no difference with the zero AMM case, in which one obtains the usual magnetic catalysis effect. For the set 1, with sizable quark AMM, a small window in which the pseudocritial temperature decreases with the magnetic field is observed \cite{Farias:2021fci}.

 \begin{figure}[H]
 \center
\includegraphics[width=0.43\textwidth]{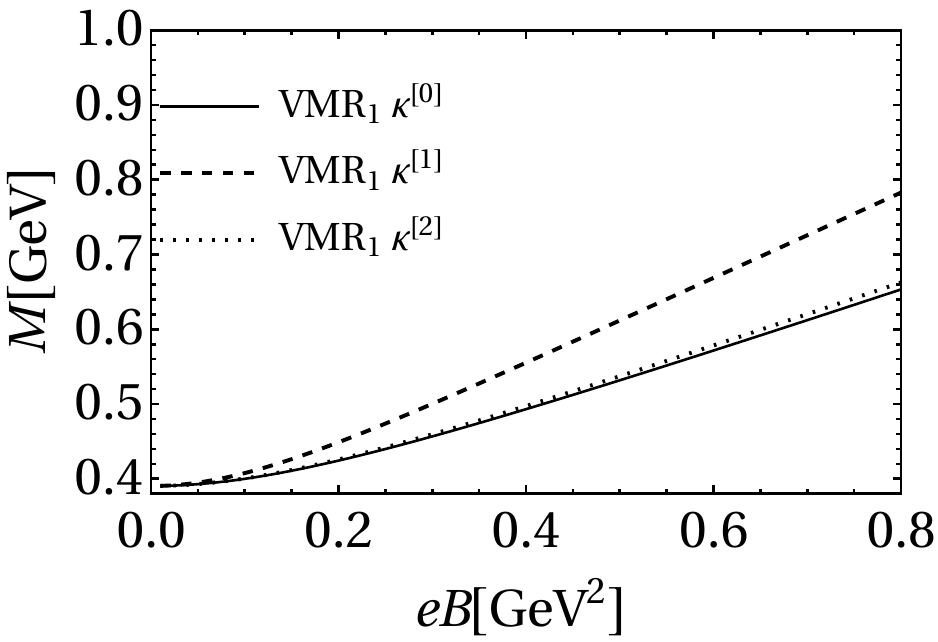}
\includegraphics[width=0.45\textwidth]{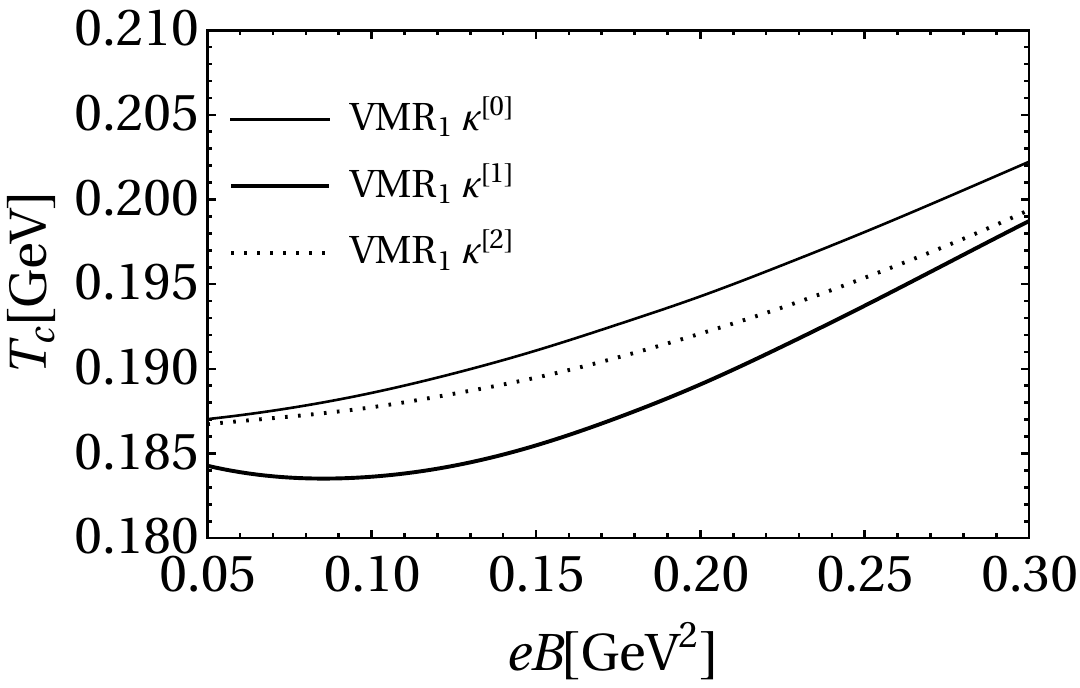}
 \caption{Left: Effective quark masses as a function of the magnetic field for different values of quark AMM. Right: Pseudocritical temperature as a function of the magnetic field for different values of quark AMM. We have defined the zero AMM case as $\kappa^{[0]}$.}
  \label{DENSE_fig2}
\end{figure}
  
As it is discussed in Ref. \cite{Tavares:2023oln},  in FF regularization there are due mass-dependent (MD) terms in the ultraviolet limit of the model, as described in Eq. (\ref{Ftau0}). These terms could change the gap equation, $\partial \Omega/\partial M=0$, and, therefore, induce artificial first-order phase transitions. One way to circumvent this problem is to apply the VMR prescription. However, it is not easy to treat the regularization of these mass-dependent terms in a general way. In fact, in the context of effective QED, the Heisenberg-Weisskopf lagrangian has the Schwinger \textit{ansatz}, which is valid in the region where $eB/m_e^2 \ll 1$, with m$_e$ being the electron mass. In our case, therefore, the limit is given by $eB/M_0^2\ll 1$, which is the first term, $n=0$, in the series expansion Eq. (\ref{F2}). When considering this situation, we obtain a mass-independent (MI) regularization, given by Eq. (\ref{vmr}) by using the $F^0_f(\tau)$ function given in Eq. (\ref{Ftau0_MI}), and the resulting effective potential is analogous with that obtained in Ref. \cite{Dittrich:1977ee}, but applied in the context of the two-flavor NJL model \cite{Farias:2021fci}. The full analysis is given in Ref. \cite{Tavares:2023oln} and in the Figure \ref{pot} we have the comparison between the thermodynamic potential, Eq. (\ref{omega}), with MD (left) regularization and MI regularization (right) scenarios at $eB=0.3 \text{GeV}^2$.

 \begin{figure}[H]
 \center
\includegraphics[width=0.4\textwidth]{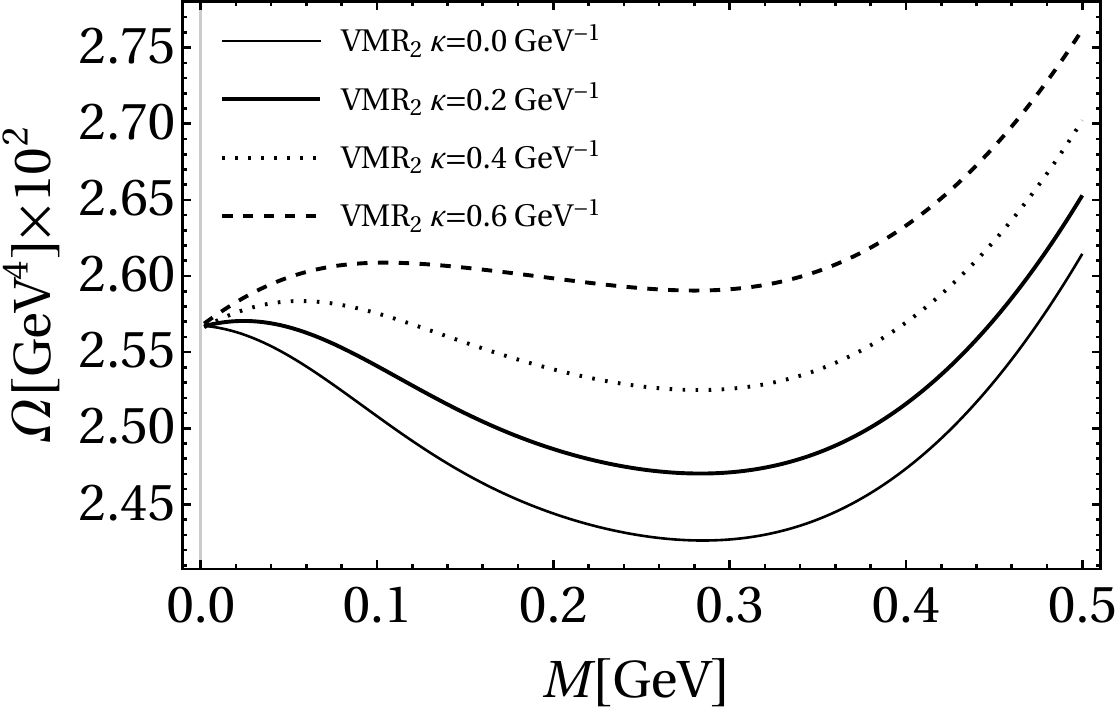}
\includegraphics[width=0.4\textwidth]{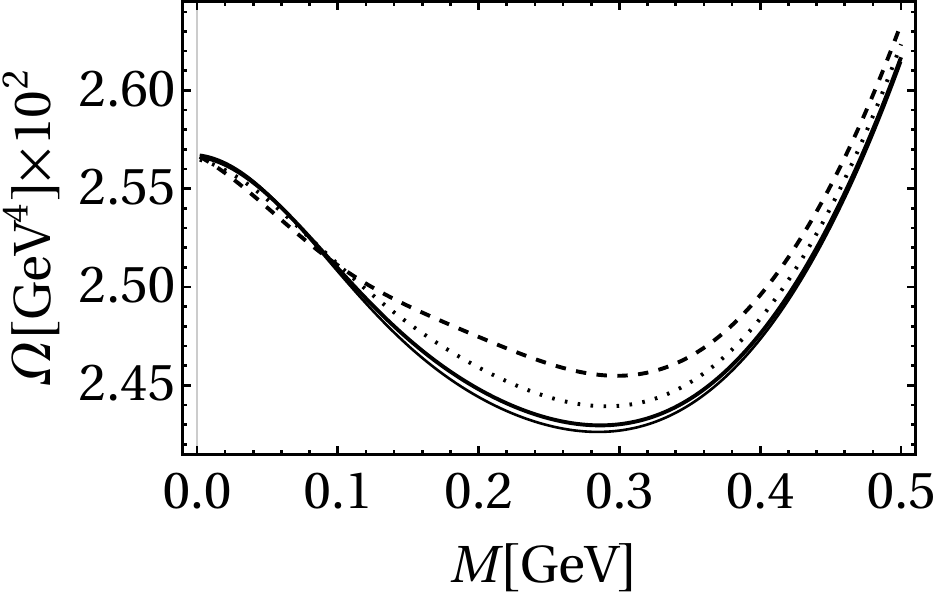}
 \caption{Effective potential as a function of the effective quark mass at $eB=0.3
  \text{GeV}^2$. Left: Mass-dependent regularization. Right: Mass-independent regularization.}
  \label{pot}
\end{figure}

There are some attempts similar to the VMR/MFIR procedures that are based in the subtraction of divergences, as explored in Ref. \cite{Aguirre:2020tiy,Aguirre:2021ljk}. Recently, the VMR/MFIR prescription has been applied to some aspects of the chiral phase transition \cite{Wen:2021mgm}, at zero temperature \cite{He:2022inw} and to investigate the anisotropy structure of magnetization in the context of Polyakov-Nambu--Jona-Lasinio model with quark AMM \cite{He:2024gnh}.

\subsection{Conclusions}\label{secV}

In this section we have analyzed the basic aspects chiral phase transitions in a magnetized two-flavor Nambu--Jona-Lasinio model with the inclusion of quark AMM. The Schwinger {\it ansatz} is used in order to include the quark AMM as a set of new parameters in the model, which is studied in the mean field approximation at zero temperature. Then, we explore two central groups of regularization, the MFIR/VMR procedures and the non-MFIR procedures with FF regularizations. We conclude that FF regularizations can induce non-physical results as, oscillations in the effective quark masses, first-order phase transitions and IMC at low temperatures. These effects can be avoided by the application of VMR procedures in the mass-independent regularization case. 

It is possible that further studies will be necessary to understand how to apply properly the FF regulators in order to avoid possible nonphysical results. Also, some non-MFIR procedures also need more specific and detailed analysis to test the consistence of the results. Future studies can be made in order to understand the dependence of the quark AMM as a function of magnetic field, temperature and density, as well as the thermodynamics, meson masses and equations of state in the context of MFIR/VMR procedures. The quark AMM connection to the IMC phenomena is an important topic to forthcoming work.

	\newpage
	\section{Summary and outlook}\label{DENSE_sec:sum}

In conclusion, this Part elucidates significant theoretical advancements in the exploration of magnetic field effects on the QCD phase diagram and dense magnetized matter. We have introduced a novel mechanism for the acceleration of proto-neutron stars through the chiral separation effect, underscoring the critical role of anisotropy in generating observable recoil effects. The examination of ring diagrams has demonstrated their essential function in modulating the chiral transition temperature and has indicated potential instances of inverse magnetic catalysis.

Furthermore, our analysis employing finite energy sum rules in dense nuclear matter has facilitated the extraction of pivotal parameters related to nucleon interactions, thereby enhancing our understanding of dense matter behavior in strong magnetic fields. The investigation into momentum diffusion coefficients for heavy quarks in a hot, magnetized medium has broadened our comprehension of quark dynamics under varying magnetic field strengths.

The implications of these findings are profound, as they bridge theoretical predictions with the physical conditions characteristic of magnetars, the primordial universe, and ultrarelativistic heavy-ion collisions. Additionally, the exploration of quark anomalous magnetic moments provides further insight into emergent phenomena within the magnetized QCD phase diagram, emphasizing the necessity for effective models to navigate the complexities of non-perturbative QCD.

Collectively, these results contribute significantly to the ongoing discourse regarding the interplay between magnetic fields and QCD, advancing our understanding of fundamental properties of matter in extreme environments.

	

    \newpage

\newcommand{\apj}{Astrophys. J.}
\newcommand{\apjl}{Astrophys. J. Lett.}
\newcommand{\prl}{Phys. Rev. Lett}
\newcommand{\araa}{Ann. Rev. Astron. and Astrophys.}
\newcommand{\prd}{Phys. Rev. D}
\newcommand{\nphysa}{Nucl. Phys. A}
\newcommand{\prc}{Phys. Rev. C}
\newcommand{\mnras}{Monthly Notices of the RAS}
\newcommand{\aap}{Astron. and Astrophys.}
\newcommand{\physrep}{Phys. Repts.}

\graphicspath{{./Figures_NS/}}

\part{Neutron Stars}\label{VI}

\section{Introduction}

In this Part,  we report on selected topics, organized in eight sections, on neutron star (NS) physics for which the consideration of strong magnetic fields is essential. 
For magnetic fields to have a direct effect on QCD properties, i.e. the hadron structure and the equation of state (EOS) of dense nuclear matter, the scale of $\Lambda_{QCD}\sim 200$ MeV is relevant which corresponds to field strengths of $\sim 10^{18}$ G. 
These field strengths influence integral properties of NSs such as masses, radii and tidal deformabilities. They may explicitly occur in the general relativistic equations of hydrodynamic stability of neutron star configurations and thus modify the standard Tolman-Oppenheimer-Volkoff (TOV) equations.
More subtle effects are occurring on the scale of nucleonic pairing gaps $\Delta\lesssim 1$ MeV for field strengths of the order $\sim 10^{15}$ G and influence transport phenomena like neutron star cooling. 
For the occurrence of these extreme field strengths, observational evidence is reported for a special class of neutron stars, the magnetars. 
But also for the more typical populations of young pulsars like Crab or Vela and old, recycled millisecond pulsars in the period-derivative vs. period diagram, with surface magnetic fields of $\sim 10^{12}$ G and $\sim 10^{8}$ G, respectively, there may prevail superstrong magnetic fields in their interiors due to a density-dependent field profile and/or collimation of field lines in superfluid or superconducting vortices. 

We begin this part with Sec.~\ref{sec:dutra} where the stage is set for discussing the present status of observational constraints for observational properties of NSs and their relation to properties of the equation of state of dense QCD matter described by relativistic (Walecka-type) or nonrelativistic (Skyrme-type) density functionals.
These models parametrize the phenomenology of nuclear saturation properties and allow for an extrapolation to supranuclear densities and isospin asymmetry as they are relevant for the description of NS interiors in $\beta$-equilibrium with electrons and muons. 
These constraints concern also the symmetry energy and its slope at the nuclear saturation density. 
Using the one-to-one relationship between the EOS of NS matter and the mass-radius relation obtained by integrating the TOV equations one may discuss constraints on models for the EOS from a comparison to the actual status of NS mass and radius measurements.
Starting from this baseline, in the following section \ref{sec:marquez} the NS matter model is extended to include heavier baryon states such as hyperons and resonances as well as the effect of a strong B-field on the EOS and thus the structure of the NS and its global properties as mass and radius when field strengths exceed $10^{17}$ G.
Spherical harmonics of the B-field are considered in sect. \ref{sec:marquez}, while in Sects. \ref{sec:tolos} and \ref{sec:zuraiq} a density-dependent profile of the B-field inside the NS are considered.
In Sec.~\ref{sec:zuraiq} the modifications of the TOV equations due to extremely strong radially and transversely oriented B-fields are presented and the possibility to see effects on gravitational wave signals in present and future laser interferometer instruments are discussed.
Turning to more subtle effects, in Sec.~\ref{sec:pais} the influence of string B-fields on nuclei in the inner crust and pasta structures at the crust-core interface are considered. 
While modifications of integral NS properties (mass, radius, tidal deformability etc.) require extremely strong magnetic fields, phenomena like cooling and glitches
as well as neutrino and electromagnetic neutron star kicks are obtained in less extreme fields.
Section \ref{sec:sedrakian} discusses that strong B-fields destroy the superfluidity and superconductivity in the neutron and proton pairing channels which results in astrophysical implications not only for neutron star cooling, but also for pulsar precession and fast radio bursts. 
In Sec.~\ref{sec:yasui} the topological surface effects are explained, which occur in the neutron p-wave superfluids structures of NSs. 
The final section,~\ref{sec:blaschke} considers the complex phenomenon of NS kicks that may be inherently related to the presence of strong magnetic fields as a precondition. Two types of kicks are distinguished: the natal kicks which NS obtain at birth in a supernova due to a neutrino rocket effect and the less strong kicks of cold recycled pulsars which are subject to an electromagnetic rocket effect due to radiation bursts in off-center magnetic fields.  
The latter may play a role in explaining why only a fraction of millisecond pulsars (MSPs) in binaries have eccentric orbits and how isolated MSPs originate.

Future observational campaigns may provide more precise dat, but may also discover more interesting compact stars that challenge the scenarios which have been developed in order to explain the rich NS phenomenology largely based on the interplay of strong and electroweak matter with extreme magnetic fields.

\section{Effective hadronic models applied to compact stars}
\label{sec:dutra}

\subsection{Overview}
Nuclear matter can exist in various temperature and density regimes, and the equation of state (EoS) is critical for describing macroscopic nuclear and astrophysical phenomena. As a result, it is essential to improve our understanding of this concept. To study nuclear matter beyond saturation density ($\rho_0 \approx 0.16$ fm$^{-3}$) in laboratory settings, nuclear reactions are necessary. Consequently, the improvement of the EoS has been one of the studies of the heavy-ion collisions experiments at intermediate energy~\cite{fopi,hades,spirit,kaos,Sorensen_2024}. In recent years,  the determination of bulk properties of nuclear matter such as the symmetry energy (isovector component of the EoS) received a lot of attention from researchers, propelled by advancements in multimessenger astronomy with the NICER (Neutron star Interior Composition ExploreR)~\cite{Miller_2019,Riley_2019,Miller_2021,Riley_2021} and the Ligo-Vigo colaboration~\cite{Abbott_2017,Abbott_2020_1,Abbott_2020_2}. These advances have significantly improved our ability to measure the properties of neutron stars, offering new approaches to studying the matter inside these compact objects. However, the manifestation of exotic degrees of freedom in the interior of massive neutron stars, where the density exceeds several times the saturation density, appears theoretically feasible but remains an unresolved question. For example, the existence of hyperons \cite{Raduta_2017,Lopes_2021} and the possibility of the presence of a quark matter core~\cite{Annala_2020,Carlomagno_2024}.

In finite systems, properties are very well established for stable nuclei, in which the number of protons and neutrons is not that different. With the current experiments, measurements of nuclear masses and radii of unstable heavy nuclei with large excess neutrons will help in a better understanding of the theory for finite nuclei and neutron stars crust~\cite{Kazuhiro_2004,Davis_2024}. In this sense, neutron stars represent one of the densest and most intriguing objects in astrophysics, with formed from the remnants of massive stars undergoing gravitational collapse after a supernova explosion. These stellar remnants pack an extraordinary amount of mass around $2M_\odot$~\cite{Demorest_2010,Antoniadis_2013,NANOGrav_2017,NANOGrav_2019,Fonseca_2021}. They also are known to be associated with strong magnetic field, which has a maximum value, estimated by the virial theorem $\approx 10^{18}$~G~\cite{Chatterjee_2015}. Neutron stars serve as critical laboratories for nuclear physics under extreme conditions, providing insights into the behavior of matter under pressures and densities unattainable in laboratories. 

Theoretically, such systems can be studied using two distinct phenomenological approaches: (i) The microscopic approach, where the description of nucleon-nucleon ($NN$) interaction is carried out through a potential whose parameters are adjusted based on few-nucleon nuclear physics observables, such as deuteron properties and scattering data~\cite{Machleidt_1989,Akmal_1997}. To account for nuclear medium effects, many-body formalisms such as the (nonrelativistic) (Dirac-)Brueckner-Hartree-Fock approximation~\cite{TerHaar_1986,vanDalen_2004,VanGiai_2010,Cugnon_1987,Zuo_2002} or variational methods~\cite{Akmal_1998} are applied. (ii) The macroscopic approach, where the description of nuclear matter is conducted using relativistic and non-relativistic effective models. 

The relativistic mean-field models (RMF) have their simplest version given by the Walecka model~\cite{Serot_1984}, and other nonlinear versions (NLW)\cite{Boguta_1977}. Another class of relativistic models is the point-coupling models\cite{Nikolaus_1992}, assuming zero-range nucleon-nucleon interactions (fermions do not interact via meson exchange). Its linear version is directly derived from the Walecka model~\cite{Gelmini_1995, Delfino_1996}, and the nonlinear version from NLW~\cite{Thomas_1935,Nikolaus_1992,Lourenco_2012}. Although these models have different EoS structures compared to the Walecka model, they yield the same relevant physics for the description of nuclear matter~\cite{Sulaksono_2003,Lourenco_2009,Dutra_2014} and neutron stars~\cite{Burgio_2021}. 

For the nonrelativistic case, a representative class of models is the Skyrme model, initially developed by T. H. R. Skyrme and further studied by D. Vautherin and D. M. Brink~\cite{Skyrme_1956,Vautherin_1971}. Later, this type of model was developed with different extensions~\cite{Goriely_2010,Margueron_2009,Bender_2009,Chamel_2009}. The Skyrme interaction, initially formulated for finite nuclei and nuclear matter at saturation density, represents a low-momentum expansion of the effective two-body nucleon-nucleon (NN) interaction in momentum space. The precise boundaries of its applicability—both lower and upper limits—remain to be definitively determined. A significant aspect of the Skyrme interaction is that it incorporates certain correlation effects via its parameters. Consequently, while it is theoretically represented as a zero-range interaction in coordinate space~\cite{Vautherin_1971}, it manifests certain finite-range characteristics~\cite{Pethick_1995}. For more details on the behavior of these models, see~\cite{Stone_2003,Dutra_2008,Dutra_2012}.

\subsection{Symmetry energy}

The EoS used to describe matter in neutron stars is strongly influenced by symmetry energy $\mathcal{S}$, a crucial term in the nuclear equation of state that represents the energy difference between symmetric nuclear matter (equal number of protons and neutrons) and matter with excess neutrons~\cite{Lattimer_2006,Chatziioannou_2024_11153}.

\begin{equation}
  \mathcal{S}(\rho) = E(\rho,y=0.5) - E(\rho,y=0), 
\end{equation}
where $E(\rho,y)$ is the energy per particle in asymmetric matter, $y = Z/(Z+N)$ the proton fraction, and $n$ the barionic density.

The nuclear symmetry energy, and particularly its dependence on nuclear density, has been the focus of extensive research over the past decade. This quantity provides critical insights into the isospin dependence of nuclear forces, which are significant both in the context of nuclear matter and in finite nuclei. Its quadratic form can be right as
\begin{equation}
    \mathcal{S}_{\rm{2}}(\rho) = \frac{1}{8}\frac{\partial^2 E(\rho,y)}{\partial y^2}\Bigg|_{y=0.5},
\end{equation}
The symmetry energy can be also expanded in terms of the density parameter $x = (\rho - \rho_0 )/(3\rho_0)$ as
\begin{equation}
    \mathcal{S}_{\rm{2}} = E_{\rm{sym,2}} + L_{\rm{sym,2}}\, x + \frac{1}{2} K_{\rm{sym,2}} \, x^2 +\cdots.
\end{equation}
where the $E_{\rm{sym,2}}$, $L_{\rm{sym,2}}$, $K_{\rm{sym,2}}$ are the quadratic nuclear empirical parameters. 

Several experimental constraints on the symmetry energy in finite nuclei have been established~\cite{Lattimer_2012,Wei_2019}. The Fig.~\ref{plot_EsymLsym_exp} illustrates a selection of these constraints. 

\begin{figure}[tb]
\centering
\includegraphics[scale=0.6]{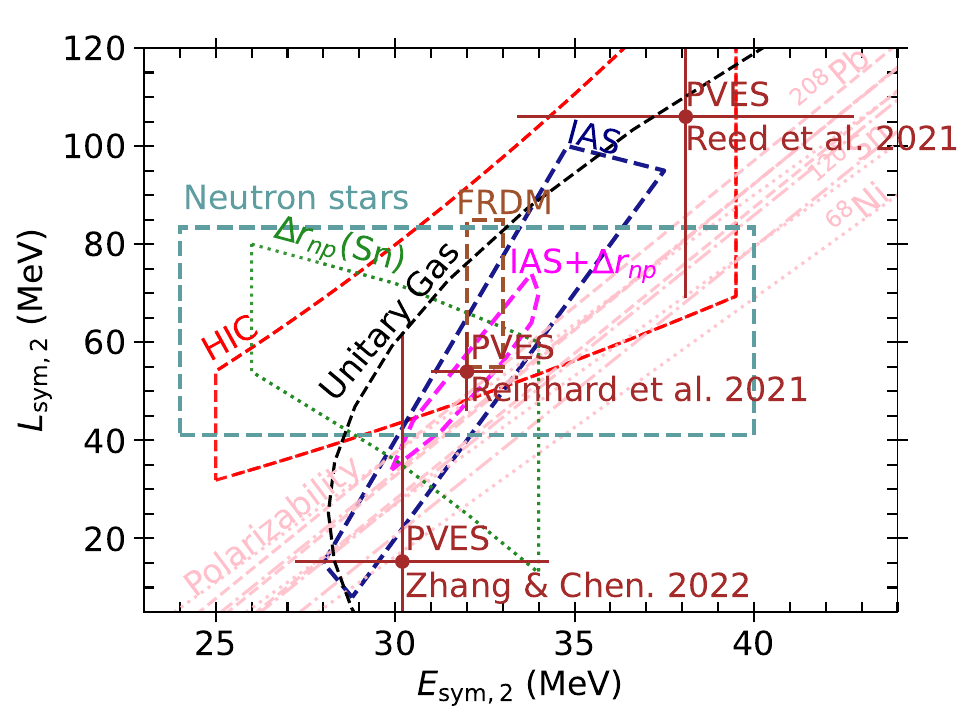}
\caption{Correlation between symmetry energy and its slope at saturation density from different constraints: “HIC”~\cite{Tsang_2008}, “Polarizability”~\cite{Roca-Maza_2015}, "$\Delta r_{np}(Sn)$" (neutron skin thickness)~\cite{Chen_2010}, "FRDM" (constraint from the finite-range droplet mass model calculations~\cite{Moller_2012}; “IAS” (isobaric analog state)~\cite{Danielewicz_2013}, “IAS + $\Delta r_{np}$"~\cite{Danielewicz_2013}, “Neutron Stars” (constraint obtained from a Bayesian analysis of mass and radius observations of neutron star by considering the $95\%$ confidence values for the slope of symmetry energy)~\cite{Steiner_2012}, "Unitary Gas"~\cite{Tews_2016}, and PVES (derived from the PREX-II~\cite{PREX_2021} and CREX~\cite{CREX_2022}). Figure extracted from Ref.~\cite{Carlson_2023}.}
\label{plot_EsymLsym_exp}
\end{figure}

In Ref.~\cite{Carlson_2023} the study, using $415$ relativistic mean field and nonrelativistic Skyrme-type interactions, finds that the better reproduction of low-energy nuclear physics data by nuclear models only weakly impacts the global properties of canonical mass neutron stars. The experimental constraint on the symmetry energy is identified as the most effective means for reducing uncertainties in neutron star matter. The authors isolate a group of interactions ($D_{\rm{4sym}}$) that exhibit a strong correlation between the symmetry energy ($E_{\rm{sym,2}}$) and its slope ($L_{\rm{sym,2}}$). The estimate found was $E_{\rm{sym,2}} = 31.8 \pm 0.7$~MeV and $L_{\rm{sym,2}} = 58.1 \pm 9.0$~MeV, indicating a significant relationship between these parameters.

\subsection{Neutron stars macroscopic properties}
The macroscopic properties of neutron stars, including mass, radius, angular momentum, and tidal deformability, are intricately linked to nuclear data. While low-energy nuclear physics provides essential constraints, the complexities of high-density matter and the density dependence of the EoS introduce significant challenges in accurately predicting these properties~\cite{Carlson_2023}. The density dependence of the EoS plays a crucial role in determining the mass and radius. To evaluate these quantities it is necessary to solve the Tolman-Oppenheimer-Volkoff (TOV) equations~\cite{Tolman_1939,Oppenheimer_1939}. The TOV equations describe the structure of a static, spherically symmetric star in hydrostatic equilibrium. They are essential for understanding how matter behaves under extreme gravitational forces, such as those found in neutron stars.

In the description of neutron stars, the RMF model and the Skyrme-type model are widely used in the literature. The EoS for pressure and energy density for both are given, respectively, by

\begin{eqnarray}
\epsilon_{\mbox{\tiny RMF}} &=& \frac{1}{2}m^2_\sigma\sigma^2 + \frac{A}{3}\sigma^3 + \frac{B}{4}\sigma^4 - \frac{1}{2}m^2_\omega\omega_0^2 
- \frac{C}{4}(g_\omega^2\omega_0^2)^2 + g_\omega\omega_0\rho - \frac{1}{2}m^2_\rho\bar{\rho}_{0(3)}^2 +\frac{g_\rho}{2}\bar{\rho}_{0(3)}\rho_3 - \frac{1}{2}\alpha'_3g_\omega^2 g_\rho^2\omega_0^2\bar{\rho}_{0(3)}^2
\nonumber\\
&-& g_\sigma g_\omega^2\sigma\omega_0^2\left(\alpha_1+\frac{\alpha'_1g_\sigma\sigma}{2}\right) - g_\sigma g_\rho^2\sigma\bar{\rho}_{0(3)}^2 
\left(\alpha_2+\frac{\alpha'_2 g_\sigma\sigma}{2}\right) + \epsilon_{\mbox{\tiny kin}}^p + \epsilon_{\mbox{\tiny kin}}^n,
\label{denerg}
\end{eqnarray}
and
\begin{eqnarray}
p_{\mbox{\tiny RMF}} &=& - \frac{1}{2}m^2_\sigma\sigma^2 - \frac{A}{3}\sigma^3 - \frac{B}{4}\sigma^4 + \frac{1}{2}m^2_\omega\omega_0^2 
+ \frac{C}{4}(g_\omega^2\omega_0^2)^2 + g_\sigma g_\omega^2\sigma\omega_0^2\left(\alpha_1 + \frac{\alpha'_1g_\sigma\sigma}{2}\right) 
+ g_\sigma g_\rho^2\sigma\bar{\rho}_{0(3)}^2 \left(\alpha_2+\frac{\alpha'_2g_\sigma\sigma}{2}\right) 
\nonumber \\
&+& \frac{1}{2}m^2_\rho\bar{\rho}_{0(3)}^2+ \frac{1}{2}{\alpha_3}'g_\omega^2 g_\rho^2\omega_0^2\bar{\rho}_{0(3)}^2 + p_{\mbox{\tiny kin}}^p + p_{\mbox{\tiny kin}}^n,
\label{pressure}
\end{eqnarray}
with
\begin{eqnarray}
\epsilon_{\mbox{\tiny kin}}^{p,n} = \frac{\gamma}{2\pi^2}\int_0^{{k_F}_{p,n}}k^2\quad \mbox{and} \quad p_{\mbox{\tiny kin}}^{p,n} = 
\frac{\gamma}{6\pi^2}\int_0^{{k_F}_{p,n}}\frac{k^4dk}{(k^2+M^{*2})^{1/2}},
(k^2+M^{*2})^{1/2}dk
\end{eqnarray}
where the Fermi momentum for proton/neutron is given by ${k_F}_{p,n}$. The zero component ($\sigma$, $\omega_0$) and isospin space third component ($\bar{\rho}_{0_{(3)}}$) are the expectation values of the mesons fields in the expressions above. $M^*= M_{\mbox{\tiny nuc}}-g_\sigma\sigma$ and the is the effective nucleon mass and $\gamma=2$ is the degeneracy factor for asymmetric matter. The self-consistency of the model imposes to $M^*$ the condition of 
\begin{eqnarray}
M^*- M_{\mbox{\tiny nuc}} + \frac{g_\sigma^2}{m_\sigma^2}({\rho_s}_p + {\rho_s}_n) - \frac{A}{m_\sigma^2}\sigma^2 - \frac{B}{m_\sigma^2}\sigma^3 = 0,\quad \mbox{with} \quad {\rho_s}_{p,n} = \frac{\gamma M^*}{2\pi^2}\int_0^{{k_F}_{p,n}}\frac{k^2dk}{(k^2+M^{*2})^{1/2}}
\end{eqnarray}
and
\begin{eqnarray}
\epsilon{\mbox{\tiny Sky}} &=& \frac{3}{10M_{\rm nuc}}\left(\frac{3\pi^2}{2}\right)^{2/3}\rho^{5/3}H_{5/3}(y)+ \frac{t_0}{8}\rho^2[2(x_0+2)-(2x_0+1)H_2(y)] \nonumber \\
&+& \frac{1}{48}\sum_{i=1}^{3}t_{3i}\rho^{\sigma_{i}+2}[2(x_{3i}+2)-(2x_{3i}+1)H_2(y)]
+\frac{3}{40}\left(\frac{3\pi^2}{2}\right)^{2/3}\rho^{8/3}[aH_{5/3}(y) + bH_{8/3}(y)],
\label{edsk}
\end{eqnarray}
with
\begin{eqnarray}
a = t_1(x_1+2)+t_2(x_2+2),\quad b = \frac{1}{2}\left[t_2(2x_2+1)-t_1(2x_1+1)\right],\quad \mbox{and} \quad H_l(y)=2^{l-1}[y^l+(1-y)^l].
\end{eqnarray}
The pressure of the model as
\begin{eqnarray}
p {\mbox{\tiny Sky}} &=& \frac{1}{5M_{\rm nuc}}\left(\frac{3\pi^2}{2} \right)^{2/3}\rho^{5/3}H_{5/3}(y) +\frac{t_0}{8}\rho^2[2(x_0+2)-(2x_0+1)H_2(y)]
\nonumber \\
&+&\frac{1}{48}\sum_{i=1}^{3}t_{3i}(\sigma_i+1)\rho^{\sigma_i+2}[2(x_{3i}+2)-(2x_{3i}+1)H_2(y)] +\frac{1}{8}\left(\frac{3\pi^2}{2}\right)^{2/3}\rho^{8/3}[aH_{5/3}(y) + bH_{8/3}(y)],
\label{prsk}
\end{eqnarray}
The nucleon rest mass $M_{\rm nuc} = 939$~MeV is the same for both models used. A particular parametrization is defined by a specific set of the following free parameters.

In order to determine the macroscopic properties of neutron stars, a series of essential steps must be undertaken~\cite{livro-Glendenning:2012,Dutra_2016}: (i) the combination of the EoS for hadronic matter with the EoS for free leptons; (ii) the imposition of conditions regarding charge neutrality and chemical equilibrium; (iii) the incorporation of the crust description, for exeample, the Baym–Pethick–Sutherland (BPS) EoS~\cite{Baym_1971} for low densities into the EoS for hadrons and leptons; (iv) the utilization of the resulting EoS as an input for the TOV, which represent the differential equations governing the structure of a static, spherically symmetric star in hydrostatic equilibrium. Subsequently, the leptonic EoS only takes into account electrons and muons, as we specifically focus on the zero-temperature deleptonized phase of stellar evolution.

An example of the results for the mentioned EoSs is shown in Fig.~\ref{mass-radius}. In this figure, were used the parametrizations SLy4~\cite{Chabanat_1997} and BRS8~\cite{Dhiman_2007} ($D_{\rm{4sym}}$), are utilized for the hadronic component of the system. These specific parameterizations have undergone recent scrutiny~\cite{Carlson_2023}, revealing their compatibility with neutron stars and the properties of finite nuclei. In the latter context~\cite{Carlson_2023}, delved into an analysis of data on various aspects of spherical nuclei, such as ground state binding energies, charge radii, and giant monopole resonances. The nuclei studied included 16O, 34Si, 40Ca, 48Ca, 52Ca, 54Ca, 48Ni, 56Ni, 78Ni, 90Zr, 100Sn, 132Sn, and 208Pb.

\begin{figure}[tb]
\centering
\includegraphics[scale=0.1]{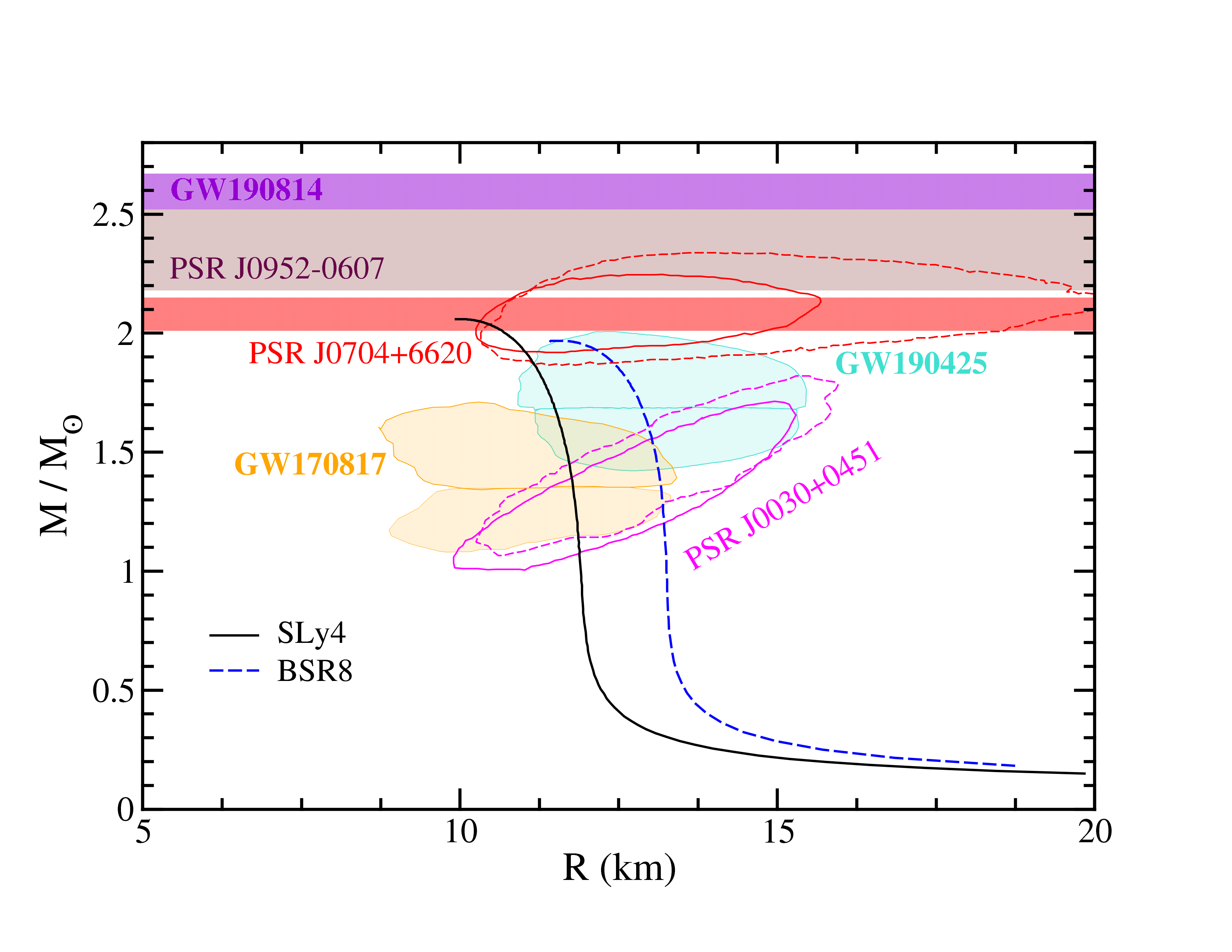}
\caption{Mass-radius relations for SLy4 and BSR2 parametrizations.The contours are related to data from the NICER mission, namely, PSR~J0030+0451~\cite{Miller_2019,Riley_2019} and PSR~J0740+6620~\cite{Miller_2021,Riley_2021}, the GW170817 event~\cite{Abbott_2017,Abbott_2018}, and the GW190425 event~\cite{Abbott_2020_2}, all of them at $90\%$ credible level. The red (brown) horizontal lines are related to PSR~J0740+6620~\cite{Fonseca_2021} (PSR~J0952+0607~\cite{Romani_2022}). The recent observational constraint on neutron star mass, GW190814~\cite{Abbott_2020_1}, is shown as the violet horizontal lines.}
\label{mass-radius}
\end{figure}

It is important to mention that different studies present in the literature provide information about the contribution of the slope of the symmetry energy in the mass-radius diagram~\cite{Lopes_2024,Carlson_2023,Lattimer_2012}, with some additional contribution from the second-order derivative of the symmetry energy. The value of $K_{\rm{sym,2}}$ determines the stiffness of the EoS, and the softer the EoS, the smaller the radius and the greater the central density for a given mass of the star. To some extent, the stiffness of the EoS is determined by the value of $L_{\rm{sym,2}}$. The values of these quantities for these two parameterizations are indicated in Table~\ref{tab:prperties}.

\begin{table}[tb]
\centering
\begin{tabular}{l|c|c|c|c|c|c|c|c}
\hline\hline
Parametrization & $n_0$ & $E_0$ & $E_{\rm{sym,2}}$ & $L_{\rm{sym,2}}$ & $K_{\rm{sym,2}}$ & $M_{\rm max}$ & $R_{\rm max}$ & $R_{\rm 1.4}$\\ \hline
SLy4~\cite{Chabanat_1997} & 0.160 & -15.97 & 32.00 & 45.94 & -119.73 & 2.05 & 10.01 & 11.83 \\ 
BSR8~\cite{Dhiman_2007} & 0.147 & -16.04 & 31.08  & 60.25 & -0.74 &  1.97 & 11.56 & 13.15 \\ 
\hline\hline
\end{tabular}
\caption{Some properties for the parametrizations showed. $E_0$ (binding energy), $E_{\rm{sym,2}}$, $L_{\rm{sym,2}}$, and $K_{\rm{sym,2}}$ are given in MeV, except for $n_0$ given in fm$^{-3}$, the maximum NS mass, $M_{\rm{max}}$ is given in units of solar mass ($M_\odot$). Both radius are given in km.}
\label{tab:prperties}
\end{table}

As a remark, future research should focus on refining nuclear models, incorporating more experimental data, exploring the density dependence of the energy density functional, investigating phase transitions, and utilizing advanced statistical methods to enhance the understanding of neutron star properties. These efforts could lead to more accurate predictions and a deeper understanding of the fundamental physics governing neutron stars~\cite{Carlson_2023}. Also, the authors from Ref.~\cite{Burgio_2021} call for significant efforts to address the current theoretical uncertainties and to explore the implications of their findings for various astrophysical phenomena, including the dynamics of neutron star binary mergers and the nucleosynthesis of heavy elements.

\section{Neutron-star matter under strong magnetic fields and magnetars}
\label{sec:marquez}

\subsection{Introduction}

The description of the macroscopic structure of compact stars -- including neutron stars and their various subclasses -- begins at the microscopic level, with the choice of the composition of dense stellar matter to be considered and of the microscopic model to accurately characterize the behavior of these components. These two ingredients are the starting point for obtaining the equation of state (EoS) of dense matter. The extraordinarily high densities within neutron star lead to the creation of heavier particle species, beyond the conventional proton-neutron-electron composition assumed in earlier models of pulsars, that are governed by some model of the strong force interactions. The calculation of the stellar matter EoS subsequently defines the distribution of baryons and leptons, constrained by equilibrium conditions such as $\beta$-stability and charge neutrality, while establishing relationships between energy density, pressure, and baryon density.
Going from micro to macrophysics requires applying the EoS that describes dense matter to conditions of mechanical (or hydrostatic) equilibrium. Compact stars are characterized by their exceptionally strong gravitational fields, necessitating the establishment of equilibrium within the framework of general relativity \cite{glendenning2012compact, meulivro}.

Magnetars are a unique class of neutron stars distinguished by having the most intense stable magnetic fields known in the nature. Surface field strengths, primarily of the poloidal component, are estimated to range from $10^{11}$ to $10^{15}$  G, with interior values surpassing these by more than an order of magnitude \citep{haensel2006neutron}. The magnetic field intensity in the central regions remains uncertain but could potentially reach up to $10^{18}$ G according to the scalar virial theorem \citep{1991ApJ...383..745L,1993A&A...278..421B}. 

The extreme magnetic field strength $B$ found in magnetars impacts the formalism describing compact stars in two significant ways. At the microscopic level, we must consider how the particles interact with the external electromagnetic field in addtition on how they interact with each other. So, the energy spectra of the particles will be altered by their interaction with $B$.
At the macroscopic level, magnetic fields exceeding $\sim10^{16}$ G, lead to deformationns in the stellar geometry away from spherical symmetry \citep{Gomes:2019paw}. Consequently, the standard relativistic hydrostatic equations used for describing non-magnetized stars, such as the Tolman-Oppenheimer-Volkoff (TOV) equations \citep{Tolman:1939jz,Oppenheimer:1939ne}, which assume spherical symmetry in their derivation from general relativity equations, become inadequate. Addressing these non-spherical configurations and the anisotropies introduced by magnetic fields also leads to the determination of the magnetic field profile in the star interior, linking the surface value of $B$ with the higher values expected deeper within the star.

In the following, we summarize the formalism developed for the description of magnetars in our previous studies -- for more detailed expositions and discussions, we refer to \citep{dexheimer2021delta}, \citep{backes2021effects} and, mainly, \citep{marquez2022}.
The formalism employed in the microscopic description of magnetized neutron-star matter is presented in Section \ref{sec1}, and the procedure of going from the EoS to the macroscopic description of a compact star through General Relativity is discussed in in Section \ref{Marquez_sec2}. The main conclusions are drawn in Section \ref{Marquez_sec3}. 

\subsection{Microphysics} \label{sec1}

The high energies found in neutron star cores can yield heavier particle species beyond the conventional proton-neutron-electron mix. The spin-1/2 baryon octet has been extensively studied in literature to the point of being almost standard procedure to include them in the stellar matter composition, and recently attention has also been directed towards the spin-3/2 decuplet, also called baryon ressonances -- see, e.g., Refs. \citep{Li:2018qaw,Yang:2018idi,marquez2022delta,lopes2023baryon}. The lightest members of this decuplet, the $\Delta$ baryons, are only about 30\% heavier than nucleons and lighter than the heaviest spin-1/2 baryons, as shown in Tab. \ref{tab:baryons}. 
The relative strengths of the coupling constants are determined by SU(3) symmetry arguments proposed in \citep{hyperonchi}, which sets the complete hyperon-meson coupling scheme from a single free parameter $\alpha_v$. 
Despite the value of $\alpha_v$, hyperons are always present in the neutron-star matter and the sequence of hyperon thresholds are always the same, with an inversely proportional relationship between $\alpha_v$ and the stiffness of the EoS. 
In this work, we choose to use $\alpha_v=0.5$, which results in hyperon potential depths of $U_{\Lambda}= -28$ MeV, $U_{\Sigma}= +21.8$ MeV and $U_{\Xi}= +35.3$ MeV.
On the other hand, the $\Delta$ baryon couplings are less constrained due to our limited knowledge about their interactions. From available literature, they are anticipated to exhibit an attractive potential of approximately 2/3 to 1 times the potential of nucleons \citep{Drago:2014oja,Raduta:2021xiz}. 
We analyse the scenarios with {$x_{\sigma \Delta} = x_{\omega \Delta} = 1.0$ and $x_{\sigma \Delta} = x_{\omega \Delta} =1.2$}, keeping $x_{\rho \Delta} = 1.0$, that generates, respectively, potentials $U_\Delta=-66.25$ MeV (equal to the nucleon potential) and $-79.50$ MeV.

Furthermore, investigating the impact of strong magnetic fields on baryon ressonances is of special interest due to the possibility of them having large electric charge and additional spin projections. The effects of Landau levels in dense stellar matter containing $\Delta$ baryons were initially explored in neutron-star contexts and subsequently in $\Delta$-admixed hypernuclear stellar matter \citep{dexheimer2021delta}. Different properties of baryons considered in this study are shown in Tab.~\ref{tab:baryons}. 

\begin{table}[!t]
    \centering
    \caption{Vacuum mass, electric charge, isospin $3^{\rm{rd}}$ component, spin, and normalized anomalous magnetic moment of baryons considered in this work.  Electric charge is shown in units of the electron charge and $\mu_N$ is the nuclear magneton.}
    \begin{tabular}{l|ccccc}\hline
          & $M_b$ (MeV) & $q_b (e)$     & $I_{3\, b}$      &$S_b$  & $\kappa_b/\mu_N$ \\ \hline
          $p$           & 939   & $+1$  & $+1/2$ &1/2    & $1.79$\\              
          $n$           & 939   & 0     & $-1/2$ &1/2    & $-1.91$\\\hline
          $\Lambda$     & 1116  & 0     & 0              &1/2    & $-0.61$\\ 
          $\Sigma^+$    & 1193  & $+1$  & $+1$           &1/2    & $1.67$\\
          $\Sigma^0$    & 1193  & 0     & 0              &1/2    & $1.61$\\
          $\Sigma^-$    & 1193  & $-1$  & $-1$           &1/2    & $-0.37$\\
          $\Xi^0$       & 1315  & 0     & $+1/2$ &1/2    & $-1.25$\\
          $\Xi^-$       & 1315  & $-1$  & $-1/2$ &1/2    & $0.06$\\\hline
          $\Delta^{++}$ & 1232  & $+2$  & $+3/2$ &3/2    & $3.47$\\
          $\Delta^{+}$  & 1232  & $+1$  & $+1/2$ &3/2    & $1.73$\\
          $\Delta^{0}$  & 1232  & $0$   & $-1/2$ &3/2    & $0.06$\\
          $\Delta^{-}$  & 1232  & $-1$  & $-3/2$ &3/2    & {$-1.69$}\\ \hline
    \end{tabular}
    \label{tab:baryons}
\end{table}

To model hadronic dense matter under magnetic fields, it is necessary to propose a Lagrangian density that describes the particle interactions with each other and with the external electromagnetic field. To adress the strong force interaction, we employ two relativistic models to have two distinct frameworks for comparisson.
The first model is a nonlinear extension of the Walecka model, where baryon interactions are mediated by mesons like $\sigma$, $\omega$, $\rho$, and $\phi$ within the mean field approximation. Specifically, we adopt the L3$\omega\rho$ parametrization \citep{l3wr}, which includes an additional $\omega\rho$ interaction crucial for predicting the symmetry energy slope, neutron-star radii, and tidal deformabilities. 
The $\phi$ meson, with hidden strangeness, interacts exclusively with hyperons, addressing the hyperon puzzle and enabling the reproduction of higher maximum masses for neutron stars. 
The second model utilized is the chiral mean-field (CMF) model, based on a nonlinear realization of the chiral sigma model. Baryon masses are dynamically generated by the medium, allowing for chiral symmetry restoration at high temperatures or densities, consistent with lattice QCD results. Here, we focus on the hadronic version developed in Ref.~\cite{Dexheimer:2008ax}, excluding phase transitions to quark matter. An additional $\omega\rho$ interaction is introduced to enhance agreement with symmetry energy slope, neutron-star radii, and tidal deformability data, while a higher-order $\omega^4$ interaction is included to model more massive neutron stars \citep{Dexheimer:2018dhb,Dexheimer:2020rlp}.

\begin{figure*}[!t]
    \centering
    \includegraphics[width=\linewidth]{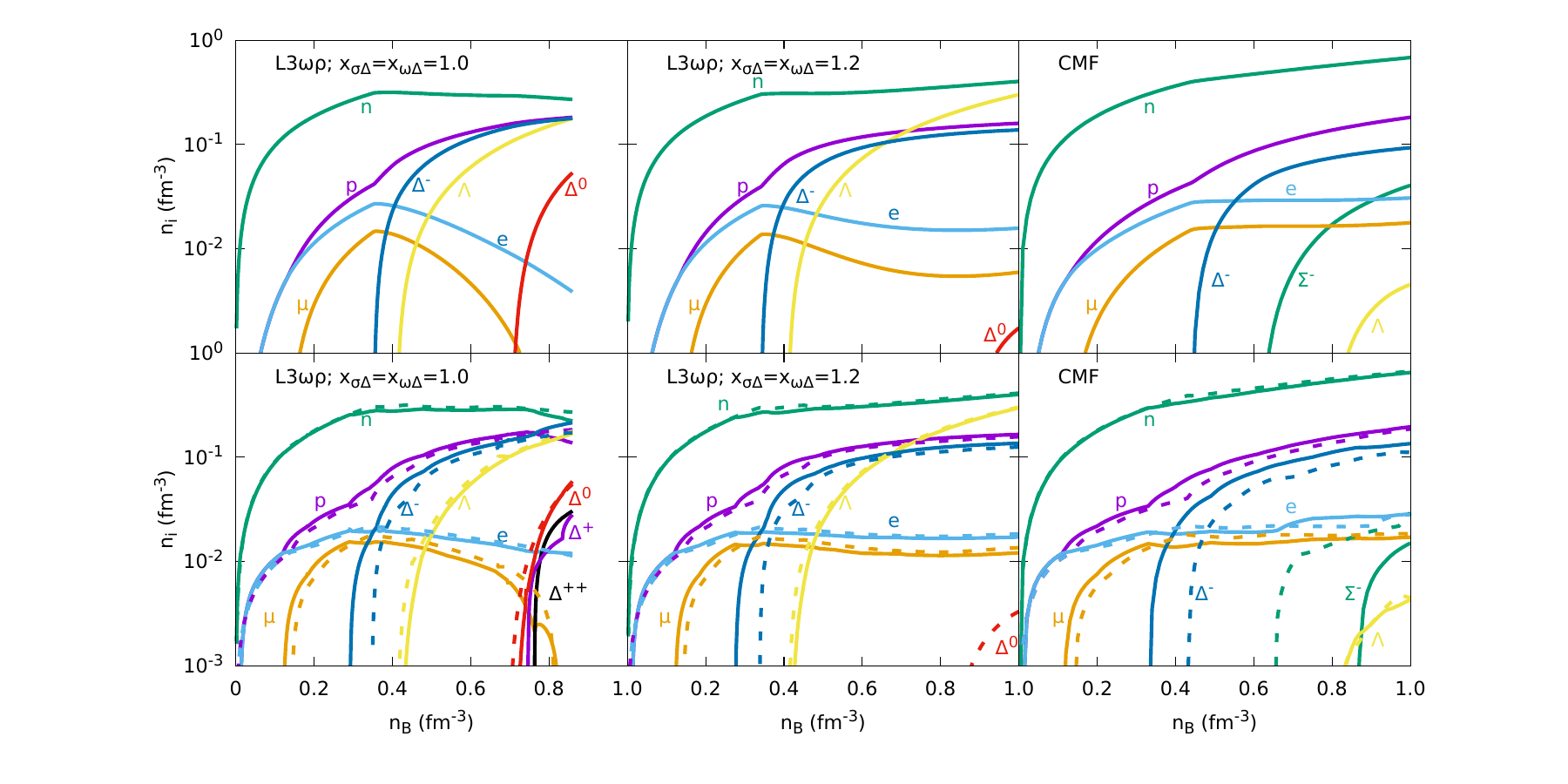}
    \caption{Particle composition of neutron-star matter with $\Delta$s, with $B=0$ (top panels) and magnetic field $B=3\times 10^{18}$ G (bottom panels), when considering (solid lines) or disregarding (dashed lines) the effects of the AMMs. Reproduced from \citep{marquez2022}.}
    \label{fig:pop}
\end{figure*}

In order to study modifications introduced by an external magnetic field $B$ on fermions, we modify the calculation of thermodynamical quantities of each (hadronic and leptonic) particle species with non-zero electric charge $q$ according to {the general procedure of taking}
\begin{eqnarray}
\sum_{\rm{spin}}\int d^3k \to \frac{|q| B}{(2\pi)^2} \sum_{\rm{spin}} \sum_{n} \int dk_z ,
\end{eqnarray}
where $k$ is the momentum, $z$ is the local direction of the magnetic field and $n$ is the discretized orbital angular momentum that the charged particle acquires in the plane transverse to $B$. The sum in $n$, at zero temperature, goes until a maximum (integer) corresponding to Landau level $\nu$ for which {{${k_z}^2\ge0$}}, i.e.,
\begin{equation}
    \nu\leq{\nu_{\rm max}}_b (s) = \left\lfloor\frac{\left( E_{F\, b}^\ast + s \kappa_b B\right)^2 - {M_b^\ast}^2} {2 |q_b| B}\right\rfloor~.
\end{equation}
where $\nu=n +\frac{1}{2}-\frac{s}{2}\frac{q}{|q|}$ depends on spin and electric charge, and the total effective energy of a particle with effective mass $M^*_b$ is given by
\begin{eqnarray}
{E_{F\, b}^\ast}^2=k_{F , b}^2 (\nu, s) + \left( \sqrt{ M_b^{\ast 2}+2 \nu |q_b| B } - s \kappa_b B \right)^2.
\end{eqnarray}

The number and scalar densities are defined for electrically neutral particles as
\begin{align}
     n_b={}&\frac{1}{2\pi^2}\sum_s \left\{\frac{k_{F\,b}^3(s)}{3} - \frac{s\kappa_b B}{2}  \left[\vphantom{\frac{ M^\ast_b }{ E_{F \,b}^\ast}}  \left(M^\ast_b - s \kappa_b B\right){k_F}_b(s)\right.\right.\nonumber\\&\left.\left.\times \,E_{F\, b}^{\ast \, 2}\left(\arcsin{\left(\frac{ M^\ast_b - s \kappa_b B}{ E_{F \,b}^\ast}\right)}-\frac{\pi}{2}\right)\right]\right\},
\end{align}
 and
\begin{align}
    n_{s\, b}={}&\frac{M_b^\ast}{4\pi^2}\sum_s\left[\vphantom{\frac{ M^\ast_b }{ E_{F \,b}^\ast}} 
    E_{F\,b}^\ast {k_F}_b(s)\right.\nonumber\\ &{}-\left. {\left( M^\ast_b - s \kappa_b B\right)^2}\ln\left|\frac{k_{F\,b}(s)+E_{F\,b}^\ast}{ M^\ast_b - s \kappa_b B}\right|\right],
\end{align}
respecively, and, for charged particles as
\begin{equation}
 n_{b}={}\frac{|q_b|B}{2\pi^2}\sum_{\nu, s} ,{k_F}_b(\nu,s)~
\end{equation}
and
\begin{align}
        n_{s\, b}={}&\frac{|q_b|B M_b^\ast}{2\pi^2}\sum_{s, \nu} \frac{\sqrt{{M_b^\ast}^2+2\nu|q_b|B} - s \kappa_b B}{\sqrt{{M_b^\ast}^2+2\nu|q_b|B}}\nonumber\\&\times\ln\left|\frac{{k_F}_b(\nu,s)+E_{F\,b}^\ast}{\sqrt{{M_b^\ast}^2+2\nu|q_b|B} - s \kappa_b B}\right|.
\end{align}
The expressions for energy density and pressure are different for each model and can be obtained from the energy-momentum tensor for matter in the usual manner.

In Fig. \ref{fig:pop}, we present the particle composition in neutron-star matter containing $\Delta$ baryons under the influence of strong magnetic fields, considering both scenarios with and without anomalous magnetic moment (AMM) corrections. The enforcement of charge conservation and chemical equilibrium results in diverse behavior among particles, depending on the sign and strength of their AMMs. Overall, the inclusion of AMMs generally leads to an increase in charged particle populations, whereas the populations of neutral particles tend to decrease.

\subsection{Macrophysics} \label{Marquez_sec2}

In addressing the influence of magnetic field effects on the mechanical equilibrium of compact stars, several considerations arise due to deviations from ideal conditions induced by $B$.
The TOV equations of relativistic hydrostatic equilibrium are derived imposing isotropic matter and spherical symmetry, both assumptions broken by extreme magnetic fields.
Various approaches have been suggested in the literature to address the effects of magnetic fields on compact stars. These include disregarding the anisotropy altogether \cite{dexheimer2021delta,backes2021effects}, employing perturbative solutions applicable only to weak magnetic fields \citep{Konno:1999zv}, which might not influence matter properties significantly, or assuming chaotic distributions of magnetic fields \citep{menezes2016quark}.

Also, magnetars exibit $B=10^{15}$ G in the star surface, escalating to approximately $10^{18}$ G in its core. Several studies try to delineate the magnetic field profile interpolating these extremes, typically involving the \textit{ad hoc} proposition of profiles for the magnetic field within the  TOV framework. Typically, these ansatz are some variation on the parameterization introduced by Ref. \cite{PhysRevLett.79.2176} in the late 90s. However, \cite{doi:10.1142/S201019451760031X} highlighted that this kind of magnetic field profiles are physically inaccurate because they do not adhere to Maxwell's equations, not preserving the cobdition $\nabla\cdot\vec B=0$. Assuming such profiles in a spherically symmetric star implies a purely monopolar distribution of the magnetic vector field.

\begin{figure}[!t]
    \centering
        \includegraphics[angle=0,width=0.5\linewidth]{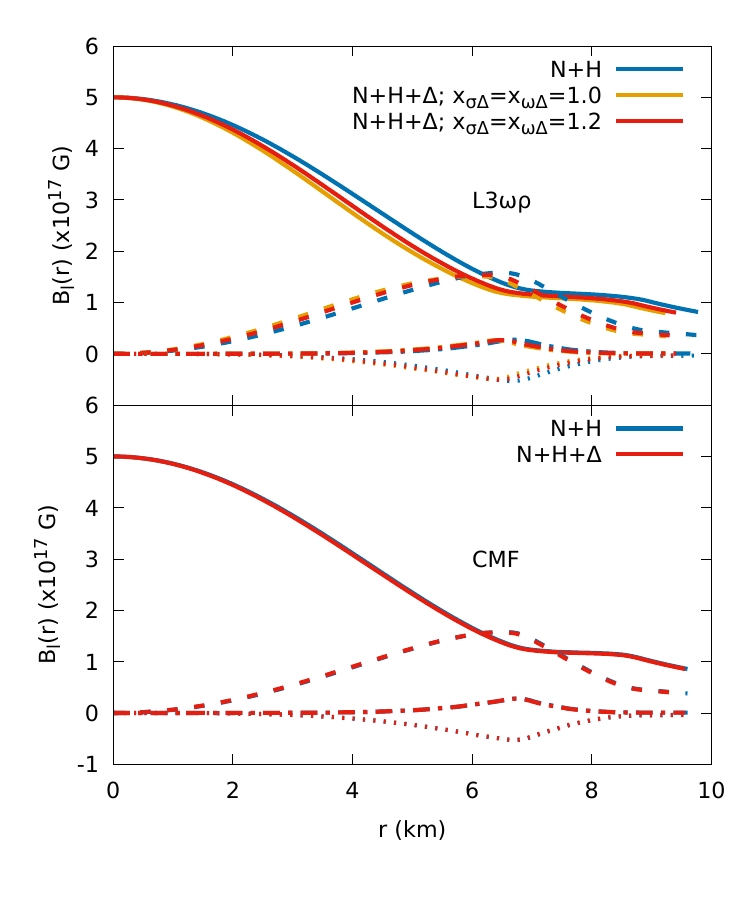}
    \caption{Magnetic field distribution inside a neutron star of mass $1.8M_{\odot}$ and central magnetic field of $B = 5\times 10^{17}$ G. Solid, dashed, dashed-dotted and dotted are $l=0,2,4,6$. Reproduced from \citep{marquez2022}.}
    \label{fig:profile1}
\end{figure}

Hence, the usual TOV equations are unsuitable for describing magnetars due to their assumption of spherical symmetry, which does not hold under strong magnetic fields. These fields cause highly deformed stellar shapes, requiring the stellar structure to be determined by solving the Einstein equations together with the Maxwell equations.
To compute the effect of the strong magnetic fields on the structure of the magnetars, then, we solve the Einstein–Maxwell equations within the numerical library LORENE \footnote{\url{http://www.lorene.obspm.fr}} using a multi-domain spectral method.

To explore througoutly effects of Landau quantization and AMMs on the star macroscopic properties, one must solve these equations with a magnetic field-dependent EoS. Refs. \cite{Chatterjee2015,Franzon:2015sya} used this approach with magnetic field-dependent quark EoS but found that the maximum neutron star mass is minimally affected by magnetic fields. Therefore, in this section of the study, we assume a non-magnetic matter contribution to the EoS, incorporating the magnetic field only through the pure electromagnetic field. However, these effects are relevant for the microscopic properties of matter, as discussed in the previous section.

The numerical routine employed in the solution of the Einstein–Maxwell equations decompose the magnetic field norm in terms of spherical harmonics,
\begin{equation}
    B(r,\theta)\simeq\sum B_l(r)Y_l^0(\theta).
\end{equation}
In Fig. \ref{fig:profile1}, we plot the first four even multipoles ($l = 0,2,4,6$) as function of coordinate radius for both the EoS models and coupling strengths. We also plot the profile of the dominant monopolar, spherically symmetric, term ($l = 0$) inside the star in Fig.~\ref{fig:profile2}. 

\begin{figure}[!t]
    \centering
        \includegraphics[angle=0,width=0.5\linewidth]{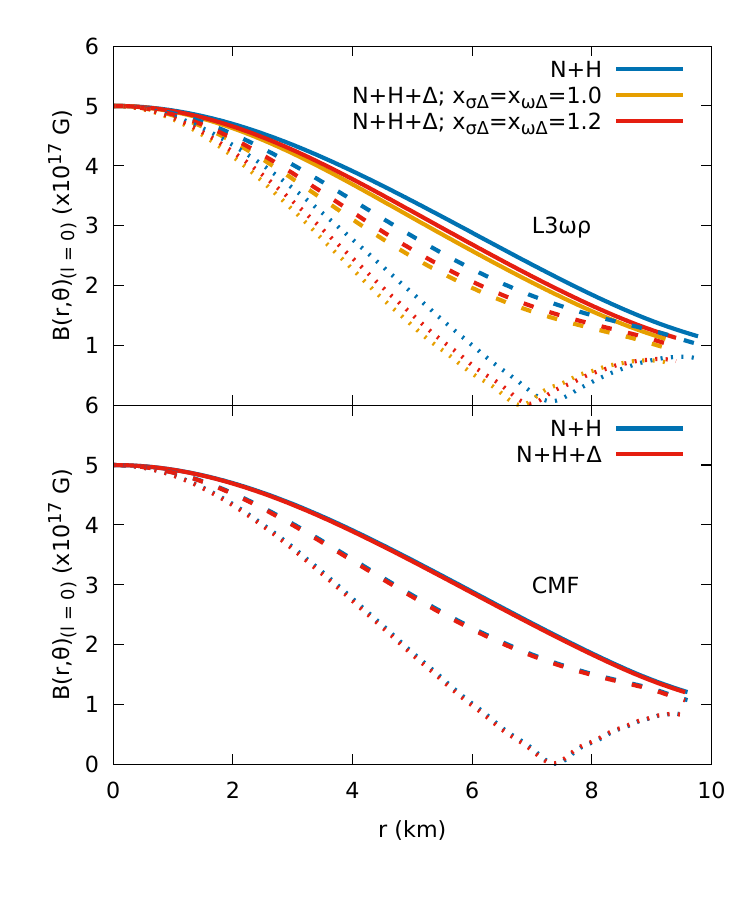}
    \caption{Magnetic field distribution inside a neutron star of mass $1.8M_{\odot}$ and central magnetic field of $B = 5\times 10^{17}$ G. Solid, dashed and dotted are the $l = 0$ term at the $\theta=0, \pi/4,\pi/2$ orientations. Reproduced from \citep{marquez2022}.}
    \label{fig:profile2}
\end{figure}

 In Fig.~\ref{fig:mxr}, we show the mass radius relations as a function of equatorial radius for a given stellar central magnetic field. The differences between the mass-radius curves for the same $B$ case (the same line dash pattern) arise from variations in the non-magnetic EoS, while differences between the same EoS with different magnetic fields (the same line color) result from the pure electromagnetic field contributions. The Lorentz force from the electromagnetic field affects the low-density part of the EoS, which is why the maximum mass of very massive stars remains unchanged with increasing magnetic field strength, while the mass and radius of less massive stars increase significantly. We observe that as the radius of the neutron star remains fixed, increasing the strength of the central magnetic field raises both the central baryon and energy densities. This occurs because greater matter pressure is needed to counterbalance the Lorentz force.

\begin{figure}[!t]
    \centering
    \includegraphics[width=0.5\linewidth]{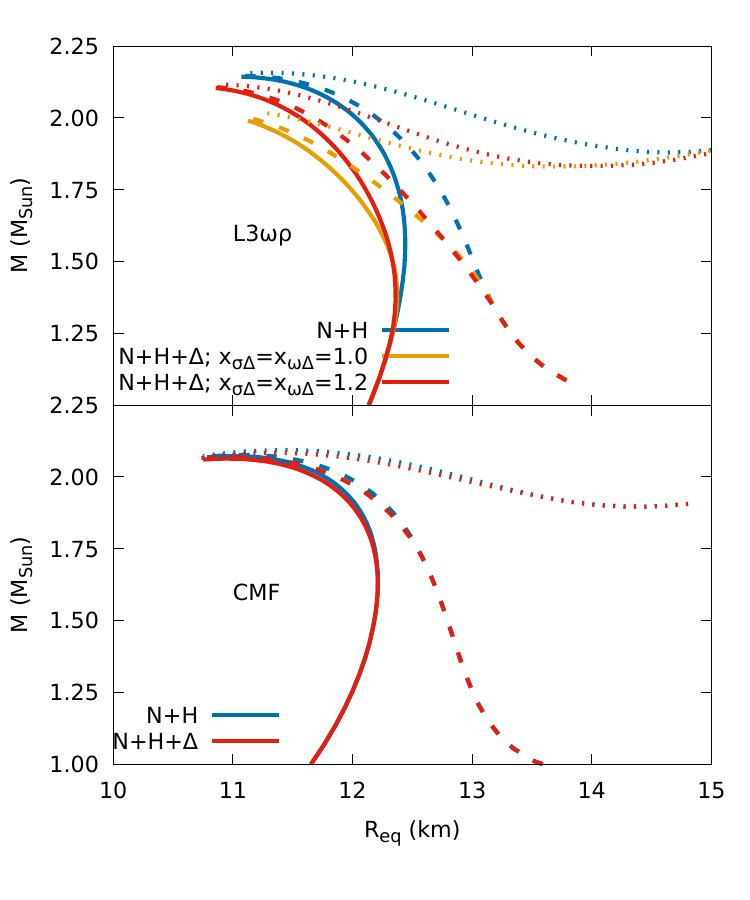}
    \caption{Stellar mass as a function of equatorial radius for different compositions and interaction strengths, for central magnetic fields $B = 0$ (solid lines), $B = 5\times 10^{17}$ G (dashed lines), and $B = 10^{18}$ G (dotted lines). Reproduced from \citep{marquez2022}.}
    \label{fig:mxr}
\end{figure}

\subsection{Conclusion} \label{Marquez_sec3}

This research aimed to shed light on the intricate interplay between strong magnetic fields, exotic particles like hyperons and baryon resonances, and the overall properties of magnetars. By investigating the microphysics stellar hadronic matter under extreme magnetic conditions, we highlighted significant alterations in the energy spectra and particle population distributions. These effects are crucial for understanding the composition and behavior of neutron star interiors under magnetic fields that can exceed $10^{18}$ G.

At the macroscopic level, we addressed the challenges posed by strong magnetic fields to the conventional models of compact stars. The spherical symmetry assumptions underlying traditional models like the TOV equations are no longer valid under such conditions, necessitating the solution of the Einstein and Maxwell equations together. This approach reveals how magnetic fields reshape the internal structure and macroscopic properties of magnetars, influencing their magnetic field profiles and mass-radius relations.

In conclusion, the comprehensive analysis presented in this paper enhances our understanding of magnetars as natural laboratories for testing the limits of matter under extreme magnetic fields. Future research directions may explore additional complexities introduced by Landau quantization and AMMs, providing further insights into the interplay between magnetic fields, exotic particles, and the overall properties of neutron stars.
\section{Hyperons in Magnetized Neutron Stars}
\label{sec:tolos}

Anomalous X-ray pulsars and soft $\gamma$-ray repeaters are  
described as magnetized neutron stars with a surface magnetic field of $ 
\sim 10^{14}-10^{15}$ G \citep{Vasisht:1997je, Kouveliotou:1998ze,Woods:1999wa}. 
This type of compact stars are identified as magnetars, that is, neutron stars with 
magnetic fields much larger than the canonical surface dipole 
magnetic fields B $ \sim 10^{12}-10^{13}$ G of the bulk of the pulsar population 
\citep{Mereghetti:2008je,Turolla:2015mwa}. Early works on dense matter in magnetars found out that the equation of state (EoS) of dense nuclear matter will be modified by magnetic fields larger than $B/B_c^e=10^5$, with $B_c^e=4.414 \times 10^{13}$ G  
defined as the critical magnetic field at which the electron cyclotron energy is equal to 
the electron mass \citep{Chakrabarty:1997ef,Bandyopadhyay:1998aq,Broderick:2000pe,Suh:2000ni,
Harding:2006qn,Chen:2005dx,Rabhi:2008je}. 

The study of the effects of very strong magnetic fields upon the EoS of matter made of hyperons was initiated in 
Ref.~\citep{Broderick:2001qw} and has been addressed later on  by several groups 
\citep{Rabhi:2009ii,Sinha:2010fm,Lopes:2012nf,Tolos:2016hhl,Gomes:2017zkc,Thapa:2020ohp,Dexheimer:2021sxs,Marquez:2022fzh,Rather:2022bmm,Sedrakian:2022ata} within different relativistic mean field (RMF) theories, where baryons interact through the exchange of  mesons, providing a covariant description of the EoS.

Within these approaches, the introduction of magnetic fields induces some important modifications in the lagrangian density. Under the assumption of minimal coupling of electromagnetic fields, the
covariant derivative  becomes $D_{\mu} = \partial_{\mu} +i q A_{\mu}$, where $A_{\mu}$ is the electromagnetic vector potential and $q$ is
the charge of the particle. Also, the coupling of the particles to the electromagnetic field tensor via the baryon anomalous magnetic moments could be considered, as done in Refs.~\cite{Rabhi:2009ii,Marquez:2022fzh}. Moreover, strong magnetic fields modify the orbits of the charged particles (leptons and charged baryons) in the direction perpendicular to the field becoming Landau quantized. Therefore, the single-particle energies of the charged baryons and leptons  as well as the scalar and vector densities have to be modified accordingly, where the integration in the orthogonal to the B-field plane must be replaced by a summation over the Landau levels up to a maximum value  (for a review see Ref.~\cite{Sedrakian:2022ata} and references therein). 

With all these modifications, the EoS for hyperonic matter together with the global properties of neutron stars, such as mass and radius, can be obtained under strong magnetic fields.  In order to illustrate the behaviour of hyperonic matter in magnetized neutron stars, we consider the approach of Ref.~\cite{Tolos:2016hhl}.  In  Ref.~\cite{Tolos:2016hhl} two slightly different parametrizations for baryonic matter within the RMF approach were developed, the FSU2R and FSU2H. The FSU2R was fitted to nuclear matter and finite nuclei properties while fulfilling some restrictions on high-density matter deduced from heavy-ion collisions as well as reproducing $2M_{\odot}$ observations and radii below 13 km for pure nucleonic matter,  whereas the FSU2H emerged by slightly refitting the parameters of the FSU2R  so that FSU2H fulfills the $2M_{\odot}$ limit with hyperons. These parametrizations were improved in Refs.~\cite{Tolos:2017lgv,Ribes:2019kno} and extended to finite temperature in Refs.~\cite{Kochankovski:2022rid,Kochankovski:2023trc}. 

The effects of strong magnetic fields were introduced in Ref.~\cite{Tolos:2016hhl} by considering the following density dependent magnetic field profile 
\begin{equation}
B(n) = B_s + B_c\left\{1 - {\rm exp}\left[-\beta\left(n/n_0\right)^\gamma\right] \right\} \ ,
\label{eq:B}
\end{equation}
with $n$ the baryonic density, following Ref.~\citep{Chakrabarty:1997ef}. The surface magnetic field value of $B_s=10^{15}$ G was taken, which is consistent with the surface magnetic fields of observed magnetars \citep{Vasisht:1997je,Kouveliotou:1998ze,Woods:1999wa} and a strong core magnetic field value of $B_c=2\times 10^{18}$ G. The parameter $\beta$ determines the density where the magnetic field saturates, whereas  $\gamma$ gives the steepness of the transition from the surface to the core field.
Changes of these parameters were allowed, although the magnetic field configuration with $\beta=0.0065$ and $\gamma=3.5$ was the preferred one, as  the magnetic field has practically saturated to its maximum value at the densities inside the core ($n \sim 5-6n_0$, with $n_0$ the nuclear saturation density). With this magnetic field profile, the  Tolman-Oppenheimer-Volkov (TOV) system was solved in order to determine the mass and radius of  magnetized neutron stars.

\begin{figure}[!ht]
\begin{center}
\includegraphics[width=0.55\textwidth]{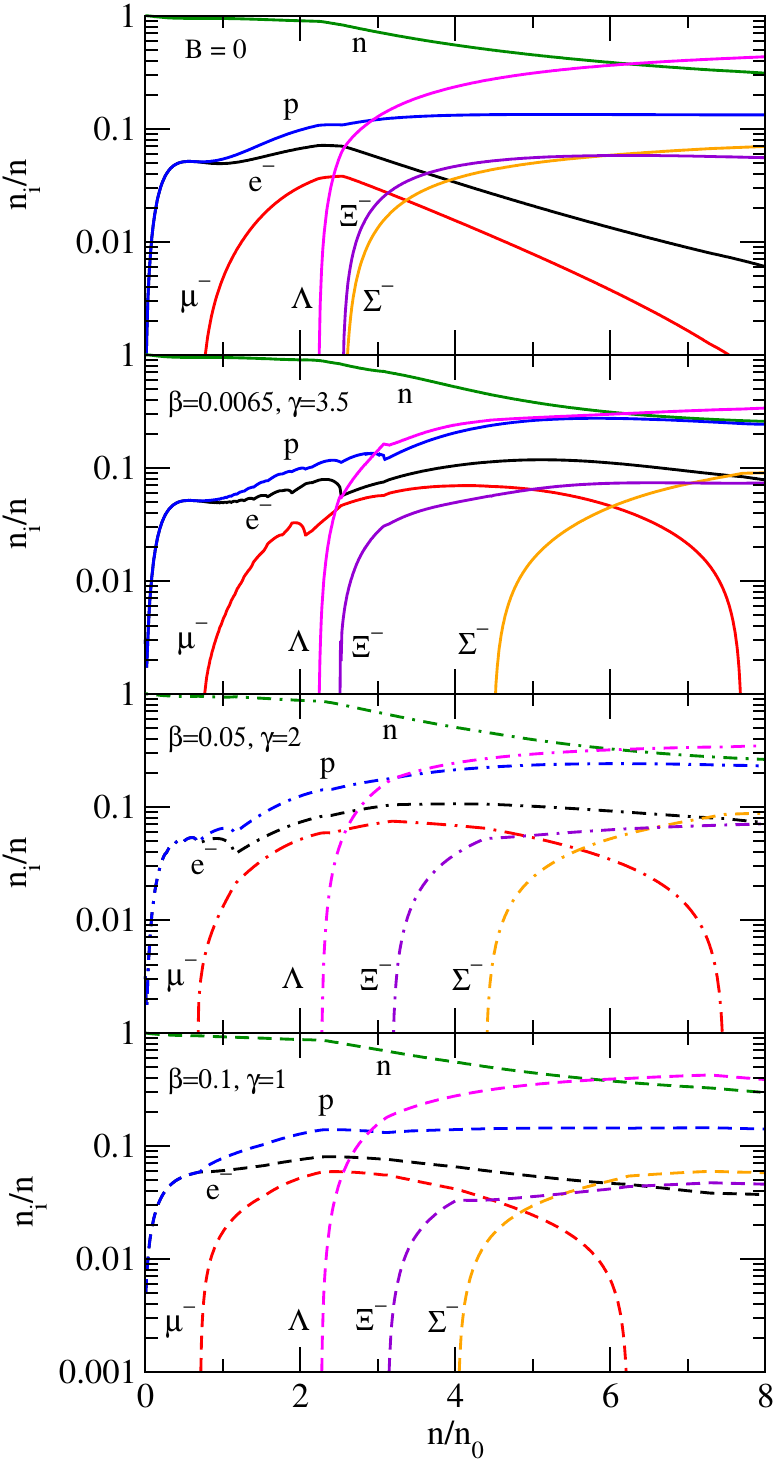}
\caption{Particle fractions as functions of the baryonic density for the FSU2H without magnetic field (upper panel) and including different magnetic field configurations (second, third and fourth lower panels), as described in the text. Figure taken from \cite{Tolos:2016hhl}.}
\label{fig:fractions}
\end{center}
\end{figure}

In Fig.~\ref{fig:fractions} the particle fractions for the FSU2H parametrization are shown for neutron star matter as functions of the baryon density. The upper panel displays the fractions in the absence of magnetic field, whereas the other panels include the density dependent magnetic field profile previously described for three different values for $\beta$ and $\gamma$. Landau oscillations are seen in the charged particle fractions when a magnetic field is applied, reflecting the successive filling of the Landau levels. For a given density, the smaller the magnetic field, the more Landau level can be accommodated and, hence, the more oscillations observed. As density increases, the magnetic field increases for all cases, and this leads to the need of only one Landau level to fit in the population of the charged particles.  Thus, the oscillations smoothen out and disappear with increasing density.  Also,  it is observed that the magnetic field affects the charged particles mostly, increasing their population with respect the $B=0$ case. At low and intermediate densities up to $n\sim 4n_0$,  an increase in the occupation of negatively charged electrons and muons is clearly seen, which delays the appearance of the negatively charged hyperons. Therefore, it can be concluded that hyperonic magnetars re-leptonize and de-hyperonize with respect to zero-field stars. At the same time, the proton abundance increases substantially, which might facilitate direct Urca processes and affect the cooling evolution of a magnetized neutron star.

\begin{figure}[!ht]
\begin{center}
\includegraphics[width=0.7\textwidth]{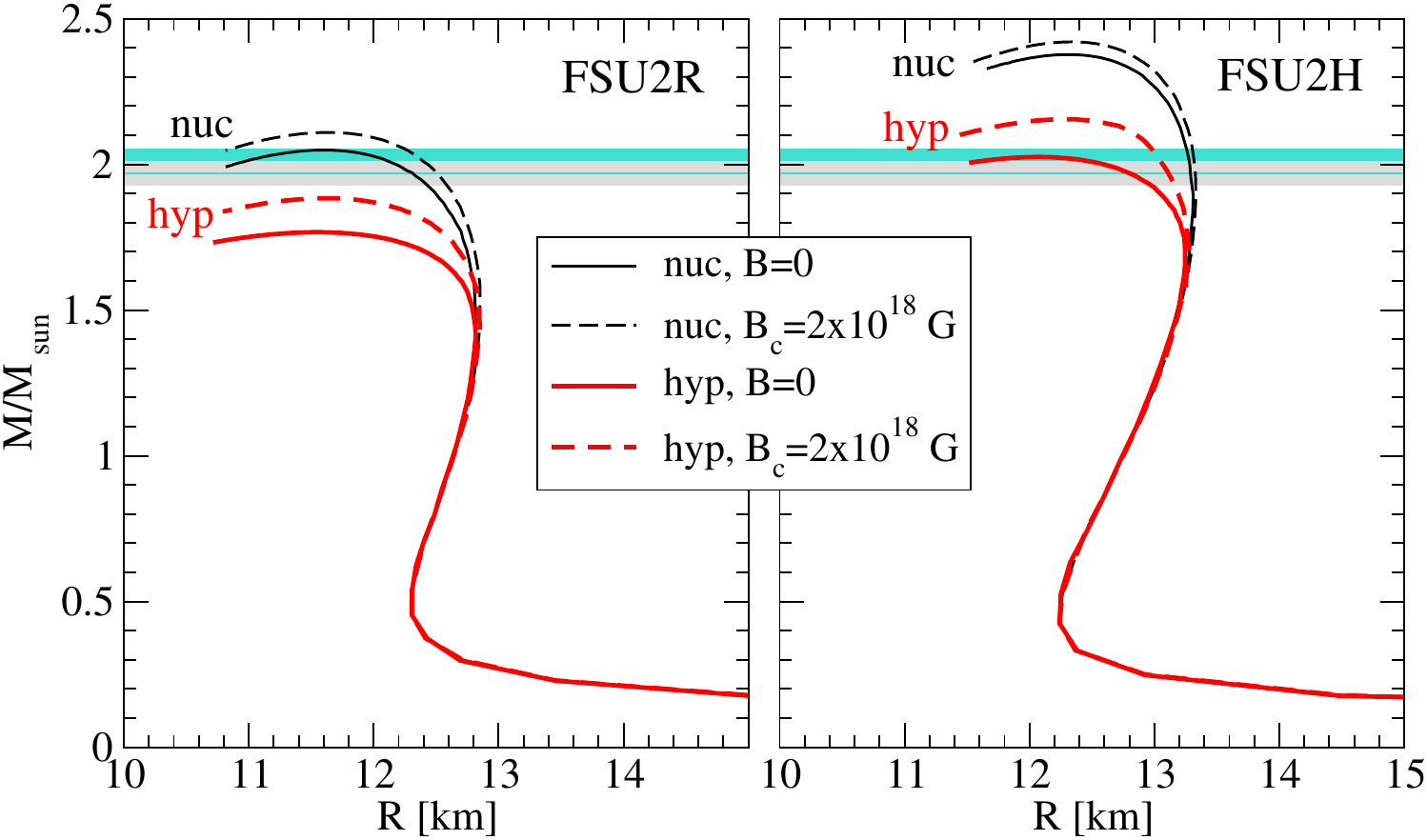}
\caption{Mass-radius relationship  of neutron stars for FSU2R (left panel) and FSU2H (right 
panel), with (thick red lines) or without (thin black lines) hyperons, and without 
(solid lines) or with (dashed lines) a magnetic field with the configuration described in the text. The two shaded bands indicate
the masses $M=1.97 \pm 0.04 M_\odot$ in the pulsar PSR J1614--2230  (grey 
band) \citep{Demorest:2010bx} and $M=2.01 \pm 0.04 M_\odot$ in the pulsar PSR 
J0348+0432 (turquoise band) \citep{Antoniadis:2013pzd}. Figure taken from \cite{Tolos:2016hhl}.}
\label{fig:mass-radius-hyp-b}
\end{center}
\end{figure}

With regards to the mass and radius of highly-magnetized neutron stars that contain hyperons, in Fig.~\ref{fig:mass-radius-hyp-b}  the mass and radius are displayed for the  FSU2R parametrization (left panel) and the FSU2H one (right panel), both considering only nucleons  as well as  nucleons and hyperons.  Whereas the solid lines correspond to vanishing magnetic field, the dashed lines include the effects of the magnetic field configuration with $\beta=0.0065$ and $\gamma=3.5$. The thin black lines indicate nucleonic neutron stars, denoted by {\it nuc}, and the thick red lines correspond to the hyperonic stars, indicated by {\it hyp}.  The inclusion of magnetic field produces stars with larger maximum masses, due to the increase in the total pressure. This enhancement is larger for the hyperonic stars than for the nucleonic ones, which is essentially due to the additional effect of de-hyperonization that takes place in the presence of a magnetic field. The reduction of hyperons is responsible for enhancing the value of the pressure, since the Fermi contributions of the other species are larger than in the $B=0$ case. It is observed, however, that large magnetic fields do not lead to large changes in the mass (or the radius).

We note that the work of Ref.~\cite{Tolos:2016hhl} as well as early works on hyperonic matter under magnetic fields \citep{Rabhi:2009ii,Sinha:2010fm,Lopes:2012nf} assumed ad hoc profiles of the magnetic field within the TOV system (that results from solving the Einstein equations for a spherically non-rotating neutron star) without solving Maxwell’s equations. However, it is nowadays clear the need of solving the coupled Einstein-Maxwell field equations in order to determine stable anisotropic magnetic-star configurations \cite{Chatterjee:2018prm,Dexheimer:2021sxs,Marquez:2022fzh,Rather:2022bmm}. Nevertheless, the general conclusions presented here regarding the behaviour of hyperons under strong magnetic fields are expected to remain unchanged.

\section{Massive, magnetized neutron stars}
\label{sec:zuraiq}

\subsection{Introduction}\label{Zuraiq_intro}

Compact stars, particularly neutron stars (NSs) and white dwarfs are incredibly rich objects from a theoretical perspective, integrating various fields of physics. One of the most fundamental results of theoretical astrophysics involving compact stars is the Chandarasekhar limit \cite{chandra}, which defines the maximum mass of a white dwarf. This concept synthesizes astrophysics, gravity and quantum statistics in one single formula. However, despite over five decades since the detection of the first NS, no analogous limit has been established for NSs due to the extreme densities at their cores,  which can far exceed the nuclear saturation density. Understanding the behavior of matter at these high densities remains a significant challenge. 

NSs serve as unique laboratories in the universe where such extreme densities are found. The exact behavior of matter under these conditions is still unknown,
which multiple candidate equation(s) of state (EOS) proposed for the NS interior.
 These EOS are constrained by both nulcear physics and astrophysical obvervations,
 making the study of NS structure vital for unraveling the mysteries of supranuclear matter. 

Recent observations have made the question of the maximum mass of an NS extremely interesting. Specifically, the discovery of “mass gap” candidates through gravitational waves (GW) and other means has intensified interest in this topic. 
 The lower mass gap, an observational range roughly between $2.5 - 5 M_\odot$, 
 was previously void of observed objects. However, events like GW190814 \cite{GW190814}, which detected the merger of a $22.2 - 24.3  M_\odot$ black hole with an object precisely in this mass gap ($2.5 - 2.67 M_\odot$), challenged this notion. Other GW observations, such as  GW200210\_092254 \cite{GW200210} and recent pulsar timing results \cite{science_massgap} further support the existence of mass gap objects.  

Thus, the above-mentioned observations suggest that the mass gap is not a theoretical inevitability, prompting the investigation of whether these objects are
massive NSs or light black holes. Thus, the critical question arises: How massive can a NS be?

In this section, we explore the possibility that massive NSs could populate the mass gap. It turns out that pure EOS effects might not suffice to support mass gap NSs due to softening effects at high densities. Therefore, we investigate additional physics, specifically the roles of magnetic fields and anisotropy, in enhancing the masses of NSs. 

Magnetic fields of compact stars represent another rich area of study with many open questions. It has been shown that strong magnetic fields can significantly affect the structure of NSs. For example, in white dwarfs, they might even be responsible for violation of the Chandrasekhar limit \cite{upasana}. Observationally, NSs are known to have significant magnetic fields, with typical surface fields ranging from $10^8 - 10^{13} \rm{ G}$. Magnetars, a subset of NSs with extreme magnetic fields,
constitute about 10\% of the NS population, with surface fields up to  $10^{15} \rm{G}$
and potentially higher central fields \cite{magnetars,magNS}. These magnetic fields modify the structure of NSs through magnetic pressure and stress. Additionally, for high enough fields, the magnetic fields can alter the microphysics of the system via Landau quantization. However, this requires quite high fields, higher than $3 \times 10^{18} \rm{G}$ \cite{sinha} for hyperonic matter. We find that magnetized NSs with strong fields ($\simeq 10^{18}$ in the core) can serve as candidates to explain mass gap objects. We put bounds on their magnetic to gravitational energy ratio ($E_{mag}/E_{grav}$) to ensure the physicality of our results. 

Another recent constraint on the NS EOS, coming from GW observations, is that of the tidal deformability, which measures how NS matter deforms in an external 
gravitational field. The tidal deformability is essentially the ratio of the quadrupole moment ($Q_{ij}$) developed in an external field to that external field 
$\epsilon_{ij}$. Thus, $Q_{ij} = -\lambda\epsilon_{ij}$, where $\lambda$ is the tidal deformability of the star. In dimensionless form, $\Lambda = \lambda/M^5 = (2/3)k_2C^{-5}$, where $C=M/R$ is the compactness of the star. Tidal deformability can further be linked to the dimensionless second love number $k_2$ arising from multipole expansion as shown in previous work \cite{hinderer}. Constraints on $\Lambda$ for a $1.4M_\odot$ star have been computed from GW events GW170817 ($\Lambda_{1.4} < 580$ and $\Lambda_{1.4} < 800$), and GW190814 ($458 < \Lambda_{1.4} < 889$) \cite{ligo1,ligo2,GW190814}. 
However, these constraints are not all independent. In fact the constraint from GW190814 was obtained by re-weighting the spectral EOS distribution from GW170817 with the probability that the maximum mass aligns with the mass of the secondary object observed in GW190814. Thus, the constraint obtained from GW190814 ended up favoring stiffer EOS as compared to the constraints from GW170817. Hence, tidal deformability is an evolving constraint. We require more such events to put better constraints on this quantity. We have, in the current article, ensured that our results are in line with the current constraints.

Massive, magnetized NSs have many interesting implications, including the potential to emit continuous GWs (CGWs). They also could be 
pulsar candidates emitting electromagnetic radiation.
This paper aims to summarize the key features of these NSs and their observational implications, contributing to the broader understanding of compact star physics and the nature of mass gap objects. 

This section is organized as follows. In subsection~\ref{ch2}, we set up the formalism required to describe the structure of our massive NSs. In subsection~\ref{ch3}, we describe our results and their possible implications. We explore the plausible detection of massive, magnetized NSs through CGWs in subsection~\ref{ch4}. We end by summarizing and providing an outlook based on our results in subsection~\ref{sec:Zuraiq_sum}.


\subsection{Massive neutron star structure}\label{ch2}

We obtain the structure of massive NSs by solving the general relativistic, hydrostatic equilibrium equations, i.e., the Tolman-Oppenheimer-Volkoff (TOV) equations \cite{tov}. The isotropic TOV equations are modified due to the inclusion of magnetic fields and anisotropic effects (both from the magnetic field and general matter effects). We follow the same modification of the TOV equations outlined in previous work by our group \cite{deb, zuraiq}, given by
\begin{equation} \label{tov}
    \frac{dm}{dr} = 4 \pi r^2 \left(\rho + \frac{B^2}{8\pi}\right), \ 
    \frac{dp_r}{dr} = 
    \begin{cases}
    \frac{- (\rho + p_r)\frac{\left(4\pi r^3\left(p_r - \frac{B^2}{8\pi}\right) + m \right)}{r(r-2m)} + \frac{2}{r}\Delta}{\left(1 - \frac{d}{d\rho}\left(\frac{B^2}{8\pi}\right)\left(\frac{d\rho}{dp_r}\right)\right)} & \text{for RO} \\
    \frac{- (\rho + p_r + \frac{B^2}{4\pi})\frac{\left(4\pi r^3\left(p_r + \frac{B^2}{8\pi}\right) + m \right)}{r(r-2m)} + \frac{2}{r}\Delta}{\left(1 + \frac{d}{d\rho}\left(\frac{B^2}{8\pi}\right)\left(\frac{d\rho}{dp_r}\right)\right)} & \text{for TO}.
    \end{cases}
\end{equation}
Here, $m$ denotes the mass, $\rho$ the density, and $B$ the magnitude of the magnetic field at a given radial distance $r$ within the star. Additionally, we have considered two different orientations for our magnetic field - radially oriented (RO) and transversely oriented (TO). These can be considered as one-dimensional analogues of poloidal and toroidal geometries, respectively. Due to the presence of anisotropy, there is a difference between the pressure along the radial direction, $p_r$, and the pressure along the transverse direction, $p_t$. This difference is expressed in the effective anisotropy factor $\Delta$ defined as 
\begin{equation} \label{delta}
    \Delta = 
     p_t - p_r + B^2/4\pi ~\text{for RO} ;~ \Delta = p_t - p_r - B^2/8\pi ~ \text{for TO}.
\end{equation}

\subsubsection{Modified Bowers-Liang model for anisotropy}

In order to close the system of equations defined above, we use the general parametric form first introduced by Bowers and Liang \cite{bl}. As done previously in our group \cite{deb, zuraiq}, we modify the Bowers-Liang form to further include the effects of the magnetic field. The anisotropy factor $\Delta$ is then given by
\begin{equation}
\label{aniso}
  \Delta =
  \kappa r^2\frac{(\rho+p_r)\left(\rho+3p_r - \frac{B^2}{4\pi}\right)}{1 - \frac{2m}{r}} \ \text{for RO; }~ \Delta =
  \kappa r^2\frac{(\rho+p_r+\frac{B^2}{4\pi})\left(\rho+3p_r + \frac{B^2}{2\pi}\right)}{1 - \frac{2m}{r}} \ \text{for TO}.
\end{equation}
This model is derived keeping in mind the following key assumptions:
\begin{itemize}
    \item The anisotropic force must vanish at the center, leading to the anisotropy term vanishing quadratically at the center.
    \item Anisotropy varies with position inside the star.
    \item $\Delta$ includes the effects due to local fluid anisotropy as well as the anisotropy due to the magnetic field (both its magnitude and orientation).
\end{itemize}

We restrict $\kappa$ to the range $[-2/3, 2/3]$ \cite{silva}. This ensures the physicality of the solution, i.e., that we do not obtain a positive $dp/dr$.

\subsubsection{Magnetic field profile}
The magnetic field is introduced in the star through a density-dependent profile \cite{bandopadhyay}, given by 
\begin{equation}
\label{mag_bando}
    B(\rho) = B_s + B_0\left[1 - \exp\left\{-\eta{\left(\frac{\rho}{\rho_0}\right)^\gamma}\right\}\right].
\end{equation}
Through this profile, we obtain the magnitude of the field as a function of the density and, thus, the radius of the star. The parameters here are: $B_s$, the surface field of the star; $B_0$, $\rho_0$, which control the field at the stellar center; $\eta$ and $\gamma$ control how the field decays from the center to the surface. Within our one-dimensional formalism and for the two orientations considered, this profile is consistent with the Maxwell equations \cite{deb}.

In this work, we choose $B_s$ to be $10^{15} \ {\rm G}$ throughout. Our results are not particularly sensitive to this value, as long as $B_s$ and $B_0$ are not comparable. We also restrict ourselves to maximum central fields of $\simeq 10^{18} \rm{ G}$. At such values, the field is not strong enough the affect the microphysics \cite{sinha}.

\subsubsection{Equations of state}

The final crucial ingredient in our depiction of massive NSs is the EOS. Terrestrial analogs for the matter at supranuclear densities found within NS cores do not exist.
There are a number of proposed EOS for high-density matter, encompassing both phenomenological and microscopic approaches. In the current work, we have considered multiple phenomenological EOS. 
We have considered several relativistic mean field (RMF) EOS: GM1L \cite{gm1l}, SWL \cite{swl}, DD2 \cite{dd2}, DD-ME1 \cite{ddme1}, and DD-ME2 \cite{ddme2}. These EOS are chosen because they best satisfy the constraints derived from both NS observations and the properties of nuclear matter at saturation density.
 
In the RMF approach, matter is modeled at the hadron level (quantum hadrodynamics). Baryon-baryon interactions present in NS matter are mediated by meson fields, which are then set to their mean values. Three such meson fields have been included in this approach: the scalar meson $\sigma$ describing attraction between baryons; the vector meson $\omega$ describing repulsion; and the isovector meson $\rho$ which is required to properly model isospin-asymmetric matter.

At the high densities found within NS cores, another source of uncertainty is the potential presence of exotic particles, i.e., particles that are not energetically favorable at the densities found in atomic nuclei.
We have considered in this work, pure nucleonic ($npe\mu$) EOS, as well as hyperon and $\Delta$ admixed ($npe\mu-Y\Delta$) EOS. 
The inclusion of hyperonic matter is achieved through meson-hyperon couplings based on the SU(3) ESC08 model.
The inclusion of $\Delta$ particles is done by including a near-universal meson-$\Delta$ coupling ($x_{i\Delta}$), with $x_{i\Delta} = g_{i\Delta}/g_{iN} = 1.2$. Here, $g$ represents the coupling constants; the subscript $\Delta$ indicates the $\Delta$ particles, $N$ the nucleons, and $i$ represents the mesons mentioned above. All EOS are shown in Fig. \ref{eos}.

\begin{figure}[!htpb]
	\centering
	\includegraphics[scale=0.3]{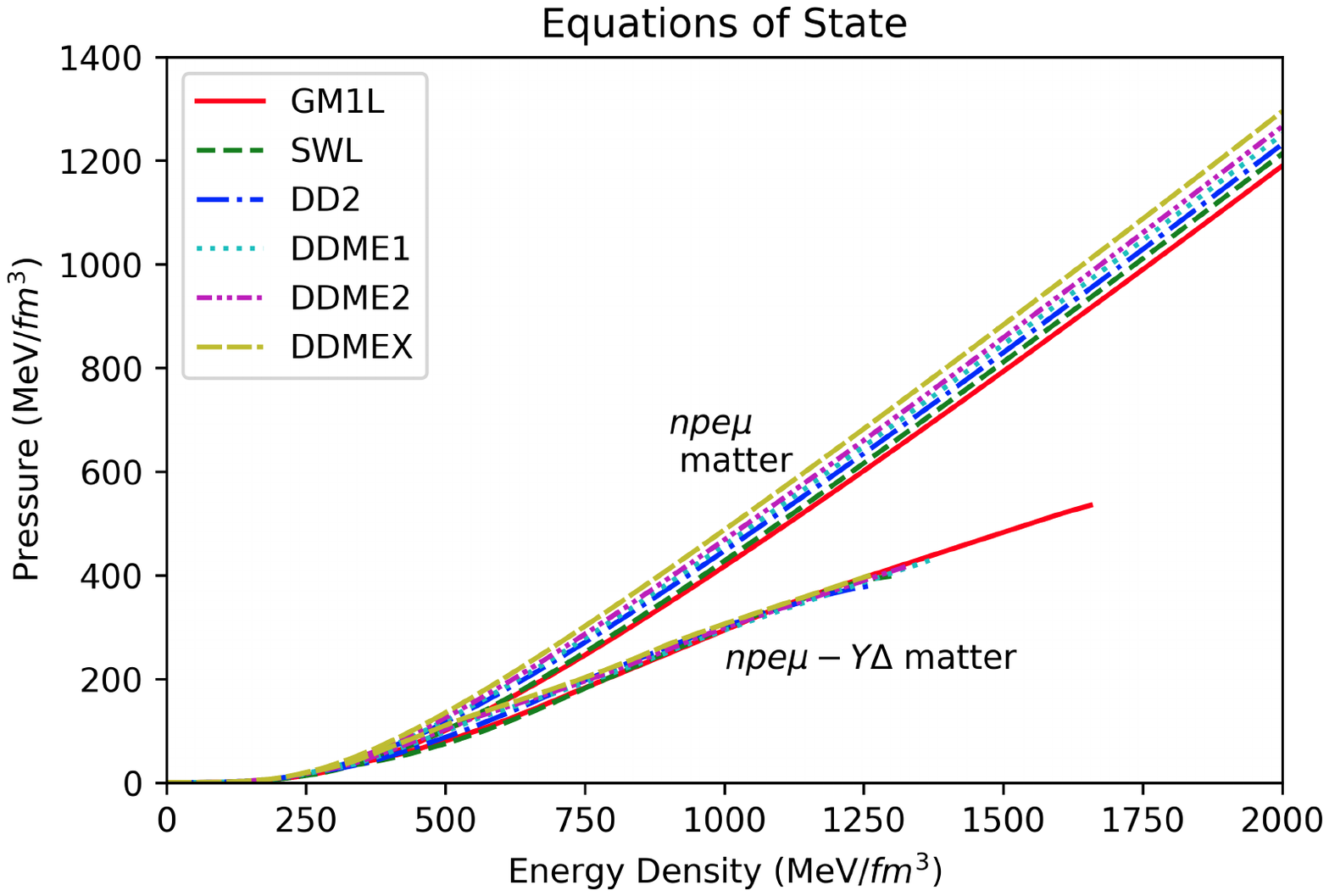}
	\caption{Different RMF EOS: Here, the upper branch denotes the $npe\mu$ EOS, while the lower branch represents the hyperon-$\Delta$ admixed $npe\mu-Y\Delta$ EOS.}
	\label{eos}
\end{figure}

\subsection{Results}\label{ch3}
One can first investigate the pure EOS effect in massive NSs by examining the stars constructed with $\kappa = 0$ and $B = 0$. As shown in the left panel of Fig. \ref{MR_iso_plot}, all the EOS considered in this study give rise to NSs with maximum-masses larger than $2M_\odot$. In the pure nucleonic cases, one obtains masses that are well in the mass gap. However, the presence of hyperon softening reduces the maximum masses to values below $\simeq 2.2M_\odot$. 
Nevertheless, by examining the tidal deformability of these NSs, shown in the right panel of Fig. \ref{MR_iso_plot}, we observe that the $npe\mu-Y\Delta$ EOS are more likely to meet the strictest observational constraints. Thus, the inclusion of exotic particles is well justified from both energetic and observational perspectives.

\FloatBarrier

\begin{figure}[!htb]
     \centering
         \centering
         \includegraphics[width=0.475\textwidth]{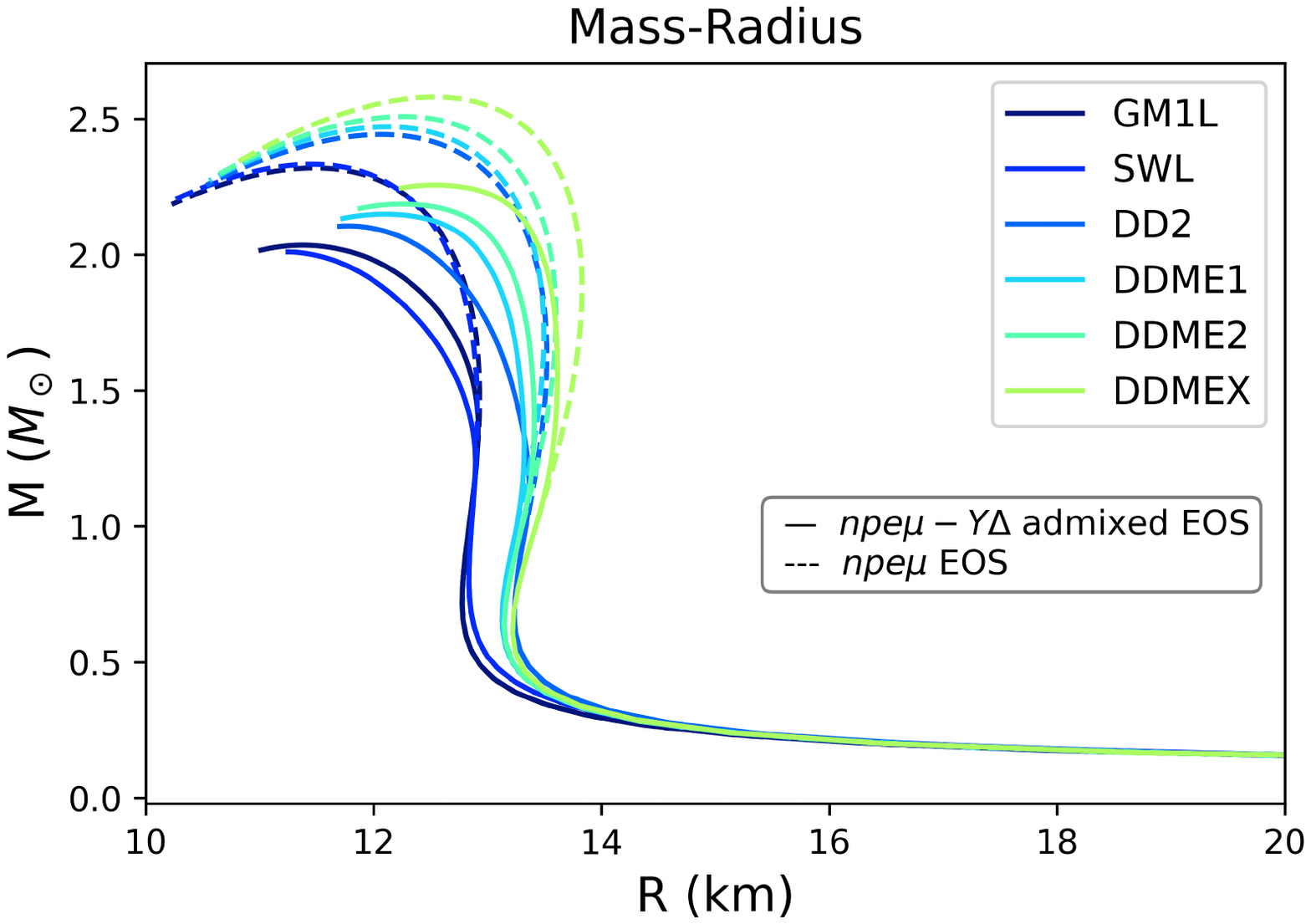}
         \centering
         \includegraphics[width=0.475\textwidth]{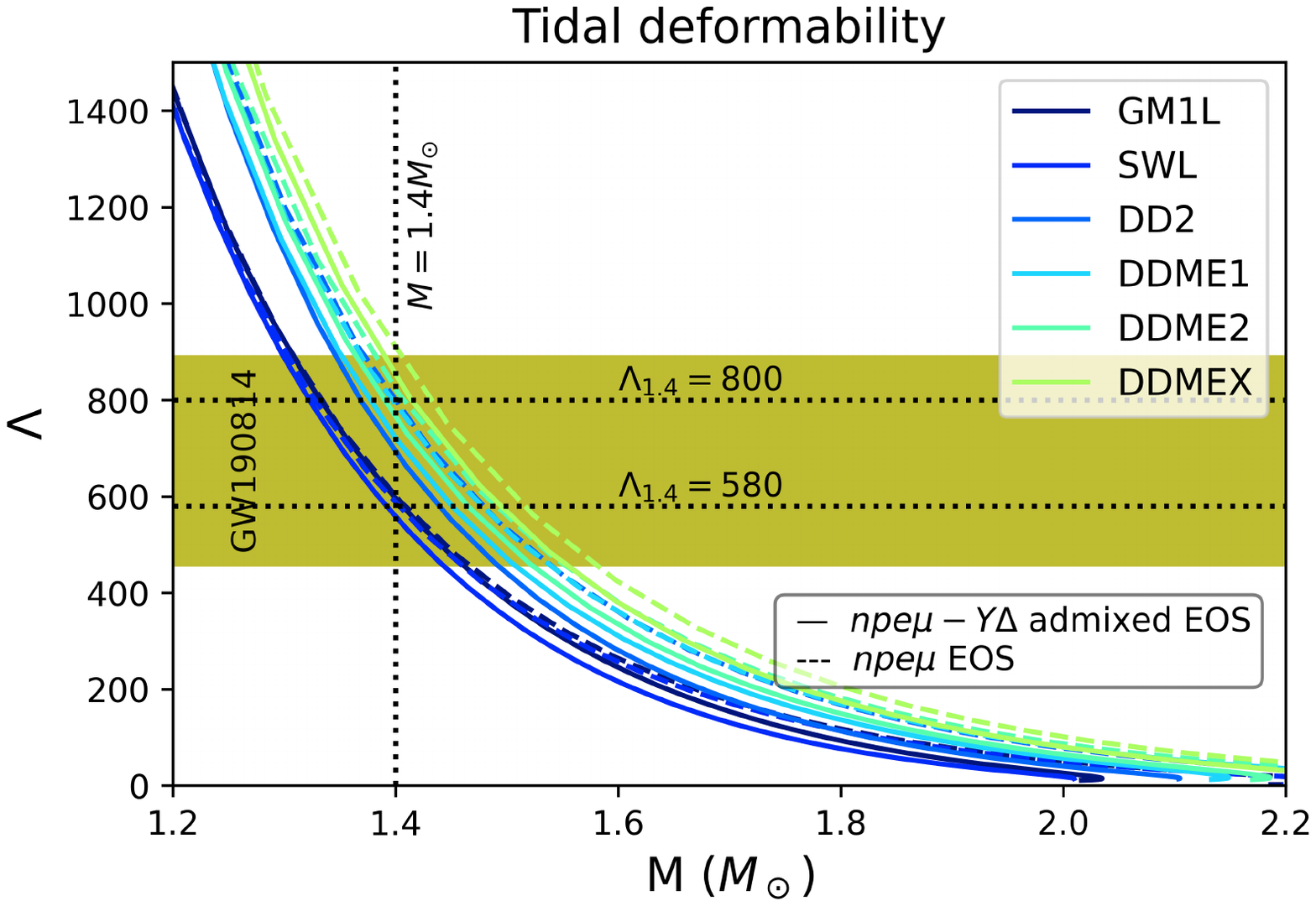}
	\caption{Mass-radius curves and tidal deformability for the isotropic cases. The dashed lines represent nucleonic cases, while solid lines represent the $npe\mu-Y\Delta$ cases.}
	\label{MR_iso_plot}
\end{figure}

\FloatBarrier
We are now thus motivated to investigate massive NSs and potential mass gap candidates by incorporating magnetic fields and/or anisotropy. To reiterate, we control the field values and their decay throughout the star using density-dependent magnetic field controlled by some parameters. However, as it is impossible to probe the magnetic field within a star, the exact magnetic field profile remains an open question.

To investigate these effects, we begin by introducing a magnetic field to the star using two different profiles/variations within the star, as shown in Fig. \ref{prof}, with fixed $\kappa = 0.5$. Profile ``1'' has a broad variation with $\eta = 0.2, \gamma = 2$, whereas profile ``2'' has a more shallow variation with $\eta = 0.01, \gamma = 2$. 
We introduce the field and anisotropy while maintaining the stiffest EOS, namely DDMEX. The resulting mass-radius curves for profile ``1'', considering different values of $B_0$ and various field orientations, are depicted on the left side of Fig. \ref{MR_ddmex}. Similar results for profile ``2'' are shown on the right side. The corresponding results are presented in Table \ref{tab_mag}.

\begin{figure}[!htpb]
	\centering
	\includegraphics[scale=0.7]{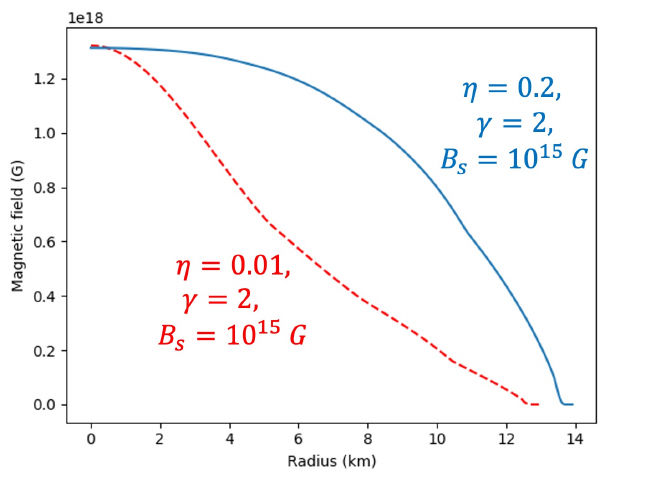}
	\caption{Two different magnetic field profiles considered in this work.}
	\label{prof}
\end{figure}

\begin{figure}[!htb]
     \centering
         \centering
         \includegraphics[width=0.48\textwidth]{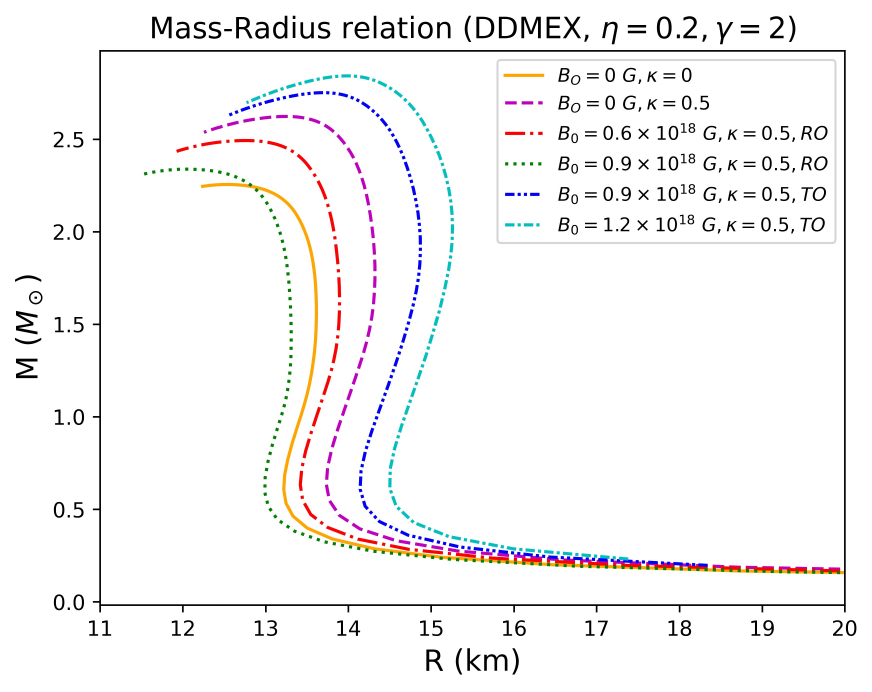}
         \centering
         \includegraphics[width=0.48\textwidth]{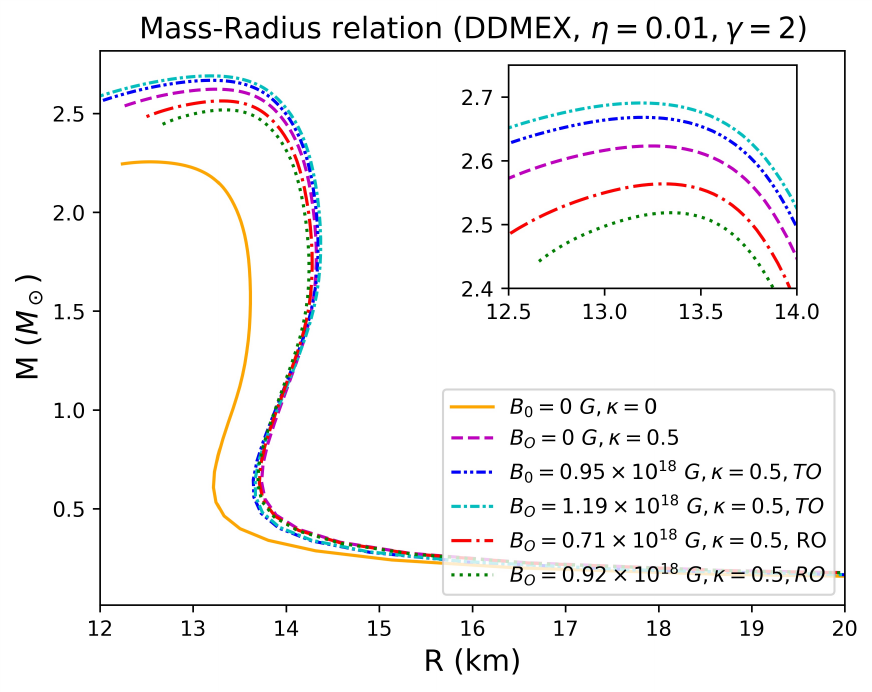}
	\caption{Mass-radius relations for anisotropic, magnetized NSs with the DDMEX EOS for two different profiles, with $\kappa = 0.5$.}
	\label{MR_ddmex}
\end{figure}

\begin{table}[]
\begin{tabular}{llll|llll}
\hline
\multicolumn{4}{l|}{Field profile: $\eta =0.2, \gamma =2$}                                                                                      & \multicolumn{4}{l}{Field profile: $\eta =0.01, \gamma =2$}                                                                                      \\ \hline
\multicolumn{1}{l|}{$B_{c} \ (10^{18} G)$} & \multicolumn{1}{l|}{$M_{max} \ (M_\odot)$} & \multicolumn{1}{l|}{$R \ (km)$} & $E_{mag}/E_{grav}$ & \multicolumn{1}{l|}{$B_{c} \ (10^{18} G)$} & \multicolumn{1}{l|}{$M_{max} \ (M_\odot)$} & \multicolumn{1}{l|}{$R \ (km)$} & $E_{mag}/E_{grav}$ \\ \hline
\multicolumn{1}{l|}{$0.9$ (RO)}              & \multicolumn{1}{l|}{2.34}                & \multicolumn{1}{l|}{12.05}     & 0.192              & \multicolumn{1}{l|}{$0.92$ (RO)}                     & \multicolumn{1}{l|}{2.52}                & \multicolumn{1}{l|}{13.32}     &  0.031                  \\ \hline
\multicolumn{1}{l|}{$0.6$ (RO)}              & \multicolumn{1}{l|}{2.49}                & \multicolumn{1}{l|}{12.76}     & 0.057             & \multicolumn{1}{l|}{$0.71$ (RO)}             & \multicolumn{1}{l|}{2.56}                & \multicolumn{1}{l|}{13.3}     & 0.019             \\ \hline
\multicolumn{1}{l|}{$0$}                     & \multicolumn{1}{l|}{2.62}                 & \multicolumn{1}{l|}{13.26}     & -                  & \multicolumn{1}{l|}{$0$}             & \multicolumn{1}{l|}{2.62}                & \multicolumn{1}{l|}{13.26}     & -             \\ \hline
\multicolumn{1}{l|}{$0.9$ (TO)}              & \multicolumn{1}{l|}{2.75}                & \multicolumn{1}{l|}{13.69}     & 0.115              & \multicolumn{1}{l|}{$0.95$ (TO)}             & \multicolumn{1}{l|}{2.67}                & \multicolumn{1}{l|}{13.22}     & 0.035              \\ \hline
\multicolumn{1}{l|}{$1.2$ (TO)}              & \multicolumn{1}{l|}{2.84}                & \multicolumn{1}{l|}{13.99}     & 0.192              & \multicolumn{1}{l|}{$1.18$ (TO)}             & \multicolumn{1}{l|}{2.69}                & \multicolumn{1}{l|}{12.18}     & 0.056              \\ \hline
\end{tabular}
\captionsetup{justification=raggedright}
\caption{\label{tab_mag}Numerical values for the physical parameters of magnetized, anisotropic ($\kappa = 0.5$) NSs from the DDMEX EOS.}
\end{table}

Some trends emerge. We see that the presence of anisotropic effects itself can enhance the masses of these objects from their isotropic values. However, we must note that in this treatment, the anisotropic and magnetic effects are not strictly independent of each other, as one of the reasons behind anisotropy could be the presence of the magnetic field itself. It is also seen that the presence of a magnetic field does not necessarily increase mass. TO fields increase the mass while RO fields decrease the mass. Thus, the  geometry of the field, not just its magnitude, matters. It seems that for addressing the mass gap agenda, it is the TO fields that play an important role.

Additionally, on comparing across profiles, one sees that the variation within the star also ends up playing a significant role. This is shown in Table \ref{tab_mag}. Although profile ``1'' gives us quite high maximum mass ($M_{max}$), even as high as $\simeq 2.8 M_\odot$ in the highly TO magnetized case, the $E_{mag}/E_{grav}$ ends up being high enough that it could destabilize the star. On the other hand, for the same central field value, profile ``2'' gives a slightly lower maximum mass, however with a much better value for $E_{mag}/E_{grav}$.

This difference is further established when we look at the tidal deformability of these cases, as shown for profile ``1'' in Fig. \ref{tidal_mag} (left) and profile ``2'' in Fig. \ref{tidal_mag} (right). One can see that the TO magnetized stars of profile ``1'', particularly the $\simeq 1.2 \times 10^{18} \ \rm{G}$ case mentioned earlier, violates both the limits of tidal deformability set out by GW170817. However, for the similar central values of field in profile ``2'', we see that even the strictest constraints of $\Lambda_{1.4}$ are comfortably met by profile ``2''.

\begin{figure}[!htb]
     \centering
         \centering
         \includegraphics[width=0.48\textwidth]{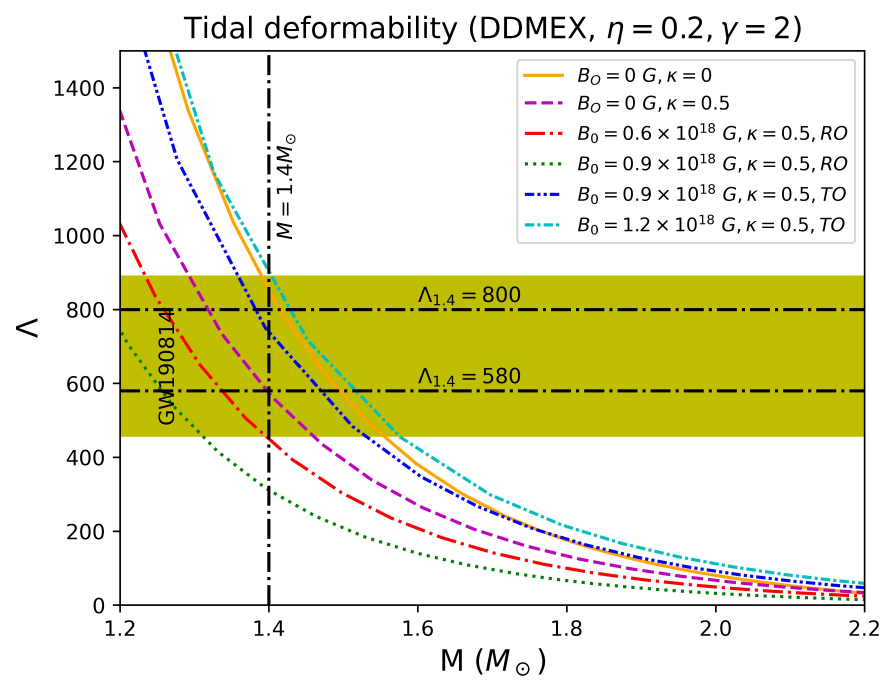}
         \centering
         \includegraphics[width=0.48\textwidth]{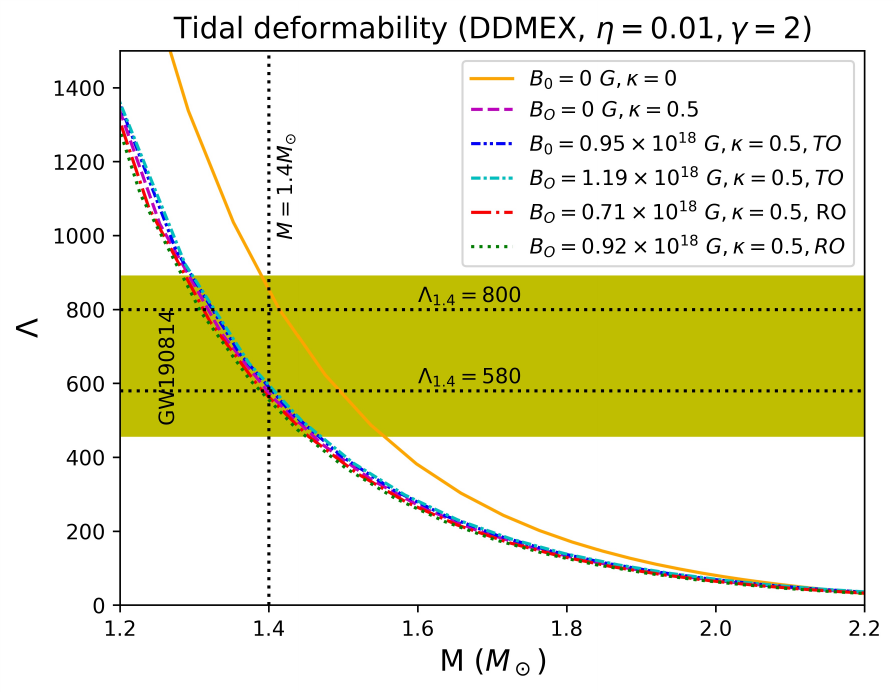}
	\caption{Tidal deformability for anisotropic, magnetized NSs with the DDMEX EOS for two different profiles, with $\kappa = 0.5$.}
	\label{tidal_mag}
\end{figure}

The presence of anisotropy also has interesting implications for the tidal deformability constraint. As shown in Fig. \ref{tidal_k}, higher degree of anisotropy ($\kappa$) leads to lower values of tidal deformability. Stiffer EOS, which might have been ruled out by purely isotropic analyses, may still be good candidates for the NS EOS if one considers this additional effect. This was also reported in other work, including \cite{biswas}.

\begin{figure}[!htpb]
	\centering
	\includegraphics[scale=0.3]{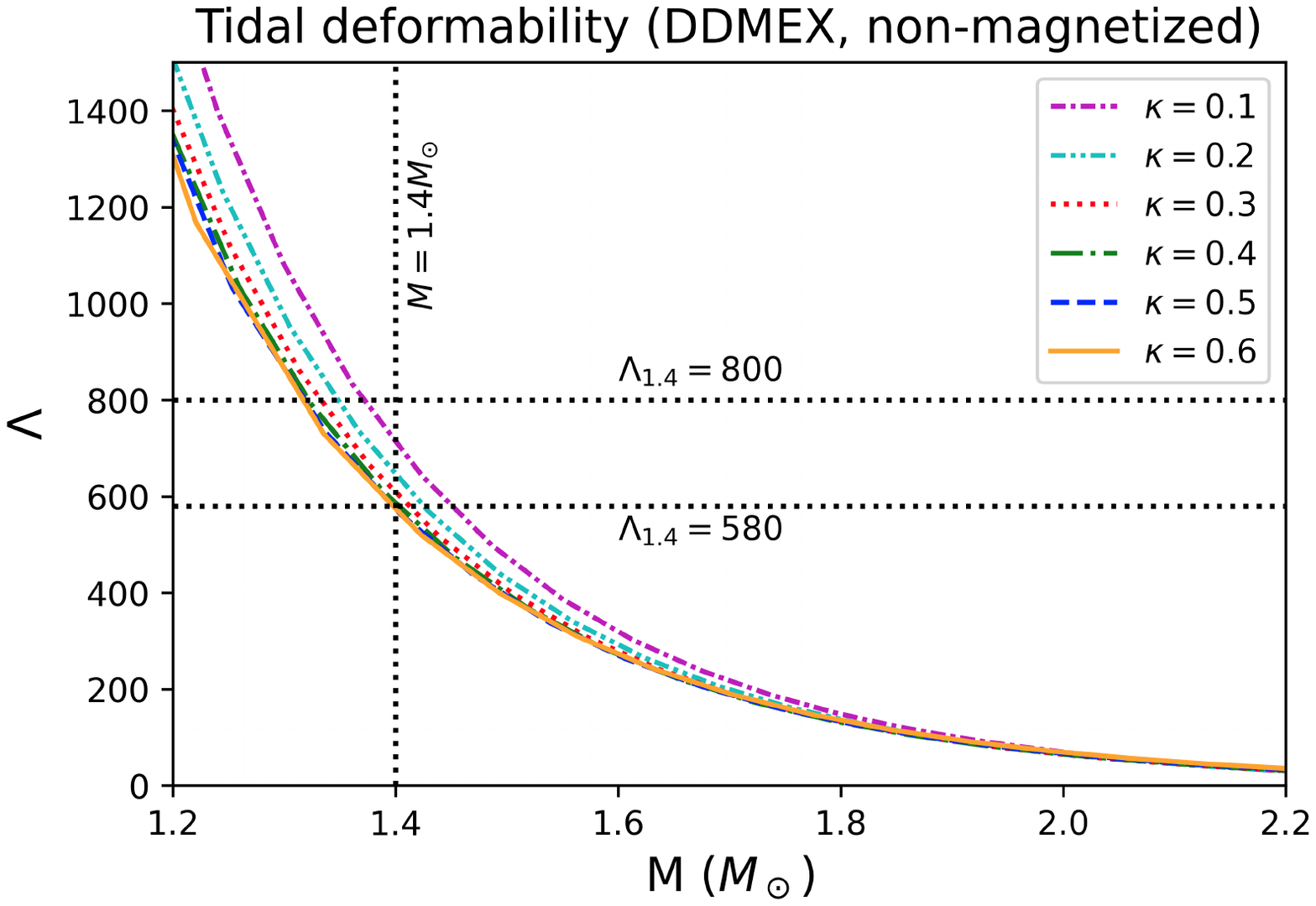}
	\caption{Change in $\Lambda$ with change of $\kappa$.}
	\label{tidal_k}
\end{figure}

\FloatBarrier

\subsection{Plausible detection through continuous gravitational waves}\label{ch4}

The detection of NSs is possible in many ways, e.g., by detecting their radio pulses (pulsars), measuring the X-ray counterpart of accreting NSs, and detecting GW from a binary system. However, the detection of isolated NSs remains a challenge due to their faintness and small size. Those isolated rotating NSs, if magnetized, can emit GWs for a very long time (many years) as long as they spin fast with the spin axis misaligned with the magnetic axis, i.e. non-zero obliquity angle. 
Such GWs are called CGWs because they are emitted over a much longer timescale compared to the transient waves radiated by binary systems. Here, we explore the tantalizing possibility of directly detecting these isolated NSs via CGWs, which would enable the measurement of the magnetic field and provide stronger constraints on the EOS.

The essential source of any system emitting GWs is non-zero, time varying quadrupole moment. If a NS consists of high magnetic field, the magnetic pressure can deform the star. The poloidal and toroidal fields deform the star to oblate and prolate spheroids, respectively. Similarly, rotation can also deform the star, resulting in the oblate spheroid. For non-zero obliquity angle, the system generates a non-zero time-varying quadrupole moment and hence becomes a source of CGWs. The emerging GW signal strength is

\begin{equation}
\begin{aligned}
h(t) &= \frac{4G}{c^4} \frac{\Omega^2(t) \epsilon I_{xx}}{d} f(\chi(t),\Omega(t), t) \\
\end{aligned}
\label{eq:gwstrain}
\end{equation}
where $c$ is the speed of light, $G$ is Newton's gravitational constant, $\chi$ is the obliquity angle, 
$\Omega$ is the angular frequency of the object, and $d$ is the distance between the detector and the source object. The ellipticity is defined as $\epsilon=|I_{zz}-I_{xx}|/I_{xx}$, where $I_{xx}$ and $I_{zz}$ are the principal moments of inertia of the star about the $x$- and $z$-axes, respectively. The ellipticity directly depends on the EOS and the magnetic field strength. The function $f$ has a (maximum) value of 0.0110297 for $t=0$, $d=10$ kpc, and $\chi=3^\circ$ (small angle approximation).
To understand the detection viability, we need to calculate the GW signal strength and compare it with the sensitivity of current and future GW detectors in similar frequency ranges. Measuring parameters such as the mass and radius is essential for calculating $h$.

In the previous sections, for ease of discussion, we neglected the rotation of NSs and focused on one-dimensional calculations. However, for the discussion of CGWs, rotation is indispensable. Therefore, for the present purpose, we perform several simulations solving the Einstein-Maxwell equations using the publicly available XNS code, without restricting the analysis to one dimension. Also unlike the previous sections, we consider here a polytropic EOS of $P=K\rho^{1.95}$ with $K$ being the polytropic constant. Based on the output of the simulation run, we calculate $h$. However, it is very important to note  that the GW signal strength is a function of time as $\Omega$ and $\chi$ are functions of time. $\Omega$ and $\chi$ can decrease with time due to angular momentum extraction from a NS due to electromagnetic and gravitational radiation \cite{mayu}. Here, we will consider a NS having a toroidally dominated internal magnetic field; hence, only gravitational radiation plays an important role in decay. The equations governing the decay are
\begin{equation}
\begin{split}
\frac{d(\Omega I_{z^{\prime}z^{\prime}})}{dt}=-\frac{2G}{5c^5} (I_{zz}-I_{xx})^2 \Omega^5 f_1(\chi)
\end{split},~~
\label{eq:dwdt}
\begin{split}
I_{z^{\prime}z^{\prime}}\frac{d\chi}{dt}=-\frac{12G}{5c^5} (I_{zz}-I_{xx})^2 \Omega^4 f_2(\chi),
\end{split}
\end{equation}
where $f_1(\chi)$ and $f_2(\chi)$ are some some complicated functions of $\chi$.
Due to the surface poloidal field of the NS, there will be an extra term due to electromagnetic radiation. However, due to the dominance of the central field, the surface field hardly would change the angular momentum extraction (the case would be different if the NS is centrally poloidally dominated).  
Note, throughout the decays of $\Omega$ and $\chi$, one may consider the magnetic field practically constant, as the field decay time scale is at least a thousand times larger than the other two. Hence, it is generally expected that the CGW strength emitted by the NSs just after birth will decrease with time and might become undetected by some detectors. Here, we show a case in Fig. \ref{fig:torwxgwdecaysen} where the NS can be detected initially, but after some years, it becomes undetectable even by Cosmic Explorer.  

\begin{figure}
\begin{center}
\includegraphics[scale=1.0]{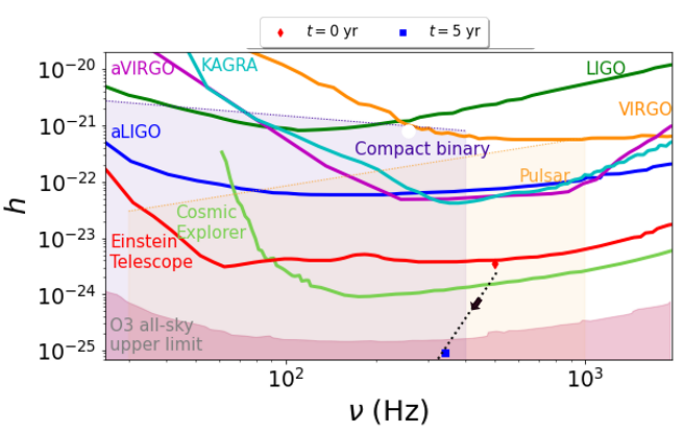}
\caption{Dimensionless GW amplitude for NSs before and after $\nu$ and $\chi$ decay along with the sensitivity curves of various detectors for $B_{max}=1.4\times 10^{17}$ G, initial $\nu=500$ Hz, $\chi=3^\circ$.}
\label{fig:torwxgwdecaysen}
\end{center}
\end{figure}

We may try to increase the detection possibility by calculating the integrated signal-to-noise ratio (SNR) for a year. See other work \cite{mayu} for details.
In some cases, such as in Fig. \ref{fig:snrtorrot}, it can be seen that the NS, which was not detectable by any detectors, might come into detection after some months of integrating the signal when it crosses above an SNR threshold limit of detection. Nevertheless, we have to keep in mind that the SNR carries the disadvantage of non-coherent GW search as the rotation rate and, hence, phase of GW changes with time, and hence, the SNR is some factor lower than that of coherent search. Also, in such a long time, the antenna pattern
may change with time, and there might be movement of the 
antenna which may change SNR, making the setup even more challenging.

\begin{figure}
\begin{center}
\includegraphics[scale=0.35]{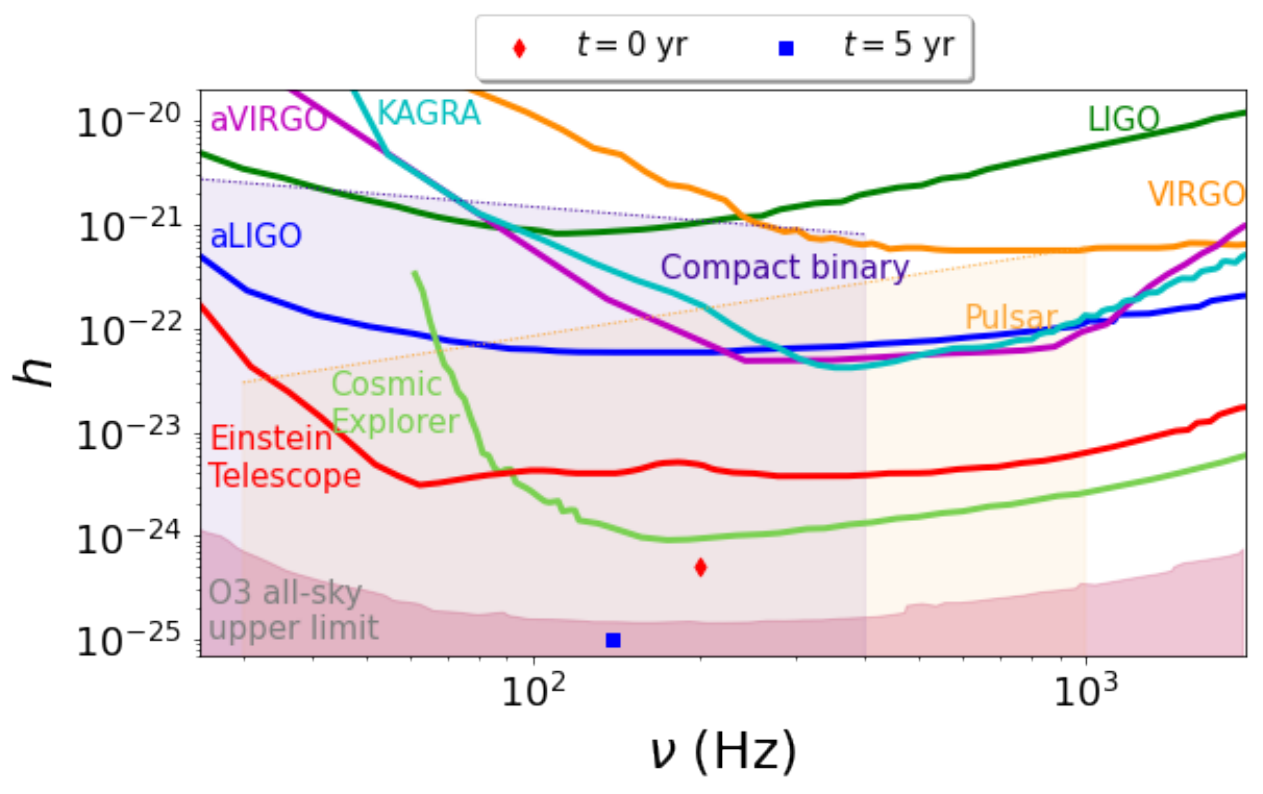}
\includegraphics[scale=0.45]{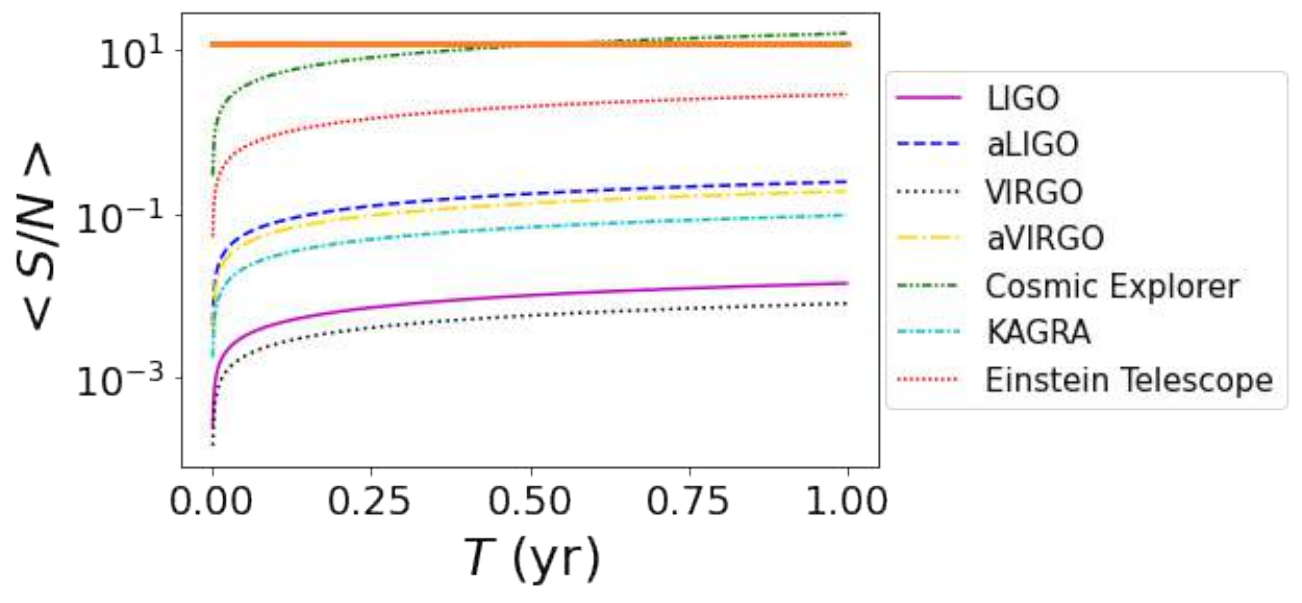}
\caption{GW amplitude for NSs before and after $\nu$ and $\chi$ decay along with the sensitivity curves of various detectors (Left) and SNR for various detectors as a function of integration time for a toroidal magnetic field dominated NS with $B_{max}=9.5\times 10^{16}$ G, initial $\chi=3^\circ$ and $\nu=200$ Hz (Right). The orange line corresponds to $SNR_{threshold}=12$.}
\label{fig:snrtorrot}
\end{center}
\end{figure}

The decay of GW strength sets a robust upper bound in the detection timescale. Although not a single isolated NS has been detected via CGW so far with LIGO-VIRGO-KAGRA detectors, with the knowledge of the timescales, future detections can be targeted carefully with Cosmic Explorer and Einstein Telescope. If the detection is successful, one can directly estimate the central magnetic field and the EOS property, as the $\epsilon$ is a function of the magnetic field and EOS type and is easily measurable once the GW is detected. The question remains, however, in their 'active timescale' (as long as some detectors detect them): what is the probability of detection? We can calculate the number of  NSs to be detected in a galactic volume in 5-year observational timescales by multiplying the birthrate of the NSs (0.002 \cite{ns_birthrate}) in galaxy per year with the observational timescale (5 yr in this case). This number comes out to be $0.01$, much smaller than unity. To detect a considerable number of NSs in the possible observational time, we may need to strengthen the sensitivity of detectors and, hence, we will be able to catch more distant sources (at least upto 500kpc). This will dramatically change the number of detectable NSs as it increases by $d^3$, making the number of NSs to be detected greater than 10 in observational timescale.

\subsection{Summary and outlook}\label{sec:Zuraiq_sum}
The observations of objects with masses in the lower mass gap range raises many interesting questions. The true nature of these mass gap objects can have many implications to our understanding of compact star physics. One of the most prominent examples of these objects is the secondary object in GW190814, which has been inferred to have a mass $2.5 - 2.67M_\odot$. As we report in the present article, we find that the masses of NSs from pure EOS are not high enough to be in this range, mainly because of the competition from hyperon softening. Hence, we are motivated to turn to the additional effects of mangetic fields and/or anisotropy.

In this section, we have investigated these objects by probing the possible maximum masses for a NS, under a combination of different effects including EOS, magnetic field and anisotropy/deformation. We initially consider the deformation effects to be sufficiently small so as to restrict ourselves to a spherically symmetric framework. On introducing magnetic fields along with a model anisotropy, we find that there is enhancement of the masses of NSs to mass gap values. However, this result is very much dependent on not only the magnitude of the magnetic field considered, but also the geometry/profile that it follows within the star itself. It seems that anisotropy has an essential part to play in all this as well. The anisotropy considered here, however, is not completely independent of the magnetic field, as we consider anisotropy to be partly due to the magnetic field effects, along with the matter effects.

Other authors tackled the question of enhancement of maximum masses through inclusion of alternate effects. Specifically for the secondary object in GW190814, the question of it being a quark/hybrid star has been investigated \cite{han_steiner}. Some other authors considered modifications to gravity \cite{mod_grav}, or to the coupling constant arising in the RMF theories \cite{lopes_GW190814, Sedrakian:2020PRD}. Another important effect that can significantly effect the eventual masses of NSs is rotation \cite{dex_prc, li}.

When considering rotation, an interesting implication is the possibility of CGW emission. These waves arise due to the non-zero, time-varying quadrupole moment generated by a deformed, rotating, magnetized massive NS.
To investigate these CGWs, simulations were set up using the XNS code with polytropic EOS. However, the plausible detection of the CGWs is complicated by the fact that there are decays of $\Omega$ and $\chi$ on the timescale at most of a few years. This decay leads to the decrease in GW amplitude, and leads to a strict upper bound on detection timescales for these objects. 
We must consider advanced, highly sensitive future detectors, such as the Cosmic Explorer and the Einstein Telescope, which will be capable of probing NSs at greater distances. Additionally, integrating the signal can help increase the SNR, thereby enhancing the possibility of detection. Therefore, the study of massive NSs, facilitated by mass gap observations, holds significant implications for both the theoretical understanding of compact objects and their potential observational evidence.

\section{Strong magnetic fields and the inner crust of neutron stars}
\label{sec:pais}

\subsection{Introduction}

Neutron stars (NS) are very dense and compact objects, with a typical radius of about $10-15$ km and a mass of about 1.4$M_\odot$ \cite{Glendenning}. They are constituted by cold catalysed stellar matter, their composition is strongly asymmetric, very neutron-rich, and their central densities can reach several nuclear saturation densities, $n_{sat}\sim 0.15$ fm$^{-3}$. They are structured in layers, with a surface constituted by an $^{56}$Fe grid, where the pressure is zero. In the outer crust, we have very neutron-rich nuclei embedded in an electron sea. As the density further increases towards the interior of the star, neutrons start dripping out of the nuclei. This defines the transition to the inner crust. In this region of the star, the nucleons can form heavy clusters with different geometrical structures, termed ''pasta phases`` \cite{ravenhall83,hashimoto84,horowitz05,watanabe05,maruyama05,avancini08,avancini10,pais12PRL,bao14,Newton22}, due to the competition between the Coulomb and the strong forces. Eventually, these structures will melt, and the core of the star starts. In there, purely homogeneous neutral matter exists, and in the very center, hyperons or even deconfined quark matter may occur.

These heavy clusters may form not only in the inner crust of neutron stars, but also in other astrophysical systems, like proto-neutron stars or neutron star mergers, that are under different thermodynamical conditions. In these kind of systems, where the temperatures can reach several MeV \cite{oertel17}, other clusters, like deuterons or $\alpha-$particles, can also form. These light clusters have also been observed in heavy-ion collisions at GANIL \cite{bougault,pais2020} or NIMROD \cite{qin}, in a range of temperatures from $\sim 5$ to $\sim 10$ MeV. 
 
The appearance of these clusters will modify the neutrino transport, and, therefore, consequences on the dynamical evolution of supernovae and on the cooling of proto-neutron stars are expected \cite{arcones}. Magnetars, in particular, may have an inner crust even more complex than non-magnetized stars \cite{fang16,fang17,fang17a,Ferreira21} due to the presence of these clusters, as we will see in the course of this paper. In Pons et al \cite{pons2013}, it was shown that a fast decay of the magnetic field could be the reason for the absence of stars with periods of rotation higher than 12 s, and this may be related to the existence of an amorphous inner crust, i.e pasta phases. Regarding the macroscopic structure of the star, the explicit inclusion of clusters in the inner crust affects the radius of intermediate mass stars: for 1.4$M_\odot$, calibrated relativistic mean-field (RMF) models predicted a radius of $13.6\pm0.3$km with a crust thickness of $\Delta R=1.36\pm 0.06$km \cite{paisVlasov}.

In order to obtain the structure of the star, we need the equation of state (EoS) to solve the TOV equations. In the available literature, one can find plenty of models, like the phenomenological ones, whose parameters are fitted to finite nuclei, and are constrained by different observables, like \textit{ab initio} calculations, experiments or observations. The relativistic mean-field models and the non-relativistic Skyrme are examples.
A collection of several of these models is gathered in the public, free online repository CompOSE \cite{compose,ComposePaper}. One example of the several observational constraints available  is the simultaneous measurement of the NS mass and radius by the NICER telescope. Their most recent measurements have put the radius and mass of the pulsar PSR J0740+6620 with 12.49$^{+1.28}_{-0.88}$ km and 2.073$^{+0.069}_{-0.069}$ M$_\odot$ \cite{Salmi24}, and the brightest rotation-powered millisecond pulsar PSR J0437-4715 with $M=1.418\pm 0.037$ M$_\odot$ and $R=11.36^{+0.95}_{-0.63}$km \cite{Choudhury24}.

Magnetars \cite{duncan,thompson,usov,paczynski}, mainly Soft Gamma Repeaters (SGRs) and Anomalous X-ray Pulsars (AXPs), belong
to a kind of neutron stars with very strong magnetic fields at the
surface, up to $10^{13} \sim 10^{15}$ G \cite{olausen,mcgill}, and quite long spin periods,
of the order of $2 \sim 20$ s. Nowadays, about thirty of
such objects have been observed \cite{mcgill}.  Moreover, magnetars are
good candidates to be a source of continuous gravitational wave emission, and we expect that in the future it will be possible to detect this type of gravitational waves.

In this section, we will briefly introduce the formalism used to calculate the inner crust EoS under the effect of a strong external magnetic field $B$ within a RMF framework, then we will show some of the results obtained, and finally some conclusions will be drawn.

\subsection{Theoretical Framework}

Here, we describe NS matter within the non-linear Walecka model, where the mesons are responsible for mediating the nuclear force. We include the isoscalar-scalar meson $\sigma$, the isoscalar-vector meson $\omega$ and the isovector-vector meson $\rho$. Electrons are also included to achieve electrical neutrality. An external electromagnetic field, oriented along the $z-$axis, is considered, $A^\mu=(0,0,Bx,0)$.
Throughout the results, we define $B^*$ as $B^* = B/B^c_e$, with $B^c_e= 4.414 \times 10^{13}$ G the critical field at which the electron cyclotron energy is equal to the electron mass.

The Lagrangian density of our system is given by
\begin{equation}
    \mathcal{L}=\sum_{i=p,n}\mathcal{L}_i+\mathcal{L}_e+\mathcal{L}_\sigma+\mathcal{L}_\omega+\mathcal{L}_\rho+\mathcal{L}_{\omega\rho}+\mathcal{L}_A \, .
\end{equation}
$\mathcal{L}_e$ and $\mathcal{L}_A$ are the electron Lagrangian density and electromagnetic term, given by
\begin{equation}
    \mathcal{L}_e=\bar{\psi}_e\big[\gamma_\mu\big(i\partial^\mu+eA^\mu\big)-m_e\big]\psi_e,
\end{equation}
\begin{equation}
    \mathcal{L}_A=-\frac{1}{4}F_{\mu\nu} F^{\mu\nu} \, ,
\end{equation}
with $F_{\mu\nu}=\partial_\mu A_\nu-\partial_\nu A_\mu$ .
The nucleon Lagrangian density is given by 
\begin{equation}
    \mathcal{L}_i=\bar{\psi}_i\big[\gamma_\mu iD^\mu-M^*-\frac{1}{2}\mu_Nk_b\sigma_{\mu\nu}F^{\mu\nu}\big]\psi_i \, ,
\end{equation}
with
\begin{equation}
    M^*=M-g_\sigma\phi \, ,
\end{equation}
the nucleon effective mass and
\begin{equation}
    iD^\mu=i\partial^\mu-g_\omega V^\mu-\frac{g_\rho}{2}\mathbf{\tau}\cdot\mathbf{b}^\mu-e\frac{1+\tau_3}{2}eA^\mu \, ,
\end{equation}

The Lagrangian density for the meson fields are given by
\begin{equation}
    \mathcal{L}_\sigma=\frac{1}{2}\bigg(\partial_\mu\phi\partial^\mu\phi-m_\sigma^2\phi^2 -\frac{1}{3}\kappa\phi^3-\frac{1}{12}\lambda\phi^4  \bigg) \, ,
\end{equation}
\begin{equation}
    \mathcal{L}_\omega=-\frac{1}{4}\Omega_{\mu\nu}\Omega_{\mu\nu}+\frac{1}{2}m_\omega^2V_\mu V^\mu +\frac{\xi}{4!}\xi g_\omega^4(V_\mu V^\mu)^2 \, ,
\end{equation}
\begin{equation}
    \mathcal{L}_\rho=-\frac{1}{4}\mathbf{B}_{\mu\nu}\cdot\mathbf{B}^{\mu\nu}+\frac{1}{2}m_\rho^2\mathbf{b}_\mu\cdot \mathbf{b}^\mu \, ,
\end{equation}
with the tensors written as
\begin{eqnarray}
\Omega_{\mu\nu}&=&\partial_\mu V_\nu - \partial_\nu V_\mu \, , \\
\mathbf{B}_{\mu\nu}&=&\partial_\mu \mathbf{b}_\nu - \partial_\nu \mathbf{b}_\mu - g_\rho \left(\mathbf{b}_\mu \times \mathbf{b}_\nu \right)\, .
\end{eqnarray}

The parameters include the couplings of the mesons to the nucleons, $g_\sigma$, $g_\omega$, $g_\rho$, the nucleon and electron masses $M$ and $m_e$, respectively, and the higher-order coupling constants $k$, $\lambda$, and $\xi$. The electromagnetic coupling constant is given by $e=\sqrt{4\pi/137}$, and $\tau_3=\pm 1$ is the third component of the Pauli matrices for protons ($+1$) and neutrons ($-1$). We also introduce in the model the anomalous magnetic moment (AMM) of the nucleons with $\sigma_{\mu\nu}=\frac{i}{2}[\gamma_\mu,\gamma_\nu]$ and strength $k_b$, with $k_n=-1.91315$ for the neutron and $k_p=1.79285$ for the proton. $\mu_N$ is the nuclear magneton. We neglect the AMM contribution for the electrons as it is negligible \cite{duncan}. 

\subsubsection{Equation of state}

Here, we consider two EoS models, NL3~\cite{Lalazissis1997}, and~NL3$\omega\rho$~\cite{Horowitz2001a,Horowitz2001b,paisVlasov}, that share the same isoscalar properties. In the NL3$\omega\rho$ model, an interaction term between the $\omega$ and the $\rho$ meson, $\mathcal{L}_{\omega\rho}$, is added to model the density dependence of the symmetry energy, since NL3 has a very large slope of the symmetry energy at saturation, $L=118$ MeV. NL3$\omega\rho$, on the other hand, has $L=55$ MeV.  Some of the symmetric nuclear matter properties at saturation density for these models are shown in Table~\ref{tab1}.
\begin{table*}
    \begin{tabular}{ccccccc}
    \hline
    \hline
          & $E/A$ (MeV) & $\rho_0$ (fm$^{-3}$) & $M^*/M$ & $K$ (MeV) & $\mathcal{E}_{sym}$ (MeV) & $L$ (MeV)  \\
    \hline
        NL3 & 16.24 & 0.148 & 0.60 & 270 & 37.34 & 118 \\
        NL3$\omega\rho$ & 16.24 & 0.148 & 0.60 & 270 & 31.66 & 55\\
    \hline
    \hline
    \end{tabular}
    \caption{Symmetric nuclear matter properties at saturation density for the NL3 and NL3$\omega\rho$ models: binding energy $E/A$, saturation density $\rho_0$, normalized nucleon effective mass, $M^*/M$, incompressibility $K$, symmetry energy $\mathcal{E}_{sym}$, and its slope $L$.}
    \label{tab1}
\end{table*} 
This extra term is given by
\begin{align}
    \mathcal{L}_{\omega\rho}=&\Lambda_{\omega\rho}g_\omega^2g_\rho^2V_\mu V^\mu\mathbf{b}_\mu\cdot \mathbf{b}^\mu \, . \notag
\end{align}

The fields equations follow from the Euler-Lagrange equations, and in the mean-field approximation, these fields are given by their constant expectation values, $\phi_0$, $V_0$, and $b_0$. In the following, for simplicity, we omit AMM in the equations. The interested reader can consult e.g. Ref.~\cite{wang22} for the equations with that term.
The scalar and vector densities for nucleons, and the electron density, are given by

\begin{eqnarray}
    \rho_{s,p}&=&\frac{q_pBM^*}{2\pi^2}\sum_{\nu=0}^{\nu_{\rm max}^p}g_s \ln\bigg|\frac{k^p_{F,\nu}+E^p_F}{\sqrt{M^{*2}+2\nu q_pB}}\bigg| \, , \label{eq:rhosp}
\end{eqnarray}
   
 \begin{eqnarray}
    \rho_{s,n}&=&\frac{M^*}{2\pi^2}\bigg[E^n_Fk^n_{F}-M^{*2}\ln\bigg|\frac{k^n_{F}+E_F^n}{M^*}\bigg|
    \bigg] \, , 
 \end{eqnarray}    

\begin{eqnarray}
    \rho_p&=&\frac{q_pB}{2\pi^2}\sum_{\nu=0}^{\nu_{\rm max}^p} g_s k^p_{F,\nu} \, , \label{eq:rhop} 
\end{eqnarray}

\begin{eqnarray}
    \rho_n&=&\frac{{k^n_{F}}^3}{3\pi^2} \, ,
 \end{eqnarray}

\begin{eqnarray}
    \rho_e&=&\frac{|q_e|B}{2\pi^2}\sum_{\nu=0}^{\nu_{\rm max}^e} g_s k^e_{F,\nu} \,  , \label{eq:rhoe}
\end{eqnarray}
where $\nu=n+\frac{1}{2}-\frac{1}{2}\frac{q}{|q|}s=0,1,\cdots,\nu_{\rm max}$ are the Landau levels (LL) for fermions with electric charge $q$, $q_e=-e$ for electrons and $q_p=e$ for protons. $s$ is the spin quantum number, $+1$ for spin up cases and $-1$ for spin down cases. The spin degeneracy factor of the Landau levels, $g_s$, is equal to $g_s=1$ for $\nu=0$ and $g_s=2$ for $\nu>0$, and $\nu_{\rm max}$ is the maximum number of LL, for which the square of the Fermi momentum of the particle is still positive, given by
\begin{align}
    \nu_{\rm max}^e=\frac{E^{e2}_F-m_e^{2}}{2 |q_e| B} \, , \label{eq:numaxe} \\
    \nu_{\rm max}^p=\frac{E^{p2}_F-M^{*2}}{2 q_p B} \, .  \label{eq:numaxp}
\end{align}
$k_{F,\nu}^q$ and $E_F^q$ are the Fermi momenta and energies of the particles, defined as 
\begin{align}
    k^{p}_{F,\nu}=&\sqrt{E^{p2}_F-M^{*2} - 2\nu q_p B} \, , \\
    k^{n}_F=&\sqrt{E^{n2}_F-M^{*2}} \, , \\
    k^{e}_{F,\nu}=&\sqrt{E^{e2}_F-m_e^{2}- 2\nu |q_e| B} \,.
\end{align}

The bulk energy density is given by:
\begin{equation}
    \mathcal{E}=\mathcal{E}_f+\mathcal{E}_p+\mathcal{E}_n \, ,
\end{equation}
where 
\begin{align} 
    \mathcal{E}_f=&\frac{m_\omega^2}{2}V_0^2 +\frac{\xi g_v^4}{8}V_0^4  + \frac{m_\rho^2}{2}b_0^2+\frac{m_\sigma^2}{2}\phi_0^2+\frac{\kappa}{6}\phi_0^3 \notag\\
    &+\frac{\lambda}{24}\phi_0^4+ 3 \Lambda_{\omega\rho}g_\rho^2 g_\omega^2 V_0^2 b_0^2 \, , 
\end{align}
\begin{align}
    \mathcal{E}_n=&\frac{1}{4\pi^2}\left[k_{F}^nE_F^{n3}-\frac{1}{2}M^*\bigg(M^*k_{F}^nE_F^{n} \right.\notag \\
    &+M^{*3} \ln\bigg|\frac{k_{F}^n+E_F^{n}}{M^*}\bigg|\bigg)\bigg] \, , 
\end{align}    

\begin{align}
    \mathcal{E}_p=&\frac{q_pB}{4\pi^2}\sum_{\nu=0}^{\nu_{\rm max}}g_s\bigg[k_{F,\nu}^pE_F^{p}+\bigg(M^{*2}+2\nu q_pB \bigg) \notag \\
    &\cdot \ln\bigg|\frac{k_{F,\nu}^p+E_F^{p}}{\sqrt{M^{*2}+2\nu q_pB}}\bigg|\bigg] \, . \label{eq:enp}
\end{align}

The chemical potentials for protons, neutrons, and electrons are given by 
\begin{align}
    \mu_p=&E^p_F+g_\omega V_0+\frac{1}{2}g_\rho b_0 \, , \\
    \mu_n=&E^n_F+g_\omega V_0-\frac{1}{2}g_\rho b_0 \, , \\
    \mu_e=&E_F^e=\sqrt{k^{e2}_{F,\nu}+m_e^2+2\nu|q_e|B} \, .
\end{align}
and the  pressure is
\begin{equation}
   P=\mu_p\rho_p+\mu_n\rho_n-\mathcal{E}.
\end{equation}

For neutron star matter, the $\beta-$equilibrium and charge-neutral conditions are imposed:
\begin{eqnarray}
   \mu_n&=&\mu_p+\mu_e \, \\
    \rho_p&=&\rho_e \, .
\end{eqnarray}

\subsubsection{Pasta structures in the CP and CLD approximations}

Here, we consider the coexistence-phase (CP) \cite{avancini12,wang22} and the  compressible liquid drop (CLD) \cite{pais15,scurto23} models to calculate the inner crust structures in $\beta-$equilibrium magnetized matter. 

In the CP approximation, separated regions of high (heavy clusters) and low (background nucleon gas) densities are considered. Gibbs equilibrium conditions are imposed, together with the charge-neutrality condition:

\begin{eqnarray}
\mu_p^I&=&\mu_p^{II} \, , \nonumber \\ 
\mu_n^I&=&\mu_n^{II} \, , \nonumber \\ 
P^I&=&P^{II} \, , \nonumber \\ 
\rho_e&=&\rho_p=f\rho_p^I+(1-f)\rho_p^{II} \, .
\end{eqnarray}
Here, $I$ labels the cluster phase and $II$ the gas phase. In this approximation, the finite size effects are taken into account by a surface and a Coulomb terms in the energy density, that is only added after the coexistence phases are achieved.

In the CLD model, the total energy density is minimized, including the surface and Coulomb terms. The equilibrium conditions become
\begin{align}
    &\mu_n^I=\mu_n^{II} \, , \\
    &\mu_p^I=\mu_p^{II}-\frac{\mathcal{E}_{surf}}{(1-f)f(\rho^I_p-\rho^{II}_p)} \, , \\
    &P^I=P^{II}+\mathcal{E}_{surf}\bigg[\frac{3}{2\alpha}\frac{\partial\alpha}{\partial f}+\frac{1}{2\Phi}\frac{\partial\Phi}{\partial f}-\frac{((1-f)\rho_p^I+f\rho_p^{II})}{(1-f)f(\rho^I_p-\rho^{II}_p)}\bigg] \, .
\end{align}

The total energy density of the system is given by

\begin{equation}
\mathcal{E}=f\mathcal{E}^I+(1-f)\mathcal{E}^{II}+\mathcal{E}_{Coul}+\mathcal{E}_{surf}+\mathcal{E}_e \, ,
\label{En_dens}
\end{equation}
where $f$ is the fraction of volume occupied by the dense phase.
The surface and Coulomb terms are given by:
\begin{eqnarray}
\mathcal{E}_{Coul}&=&2\alpha e^2\pi\Phi R_d^2 \left(\rho_p^I-\rho_p^{II}\right)^2 \, , \label{eq:ecoul} \\
\mathcal{E}_{surf}&=&\frac{\sigma\alpha D}{R_d} \label{eq:esurf}
\end{eqnarray}
where $\alpha=f$ for droplets, rods and slabs and $\alpha=1-f$ for tubes and bubbles, and $R_d$ is the size of the cluster. $\Phi$ is given by
\begin{eqnarray}
\Phi&=&\left(\frac{2-D\alpha^{1-2/D}}{D-2}+\alpha\right)\frac{1}{D+2} \, , \qquad D=1,3 \, ,  \nonumber \\
\Phi&=&\frac{\alpha-1-\ln\alpha}{D+2} \, , \qquad  D=2 \, .
\end{eqnarray}
The surface tension  parameter $\sigma$ was obtained from a fit to a relativistic Thomas-Fermi calculation \cite{avancini12}. 
The following relation is obtained, when minimizing the surface and Coulomb terms with respect to $R_d$
\begin{eqnarray}
\mathcal{E}_{surf}&=&2\mathcal{E}_{Coul} \, ,   \\
R_d&=&\left[\frac{\sigma D}{4\pi e^2\Phi\left(\rho_p^I-\rho_p^{II}\right)^2 }\right]^{1/3} \, .
\end{eqnarray}

\begin{figure}
		\begin{tabular}{c}
			\includegraphics[width=0.5\textwidth]{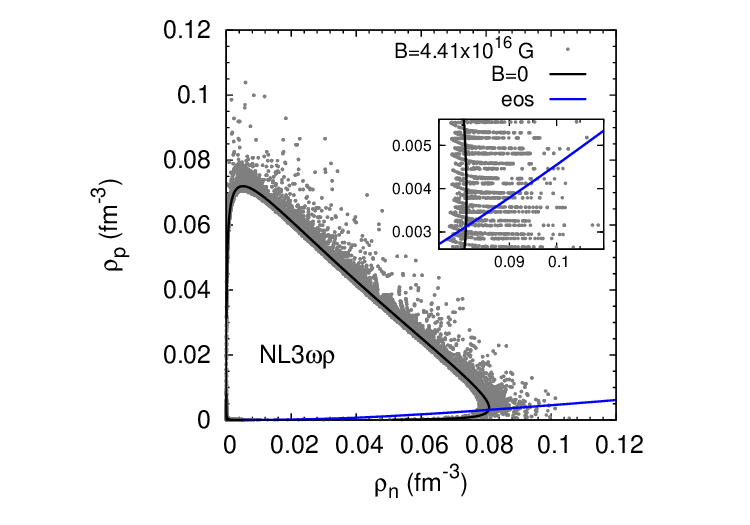}
			\includegraphics[width=0.5\textwidth]{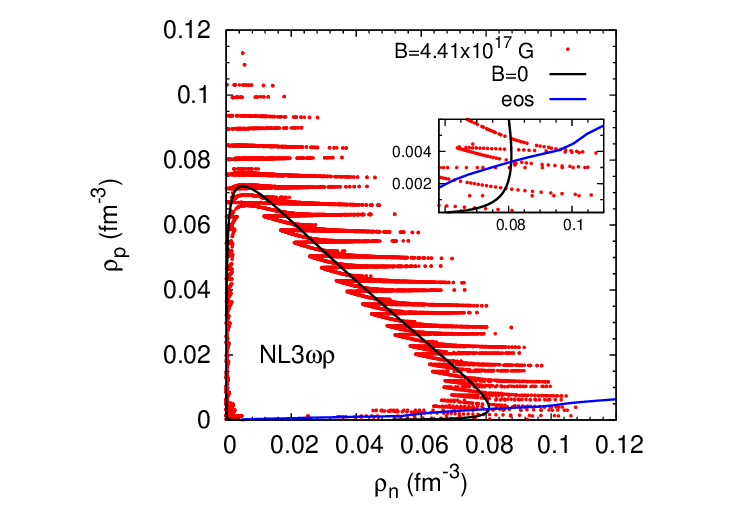}
		\end{tabular}	
	\caption{Dynamical spinodal regions for NL3$\omega\rho$, for a momentum transfer of $k = 75$ MeV, with anomolous magnetic moment  for $B = 4.41\times 10^{16}$ G (left), and $B= 4.41\times 10^{17}$G (right). A comparison with the $B = 0$ (black lines) results is also made. The EoS for $\beta-$equilibrium matter is also shown (blue). Figure adapted from Ref.~\cite{fang17}.} 
	\label{Pais_fig1}
 \end{figure}

\subsection{Results and discussion}

Here, we start by addressing the results of the estimation of the crust-core transition density from a dynamical spinodal calculation within the Vlasov formalism \cite{nielsen91,providencia06,paisVlasov} for magnetized nuclear matter. For $\beta-$equilibirum and zero temperature, i.e. NS conditions, this calculation is in very good agreement with more sophisticated calculations, like Thomas-Fermi \cite{avancini10}. In this calculation, the instability region is determined from the collective modes of nuclear matter that correspond to small oscillations around equilibrium. Only longitudinal modes are considered, and the boundary of this region is defined by the frequency of these modes to be zero. Inside this (unstable) region, the mode with the largest frequency drives the system to the formation of the instabilities. To calculate the crust-core transition, we cross the EoS with these spinodal surfaces, in the ($\rho_p,\rho_n$) space. In Ref.~\cite{paisVlasov}, it was seen that the larger the slope of the symmetry energy, the smaller the spinodal regions. This was then reflected in an anti-correlation between $L$ and the crust-core transition density: the larger the $L$, the smaller the density \cite{paisVlasov}.

In Fig.~\ref{Pais_fig1}, we show the dynamical spinodal regions for the NL3$\omega\rho$ model, for two different values of the magnetic field, and we also compare with the $B=0$ result. We observe that the magnetic field is giving rise to alternate bands of homogeneous and non-homogeneous matter, that appear due to the Landau levels. The stronger the magnetic field is, the greater the size of the spinodal region. Also, with the increase of the magnetic field, the number of the bands becomes smaller and the bands become wider. This happens because as the $B-$field increases, there is a decrease in the number of Landau levels. We also observe that the crust-core transition extends to a larger range of densities, as opposed to what happens at $B=0$.

\begin{figure}[htbp]
		\begin{tabular}{c}
			\includegraphics[width=0.7\textwidth]{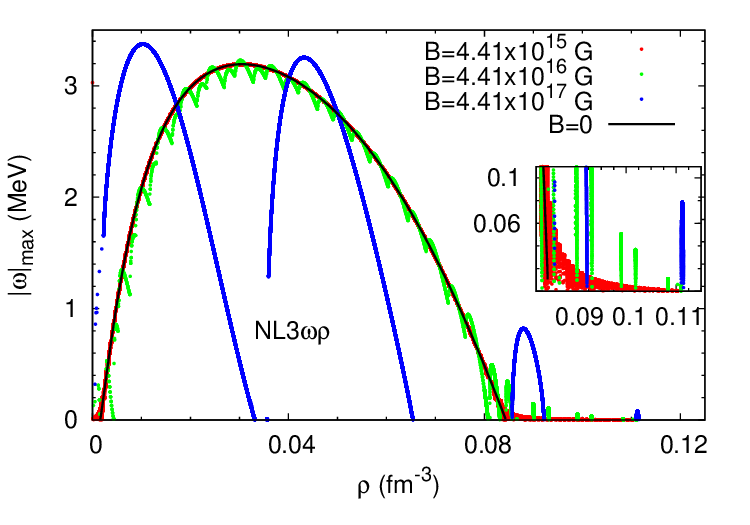}
		\end{tabular}
	\caption{Largest growth rate versus density for NL3$\omega\rho$, with a proton fraction of yp = 0.035, with anomalous magnetic moment. Three different values of B are considered: $B=4.41\times 10^{15}$G (red), $B = 4.41\times 10^{16}$ (green), and $B = 4.41\times 10^{17}$G (blue). A comparison with the $B = 0$ (black lines) results is also made. Figure adapted from Ref.~\cite{fang17}.} 
	\label{Pais_fig2}
 \end{figure}

In Fig.~\ref{Pais_fig2}, we plot the largest growth rate as a function of the density for NL3$\omega\rho$, and three different values of the magnetic field. We observe the appearance of oscillations around the $B=0$ results, below the $B=0$ crust-core transition density, that is given when the mode goes to zero, and above this value, we get this alternate regions of clusterized and non-clusterized matter, as already observed in Fig.~\ref{Pais_fig1}. Since the magnetic field is giving rise to larger crust-core transition densities, the correspondent pressures also become larger, and this has a direct influence in the fractional moment of inertia of the crust (see Table I of \cite{fang17}), that also becomes larger.

 \begin{figure}[htbp]
			\includegraphics[width=\textwidth]{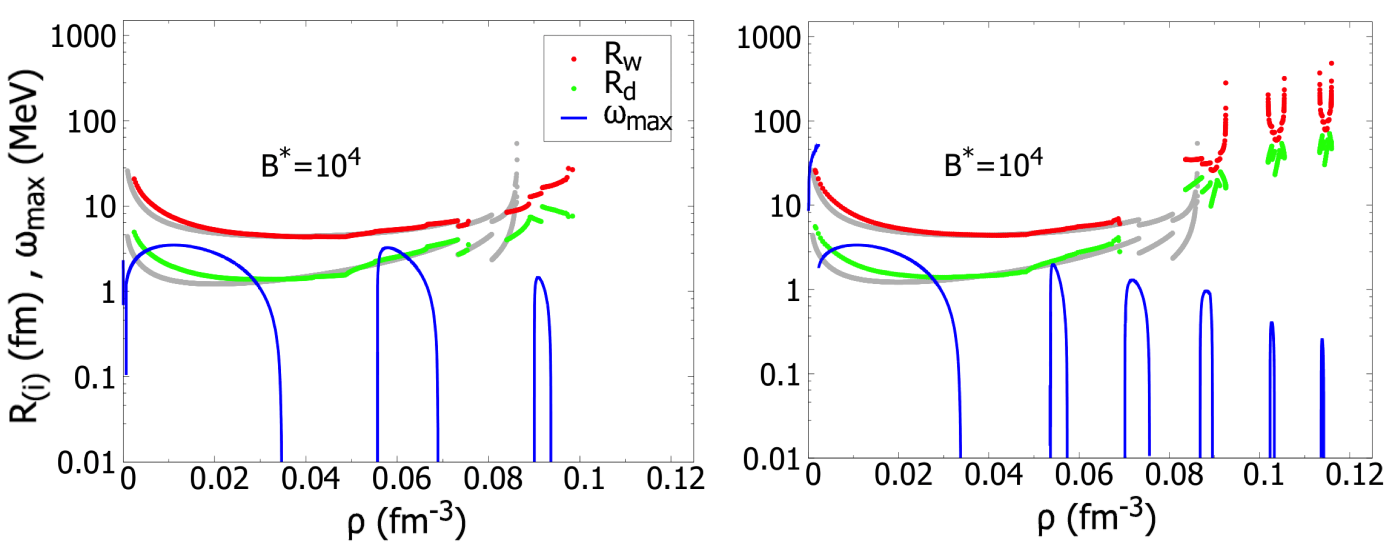}
	\caption{Radii of the WS cell (red) and nucleus (green) for $\beta-$equilibrium matter using the NL3$\omega\rho$ parametrization without (left) and with (right) the inclusion of AMM for $B=4.41\times 10^{17}$G. The no-field case is also shown with gray points as a reference. Growth rates obtained with a dynamical spinodal calculation in \cite{fang17} are plotted with blue lines. Figure adapted from Ref.~\cite{wang22}.} 
	\label{fig3}
 \end{figure}   

In the following, we show the results obtained from a CP and CLD calculations for the NS magnetized inner crust. We start with the CP calculation.

In Fig.~\ref{fig3}, we show the radii of the Wigner-Seitz (WS) cell (red) and of the nucleus (green) as a function of the density for the same model as above and considering results with (bottom) and without (top) AMM. We consider the highest $B-$field strength in our calculations to be $B^*=10^4$, or $B=4.41\times 10^{17}$G. As for comparison, the blue lines represent the maximum growth rates as shown in Fig.~\ref{Pais_fig2}. These results seem to be in agreement with what we previously found with a dynamical spinodal calculation: several disconnected regions of non-homogeneous matter appear above the $B=0$ region. If AMM is considered, these regions are more numerous (the double) and narrower, because the spin polarisation degeneracy is removed.
 
\begin{figure}[htbp]
			\includegraphics[width=\textwidth]{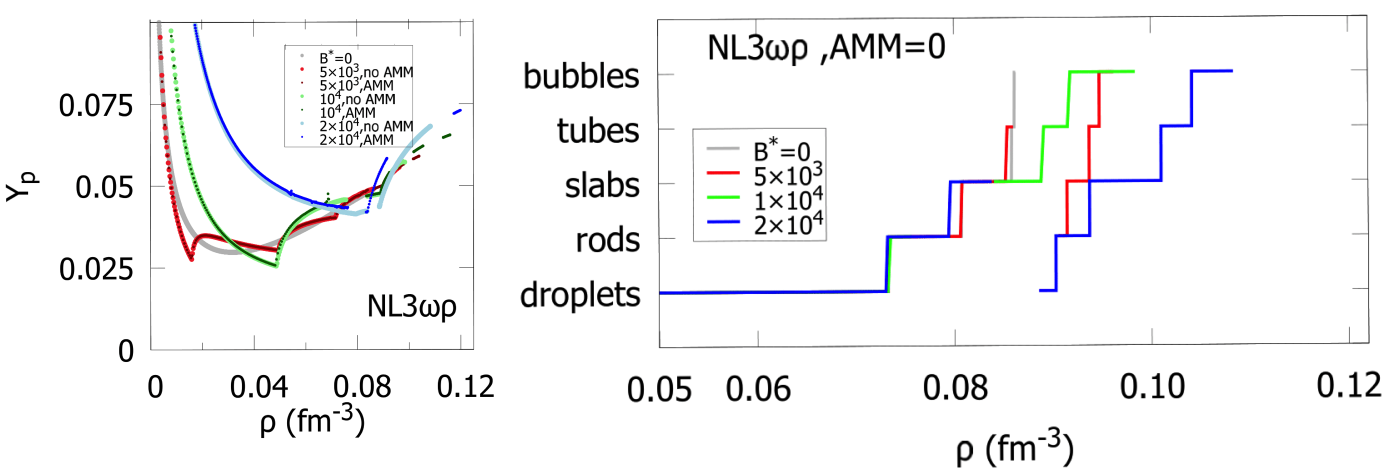}
	\caption{ The proton fraction (left) and the evolution of the pasta phases (right) as a function of the baryonic density for $\beta-$equilibrium matter using the NL3$\omega\rho$ model, considering different magnetic field strengths. The results consider calculations with (dark colors) and without (light colors) AMM. In the case of the shapes, only the results without AMM are shown. Figure adapted from Ref.~\cite{wang22}.} 
	\label{fig4}
 \end{figure}

In Fig.~\ref{fig4}, we show the proton fraction and the evolution of the shapes as a function of the density. We observe that the larger the magnetic field, the larger the proton fraction, showing fluctuations due to the opening of new Landau levels. For NL3$\omega\rho$ model, these disconnected pasta regions that appear above the main $B=0$ region contain all types of geometric configurations in their narrow density range. For the NL3 model, that only shows the the droplet configuration in the $B=0$ case, the finite magnetic field induces the appearance of all geometric structures in the first region (see Fig.~3 of Ref.~\cite{wang22}), as well as in the narrow disconnected regions.

\begin{figure}[htbp]
		\begin{tabular}{c}
			\includegraphics[width=0.5\textwidth]{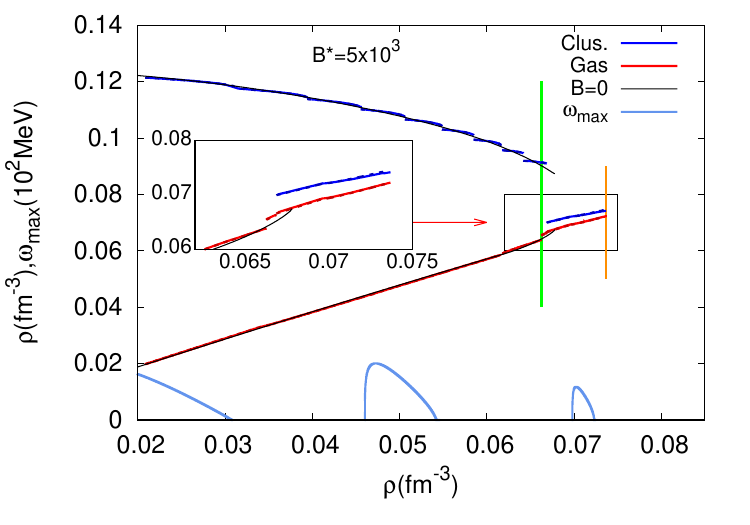}
			\includegraphics[width=0.5\textwidth]{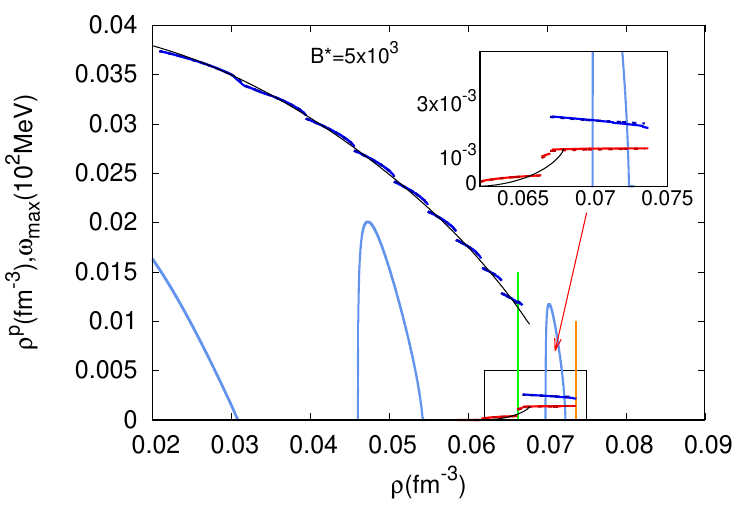}
		\end{tabular}
	\caption{Baryonic (left) and proton (right) densities of liquid (1, blue) and gas (2, red) phases as function of the total baryon density for the NL3 model in a CP (dashed lines) and CLD (solid line) calculations, with $B^*= 5 \times 10^3$. We also plot the magnetized growth rates divided by a factor $10^2$, $|\omega_{max}|$ (light blue), as well as the densities in the $B = 0$ case (black). The green and orange segments indicate, respectively, $\rho_{1\rightarrow 2}$ and $\rho_{cc}$, both defined in the text.  Figure adapted from Ref.~\cite{scurto23}.} 
	\label{fig5}
 \end{figure}

Now we focus on the CLD results. In Fig.~\ref{fig5}, we show the baryonic and proton densities in the gas and liquid phases as a function of the density. Also shown are the maximum growth rates in a dynamical spinodal calculation (see Fig.~\ref{Pais_fig2}). These results are for the NL3 model. A comparison with a CP calculation from Wang et al \cite{wang22} is also made, though one can almost not distinguish except for the inset panel in the bottom panel.We see that the crust-core transition density (orange lines, referred as $\rho_{cc}$) gets shifted to higher values with respect to the $B=0$ case (green lines, referred as $\rho_{1\rightarrow 2}$). As before, in the CP case, this results are in line with the previous studies using the dynamical spinodal calculations \cite{fang16,fang17,fang17a}. It is interesting to notice that in this extra region that appears due the magnetic field, the proton and baryonic densities of the liquid and gas become very similar.

\begin{figure}[htbp]
		\begin{tabular}{c}
			\includegraphics[width=0.5\textwidth]{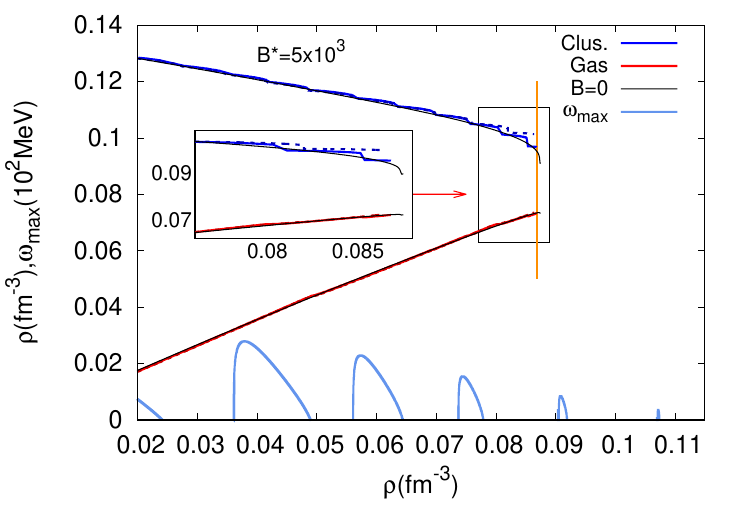}
			\includegraphics[width=0.5\textwidth]{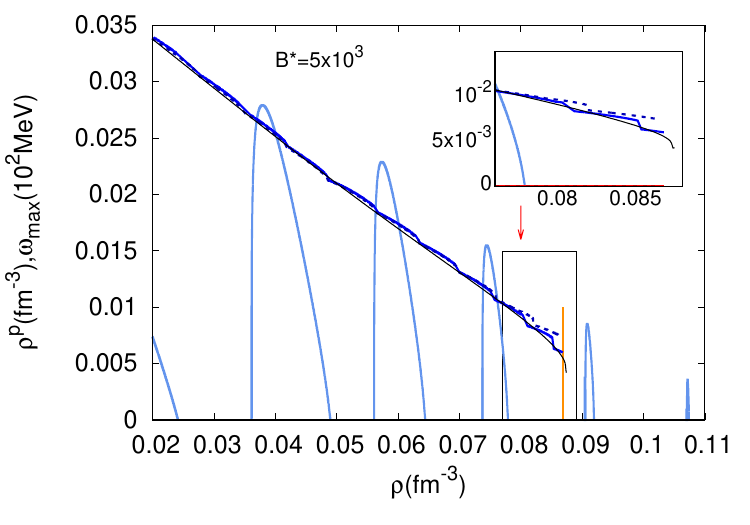}
		\end{tabular}
	\caption{Baryonic (left) and proton (right) densities of liquid (1, blue) and gas (2, red) phases as function of the total baryon density for the NL3$\omega\rho$ model in a CP (dashed lines) and CLD (solid line) calculations, with $B^*= 5 \times 10^3$. We also plot the magnetized growth rates divided by a factor $10^2$, $|\omega_{max}|$ (light blue), as well as the densities in the $B = 0$ case (black). The orange segments indicate the $\rho_{cc}$. Figure adapted from Ref.~\cite{scurto23}.} 
	\label{fig6}
 \end{figure}

In Fig.~\ref{fig6}, we show the same quantities as above but this time for the NL3$\omega\rho$ model. Unlike for the NL3 model, here we see that with increasing $B$, the crust-core transition decreases and the extra region does not appear, unlike with the CP calculation. In Wang et al \cite{wang22}, an extra region of non-homogeneous matter was found for this model, although smaller than for the NL3 model. However, in this work, the energy criterium, according to which the stable configuration has the lowest free energy, was not applied as in \cite{scurto23}, and some non-homogeneous configurations for the larger densities have an energy above homogeneous matter. Notice, however, that the calculation is not self-consistent as it would be, for instance, a Thomas-Fermi calculation that tends smoothly to the homogeneous solution, and, therefore, the crust-core transition should be further analysed. The different behaviour of the two models can be explained by the different behaviour of their symmetry energy: even though the slope of the symmetry energy at saturation is higher for the NL3 model than for NL3$\omega\rho$ (see Tab.~\ref{tab1}), for densities below $\sim 0.1$fm$^{-3}$, the symmetry energy of NL3$\omega\rho$ is higher than the one of NL3, as we can see from Fig.~\ref{fig7}. This means that NL3$\omega\rho$ (NL3) will have a larger (smaller) proton fraction, that will translate into a smaller (larger) effect of the magnetic field, and therefore a smaller (larger) extension of the crust.
The symmetry energy behaviour favours larger proton fractions for NL3$\omega\rho$, and smaller $B-$field effects, when compared to NL3.

We also need to point out that, even though both these calculations tend to give similar results, they are not self-consistent, since the surface tension is parametrised from a fit to a Thomas-Fermi calculation without magnetic field. This quantity influences the crust-core transition density so in a near future it would be interesting to obtain a calculation for a magnetized surface tension, and to analyze where the crust-core transition occurs.

\begin{figure}
		\begin{tabular}{c}
			\includegraphics[width=0.5\textwidth]{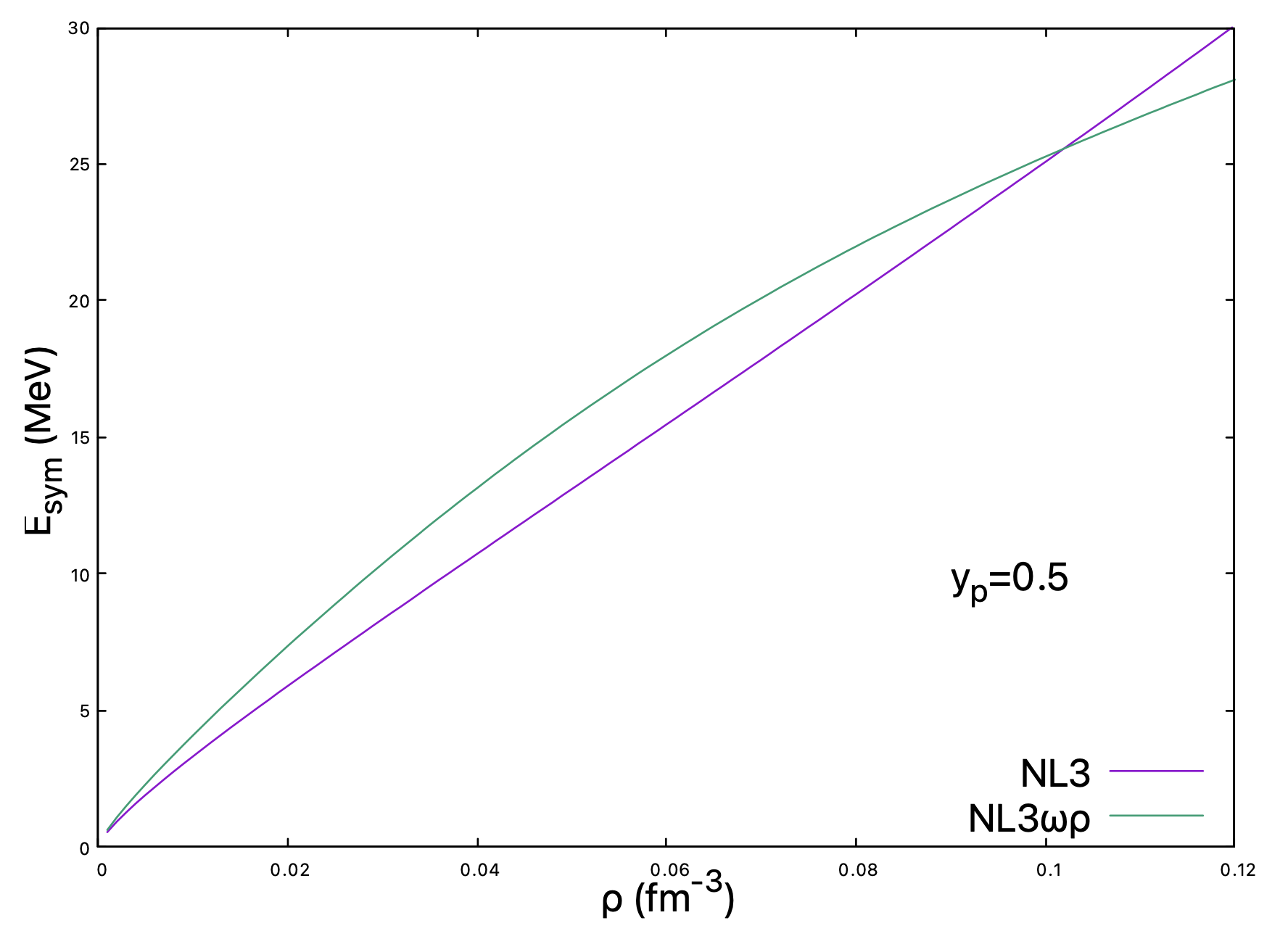}
		\end{tabular}
	\caption{Symmetry energy as a function of the density for the NL3$\omega\rho$ and NL3 models.} 
	\label{fig7}
 \end{figure}

\subsection{Conclusions}	

In this section, the structure of the inner crust of a neutron star in the presence of a strong magnetic field, within a relativistic mean-field framework, using the coexistence phase and compressible liquid drop models for the calculation of the pasta phases, was addressed. These results were compared to a dynamical spinodal method \cite{fang16,fang17,fang17a}. Two RMF models, NL3 \cite{Lalazissis1997} and NL3$\omega\rho$ \cite{Horowitz2001a,Horowitz2001b}, were considered.

We found that an extended region of clusters appears due to the presence of the magnetic field. This region contains matter in different geometric structures, and, in there, the cluster and gas densities are very close, for both neutrons and protons. This extra region seems to depend on the behaviour of the symmetry energy in the crustal EoS. We found that the transition densities given by the CLD calculation 
are in good agreement with the CP approximation, and also in agreement with the dynamical spinodal calculation. These heavy clusters are very dependent on the surface tension, therefore a calculation of a magnetized surface tension should be explored in a near future.

\section{Superconductivity and superfluidity in strong magnetic fields}
\label{sec:sedrakian}

\subsection{Introductory remarks}

We have observed that extremely large magnetic fields, on the order of
$B\sim 10^{18}-10^{19}$~G, can impact the hydrostatic equilibrium and,
consequently, the integral parameters (mass, radius, tidal
deformability, etc.) of compact stars. However, much
smaller magnetic fields are needed to influence properties governed by
the quasiparticle spectrum near the Fermi surface. These properties
include the superfluidity and superconductivity of nucleonic fluids
(neutrons and protons) and transport phenomena associated with
fermionic (primarily electrons, protons, and neutrons) and bosonic
degrees of freedom (such as pion and kaon condensates). The changes in
these properties due to magnetic fields can significantly affect the
macroscopic behavior of magnetars, as discussed in this subsection.

There are two mechanisms of suppression of nucleonic superfluids in
strong magnetic fields~\cite{Stein2016PhRvC,Sinha2015PhRvC}. The
interaction of the magnetic field with the neutron or proton spin
induces an imbalance in the number of spin-up and spin-down particles,
which implies that the Cooper pairing will be suppressed because not
all spin-up particles will find spin-down
``partners''~\cite{Stein2016PhRvC}. This so-called {\it Pauli
  paramagnetic suppression} acts for both proton and neutron
condensates, but is the sole suppression mechanism only in the case of
neutrons. The proton condensate is also suppressed due to the Larmor
motion of protons in the magnetic field (Landau
diamagentism)~\cite{Sinha2015PhRvC}. We now discuss these effects on a more quantitative level.

An important threshold is reached when electromagnetic interactions
approach the nuclear energy scale for cold neutron stars, typically,
beteween 0.1 and 1 MeV. The first suppression mechanism that acts
independent of particle charges arises when the interaction energy
between the magnetic field and the nucleon spin is given by $\mu_NB$,
where $\mu_N = e\hbar/2m_p$ is the nuclear magneton become of the
order of scales indicated above.  A simple comparison of these scales
shows that the magnetic fields on the order of $10^{16}-10^{17}$~G
would significantly influence 0.1 to 1 MeV-scale physics through the
nucleonic spin--$B$-field interaction.

An additional interaction channel arises for charged particles due to
the coupling of the particle charge to the magnetic field. Physically,
the suppression becomes effective when the Larmor radius quantifies
the winding of the particle trajectory in the magnetic field becomes
small enough (once the field is large enough) to be comparable to the
condensate coherence length. The latter defines the size of the Cooper pairs in the limit of weak-coupling BCS theory, which is appropriate
for neutron and proton fluids in neutron stars.  At this threshold the
coherence of Cooper pairing is destroyed by the field. To make a
simple estimate of the effect, we use the Landau criterion for the
critical velocity is given by $v_s \sim \Delta/p_\perp$, where
$\Delta$ is the pairing gap and $p_\perp$ is the characteristic
momentum in the plane orthogonal to the field. Since this momentum is
smaller than the Fermi momentum $p_\perp \le p_F$, we find that 
the relevant energy scale is given by
$v_sp_F \sim \pi (v_s/c) (\xi/10~\textrm{fm})
(B/10^{16}~\textrm{Gauss})$ MeV by equating the particle's Larmor radius
$p_F c/eB$ to the coherence length of the condensate, which for
neutrons and protons are of the order of $\xi\simeq 10$~fm. In our
estimate we used that $v_s/c\le 0.3$. We conclude that the fields
needed to suppress the pairing by magnetic fields are significantly
smaller than those that would affect their equation of state and
structure.

\subsection{Destruction of $S$-wave neutron superfluidity by strong magnetic fields }

In the case of $S$-wave pairing the condensate consists of Cooper
pairs of neutrons with opposite spins and momenta. When magnetic field
is applied, the spin-up and spin-down populations become unequal
because of the interaction of the neutron's spin magnetic moment with the
magnetic field - there are more neutrons aligned with the magnetic
field than neutrons that are anti-aligned. This imbalance has
a disruptive effect on $S$-wave neutron Cooper pairs, as has been
first established in the context of ordinary superconductors with
magnetic impurities~(for a review see~\cite{Sedrakian2019EPJA}).  A
sufficiently strong magnetic field can completely suppress this
pairing as neutrons with spins along the field will not find opposite
spin partners to pair. This critical field will be referred to as
$H^n_{c2}$ is the analog of the Chandrasekhar-Clogston limit in
ordinary superconductors~\cite{Abrikosov:Fundamentals}.

As the density is lowered there is also a possibility of BCS-BEC
crossover from the weakly coupled pairs of neutrons (BCS regime) to
the strongly coupled regime (BEC regime). Such a transition in nuclear
systems was studied first in the context of proton-neutron pairing
where in the vacuum there is a bound state - the deuteron. Naturally
enough, one would expect in this case formation of Bose gas of
deuterons at very low densities and a smooth transition from the BCS
regime to the BEC regime as the density is
lowered~\cite{Alm1993NuPhA,Baldo1995PhRvC,Sedrakian2006PhRvC,Huang2010PhRvC}. However,
in this case, the isospin asymmetry - a feature common in the finite
and infinite nuclear systems - has a destructive effect, analogous to
the spin-polarization, and the phase diagram is not limited to BCS-BEC
crossover but various novel phases may arise~\cite{Lombardo2001PhRvC,Muther2003PhRvC}. Since two
neutrons do not form a bound pair in free space, there is no initial
basis for a Bose-Einstein condensate (BEC) of neutron-neutron
pairs. However, by applying Nozi\'eres-Schmitt-Rink theory~\cite{Nozieres1985JLTP}, it
was possible to identify several signatures in neutron matter that
can be viewed as precursors to the BCS-BEC crossover. In the case of
zero polarization, similar results were
obtained~\cite{Bertsch1991AnPhy,Margueron2007PhRvC,Abe2009PhRvC,Sun2013NuPhA}.

Ref.~\cite{Stein2016PhRvC} have determined the critical field for un-
pairing of the neutron condensate to be in the range
$B \sim 10^{17}$~G.  In Fig.~\ref{fig:Hc2neutrons} we
plot values of $H_{c2}$ for the neutron condensate as a function of
density, determined based on a phase-shift-equivalent nucleon-nucleon
interaction and numerical solutions of the BCS equations in the case
of spin-polarized neutron matter~\cite{Stein2016PhRvC}. The shape of
the curve reflects the corresponding density dependence of the pairing
gap and its temperature dependence follows the BCS prediction: it is
largest at $T =0$ and decreases as the pairing gap decreases with
increasing temperature.
\begin{figure}[t]
\centering
 \includegraphics[width=10cm]{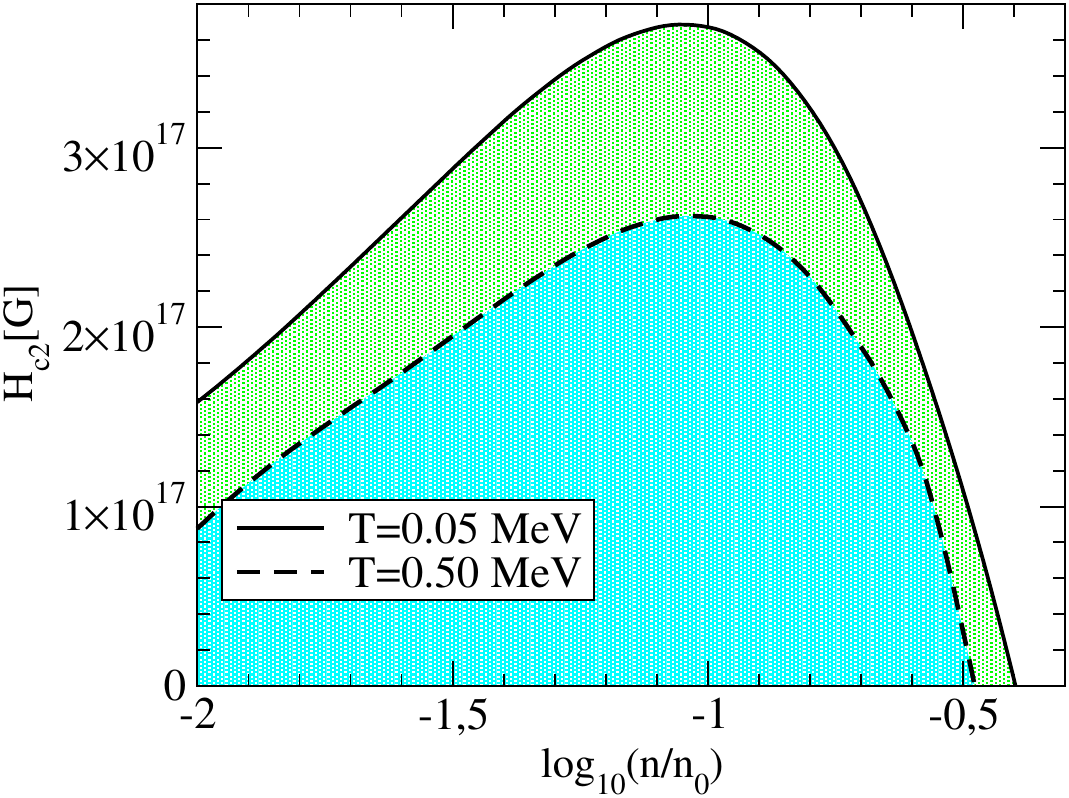}
 \caption{ Critical unpairing magnetic field $H_{c2}$ for the neutron
   condensate due to spin alignment, shown for two temperatures. The
   density is expressed in units of the nuclear saturation density.  }
 \label{fig:Hc2neutrons}
\end{figure}
Thus, if the local field in a magnetar crust exceeds the value
$H^n_{c2}$, the magnetic field will destroy the condensate. Note that
according to Fig.~\ref{fig:Hc2neutrons} the critical field is larger for
lower temperatures because (a) the pairing gap increases with
temperature and (b) the magnetic field required to produce a certain
polarization increases with decreasing temperature.

From a phenomenological perspective, $S$-wave pairing is important in
the crust of magnetars, which extends below the density of half 
the saturation density of symmetric nuclear matter. At higher
densities, the dominant pairing state in neutron matter shifts to the
$^3P_2$-$^3F_2$ channel, where neutrons pair in a total spin-1 state
(see~\cite{Khodel2014PAN,Zverev2003NuPhA}). In
this situation, the spin-polarizing effect of the magnetic field on
the internal structure of spin-1 pairs does not break the
pairing~\cite{Muzika1980PhRvD,Tarasov2014,Mizushima2021PhRvC}.

\subsection{Destruction of $S$-wave superconductivity of protons in
  strong magnetic field}
\label{sec:Protons}

BCS superconductors are characterized by at least three distinct
length scales: (i) the {\it London penetration depth} $\lambda$, (ii)
the {\it coherence length} $\xi$, and (iii) the interparticle distance
$d$. For this discussion, it is assumed that the
interparticle distance $d$ is much smaller than the other two scales,
indicating that the superconductor is in the weakly coupled
regime. The ratio of the remaining two scales, $\lambda$  and $\xi$,
defines the type of superconductivity through the Ginzburg-Landau
parameter $\kappa = \lambda/\xi$ (see, e.g., \cite{Abrikosov:Fundamentals}). If
$1/\sqrt{2} < \kappa < \infty$, the material is classified as a
type-II superconductor; otherwise, it is type-I. In type-II
superconductors, the magnetic field is carried by electromagnetic
vortices with quantum flux $\phi_0 = \pi/e$ (with $\hbar = c = 1$
assumed hereafter), while in type-I superconductors, the magnetic
field forms domain structures~\cite{Bruk1973Afz,Sedrakian1997MNRAS,Alford2005PhRvC,Sedrakian2005PhRvD}.

These two scales, $\lambda$ and $\xi$, also define three distinct magnetic field scales when combined with the flux quantum:
\begin{equation}
H_{c 1} \simeq \frac{\phi_0}{\lambda^2}, \quad H_{c m} \simeq \frac{\phi_0}{\xi \lambda}, \quad H_{c 2} \simeq \frac{\phi_0}{\xi^2}
\end{equation}
In type-II superconductors, the hierarchy of these fields is
$H_{c1}\le H_{cm}\le H_{c2}$ when $\kappa \ge 1$. At and above
$H_{c1}$, the formation of a single flux tube (Abrikosov quantum
vortex) becomes energetically favorable. The field $H_{cm}$ is the
thermodynamic magnetic field, where the energy density equals the
difference between the energy densities of the superconducting and
normal states. Finally, $H_{c2}$ is the field strength at which
superconductivity vanishes, as the density of flux tubes becomes so
high that the normal vortex cores overlap.

The Ginzburg-Landau  theory of neutron-proton superfluid mixtures was used to
compute the upper critical magnetic field in
Refs.~\cite{Sinha2015PhRvC} and is extended to account for coupling
between the neutron and proton currents in Ref.~\cite{Haber2017PhRvD}.

\begin{figure}[t]
\centering 
\includegraphics[width=8cm]{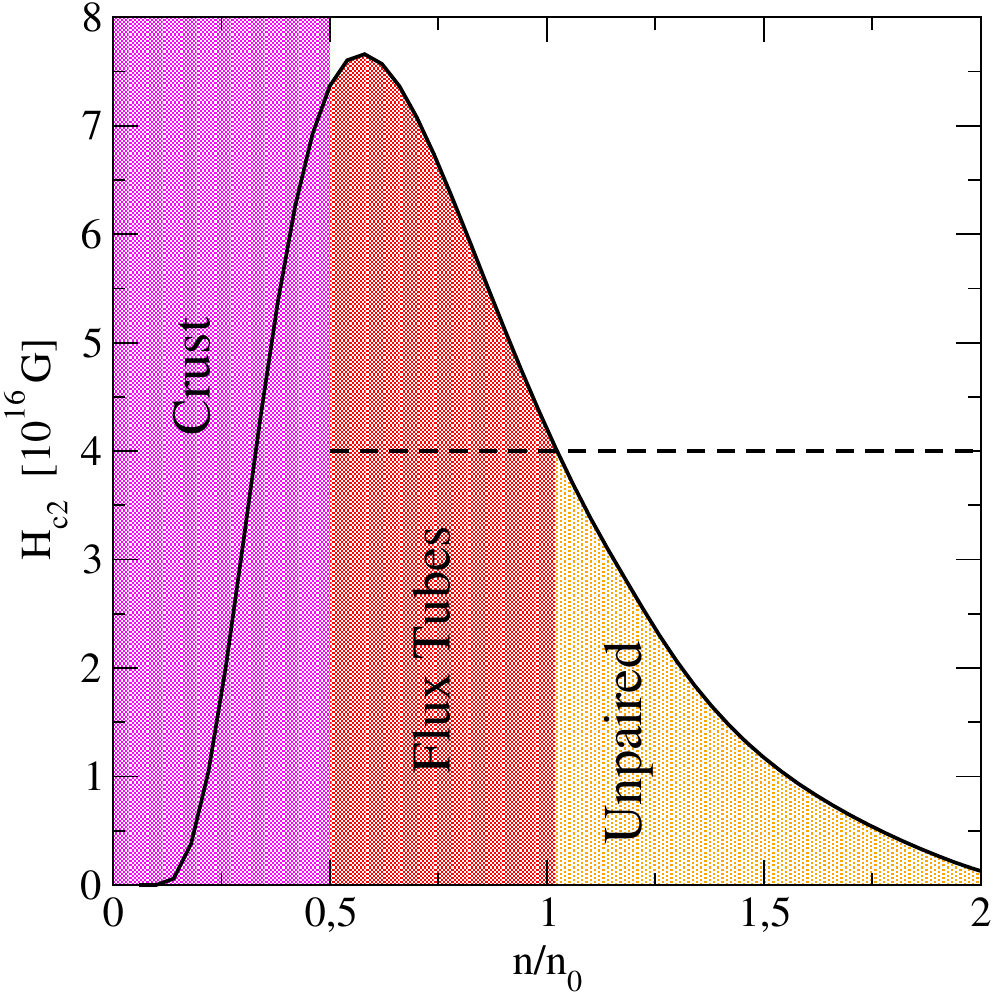}
 \caption{The proposed layered structure of a magnetar features a
   constant magnetic field in the core (represented by the dashed
   line)~\cite{Sedrakian2016AA}. To the right of the point where this field intersects with
   $H_{c2}$, the proton fluid is non-superconducting; to the left,
   where $B \leq H_{c2}$, it remains superconducting. The crust
   contains a uniform magnetic field $B$. The density is expressed in
   units of the nuclear saturation density.}
 \label{fig:Hc2_protons}
\end{figure}
The dependence of the $H_{c2}$ field on density is depicted in
Fig.\ref{fig:Hc2_protons}. Several key features are worth
mentioning. The $H_{c2}$ field reaches its maximum near the crust-core
interface at a density of $n_{\rm b} = 0.5 n_0$, where $n_0$ is the
nuclear saturation density. As seen from the figure, a magnetar is
likely to have a layered structure, consisting of (a) an inner core
devoid of superconductivity, (b) an outer core threaded by flux tubes,
and (c) a crust containing a homogeneous magnetic field of magnitude
$B$. If the field $B$ exceeds the maximum $H_{c2}$, the flux-carrying
region in the outer core vanishes, resulting in the complete
destruction of superconductivity within the magnetar.

\subsection{Astrophysical implications: Precessing magnetars and fast radio bursts (FRBs)}
\label{sec:astro_implicationss}
\label{sec:Dynamics}

The absence or partial presence of superfluidity in
magnetars will significantly impact various both macrophysical and
microphysical properties of magnetars. Below, we illustrate the key
modifications without the intent to give a complete picture.

The possibility of neutron star precession remains an intriguing unsolved
issue, the main obstacle being the fact that the precession is
strongly suppressed by the coupling of the crust to the superfluid
phases inside the neutron stars, see~\cite{Sedrakian1999ApJ}. The
precession in magnetars may not have the same fate if the fields are
in the range where the suppression of proton superconductivity is
effective~\cite{Sedrakian2016AA} and it has been suggested that free
precession can account for FRBs~\cite{Zanazzi2020ApJ}. The reason is the following: when
protons become unpaired, they can scatter off neutron quasiparticles
in the vortex cores, a process that is significantly more effective than the scattering of electrons off magnetized neutron vortices via
electromagnetic forces, which is the dominant mechanism in type-I
proton
superconductors~\cite{Alpar1984ApJ,Sedrakian2005PhRvD}. However, in
the more realistic scenario of type-II superconductivity, which would
be absent in magnetars due to unpairing, the coupling mechanism is
likely more complex due to significant interactions between protonic
flux tubes and neutron
vortices~\cite{Haskell2013ApJ,Drummond2018MNRAS,Drummond2017MNRAS,Sedrakian1995ApJ}. It
has been proposed that FRBs are associated with magnetars and the
subclass of repeated FRBs may originate from the precession of
magnetars\cite{Wasserman2022ApJ,Cordes2022ApJ,Gao2023MNRAS,Desvignes2024NatAs}. The  fact that only a small subclass of FRBs shows periodicity might be
interpreted as these objects having a favorable magnetic field
configuration admitting weak coupling of the neutron superfluid to the
crust. If the magnetic field in the core of a magnetar completely
unpairs the proton fluid, it will interact with the electron fluid on
plasma time scales, which are significantly shorter than
hydrodynamical time scales. As a result, the unpaired core of a
magnetar can be described as a two-fluid system, consisting of a
superfluid neutron condensate and a normal component made up of the
proton and electron fluids. 
The mutual coupling between these components is expected to be weak in this case.

Finally, it is important to note that, in addition to precession, the
dynamic coupling between the superfluid and the normal plasma plays a
crucial role in understanding the rotational irregularities of
magnetars. This includes phenomena such as glitches, anti-glitches,
post-glitch relaxation, and oscillations.

\subsection{Neutrino radiation from magnetars}
\label{sec:Neutrinos}

The suppression of pairing by MeV-scale magnetic fields will also have
profound consequences on the thermal evolution of magnetars, because
the dominant processes of neutrino radiation will not be suppressed by
the Boltzmann factor containing the gap in the quasiparticle spectrum
of baryons. At the same time, the processes that are intrinsic to
condensates, such as the pair-breaking emission of
neutrino-anti-neutrino pairs, will not operate by definition.

Let us start with the direct Urca process: $n\to p + e + \bar\nu_e$
where $n$ stands for neutron, $p$ - proton, $e$ - elections and
$\bar\nu_e$ for electron antineutrino. 
It is well known that proton and neutron pairing suppresses the Urca
process when nucleons transition into a superconducting or superfluid
state. At very low temperatures, emissivity is reduced by a factor of
$\exp(-\Delta/T)$ for each participating nucleon, where $\Delta$
represents the relevant pairing gap and $T$ is the
temperature. The suppression is evidently reduced when neutron
and proton $S$-wave pairing is terminated by the magnetic field. Since
the neutron pairing gap in the $P$-wave channel is smaller than the
proton pairing gap in the $S$-wave channel, the disruption of proton
superconductivity by a MeV-scale magnetic field will significantly
impact Urca emissivity. Numerical examples are provided in
Ref.~\cite{Sinha2015PhRvC}.

Next, consider pair-formation and breaking (PFB) processes:
$[NN] \to N+N + \nu_f+\bar\nu_f$ where $N$ stands for either neutron
or proton, $\nu_f$ and $\bar\nu_f$ for neutrino and anti-neutrino of
flavor $f$~(for a review see \cite{Sedrakian2019EPJA}).  The rates of
neutrino emission through PBF processes follows the relation
$\epsilon \propto \Delta^2T^7$, where $\Delta$ is the pairing gap. As
a result, the unpairing of $S$-wave condensates will effectively
eliminate PBF processes in regions where the magnetic field locally
surpasses the unpairing thresholds for protons and
neutrons. Consequently, the overall neutrino emission rate will
asymptotically decrease to the level associated with PBF emission by
the $P$-wave condensate.

Apart from emission rates, the thermal evolution of magnetars depends
on the specific heat of the interior matter, i.e. star's thermal
inertia. In neutron stars that are fully superconducting or superfluid
and have low magnetic fields, the heat capacity is predominantly
determined by electrons because of the exponential suppression of specific
heat of nucleons. Once magnetic unpair nucleons, their specific heat
will contribute to the total budget of the star's thermal inertia. An
increase in the specific heat of the interior matter will lead to a
longer cooling time scale for a magnetar.  For the current state of
the simulations of coupled magnetic and thermal evolution
see Refs.~\cite{Ascenzi2024MNRAS} and references therein.

\section{Neutron P-wave superfluids in neutron stars: topological surface defects}
\label{sec:yasui}

\subsection{Introduction}
\label{sec:introduction}

Neutron stars are important subjects which are studied by many researchers in astrophysics, nuclear physics and condensed matter physics (see recent reviews, e.g., Refs.~\cite{Graber2017,Baym2018}).
Neutron superfluids have been interested since the early stage of researching neutron stars.
It is considered that the neutron $^{1}S_{0}$ superfluids exist at low density region near the surface of neutron stars.
This superfluid is produced by spin-singlet neutron pairings with $S$-wave by the low-energy scattering process.
At higher density, there appears another type of the pairing structure: spin-triplet and P-wave denoted by $^{3}P_{J}$.
In this pairing, the total angular-momentum $J=2$ is favored, while the $J=0$ and $J=1$ channels are 
disfavored, because the latter channels are repulsive.
Thus, the neutron matter is described as the neutron $^{3}P_{2}$ superfluids.
The neutron $^{3}P_{2}$ pairings exhibit interesting properties induced by the rich structures of the internal degrees of freedom combined by spin and angular momentum.
In this section, we review our recent studies on the neutron $^{3}P_{2}$ superfluids.

We summarize briefly the history of the researches of the neutron $^{1}S_{0}$ and $^{3}P_{2}$ superfluids.
In the condensed matter physics, there are rapid development of the researches of fermion superfluids.
In 1972, $^{3}$He atom (fermion) superfluids were discovered in experiments~\cite{PhysRevLett.28.885,PhysRevLett.29.920}, where it was studied that spin-triplet P-wave Cooper pairs stem from the spin-fluctuation interactions.
In 1995, Ru atom (boson) superfluids were discovered in experiments~\cite{doi:10.1126/science.269.5221.198,PhysRevLett.75.3969,PhysRevLett.75.1687}.
In 2003, the fermionic condensate was found in 
an experiment by using $^{40}$K (fermion) atom~\cite{PhysRevLett.92.040403}.
In parallel to the development in the condensed matter physics, the possibility of superfluids has been pursuit in nuclear physics.
In 1960, Migdal first proposed the existence of neutron $^{1}S_{0}$ superfluids at low density~\cite{Migdal1960}.
In 1966, however, Wolf discussed that strong repulsion core makes the interaction to attractive one at high density, and it makes the neutron $^{1}S_{0}$ pairings unstable~\cite{Wolf1966}.

As a new type of pairings, in 1968, Tabakin proposed that the neutron $^{3}P_{2}$ superfluids can be realized by an attraction in this channel due to the nuclear LS potential at high energy scatterings~\cite{Tabakin1968}.
The LS potential exhibits the dependence on both relative angular momentum ($L$) and total spin ($S$) of the scattering two neutrons.
After this pioneering work, further studies have been developed by many authors, see, e.g., Refs.~\cite{Hoffberg1970,Tamagaki1970,Takatsuka1971,Takatsuka1972,Fujita1972,Richardson1972} as early works.
For example, it was presented that the $^{3}P_{2}$ channel is coupled to the $^{3}F_{2}$, and the mixing effect can enhance the gap size in the neutron $^{3}P_{2}$ superfluids leading to a few MeV order~\cite{Tamagaki1970,Takatsuka1971,Takatsuka1972}.
Recent calculations, however, indicate the pairing size may be smaller due to many-body effects in nuclear matter.
The neutron $^{3}P_{2}$ pairings are also concerned with the cooling process of neutron stars~\cite{Shternin2011,Page2011,Heinke2010} (see also Refs.~\cite{Blaschke2012,Blaschke2013,Grigorian2016}).

The P-wave pairings show the topological properties. 
For example, the Weyl and/or Majorana fermions can exist in the neutron $^{3}P_{2}$ superfluids~\cite{Mizushima2017,Mizushima2018,Masaki:2023rtn}.
There are also rich bosonic excitations in the the neutron $^{3}P_{2}$ superfluids~\cite{Bedaque:2003wj,Leinson:2011wf,Leinson:2012pn,Leinson:2013si,Bedaque:2012bs,bedaquePRC14,Bedaque:2013fja,Bedaque:2014zta,Leinson:2009nu,Leinson:2010yf,Leinson:2010pk,Leinson:2010ru,Leinson:2011jr} which may be relevant to the cooling process.
The exotic topological vortices also exhibit many interesting properties such as the spontaneous magnetizations~\cite{Fujita1972,muzikarPRD80,Sauls:1982ie,Masuda:2015jka} and vortices with Majorana fermions~\cite{Masaki:2019rsz,Masaki:2021hmk,Masaki:2023rtn}, half-quantized non-Abelian vortices~\cite{Masuda:2016vak,Masaki:2021hmk,Kobayashi:2022moc,Kobayashi:2022dae,Masaki:2023rtn}, 
solitonic excitations on 
a vortex~\cite{Chatterjee:2016gpm}, 
and domain walls~\cite{Yasui:2019vci}.
The neutron $^{3}P_{2}$ superfluids also exhibit interesting topological defects at the surface on the neutron stars~\cite{Yasui:2019pgb}, as discussed in details in Sec.~\ref{sec:surface_defects}.
Recently, it was shown that 
since a singly quantized vortex 
is split into two half-quantized vortices
\cite{Masaki:2021hmk,Kobayashi:2022moc}, 
vortex junctions 
are produced at the interface between the neutron $^{1}S_{0}$ superfluid and the neutron $^{3}P_{2}$ superfluid, 
thereby inducing the vortex networks which can explain the power-law of the glitches in neutron stars~\cite{Marmorini:2020zfp}.
The vortex junctions at the boundary of the superfluids are called the boojums which were discussed originally for $^{3}$He superfluids by N.~D.~Mermin, see, e.g., Ref.~\cite{RevModPhys.51.591}.
Thus, the  $^{3}P_{2}$ superfluids are important subjects, not only in the nuclear physics and the astrophysics, but also in the condensed matter physics.

The neutron $^{3}P_{2}$ superfluids especially play important roles in magnetars.
The magnetars are the special neutron stars with very strong magnetic fields which reach $10^{15}$ Gauss~\cite{Kaspi:2017fwg}.
This is about hundred times larger than the magnetic fields in neutron stars $10^{13}$ Gauss.
The neutron $^{3}P_{2}$ superfluids are stable against the strong magnetic fields.
This can be intuitively understood: the spin-parallel (spin-one) pairs in the neutron $^{3}P_{2}$ superfluids are irrelevant to the Zeeman splitting.
This property is sharply contrasted to the fragility of the spin-antiparallel (spin-zero) pairings in the neutron $^{1}S_{0}$ superfluids.
Therefore, it is expected that the neutron matter in the magnetars is dominated by the neutron $^{3}P_{3}$ superfluids.

In the followings, we introduce the basic properties of the neutron $^{3}P_{2}$ superfluids in Sec.~\ref{sec:BdG_GL_phase_diagam}. After summarizing the symmetries of the neutron $^3{P}_{2}$ superfluids in Sec.~\ref{sec:symmetry}, we explain the Bogoliubov-de Gennes (BdG) equation and the Ginzburg-Landau (GL) theory in Sec.~\ref{sec:BdG_GL_phase_diagam}, and present that they give the phase diagram with rich structures.
In Sec.~\ref{sec:surface_defects}, based on the GL theory, we show our study on the topological surface defects on the boundary of  the neutron $^{3}P_{2}$ superfluids.
We summarize our presentation in Subsec.~\ref{sec:summary}.
This section is dedicated to the summary of our previous studies performed partly by the author. The relevant references are shown in the text.

\subsection{Phase diagram of neutron $^{3}P_{2}$ superfluids} \label{sec:BdG_GL_phase_diagam}

\subsubsection{Symmetry}
\label{sec:symmetry}

Let us describe some basic properties of the neutron $^{3}P_{2}$ pairings in neutron matter.
The neutron $^{3}P_{2}$ pairs have the total spin one, $\vec{s}=(s^{1},s^{2},s^{3})$, as a sum of two neutrons.
Besides, the neutron $^{3}P_{2}$ pairings have also dependence on three-dimensional momentum, $\vec{q}=(q^{1},q^{2},q^{3})$, transferred between two neutrons.
Those properties stem from the LS potential which is dominantly important in the interaction between two neutrons at high-energy scatterings.
They induce the tensor-type condensate for the neutron $^{3}P_{2}$ superfluids, whose pairing form denoted by $A^{ab}$ is given by
\begin{align}
   A^{ab} \propto \frac{1}{2} (s^{a}q^{b}+s^{b}q^{a}) - \frac{1}{3} \delta^{ab} \vec{s}\cdot\vec{q},
\label{eq:A_indices}
\end{align}
as a symmetric and traceless tensor over the indices $a,b,c=1,2,3$.
The pairing matrix~\eqref{eq:A_indices} is described essentially by five complex components as independent degrees of freedom.
Including the overall global phase, the neutron $^{3}P_{2}$ gap can be parametrized by the matrix form.
In general, the matrix $A$ in Eq.~\eqref{eq:A_indices} can be expressed by
\begin{align}
   A(\vec{x})
= \left(
   \begin{array}{ccc}
    f_{1}(\vec{x}) & g_{3}(\vec{x}) & g_{2}(\vec{x}) \\
    g_{3}(\vec{x}) & f_{2}(\vec{x}) & g_{1}(\vec{x}) \\
    g_{2}(\vec{x}) & g_{1}(\vec{x}) & -f_{1}(\vec{x})-f_{2}(\vec{x})  
   \end{array}
   \right),
\label{eq:order_parameter_general}
\end{align}
where $f_{i}(\vec{x})$ ($i=1,2$) and $g_{j}(\vec{x})$ ($j=1,2,3$) are complex functions with the three-dimensional coordinate $\vec{x}$.

\begin{table}[tb]
\caption{The bulk phases and symmetries of the uniform neutron $^{3}P_{2}$ superfluids are shown for the different strengths of magnetic fields (cf.~Fig~\ref{fig:240804_phase_diagram}). $r$ is an internal parameter in the pairing matrix, see Eq.~\eqref{eq:order_parameter_diagonal}. More detailed classification of phases is found in Ref.~\cite{Mizushima:2021qrz}.}
\begin{center}
\begin{tabular}{cccc}
\hline
   magnetic field & zero & weak & strong \\
\hline
   $r$ & $-1$ & $(-1,-1/2)$ & $-1/2$ \\
   bulk phase & UN & $\mathrm{D}_{2}$-BN & $\mathrm{D}_{4}$-BN \\
   symmetry & $\mathrm{O}(2)$ & $\mathrm{D}_{2}$ & $\mathrm{D}_{4}$ \\
\hline
\end{tabular}
\end{center}
\label{tbl:phase_magnetic_field}
\end{table}%

When the neutron $^{3}P_{2}$ superfluid is uniform in space, the pairing matrix $A$ can be expressed simply by a diagonal form as
\begin{align}
   A
= A_{0}
   \left(
   \begin{array}{ccc}
    r & 0 & 0 \\
    0 & -1-r & 0 \\
    0 & 0 & 1   
   \end{array}
   \right).
\label{eq:order_parameter_diagonal}
\end{align}
Here $A_{0}$ is a complex scalar including a complex phase, and
$r$ indicates the internal parameter in the neutron $^{3}P_{2}$ gap which is given by the range of $r$ can be restricted to be $-1 \le r \le -1/2$ without loss of generality.
According to the value of $r$, the neutron $^{3}P_{2}$ pairing exhibits various phases with different (continuous or discrete) symmetries:
the $\mathrm{O}(2)$ symmetry for $r=-1$,
the $\mathrm{D}_{2}$ symmetry (a discrete symmetry for rectangular form) for $-1<r<-1/2$ and
the $\mathrm{D}_{4}$ symmetry (a discrete symmetry for square prism) for $r=-1/2$.
These states are called the UN (uniaxial nematic) phase, the $\mathrm{D}_{2}$-BN ($\mathrm{D}_{2}$ biaxial nematic) phase and the $\mathrm{D}_{4}$-BN ($\mathrm{D}_{4}$ biaxial nematic) phase, respectively.
The phases are determined by minimizing the free energy with respect to $r$ at different magnetic fields, see Table~\ref{tbl:phase_magnetic_field}.

In each phase, the symmetry breaking patters are given as redundant information with the above
\begin{align}
   \mathrm{U}(1) \times \mathrm{SO}(3)_{\mathrm{L}+\mathrm{S}} \rightarrow \mathrm{O}(2),
\label{eq:symmetry_breaking_O2}
\end{align}
in the UN phase,
\begin{align}
   \mathrm{U}(1) \times \mathrm{SO}(3)_{\mathrm{L}+\mathrm{S}} \rightarrow \mathrm{D}_{2},
\label{eq:symmetry_breaking_D2}
\end{align}
in the $\mathrm{D}_{2}$-BN phase and
\begin{align}
   \mathrm{U}(1) \times \mathrm{SO}(3)_{\mathrm{L}+\mathrm{S}} \rightarrow \mathrm{D}_{4},
\label{eq:symmetry_breaking_D4}
\end{align}
in the $\mathrm{D}_{4}$-BN phase.
Here $\mathrm{U}(1)$ is the global symmetry and $\mathrm{SO}(3)_{\mathrm{L}+\mathrm{S}}$ is the symmetry for simultaneous rotation in three-dimensional space and spin stemming from the LS potential.

The symmetry breaking patters in Eqs.~\eqref{eq:symmetry_breaking_O2}, \eqref{eq:symmetry_breaking_D2} and \eqref{eq:symmetry_breaking_D4} are important for the properties of the neutron $^{3}P_{2}$ superfluids.
We remind us that, in the symmetry breaking from the group $G$ to the subgroup $H$, the coset space $G/H$ exhibits nontrivial homotopy groups $\pi_{n}(G/H)$, see, e.g., Refs.~\cite{Mizushima2017,Mizushima:2021qrz}.
In two-dimensional space, a variety of homotopy groups in $\pi_{1}(G/H)$ leads to the rich phenomena of quantum vortices in neutron stars in rotations and/or magnetic fields.


\subsubsection{BdG theory and phase diagram}

The neutron $^{3}P_{2}$ superfluids are described in terms of the non-relativistic Hamiltonian with the inter-neutron interaction.
The relativistic corrections are supplied by the LS potential which is proportional to $\vec{L}\cdot\vec{S}$.
Here, the angular momentum $\vec{L}$ is coupled to the total spin $\vec{S}$ in the two neutron pairs, where the total spin is defined by $\vec{S}=\vec{s}_{1}+\vec{s}_{2}$ with $\vec{s}_{i}$ the spin operator for neutrons labeled by $i=1,2$.
The interaction with the magnetic field ($\vec{B}$) is introduced through the term $-\gamma_{\mathrm{n}}\vec{s}_{i}\cdot\vec{B}$, where $\gamma_{\mathrm{n}}=1.2 \times 10^{-13}$ MeV/T (T for unit of tesla) is the gyromagnetic ratio of a neutron.

The superfluid state can be analyzed by solving the BdG equation for the given Hamiltonian for neutrons.
In the BdG equation, the gap function is obtained by the self-consistent solution in the mean-field approximation, see, e.g., Ref.~\cite{Mizushima2017,Mizushima:2021qrz} as recent studies.
This is equivalent to the evaluation of the effective action in the one-loop approximation in the field theory.
The BdG equation gives the precise solution of the gap in terms of the fermionic degrees of freedom.

\begin{figure}[htb]
\includegraphics[keepaspectratio, scale=0.5]{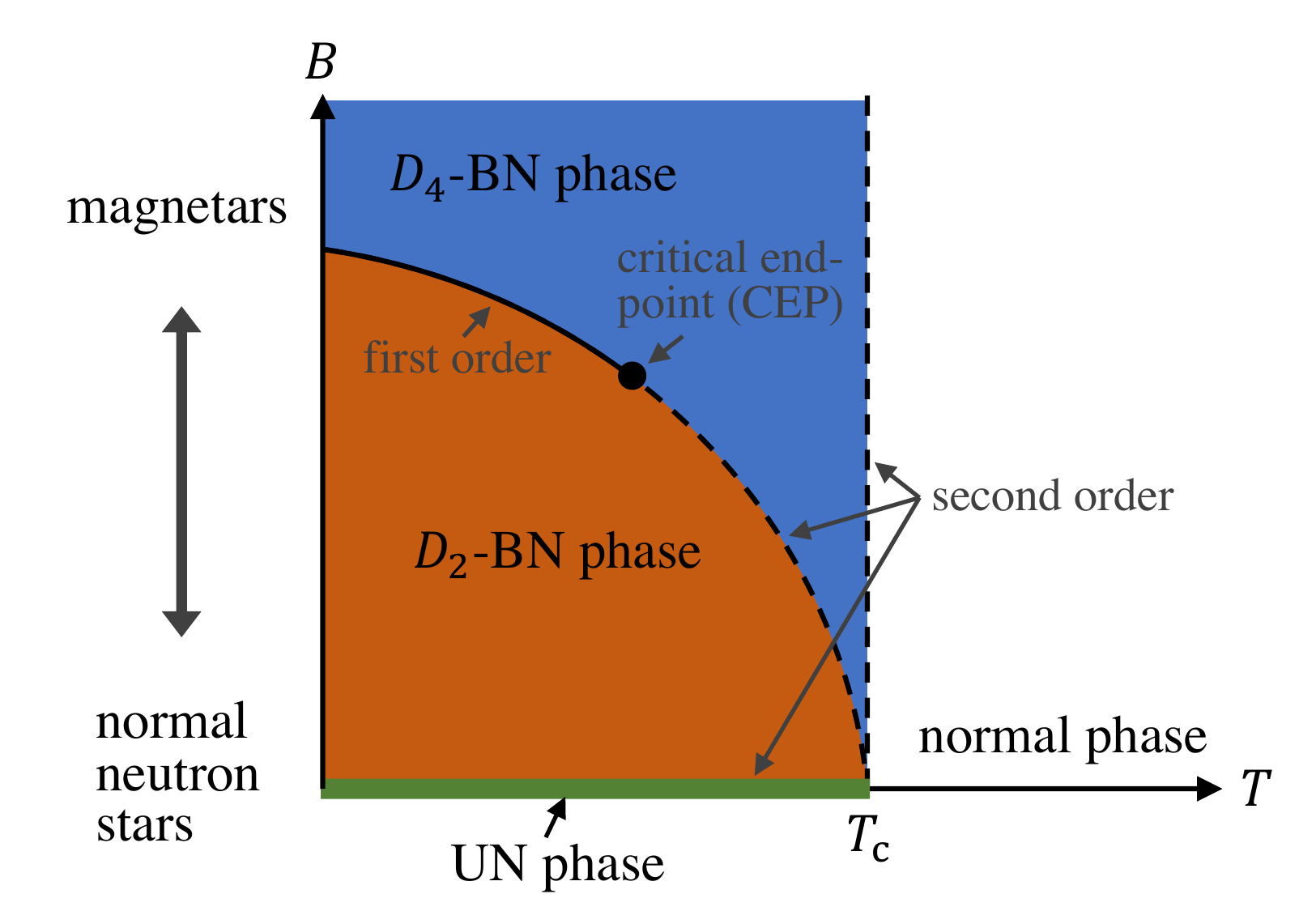} %
\caption{
The schematic figure for the phase diagram of the neutron $^{3}P_{2}$ superfluids is shown on the plane spanned by temperature ($T$) and magnetic field ($B$) (cf.~Table~\ref{tbl:phase_magnetic_field}). More precise information is found in Refs.~\cite{Mizushima2017,Mizushima:2021qrz}. See the text for the details.
\label{fig:240804_phase_diagram}}
\end{figure}

The phase diagram of the neutron $^{3}P_{2}$ superfluids is obtained by the BdG equation as shown schematically in Fig.~\ref{fig:240804_phase_diagram} (see also Table~\ref{tbl:phase_magnetic_field}).
In the figure, $T_{\rm c}$ is the critical temperature for phase transition from superfluid phase to normal phase.
The properties of the phase diagram is explained in the followings.
At zero magnetic field, the UN phase is realized on the temperature axis below $T_{\rm c}$.
This phase is, however, unstable against the nonzero magnetic field, and it changes to the $\mathrm{D}_{2}$-BN phase at nonzero but small magnetic fields.
At stronger magnetic fields, instead of the $\mathrm{D}_{2}$-BN phase, the $\mathrm{D}_{4}$-BN phase appears as the most stable state.
The $\mathrm{D}_{4}$-BN phase survives in the strong magnetic fields due the parallel spin in the neutron pairings as explained in Sec.~\ref{sec:symmetry}.

More detailed information is explained for the phase diagram in Fig.~\ref{fig:240804_phase_diagram}.
The transitions between the UN phase and the $\mathrm{D}_{2}$-BN phase is the second-order transition, while the transitions between the $\mathrm{D}_{2}$-BN phase and the $\mathrm{D}_{4}$-BN phase exhibit the first-order transition at lower temperatures and the second-order transitions at higher temperatures.
 It is important to note the existence of the critical end point (CEP) at the meeting point of the lines from the first-order and the second-order phase transitions.
The CEP shows the power-law divergences for several physical quantities, such as the heat capacity, the magnetization, and the spin susceptibility~\cite{Mizushima:2019spl}.
Interestingly, it leads to the new universality class for powers which has not been known for the other quantum systems.

\subsubsection{GL theory: bosonic effective theory}

Near the critical temperature ($T \lesssim T_{\rm c}$), 
 the relevant degrees of freedom are given basically in terms of the the bosons, i.e., the pairs of fermions (nucleons).
In this temperature region, the bosons are regarded as the low-energy effective degrees of freedom.
The effective theory for bosons is given by the GL theory as the effective theory of the BdG equation for fermions~\cite{Fujita1972,Richardson:1972xn,Sauls:1978lna,Muzikar:1980as,Sauls:1982ie,Vulovic:1984kc,Masuda:2015jka,Masuda:2016vak,Yasui:2018tcr,Yasui:2019tgc,Yasui:2019unp,Yasui:2019pgb,Yasui:2020xqb,Mizushima:2019spl,Mizushima:2021qrz}.

Noting the traceless and symmetric $3 \times 3$ tensor, $A$, for the order parameter in Eq.~\eqref{eq:order_parameter_general}, whose diagonal form was shown in Eq.~\eqref{eq:order_parameter_diagonal}, 
the GL free energy is given by
\begin{align}
   f_{8}[{A}] = f_{8}^{(0)}[{A}] + f_{2}^{(\le4)}[{A}] + f_{4}^{(\le2)}[{A}],
\label{eq:GL_free_energy}
\end{align}
as a truncated series of the power function of $A$ up to ${\cal O}(A^{8})$, see, e.g., Refs.~\cite{Yasui:2018tcr,Yasui:2019unp,Mizushima:2019spl,Mizushima:2021qrz}.
Each term is defined by
\begin{eqnarray}
 f_{8}^{(0)}[{A}]
&=&
  K^{(0)}
  \sum_{i,j,\mu=1,2,3}
  \Bigl(
        \nabla_{j} {A}_{i \mu}^{\ast} \nabla_{j} {A}_{\mu i}
     + \nabla_{i} {A}_{i \mu}^{\ast} \nabla_{j} {A}_{\mu j}
     + \nabla_{i} {A}_{j \mu}^{\ast} \nabla_{j} {A}_{\mu i}
  \Bigr)
\nonumber \\ &&
+ \alpha^{(0)}
   \bigl(\mathrm{tr} {A}^{\ast} {A} \bigr)
+ \beta^{(0)}
   \Bigl(
        \bigl(\mathrm{tr} \, {A}^{\ast} {A} \bigr)^{2}
      - \bigl(\mathrm{tr} \, {A}^{\ast 2} {A}^{2} \bigr)
   \Bigr)
\nonumber \\ &&
+ \gamma^{(0)}
   \Bigl(
         - 3  \bigl(\mathrm{tr} \, {A}^{\ast} {A} \bigr) \bigl(\mathrm{tr} \, {A}^{2} \bigr) \bigl(\mathrm{tr} \, {A}^{\ast 2} \bigr)
        + 4 \bigl(\mathrm{tr} \, {A}^{\ast} {A} \bigr)^{3}
        + 6 \bigl(\mathrm{tr} \, {A}^{\ast} {A} \bigr) \bigl(\mathrm{tr} \, {A}^{\ast 2} {A}^{2} \bigr)
      + 12 \bigl(\mathrm{tr} \, {A}^{\ast} {A} \bigr) \bigl(\mathrm{tr} \, {A}^{\ast} {A} {A}^{\ast} {A} \bigr)
              \nonumber \\ && \hspace{3em} 
         - 6 \bigl(\mathrm{tr} \, {A}^{\ast 2} \bigr) \bigl(\mathrm{tr} \, {A}^{\ast} {A}^{3} \bigr)
         - 6 \bigl(\mathrm{tr} \, {A}^{2} \bigr) \bigl(\mathrm{tr} \, {A}^{\ast 3} {A} \bigr)
       - 12 \bigl(\mathrm{tr} \, {A}^{\ast 3} {A}^{3} \bigr)
      + 12 \bigl(\mathrm{tr} \, {A}^{\ast 2} {A}^{2} {A}^{\ast} {A} \bigr)
        + 8 \bigl(\mathrm{tr} \, {A}^{\ast} {A} {A}^{\ast} {A} {A}^{\ast} {A} \bigr)
   \Bigr)
\nonumber \\ &&
 + \delta^{(0)}
\Bigl(
       \bigl( \mathrm{tr}\,A^{\ast 2} \bigr)^{2} \bigl( \mathrm{tr}\, A^{2} \bigr)^{2}
 + 2 \bigl( \mathrm{tr}\,A^{\ast 2} \bigr)^{2} \bigl( \mathrm{tr}\, A^{4} \bigr)
  - 8 \bigl( \mathrm{tr}\,A^{\ast 2} \bigr) \bigl( \mathrm{tr}\,A^{\ast}AA^{\ast}A \bigr) \bigl( \mathrm{tr}\,A^{2} \bigr)
  - 8 \bigl( \mathrm{tr}\,A^{\ast 2} \bigr) \bigl( \mathrm{tr}\,A^{\ast}A \bigr)^{2} \bigl( \mathrm{tr}\,A^{2} \bigr)
       \nonumber \\ && \hspace{3em}
 - 32 \bigl( \mathrm{tr}\,A^{\ast 2} \bigr) \bigl( \mathrm{tr}\,A^{\ast}A \bigr) \bigl( \mathrm{tr}\,A^{\ast}A^{3} \bigr)
 - 32 \bigl( \mathrm{tr}\,A^{\ast 2} \bigr) \bigl( \mathrm{tr}\,A^{\ast}AA^{\ast}A^{3} \bigr)
 - 16 \bigl( \mathrm{tr}\,A^{\ast 2} \bigr) \bigl( \mathrm{tr}\,A^{\ast}A^{2}A^{\ast}A^{2} \bigr)
       \nonumber \\ && \hspace{3em}
  + 2 \bigl( \mathrm{tr}\,A^{\ast 4} \bigr) \bigl( \mathrm{tr}\,A^{2} \bigr)^{2}
  + 4 \bigl( \mathrm{tr}\,A^{\ast 4} \bigr) \bigl( \mathrm{tr}\,A^{4} \bigr)
  - 32 \bigl( \mathrm{tr}\,A^{\ast 3}A \bigr) \bigl( \mathrm{tr}\,A^{\ast}A \bigr) \bigl( \mathrm{tr}\,A^{2} \bigr)
       \nonumber \\ && \hspace{3em}
  - 64 \bigl( \mathrm{tr}\,A^{\ast 3}A \bigr) \bigl( \mathrm{tr}\,A^{\ast}A^{3} \bigr)
  - 32 \bigl( \mathrm{tr}\,A^{\ast 3}AA^{\ast}A \bigr) \bigl( \mathrm{tr}\,A^{2} \bigr)
  - 64 \bigl( \mathrm{tr}\,A^{\ast 3}A^{2}A^{\ast}A^{2} \bigr)
  - 64 \bigl( \mathrm{tr}\,A^{\ast 3}A^{3} \bigr) \bigl( \mathrm{tr}\,A^{\ast}A \bigr)
       \nonumber \\ && \hspace{3em}
  - 64 \bigl( \mathrm{tr}\,A^{\ast 2}AA^{\ast 2}A^{3} \bigr)
  - 64 \bigl( \mathrm{tr}\,A^{\ast 2}AA^{\ast}A^{2} \bigr) \bigl( \mathrm{tr}\,A^{\ast}A \bigr)
 + 16 \bigl( \mathrm{tr}\,A^{\ast 2}A^{2} \bigr)^{2}
 + 32 \bigl( \mathrm{tr}\,A^{\ast 2}A^{2} \bigr) \bigl( \mathrm{tr}\,A^{\ast}A \bigr)^{2}
       \nonumber \\ && \hspace{3em}
 + 32 \bigl( \mathrm{tr}\,A^{\ast 2}A^{2} \bigr) \bigl( \mathrm{tr}\,A^{\ast}AA^{\ast}A \bigr)
 + 64 \bigl( \mathrm{tr}\,A^{\ast 2}A^{2}A^{\ast 2}A^{2} \bigr)
  -16 \bigl( \mathrm{tr}\,A^{\ast 2}AA^{\ast 2}A \bigr) \bigl( \mathrm{tr}\,A^{2} \bigr)
   + 8 \bigl( \mathrm{tr}\,A^{\ast}A \bigr)^{4}
       \nonumber \\ && \hspace{3em}
 + 48 \bigl( \mathrm{tr}\,A^{\ast}A \bigr)^{2} \bigl( \mathrm{tr}\,A^{\ast}AA^{\ast}A \bigr)
 +192 \bigl( \mathrm{tr}\,A^{\ast}A \bigr) \bigl( \mathrm{tr}\,A^{\ast}AA^{\ast 2}A^{2} \bigr)
 + 64 \bigl( \mathrm{tr}\,A^{\ast}A \bigr) \bigl( \mathrm{tr}\,A^{\ast}AA^{\ast}AA^{\ast}A \bigr)
       \nonumber \\ && \hspace{3em}
  -128 \bigl( \mathrm{tr}\,A^{\ast}AA^{\ast 3}A^{3} \bigr)
 + 64 \bigl( \mathrm{tr}\,A^{\ast}AA^{\ast 2}AA^{\ast}A^{2} \bigr)
 + 24 \bigl( \mathrm{tr}\,A^{\ast}AA^{\ast}A \bigr)^{2}
 +128 \bigl( \mathrm{tr}\,A^{\ast}AA^{\ast}AA^{\ast 2}A^{2} \bigr)
       \nonumber \\ && \hspace{3em}
 + 48 \bigl( \mathrm{tr}\,A^{\ast}AA^{\ast}AA^{\ast}AA^{\ast}A \bigr)
\Bigr),
\label{eq:eff_pot_w0_coefficient02_f}
\\
   f_{2}^{(\le4)}[{A}]
&=&
      \beta^{(2)}
      \vec{B}^{t} {A}^{\ast} {A} \vec{B}
+ \beta^{(4)}
   |\vec{B}|^{2}
   \vec{B}^{t} {A}^{\ast} {A} \vec{B},
\label{eq:eff_pot_B4w2_coefficient02_f}
\\
   f_{4}^{(\le2)}[{A}]
&=&
  \gamma^{(2)}
  \Bigl(
       - 2 \, |\vec{B}|^{2} \bigl(\mathrm{tr} \, {A}^{2} \bigr) \bigl(\mathrm{tr} \, {A}^{\ast 2} \bigr)
       - 4 \, |\vec{B}|^{2} \bigl(\mathrm{tr} \, {A}^{\ast} {A} \bigr)^{2}
      + 4 \, |\vec{B}|^{2} \bigl(\mathrm{tr} \, {A}^{\ast} {A} {A}^{\ast} {A} \bigr)
      + 8 \, |\vec{B}|^{2} \bigl(\mathrm{tr} \, {A}^{\ast 2} {A}^{2} \bigr)
        + \vec{B}^{t} {A}^{2} \vec{B} \bigl(\mathrm{tr} \, {A}^{\ast 2} \bigr)
            \nonumber \\ && \hspace{2em}
       - 8 \, \vec{B}^{t} {A}^{\ast} {A} \vec{B} \bigl(\mathrm{tr} \, {A}^{\ast} {A} \bigr)
         + \vec{B}^{t} {A}^{\ast 2} \vec{B} \bigl(\mathrm{tr} \, {A}^{2} \bigr)
      + 2 \, \vec{B}^{t} {A} {A}^{\ast 2} {A} \vec{B}
      + 2 \, \vec{B}^{t} {A}^{\ast} {A}^{2} {A}^{\ast} \vec{B}
       - 8 \, \vec{B}^{t} {A}^{\ast} {A} {A}^{\ast} {A} \vec{B}
            \nonumber \\ && \hspace{2em}
       - 8 \, \vec{B}^{t} {A}^{\ast 2} {A}^{2} \vec{B}
  \Bigr),
\label{eq:eff_pot_B2w4_coefficient02_f}
\end{eqnarray}
with the derivative $\nabla_{i}$ for the spatial direction $i=1,2,3$ and the external magnetic field $\vec{B}$.
The coefficients in the GL free energy are obtained as 
\begin{gather}
  K^{(0)}
=
   \frac{7 \, \zeta(3)N_{\rm F} p_{\rm F}^{4}}{240m^{2}(\pi T_{\mathrm{c}})^{2}},
\quad
   \alpha^{(0)}
=
   \frac{N_{\rm F}p_{\rm F}^{2}}{3} \log\frac{T}{T_{\mathrm{c}}},
\nonumber \\ 
  \beta^{(0)}
=
   \frac{7\,\zeta(3)N_{\rm F}p_{\rm F}^{4}}{60\,(\pi T_{\mathrm{c}})^{2}},
\quad
   \beta^{(2)}
=
    \frac{7\,\zeta(3)N_{\rm F}p_{\rm F}^{2}\gamma_{\mathrm{n}}^{2}}{48(1+G^{({\rm n})}_0)^{2}(\pi T_{\mathrm{c}})^{2}},
\nonumber \\ 
   \beta^{(4)}
=
    - \frac{31\,\zeta(5)N_{\rm F}p_{\rm F}^{2}\gamma_{\mathrm{n}}^{4}}{768(1+G^{({\rm n})}_0)^{4}(\pi T_{\mathrm{c}})^{4}}, \quad
  \gamma^{(0)}
=
   - \frac{31\,\zeta(5)N_{\rm F}p_{\rm F}^{6}}{13440\,(\pi T_{\mathrm{c}})^{4}},
\nonumber \\ 
  \gamma^{(2)}
=
  \frac{31\,\zeta(5)N_{\rm F}p_{\rm F}^{4}\gamma_{\mathrm{n}}^{2}}{3840(1+G^{({\rm n})}_0)^{2}(\pi T_{\mathrm{c}})^{4}},
\quad
  \delta^{(0)}
=
  \frac{127\,\zeta(7)N_{\rm F}p_{\rm F}^{8}}{387072\,(\pi T_{\mathrm{c}})^{6}}. 
\label{eq:GL_coefficients}
\end{gather}
Here we define $N_{\rm F}=mp_{\rm F}/(2\pi^{2})$ for the density-of-state of the neutrons on the Fermi surface for the neutron mass $m$ and the Fermi momentum $p_{\rm F}$.
$\zeta(n)$ is the zeta function.
The GL free energy~\eqref{eq:GL_free_energy} with the coefficients in Eq.~\eqref{eq:GL_coefficients} is deduced analytically by the one-loop approximation for the effective action from the BdG equation.
In the derivation, the particle-hole symmetry is supposed as an approximation for simplicity which should be valid at sufficiently high density.
In the above expression, we consider that the neutron magnetic moment $\vec{\mu}_{n}$ in vacuum should be replaced to $\vec{\mu}_{n}^{\ast}=\gamma_{\mathrm{n}}\vec{s}/(1+G^{({\rm n})}_0)$ with a neutron spin $\vec{s}$ and the neutron gyromagnetic ratio $\gamma_{\mathrm{n}}$ due to the modification by the Landau parameter $G^{({\rm n})}_0$ in the Fermi liquid theory at finite density.


In the GL free energy~\eqref{eq:GL_free_energy}, each term has the following meanings.
At ${\cal O}(A^{2})$, the first term ($K^{0}$) is the kinetic term and the second term ($\alpha^{(0)}$) indicates the leading interaction.
At ${\cal O}(A^{4})$, the interaction term ($\beta^{(0)}$) exhibits the $\mathrm{SO}(5)$ symmetry as the dynamical symmetry which is absent in the BdG theory.
This symmetry appears accidentally up to this order, and it generates the quasi Nambu-Goldstone (qNG) bosons \cite{Uchino:2010pf}. 
The $\mathrm{SO}(5)$ symmetry is, however, explicitly broken by the term ($\gamma^{(0)}$) at ${\cal O}(A^{6})$, and hence the qNG bosons disappear finally.
Nevertheless, we cannot stop the expansion up to ${\cal O}(A^{6})$, because the GL free energy up to this order has only the local minimum but no global minimum.
This problem is solved by the term ($\delta^{(0)}$) at ${\cal O}(A^{8})$ by which the global minimum is realized successfully~\cite{Yasui:2019unp}.
It is important that the terms up to ${\cal O}(A^{8})$ reproduce qualitatively the position of the CEP that is already known in the BdG theory (cf.~Fig.~\ref{fig:240804_phase_diagram}).
Thus, the GL free energy up to ${\cal O}(A^{8})$ supplies the minimum set of terms.
The complexity of the GL equation for the neutron $^{3}P_{2}$ superfluids is qualitatively different from the case of usual S-wave superfluids with second-order phase transition in which usually the terms up the quadratic order are enough.

In Eq.~\eqref{eq:GL_free_energy}, the magnetic terms are calculated not only at the leading order (LO) ($\beta^{(2)}$)  ${\cal O}(B^{2}A^{2})$ but also at the next-to-leading orders (NLO) ($\beta^{(4)}$ and $\gamma^{(2)}$)  ${\cal O}(B^{4}A^{2})$ and ${\cal O}(B^{2}A^{4})$~\cite{Yasui:2018tcr}.
The NLO terms for the magnetic field are necessary in order for checking the convergence of the expansion series in strong magnetic fields such as in magnetars.

The minimization of the GL free energy with respect to the order parameter $A$ provides the phase diagram of the neutron $^{3}P_{2}$ superfluids as shown, e.g., in Refs.~\cite{Yasui:2018tcr,Yasui:2019unp,Yasui:2019pgb,Mizushima:2019spl,Mizushima:2021qrz}.
The GL equation up to ${\cal O}(A^{8})$ reproduces the general features of the phase diagram obtained in the BdG equation (cf.~Fig.~\ref{fig:240804_phase_diagram}).
At quantitative level, however, we notice that the positions of the first-order phase transition line and the CEP in the GL theory are different from the original ones in the BdG theory.
Nevertheless, it should be emphasized that the neutron $^{3}P_{2}$ superfluid phases changing from the UN phase to the $\mathrm{D}_{4}$-BN phase through the $\mathrm{D}_{2}$-BN phase
 can be described by the GL theory successfully.
Therefore, the GL theory can be used to understand the properties of the neutron $^{3}P_{2}$ superfluids inside neutron stars and magnetars.
We comment that the GL free energy~\eqref{eq:GL_free_energy} can be furthermore extended to include the condensate of the neutron $^{1}S_{0}$ superfluids allowing the coexistence of the neutron $^{1}S_{0}$ and $^{3}P_{2}$ phases~\cite{Yasui:2020xqb} for studying a wide range of phases in neutron stars and magnetars.

\subsection{Topological surface defects on neutron star and magnetar} \label{sec:surface_defects}

As introduced in Sec.~\ref{sec:symmetry}, there are many interesting topological properties in the neutron $^{3}P_{2}$ superfluids.
Among them, we focus on the topological defects on the surface of neutron star and magnetars~\cite{Yasui:2019pgb}.
This subject is interesting not only in the astrophysics but also in the the condensed matter physics such as the $^{3}$He superfluids and liquid crystals~\cite{Mermin:1979zz,Urbanski_2017,PhysRevResearch.6.013046}.
In this subject, we are interested in the properties of the neutron $^{3}P_{2}$ superfluids at the surface, not in bulk, and thus it is important to consider the appropriate boundary condition for the neutron $^{3}P_{2}$ faced to the outer space.

At the surface, the directions of the neutron $^{3}P_{2}$ superfluids are constrained due to the anisotropy of space.
Let us consider for simplicity that the boundary region is regarded approximately as the flat plane by neglecting the curvature~\cite{Yasui:2019pgb}.
Such approximation will be valid for the surface of a neutron star, because the scale length of the coherent length of the neutron $^{3}P_{2}$ pairings is much smaller than the scale size for curvature of the star surface.
Thus, the boundary condition is characterized by the normal vector $\vec{n}$ whose direction is perpendicular to the surface.
Here the positive direction of $\vec{n}$ is defined to be outward from the inside to the outside.

In our setting for the boundary condition, we suppose that the outer space of the neutron $^{3}P_{2}$ superfluids is simply a vacuum rather than the other matter phase, such as the neutron $^{1}S_{0}$ superfluids.
In reality, we may need to consider that the neutron $^{3}P_{2}$ superfluid may be faced to the neutron $^{1}S_{0}$ superfluids, because the $^{1}S_{0}$ superfluids should be realized at lower density region near the surface of the neutron stars (cf.~Ref.~\cite{Yasui:2020xqb}).
Thus, the present simple setting should be regarded as being to the first step to investigate the surface effect.
Regardless of the simple approximation, we will show that there are rich phenomena induced by topology on the surface.

We consider the constraint condition for the condensate matrix $A$ at the surface:
\begin{align}
   \vec{n}^{t}A\vec{n} = 0,
\label{eq:constraint_condition_surface}
\end{align}
where $\vec{n}$ is the normal vector perpendicular to the surface.
Here $\vec{n}^{t}$ is a transpose of $\vec{n}$.
In addition, we define the two-dimensional vector
\begin{align}
   \vec{A}_{\vec{n}}=\vec{n}^{t}A,
\end{align}
which satisfies the condition $\vec{A}_{\vec{n}} \!\cdot\! \vec{n} = 0$ from Eq.~\eqref{eq:constraint_condition_surface}.
Thus, $\vec{A}_{\vec{n}}$ is a two-dimensional vector field confined on the surface.
The two-dimensional vector $\vec{A}_{\vec{n}}$ exhibits an interesting behavior on the surface: it induces the topological defects as vortices.
Such defects emerge from the property that the boundary condition~\eqref{eq:constraint_condition_surface} has the symmetry $(\mathrm{S}^{1} \times \mathrm{S}^{1})/\mathbb{Z}_{2}$: the invariance under the one-axis rotation around the normal vector $\vec{n}$ and the global $\mathrm{U}(1)$ phase in the complex matrix $A$.

\begin{figure}[htb]
\includegraphics[keepaspectratio, scale=0.5]{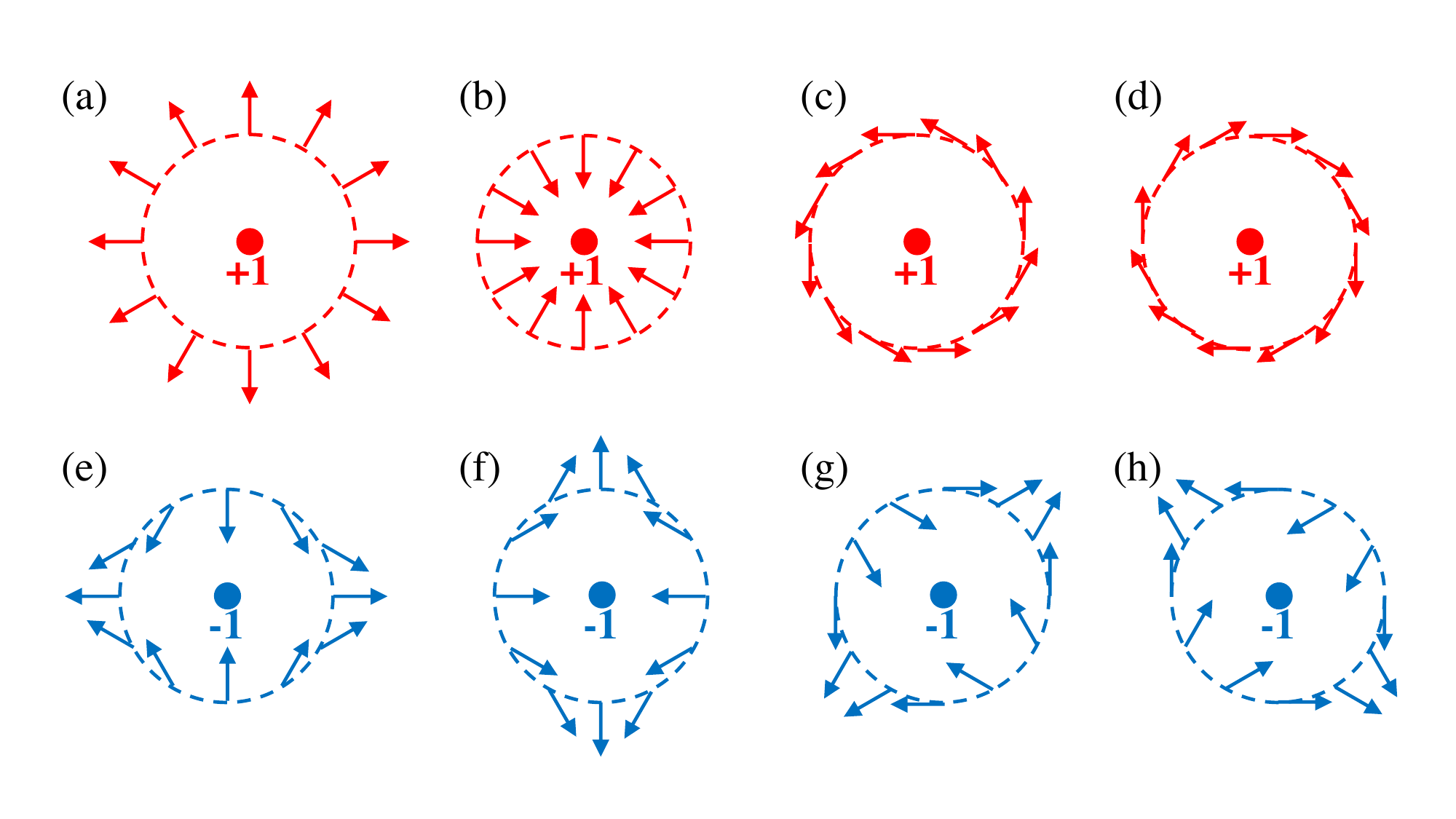} %
\caption{
The schematic figure for the topological charges is shown for the two-dimensional vector fields on the surface. The panels (a)-(d) indicate $+1$ charge, and the panels (e)-(h) indicate $-1$ charge. See the text for the details. 
\label{fig:240805_vortices}}
\end{figure}

Here we note that, in general, there are two different types of topological surface defects with topological charges $\pm1$, 
 see Fig.~\ref{fig:240805_vortices}.
The vector-field configurations with topological charges can be deformed smoothly.
As a result, the defects with the same charge are not distinguished from the topological point of view, while the configurations with different topological charges cannot be transformed without creating singular points.
Thus, the charges of topological defects on the surface should be classified to either $+1$ or $-1$.
Therefore, it should be expected that the two-dimensional vector $\vec{A}_{\vec{n}}$ should exhibit the topological defects as well.

The solutions of the condensate matrix $A$ under the boundary condition~\eqref{eq:constraint_condition_surface} can be obtained numerically.
Similarly to Eq.~\eqref{eq:order_parameter_general}, we express $A$ by
\begin{align}
   A(x)
= \left(
   \begin{array}{ccc}
    f_{1}(x) & g_{3}(x) & g_{2}(x) \\
    g_{3}(x) & f_{2}(x) & g_{1}(x) \\
    g_{2}x) & g_{1}(x) & -f_{1}(x)-f_{2}(x)  
   \end{array}
   \right),
\label{eq:order_parameter_one_dimensional}
\end{align}
where the one-dimensional axis ($x \ge 0$) is defined as the half line.
This half line is directed from the surface of the neutron star ($x=0$) to the bulk space in center of a neutron star ($x \rightarrow \infty$).
Because the neutron stars are too large in comparison with the coherence length of the neutron $^{3}P_{2}$ superfluids, the large limit of $x$ can be regarded as an infinity. 
The boundary condition at the infinity is supplied by the bulk phases, i.e., the UN, $\mathrm{D}_{2}$-BN and $\mathrm{D}_{4}$-BN phases, according to the strengths of the magnetic fields, as summarized in Table~\ref{tbl:phase_magnetic_field}.
We observe, as a matter of course, that the condensate matrix $A$ are dependent on the distance from the surface.
Some examples of the behaviors of the solutions are shown in Ref.~\cite{Yasui:2019pgb}.
We notice that the off-diagonal components, $g_{1}$, $g_{2}$ and $g_{3}$, in Eq.~\eqref{eq:order_parameter_one_dimensional} cannot be eliminated by global transformation in general due to the boundary condition~\eqref{eq:constraint_condition_surface}.
Indeed, not only the diagonal components, $f_{1}$ and $f_{2}$, but also the off-diagonal components, $g_{1}$, $g_{2}$ and $g_{3}$, play important roles for generating the topological surface defects.
They all together contribute to form the two-dimensional vector field $\vec{A}_{\vec{n}}$. 

With the above setup, the configurations of $\vec{A}_{\vec{n}}$ can be presented on an ``atlas" of the neutron star surface, where the Mercator projection is adopted with the longitude ($\varphi$) and the attitude ($\theta$)~\cite{Yasui:2019pgb}.
We consider zero, weak and strong strengths of the magnetic fields, in which the corresponding bulk phase is given by the UN phase, the $\mathrm{D}_{2}$-BN phase and the $\mathrm{D}_{4}$-BN phase at the center of neutron star, see Table~\ref{tbl:phase_magnetic_field}.
The vector stream lines on the surface are different to each magnetic field (or the bulk phase), as presented in Ref.~\cite{Yasui:2019pgb}.
We notice that there exist several topological defects with charges $\pm1$, regarded as the vortices, induced by $\vec{A}_{\vec{n}}$.
It is interesting that the spatial distributions of the topological charge on the surface are changed by tuning the strengths of the magnetic field, as summarized in Table~\ref{tbl:surface_defects}.
Correspondingly, the surface defects have $+10$ and $-8$ charges in the case of the UN phase,
$+12$ and $-10$ charges in the case of the $\mathrm{D}_{2}$-BN phase and $+8$ and $-6$ charges in the case of the $\mathrm{D}_{4}$-BN phase.
In any case, the total charges are always $+2$ irrespective to the symmetries in the bulk phases.

The total charge $+2$ is not accidental, because the total charges of defects of two-dimensional vector fields on closed manifolds should be determined solely by the Euler characteristic of the closed manifold.
This is the Poincar\'e-Hopf theorem known as the hairy ball theorem.
Precisely, this theorem is stated as
\begin{align}
   \sum_{p\in M} \mathrm{index}_{p} \, \vec{v} = \chi(M),
\end{align}
for the closed manifold $M$ and the vector field $\vec{v}$ on $M$.
In the left-hand side, $p \in M$ is the points on $M$ and $\mathrm{index}_{p} \, \vec{v}$ means the index zero-point of $\vec{v}$ at $p$, indicating charges $\pm1$ in our case.
Thus, the left-hand side presents the sum of the indices on $M$.
In the right-hand side, $\chi(M)$ is the Euler characteristic for the manifold $M$, giving $\chi=2$ in the case of a ball-like 
 neutron stars and magnetars.
Therefore, the total charges of the two-dimensional vector fields on 
the sphere
should be always two.
This theorem is applied to the case of the neutron $^{3}P_{2}$ superfluids, where the two-dimensional vector field is supplied by $\vec{A}_{\vec{n}}$, and the total charge $+2$ is assured always.
In terms of the dynamics, it is interesting to observe pair-creations and annihilations of the defects on the surface of neutron stars:
two pairs of defects and anti-defects are created from zero magnetic field to weak magnetic field, and four pairs of defects 
and anti-defects are annihilated from weak magnetic field to strong magnetic field.
Notice that a pair of defects with $+1$ and $-1$ charges should be simultaneously created and/or annihilated to conserve the total charge.

\begin{table}[tb]
\caption{The numbers of the topological surface defects are summarized for different bulk phase at center of neutron stars under different strengths of magnetic fields~\cite{Yasui:2019pgb}. See the text for the details.}
\begin{center}
\begin{tabular}{cccc}
\hline
   magnetic field strength & zero & weak & strong \\
\hline
   bulk phase at center & UN & $\mathrm{D}_{2}$-BN & $\mathrm{D}_{4}$-BN \\
   positive charge & 10 & 12 & 8 \\
   negative charge & -8 & -10 & -6 \\
\hline
\end{tabular}
\end{center}
\label{tbl:surface_defects}
\end{table}%

Summarizing this section, considering the bulk phase and the boundary condition induces the topological surface defects whose total charge is conserved by the Poincar\'e-Hopf theorem.
This also has impact on the astrophysical research: the correspondence between the distributions of defects on the surface and the bulk phase at the center makes it possible to research the internal bulk phase by observing the defects on its surface.
Such topological surface defects may be investigated by observing neutron stars and magnetars.

\subsection{Summary} 
\label{sec:summary}

In this section, we have reviewed recent progress on the neutron $^{3}P_{2}$ superfluids in neutron stars and magnetars 
focusing on the surface defects.
The research of the neutron $^{3}P_{2}$ superfluids is still producing new results and phenomena which should be shared not only with researchers in astrophysics and nuclear physics but also with those in condensed matter physics.
Especially the topology is an important key word to understand the properties of the neutron $^{3}P_{2}$ superfluids.
It is expected that neutron stars and magnetars are researched in terms of the topological stars in the coming future.

\section{Neutron star kicks: rocket effects in strong B-fields}
\label{sec:blaschke}

\subsection{Introduction}
Pulsars are rotating neutron stars with high magnetic fields causing observable
radio dipole signals.
Most of the known pulsars are born in the neighborhood of the galactic plane 
and move away from it with natal kick velocities which are typically higher 
than those of their progenitors \cite{Lyne:1994az}.
This implies that the birth process of pulsars also produces their high 
velocities and thus cannot be entirely isotropic \cite{Spruit:1998sg}.
Two examples for directly observed pulsar motion are the isolated neutron star
RX J1856.5-3754 and for the binary system B0950+08.
The spatial distribution of observed pulsar kick velocities in galactic coordinates
is given in \cite{Hobbs:2005yx}. 
Up to now, the mechanisms driving this asymmetry are far from clear.
There are several hypotheses, ranging from 
asymmetric supernova explosions \cite{Lai:2000pk} over neutron star instabilities
\cite{Colpi:2002cu,Imshennik:2003ne} and magnetorotational effects 
\cite{2003astro.ph.10142M,2004Ap.....47...37A} to the model of an electromagnetic \cite{Agalianou:2023lvv} or neutrino rocket \cite{Lai:2000pk,Berdermann:2006rk,Sagert:2007as,Kaminski:2014jda}.
For a recent overview, see \cite{Bombaci:2004nu,Ayala:2024wgb} and references therein.
It is also not clarified whether the distribution of pulsar kick 
velocities is bimodal with a lower component of $v \leq 100 {\rm km/s}$ 
(20 \% of the known objects) and a higher component of  
$v \geq 500 {\rm km/s}$ (80\%) as suggested by \cite{Arzoumanian:2001dv} or 
whether it can be explained by a one-component distribution 
\cite{Hobbs:2005yx}. 
Note, that the high velocity tail is possibly underrepresented, due to the 
fact that pulsars with high velocity move out of the observational area 
faster than lower ones \cite{Cordes:1997mm}. 
Most of the models are capable of explaining kick velocities of
$v \sim 100 ~{\rm km/s}$, but it is a nontrivial problem to explain
the highest measured pulsar velocities around $1600 ~{\rm km/s}$.

While there is evidence for the alignment of kick velocity vectors with the
rotation axis of pulsars \cite{Johnston:2005ka} the suggestion of a correlation
between magnetic field and kick velocity \cite{Spruit:1998sg,Bisnovatyi-Kogan:1993ljj}
could not be verified from observations so far.

In this section we show that beamed neutrino radiation from a
strongly magnetized strange star in a color superconducting state could lead
to an acceleration during the early cooling phase of the neutron star.
It has recently been suggested that the typical properties of
gamma ray bursts might be explained by the neutrino release off a strongly
magnetized quark core star \cite{2006sqmc.conf..377A,Drago:2005qb} since neutrinos can 
very effectively be converted into $e^+e^-$ pairs and gammas 
\cite{Haensel:1991um}.
Here we show that such a collimated neutrino release (neutrino rocket) can
lead to high  neutron
star kick velocities in qualitative agreement with the observational data.
The key ingredient to this mechanism are: (i) the mass defect of the order of 
$10^{52}$ erg in the formation of a (color superconducting) quark star
\cite{Aguilera:2002dh,Blaschke:2005uj},
(ii) the anisotropic neutrino transport (beaming) since normal quark matter 
vortices (short mean free path) aligned with the strong magnetic field are
immersed in a superconducting matrix (free neutrino propagation) and (iii)   
parity nonconservation of neutrino production in the strong magnetic field
 \cite{Vilenkin:1995um,Horowitz:1997mk} which leads to a sufficiently large 
net momentum transfer to the star resulting in a pulsar kick. 
The color superconductivity plays an essential role since those neutrinos 
generating the propulsion of the star propagate without secondary interaction 
thus circumventing the {\it no-go theorem} 
\cite{Vilenkin:1995um,Kusenko:1998yy}, that there is no parity violation in 
equilibrium. In the remainder of this contribution, we explain the ingredients 
of the model and present results of a model calculation \cite{Berdermann:2006rk}. 

\subsection{Neutrino beaming by magnetic vortices and cooling delay}
When designing a scenario for the origin of natal pulsar kicks it is natural to
make contact to other puzzling phenomena occuring in the realm of a supernova 
explosion such as gamma ray bursts (GRBs). 
It has been suspected that GRB charcteristics could find an explanation 
in the phase transition to quark matter in the protoneutron star 
(see \cite{2006sqmc.conf..377A,Drago:2005qb} and Refs. therein) 
whereby the phase with all quarks being paired with large gaps in a 
{\it color-flavor-locking (CFL)} state is
of particular interest for our concern of the neutrino propagation. 
Let us, in the following, assume that the bulk of the star is in the CFL phase
with a tiny hadronic crust and a surface magnetic field in excess of  
$B_s ~ \sim 10^{12}~ {\rm G}$. Then, one may expect by flux conservation that
the inner magnetic field at the quark core surface can reach values of
$B_{\rm in,s}= B_s(n_{in}/n_s)^{2/3}=10^{12}-10^{17}~{\rm G}$, where 
$n_s \sim 6\times 10^{-10}~ {\rm fm^{-3}}$ and $n_{in}\sim 0.3~ {\rm fm^{-3}}$ 
are the corresponding densities at the core-crust interface.
The presence of such high magnetic fields allows creation of a vortex structure
along the magnetic axis  \cite{Blaschke:1999fy} as shown in Fig. \ref{fig:db1}.

\begin{figure}[tb]
\includegraphics[width=\textwidth]{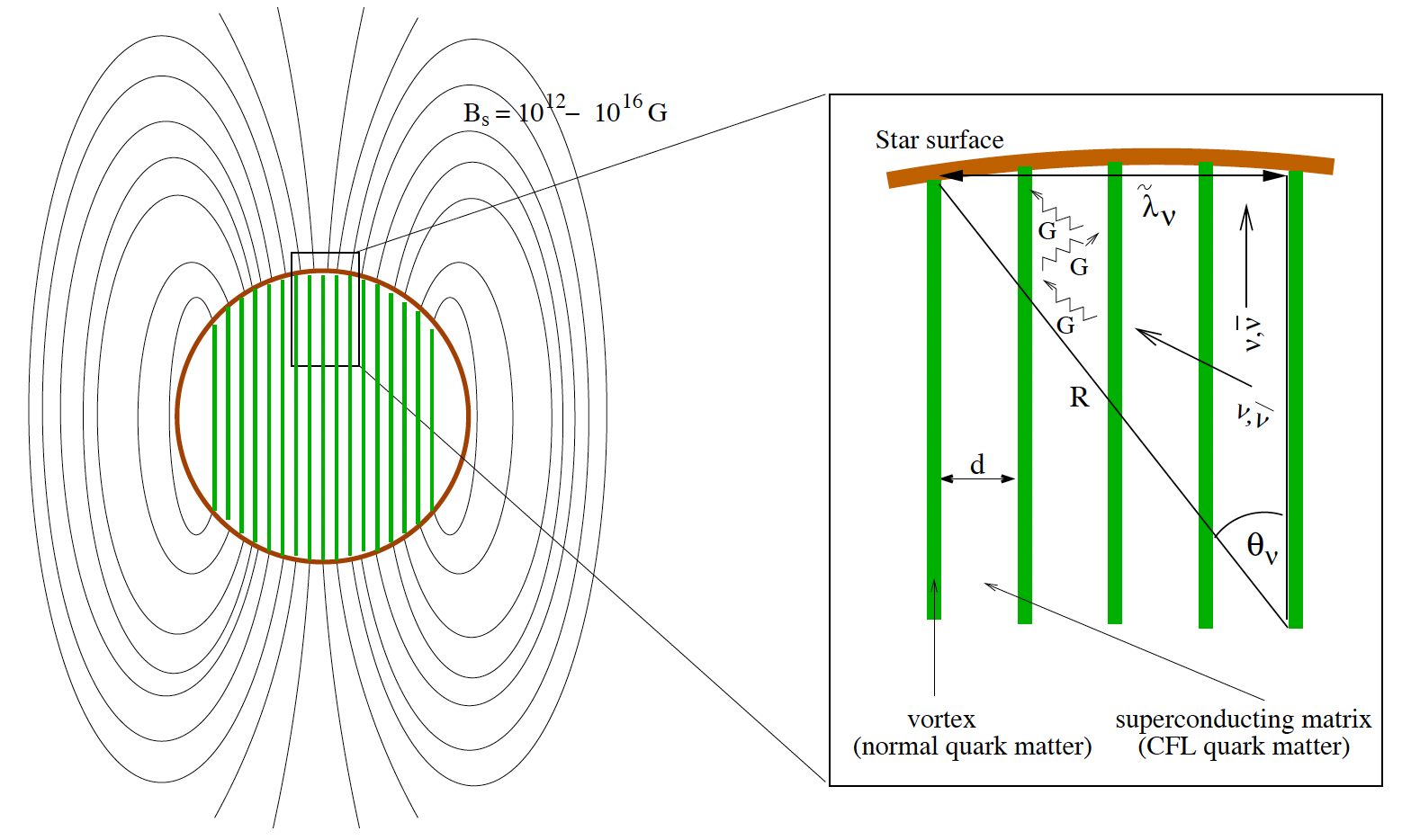}
\caption{Vortex structure in highly magnetized superconducting quark stars.
\label{fig:db1}}
\end{figure}

Due to magnetic flux conservation the  number of vortices in the quark star 
interior is
$N_{vo} = \pi B_{in,s} R^2/ (6\Phi_0)$,
where $\Phi_0 =2\cdot 10^{-7}~$G$~ \mbox{cm}^2$ is the magnetic flux quantum, 
cf. \cite{Blaschke:1999fy,Blaschke:1999qx,Sedrakian:2000kw}.
Superconductivity is expelled from the vortex interior, where the matter is 
in the state of a strongly magnetized quark-gluon plasma.
The volume of a vortex can be estimated as $V_{vo}\sim 2 \pi \lambda_D^2~R$, 
where the Debye screening length $\lambda_D$ is of the order of the 
penetration depth
\begin{eqnarray} \label{la}
\lambda_{{\rm CFL}} &\simeq&
({\mu_q}/{300~{\rm MeV}})^{-1}
(1-{T}/{T_c})^{-1/2}~{\rm fm}, 
\end{eqnarray}
and for the critical temperature the BCS relation,  
$T_c = 0.57~\Delta $, can be adopted.
The vortex volume has a large neutrino emissivity due to the quark direct 
Urca (QDU) process \cite{Iwamoto:1980eb}
\begin{equation} \label{QDU}
\epsilon^{QDU}_{\nu} \simeq 2.2 \times 10^{26}~\alpha_s~u~Y_e^{1/3}~T^6_9~ 
{\rm erg~cm^{-3}~s^{-1}},
\end{equation} 
with $\alpha_s\sim 1$ being the strong coupling constant, the compression 
$u=n_b/n_0$ is the factor by which the baryon density exceeds its value at 
saturation $n_0=0.16$ fm$^{-3}$, $Y_e$ is the electron fraction and $T_9$ the 
temperature in units of $10^9$ K. 
Correspondingly, the neutrino mean free path (MFP) in normal quark matter 
becomes very small
\begin{equation} \label{MFP}
\lambda_{\nu} = 
3.4\times 10^6~\alpha_s^{-1}~u^{-1}~Y_e^{-1/3}~T_9^{-2}~{\rm cm}. 
\end{equation}
Note that the direct Urca processes in quark matter, which is the responsible 
for the high neutrino production inside the star, remain virtually unaltered 
in the presence of magnetic fields in this range \cite{Bandyopadhyay:1998aq}.
In the region between the vortices matter is in the CFL state with pairing 
gaps $\Delta \sim 100$ MeV, so that the direct Urca process is suppressed by
the exponential factor $\zeta_{QDU}=\exp(-\Delta/T)$ and the neutrino MFP 
$\lambda_{\nu}^{CFL}=\zeta_{QDU}^{-1}\lambda_{\nu}$ exceeds the
radius of the star for temperatures below  $T_{opac}^{CFL}\approx 15$ MeV.
This prominent difference between the neutrino MFP in normal (vortex core) and
in CFL quark matter (matrix) leads to anisotropic neutrino emission from the
star within a cone of the temperature-dependent opening angle 
$\theta_{\nu}(T) \sim \tilde{\lambda}_{\nu}(T)/R$, where
$\tilde{\lambda}_{\nu}(T)=\lambda_{\nu}(T)~V /(N_{vo}~V_{vo})$, see Fig. 
\ref{fig:db2}. 

\begin{figure}[htb]
\centering
\includegraphics[width=0.8\textwidth]{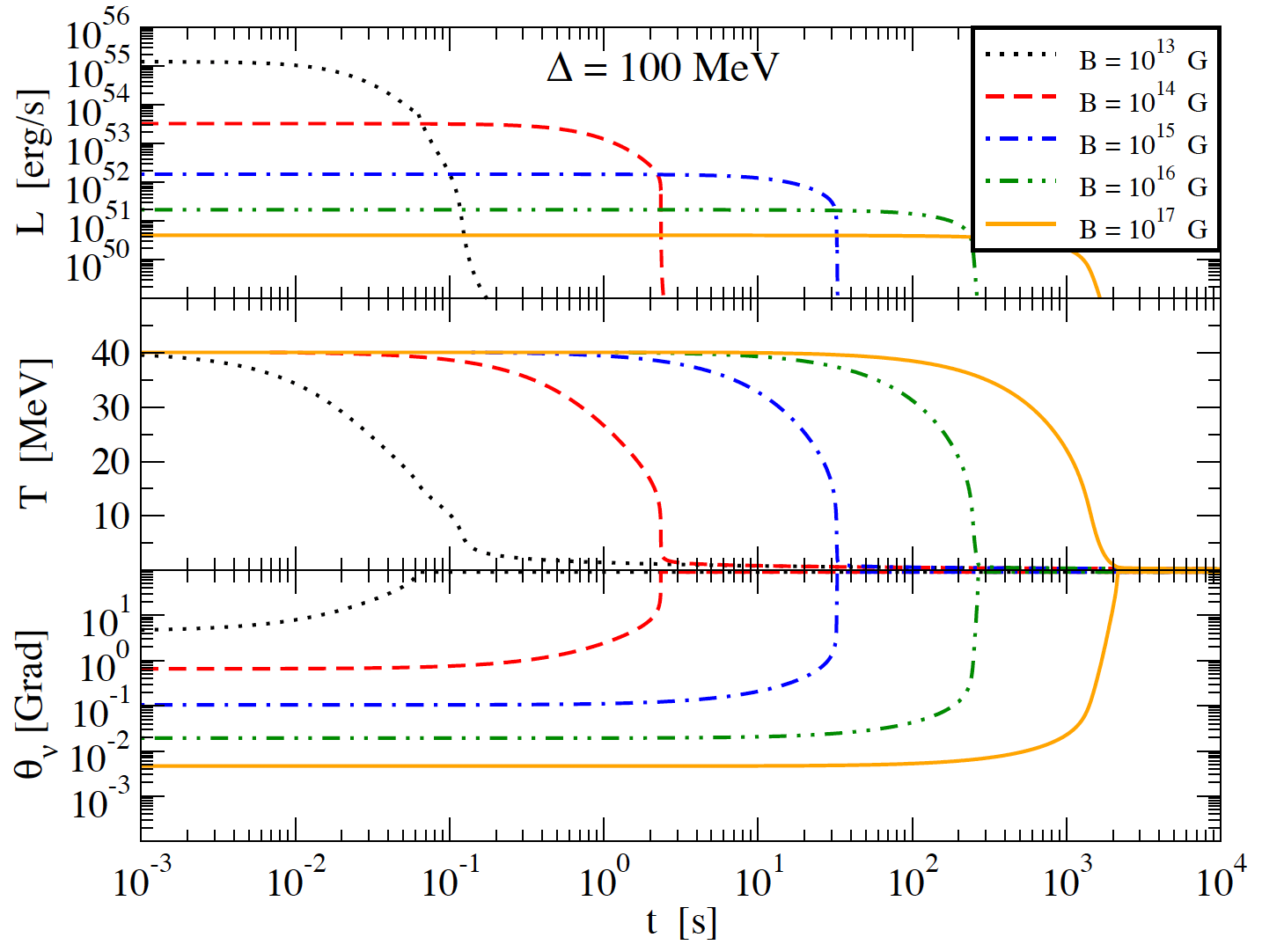}
\caption{Time evolution of neutrino luminosity, temperature and neutrino 
beaming angle.
\label{fig:db2}}
\end{figure}

The cooling evolution $T(t)$ can be described by inverting the solution of
\begin{equation} 
t = -\int\limits_{T_i}^{T(t)} \frac{C_V(T')~{\rm d}T'}{L(T')}
\end{equation}
where $T_i\sim 40$ MeV is the initial temperature of the protoneutron star,
$C_V(T) = [1-{N_{vo}V_{vo}}/{V}]~C_V^q(T)~\zeta_{DU}$
is the specific heat dominated by the normal quark matter contribution, and
the luminosity
\begin{equation}\label{solL}
L (T)= [1-{\rm cos} ~\theta_{\nu}(T)] L_0(T) \simeq
L_0(T) \theta_{\nu}(T)^2 /2 
\end{equation}
takes into account the effect of the neutrino beaming.
The isotropic luminosity
\begin{eqnarray}\label{f7.9+}
L_0(T) &=& ({N_{vo}V_{vo}}/{V}) \int dV \epsilon_{\nu}^{QDU}(T)
+ \left[1-({N_{vo} V_{vo}}/{V})\right] \int dV \epsilon_{\nu}^{QDU}(T)~\zeta_{QDU}
\end{eqnarray}
has contributions from both, the unsuppressed QDU process from the small volume
of the vortices and the suppressed contribution from the large CFL matrix 
volume.
Results for the time evolution of neutrino luminosity, temperature and neutrino
beaming angle shown in Fig. \ref{fig:db2} demonstrate the effect of cooling 
delay due to neutrino beaming in a strong magnetic field.

\subsection{Pulsar kicks from color superconductivity parity violation}
Parity nonconservation in the weak interaction QDU neutrino production process 
under the conditions of a strong magnetic field in a protoneutron star leads
to a violation of reflection symmetry since the neutrino flux is a polar vector
while the magnetic field is an axial one.
The magnitude of this asymmetry has been estimated as 
$A_\nu\sim 10^{-4}~B_{14}$, where $B_{14}=B/10^{14}$ G, see 
\cite{Vilenkin:1995um} and \cite{Horowitz:1997mk} for a similar result. 
This nonvanishing $A_\nu$ leads to a net momentum transfer from the neutrino
flux to the star of mass $M$ resulting in a time-dependent kick velocity
\begin{eqnarray}
v(t)&=&\frac{A_\nu~c}{M c^2} \int_0^t dt' L_\nu(T)~,
\end{eqnarray}
which saturates at the B-dependent asymptotic value
\begin{eqnarray}
v= 3.6 \times B_{14} (M/M_\odot)^{-1} ~{\rm km~s^{-1}}
\end{eqnarray} 
as soon as the beaming ceases. Numerical results are shown in Fig. \ref{fig:db3}
for a typical compact star with $M=1.4~M_\odot$ and $R=10$ km. 
In order to obtain pulsar kicks as large as $10^3$ km s$^{-1}$, magnetic fields
of the order of $10^{16}-10^{17}$ G in the stars interior are required. 
Such values correspond to our above estimates. 

\begin{figure}[h!]
\centering
\includegraphics[width=0.8\textwidth]{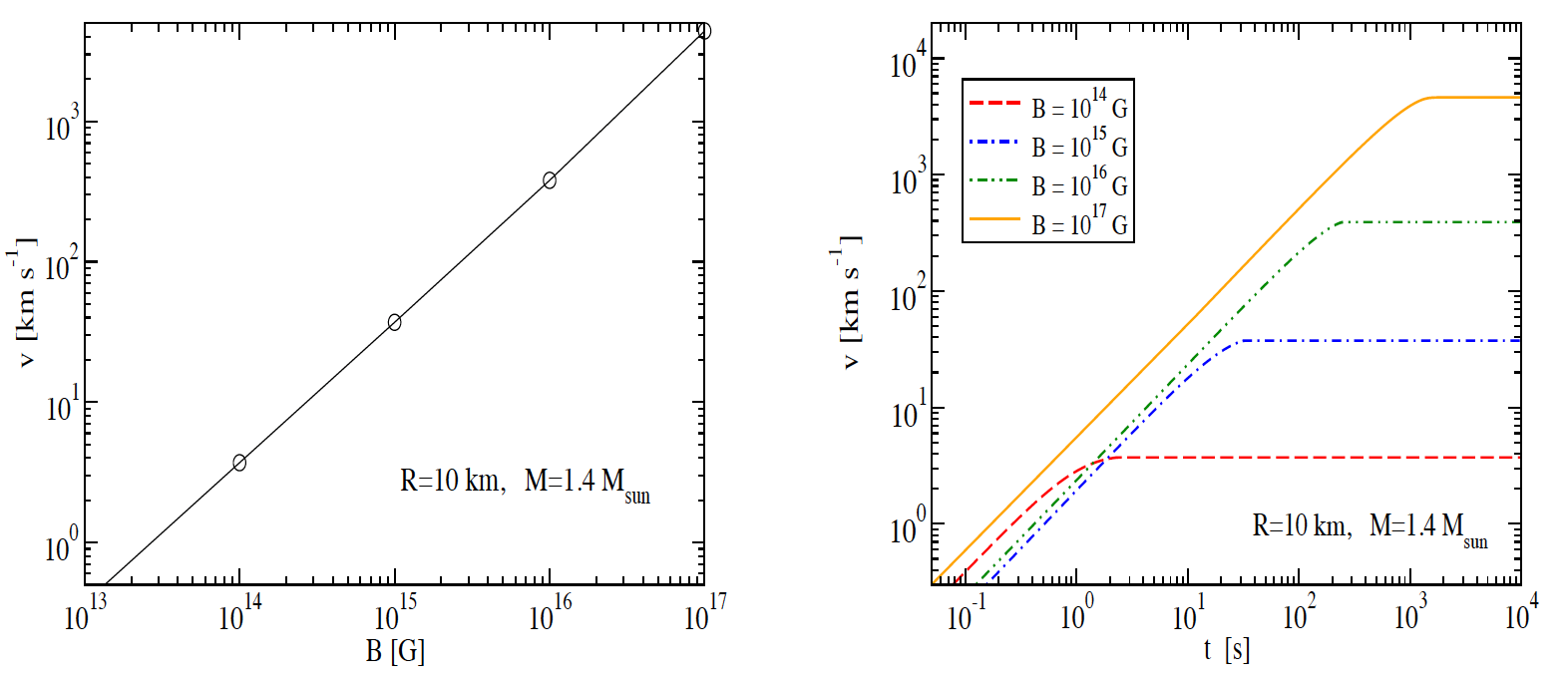}
\caption{Time evolution  of the pulsar kick velocity for different 
magnetic fields (right panel) and magnetic field dependence of the final
kick velocity (left panel).
\label{fig:db3}}
\end{figure}

We have to comment on the argument, that any initial asymmetry in neutrino 
emission processes is only a local phenomenon and can not survive at large 
distances due to strong neutrino scattering inside the neutron star
(no go theorem \cite{Vilenkin:1995um,Kusenko:1998yy}).
Note, that in our case the neutrinos move along the magnetic axis between the 
normal quark matter vortices in areas where the quark matter is in a 
CFL phase and therefore all neutrino interactions are exponentially suppressed 
with the gap in the energy spectrum.
This leads to a strong suppression of scattering processes in this direction.
The initial asymmetry in neutrino emission can therefore survive and lead to 
the observable pulsar kick.

Finally, one can test the predictive power of the neutrino rocket scenario
by testing the correlation between the pulsar mass distribution and the 
kick velocity distribution.

\subsection{Electromagnetic rocket effect}
\label{ssec:em-rocket}

The electromagnetic rocket effect was introduced in \cite{1975ApJ...201..447H} and has recently been revisited by \cite{Agalianou:2023lvv}.
It is based on a displacement $\textbf{s}$ of a strong dipolar magnetic field from the center of the NS (see Fig. \ref{fig:off-center}) and the sufficiently fast rotation of the latter.
As it was discussed in \cite{Agalianou:2023lvv}, observational evidence provided by the \texttt{NICER} (Neutron star Interior Composition ExploreR) X-ray observatory, particularly from investigating the location of hot spots on the surfaces of MSPs \citep{Miller:2019cac, Riley:2019yda}, indicates that the magnetic field diverges from a centered dipole configuration.
Based on the results for PSR J0030+0451, \cite{Kalapotharakos:2020rmz} estimated the magnetic field structure and found an off-centered magnetic field consisting of dipole and quadruple components. 
\begin{figure}[htb]
    \centering
    \includegraphics[width=0.5\linewidth]{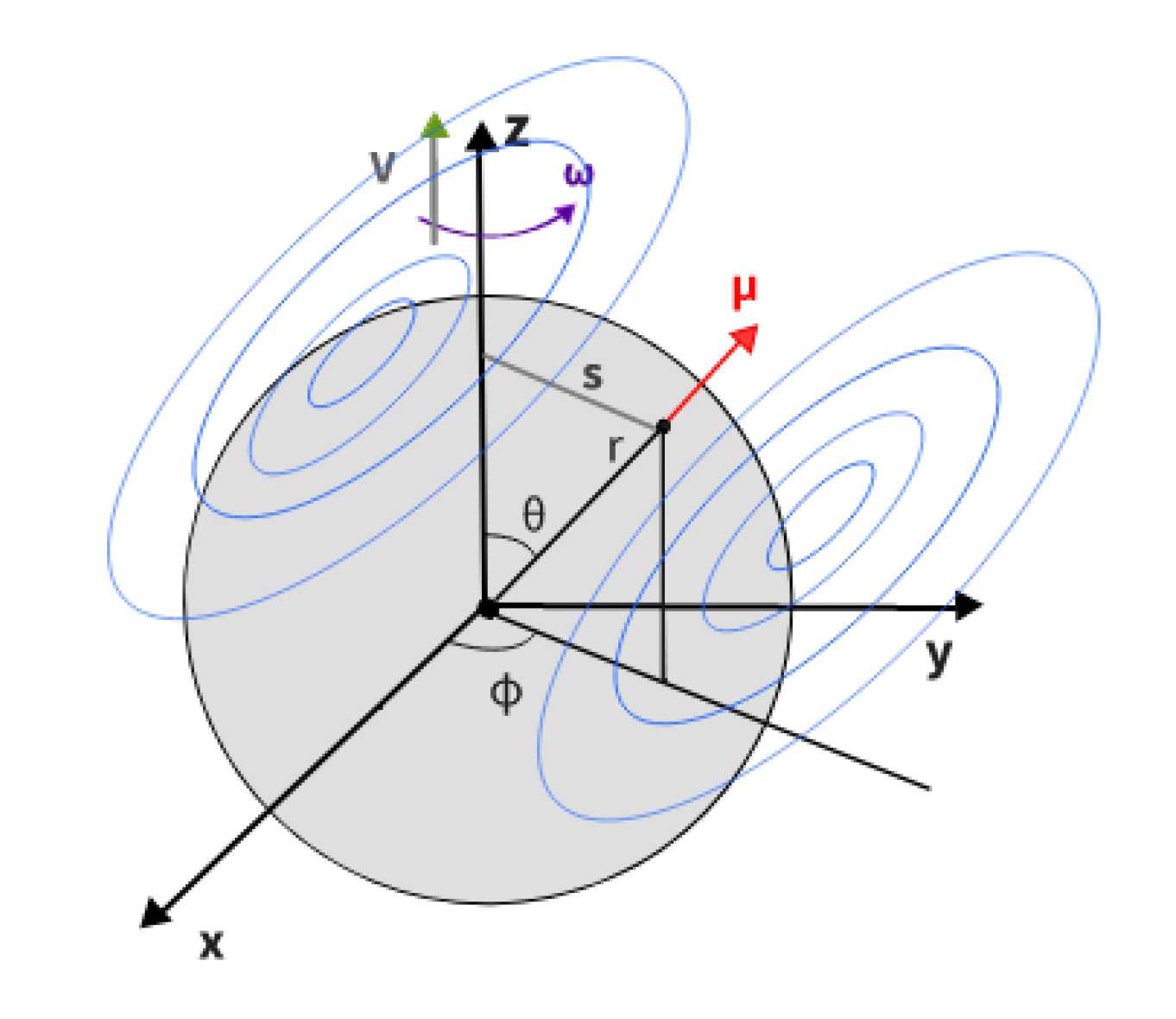}
    \caption{Displacement of the strong dipolar magnetic field inside the NS as a condition for the electromagnetic rocket effect. Figure from Ref. \cite{Agalianou:2023lvv}.}
    \label{fig:off-center}
\end{figure}

Since both the neutrino and electromagnetic rocket mechanisms require strong magnetic fields of at least the young pulsar field strength $B\sim 10^{12}$ Gauss, such effects seem at first glance unlikely for ``old'' MSPs with low magnetic fields of
$B\sim 10^{8}$ Gauss.
However, in systems with mass accretion from a companion star like in the case of LMXBs, the argument of \cite{1989ApJ...346..847C} can be applied that the strong magnetic field of the interior gets buried in the crust of the pulsar \cite{Bhattacharya:1991pre, 2001ApJ...557..958C, 10.1111/j.1365-2966.2004.07397.x}.
Thus, a small surface magnetic field is compatible with a super-strong magnetic field in the interior. 
Due to such a magnetic field profile, kick velocities of a few dozens of km s$^{-1}$ can be produced in MSPs, as it has been demonstrated in Ref. \cite{Agalianou:2023lvv} by applying the electromagnetic rocket effect to the case of PSR J0030+0451 with its true transverse velocity not exceeding $17$ km/s that is obtained by the electromagnetic rocket effect for a displacement $s=3.02$ km and the initial spin period $3.52$ ms.

\subsection{Eccentric and isolated MSPs from accretion-induced twin star transition}

In concluding this section, we discuss a possible application of neutron star kick mechanisms by electromagnetic or neutrino rocket effects for the explanation of the puzzle why a few millisecond pulsars in low-mass X-ray binaries with periods in the range of 20-50 days have orbits with large eccentricities $e\sim 0.1$, while generally the orbits are rather circular with eccentricities following the Phinney line \cite{Phinney:1992}, see Fig. \ref{fig:eccentric}. 

\begin{figure}
    \centering
    \includegraphics[height=0.33\linewidth]{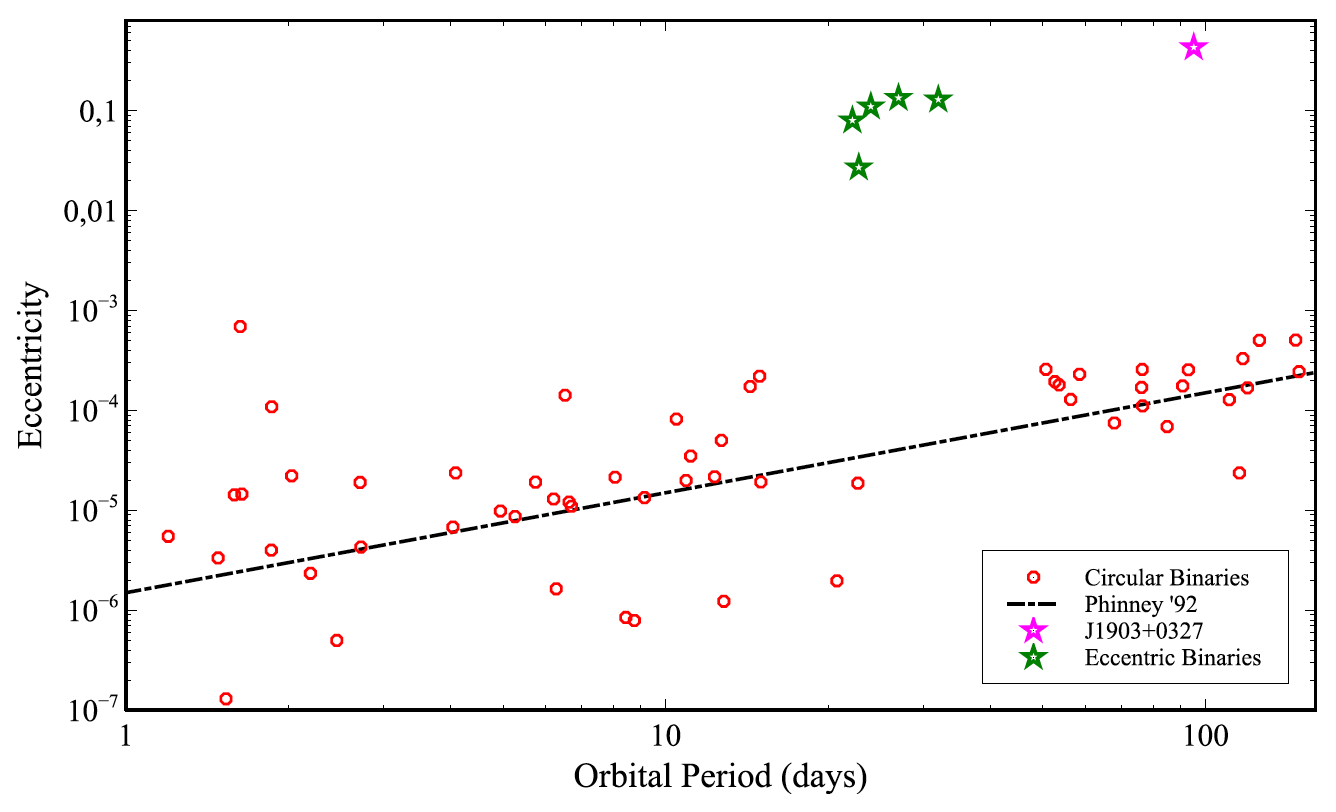}\hfill
    \includegraphics[height=0.34\linewidth]{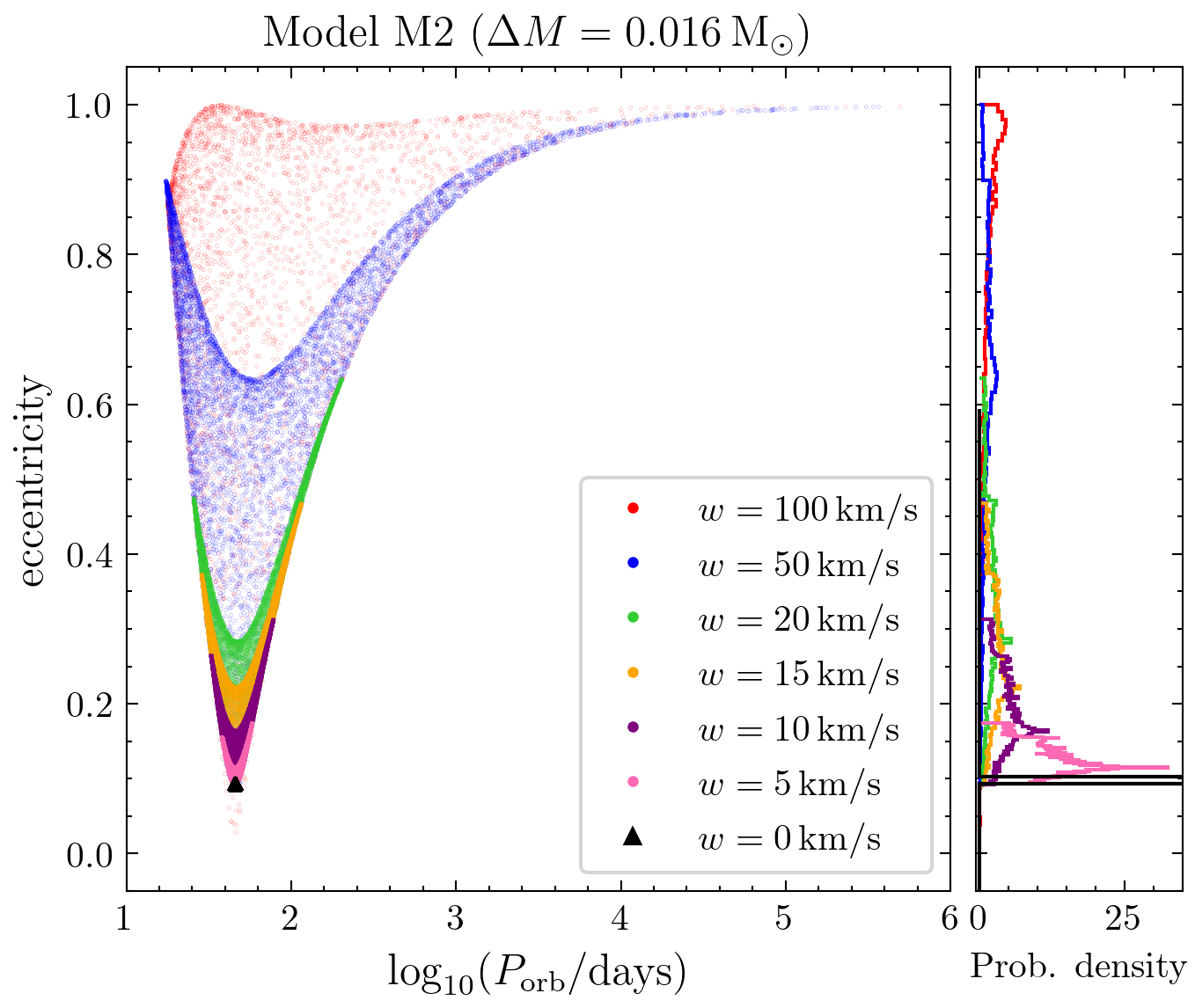}
    \caption{Eccentricity vs. orbital period for millisecond pulsars in binaries with white dwarf companions. Left panel: observations compared with the Phinney line \cite{Phinney:1992}, from Ref. \cite{Alvarez-Castillo:2019apz}. Right panel: Monte-Carlo simulation for the post-transition distribution of orbital eccentricities resulting from a transition-triggered mass loss of $\Delta M = 0.016~M_\odot$ and a fixed magnitude kick velocity $w$ corresponding to model \textsc{m2} of Ref. \cite{Chanlaridis:2024rov}.}
    \label{fig:eccentric}
\end{figure}

In Ref. \cite{Alvarez-Castillo:2019apz} it has been suggested that an accretion-induced phase transition from nuclear to quark matter in the inner core of the neutron star, transforming it to a hybrid star on the disconnected third family branch \cite{Gerlach:1968zz} entails a mass defect that is sufficient to cause an orbital reconfiguration that explains the observed large eccentricities $e\sim 0.1$. 
Earlier, in refs. \cite{Jiang:apj15,Jiang:raa2021} the authors have already discussed that an eccentricity of $\sim 0.11 - 0.15$ will be induced just by instantaneous gravitational mass loss of the NS due to a phase transition in its interior, even without invoking an explicit kick mechanism.
However, in their scenario, it may not be possible to explain the ``instantaneity'' of the transition when the NS spin is included. As has been pointed out by Glendenning et al. in \cite{Glendenning:1997fy} in the context of their scenario for an accretion-induced phase transition to a hybrid star branch connected with the NS one, the reconfiguration of the mass distribution in the NS interior is accompanied by a change in the moment of inertia. This would entail a ``pirouette effect'' (a spin-up), which in turn would inhibit the completion of the transition due to the sensible dependence of the density profile inside the star on its spin state caused by centrifugal forces. The authors  obtain in their scenario a time scale of $10^5$ years for the completion of the phase transition, which is in stark contrast with the requirement that the mass defect has to occur on a timescale much smaller than the orbital period.

This problem is circumvented when the phase transition leads to a member of the third family of compact stars (so-called ``twin stars'') beyond the conventional  white dwarf (WD) and NS classifications, which could emerge as the result of a {sufficiently strong} phase transition in the NS core \cite{Schertler:2000xq,Blaschke:2013ana,Benic:2014jia}. 
In that case, baryon number and angular momentum can be conserved simultaneously during the transition so that this process can occur almost instantaneously, on a dynamic timescale, which is insignificant compared to the orbital period.
This twin transition scenario has been worked out in detail recently in Ref. \cite{Chanlaridis:2024rov}, where also a Monte-Carlo simulation was performed which resulted in a distribution of post-transition eccentricities as a function of the orbital period which is shown in Fig. \ref{fig:eccentric} (right panel), including an additional explicit kick velocity parameter $w$, varied between 0 and 100 km/s.

The evolution of the orbit and the NS spin were followed until the NS reached the phase transition threshold (see Fig.~\ref{fig:twin-trans}, left panel), when the phase transition occurred instantaneously (i.e., on a time scale that is much shorter than the orbital period). To investigate the impact of the phase transition on the orbit, the prescriptions of \cite{1983apj:Hills} and \cite{Tauris:2017apj} were used, in which the ratio of the post-transition semi-major axis $a_\mathrm{f}$ to the pre-transition one ($a_\mathrm{i}$) is given by:
\begin{equation}
    \label{eq:major_axis_posttrans}
    \frac{a_\mathrm{f}}{a_\mathrm{i}} = \frac{1 - \Delta M/M}{1 - 2\Delta M/M - (w/v_\mathrm{rel})^2 - 2\cos\theta(w/v_\mathrm{rel})}.
\end{equation}
Here $\Delta M$ is the instantaneous mass defect corresponding to the released gravitational binding energy during the phase transition, $M$ is the total mass of the pre-transition system, $v_\mathrm{rel}$ is the relative velocity between the two stars ($v_\mathrm{rel} = \sqrt{GM/a_i}$), $w$ is the magnitude of the kick velocity, and $\theta$ is the kick angle between the kick velocity vector and the pre-transition orbital velocity vector.
The eccentricity of the post-transition binary system is given by:
\begin{equation}
    \label{eq:eccentricity_posttrans}
    e = \sqrt{1 + \frac{2E_\mathrm{orb,f}L_\mathrm{orb,f}^2}{\mu_\mathrm{f} G^2 M_\mathrm{f,1}^2 M_\mathrm{f,2}^2}},
\end{equation}
where $L_\mathrm{orb,f} = a_\mathrm{i} \mu_\mathrm{f} \sqrt{(v_\mathrm{rel} + w\cos\theta)^2 + (w\sin\theta \sin\phi)^2}$ is the post-transition orbital angular momentum, with $\phi$ being the kick angle on the plane perpendicular to the pre-transition velocity vector, $\mu_\mathrm{f}$ is the post-transition reduced mass, and $E_\mathrm{orb,f} = -GM_\mathrm{f,1}M_\mathrm{f,2}/2a_\mathrm{f}$ is the post-transition orbital energy as a function of the post-transition masses $M_\mathrm{f,1}$ and $M_\mathrm{f,2}$ in the binary. 

To investigate the post-transition orbital configurations, a mass defect of $\Delta M = 0.016~M_\odot$ obtained for the considered hybrid EOS (see Fig. \ref{fig:twin-trans}, right panel) and secondary kicks with magnitudes $w$ up to $100$ km s$^{-1}$ were considered. It was assumed that the kick lacks a preferential orientation, so that the kick angles $\theta$ and $\phi$ were modeled as uniformly distributed variables. 
The range of kick velocities is consistent with results for the electromagnetic rocket effect discussed in subsect. \ref{ssec:em-rocket}. 

\begin{figure}
    \centering
    \includegraphics[height=0.36\linewidth]{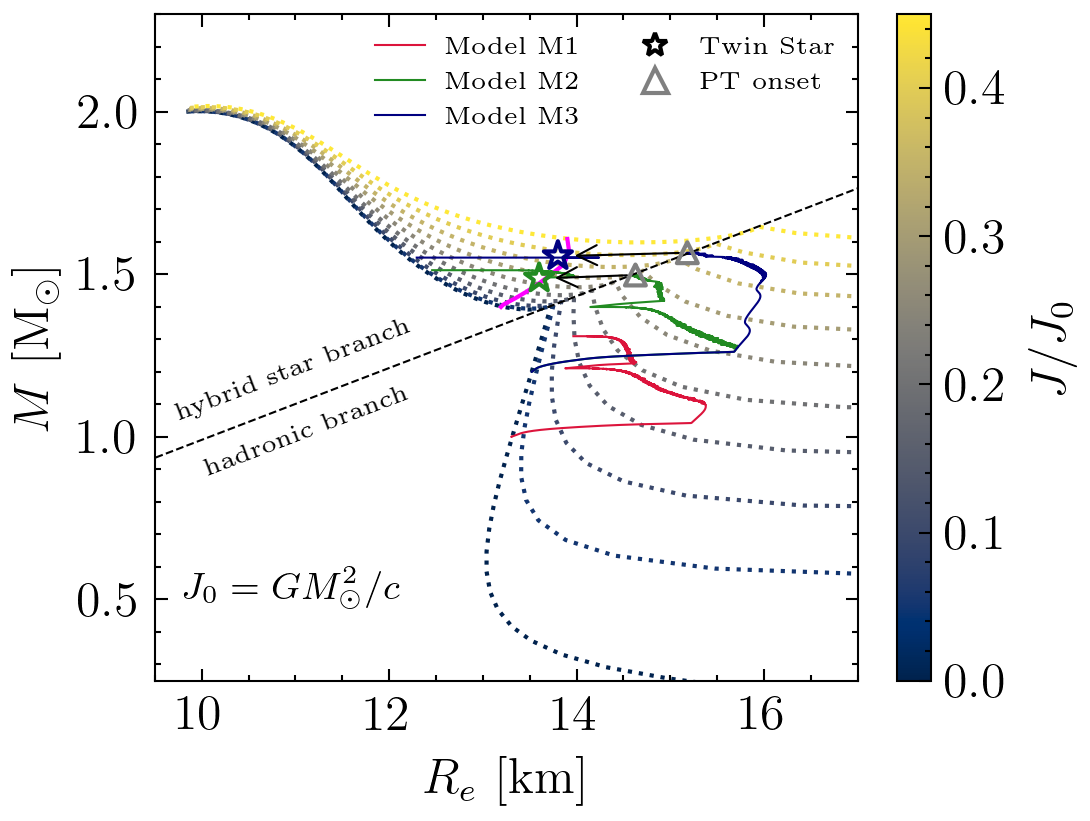}\hfill
    \includegraphics[height=0.35\linewidth]{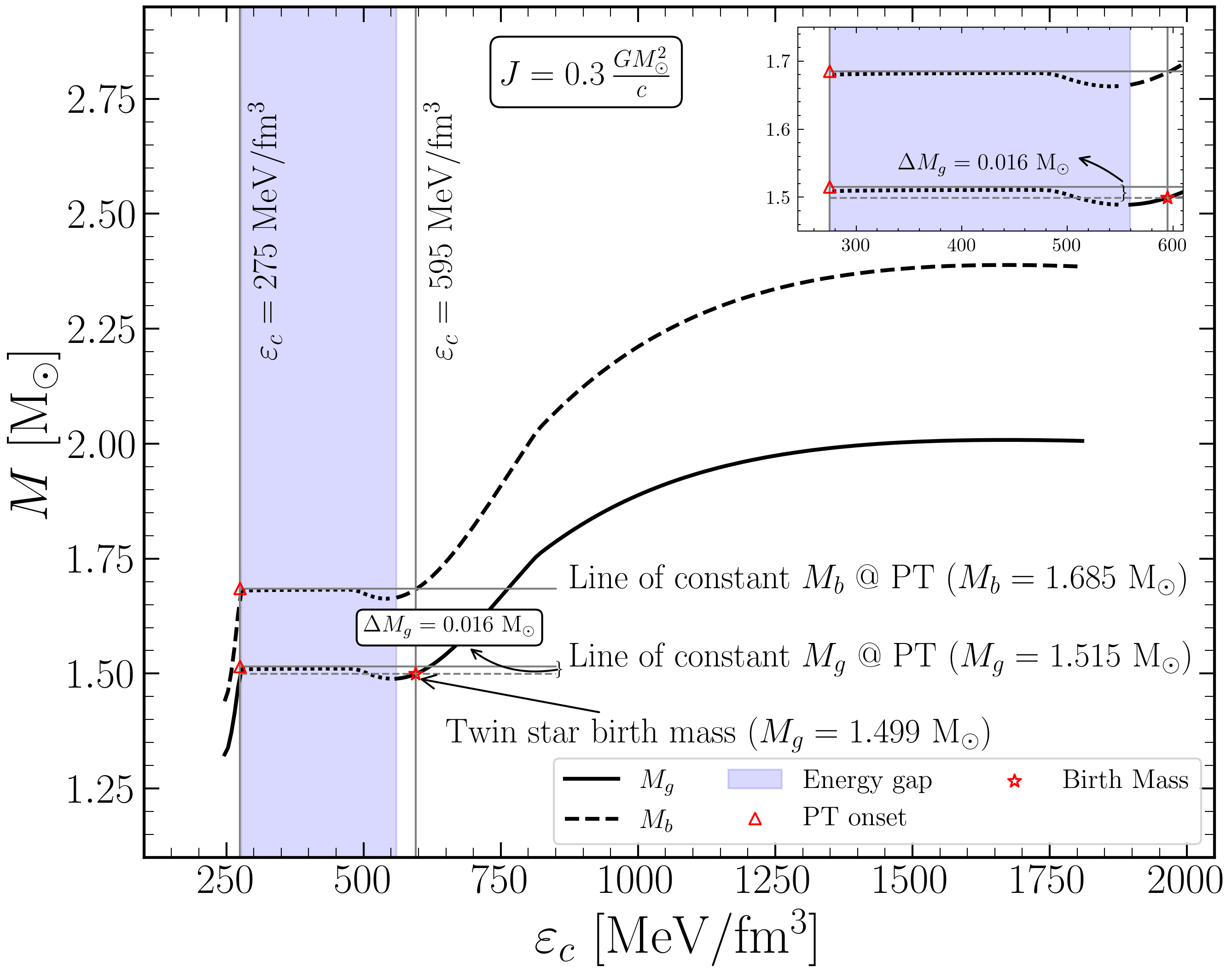}
    \caption{Left panel: Gravitational mass vs. equatorial radius for constant angular momentum values $J=0,...,0.45~J_0$ (dotted lines). The black dashed line highlights the points of maximum stability in the hadronic branches, signaling the onset of a phase transition (PT). 
    The thick magenta curve connects points on the hybrid star branches that can be reached through a collapse conserving both angular momentum and baryon mass.
   Star markers denote the endpoint trajectories for direct (\textsc{m2}) and delayed (\textsc{m3}) collapse models. Here, the arrows are inclined because gravitational mass, unlike baryonic mass, is not conserved during the PT. Right panel: Baryonic mass ($M_b$) and gravitational mass ($M_g$) as functions of the central energy density ($\varepsilon_c$) for a fixed angular momentum $J$. Vertical gray lines mark the central energy densities at the onset ($\varepsilon_c = 275\,\text{MeV/fm}^3$) and completion ($\varepsilon_c = 595\,\text{MeV/fm}^3$) of the phase transition (PT). Black dotted lines and the blue-shaded region denote unstable configurations.
    The inset zooms into the PT region, illustrating the mass defect $\Delta M_g=0.016~M_\odot$ associated with the transition from hadronic to quark matter. From Ref. \cite{Chanlaridis:2024rov}.  }
    \label{fig:twin-trans}
\end{figure}

The results for the model \textsc{m2} are shown in the right panel of Fig. \ref{fig:eccentric}
and compare well with the observed effect shown in the left panel of that figure, even without an explicit kick mechanism like electromagnetic or neutriino rocket effect. 
However, when such rocket effects in the strong magnetic field of a pulsar are considered, the model \cite{Chanlaridis:2024rov} for accretion-induced twin star phase transitions in low-mass X-ray binaries allows to explain the disruption of the binary and thus provides a scenario for the origin of isolated MSPs.

\subsection{Summary and outlook}

In this section, we have demonstrated on the example of neutrino and electromagnetic propulsion (rocket) effects for neutron stars that if these mechanisms can serve as the explanation of kick velocities up to 1000 km/s, they require strong magnetic fields in excess of $10^{12}$ G.
The neutrino rocket effect we outlined here requires a channeling and thus collimation of the neutrino emission along the magnetic field axis where magnetic vortex lines are formed in color superconducting (CFL phase) quark matter. Perpendicular to the vortex lines, the neutrino propagation is in the diffusion regime because of the frequent scattering on the dense network of these lines while parallel to them the neutrinos are free streaming along the B-field axis because of their long mean free path in CFL quark matter where scattering processes are exponentially suppressed by large pairing gaps. Within the cores of the vortex lines quark matter is normal (non-superconducting) and therefore copious neutrino production by the unsuppressed direct Urca process can take place, in particular at the early (hot) stages of the evolution.
The neutrino rocket scenario seems to be most appropriate for NS at their birth in supernova explosions, where velocities up to 1000 km/s  can be reached for young pulsars with high, magnetar-like B-fields.

For the late evolution phases of cold, recycled MSPs in LMXBs, the electromagnetic rocket effect seems to be appropriate which produces moderate kick velocities of $\sim 10 ... 100$ km/s. Those kick velocities are sufficient to provide a possible scenario for the origin of both eccentric and isolated MSPs, as has been demonstrated in recent Monte-Carlo simulations discussed in this section. It has been pointed out that for the accretion induced phase transition scenario of explaining MSPs with eccentric orbits it is important that a disconnected third family branch of of hybrid stars exists. This requires that the QCD deconfinement transition is strongly first order and occurs at moderate densities in NS with typical masses of 1.4 $M_\odot$.

Future observations and further theoretical studies have to contribute to a further clarification of the role of the phase transition to quark matter, color superconducting phases and anomalous neutrino transport in QCD at strong magnetic fields as they can be met in neutron stars.

    \newpage
    \part{Conclusions}\label{concl}

In this work we have presented a comprehensive review of the state-of-the-art in the investigation of the properties of particle systems subject to the influence of strong magnetic fields. The review was motivated by discussions and presentations  during the ``Workshop on Strongly Interacting Matter in Strong Electromagnetic Fields'' that took place in the European Centre for Theoretical Studies in Nuclear Physics and Related Areas (ECT*) in the city of Trento, Italy, September 25-29, 2023. We have divided the work into five parts, each accounting for the main subject discussed during one of the five days of the workshop. Based on the different frontier topics that are hereby discussed, we can conclude that the field is very  
active and  rapidly evolving. Nevertheless, the present work captures a large enough cross section of the main subjects of current interest, and provides also a very extensive list of references. The prospective for future developments will benefit from this kind of crosstalk exercises where different methods and techniques are used across seemingly different fields but in reality closely related by a driving subject, in this case, the physics of systems of particles in the presence of magnetic fields.

	\clearpage
	\section*{Acknowledgements}
	This work has been supported by STRONG-2020 ``The strong interaction at the frontier of knowledge: fundamental research and applications'' which received funding from the European Union's Horizon 2020 research and innovation programme under grant agreement No 824093.
	M.K. was supported, in part, by the U.S.~Department of Energy grant DE-SC0012447.
	The material presented in this article is based in part upon work done while C.C. was supported by the U.S. Department of Energy, Office of Science, Office of Workforce Development for Teachers and Scientists, Office of Science Graduate Student Research (SCGSR) program. The SCGSR program is administered by the Oak Ridge Institute for Science and Education (ORISE) for the DOE. ORISE is managed by ORAU under contract number DE‐SC0014664. All opinions expressed in this paper are the author’s and do not necessarily reflect the policies and views of DOE, ORAU, or ORISE. CC is currently supported in part by the Netherlands Organisation for Scientific Research (NWO) under the VICI grant VI.C.202.104.
	S.G. was supported in part by the Office of Science, Office of Nuclear Physics, U.S. Department of Energy under Contract No. DE- FG88ER40388 and in part by a Feodor Lynen Research fellowship of the Alexander von Humboldt foundation. Part of the material presented in this work is based upon work done while S.G. was supported by the Fulbright Visiting Scholar Program which is sponsored by the US Department of State and the German-American Fulbright Commission and the DAAD. 
	J.H. is supported by FWO-Vlaanderen through a Junior Postdoctoral Fellowship. Research by J.H. is also supported by FWO-Vlaanderen project G012222N, and by Vrije Universiteit Brussel through the Strategic Research Program High-Energy Physics.
	A.M. was financed, in part, by the São Paulo Research Foundation (FAPESP), Brasil, Process Number 
2023/08826-7. 
	The work of G.E., E.G.V. and G.M. was funded by the DFG (Collaborative Research Center CRC-TR 211 ``Strong-interaction matter under extreme conditions'' - project number 315477589 - TRR 211). G.E. acknowledges support by the Hungarian National Research, Development and Innovation Office (Research Grant Hungary). 
	E.G.V. acknowledges support by the Helmholtz Graduate School for Hadron and Ion Research (HGS-HIRe for FAIR).
	 The work of P.B. was funded in part by the UK STFC Consolidated Grant ST/T000988/1.
  This work was partially supported by Conselho Nacional de
Desenvolvimento Cient\'ifico  e Tecno\-l\'o\-gico  (CNPq),
312032/2023-4 (R.L.S.F.);  Funda\c{c}\~ao de Amparo \`a
Pesquisa do Estado do Rio Grande do Sul (FAPERGS), Grants
Nos. 19/2551- 0000690-0  and 19/2551-0001948-3 (R.L.S.F.); The work is
also part of the  project Instituto Nacional de Ci\^encia e Tecnologia
- F\'isica Nuclear e Aplica\c{c}\~oes  (INCT - FNA), Grant
No. 464898/2014-5, by Fundação Carlos Chagas Filho de Amparo \`a Pesquisa do Estado do Rio de Janeiro (FAPERJ) under Grant No.~SEI-260003/019544/2022 (W.R.T). This study was financed in part by the Coordena\c{c}\~ao de Aperfei\c{c}oamento de Pessoal de N\'ivel Superior - Brasil (CAPES) – Finance Code  001 (R.M.N.) and Grant no. 88887.826087/2023-00 (R.P.C). A.B. acknowledges the support from the Alexander von Humboldt foundation. C.V. acknowledge financial support from ANID/FONDECYT under
Grant No. 1190192. P.M.L. acknowledges partial support from the Polish
National Science Center (NCN) under the Opus grant no.
2022/45/B/ST2/01527. The work of G.K. is a part of the project INCT-FNA proc. No. 464898/2014-5. G.K. is also supported by Conselho Nacional de Desenvolvimento Cient\'ifico e Tecnol\'ogico (CNPq) under Grant No. 309262/2019-4 and by Funda\c{c}\~ao de Amparo \`a Pesquisa do Estado de S\~ao Paulo (FAPESP) under Grant No. 2018/25225-9. The work of V.S.T. is also supported by Conselho Nacional de Desenvolvimento Cient\'ifico e Tecnol\'ogico (CNPq) under Grant No. 305004/2022-0 and by Funda\c{c}\~ao de Amparo \`a Pesquisa do Estado de S\~ao Paulo (FAPESP) under Grant No. 2019/10889-1. The work of A.A. was supported by UNAM-PAPIIT grant number IG100322 and by Consejo Nacional de Humanidades, Ciencia y Tecnolog\'ia grant number CF-2023-G-433. The work of L.T. is supported under contract  No.\,PID2022-139427NB-I00 financed by the Spanish MCIN/AEI/10.13039/501100011033/FEDER, UE; by the Generalitat de Catalunya under contract 2021 SGR 00171; by Generalitat Valenciana under contract PROMETEO/2020/023;  from the project CEX2020-001058-M (Unidad de Excelencia ``Mar\'{\i}a de Maeztu''); from the European Union Horizon 2020 research and innovation programme under the program H2020-INFRAIA-2018-1, grant agreement No.\,824093 of the STRONG-2020 project; and from the CRC-TR 211 'Strong-interaction matter under extreme conditions'- project Nr. 315477589 - TRR 211. A.S. is supported by Deutsche Forschungsgemeinschaft (DFG) Grant No. SE 1836/5-3 and the Polish National Science Center (NCN) Grant No. 2023/51/B/ST9/02798. K.D.M. thanks the financial support from the São Paulo State Research Foundation (FAPESP), under Grant No. 2024/01623-6. The work of M.C., D.G.D. and N.N.S. has been partially funded by CONICET (Argentina) under
Grant No. PIP 2022 GI-11220210100150CO, by ANPCyT (Argentina) under
Grant No.~PICT20-01847. The work of D.G.D. has also been funded by the National University of La Plata (Argentina),
Project No.~X824. S.N. acknowledges support by Ministerio de Ciencia e Innovaci\'on and Agencia
Estatal de Investigaci\'on (Spain), and European Regional Development Fund
Grant No.~PID2019-105439GB-C21, by EU Horizon 2020 Grant No.~824093
(STRONG-2020), and by Conselleria de Innovaci\'on, Universidades, Ciencia y
Sociedad Digital, Generalitat Valenciana, GVA PROMETEO/2021/083. The work was also partially supported by national funds from FCT (Fundação para a Ciência e a Tecnologia, I.P, Portugal) under projects 
UIDB/04564/2020 and UIDP/04564/2020, with DOI identifiers 10.54499/UIDB/04564/2020 and 10.54499/UIDP/04564/2020, respectively, and the project 2022.06460.PTDC with the associated DOI identifier 10.54499/2022.06460.PTDC. H.P. acknowledges the grant 2022.03966.CEECIND (FCT, Portugal) with DOI identifier 10.54499/2022.03966.CEECIND/CP1714/CT0004. L.S. acknowledges the PhD grant 2021.08779.BD (FCT, Portugal) with DOI identifier 10.54499/2021.08779.BD. P.A. is supported by the U.S. Department of Energy, Office of Science, Office of Nuclear Physics and Quantum Horizons Program under Award Number DE-SC0024385. 
P.A. also acknowledges the hospitality of the the Kavli Institute for Theoretical Physics, Santa Barbara, through which the research was supported in part by the National Science Foundation under Grant No. NSF PHY-1748958. M.E.T-Y acknowledges support by Consejo Nacional de Humanidades, Ciencia y Tecnologia grant number A1-S-7655. The work of O.L. is a part of the project INCT-FNA proc. No. 464898/2014-5. It was also supported by Conselho Nacional de Desenvolvimento Cient\'ifico e Tecnol\'ogico (CNPq) under Grants No. 307255/2023-9 and 401565/2023-8 (Universal) and by Funda\c{c}\~ao de Amparo \`a Pesquisa do Estado de S\~ao Paulo (FAPESP) under Grant No. 2022/03575-3 (BPE). The work of M.D. is a part of the project INCT-FNA proc. No. 464898/2014-5. It is also
supported by Conselho Nacional de Desenvolvimento Cient\'ifico e Tecnol\'ogico (CNPq) under Grant No. 308528/2021-2 and No. 401565/2023-8 (Universal). This study was also financed in part by the Coordena\c{c}\~ao de Aperfei\c{c}oamento de Pessoal de N\'ivel Superior - Brazil (CAPES) - Finance Code 001 - Project number 88887.687718/2022-0. F.L.B. is a member of INCT-FNA, 464898/2014-5 and he acknowledges partial support also from
CNPq-312750/2021-8 and CNPq-407162/2023-2. 

	\newpage
	\appendix
	\renewcommand*{\thesection}{\Alph{section}}
	
	
	\bibliography{biblio_filtered}
    

\end{document}